\def\xybibbackend{bibtex}
\documentclass[11.5pt]{scrartcl}
\PassOptionsToPackage{dvipsnames}{xcolor}   % RoyalBlue etc., before xcolor is pulled in

\usepackage{xy-format}     % KOMA scrartcl style: newtx fonts, a4/32mm, authblk, biblatex(alphabetic)
\usepackage{xy-theorem}    % amsthm theorem styles

\usepackage{dsfont}        % \mathds{1} (identity, distinct from \mathbf irreps)
\usepackage{mathtools}
\usepackage{array}         % wrapped p{} table columns
\usepackage{framed}        % keybox
\usepackage{tikz}
\usetikzlibrary{arrows.meta,positioning,decorations.markings,calc}
\tikzset{
  gaugenode/.style={circle,draw,thick,minimum size=8mm,inner sep=1pt},
  flavnode/.style={rectangle,draw,thick,minimum size=8mm,inner sep=1pt},
  bifund/.style={-{Stealth[length=2mm]},thick},
}

\newcommand{\ec}[1]{}
\newcommand{\conv}[1]{}
\newcommand{\xw}[1]{}
\newenvironment{keybox}[1]{%
  \par\smallskip\noindent\begin{minipage}{\linewidth}%
  \begin{framed}\small\noindent\textsf{\textbf{#1}}\par\nopagebreak\smallskip}%
  {\end{framed}\end{minipage}\par\smallskip}

\let\chapter\section
\let\section\subsection
\let\subsection\subsubsection

\numberwithin{equation}{section}
\numberwithin{figure}{section}
\numberwithin{table}{section}
\usepackage{fancyhdr}
\title{A Crash Course in Supersymmetric Field Theory Across Dimensions}
\author{Xingyang Yu}
\affil[1]{Department of Physics, Virginia Tech, Blacksburg, VA 24061, USA}
\date{}
\hypersetup{
  pdftitle={A Crash Course in Supersymmetric Field Theory Across Dimensions},
  pdfauthor={Xingyang Yu},
  pdfsubject={Volume 0 of String-Theoretic Construction of Quantum Field Theory},
  pdfkeywords={supersymmetry, quantum field theory, lecture notes, dimensions}
}

\begin{document}
\maketitle

\begin{abstract}
These notes give a cross-dimensional crash course in supersymmetric field theory. The goal is to
provide enough common language to recognize the basic setups used in talks and papers, while also
working through representative computations that make the terminology concrete. We begin with
supersymmetry algebras, multiplets, protected quantities, moduli spaces, deformations, and anomalies,
then follow these ideas through examples in two to ten spacetime dimensions. The emphasis is on
recurring structures, such as holomorphy, supersymmetric vacua, dualities, extremization principles,
indices, BPS data, and anomaly constraints, rather than on a classification of theories in any one
dimension.
\end{abstract}

\newpage
\section*{Preface}
\addcontentsline{toc}{section}{Preface}

These notes grew out of a practical need and are based on the author's personal handwritten notes.
They include standard material, but also some well-known pieces of folklore and practical setup that
are not always easy to find in one place, or explained at the level needed for a first calculation.

Supersymmetric field theory has many excellent dimension-specific references, but a reader who moves
between seminars on $2d$ GLSMs, $3d$ monopole operators, $4d$ Seiberg duality, $5d$ fixed points,
and $6d$ anomaly polynomials can still feel as if every dimension starts from a different set of
instincts. The separation is not only dimensional. In a broad sense, the four-supercharge and
eight-or-more-supercharge worlds also tend to have different research styles, favorite questions,
and calculational languages. Four-supercharge theories often lean on holomorphy, superpotentials,
anomaly constraints, extremization, and chiral or twisted-chiral data. Eight-or-more-supercharge
theories more often use Coulomb and Higgs branch geometry, central charges, BPS objects, special or
hyperkahler geometry, and tensor-branch data as rigid organizing guides. These worlds are not
disjoint: dimensional reduction, duality, geometric engineering, and protected observables constantly
translate between them. The aim here is more modest and more horizontal. This is a crash course meant
to make the basic setups, protected quantities, and standard first computations recognizable across
dimensions and across these two broad languages.

The intended coherence is not uniformity. The point is not to pretend that every dimension uses the
same technology, but to ask the same first questions each time: what is the supersymmetry algebra and
R-symmetry, what local or protected data carries the theory, what branches or deformation parameters
are visible, and which quantities are rigid enough to compare across descriptions. The answers differ
from dimension to dimension, but the reading format is shared.

The goal is not to make the reader an expert in every dimension and every number of
supercharges. Rather, after reading these notes, the reader should be able to follow the
motivation and vocabulary of a typical supersymmetric field
theory talk, identify the kind of branch or protected observable being used, and work through a small
calculation that makes the words less opaque. The emphasis is therefore on reusable local tools:
supercharges and $R$-symmetries, multiplets and superpotentials, chiral and protected rings,
anomaly constraints, extremization principles, indices, Coulomb and Higgs branches, tensor branches,
and the standard examples that test these ideas.

The repeated end-of-section structure is part of the same design. The exit checklist states what the
reader should be able to do after the section. Sources and notes record provenance, conventions, and
deliberate omissions. Further reading gives a small number of next references, while the references
section lists the works cited in the section.

\medskip\noindent\textbf{Reproducibility note.}\enspace
All finite computations described in the text as checked, machine checked, or verified by check code
are reproducible from the arXiv source bundle. The Python scripts are included as ancillary files
under \path{anc/code/}: the section-level check suites are in \path{anc/code/chNN_*/checks/}, and
shared helper routines live in \path{anc/code/_engine/}. The file \path{anc/code/README.md} gives
the run commands and dependency list.

\newpage
\tableofcontents
\bigskip

\begin{refsection}\chapter{Supersymmetry algebras and the dimensional ladder}
\label{ch:V01}

\noindent\textbf{Guide to this section.}\enspace
This is the spine of these notes and the section from which the later sections read their
supersymmetry bookkeeping. It teaches one structure, the
super-Poincar\'e algebra, and one accounting question: \emph{how much} supersymmetry a
field-theory label carries in each spacetime dimension, and how that label changes under
ordinary dimensional reduction. It is not a comprehensive quantum field theory course
(\S\ref{sec:V01-intro}); it states the algebra and the bookkeeping and leaves detailed dynamics
to the dimension-specific sections and the cited literature. By the end you can read the dimensional-ladder master tables of
\S\ref{sec:V01-table}: align the supersymmetry notation across dimensions, name the R-symmetry,
the central charge, and the BPS role of an object, and decide how many real supercharges a given
field-theory supersymmetry label denotes.

\begin{keybox}{What this section delivers}
The real supercharge count and spinor reality type in every dimension $d=2,\dots,11$
(\S\ref{sec:V01-count}); the R-symmetry that rotates them (\S\ref{sec:V01-rsym}); the
dimensional ladder that carries one count across many $(d,\mathcal{N})$ labels
(\S\ref{sec:V01-ladder}); the BPS bound and the shortening that protects BPS data
(\S\ref{sec:V01-bps}); the superconformal ceiling that bounds where interacting fixed
points live (\S\ref{sec:V01-sconf}); and the master tables that collect it all
(\S\ref{sec:V01-table}).
\end{keybox}

\section{The toolkit and the algebra}
\label{sec:V01-intro}

The subject of this section is the supersymmetry algebra of ordinary relativistic quantum field
theory. Every supersymmetric field theory begins with a count of real supercharges: $2d\ (0,1)$
has one, $2d\ (0,2)$ and $3d\ \mathcal{N}=1$ have two, $4d\ \mathcal{N}=1$ has four,
$4d\ \mathcal{N}=2$ and $5d\ \mathcal{N}=1$ have eight, and
$4d\ \mathcal{N}=4$ has sixteen. The notation changes with dimension, but the algebraic count is
the invariant. Later sections use this section as a dictionary, but no extra construction is
needed to define the dictionary itself.

These notes collect that language once, organized by the dimensional ladder, so later sections do
not re-derive it. It is deliberately \emph{not} a replacement for a full quantum field theory reference.
It does not develop perturbation theory, renormalization, or the representation theory of the
conformal group. It states the supersymmetry facts used below, proves only what is needed to use
them, and leaves the rest to cited references. The boundary is worth fixing now: a crash course can
grow without end unless its contract is explicit.

\begin{keybox}{Two things these notes are not}
\textbf{(i)} These notes are not a comprehensive quantum field theory course. They teach a
supersymmetry toolkit and cite the detailed proofs and dynamics rather than reproducing all of them.
\textbf{(ii)} The label
$\mathcal{N}$ (or $(p,q)$) does not denote the same amount of supersymmetry across dimensions:
the invariant along the dimensional ladder is the real supercharge \emph{count}, not the label
(\S\ref{sec:V01-ladder}).
\end{keybox}

The organizing object is the super-Poincar\'e algebra. The rest of the section is a long unpacking
of what that algebra tells us: first how many real supercharges a spinor can carry, then how those
charges are rotated by R-symmetry, then how the same count is renamed under dimensional reduction,
and finally how central charges, BPS shortening, and conformal symmetry turn the bookkeeping into
protected physics.

\medskip\noindent\textbf{The super-Poincar\'e algebra.}\enspace
\label{sec:V01-algebra}

Supersymmetry is the unique consistent extension of the Poincar\'e symmetry of a relativistic
quantum field theory by generators that carry spin. The Coleman--Mandula theorem forbids the
Poincar\'e generators from combining nontrivially with any internal symmetry through
\emph{bosonic} charges. The supercharges evade it because they are \emph{fermionic}: they transform
as spinors of the Lorentz group. The Haag--Lopusza\'nski--Sohnius theorem then shows that the
resulting graded extension is essentially unique. We work throughout in Lorentzian signature
$\eta=\mathrm{diag}(-,+,\dots,+)$ and state dimension-dependent spinor conventions whenever they
affect a sign or normalization.

A supercharge $Q$ is a spinor of $\mathrm{Spin}(1,d-1)$. The defining relation of the algebra is
the anticommutator of two supercharges, which closes on the translations,
\begin{equation}
\label{eq:V01-QQ}
\{Q_\alpha,\,\bar Q_{\dot\beta}\}=2\,(\sigma^\mu)_{\alpha\dot\beta}\,P_\mu,
\end{equation}
written here in the four-dimensional two-component form for definiteness; in general $d$ the
right-hand side is $(C\Gamma^\mu)_{\alpha\beta}P_\mu$, with $C$ the charge-conjugation matrix.
The supercharges commute with the momentum, $[P_\mu,Q_\alpha]=0$, and rotate as
spinors under the Lorentz generators, $[M_{\mu\nu},Q_\alpha]=-(\sigma_{\mu\nu})_\alpha{}^\beta
Q_\beta$. Equation~\eqref{eq:V01-QQ} is the entire content from which the rest of the section
follows. Its right-hand side is a positive operator, $2\sigma^\mu P_\mu$, and that forces the
energy bound of \S\ref{sec:V01-bps}. Its left-hand side is built from spinors, and their reality
and chirality fix the count of \S\ref{sec:V01-count}.

\medskip\noindent\textbf{Extended supersymmetry and central charges.}\enspace
With $\mathcal{N}$ supercharges $Q^I$, $I=1,\dots,\mathcal{N}$, the algebra admits an extension
that the minimal case does not:
\begin{equation}
\label{eq:V01-central}
\{Q^I_\alpha,\,\bar Q_{\dot\beta\,J}\}=2(\sigma^\mu)_{\alpha\dot\beta}P_\mu\,\delta^I_J,
\qquad
\{Q^I_\alpha,\,Q^J_\beta\}=\epsilon_{\alpha\beta}\,Z^{IJ},
\end{equation}
where $Z^{IJ}=-Z^{JI}$ is a \emph{central charge}: it commutes with every generator of the
algebra. In higher dimensions the place of $Z^{IJ}$ is taken by a collection of antisymmetric
tensors $Z_{\mu_1\cdots\mu_p}$, the p-form charges under which extended defects are BPS. The
central charge is the second datum of every construction. It is the quantity the BPS bound of
\S\ref{sec:V01-bps} pins. In the Seiberg--Witten theory of four-dimensional $\mathcal{N}=2$ it is
the holomorphic combination of electric and magnetic charges, and it controls the entire low-energy
dynamics.

\medskip\noindent\textbf{The Nahm ceiling.}\enspace
The same positivity that bounds the energy bounds the \emph{size} of the algebra. A multiplet of
a super-Poincar\'e algebra with more than $32$ real supercharges necessarily contains fields of
spin greater than two, while $32$ real supercharges already force a spin-two field in a massless
multiplet. This is the physical ceiling used in field theory: $d=11$ is the top gravity-coupled
case with a $32$-component minimal spinor, and a formal $d=12$ spinor would carry $64$ real
components and force higher-spin states. We state this Nahm ceiling and use it, citing the proof
rather than reproducing it (\S\ref{sec:V01-gap}). It is the reason the dimensional ladder treats
$32Q$ as an ambient ceiling rather than an ordinary row of decoupled QFTs.

\section{Spinor reality, R-symmetry, and supercharge accounting}
\label{sec:V01-count}

How many real supercharges does the minimal spinor carry in dimension $d$? The answer is fixed by
the reality structure of a spinor of $\mathrm{Spin}(1,d-1)$. That structure comes from the real
Clifford algebra, and it repeats with period eight in $d$ (the Bott periodicity of the real
Clifford algebras). A spinor can carry up to two independent reductions of its components:
\begin{itemize}
\item a \emph{Majorana} (M) condition, a reality condition $\psi^*=B\psi$ compatible with the
Lorentz action, available when the charge-conjugation matrix squares the right way; it halves the
real dimension;
\item a \emph{Weyl} (W) condition, a chirality projection $\Gamma_*\psi=\pm\psi$, available in
even $d$; it also halves the dimension.
\end{itemize}
When both conditions hold at once the spinor is Majorana--Weyl (MW), and it is quartered. When the
natural reality condition pairs two spinors instead of constraining one, the spinor is
symplectic-Majorana (SM); if a chirality also survives, it is symplectic-Majorana--Weyl (SMW).
Which of $\{$M, W, MW, SM, SMW$\}$ occurs in dimension $d$ is fixed by $(d-2)\bmod 8$. The minimal
real supercharge count then follows from the type and the complex Dirac dimension $2^{\lfloor d/2\rfloor}$:
\begin{equation}
\label{eq:V01-counts}
\begin{aligned}
&\text{M:}\ 2^{\lfloor d/2\rfloor};\quad
\text{MW:}\ 2^{\lfloor d/2\rfloor-1}\ \text{(per chirality)};\\
&\text{SM:}\ 2^{\lfloor d/2\rfloor+1};\quad
\text{SMW:}\ 2^{\lfloor d/2\rfloor}\ \text{(per chirality)}.
\end{aligned}
\end{equation}
Evaluating the type and \eqref{eq:V01-counts} dimension by dimension gives the minimal real
supercharge count, the first column every later section reads:
\begin{equation}
\label{eq:V01-table-counts}
\begin{array}{l|cccccccccc}
d & 2 & 3 & 4 & 5 & 6 & 7 & 8 & 9 & 10 & 11\\\hline
\text{type} & \text{MW} & \text{M} & \text{M} & \text{SM} & \text{SMW} & \text{SM} & \text{M} & \text{M} & \text{MW} & \text{M}\\
\text{min }Q & 1 & 2 & 4 & 8 & 8 & 16 & 16 & 16 & 16 & 32
\end{array}
\end{equation}
The pattern is not a list to memorize. It is the output of the Clifford computation. Our check code
builds the gamma matrices in each dimension, solves for the reality structure $B$, and reads off
its Majorana sign and Weyl compatibility. It reproduces \eqref{eq:V01-table-counts} term by term,
and the companion source bundle records the check inventory. Two features carry through the
rest of the section. First, the count is \emph{not} monotone in $d$: it is $16$ for $d=7,8,9,10$
and jumps to $32$ only at $d=11$, the gravity-coupled ceiling. Second, $d=8$ has a
$16$-real-component minimal spinor but no Majorana--Weyl spinor. Equivalently, the Weyl spinor
there is complex, and the real Majorana package is not chiral. So the naive expectation ``even $d$
has Majorana--Weyl spinors'' is wrong: the reality type decides, not the parity of $d$. The label
records the same content either way. In four dimensions the four real components are often packaged
as a complex Weyl supercharge plus its conjugate; the table labels the equivalent real Majorana
package.

\begin{figure}[ht]
\centering
\begin{tikzpicture}
 \draw[gray!35] (0,0) circle (1.8);
 \foreach \r/\type/\dims in {%
 0/MW/{2,\,10}, 1/M/{3,\,11}, 2/M/{4}, 3/SM/{5},
 4/SMW/{6}, 5/SM/{7}, 6/M/{8}, 7/M/{9}}{
 \node[draw,rounded corners,fill=white,inner sep=2.2pt,font=\scriptsize,align=center]
 at ({90-45*\r}:1.8) {$\type$\\[-2pt]{\tiny $d{=}\dims$}};
 \node[font=\tiny,gray] at ({90-45*\r}:2.62) {$r{=}\r$};
 }
 \node[align=center,font=\scriptsize,text width=1.8cm] at (0,0)
 {reality type by\\ $(d{-}2)\bmod 8$};
\end{tikzpicture}
\caption{The spinor reality type is periodic in $(d-2)\bmod 8$ (Bott periodicity of the real
Clifford algebra), so it is fixed by the residue $r$ alone: the engine computes it at the eight
representatives $d=2,\dots,9$ and the higher dimensions inherit it ($d=10$ shares $r=0$ with $d=2$,
$d=11$ shares $r=1$ with $d=3$). The minimal real \emph{count}, the type applied to the Dirac
dimension $2^{\lfloor d/2\rfloor}$, is \emph{not} periodic: it grows with $d$ even where the type
repeats.}
\label{fig:V01-bott}
\end{figure}
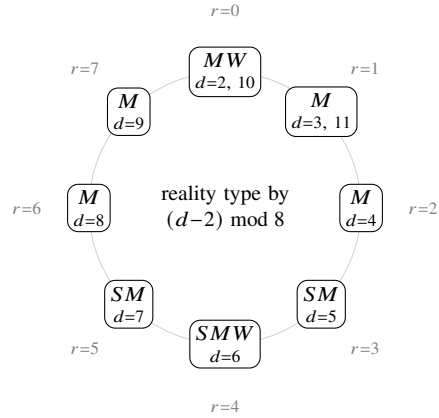

An amount of supersymmetry with $\mathcal{N}$ copies of the minimal spinor, or with chirality
content $(p,q)$ in the even dimensions that admit it, carries $\mathcal{N}$ (respectively $p+q$)
times the minimal count of \eqref{eq:V01-table-counts}. That product is the single number the
next two sections organize.

\medskip\noindent\textbf{R-symmetry.}\enspace
\label{sec:V01-rsym}

The $\mathcal{N}$ supercharges of an extended algebra can be rotated into one another by a global
symmetry that commutes with the Lorentz group and acts on the index $I$ of \eqref{eq:V01-central}.
This is the \emph{R-symmetry}: the automorphism group of the supersymmetry algebra,
\begin{equation}
\label{eq:V01-Raction}
[\,R_a,\,Q^I_\alpha\,]=(t_a)^I{}_J\,Q^J_\alpha,
\end{equation}
with $t_a$ the generators in the representation carried by the supercharges. As a first mnemonic,
real, complex, and symplectic supercharge packages are rotated respectively by orthogonal, unitary,
and compact symplectic automorphism groups. The actual R-symmetry of a physical or superconformal
theory is dimension-dependent, and can be reduced by chirality, central charges, or projectivization.
The master R-symmetry table below (Table~\ref{tab:V01-master-rsym}) is the authoritative lookup used
below. At a fixed point its preserved current
becomes the superconformal R-current of \S\ref{sec:V01-sconf}. Identifying the R-symmetry correctly
is therefore the prerequisite for the extremization principles ($a$-maximization, $F$-maximization,
$c$-extremization) used in later examples to compute exact central charges.

The extremes are as important as the middle of the ladder. At the very bottom, a purely
two-dimensional chiral algebra can have just one real supercharge: $2d\ (0,1)$ or $(1,0)$ has a
single Majorana--Weyl generator and no continuous R-symmetry, only the discrete sign automorphism
of that generator. With two real supercharges the R-symmetry can still be small or absent. The $3d\ \mathcal{N}=1$ and
$2d\ (1,1)$ algebras have no continuous R-symmetry that acts as a superconformal R-current. The
chiral $2d\ (0,2)$ algebra does: it has a right-moving $U(1)_R$. With sixteen real supercharges the
R-symmetry is large enough to dominate the representation theory. Here $4d\ \mathcal{N}=4$ has
$SU(4)_R\simeq \mathrm{Spin}(6)$, and $3d\ \mathcal{N}=8$ superconformal theories have
$\mathrm{Spin}(8)_R$ (the maximal-SYM Lagrangian shows only the scalar-rotation $\mathrm{Spin}(7)$,
enhanced in the infrared). The high-dimensional rows are smaller. The $8d\ \mathcal{N}=1$ row has
$U(1)_R$, from the two transverse directions to a 7-brane, and $7d\ \mathcal{N}=1$ has $SU(2)_R$.
The $5d$ maximal super Yang--Mills and $6d\ (2,0)$ rows have $USp(4)_R\simeq\mathrm{Spin}(5)_R$, and
$6d\ (1,1)$ has the $\mathrm{Spin}(4)_R$ of its nonchiral sixteen supercharges. These high-symmetry
rows are not
decorative. They are the reason the maximally supersymmetric constructions are so rigid. They are
also the rows from which maximal-SYM and sixteen-supercharge fixed-point discussions read their
protected data.

Three identifications recur and are worth fixing against their common confusions. In four
dimensions, $\mathcal{N}=2$ has R-symmetry $SU(2)_R\times U(1)_R$, not $SU(3)$: the
two Weyl supercharges $Q^I_\alpha$, carrying eight real components in total, form an $SU(2)_R$ doublet and
carry a $U(1)_R$ charge. That automorphism algebra has dimension four and rank two; $SU(3)$ has
dimension eight, too large to act on a doublet. In two dimensions, the $(0,2)$ algebra has a single
$U(1)_R$, the right-moving R-symmetry. The left-moving $U(1)_L$ that often accompanies it is a
\emph{flavor} current, used in $c$-extremization, not an R-symmetry factor. In five dimensions,
$\mathcal{N}=1$ has R-symmetry $SU(2)_R$. The $U(1)_I$ instanton symmetry that enhances to $E_n$ at
the ultraviolet fixed point is a \emph{flavor} symmetry, not an R-symmetry. Each of these is a
row of the R-symmetry table (Table~\ref{tab:V01-master-rsym}), and each is pinned in the check code by computing
the Lie-algebra dimension and rank from the Cartan type and matching the tabulated group; the
companion source bundle records the falsifiers. The remaining cells follow the same rule:
$3d\ \mathcal{N}=4$ has
$SU(2)\times SU(2)$ (the algebra $\mathfrak{so}(4)$), and $6d\ (2,0)$ has $USp(4)$, the spin cover
$\mathrm{Spin}(5)$ of $SO(5)$, of dimension ten and rank two.

\section{The dimensional ladder}
\label{sec:V01-ladder}

Dimensional reduction does not change the number of supercharges. Compactifying a theory on a
circle keeps every component of every spinor; what changes is the Lorentz group the spinor is a
representation of, and therefore the \emph{label} $\mathcal{N}$, not the count. The same number of
real supercharges thus wears different $(d,\mathcal{N})$ names along a ladder. Reading the ladder
is the act of recognizing one amount of supersymmetry under many names, and also of noticing when
two theories have the same count but a different chiral distribution, R-symmetry, or amount of
protection. The four- and eight-supercharge rows appear most often in applications, but a
cross-dimensional crash course has to display the whole spine.

\begin{table}[ht]
\centering
\scriptsize
\setlength{\tabcolsep}{4pt}
\renewcommand{\arraystretch}{1.18}
\begin{tabular}{@{}c >{\raggedright\arraybackslash}p{7.2cm} >{\raggedright\arraybackslash}p{4.2cm}@{}}
\toprule
real $Q$ & balanced reduction ladder & chiral / same-count variants\\
\midrule
1 & purely $2d$ & $2d\ (0,1)$ or $(1,0)$; the minimal chiral field-theory row\\
\midrule
2 & $3d\ \mathcal{N}{=}1 \to 2d\ (1,1)$ & $2d\ (0,2)$ or $(2,0)$; the minimal chiral gauge-theory row\\
\midrule
4 & $4d\ \mathcal{N}{=}1 \to 3d\ \mathcal{N}{=}2 \to 2d\ (2,2)$ & $2d\ (0,4)$ is same count, different chirality\\
\midrule
8 & $6d\ (1,0) \to 5d\ \mathcal{N}{=}1 \to 4d\ \mathcal{N}{=}2 \to 3d\ \mathcal{N}{=}4 \to 2d\ (4,4)$ & $2d\ (0,8)$ chiral cousins; hypermultiplet row\\
\midrule
16 & \begin{tabular}[t]{@{}l@{}}
$10d\ \mathcal{N}{=}1\ \mathrm{SYM} \to 9d\ \mathcal{N}{=}1 \to 8d\ \mathcal{N}{=}1 \to 7d\ \mathcal{N}{=}1$\\
$\to 6d\ (1,1)$
$\to 5d\ \mathcal{N}{=}2 \to 4d\ \mathcal{N}{=}4 \to 3d\ \mathcal{N}{=}8 \to 2d\ (8,8)$
\end{tabular} & $6d\ (2,0)$ is chiral 16Q, not an ordinary Yang--Mills Lagrangian\\
\midrule
32 & $11d$ super-Poincar\'e / ten-dimensional type-II algebra & ceiling for gravity-coupled algebras, not a decoupled-QFT row here\\
\bottomrule
\end{tabular}
\caption{The dimensional ladder by real supercharge count. The one-supercharge row is genuinely
two-dimensional; later arrows are ordinary circle reductions, so the count is conserved while the
dimension-dependent label changes. Same-count rows with different chirality are not
interchangeable.}
\label{tab:V01-ladder-counts}
\end{table}

\begin{figure}[ht]
\centering
\begin{tikzpicture}[
 x=1.7cm,y=0.85cm,
 box/.style={draw,rounded corners,fill=white,inner sep=1.5pt,font=\scriptsize,
 align=center,minimum width=1.15cm,minimum height=0.62cm},
 red/.style={->,gray!70,shorten >=1.5pt,shorten <=1.5pt}]
 \foreach \c/\lab in {0/{$1Q$},1/{$2Q$},2/{$4Q$},3/{$8Q$},4/{$16Q$}}
 \node[font=\footnotesize\bfseries] at (\c,8.7) {\lab};
 % 16Q column
 \node[box](a10) at (4,8) {$10d$\\$\mathcal{N}{=}1$};
 \node[box](a9) at (4,7) {$9d$\\$\mathcal{N}{=}1$};
 \node[box](a8) at (4,6) {$8d$\\$\mathcal{N}{=}1$};
 \node[box](a7) at (4,5) {$7d$\\$\mathcal{N}{=}1$};
 \node[box](a6) at (4,4) {$6d$\\$(1,1)$};
 \node[box](a5) at (4,3) {$5d$\\$\mathcal{N}{=}2$};
 \node[box](a4) at (4,2) {$4d$\\$\mathcal{N}{=}4$};
 \node[box](a3) at (4,1) {$3d$\\$\mathcal{N}{=}8$};
 \node[box](a2) at (4,0) {$2d$\\$(8,8)$};
 \foreach \hi/\lo in {a10/a9,a9/a8,a8/a7,a7/a6,a6/a5,a5/a4,a4/a3,a3/a2} \draw[red](\hi)--(\lo);
 \node[box,fill=gray!10](a6b) at (4.92,4) {$6d$\\$(2,0)$};
 % 8Q column
 \node[box](b6) at (3,4) {$6d$\\$(1,0)$};
 \node[box](b5) at (3,3) {$5d$\\$\mathcal{N}{=}1$};
 \node[box](b4) at (3,2) {$4d$\\$\mathcal{N}{=}2$};
 \node[box](b3) at (3,1) {$3d$\\$\mathcal{N}{=}4$};
 \node[box](b2) at (3,0) {$2d$\\$(4,4)$};
 \foreach \hi/\lo in {b6/b5,b5/b4,b4/b3,b3/b2} \draw[red](\hi)--(\lo);
 % 4Q column
 \node[box](e4) at (2,2) {$4d$\\$\mathcal{N}{=}1$};
 \node[box](e3) at (2,1) {$3d$\\$\mathcal{N}{=}2$};
 \node[box](e2) at (2,0) {$2d$\\$(2,2)$};
 \foreach \hi/\lo in {e4/e3,e3/e2} \draw[red](\hi)--(\lo);
 % 2Q column
 \node[box](f3) at (1,1) {$3d$\\$\mathcal{N}{=}1$};
 \node[box](f2) at (1,0) {$2d$\\$(1,1)$};
 \draw[red](f3)--(f2);
 % 1Q column
 \node[box](g2) at (0,0) {$2d$\\$(0,1)$};
 % 32Q ceiling
 \node[draw,densely dashed,rounded corners,fill=gray!12,font=\scriptsize,align=center,inner sep=3pt]
 at (2.35,9.7) {$32Q$ ceiling: $11d$ and $10d$ type II (gravity-coupled, not a QFT column)};
\end{tikzpicture}
\caption{The dimensional ladder by real supercharge count. Each column is one count; a circle
reduction moves \emph{down} a column (arrows), conserving the count while the $(d,\mathcal{N})$
label changes. Chiral cousins carry the same count by a different chirality: the $6d\,(2,0)$ theory
beside the non-chiral $6d\,(1,1)$, and the $2d\,(p,q)$ rows. The $32Q$ row is the gravity-coupled
ceiling, not a decoupled-QFT column. This is the picture behind Table~\ref{tab:V01-ladder-counts}.}
\label{fig:V01-ladder}
\end{figure}
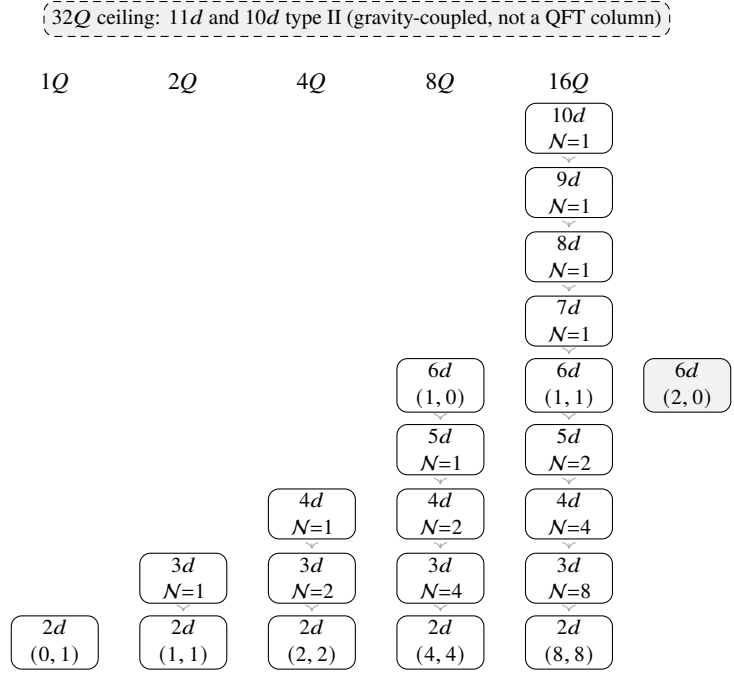

Table~\ref{tab:V01-ladder-counts} displays the named rungs used most often in these notes, not every
possible label in every dimension. The exceptional first row is possible because a $2d$
Majorana--Weyl spinor has one real component per chirality: a $(0,1)$ or $(1,0)$ theory is a
one-supercharge theory, not a shorthand for a hidden two-supercharge algebra. The complete
low-dimensional field-theory list has the same simple arithmetic behind it. In three dimensions
$\mathcal{N}=1,\dots,8$ carry $2,4,\dots,16$ real supercharges. In four dimensions
$\mathcal{N}=1,2,3,4$ carry $4,8,12,16$. In five dimensions $\mathcal{N}=1,2$ carry $8,16$. In six
dimensions the standard rows are $(1,0)$, $(1,1)$, and $(2,0)$. Seven through ten dimensions each
have the $\mathcal{N}=1$ sixteen-supercharge pure-SYM row. Two dimensions are more flexible,
because the left- and right-moving supercharges can be chosen independently: $(p,q)$ has $p+q$ real
supercharges. The count table below records the compact version of this list; the count itself
always comes from \eqref{eq:V01-table-counts}.

The one-supercharge row is the first place where the count is too small to hide behind any
higher-dimensional notation. It is a genuinely two-dimensional possibility: a $2d\ (0,1)$ theory
has one right-moving supercharge, and the left-moving $(1,0)$ version is its parity mirror. The
two-supercharge row is the first row connected to ordinary higher-dimensional reduction. In three
dimensions $\mathcal{N}=1$ means two real supercharges; reducing it on a circle gives
$2d\ (1,1)$, again two real supercharges. The chiral
$2d\ (0,2)$ algebra also has two real supercharges, but it is not just the reduction of the
balanced $(1,1)$ theory. Its right-moving chirality, its Fermi multiplets, and its $U(1)_R$ are
genuine two-dimensional field-theory data, which is why a $(0,2)$ gauge theory is not a disguised
$(2,2)$ theory. The lesson is already visible here: a
supercharge count is necessary bookkeeping, but it is not a complete theory label.

The four-supercharge ladder collects
\begin{equation}
\label{eq:V01-track4}
4d\ \mathcal{N}{=}1\ \cong\ 3d\ \mathcal{N}{=}2\ \cong\ 2d\ (2,2)
\ =\ 4\ \text{real supercharges},
\end{equation}
This is the standard four-supercharge field-theory row: four supercharges are just enough for
chiral multiplets, holomorphic superpotentials, chiral rings, and $a$-maximization in four
dimensions, but not enough for the rigidity of the eight- or sixteen-supercharge rows.

The eight-supercharge ladder collects
\begin{equation}
\label{eq:V01-track8}
6d\ (1,0)\ \cong\ 5d\ \mathcal{N}{=}1\ \cong\ 4d\ \mathcal{N}{=}2
\ \cong\ 3d\ \mathcal{N}{=}4\ \cong\ 2d\ (4,4)
\ =\ 8\ \text{real supercharges},
\end{equation}
This is the hypermultiplet row: hypermultiplets, Coulomb-Higgs factorization, four-dimensional
Coulomb-branch low-energy data, and enough protection that many statements about the low-energy
effective theory become exact.

The sixteen-supercharge row is a different animal. One branch is the maximally supersymmetric
Yang--Mills ladder,
\begin{equation}
\label{eq:V01-track16}
\begin{aligned}
10d\ \mathcal{N}{=}1\ \mathrm{SYM}
&\to 9d\ \mathcal{N}{=}1
\to 8d\ \mathcal{N}{=}1
\to 7d\ \mathcal{N}{=}1
\to 6d\ (1,1)\\
&\to 5d\ \mathcal{N}{=}2
\to 4d\ \mathcal{N}{=}4
\to 3d\ \mathcal{N}{=}8
\to 2d\ (8,8),
\end{aligned}
\end{equation}
all with $16$ real supercharges. Another branch with the same count is the chiral
$6d\ (2,0)$ theory: it has no ordinary Yang--Mills Lagrangian in six dimensions, yet it carries
sixteen real supercharges and sits on the same count ladder, its circle reduction landing on the
five-dimensional maximal theory. It is not ``more of the same'' as the eight-supercharge row; the
larger R-symmetry and the absence or presence of a Lagrangian are load-bearing field-theory data.

The high-dimensional end of this ladder is still field theory. Ten-dimensional
$\mathcal{N}=1$ SYM is the parent pure gauge theory; reducing it gives legitimate
non-gravitational $9d$, $8d$, and $7d$ $\mathcal{N}=1$ SYM theories. None of these
sixteen-supercharge rows is forced to contain a graviton. The reason $8d$ matters concretely for
string-theoretic constructions is F-theory: one uses the eight-dimensional gauge theory living on
7-branes, especially in the elliptic K3 / heterotic dual interface. The neighboring
$7d\ \mathcal{N}=1$ row similarly appears in $G_2$-holonomy constructions. Section~1 treats the $7d$-$10d$ pure-SYM
rows uniformly at the algebraic level; later sections may group them because their
dimension-specific field-theory structures are lighter than the $2d$-$6d$ cases.

There is one useful field-theory check on this row, and it is a short derivation rather than a
slogan. A propagating minimal pure super Yang--Mills Lagrangian has only a gauge field and one
gaugino, so it has the familiar on-shell degree-of-freedom match only where their physical
polarizations agree. A massless vector in $d$
dimensions has $d-2$ physical polarizations, while the minimal gaugino carries half the minimal
real supercharge count of \eqref{eq:V01-table-counts} on shell (the Dirac equation halves the
off-shell components); equating the two,
\begin{equation}
\label{eq:V01-symmatch}
d-2\ =\ \tfrac12\,Q_{\min}(d)
\quad\Longrightarrow\quad
d-2\in\{1,2,4,8\},\qquad d\in\{3,4,6,10\},
\end{equation}
since $Q_{\min}=2,4,8,16$ at $d=3,4,6,10$ gives $\tfrac12 Q_{\min}=1,2,4,8=d-2$, and no other
dimension matches. These four are the \emph{minimal} super-Yang--Mills theories; reducing the
ten-dimensional one on a torus generates the entire maximal-SYM ladder \eqref{eq:V01-track16}.
The reduction also fixes each rung's R-symmetry. Reducing $10d\ \mathcal{N}=1$ SYM to $D$
dimensions leaves $10-D$ adjoint scalars. The $SO(10-D)$ that rotates them is the R-symmetry of the
maximal theory. It acts directly on the scalars; on the gaugino, and hence on the supercharges, it
acts through its spin cover. The group acting on the \emph{supercharges} is therefore $\mathrm{Spin}(10-D)$.
For $D=8$ this is $SO(2)\simeq U(1)_R$. For $D=7$ the scalars rotate by $SO(3)$, and the
supercharges feel its double cover $SU(2)_R=\mathrm{Spin}(3)$. This is the $SU(2)_R$ of the master
table, \emph{not} a quotient of it. The pattern continues down to $D=3$, where the Lagrangian shows
only $\mathrm{Spin}(7)$ and the infrared fixed point enhances it to $\mathrm{Spin}(8)$.

Finally, the thirty-two-supercharge ceiling is the algebraic upper edge. It is the spinor size of
eleven-dimensional super-Poincar\'e symmetry and of the type-II ten-dimensional algebras, and it
tells us why the ladder stops. It is not a row of ordinary decoupled QFTs in these notes. With
that many supercharges one is at the gravity-coupled ceiling, not at a stand-alone supersymmetric
field theory in the sense used here. Going to $64Q$, as in a formal twelve-dimensional
spinor algebra, would force higher-spin states.

This is the single most useful fact in the toolkit, and the single most misread. The invariant is
the count, not the symbol $\mathcal{N}$. Four examples make the point. Four-dimensional
$\mathcal{N}=2$ and three-dimensional $\mathcal{N}=2$ are \emph{different} amounts, eight against
four, because the minimal spinor halves between the two dimensions. Five-dimensional
$\mathcal{N}=1$ and four-dimensional $\mathcal{N}=2$ are the \emph{same} amount. Two-dimensional
$(0,1)$ has no three-dimensional parent with its single count. Two-dimensional $(0,2)$ and
three-dimensional $\mathcal{N}=1$ share a count but not a chirality. A reader who carries
$\mathcal{N}$ across a reduction without recounting will mis-assign every theory downstream. So
state the count first, then the dimension, then the chirality and R-symmetry data.

\begin{keybox}{Worked fixture: embeddings preserve Q}
The downstream engines pick a rung of this ladder from a holonomy embedding, and the rule is pure
supercharge bookkeeping. Start with the D3-brane parent, which has sixteen real supercharges. The
transverse orbifold group $\Gamma$ decides how many survive. The surviving supercharges are those
left invariant by $\Gamma$ acting on the four-component transverse spinor (the $\mathbf{4}$ of the
transverse $\mathrm{Spin}(6)\simeq SU(4)$), and each invariant component carries four real
supercharges. If $\Gamma\subset SU(2)$, one complex transverse direction is inert and two of those
components stay invariant ($\mathbf{4}\to\mathbf{2}+\mathbf{1}+\mathbf{1}$): the four-dimensional
probe theory keeps eight of the sixteen supercharges, namely $4d\ \mathcal{N}=2$. This is the
$\mathbb{C}^2/\Gamma$ ADE block. If $\Gamma$ lies generically in $SU(3)$, only one component stays
invariant ($\mathbf{4}\to\mathbf{3}+\mathbf{1}$): the probe theory keeps four real supercharges,
namely $4d\ \mathcal{N}=1$. This is the $\mathbb{C}^3/\Gamma$ chiral-quiver block. If an $SU(3)$ embedding
accidentally factors through an $SU(2)$ subgroup, the count jumps back to eight real supercharges.
The embedding thus chooses the row of the ladder before any quiver or superpotential data is read.
The classification is by complex dimension, not by notation: $\mathbb{C}^2/\Gamma$ orbifold probes give
the $8Q$, $4d\ \mathcal{N}=2$ ADE block; $\mathbb{C}^3/\Gamma$ orbifold probes give the $4Q$,
$4d\ \mathcal{N}=1$ chiral-quiver block; general toric CY$_3$ probes require dimer technology.
\end{keybox}

\section{Central charges, the BPS bound, and shortening}
\label{sec:V01-bps}

The positivity of the anticommutator \eqref{eq:V01-QQ} is not a formality: it is a bound on the
mass of every state, and the states that saturate it are the protected spectrum the constructions
compute. Acting on a one-particle state of mass $M$ and central charge $Z$, the supersymmetry
algebra reduces, in the relevant sector, to a Hermitian block built from the rest mass and the
central charge,
\begin{equation}
\label{eq:V01-block}
A=\begin{pmatrix} M & Z\\ \bar Z & M\end{pmatrix},
\qquad
\text{eigenvalues}\quad M\pm|Z|.
\end{equation}
Because $A=\{Q,Q^\dagger\}\ge 0$ is a sum of squares, both eigenvalues are nonnegative, which is
the \emph{BPS bound}
\begin{equation}
\label{eq:V01-bpsbound}
M\ \ge\ |Z|.
\end{equation}
A state saturating the bound, $M=|Z|$, makes $\det A=M^2-|Z|^2$ vanish: the block drops rank, and
half of the supercharges (the kernel of $A$) annihilate the state. These \emph{preserved}
supercharges are the ones for which $\{Q,Q^\dagger\}=M-|Z|=0$; the remaining \emph{broken}
supercharges, with eigenvalue $2M$, act nontrivially and build a multiplet that is
\emph{shortened}, with half the states of a generic massive multiplet. The shortening is what
protects BPS quantities. A short multiplet has fewer states than a long one, so its mass cannot
drift off $M=|Z|$ under a deformation that keeps the multiplet short. The protection is robust but
not absolute. What the shortening protects is the algebraic mass formula for a state that stays in
that short representation. It does not, by itself, guarantee that the state survives as a stable
one-particle state everywhere in moduli space. Even strongly shortened BPS sectors can meet walls of
marginal stability, singular loci, or decay channels. What is protected is the shortening relation
and the appropriate charge or index data in its chamber. The protection is graded by the
supersymmetry fraction. A low fraction, such as a quarter-BPS dyon degeneracy, is more susceptible
to recombination and to wall-crossing, where the protected number jumps between chambers; but the
caveat is not limited to the low fractions, since a short multiplet can \emph{recombine} into a long
one whenever the spectrum supplies the missing states. This graded protection is what lets
supersymmetric field theory compute BPS masses, indices, and central charges through controlled
chambers and deformations, surviving deformations that would otherwise leave perturbative control.

The canonical instance comes from the Seiberg--Witten solution of four-dimensional $\mathcal{N}=2$
super Yang--Mills, in the normalization used throughout these notes. There the central charge is the
holomorphic combination of the electric and magnetic charges $(n_e,n_m)$,
\begin{equation}
\label{eq:V01-SW}
Z\ =\ a\,n_e\ +\ a_D\,n_m,
\qquad
M_{\mathrm{BPS}}\ =\ \sqrt{2}\,|Z|,
\end{equation}
with $a$ and $a_D$ the special coordinate and its dual on the Coulomb branch. A purely electric
state, $n_m=0$, has $M=\sqrt2\,|a|\,|n_e|$, the W-boson tower; a monopole has $n_m\ne0$. Section~1
states the bound, the shortening mechanism, and this canonical central charge. It computes no
specific spectrum. Which states are present, and how $a$ and $a_D$ vary, belongs to the
Seiberg--Witten dynamics cited in the further-reading references. The block \eqref{eq:V01-block}, its
eigenvalues, and the kernel-at-saturation are the elementary algebraic part used here.

\section{Superconformal algebras and the unitarity bound}
\label{sec:V01-sconf}

The constructions usually flow to fixed points. There the Poincar\'e symmetry enhances to the
conformal group $SO(d,2)$, and the supersymmetry algebra enlarges with it. This section does for
the fixed point what \S\ref{sec:V01-bps} did for the massive spectrum. The same positivity that
turned $\{Q,Q^\dagger\}\ge 0$ into a mass bound now bounds scaling dimensions. The operators that
saturate the bound are the protected, short spectrum, and a superconformal theory computes them
exactly. This structure recurs across the notes, so we state it once here.

\medskip\noindent\textbf{From super-Poincar\'e to superconformal.}\enspace
The conformal group adds to the Poincar\'e generators $(P_\mu,M_{\mu\nu})$ the dilatation $D$ and
the special conformal generators $K_\mu$. Supersymmetrizing it cannot stop at the Poincar\'e
supercharges $Q$: closing $Q$ with $K$ forces a second fermionic generator, the
\emph{superconformal supercharge} $S$, generated by the commutator $[K_\mu,Q]$. It is the
dimension-lowering partner of $Q$,
\begin{equation}
\label{eq:V01-Sgen}
[\,D,\,Q\,]\ =\ +\tfrac12\,Q,\qquad\qquad [\,D,\,S\,]\ =\ -\tfrac12\,S,
\end{equation}
so $Q$ raises the scaling dimension by half a unit and $S$ lowers it. The three fermionic
brackets then close on the three bosonic sectors; in the standard conventions they are
\begin{equation}
\label{eq:V01-scfclose}
\{Q,\bar Q\}=2\sigma^\mu P_\mu,\qquad
\{S,\bar S\}=2\sigma^\mu K_\mu,\qquad
\{Q_\alpha,S^\beta\}=(\sigma^{\mu\nu})_\alpha{}^\beta\,M_{\mu\nu}+\delta_\alpha^\beta\,(D-c\,R),
\end{equation}
The first is the super-Poincar\'e relation \eqref{eq:V01-QQ} itself, and the second is its
conformal conjugate ($P\to K$). The third is the decisive one. Its Lorentz part
$(\sigma^{\mu\nu})_\alpha{}^\beta M_{\mu\nu}$ annihilates a scalar primary. What remains on the
diagonal is the combination $D-cR$ of the dilatation and the R-symmetry, with $c$ the algebra
constant ($c=\tfrac32$ for $4d\ \mathcal{N}=1$). At a fixed point the R-symmetry is therefore no
longer an external automorphism of the algebra (\S\ref{sec:V01-rsym}) but a \emph{generator inside
it}, graded next to the dilatation. That single fact makes the R-symmetry load-bearing at a fixed
point, and it is the engine of every exact-dimension statement that follows.

\medskip\noindent\textbf{The unitarity bound and BPS shortening at the fixed point.}\enspace
Take a superconformal primary $|\mathcal{O}\rangle$, an operator annihilated by every $S$ (the
conformal analog of a lowest-weight state), with scaling dimension $\Delta$ and R-charge $R$.
Exactly as in \S\ref{sec:V01-bps}, the relevant supercharge pairing is a positive Hermitian
operator. In radial quantization the conformal conjugate of $Q$ is precisely $S=Q^\dagger$, and a
superconformal primary is annihilated by every $S$, so the diagonal matrix element collapses to a
single norm,
\begin{equation}
\label{eq:V01-scfblock}
\langle\mathcal{O}|\,\{Q,\,S\}\,|\mathcal{O}\rangle
\ =\ \langle\mathcal{O}|\,Q^\dagger Q\,|\mathcal{O}\rangle
\ =\ \big\|\,Q|\mathcal{O}\rangle\,\big\|^{2}\ \ge\ 0,
\end{equation}
where the $QS=QQ^\dagger$ term dropped because $S|\mathcal{O}\rangle=0$ on a primary. By
\eqref{eq:V01-scfclose}, the left-hand side equals the eigenvalue $\Delta-c\,R$ (the Lorentz part
vanishes on a scalar). Reading it off gives a \emph{superconformal shortening bound}, the
fixed-point counterpart of $M\ge|Z|$,
\begin{equation}
\label{eq:V01-unitarity}
\Delta\ \ge\ c\,R\qquad(\text{scalar primary, the chosen }\{Q,S\}\text{ branch}),
\end{equation}
with $c$ the same algebra constant as in \eqref{eq:V01-scfclose}. This is one BPS branch. The full
unitarity bounds for a generic (long) primary carry extra spin-dependent offsets; those are the
representation-theoretic statement classified in the cited literature. A primary that
\emph{saturates} the bound, $\Delta=cR$, is annihilated by the corresponding supercharge. It sits
in a \emph{short} (BPS) representation, the exact analog of the rank drop of \eqref{eq:V01-block}.
In $4d\ \mathcal{N}=1$ the saturating primaries are the \emph{chiral} operators, annihilated by all
$\bar Q$; the conjugate positivity $\{\bar Q,\bar Q^\dagger\}\ge 0$ then locks their dimension to
the R-charge,
\begin{equation}
\label{eq:V01-chiral}
\Delta\ =\ \tfrac32\,R\qquad(4d\ \mathcal{N}=1\ \text{chiral primary}),
\end{equation}
in the normalization where a chiral superpotential carries $R=2$ (so that $R>0$ on a chiral
primary).
This is the fixed-point version of $M=|Z|$. The dimension of a chiral/BPS operator is fixed by an
algebraic charge, and it stays fixed under any deformation that preserves the short representation.
That is what lets a superconformal theory pin anomalous dimensions perturbation theory cannot
reach. The same graded caution as in \S\ref{sec:V01-bps} applies: the protection is the
representation-theoretic shortening, and weaker short multiplets can recombine.

\medskip\noindent\textbf{A worked instance and the extremization principle.}\enspace
The smallest check of \eqref{eq:V01-chiral} fixes the R-normalization from a marginal cubic.
Requiring a cubic superpotential $W=\Phi^3$ to carry R-charge two (so that $\int\!d^2\theta\,W$ is
R-neutral) forces $R(\Phi)=\tfrac23$, hence
\begin{equation}
\label{eq:V01-freechiral}
\Delta(\Phi)\ =\ \tfrac32\,R(\Phi)\ =\ \tfrac32\cdot\tfrac23\ =\ 1,
\end{equation}
the free-field dimension, consistent with the cubic having classical dimension three
($\Delta(W)=3$, marginal at the free point). A free chiral primary thus saturates the bound
\eqref{eq:V01-unitarity}, while a non-chiral operator (a real scalar of $R=0$ and $\Delta>0$) does
not. This is a normalization fixture for the $4d\ \mathcal{N}=1$ chiral-primary relation. It is not
a claim that the single-field cubic Wess--Zumino model defines a nontrivial four-dimensional
interacting fixed point. In an interacting theory the superconformal $R$ is not put in by hand. It is a specific
combination of all the abelian symmetries, \emph{selected} by an extremization principle:
$a$-maximization in $4d$, $F$-maximization in $3d$, $c$-extremization in $2d$. Section~1 states the
principle and its place in the algebra. The per-dimension derivations and worked computations are
introduced in the dimension sections below. The
chiral relation \eqref{eq:V01-chiral}, the
free-chiral saturation \eqref{eq:V01-freechiral}, and the non-saturation of a non-chiral operator
are verified symbolically in the check code recorded at section end. The whole derivation is one
positivity argument used twice, at a massive state in \S\ref{sec:V01-bps} and at a fixed point
here; Table~\ref{tab:V01-bpsscf} reads the two off side by side.

\begin{table}[ht]
\centering
\small
\setlength{\tabcolsep}{6pt}
\renewcommand{\arraystretch}{1.25}
\begin{tabular}{@{}>{\raggedright\arraybackslash}p{3.0cm} >{\raggedright\arraybackslash}p{4.55cm} >{\raggedright\arraybackslash}p{4.55cm}@{}}
\toprule
 & massive (super-Poincar\'e, \S\ref{sec:V01-bps}) & fixed point (superconformal, \S\ref{sec:V01-sconf})\\
\midrule
positive operator & $\{Q,Q^\dagger\}=A\ge 0$ & $\langle\mathcal{O}|\{Q,S\}|\mathcal{O}\rangle\ge 0$\\
\midrule
bound & $M\ge|Z|$ & $\Delta\ge cR$\\
\midrule
saturation & $M=|Z|$ (half-BPS) & $\Delta=cR$ (chiral, short)\\
\midrule
mechanism & preserved $Q$ annihilate the state & $Q|\mathcal{O}\rangle=0$, a short representation\\
\midrule
protected datum & BPS mass / index & dimension, $\Delta=\tfrac32 R$ at $4d\ \mathcal{N}{=}1$\\
\bottomrule
\end{tabular}
\caption{The BPS bound and the superconformal shortening bound are the same positivity argument,
applied to a massive one-particle state (\S\ref{sec:V01-bps}) and to a conformal primary
(\S\ref{sec:V01-sconf}). The eigenvalues, the saturation, and the kernel-at-saturation match row by
row.}
\label{tab:V01-bpsscf}
\end{table}

\medskip\noindent\textbf{The classification ceiling \texorpdfstring{$d\le 6$}{d<=6}.}\enspace
Superconformal algebras are far more constrained than super-Poincar\'e ones. Requiring the
fermionic generators to close into the conformal group as in \eqref{eq:V01-scfclose}, and the
whole to admit unitary representations, the Nahm classification of superconformal algebras leaves
only a finite list, and it terminates: \emph{interacting} superconformal field theories exist only
for $d\le 6$, the four families and their boundary collected in Table~\ref{tab:V01-sca}. The
two-dimensional case is special: local superconformal symmetry there is governed by the
super-Virasoro algebras, not by one of the finite-dimensional Nahm families below.
Table~\ref{tab:V01-sca} records the finite-dimensional $d\ge 3$ superconformal algebras and the
$d\ge 7$ ceiling; the two-dimensional superconformal toolkit is treated in the $2d$ section.

\begin{table}[htb]
\centering
\small
\setlength{\tabcolsep}{6pt}
\renewcommand{\arraystretch}{1.2}
\begin{tabular}{@{}c >{\raggedright\arraybackslash}p{2.5cm} >{\raggedright\arraybackslash}p{3.2cm} >{\raggedright\arraybackslash}p{1.9cm} >{\raggedright\arraybackslash}p{3.3cm}@{}}
\toprule
$d$ & superconformal algebra & bosonic part (conformal $\times$ R) & max interacting $Q$ & note\\
\midrule
$3$ & $\mathrm{OSp}(\mathcal{N}|4)$ & $Sp(4)\times SO(\mathcal{N})$ & $16$ $(\mathcal{N}{=}8)$ & $\mathfrak{so}(3,2){=}\mathfrak{sp}(4)$; $\mathcal{N}{>}8$ free\\
\midrule
$4$ & $SU(2,2|\mathcal{N})$ & $SU(2,2)\times U(\mathcal{N})$ & $16$ $(\mathcal{N}{=}4)$ & $\mathfrak{su}(2,2){=}\mathfrak{so}(4,2)$; $\mathcal{N}{=}4$ projectivized\\
\midrule
$5$ & $F(4)$ & $SO(5,2)\times SU(2)$ & $8$ $(\mathcal{N}{=}1)$ & the unique $5d$ superconformal algebra\\
\midrule
$6$ & $\mathrm{OSp}(8^*|2\mathcal{N})$ & $SO(6,2)\times USp(2\mathcal{N})$ & $16$ $\big((2,0)\big)$ & chiral $(\mathcal{N},0)$; $(2,0)$ is $\mathcal{N}{=}2$\\
\midrule
$\ge 7$ & none & $-$ & $-$ & no interacting superconformal field theory\\
\bottomrule
\end{tabular}
\caption{Interacting superconformal algebras by dimension (Nahm's classification). The conformal
factor is $\mathfrak{so}(d,2)$ and the R-symmetry factor is the automorphism of
\S\ref{sec:V01-rsym}, now graded inside the algebra. The maximal interacting supercharge count is
checked against the ladder of \S\ref{sec:V01-ladder}; the algebra names and
bosonic parts are the stated classification. There is no interacting superconformal field theory
above six dimensions.}
\label{tab:V01-sca}
\end{table}

Table~\ref{tab:V01-sca} makes the consequences sharp. The $5d$ algebra has only eight supercharges,
so the $5d\ \mathcal{N}=2$ sixteen-supercharge row is a maximal-SYM row, not a superconformal one.
In $4d$ and $6d$ no superconformal theory has more than sixteen supercharges, and in $3d$ a theory
with more than sixteen is free. Above six dimensions there is no interacting superconformal field
theory at all. We state this classification and cite its proof (\S\ref{sec:V01-gap}); the $d\le6$
ceiling is Nahm's. The argument has the same spirit as the positivity behind the super-Poincar\'e
spin ceiling of \S\ref{sec:V01-algebra}. The gap between the two ceilings is the room in which the
gravity-coupled and the non-conformal supersymmetric theories live.

\section{Master tables and section boundary}
\label{sec:V01-table}

The section's deliverable is the navigation map in Tables~\ref{tab:V01-master-counts}--\ref{tab:V01-master-tools}.
Every later section reads its supersymmetry notation, R-symmetry, multiplet content,
moduli-branch structure, and extremization principle off these tables, rather than redefining the
notation locally. The split is intentional:
Table~\ref{tab:V01-master-counts} answers ``how many real supercharges?'';
Table~\ref{tab:V01-master-rsym} answers ``what rotates them?''; and
Table~\ref{tab:V01-master-tools} answers ``what field-theory data and tools come with the
row?'' The rows run from two to ten dimensions, the range in which pure supersymmetric
Yang--Mills field theories exist without coupling to gravity. The $2d$ through $6d$ rows carry the
main fixed-point toolkit; the $7d$ through $10d$ rows are the high-dimensional
sixteen-supercharge SYM interface, including the F-theory 7-brane row at $8d$. The $32Q$ ten- and
eleven-dimensional algebras are the gravity-coupled ceiling, not ordinary pure-field-theory rows.

\begin{center}
\refstepcounter{table}\label{tab:V01-master-counts}
\centering
\scriptsize
\setlength{\tabcolsep}{5pt}
\renewcommand{\arraystretch}{1.03}
\begin{tabular}{@{}l >{\raggedright\arraybackslash}p{4.1cm} >{\raggedright\arraybackslash}p{3.1cm} >{\raggedright\arraybackslash}p{4.2cm}@{}}
\toprule
dim & labels used in this toolkit & real count & reading note\\
\midrule
$2d$ & $(0,1)/(1,0)$; $(0,2)/(1,1)$; $(2,2),(4,4),(8,8)$ &
$1,2,4,8,16$ & $(p,q)$ has $p+q$ real supercharges; chirality is part of the label.\\
\midrule
$3d$ & $\mathcal{N}{=}1,\ldots,8$ &
$Q=2\mathcal{N}$ & The interacting superconformal list stops at $\mathcal{N}\le 8$.\\
\midrule
$4d$ & $\mathcal{N}{=}1,2,3,4$ &
$Q=4\mathcal{N}$ & $\mathcal{N}=3$ has the same count as twelve real supercharges, not an
extra reduction rung.\\
\midrule
$5d$ & $\mathcal{N}{=}1,2$ &
$Q=8,16$ & The sixteen-supercharge row is maximal SYM, not a 5d SCFT row.\\
\midrule
$6d$ & $(1,0),(1,1),(2,0)$ &
$Q=8,16,16$ & The two sixteen-supercharge labels differ by chirality and field content.\\
\midrule
$7d$ & $\mathcal{N}{=}1$ &
$Q=16$ & The 7d SYM row used by the $G_2$ roadmap interface.\\
\midrule
$8d$ & $\mathcal{N}{=}1$ &
$Q=16$ & The 7-brane gauge-theory row used by the F-theory engine.\\
\midrule
$9d$ & $\mathcal{N}{=}1$ &
$Q=16$ & Pure SYM exists; no current engine needs a separate 9d section.\\
\midrule
$10d$ & $\mathcal{N}{=}1$ &
$Q=16$ & The parent pure-SYM row; 32Q type-II algebras are gravity-coupled.\\
\bottomrule
\end{tabular}
\par\vspace{2pt}
\begin{minipage}{0.98\textwidth}
\footnotesize
Table~\thetable: Master count table: quick lookup from each dimension-dependent SUSY label to its
real-supercharge count. The ambient $32Q$ ceiling is tracked in
Table~\ref{tab:V01-ladder-counts} and \eqref{eq:V01-table-counts}, not as a decoupled-QFT row.
\end{minipage}
\end{center}

\clearpage

\begin{center}
\refstepcounter{table}\label{tab:V01-master-rsym}
\centering
\footnotesize
\setlength{\tabcolsep}{3pt}
\renewcommand{\arraystretch}{1.14}
\begin{tabular}{@{}>{\raggedright\arraybackslash}p{0.65cm} >{\raggedright\arraybackslash}p{3.1cm} >{\raggedright\arraybackslash}p{4.0cm} >{\raggedright\arraybackslash}p{3.55cm}@{}}
\toprule
dim & label & R-symmetry & caution\\
\midrule
$2d$ & $(0,1)/(1,0)$; $(0,2)$; $(2,2)$; $(4,4)$ &
none; $U(1)_R$; $U(1)_V{\times}U(1)_A$; $SU(2){\times}SU(2)$, respectively &
$U(1)_L$ in $(0,2)$ is flavor, not the superconformal R-current.\\
\midrule
$3d$ & $\mathcal{N}{=}1$; $\mathcal{N}{=}2$; $\mathcal{N}\ge 3$ &
none; $U(1)_R$; $\mathrm{Spin}(\mathcal{N})$, respectively &
Maximal SYM has UV scalar-rotation $\mathrm{Spin}(7)$; the IR superconformal algebra has $\mathrm{Spin}(8)_R$.\\
\midrule
$4d$ & $\mathcal{N}{=}1$; $\mathcal{N}{=}2$; $\mathcal{N}{=}3$; $\mathcal{N}{=}4$ &
$U(1)_R$; $SU(2)_R{\times}U(1)_R$; $SU(3){\times}U(1)$; $SU(4)_R$, respectively &
The $\mathcal{N}=2$ row is $SU(2)_R{\times}U(1)_R$, not $SU(3)$.\\
\midrule
$5d$ & $\mathcal{N}{=}1$; $\mathcal{N}{=}2$ &
$SU(2)_R$; $USp(4)_R$, respectively &
$U(1)_I/E_n$ is flavor, not R-symmetry; the $16Q$ row is maximal SYM, not a $5d$ SC algebra.\\
\midrule
$6d$ & $(1,0)$; $(1,1)$; $(2,0)$ &
$SU(2)_R$; $\mathrm{Spin}(4)_R$; $USp(4)_R{=}\mathrm{Spin}(5)_R$, respectively &
The two $16Q$ labels have different chirality and different R-symmetry.\\
\midrule
$7d$ & $\mathcal{N}{=}1$ & $SU(2)_R$ & This is a sixteen-supercharge SYM row, not a separate fixed-point family.\\
\midrule
$8d$ & $\mathcal{N}{=}1$ & $U(1)_R$ & The $U(1)_R$ is the $SO(2)$ transverse rotation of the 7-brane gauge theory.\\
\midrule
$9d$ & $\mathcal{N}{=}1$ & none continuous & The $SO(1)$ transverse rotation of the SYM parent is trivial.\\
\midrule
$10d$ & $\mathcal{N}{=}1$ & none continuous & The parent pure-SYM row has no transverse rotation R-symmetry.\\
\bottomrule
\end{tabular}
\par\medskip
\begin{minipage}{0.98\textwidth}
\small
Table~\thetable: The master R-symmetry table. It isolates the automorphism group that rotates the
supercharges and the flavor-versus-R distinctions that later exact computations depend on. Within
each row, the entries in the label column and the R-symmetry column are read in the same
order. The two-dimensional entries shown are the rows operated downstream in these notes; extended
chiral variants not displayed separately are discussed in the $2d$ section. The table is a downstream-use
lookup, not a full classification of every $2d$ superconformal variant.
\end{minipage}
\end{center}

\clearpage

\begin{center}
\refstepcounter{table}\label{tab:V01-master-tools}
\centering
\footnotesize
\setlength{\tabcolsep}{5pt}
\renewcommand{\arraystretch}{1.18}
\begin{tabular}{@{}l >{\raggedright\arraybackslash}p{4.15cm} >{\raggedright\arraybackslash}p{3.55cm} >{\raggedright\arraybackslash}p{3.25cm}@{}}
\toprule
dim & multiplets and branches & protected tools & where they reappear\\
\midrule
$2d$ & chiral, twisted-chiral, vector, Fermi; Higgs, Coulomb, LG phases &
$c$-extremization for $(0,2)+$; elliptic genus & worldsheet, GLSM, and brane-brick examples\\
\midrule
$3d$ & scalar, chiral, vector, hyper; Higgs, Coulomb; mirror pairs &
$F$-maximization; $S^3$ index; $\mathcal{N}>8$ free & monopoles, mirrors, and Chern--Simons matter\\
\midrule
$4d$ & chiral, vector, hyper; Higgs and Coulomb branches; Coulomb-branch low-energy data &
$a$-maximization; superconformal index & quivers, dualities, and geometric-engineering examples\\
\midrule
$5d$ & vector, hyper; real Coulomb branch &
5d index only for $\mathcal{N}=1$; no $\mathcal{N}=2$ SC algebra & UV fixed points and instanton particles\\
\midrule
$6d$ & tensor, hyper, vector; tensor branch &
anomaly 8-form and Green--Schwarz factorization & tensor branches and anomaly matching\\
\midrule
$7d$ & $\mathcal{N}=1$ vector multiplet; Coulomb branch &
sixteen-supercharge SYM bookkeeping & $G_2$-holonomy gauge-theory examples\\
\midrule
$8d$ & $\mathcal{N}=1$ vector multiplet; two adjoint scalars &
7-brane gauge-theory bookkeeping; no SC extremization & F-theory 7-brane gauge sectors\\
\midrule
$9d$ & $\mathcal{N}=1$ vector multiplet; one adjoint scalar &
reduction-parent bookkeeping; no SC extremization & high-dimensional SYM interface\\
\midrule
$10d$ & $\mathcal{N}=1$ vector multiplet &
parent pure-SYM bookkeeping; no SC extremization & dimensional-reduction parent theory\\
\bottomrule
\end{tabular}
\par\medskip
\begin{minipage}{0.98\textwidth}
\small
Table~\thetable: The master tools table. It records the field-theory objects, protected
computations, and recurring uses associated with each dimension after the count and
R-symmetry have already been fixed.
\end{minipage}
\end{center}

\medskip\noindent\textbf{What this section states but does not prove.}\enspace
\label{sec:V01-gap}

Section~1 gives the algebra and the bookkeeping, not the dynamics, and the boundary is first-class
content. It states two classification theorems and cites them rather than proving them. The first
is the Nahm classification of super-Poincar\'e algebras, with its $d\le 11$ and $32$-supercharge
ceiling (\S\ref{sec:V01-algebra}). The second is the classification of superconformal algebras,
with its $d\le 6$ interacting ceiling (\S\ref{sec:V01-sconf}). They are cited mathematics, not
proved here. The section also states the BPS bound and its
shortening, but computes no specific BPS spectrum. Which states saturate it, and how the central
charge varies over a moduli space, is the dynamics of the dimension-specific sections.
Finally, the section does not derive supersymmetry preservation from external construction data,
and it does not develop the Euclidean-spinor and localization-contour conventions used in index
and partition-function computations. The intended takeaway is: align the supersymmetry notation through the
master tables, read off the R-symmetry, central-charge, and BPS roles, and translate a field-theory
SUSY label into its real-supercharge count.

\medskip\noindent\textbf{Exit checklist.}\enspace
After this section the reader can
\begin{enumerate}
\item translate any field-theory supersymmetry label in the master tables into a real supercharge
count;
\item follow the $4Q$, $8Q$, and $16Q$ dimensional ladders under ordinary circle reduction;
\item identify the R-symmetry, central-charge, and BPS role attached to a row;
\item use the $SU(2)$/$SU(3)$ embedding fixture to decide whether the standard D-brane orbifold
probe has $8Q$ / $4d\ \mathcal{N}=2$ or $4Q$ / $4d\ \mathcal{N}=1$;
\item state which classification results are being cited rather than proved.
\end{enumerate}

\bigskip
\section*{Sources and notes}
\addcontentsline{toc}{subsection}{Sources and notes}
\markboth{Sources and notes}{Sources and notes}
{\small

\noindent\textsf{\textcolor{RoyalBlue}{Sources and notes.}}\enspace
This is the spine section and template for the remaining sections; it makes no constructive
claim beyond the algebraic dictionary.

\medskip\noindent\textsf{\textcolor{RoyalBlue}{\textbf{\S\ref{sec:V01-algebra}\enspace The super-Poincar\'e algebra.}}}\enspace
The graded extension of Poincar\'e \eqref{eq:V01-QQ}, its central extension \eqref{eq:V01-central},
and the Nahm spin ceiling: $32Q$ forces gravity and $64Q$ forces higher spin, stated and cited.
(\textcite{Coleman:1967ad} the bosonic no-go; \textcite{Haag:1974qh} the supersymmetric extension;
\textcite{Nahm:1977tg} the classification + ceiling).

\medskip\noindent\textsf{\textcolor{RoyalBlue}{\textbf{\S\ref{sec:V01-count}\enspace The minimal spinor and the count.}}}\enspace
The reality table \eqref{eq:V01-table-counts} (types M/MW/SM/SMW and minimal real counts,
$d=2$--$11$) and the count formula \eqref{eq:V01-counts}, fixed by the real Clifford algebra and
$(d-2)\bmod8$. (\textcite{Kugo:1982bn}). 

\medskip\noindent\textsf{\textcolor{RoyalBlue}{\textbf{\S\ref{sec:V01-rsym}\enspace R-symmetry.}}}\enspace
The R-symmetry as the automorphism rotating the supercharges \eqref{eq:V01-Raction}, with the
$2d\,(0,1)$ (no continuous R), $4d\,\mathcal{N}{=}2$ ($SU(2){\times}U(1)$, not $SU(3)$),
$2d\,(0,2)$ ($U(1)_R$ only, $U(1)_L$
flavor), $5d$ ($SU(2)_R$, $U(1)_I/E_n$ flavor), $8d\ \mathcal{N}=1$ ($U(1)_R$ from
$SO(2)$ transverse rotations), and 3d maximal-SYM UV $\mathrm{Spin}(7)$ versus IR
$\mathrm{Spin}(8)$ identifications. 

\medskip\noindent\textsf{\textcolor{RoyalBlue}{\textbf{\S\ref{sec:V01-ladder}\enspace The dimensional ladder.}}}\enspace
The full count ladder Table~\ref{tab:V01-ladder-counts}: $1Q$, $2Q$, $4Q$ \eqref{eq:V01-track4}, $8Q$
\eqref{eq:V01-track8}, $16Q$ \eqref{eq:V01-track16} (including the $10d$ parent and the
$9d$ / $8d$ / $7d\ \mathcal{N}=1$ pure-SYM rows), and the $32Q$ ambient ceiling, each as
count $=$ $\mathrm{min}(d)\times\mathcal{N}$ (or $\mathrm{min}(d)\times(p+q)$ in chiral even
dimensions).

\medskip\noindent\textsf{\textcolor{RoyalBlue}{\textbf{\S\ref{sec:V01-bps}\enspace Central charges and the BPS bound.}}}\enspace
The BPS block \eqref{eq:V01-block}, the bound \eqref{eq:V01-bpsbound}, the $\tfrac12$-BPS
shortening (the preserved supercharges annihilate the saturating state), and the Seiberg--Witten
central charge \eqref{eq:V01-SW} in the normalization used below
(\textcite{Seiberg:1994rs}).
The graded protection (shortening protects the algebraic relation and the relevant charge/index
data in its chamber, but not global one-particle stability across the whole moduli space;
recombination and wall-crossing are dimension-specific dynamical questions) is stated
qualitatively in the body. Machine-checked symbolically: the block's eigenvalues $M\pm|Z|$, the bound
from $\det A=M^2-|Z|^2$, the rank-one saturation with the preserved kernel annihilating and the
broken combination at eigenvalue $2M$, and $Z=an_e+a_Dn_m$, $M=\sqrt2|Z|$.

\medskip\noindent\textsf{\textcolor{RoyalBlue}{\textbf{\S\ref{sec:V01-sconf}\enspace Superconformal algebras and the unitarity bound.}}}\enspace
The superconformal closure \eqref{eq:V01-scfclose} (the conformal supercharge $S\sim[K,Q]$ of
\eqref{eq:V01-Sgen}, the R-symmetry inside the algebra), the superconformal unitarity bound
\eqref{eq:V01-unitarity} from the positivity \eqref{eq:V01-scfblock}, the $4d\ \mathcal{N}=1$
chiral relation $\Delta=\tfrac32 R$ \eqref{eq:V01-chiral} as the fixed-point analog of $M=|Z|$, the
Nahm classification of superconformal algebras with its $d\le6$ interacting ceiling
(Table~\ref{tab:V01-sca}), and the row-specific constraints (especially $5d$ only has the $8Q$
$F(4)$ algebra): derived where elementary, otherwise stated and cited. (\textcite{Nahm:1977tg} the classification; \textcite{Minwalla:1997ka} the unitarity bounds). The
BPS-to-superconformal parallel is collected in Table~\ref{tab:V01-bpsscf}. Machine-checked
symbolically and arithmetically: the free-chiral R-charge $R(\Phi)=\tfrac23$ solved from
$R(W{=}\Phi^3)=2$ (with the solver tested on quartic/quintic fixtures so the provenance is
executable), the saturation $\Delta(\Phi)=\tfrac32 R=1$ \eqref{eq:V01-freechiral}, the
non-saturation of a non-chiral ($R=0$) operator, and the Table~\ref{tab:V01-sca} maximal-interacting-$Q$
column plus the $d\le6$ ceiling against the ladder. The extremization
principle ($a$-/$F$-maximization, $c$-extremization) that selects the superconformal R is stated;
its per-dimension realizations appear in the later sections.

\medskip\noindent\textsf{\textcolor{RoyalBlue}{\textbf{\S\ref{sec:V01-table}\enspace The master tables.}}}\enspace
Tables~\ref{tab:V01-master-counts}--\ref{tab:V01-master-tools}, the navigation map;
consistent term by term, with the per-row content owned by the named sections.
}

\subsection*{Further reading}
\addcontentsline{toc}{subsection}{Further reading}
For the super-Poincar\'e algebra and the uniqueness theorems, standard references are
\textcite{Weinberg:2000cr,Terning:2006bq} and the review \textcite{Sohnius:1985qm}; the founding papers
are \textcite{Golfand:1971iw,Volkov:1973ix,Wess:1974tw}. Central charges and the BPS bound originate
with \textcite{Witten:1978mh}. Compact pedagogical introductions at the level of these notes are
\textcite{Martin:1997ns,Bilal:2001nv,Lykken:1996xt,Argyres:2001eva}, and the superspace apparatus is
catalogued in \textcite{Gates:1983nr}.

Modern treatments of superconformal algebras and their representations across dimensions are given
in \textcite{Eberhardt:2020cxo}; the geometry of square-zero supercharges and the resulting
classification of twists are developed in \textcite{Eager:2018dsx,Elliott:2020ecf}.
A recent computation-oriented survey of large-$N$ limits across $3d$, $4d$, and $5d$ supersymmetric
field theories is \textcite{Santilli:2025zum}.

\section*{References}
\addcontentsline{toc}{subsection}{References}
\markboth{References}{References}
\printbibliography[heading=none]
\end{refsection}
\begin{refsection}\chapter{Common SUSY-QFT grammar: multiplets, local terms, vacua, anomalies}
\label{ch:V02}

\noindent\textbf{Guide to this section.}\enspace
This section is a compact bridge. Section~1 fixed the
algebraic spine: supercharges, dimensional reduction, R-symmetry labels,
central charges, and holonomy projections. The dimension sections that
follow need a different but equally common language. They need to say what
kind of multiplet carries a field, what kind of supersymmetric local term is
being written, what it means to solve for a vacuum, and how an anomaly or
protected current constrains the answer.

The goal here is deliberately modest. We will not teach full superspace
technology, not classify moduli spaces, not run extremization, and not prove
duality. Instead we build the shared dictionary that lets the later sections
move quickly without redefining every word. Whenever a topic becomes
dimension-specific, this section stops and points forward.

\begin{keybox}{What this section delivers}
After this section the reader should be able to parse the local vocabulary of
a supersymmetric field theory: a multiplet assignment, an F-term, a D-term, a
superpotential equation, a moment-map equation, a quotient by a gauge group,
a branch label, a protected chiral operator, a fermion anomaly trace, and
deformations/background parameters. This section also fixes these notes'
navigation rule: E/J systems and GLSMs live in $2d$, Chern--Simons terms and
monopoles in $3d$, a-maximization and Seiberg-type technology in $4d$,
prepotentials and instanton operators in $5d$, tensor-branch and
Green--Schwarz anomaly technology in $6d$.
\end{keybox}

\section{Multiplets are representations with local descendants}
\label{sec:V02-multiplets}

A supersymmetric field theory begins with an ordinary spacetime and a
super-Poincar\'e algebra, but the fields do not appear one at a time. They
appear in representations of the algebra. We call such a representation a
\emph{multiplet}. The word is intentionally neutral: a multiplet can be
off-shell in a convenient formalism, on-shell in a more economical one, or
non-Lagrangian if the theory is known by protected data rather than by a
field list. What survives across all these cases is that supersymmetry
relates bosons and fermions and packages their local descendants into a
common object.

The smallest useful distinction is between three pieces of data.
\begin{enumerate}
\item The \emph{spacetime dimension} and the amount of supersymmetry,
fixed by Section~1.
\item The \emph{multiplet type}, such as chiral, vector, hyper, tensor,
or Fermi.
\item The \emph{local terms} allowed by supersymmetry: kinetic terms,
superpotential-type terms, moment-map or D-type terms, and in special
dimensions topological couplings.
\end{enumerate}
For example, a four-dimensional \(\mathcal N=1\) chiral multiplet is not
the same representation as a two-dimensional \((0,2)\) chiral multiplet,
even though the same word ``chiral'' is used. The label is meaningful only
after the algebra and dimension have been fixed.

\begin{table}[ht]
\centering
\small
\begin{tabular}{@{}p{0.16\linewidth}p{0.24\linewidth}p{0.47\linewidth}@{}}
\toprule
\textbf{Multiplet} & \textbf{Typical use} & \textbf{Reading rule} \\
\midrule
Chiral & complex scalar plus fermions in $4d$ \(\mathcal N=1\), and related lower-dimensional systems
& Carries holomorphic local data. Its detailed component content is dimension-dependent. \\
\midrule
Vector & gauge field plus superpartners & Carries gauge redundancy and a moment-map/D-type constraint when a Lagrangian description exists; a $3d$ \(\mathcal N=2\) vector also contains a real scalar after reduction. \\
\midrule
Hyper & extended-SUSY matter & Usually best viewed as paired chiral data from a smaller-SUSY perspective; its quaternionic structure is dimension-specific. \\
\midrule
Tensor & higher-form field strength plus partners & Essential in $6d$. Tensor branches and Green--Schwarz terms are owned by the $6d$ section. \\
\midrule
Fermi & left-moving fermionic multiplet in $2d$ \((0,2)\) language & Carries \(E\)- and \(J\)-type data. This bridge only names the words; the $2d$ section does the real work. \\
\midrule
Current multiplet & conserved currents and stress tensor & Encodes global symmetries, R-currents, anomalies, and protected observables. \\
\bottomrule
\end{tabular}
\caption{Multiplet types and the reading rule used here. The table is a reading aid, not a classification theorem.}
\label{tab:V02-multiplets}
\end{table}

Table~\ref{tab:V02-multiplets} is not a classification theorem. It is a reading aid. A dimension
section may refine a row, split it, or replace it by a more natural
language. The bridge rule is that a multiplet name is not a promise of a
universal component formula. It is a promise that the later section will
state which algebra is acting and what local data the multiplet carries.

\subsection*{A minimal superspace slogan}

Superspace is a powerful notation, but we will use only its slogan here.
One introduces anticommuting coordinates \(\theta\) so that some
supersymmetry transformations become translations in \(\theta\). A
constraint on a superfield then defines a multiplet, and an integral over
part or all of the \(\theta\)-space defines a supersymmetric local term.

For a four-dimensional \(\mathcal N=1\) Lagrangian, the familiar schematic
form is
\begin{equation}
\label{eq:V02-4dN1-schematic}
 \mathcal L
 =
 \int d^4\theta\, K(\Phi^\dagger e^V,\Phi)
 +
 \left(
 \int d^2\theta\, W(\Phi)
 +
 \int d^2\theta\, \frac{\tau}{16\pi i}\,
 \operatorname{Tr} W^\alpha W_\alpha
 + \text{c.c.}
 \right)
.
\end{equation}
Here \(\Phi\) denotes chiral multiplets, \(V\) vector multiplets,
\(K\) a Kähler potential, \(W\) a holomorphic superpotential, and
\(\tau\) a complexified gauge coupling. Equation~\eqref{eq:V02-4dN1-schematic}
is not the master equation for these notes. It is a compact example of the
grammar: full superspace integrals behave like D-type terms, chiral
superspace integrals behave like F-type terms, and gauge kinetic terms can
also be protected holomorphic data.

This one line already explains why the bridge stays short. In $2d$
\((0,2)\), the natural language uses chiral and Fermi multiplets with
\(E\)- and \(J\)-functions. In $3d$, Chern--Simons couplings and monopole
operators become central. In $5d$, a cubic prepotential controls the Coulomb
branch. In $6d$, tensor multiplets and anomaly polynomials organize the
story. The common word is ``local supersymmetric term''; the implementation
belongs to the corresponding dimension.

\subsection*{Off-shell convenience versus physical content}

The reader will often meet phrases such as ``off-shell multiplet'' and
``auxiliary field''. The bridge uses them in a pragmatic way. An off-shell
description realizes the supersymmetry transformations without imposing the
equations of motion, usually by adding auxiliary fields. This can make
local terms compact to write and vacuum equations mechanically visible. It
is not, by itself, the physical definition of the theory.

The Wess--Zumino example in
Equation~\eqref{eq:V02-wz-superpotential} has an auxiliary field \(F\).
Before eliminating it, the bosonic part contains
\begin{equation}
\label{eq:V02-offshell-F}
 |F|^2
 +
 F\,\frac{\partial W}{\partial\phi}
 +
 \overline F\,\frac{\partial\overline W}{\partial\overline\phi}.
\end{equation}
The equation of motion sets \(F=-\partial\overline W/\partial\overline\phi\),
equivalently \(\overline F=-\partial W/\partial\phi\), and the positive
potential \(V_F=|\partial W|^2\) follows. The auxiliary field is a
bookkeeping device for the local supersymmetric term. It is not a
propagating particle.

This distinction becomes useful later because different dimensions have
different off-shell technologies. Four-dimensional \(\mathcal N=1\)
superspace is economical. Extended supersymmetry can force one to use a
smaller visible subalgebra, harmonic or projective superspace, or no compact
off-shell formalism at all. These notes will not standardize those choices.
It will standardize the questions: what are the multiplets, what local
terms are allowed, which auxiliary or protected equations follow, and what
physical data remain after quotienting redundancies?

\subsection*{Smaller-SUSY language is often a chart}

A common move is to describe a theory with a smaller visible supersymmetry
than the theory actually has. A four-dimensional \(\mathcal N=2\) vector
multiplet can be described, in \(\mathcal N=1\) language, by a vector
multiplet plus an adjoint chiral multiplet. A hypermultiplet can be
described by a pair of chiral multiplets in conjugate representations.
This is a chart on the local grammar, not a statement that the extra
supersymmetry has disappeared.

The same idea appears after dimensional reduction. A component that was a
gauge field in a higher dimension can become a scalar in a lower dimension.
The multiplet name then changes with the algebra. This is why these notes
keep Section~1 and the present bridge separate: Section~1 says how the
supercharges and R-symmetries reduce; this section says how the local
objects are named once a dimension and visible algebra have been chosen.

\section{F-terms, D-terms, and the first vacuum equations}
\label{sec:V02-local-terms}

In a Lagrangian frame, supersymmetric vacua are often found by eliminating
auxiliary fields. This is the most useful entry point for the later
sections, because it turns a local term into an equation. We record the
minimal pattern.

Consider a single four-dimensional \(\mathcal N=1\) chiral multiplet
\(\Phi\) with scalar component \(\phi\) and superpotential
\begin{equation}
\label{eq:V02-wz-superpotential}
 W(\Phi) = \frac{m}{2}\Phi^2 + \frac{\lambda}{3}\Phi^3.
\end{equation}
The auxiliary field \(F\) is eliminated by
\begin{equation}
\label{eq:V02-F-elim}
 F = -\frac{\partial \overline W}{\partial \overline\phi},
 \qquad
 V_F(\phi)=\left|\frac{\partial W}{\partial\phi}\right|^2
 =
 \left|m\phi+\lambda\phi^2\right|^2.
\end{equation}
A supersymmetric vacuum in this sector solves
\begin{equation}
\label{eq:V02-F-flat}
 \frac{\partial W}{\partial\phi}=0.
\end{equation}
This elementary example is enough to fix the usage here: an
\emph{F-flat equation} is the vanishing of the holomorphic derivative of the
superpotential, or of its dimension-specific analogue.

For a gauge theory with chiral fields \(\phi_i\) of charges \(q_i\) under a
U(1), a D-type auxiliary field gives
\begin{equation}
\label{eq:V02-D-term}
 D = \sum_i q_i |\phi_i|^2 - \zeta,
 \qquad
 V_D = \frac{g^2}{2}D^2.
\end{equation}
Here \(\zeta\) is a Fayet--Iliopoulos parameter when the dimension and
supersymmetry allow it. The D-flat equation is
\begin{equation}
\label{eq:V02-D-flat}
 \sum_i q_i |\phi_i|^2 = \zeta.
\end{equation}
For a general compact gauge group the same statement becomes a moment-map equation
\begin{equation}
\label{eq:V02-moment-map}
 \mu(\phi)=\zeta_{\text{central}}.
\end{equation}
The level \(\zeta_{\text{central}}\) must lie in the gauge-invariant
central part of the Lie algebra dual. For a semisimple nonabelian factor
the corresponding level is zero. Thus one often writes \(\mu^a=0\) for
nonabelian generators and \(\mu_{U(1)}=\zeta\) for abelian trace factors.
The exact normalization of \(\mu\) depends on conventions. The invariant
content is that D-type terms impose real moment-map equations before one
divides by gauge equivalence.

\begin{keybox}{Common misconception: F-terms and D-terms are specific local terms}
``F-term'' and ``D-term'' do not mean ``everything holomorphic'' and
``everything non-holomorphic'' in a dimension-independent way. They mean
specific supersymmetric local terms in a specified algebra. This bridge
uses the four-dimensional \(\mathcal N=1\) notation only as a compact
example, not as a universal formalism.
\end{keybox}

\subsection*{The quotient is part of the equation}

Solving the equations is not yet the vacuum space. Gauge-equivalent
solutions represent the same physical vacuum. A basic U(1) example makes
the point. Let \(Q\) and \(\widetilde Q\) have charges \(+1\) and \(-1\),
with no superpotential and \(\zeta=0\). The D-flat equation is
\begin{equation}
\label{eq:V02-U1-plus-minus}
 |Q|^2-|\widetilde Q|^2=0.
\end{equation}
Modulo the U(1) action
\begin{equation}
\label{eq:V02-U1-action}
 (Q,\widetilde Q)\mapsto
 (e^{i\alpha}Q,e^{-i\alpha}\widetilde Q),
\end{equation}
the gauge-invariant coordinate is
\begin{equation}
\label{eq:V02-meson}
 M=Q\widetilde Q.
\end{equation}
Thus the quotient is parameterized by the complex number \(M\). This is a
toy model, but it contains the rule used throughout these notes: the physical
vacuum space is described by equations plus equivalence, often translated
into gauge-invariant operators and their relations.

A slightly different U(1) example explains why the sign of a real parameter
can change the description. Let \(X_1\) and \(X_2\) both have charge \(+1\)
and take \(\zeta>0\). The D-flat equation
\begin{equation}
\label{eq:V02-CP1-D}
 |X_1|^2+|X_2|^2=\zeta
\end{equation}
modulo the common phase gives \(\mathbb P^1\). At \(\zeta=0\), the quotient
collapses to the origin in the gauge-invariant affine description. The
later $2d$ GLSM section will make this kind of statement precise using its own
language. Here we only keep the reading rule: an FI-like parameter can
separate phases, and a quotient by gauge equivalence is part of the vacuum
construction.

\subsection*{Complex quotient language}

There is an equivalent algebraic way to package many of these examples.
Instead of solving the real moment-map equation and dividing by the compact
gauge group, one often solves the holomorphic F-flat equations and divides
by the complexified gauge group, with a stability condition determined by
the FI parameter. For the \(X_1,X_2\) example with \(\zeta>0\), the
complexified group is \(\mathbb C^\ast\) acting by
\begin{equation}
\label{eq:V02-Cstar-action}
 (X_1,X_2)\mapsto (\lambda X_1,\lambda X_2),
 \qquad
 \lambda\in \mathbb C^\ast.
\end{equation}
The stable locus removes the origin, and
\begin{equation}
\label{eq:V02-CP1-complex-quotient}
 (\mathbb C^2-\{0\})/\mathbb C^\ast \simeq \mathbb P^1.
\end{equation}
This is the same physical quotient as the D-flat sphere modulo U(1). The
bridge does not develop geometric invariant theory. It only records why
the same vacuum may be written in a differential form, as a moment-map
quotient, or in an algebraic form, as equations modulo a complexified
redundancy.

The distinction matters because the later sections use both languages. A
$2d$ GLSM phase is usually clearest in quotient language. A $3d$ Coulomb branch
may require quantum coordinates that are not visible in a naive classical
quotient. A $4d$ chiral ring is often presented algebraically, while its
unitary vacuum interpretation remembers D-flatness. These are not three
different theories. They are three controlled coordinate systems on the
same local problem when the assumptions of the chosen frame apply.

\subsection*{A superpotential ideal}

The simplest F-term ideal already shows how relations enter a protected
ring. Take three chiral multiplets \(X,Y,Z\) and
\begin{equation}
\label{eq:V02-XYZ-W}
 W = XYZ.
\end{equation}
The F-flat equations are
\begin{equation}
\label{eq:V02-XYZ-F}
 YZ=0,
 \qquad
 XZ=0,
 \qquad
 XY=0.
\end{equation}
The classical chiral ring in this frame is
\begin{equation}
\label{eq:V02-XYZ-ring}
 \mathcal R_{XYZ}
 =
 \mathbb C[X,Y,Z]/(XY,XZ,YZ).
\end{equation}
Geometrically this is the union of three coordinate axes. Physically it is
a reminder that F-flatness can split the vacuum set into components, and
that a branch need not be smooth even in a tiny example. We will not use
this as a model of a particular dimension's dynamics. We will use it as a
portable grammar for the phrases ``F-term ideal'', ``branch'', ``component'',
and ``ring relation''.

\section{Branches, protected rings, and local coordinates}
\label{sec:V02-branches}

Supersymmetric vacua often form families. A connected component or
irreducible piece of such a family is called a \emph{branch}. The name of a
branch is physical shorthand, not a theorem by itself. A Higgs branch is
usually visible in a frame where charged matter has expectation values. A
Coulomb branch is usually visible in a frame where vector-multiplet scalars,
dual photons, or related abelianized data parameterize the vacuum. A mixed
branch contains both kinds of data.

The word ``usually'' matters. Dualities can exchange descriptions, and
strong-coupling physics can introduce coordinates not visible in a weakly
coupled Lagrangian. A branch label therefore records how the branch is seen
in a chosen frame. It is not an invariant name that every dual frame must
use in the same way.

\begin{table}[ht]
\centering
\small
\begin{tabular}{@{}p{0.16\linewidth}p{0.31\linewidth}p{0.42\linewidth}@{}}
\toprule
\textbf{Label} & \textbf{Typical weak-frame signal} & \textbf{Bridge rule} \\
\midrule
Higgs branch & charged matter expectation values and gauge symmetry broken partly or fully
& Use the label when matter coordinates or gauge-invariant composites are the natural local coordinates. \\
\midrule
Coulomb branch & vector-multiplet scalars, abelianization, dual photons, monopole or instanton coordinates in suitable dimensions
& Use the label only after the dimension-specific section states the appropriate coordinates. \\
\midrule
Mixed branch & simultaneous matter and vector-multiplet coordinates
& Treat as an intersection or fibration when the later section supplies the local structure. \\
\midrule
Conformal manifold & exactly marginal deformations modulo redundancies
& The bridge names it; the dimension section supplies the protected marginality tests. \\
\bottomrule
\end{tabular}
\caption{Branch labels and their frame-dependent weak-coupling signals.}
\label{tab:V02-branches}
\end{table}

Protected chiral or BPS operator rings are a second way to describe vacuum
data. In a Lagrangian example, one begins with gauge-invariant polynomials
in fields and divides by relations generated by F-flat equations. If
\begin{equation}
\label{eq:V02-ring-example}
 \mathcal R = \mathbb C[x,y,z]/(xy),
\end{equation}
then the relation \(xy=0\) says that the affine variety is the union of two
coordinate planes meeting along the \(z\)-axis. A coarse Hilbert series is
\begin{equation}
\label{eq:V02-hilbert-toy}
 H_{\mathcal R}(t)=\frac{1-t^2}{(1-t)^3}.
\end{equation}
This formula is included only to normalize the vocabulary. More advanced uses of Hilbert-series and
perfect-matching technology sharpen the same idea, but these notes only need the reader to recognize
that a ring presentation consists of
generators, relations, and gradings.

\subsection*{What a Hilbert series is allowed to mean here}

A Hilbert series is a counting device for graded operators. In the toy
ring \(\mathcal R=\mathbb C[x,y,z]/(xy)\), assign
\(\deg x=\deg y=\deg z=1\). The polynomial ring has generating function
\((1-t)^{-3}\). The single degree-two relation removes the contribution
of the ideal generated by \(xy\), giving
Equation~\eqref{eq:V02-hilbert-toy}. That is the entire level of detail
needed in this section.

The later sections will be more careful. A physically meaningful Hilbert
series may refine the grading by flavor fugacities, impose gauge
invariance through an integral or a Molien sum, or include quantum
coordinates. It may also fail to capture the full geometry if the chosen
ring is not the right protected object. The displayed
\((1-t^2)/(1-t)^3\) form assumes a single complete-intersection relation;
it is not a general algorithm for arbitrary non-complete-intersection
rings. The bridge therefore uses Hilbert series only as a word in the
dictionary: it is a compact way to count graded protected operators after
generators and relations have been specified.

\begin{keybox}{Common misconception: not every vacuum component is a Higgs branch}
The bridge section does not identify every vacuum component with a Higgs
branch. A Higgs/Coulomb/mixed label depends on which multiplets provide the
coordinates in the chosen description, and in some dimensions the natural
Coulomb coordinates are intrinsically quantum.
\end{keybox}

\subsection*{When no Lagrangian is available}

Some supersymmetric field theories are not best introduced by a weakly
coupled Lagrangian. The bridge still has something to say. A theory can be
specified, partly or wholly, by protected data:
\begin{equation}
\label{eq:V02-protected-data}
 \mathfrak D_{\text{protected}}
 =
 \left\{
 \begin{array}{l}
 \text{superalgebra and R-symmetry},\\
 \text{global symmetry and anomaly data},\\
 \text{protected operators and rings},\\
 \text{deformations, defects, and boundary data}
 \end{array}
 \right\}.
\end{equation}
This list is not a replacement for dynamics. It is a way to read the later
sections without assuming that every supersymmetric QFT has a convenient
elementary-field Lagrangian. In particular, non-Lagrangian four-dimensional
\(\mathcal N=2\) theories and six-dimensional SCFTs are not exceptions to
the grammar; they are cases where the grammar is applied to protected
observables and deformation data rather than to a microscopic component
action.

There is a discipline to this. Calling a theory non-Lagrangian should not
license vague prose. One should still be able to say which supersymmetry
algebra is realized, which global symmetries act on protected operators,
which deformations are allowed, which anomalies are known, and what
evidence supports an identification with another description. A later
section may use a curve, a category of defects, a BPS spectrum, or an
anomaly polynomial as primary data. The bridge accepts all of these as
field-theoretic data, provided the local/global vocabulary is kept explicit.

\section{Symmetries, currents, and anomaly traces}
\label{sec:V02-anomalies}

Supersymmetric theories come with several kinds of symmetries. The most
important distinction is between gauge redundancy and global symmetry. A
gauge symmetry is not an ordinary physical symmetry acting faithfully on the
Hilbert space; it is a redundancy in the description. A global symmetry
acts on operators and states. An R-symmetry is a global symmetry that also
acts on the supercharges. A flavor symmetry commutes with the
supercharges.

This distinction controls anomaly language. A gauge anomaly is a
consistency obstruction for a dynamical gauge symmetry unless it is
cancelled, for example by a local counterterm, by anomaly inflow, or by a
Green--Schwarz-type mechanism when such a mechanism exists in that
dimension. A 't Hooft
anomaly of a global symmetry is protected response data: it must match
across dual descriptions and becomes an obstruction only if one tries to
gauge that global symmetry. An R-symmetry anomaly can be both a diagnostic
and, in special dimensions, part of an extremization principle.

For a four-dimensional chiral fermion in representation \(\rho\) of a
symmetry group, the perturbative cubic anomaly is measured schematically by
\begin{equation}
\label{eq:V02-cubic-anomaly}
 \mathcal A^{abc}
 =
 \operatorname{Tr}_{\text{left fermions}}
 \rho(T^a)\{\rho(T^b),\rho(T^c)\}.
\end{equation}
For a mixed R-flavor-flavor anomaly one similarly traces the R-charge of the
fermion times the two flavor generators:
\begin{equation}
\label{eq:V02-RFF-anomaly}
 \mathcal A_{RFF}
 =
 \operatorname{Tr}_{\text{left fermions}}
 R_{\psi}\, \rho(T^a)\rho(T^b).
\end{equation}
The subscript \(\psi\) is deliberate. In a four-dimensional \(\mathcal N=1\)
chiral multiplet, if the scalar has R-charge \(R_\phi\), then the fermion
has R-charge
\begin{equation}
\label{eq:V02-fermion-R}
 R_\psi = R_\phi - 1.
\end{equation}
Anomaly traces are fermion traces. Forgetting this shift is one of the most
common ways to get the wrong answer.

\subsection*{Gauge cancellation in a vectorlike fixture}

As a minimal four-dimensional fixture, take an SU\((N_c)\) gauge theory
with a chiral multiplet \(Q\) in the fundamental representation and a
chiral multiplet \(\widetilde Q\) in the antifundamental. The left-handed
fermion in \(Q\) contributes the cubic gauge anomaly coefficient
\(A(\mathbf N)\), while the left-handed fermion in \(\widetilde Q\)
contributes \(A(\overline{\mathbf N})=-A(\mathbf N)\). The pair is
therefore vectorlike with respect to the gauge group:
\begin{equation}
\label{eq:V02-vectorlike-cancel}
 A(\mathbf N)+A(\overline{\mathbf N})=0.
\end{equation}
This is not a statement about the flavor anomalies of the same fields. The
flavor symmetry can still have nonzero 't Hooft anomalies, and those
anomalies are meaningful infrared data. The same fermions participate in
both computations; what changes is whether the traced generators belong to
a gauge redundancy or to a global symmetry.

This fixture also explains why the word ``matter'' is too imprecise for an
anomaly computation. Scalars contribute to the multiplet and to the vacuum
coordinates, but perturbative chiral anomalies are fermion traces. Gauge
representations, chirality, and R-charge shifts have to be read at the
fermion level.

\begin{keybox}{Common misconception: gauge anomalies versus 't Hooft anomalies}
A gauge anomaly and a 't Hooft anomaly of a global symmetry play different
roles. Gauge anomalies must be removed for the description to be
consistent. 't Hooft anomalies are protected response data; they are
allowed because they are observable constraints, and they must match across
dual descriptions.
\end{keybox}

\subsection*{A minimal anomaly-matching fixture}

Suppose two descriptions claim to flow to the same infrared theory. If they
have the same non-anomalous global symmetry \(G_F\), then the 't Hooft
anomalies for \(G_F\) and for any preserved R-symmetry must match. In a
four-dimensional \(\mathcal N=1\) frame the two basic fermion anomaly traces are
\begin{equation}
\label{eq:V02-a-anomaly-traces}
 \operatorname{Tr} R
 =
 \sum_{\psi} R_\psi,
 \qquad
 \operatorname{Tr} R^3
 =
 \sum_{\psi} R_\psi^3.
\end{equation}
The trial central charge built from these traces is
\begin{equation}
\label{eq:V02-a-trial}
 a_{\text{trial}}(R)
 =
 \frac{3}{32}
 \left(3\operatorname{Tr} R^3-\operatorname{Tr} R\right).
\end{equation}
This equation is here only as vocabulary. The actual a-maximization
principle belongs to the $4d$ section, where four-dimensional dynamics
is the subject. The bridge records the ingredients: a candidate
R-symmetry, a list of fermions, and anomaly traces. One caution is
worth stating now. Matching anomalies is a necessary consistency check
on a proposed duality, not a proof of it. Two theories can share all
their 't Hooft anomalies and still be different; the dimension sections
supply the further evidence a duality claim requires.

The same caution applies in other dimensions. Two-dimensional gravitational
and flavor anomalies are naturally chiral and lead into c-extremization in
\((0,2)\) theories. Three-dimensional parity anomalies and contact terms
are tied to Chern--Simons levels. Six-dimensional theories organize
anomalies in an eight-form polynomial, often with tensor-branch
Green--Schwarz cancellation. These are not variants of
Equation~\eqref{eq:V02-cubic-anomaly} with the dimension label changed.
They are distinct technologies whose shared concept is anomaly as protected
local or global data.

\begin{keybox}{Common misconception: the R-symmetry is not ``whatever U(1) is convenient''}
R-symmetry is not ``whatever U(1) is convenient''. It is the symmetry that
acts on the supercharges, and in interacting theories the exact infrared
R-symmetry can mix with flavor symmetries. The bridge names this fact; the
dimension section performs the relevant extremization or replacement
principle.
\end{keybox}

\subsection*{Background fields as a unifying language}

A clean way to remember the gauge/global distinction is to couple global
symmetries to nondynamical background fields. If the background field is
made dynamical, the symmetry has become a gauge redundancy and the anomaly
must be cancelled. If the field remains a background, the anomaly is a
well-defined response of the quantum theory. This response can be encoded
as a phase of the partition function, a contact term, or an anomaly
polynomial, depending on dimension.

This language lets the same bridge sentence cover several later cases
without pretending they are identical. In $2d$, a chiral imbalance of
left-moving and right-moving fermions produces gravitational and flavor
anomalies. In $3d$, integrating out massive fermions can shift
Chern--Simons contact terms, and the parity anomaly constrains allowed
backgrounds and levels. In $4d$, cubic and mixed 't Hooft anomalies constrain
dual descriptions and feed R-symmetry tests. In $6d$, the anomaly polynomial
and Green--Schwarz terms are central data. The bridge statement is only
this: anomalies are local or global obstructions/responses attached to
background or dynamical symmetry fields, and the dimension section tells
which mathematical object represents them.

\section{Deformations and parameters}
\label{sec:V02-deformations}

The same local grammar also organizes deformations of a theory. A
\emph{relevant} deformation changes the infrared theory, a \emph{marginal}
deformation has classical dimension equal to the spacetime dimension, and
an \emph{exactly marginal} deformation remains marginal after quantum
effects and quotienting by redundancies. A \emph{mass parameter} is a
background value for a flavor multiplet. A \emph{coupling} is a coefficient
of a local term. An \emph{FI parameter}, when available, is a parameter in
a moment-map equation.

The bridge uses this vocabulary but does not assert that the same parameter
exists in every dimension. This is especially important for supersymmetric
field theories in dimensions above four, where Yang--Mills couplings are
dimensionful. With the standard normalization \((1/g^2)\int F^2\), the
coupling has \([g^2]=4-d\), while the inverse coefficient has
\([1/g^2]=d-4\). Thus the existence of a UV fixed point is a dynamical
question, not a consequence of the local term alone. It is also important
in two and three dimensions, where real masses, FI terms, twisted masses,
Chern--Simons levels, and monopole operators have their own protected
roles.

\begin{table}[ht]
\centering
\small
\begin{tabular}{@{}p{0.20\linewidth}p{0.31\linewidth}p{0.39\linewidth}@{}}
\toprule
\textbf{Word} & \textbf{Bridge meaning} & \textbf{Where details live} \\
\midrule
Superpotential deformation & Holomorphic local deformation of chiral data & $2d$, $3d$, and $4d$ sections in their own notation \\
\midrule
Mass parameter & Background value associated to a flavor symmetry & Dimension sections, especially $3d$ and $4d$ examples \\
\midrule
FI parameter & Moment-map parameter when the algebra allows it & $2d$ GLSMs and $3d$ gauge dynamics \\
\midrule
Exactly marginal coupling & Protected marginal deformation modulo redundancies & $4d$ conformal-manifold discussion \\
\midrule
Topological level & Quantized coefficient such as a Chern--Simons level & $3d$ section \\
\midrule
Tensor-branch parameter & Scalar expectation value in tensor multiplet data & $6d$ section \\
\bottomrule
\end{tabular}
\caption{Deformation and parameter vocabulary, and the sections that own the technical details.}
\label{tab:V02-deformations}
\end{table}

\begin{keybox}{Common misconception: a deformation is not specified by its name alone}
A deformation is not specified by its name alone. One must say which
operator is being added, which symmetry protects or forbids it, what its
dimension is, and whether the parameter is continuous, quantized, or
background-field data.
\end{keybox}

\subsection*{Relevant, marginal, and protected are different adjectives}

The bridge keeps three adjectives separate. A deformation is
\emph{relevant}, \emph{marginal}, or \emph{irrelevant} according to the
scaling dimension of the operator at the fixed point under discussion. It
is \emph{supersymmetric} if it can be written as an allowed supersymmetric
local term. It is \emph{protected} if supersymmetry and other symmetries
constrain its quantum behavior strongly enough that a finite computation or
index captures it.

These adjectives often correlate, but they are not synonyms. A
superpotential mass term can be relevant and supersymmetric. A gauge
kinetic term may be marginal in four dimensions but dimensionful in five.
A Chern--Simons level in three dimensions is supersymmetric and quantized,
but it is not a continuous exactly marginal parameter. A tensor-branch
coefficient in six dimensions is not a four-dimensional gauge coupling in
disguise.

The operational rule is to ask four questions:
\begin{enumerate}
\item What operator or background multiplet is being turned on?
\item What symmetries, including R-symmetry, allow the term?
\item What is the mass dimension of the parameter at the point being
studied?
\item Is the parameter continuous, discrete, quantized, or a background
choice?
\end{enumerate}
This rule prevents the bridge from importing a four-dimensional intuition
into every dimension.

\subsection*{Masses as background multiplet data}

Mass parameters are a useful place to see the local/global vocabulary at
work. A flavor symmetry current lives in a current multiplet. Coupling
that current multiplet to a nondynamical background vector multiplet and
giving the background scalar a value produces a mass parameter, when the
dimension and supersymmetry allow such a coupling. In a weak
four-dimensional \(\mathcal N=1\) frame, a holomorphic mass can appear in a
superpotential term
\begin{equation}
\label{eq:V02-holomorphic-mass}
 W_{\text{mass}} = m\, Q\widetilde Q.
\end{equation}
In lower dimensions, real masses and twisted masses are often more natural.
The bridge does not equate these. It records the shared logic: a mass is
not merely a number in a Lagrangian; it is usually tied to a flavor
symmetry, a background multiplet, and a choice of preserved supersymmetry.

This statement has dimension-specific realizations. Holomorphic masses in
a four-dimensional \(\mathcal N=1\) frame are naturally superpotential or
chiral-spurion data, whereas real masses in three dimensions and twisted
masses in two dimensions are often background-vector-multiplet scalar data.
The bridge records the shared flavor-symmetry organization without
identifying these mechanisms.

This viewpoint is especially useful when comparing two descriptions. If a
duality maps a flavor symmetry \(G_F\) on one side to the same symmetry on
the other, its mass deformations should map as background values for
\(G_F\). If an FI parameter is present, it should map according to the
topological or gauge symmetry structure of that dimension. The details are
dimension-specific, but the bridge-level sentence is stable: deformations
are organized by the multiplets and symmetries to which their parameters
belong.

\section{The navigation table for the dimension sections}
\label{sec:V02-navigation}

This is the most important boundary of the bridge. The bridge
defines common words, then gets out of the way. Table~\ref{tab:V02-navigation} records
where the real technology belongs.

\begin{table}[ht]
\centering
\small
\begin{tabular}{@{}p{0.10\linewidth}p{0.35\linewidth}p{0.43\linewidth}@{}}
\toprule
\textbf{Dim.} & \textbf{Owns} & \textbf{Bridge only says} \\
\midrule
$2d$ & \((2,2)\) and \((0,2)\) multiplets, GLSMs, \(E/J\) data, elliptic genera, c-extremization
& Chiral, vector, and Fermi multiplets are local representations; \(E\) and \(J\) are dimension-specific local data. \\
\midrule
$3d$ & Chern--Simons terms, parity anomaly, monopole operators, mirror symmetry, F-maximization
& Topological terms and disorder operators are part of the local/protected grammar in $3d$. \\
\midrule
$4d$ & \(\mathcal N=1\) chiral/vector grammar, Seiberg duality, a-maximization, \(\mathcal N=2\) Seiberg--Witten data, class S
& F-terms, D-terms, chiral rings, and anomaly traces have a familiar compact notation. \\
\midrule
$5d$ & Coulomb-branch prepotentials, instanton particles/operators, enhanced flavor symmetry, $5d$ SCFT limits
& Couplings and Coulomb coordinates are dimensionful and can signal UV fixed-point data. \\
\midrule
$6d$ & tensor branches, anomaly polynomials, Green--Schwarz cancellation, BPS strings
& Tensor multiplets and anomaly data are essential; the bridge does not compute the eight-form. \\
\midrule
$7d$--$10d$ & high-dimensional supersymmetric field theories used as boundary/interface or reduction inputs
& The common words still apply, but these sections are compact and field-theory-first. \\
\bottomrule
\end{tabular}
\caption{The navigation table: which dimension section owns which technology, and what the bridge alone fixes.}
\label{tab:V02-navigation}
\end{table}

The table also fixes a negative rule. These notes will not turn into a
standalone reference on superspace, algebraic geometry of moduli spaces,
dualities, or anomaly-polynomial technology. Those topics appear when they
are needed to understand a dimension's supersymmetric field theories.

The positive rule is just as important. The dimension sections are allowed
to be technically serious because the bridge has removed the repeated
setup. When the $2d$ section introduces \((0,2)\) \(E\)- and \(J\)-data, it
does not have to re-explain what a multiplet is or why F-type equations
enter a vacuum problem. When the $3d$ section introduces monopole operators,
it can say immediately how these operators enlarge the protected coordinate
ring. When the $4d$ section discusses anomaly matching or exact R-symmetry,
it can rely on the fermion-trace rule fixed here. When the $5d$ and $6d$
sections use prepotentials, instanton particles, tensor branches, and
anomaly polynomials, they can point back to the bridge for the local/global
vocabulary and then proceed with the actual higher-dimensional dynamics.

This is also the reason for the title ``supersymmetric field theories''
rather than a gauge-centered title. Gauge theories provide many of the most
explicit frames, but they are not the whole subject. The
dimension sections must be able to discuss sigma models, Landau--Ginzburg
models, non-Lagrangian SCFTs, boundary/interface systems, and compact
higher-dimensional field theories without pretending that every example is
best introduced by a gauge field and charged matter.

\begin{keybox}{Common misconception: non-Lagrangian does not mean non-field-theoretic}
Non-Lagrangian does not mean non-field-theoretic. A theory can be defined
or constrained by symmetry, anomalies, protected rings, defects, and
deformation data even when no weakly coupled elementary Lagrangian is the
right starting point.
\end{keybox}

\subsection*{Near-misses that the bridge forbids}

The bridge is as much about preventing false identifications as about
introducing terms. A four-dimensional F-term and a two-dimensional
\((0,2)\) \(J\)-term are related by family resemblance, but they are not
the same object with a new name. A three-dimensional monopole operator is
a protected disorder operator, not merely a scalar expectation value that
was forgotten in the elementary field list. A five-dimensional instanton
particle is tied to the topological current and UV completion data, not to
the four-dimensional instanton correction story by a change of notation.
The six-dimensional tensor branch has its own anomaly-cancellation
mechanism, not a D-term quotient written in a larger font.

These near-misses are useful because they tell us what the bridge should
not do. It should not flatten all local terms into the four-dimensional
\(\mathcal N=1\) schematic line. It should not call every vacuum family a
Higgs branch just because elementary scalars appear somewhere in the
description. It should not treat every exact R-symmetry problem as
a-maximization. It should not mistake the presence of a gauge-theory frame
for a promise that the theory itself is always best defined by that frame.

The correct bridge move is to translate only the first layer. If a later
section says ``multiplet'', ask which representation of which algebra. If
it says ``local term'', ask which supersymmetric measure or protected
coupling is available in that dimension. If it says ``vacuum'', ask for
equations and equivalences. If it says ``anomaly'', ask whether the
symmetry is dynamical or background and which anomaly object the dimension
uses. Once those questions are answered, the dimension section takes over.

There is a parallel near-miss on the reader's side. Knowing the bridge
vocabulary is not the same as knowing a theorem about every dimension. The
bridge lets the reader recognize the shape of a statement before learning
the technology that proves it. It is a grammar, not a substitute for the
later sections' dynamics.

\section{Theory cards: the bridge format}
\label{sec:V02-theory-cards}

For later use it is helpful to compress the bridge vocabulary into a
standard ``theory card''. This is not a mathematical definition of QFT. It
is a reading format. When a dimension section introduces a theory
\(\mathcal T\), the first pass should identify
\begin{equation}
\label{eq:V02-theory-card}
 \mathcal T:
 \quad
 (d,\mathcal Q,G_R;\ \mathfrak M_{\text{local}};
 \mathcal I_{\text{terms}};
 \mathcal V_{\text{vac}};
 \mathfrak D_{\text{protected}}).
\end{equation}
Here \(d\) is the spacetime dimension, \(\mathcal Q\) the supersymmetry
algebra data from Section~1, \(G_R\) the R-symmetry,
\(\mathfrak M_{\text{local}}\) the multiplet or protected local-data list,
\(\mathcal I_{\text{terms}}\) the allowed local interactions,
\(\mathcal V_{\text{vac}}\) the visible vacuum equations and quotient, and
\(\mathfrak D_{\text{protected}}\) the protected data used for comparison.

\begin{table}[ht]
\centering
\small
\begin{tabular}{@{}p{0.22\linewidth}p{0.30\linewidth}p{0.38\linewidth}@{}}
\toprule
\textbf{Card entry} & \textbf{Question} & \textbf{Examples of acceptable answers} \\
\midrule
Algebra & Which SUSY algebra and R-symmetry? & $4d$ \(\mathcal N=1\) with U(1)\(_R\); $2d$ \((0,2)\) with U(1)\(_R\); $6d$ \((1,0)\) with SU(2)\(_R\). \\
\midrule
Local data & What carries the fields or protected observables? & Chiral and vector multiplets; Fermi multiplets; tensor multiplets; current multiplets; protected operator data. \\
\midrule
Local terms & What interactions are allowed here? & Superpotential, D-term, FI term, Chern--Simons level, prepotential term, Green--Schwarz coupling, each only in its proper dimension. \\
\midrule
Vacua & What equations and equivalences are visible? & F-flat equations, moment-map equations, complex quotient, quantum Coulomb coordinates, branch decomposition. \\
\midrule
Symmetries & Which symmetries are redundancies or physical? & Gauge group, flavor group, topological symmetry, R-symmetry, higher-form symmetry when the section introduces it. \\
\midrule
Protected data & What must match or be preserved? & Anomaly traces, protected rings, BPS spectra, indices, conformal-manifold data, defect or boundary data. \\
\bottomrule
\end{tabular}
\caption{The theory-card reading format: the question each card entry answers.}
\label{tab:V02-card}
\end{table}

The card format keeps examples honest. If a theory is Lagrangian, the
local-data entry may be a literal field list. If a theory is
non-Lagrangian, the local-data entry may instead be a protected operator
spectrum, an anomaly package, or a deformation pattern. If a theory has a
gauge-theory frame in one regime, the card records that frame without
declaring it to be the unique definition of the theory. If two descriptions
are dual, their cards need not have identical local-data entries, but their
protected data and deformation responses must be compatible.

\subsection*{A sample card}

Return to the simple U(1) theory with \(Q\) and \(\widetilde Q\) of charges
\(+1\) and \(-1\), no superpotential, and \(\zeta=0\). Its bridge-level card
is
\begin{equation}
\label{eq:V02-sample-card}
 \begin{array}{ll}
 d,\mathcal Q: & \text{chosen from the dimension section},\\
 \mathfrak M_{\text{local}}: & Q,\widetilde Q \text{ plus a U(1) vector multiplet},\\
 \mathcal I_{\text{terms}}: & \text{kinetic terms, no } W, \zeta=0,\\
 \mathcal V_{\text{vac}}: & |Q|^2-|\widetilde Q|^2=0 \text{ modulo U(1)},\\
 \text{invariant coordinate}: & M=Q\widetilde Q.
 \end{array}
\end{equation}
This card does not determine all dynamics. It does determine what the
bridge is allowed to say: the vacuum quotient has a coordinate \(M\), the
branch label depends on the dimension and frame, and any anomaly statement
must specify the fermions and symmetries being traced.

The same card can be refined in several directions. A two-dimensional
version may add \(E/J\) data and GLSM phases. A three-dimensional version
may add real masses, FI parameters, and monopole operators. A
four-dimensional version may add anomaly matching and an exact R-symmetry
problem. The bridge does not perform these refinements, but it prevents
the reader from losing track of which entry of the card is being modified.

\section{Worked dictionary: one frame, many later uses}
\label{sec:V02-worked-dictionary}

We close with a small dictionary calculation that will be reused in several
dimensions. Take chiral fields \(\Phi_i\), a holomorphic superpotential
\(W(\Phi_i)\), and a gauge group \(G\). In a Lagrangian frame, the local
supersymmetric data determine the classical vacuum equations
\begin{equation}
\label{eq:V02-master-vacuum}
 \frac{\partial W}{\partial \Phi_i}=0,
 \qquad
 \mu(\phi)=\zeta_{\text{central}},
 \qquad
 \text{modulo } G.
\end{equation}
The first equation is F-flatness, the second is D-flatness or moment-map
flatness, with nonzero levels only in central or abelian directions; the
last phrase is gauge equivalence. The protected ring in this frame is
represented schematically by
\begin{equation}
\label{eq:V02-master-ring}
 \mathcal R
 =
 \frac{\mathbb C[\text{gauge-invariant chiral words}]}
 {\left(\partial W=0\right)}.
\end{equation}
Equation~\eqref{eq:V02-master-ring} is not a universal theorem. Quantum
effects can deform it, monopole or instanton operators can add coordinates,
and non-Lagrangian theories can require a different presentation. Its use
is pedagogical: it identifies the ingredients that later sections will keep
modifying.

The same frame has candidate global symmetries \(G_F\), candidate
R-symmetries, and possible anomalies. To compute an anomaly one lists the
fermions and traces their charges, not the scalar charges directly. To
test a duality one compares protected data: symmetries, anomaly traces,
operator maps, branch dimensions, and deformations. To study a family of
vacua one decides which coordinates remain light and which equivalences have
already been divided out.

\subsection*{A parsing algorithm}

When later sections present a new supersymmetric field theory, read it in
the following order. First identify the algebra: dimension, number of
supercharges, and R-symmetry. Second identify the visible multiplets or,
if the theory is non-Lagrangian, the protected data standing in for
multiplets. Third list the local supersymmetric terms. Fourth derive the
vacuum equations or protected-ring relations that are visible in the chosen
frame. Fifth separate gauge redundancies from global symmetries. Sixth
record anomalies and other protected quantities that have to survive along
renormalization-group flow.

This order is not bureaucratic. It prevents common mistakes. If one
starts from a branch name before identifying the multiplets, one can call a
quantum Coulomb coordinate a Higgs coordinate by accident. If one starts
from scalar R-charges before shifting to fermion R-charges, the anomaly
trace is wrong. If one writes a quotient before stating the stability
condition or FI chamber, one may collapse the wrong locus. If one treats a
't Hooft anomaly of a global symmetry like a gauge anomaly, one may
incorrectly discard physical data.

Table~\ref{tab:V02-card} is the static card; Table~\ref{tab:V02-parsing} is the
entry checklist. The vocabulary overlaps because the checklist acts on the
card entries, but its job is procedural: it says what has to be fixed before a
dimension-specific tool is allowed to take over.

\begin{table}[ht]
\centering
\small
\begin{tabular}{@{}p{0.20\linewidth}p{0.34\linewidth}p{0.36\linewidth}@{}}
\toprule
\textbf{Checkpoint} & \textbf{Action before moving on} & \textbf{Mistake it blocks} \\
\midrule
Algebra fixed & Name the dimension, supercharges, and R-symmetry from Section~1. & Treating the same multiplet word as dimension-independent. \\
\midrule
Local frame chosen & Say whether the data are elementary fields, protected data, or both. & Importing full superspace or a gauge-theory frame when not given. \\
\midrule
Equations and quotient & Write the visible equations together with the quotient or stability condition. & Forgetting gauge equivalence or complexified stability. \\
\midrule
Protected coordinates & State whether ring, index, BPS, monopole, or instanton data add coordinates. & Assuming all coordinates are elementary fields. \\
\midrule
Symmetry role separated & Mark each symmetry as gauge, flavor, topological, or R before tracing anomalies. & Cancelling a 't Hooft anomaly of a global symmetry or matching a gauge redundancy. \\
\midrule
Next step chosen & Move to the dimension section only after the previous entries are fixed. & Running extremization, monopole, or tensor-branch machinery too early. \\
\bottomrule
\end{tabular}
\caption{The checklist for entering a dimension section. It is the operational use of Table~\ref{tab:V02-card}, not a second theory-card schema.}
\label{tab:V02-parsing}
\end{table}

This is all the bridge needs to do. The reader fills the card, uses the
entry checklist to avoid the standard false starts, and then lets the
dimension section supply the real technology.
The answer will look different in $2d$, $3d$, $4d$, $5d$, $6d$, and $7d$--$10d$. The
grammar is common; the physics is dimensional.

\section*{Exit checklist}
\addcontentsline{toc}{subsection}{Exit checklist}
\markboth{Exit checklist}{Exit checklist}

The section has done its job if the reader can perform the following
translation without importing dimension-specific machinery too early.
\begin{center}
\small
\begin{tabular}{@{}p{0.39\linewidth}p{0.49\linewidth}@{}}
\toprule
\textbf{Prompt} & \textbf{Expected translation} \\
\midrule
``Write the F-flat equations.'' & Differentiate the holomorphic local term in the appropriate multiplet language. \\
\midrule
``Take the vacuum quotient.'' & Solve the equations and divide by gauge equivalence, or describe the invariant coordinate ring. \\
\midrule
``Name the branch.'' & State which local coordinates define the branch in this frame; do not assume the label is duality-invariant. \\
\midrule
``Check the anomaly.'' & Trace over fermions, distinguish gauge consistency from global protected data, and use the dimension's anomaly technology. \\
\midrule
``Route the topic.'' & Put \(E/J\), GLSM, Chern--Simons, monopoles, extremization, Seiberg--Witten, prepotentials, tensor branches, and anomaly polynomials in their dimension sections. \\
\bottomrule
\end{tabular}
\end{center}

\section*{Sources and notes}
\addcontentsline{toc}{subsection}{Sources and notes}
\markboth{Sources and notes}{Sources and notes}

{\small

\noindent\textsf{\textcolor{RoyalBlue}{Sources and notes.}}\enspace
This is the compact bridge section, the common-grammar layer between the algebraic spine
(Section~1) and the dimension sections.

\medskip\noindent\textsf{\textcolor{RoyalBlue}{\textbf{\S\ref{sec:V02-multiplets}\enspace Multiplets.}}}\enspace
A multiplet is a representation of the super-Poincar\'e algebra; a multiplet name is dimension- and
algebra-dependent, not a universal component formula (Table~\ref{tab:V02-multiplets}).

\medskip\noindent\textsf{\textcolor{RoyalBlue}{\textbf{\S\ref{sec:V02-local-terms}\enspace F-terms, D-terms, and vacua.}}}\enspace
The F-flat equation $\partial W/\partial\phi=0$, the D-flat / moment-map equation
$\mu(\phi)=\zeta_{\text{central}}$, and the gauge quotient, in the four-dimensional $\mathcal{N}=1$
notation used as a compact example, with the $U(1)$ meson $M=Q\widetilde Q$ and the
$\mathbb{C}^\ast$-quotient $\mathbb{P}^1$. (\textcite{Wess:1992cp} the superspace grammar).

\medskip\noindent\textsf{\textcolor{RoyalBlue}{\textbf{\S\ref{sec:V02-branches}\enspace Branches and protected rings.}}}\enspace
A branch label is frame-dependent shorthand (Table~\ref{tab:V02-branches}); a protected ring is
generators modulo F-flat relations, counted by a graded Hilbert series at the dictionary level only.
(\textcite{Benvenuti:2010pq} the Hilbert-series language). 

\medskip\noindent\textsf{\textcolor{RoyalBlue}{\textbf{\S\ref{sec:V02-anomalies}\enspace Symmetries, currents, and anomaly traces.}}}\enspace
The gauge / 't~Hooft distinction; anomaly traces are fermion traces with the
$R_\psi=R_\phi-1$ shift; the vectorlike cancellation $A(\mathbf N)+A(\overline{\mathbf N})=0$; the
trace data $\operatorname{Tr}R,\operatorname{Tr}R^3$ and $a_{\text{trial}}$ named as vocabulary only.
(\textcite{Intriligator:2003jj} the $a$-maximization ingredients;
\textcite{Seiberg:1994pq} 't~Hooft matching across duals). 

\medskip\noindent\textsf{\textcolor{RoyalBlue}{\textbf{\S\ref{sec:V02-deformations}\enspace Deformations and parameters.}}}\enspace
Relevant / marginal / exactly-marginal / mass / coupling / FI vocabulary (Table~\ref{tab:V02-deformations}),
with the dimensionful-coupling bookkeeping $[g^2]=4-d$ flagging that a UV fixed point is dynamical.
. The deformation vocabulary is sanity-checked for bridge ownership versus later dimension-specific use.

\medskip\noindent\textsf{\textcolor{RoyalBlue}{\textbf{\S\S\ref{sec:V02-navigation}--\ref{sec:V02-worked-dictionary}\enspace Navigation, theory cards, and the parsing algorithm.}}}\enspace
The navigation table (Table~\ref{tab:V02-navigation}) assigning $E/J$+GLSM+$c$-extremization to $2d$,
Chern--Simons+monopoles+$F$-maximization to $3d$, $a$-maximization+Seiberg--Witten+class~S to $4d$,
prepotentials+instanton operators to $5d$, tensor branches+anomaly polynomials to $6d$; the
theory-card format (Table~\ref{tab:V02-card}) and the parsing algorithm (Table~\ref{tab:V02-parsing}).
(\textcite{Gadde:2013wq} a further pointer for the $3d$ protected-data vocabulary). The theory card
\eqref{eq:V02-theory-card} is a reading format, not a new convention.
}

\subsection*{Further reading}
\addcontentsline{toc}{subsection}{Further reading}
The chiral, vector, and current multiplets and the superspace formalism are developed in
\textcite{Wess:1992cp}; the founding superfield papers are \textcite{Salam:1974jj,Ferrara:1974ac}. The
supercurrent and the supersymmetry multiplets of conserved currents are treated in
\textcite{Komargodski:2010rb}, and rigid supersymmetry on curved backgrounds in
\textcite{Festuccia:2011ws,Dumitrescu:2016ltq}. Non-renormalization from holomorphy is explained in
\textcite{Seiberg:1993vc}, the four-dimensional $a$-theorem constraining renormalization-group flows in
\textcite{Komargodski:2011vj}, and supersymmetry breaking as a general phenomenon in
\textcite{Luty:2005sn}.

Modern extensions of the symmetry-and-anomaly dictionary to higher-form and higher-group symmetries
are reviewed pedagogically in \textcite{Gomes:2023ahz,Bhardwaj:2023kri}.

\section*{References}
\addcontentsline{toc}{subsection}{References}
\markboth{References}{References}
\printbibliography[heading=none]
\end{refsection}
\begin{refsection}\chapter{\texorpdfstring{$4d$ $\mathcal{N}=1$}{4d N=1} supersymmetric field theories}
\label{ch:V03}

\noindent\textbf{Guide to this section.}\enspace
Sections~1 and~2 fixed the algebra and the common words. This section teaches the
four-supercharge holomorphic workhorse, $4d\ \mathcal{N}=1$ field theory, and the extremization
principle the later sections reuse. It is the anchor of the dimension sections: the $2d$ and $3d$
sections point back here for R-symmetries, chiral rings, 't~Hooft anomalies, and extremization
logic. It is a foundations section, but a working one. It states the
field-theory facts and then runs them as computations: the dynamical scale and its threshold
matching, the chiral ring, gaugino condensation and the index, the exact R-symmetry, the 't~Hooft
and Konishi anomalies, $a$-maximization with worked extrema, the full SQCD phase map with the
Affleck--Dine--Seiberg runaway, the quantum-deformed and s-confining constraints, Seiberg duality
checked trace by trace, and dynamical supersymmetry breaking. Only the deep theorems (the
$a$-theorem, the proof of Seiberg duality, the superconformal-index integral) are cited but not
proved here. By the end you can take a $4d\ \mathcal{N}=1$ theory, given by a Lagrangian or a quiver,
identify its multiplets and chiral ring, assign a trial R-symmetry, compute the 't~Hooft anomalies,
run $a$-maximization to the superconformal R, write the Seiberg dual, and check the match by hand.

\begin{keybox}{What this section delivers}
The holomorphic data of a $4d\ \mathcal{N}=1$ theory, the superpotential and the gauge coupling, and
the dynamically generated scale $\Lambda$ with its threshold matching
(\S\ref{sec:V03-multiplets}); the chiral ring and non-renormalization (\S\ref{sec:V03-chiralring});
pure super Yang--Mills, gaugino condensation, and the Witten index (\S\ref{sec:V03-sym}); the
$U(1)_R$ symmetry, the NSVZ beta function, and the superconformal R (\S\ref{sec:V03-rsymmetry});
't~Hooft anomaly matching and the Konishi anomaly as tools, with the SQCD traces run explicitly
(\S\ref{sec:V03-anomaly}); $a$-maximization, the prototype of the extremization through-line, with
SQCD, adjoint SQCD, and the dP$_0$ quiver worked (\S\ref{sec:V03-amax}); SQCD across the flavor
window, the Affleck--Dine--Seiberg runaway, the quantum-deformed moduli space, s-confinement, and
Seiberg duality with four explicit matching traces (\S\ref{sec:V03-seiberg}); and the Witten-index
criterion for dynamical supersymmetry breaking with a worked rank-condition fixture
(\S\ref{sec:V03-susybreaking}).
\end{keybox}

\section{Multiplets, holomorphy, and the dynamical scale}
\label{sec:V03-multiplets}

Before any holomorphic technology enters, fix the row in the ladder. A $4d\ \mathcal{N}=1$ theory has
four real supercharges: the same $4Q$ row that appears as $3d\ \mathcal{N}=2$ after circle reduction
and as $2d\ \mathcal{N}=(2,2)$ after torus reduction. Its four-dimensional R-symmetry is a single
$U(1)_R$. This section is therefore the prototype of the low-supercharge language, not because every
other dimension copies four dimensions, but because the other $4Q$ sections keep translating their
local terms, anomalies, and extremization principles back to this row.

A $4d\ \mathcal{N}=1$ theory is built from two multiplets fixed in Section~2. A chiral multiplet
$\Phi$ carries a complex scalar, a Weyl fermion, and an auxiliary field $F$. A vector multiplet $V$
carries a gauge field, a gaugino, and an auxiliary field $D$. The renormalizable dynamics is fixed by
holomorphic data: the superpotential $W$, a holomorphic function of the chiral superfields, and the
holomorphic gauge coupling
\begin{equation}
\label{eq:V03-tau}
 \tau \;=\; \frac{\theta}{2\pi} \;+\; \frac{4\pi i}{g^2}.
\end{equation}
The classical supersymmetric vacua solve the F-flat equations $F_i = \partial W/\partial\phi_i = 0$
and the D-flat (moment-map) equation $D=0$ modulo gauge equivalence, exactly as in Section~2. We do
not re-derive that grammar here. We use it. Throughout, the gauge group is $SU(N_c)$ unless a quiver
is named, in which case it is written as $\prod_i U(N_i)$.

Holomorphy is the organizing principle of the whole section. That the superpotential and the gauge
coupling are holomorphic constrains the quantum theory far beyond perturbation theory. We use it three
times in this section, each time as a computation: the dynamical scale, the threshold matching, and the
top of the conformal window.

\medskip\noindent\textbf{The one-loop coefficient and the dynamical scale.}\enspace
A non-abelian gauge theory is not scale invariant at one loop. Its running coupling is set by the
one-loop coefficient
\begin{equation}
\label{eq:V03-b0}
 b_0 \;=\; 3\,T(\mathrm{adj}) \;-\; \sum_i T(r_i) \;=\; 3\,N_c \;-\; \sum_i T(r_i),
\end{equation}
the gauge contribution $3T(\mathrm{adj}) = 3N_c$ for $SU(N_c)$ (the adjoint Dynkin index is
$T(\mathrm{adj}) = N_c$ in the normalization $T(\square)=\tfrac12$) minus the Dynkin indices $T(r_i)$
of the matter chiral multiplets. For $SU(N_c)$ with $N_f$ fundamental \emph{flavors} the matter is
$N_f$ quarks $Q$ in the fundamental plus $N_f$ antiquarks $\widetilde Q$ in the antifundamental, so
$2N_f$ fundamental chiral fields, each of index $T(\square)=\tfrac12$. The matter contributes
$2N_f\cdot\tfrac12 = N_f$, and
\begin{equation}
\label{eq:V03-b0sqcd}
 b_0 \;=\; 3N_c - N_f.
\end{equation}
The count is $3N_c - N_f$, not $3N_c - 2N_f$: it is the sum of Dynkin \emph{indices} (a half each),
not a count of fields. Getting this wrong by a factor of two on the matter is the most common
arithmetic slip in the whole subject, and it moves the top of the conformal window. A few worked
values fix the formula, each \eqref{eq:V03-b0} read with the right Dynkin indices,
\begin{align}
\label{eq:V03-b0values}
 b_0\big|_{SU(3),\,N_f=0} &= 9, &
 b_0\big|_{SU(3),\,N_f=6} &= 9-6 = 3, \notag\\[2pt]
 b_0\big|_{SU(3),\,N_f=9} &= 0, &
 b_0\big|_{\text{adj},\,N_f=0} &= 3N_c - N_c = 2N_c, \notag\\[2pt]
 b_0\big|_{\mathcal{N}=4} &= 3N_c - 3N_c = 0. &&
\end{align}
The running stops at $N_f=9$, and it stops for $\mathcal{N}=4$ (three adjoint chirals), where no
holomorphic one-loop scale is generated. The stronger statement that $\mathcal{N}=4$ super Yang--Mills
is exactly conformal follows from the extended supersymmetry and is not derived here.

Dimensional transmutation now trades the dimensionless coupling for a dimensionful,
renormalization-group-invariant, holomorphic scale. The one-loop running of the holomorphic coupling
is exact (the superpotential and the holomorphic $\tau$ run only at one loop), and it follows from the
definition of $b_0$:
\begin{equation}
\label{eq:V03-taurun}
 \tau(\mu) \;=\; \tau(\mu_0) \;+\; \frac{b_0}{2\pi i}\,\ln\!\frac{\mu_0}{\mu},
\end{equation}
whose imaginary part is the familiar running of the physical coupling,
\begin{equation}
\label{eq:V03-grun}
 \frac{8\pi^2}{g^2(\mu)} \;=\; b_0\,\ln\!\frac{\mu}{\Lambda}.
\end{equation}
We package this running into a single dimensionful object, written in the form that avoids a branch
ambiguity,
\begin{equation}
\label{eq:V03-lambda}
 \Lambda^{b_0} \;=\; \mu^{\,b_0}\, e^{\,2\pi i\,\tau(\mu)},
\end{equation}
with $\mu$ the sliding scale. (The equivalent $\Lambda = \mu\, e^{2\pi i\tau/b_0}$ is the same object
once a $b_0$-th root is chosen; we keep the power form \eqref{eq:V03-lambda} in the main text because
it is single-valued.) The combination is built so that the explicit $\mu$ dependence cancels the
running of $\tau$. Differentiating the logarithm of \eqref{eq:V03-lambda},
\begin{equation}
\label{eq:V03-lambdaRG}
 b_0\,\frac{d\ln\Lambda}{d\ln\mu}
 \;=\; b_0 \;+\; 2\pi i\,\frac{d\tau}{d\ln\mu}
 \;=\; b_0 \;+\; 2\pi i\cdot\left(-\frac{b_0}{2\pi i}\right)
 \;=\; 0,
\end{equation}
where the explicit $\mu^{b_0}$ contributes $+b_0$ and the running \eqref{eq:V03-taurun}, which has
$d\tau/d\ln\mu=-b_0/2\pi i$, contributes $-b_0$. So $d\ln\Lambda/d\ln\mu =
0$: along the one-loop trajectory $\Lambda$ does not move. This $\Lambda$ is the
renormalization-group-invariant strong-coupling scale, the variable in which gaugino condensation, the
Affleck--Dine--Seiberg superpotential, and Seiberg duality are all written. It is genuinely
nonperturbative data: $\Lambda \sim \mu\, e^{-8\pi^2/(b_0 g^2)}$ is exponentially small in $1/g^2$ and
invisible to any finite order of perturbation theory.

At the upper edge $N_f = 3N_c$ the coefficient $b_0$ vanishes: the one-loop running stops, no scale
is generated, and \eqref{eq:V03-lambda} degenerates ($\Lambda^0 = 1$ carries no scale). That edge is
the top of the conformal window of \S\ref{sec:V03-seiberg}. One must not read a finite dynamical
scale into a $b_0 = 0$ theory; there simply is none, and the would-be $\Lambda$ is ill-defined.

\medskip\noindent\textbf{Threshold matching: integrating out one flavor.}\enspace
The holomorphic scale obeys a simple matching rule when a heavy flavor is removed, and that rule is a
clean first computation. Take $SU(N_c)$ with $N_f$ flavors and give the last flavor a mass by adding
$\Delta W = m\, Q_{N_f}\widetilde Q_{N_f}$ to the superpotential. Below the mass $m$ the heavy quarks
decouple and the theory is $SU(N_c)$ with $N_f-1$ light flavors, governed by its own scale
$\Lambda_{N_f-1}$. Holomorphy of the gauge coupling, with the threshold at $\mu=m$, demands that the
two holomorphic scales agree across the threshold:
\begin{equation}
\label{eq:V03-threshold}
 \Lambda_{N_f-1}^{\,3N_c-(N_f-1)} \;=\; m\,\Lambda_{N_f}^{\,3N_c-N_f}.
\end{equation}
Read the exponents,
\begin{equation}
\label{eq:V03-threshold-exps}
 b_0^{\mathrm{lo}} = 3N_c - (N_f-1) = b_0^{\mathrm{hi}} + 1, \qquad
 b_0^{\mathrm{hi}} = 3N_c - N_f,
\end{equation}
so the exponent shifts up by exactly one, and that one is exactly $2T(\square) = 2\cdot\tfrac12 = 1$,
the total Dynkin index of the removed pair $(Q_{N_f},\widetilde Q_{N_f})$. The factor of $m$ on the
right carries the one extra
power of mass dimension that the higher exponent on the left needs, so the equation is dimensionally
homogeneous: $[\Lambda_{N_f-1}^{b_0^{\mathrm{hi}}+1}] = [\Lambda_{N_f-1}]^{b_0^{\mathrm{hi}}+1}$
matches $[m][\Lambda_{N_f}]^{b_0^{\mathrm{hi}}}$. The relation \eqref{eq:V03-threshold} is what makes
$\Lambda$ a bookkeeping device one can carry through the entire flavor window: removing a flavor
raises the exponent by the removed index and inserts a factor of the mass.

\emph{Worked numbers.} Take $N_c=3$. The $N_f=6$ theory has $b_0^{\mathrm{hi}} = 9-6 = 3$, and after
integrating out one flavor the $N_f=5$ theory has $b_0^{\mathrm{lo}} = 9-5 = 4$. The matching
\eqref{eq:V03-threshold} reads
\begin{equation}
\label{eq:V03-threshold-num}
 \Lambda_5^{4} \;=\; m\,\Lambda_6^{3}
 \quad\Longrightarrow\quad
 \Lambda_5 \;=\; \big(m\,\Lambda_6^{3}\big)^{1/4},
\end{equation}
a clean fourth root. As the mass $m$ is raised toward the ultraviolet, $\Lambda_5\to\infty$ relative to
$\Lambda_6$, the statement that removing matter makes a theory more strongly coupled at a fixed scale.
Iterating down to $N_f=0$ (all flavors massive) recovers pure $SU(3)$ super Yang--Mills with $b_0 = 9$,
\begin{equation}
\label{eq:V03-threshold-iter}
 \Lambda_{\mathrm{SYM}}^{9} \;=\; m_1\cdots m_6\,\Lambda_6^{3},
\end{equation}
the matching that will tie the SQCD scale to the gaugino condensate of \S\ref{sec:V03-sym}.

\emph{Sanity check.} If one forgets the threshold and writes $\Lambda_{N_f-1}=\Lambda_{N_f}\equiv\Lambda$,
\eqref{eq:V03-threshold} would read
\begin{equation}
\label{eq:V03-threshold-fail}
 \Lambda^{\,3N_c-(N_f-1)} \;\ne\; m\,\Lambda^{\,3N_c-N_f}
 \quad\text{(the two sides differ by the factor $m\,\Lambda^{-1}$),}
\end{equation}
so the equation fails dimensionally. The mass insertion is not optional. This threshold rule is the $4d\
\mathcal{N}=1$ analog of the scale-matching the $4d\ \mathcal{N}=2$ section performs for its own (different) scale,
and the two should never be conflated: that section's scale runs with an $\mathcal{N}=2$ one-loop
coefficient and is matched to this one only across an $\mathcal{N}=1$ mass deformation, as
\S\ref{sec:V03-sym} explains.

\begin{keybox}{Common misconception: the gauge algebra fixes the theory}
The Lagrangian fixes only the gauge \emph{algebra}. The global form of the gauge \emph{group} is
extra data. The algebra $\mathfrak{su}(2)$ is realized by $SU(2)$ and by $SO(3) = SU(2)/\mathbb{Z}_2$,
and these are \emph{different} quantum theories: they admit different spectra of genuine line
operators (which Wilson and 't~Hooft lines are present), equivalently a different discrete one-form
symmetry and different allowed 't~Hooft fluxes. Two theories with the same algebra and the same local
action can be globally distinct. We flag the distinction and cite the line-operator literature rather
than classifying all global forms here. These notes fix no new convention here. They only warn
that the Lagrangian, by itself, does not fix the theory.
\end{keybox}

\section{The chiral ring and non-renormalization}
\label{sec:V03-chiralring}

The protected backbone of a $4d\ \mathcal{N}=1$ theory is its chiral ring. A chiral operator is one
annihilated by all the $\bar Q$ supercharges. Two chiral operators that differ by a $\bar Q$-exact
piece have the same correlators in any supersymmetric configuration, so the chiral operators are
counted modulo $\bar Q$-exact terms. The product of two chiral operators is again chiral, and the
product is commutative in the ring (the operators can be separated and their ordering is
$\bar Q$-exact). The chiral ring is therefore a commutative ring of protected operators, graded by
the abelian charges. In a Lagrangian frame its relations are the F-term equations: the ring is
gauge-invariant chiral words modulo the ideal generated by $\partial W = 0$, exactly the schematic
$\mathcal R = \mathbb{C}[\text{chiral words}]/(\partial W)$ of Section~2.

\medskip\noindent\textbf{Generators: mesons and baryons.}\enspace
For $SU(N_c)$ SQCD with $N_f$ flavors the gauge-invariant chiral generators are built by contracting
color indices. The meson is the simplest,
\begin{equation}
\label{eq:V03-meson}
 M^i_{\ j} \;=\; Q^i_{\ a}\,\widetilde Q^a_{\ j},
\end{equation}
an $N_f\times N_f$ matrix in flavor indices $i,j$ with the color index $a$ contracted. When the ranks
allow it ($N_f\ge N_c$) one can also antisymmetrize $N_c$ quarks on their color indices with an
epsilon tensor to form baryons,
\begin{equation}
\label{eq:V03-baryon}
 B^{i_1\cdots i_{N_c}} \;=\; \epsilon^{a_1\cdots a_{N_c}}\,Q^{i_1}_{\ a_1}\cdots Q^{i_{N_c}}_{\ a_{N_c}},
\end{equation}
and an analogous $\widetilde B$ from the antiquarks. These generators are graded by the global charges;
with $B(Q)=+1$, $B(\widetilde Q)=-1$ and the R-charges adding from the constituents,
\begin{align}
\label{eq:V03-gengrading}
 &B(M)=0,\quad B(B)=N_c,\quad B(\widetilde B)=-N_c, \notag\\[2pt]
 &R(M)=2R(Q)=2\big(1-\tfrac{N_c}{N_f}\big),\quad R(B)=N_c R(Q).
\end{align}
The classical chiral
ring is the polynomial ring in $M,B,\widetilde B$ modulo the relations forced by the epsilon
identities. These relations are not optional: they encode the geometry of the classical moduli space.
A clean example is $N_f = N_c$, where the single baryon and antibaryon and the $N_c\times N_c$ meson
obey
\begin{equation}
\label{eq:V03-classring}
 \det M - B\widetilde B \;=\; 0,
\end{equation}
the statement that the meson, baryon, and antibaryon are not independent: a single holomorphic
relation cuts the chiral ring down to the moduli space of a rank-$N_c$ matrix with its baryonic
completion. We will see this relation deform quantum-mechanically in \S\ref{sec:V03-seiberg}.

A dimension count makes the chiral ring concrete. Consider $SU(N_c)$ with $N_f\ge N_c$ flavors on its
classical Higgs branch. The scalars are $2N_fN_c$ complex quark components; the D-flat condition and
the gauge quotient remove $\dim SU(N_c) = N_c^2-1$ complex directions (the broken generators), so the
generic Higgs branch has complex dimension
\begin{equation}
\label{eq:V03-higgsdim}
 \dim_{\mathbb{C}}\mathcal{M}_{\mathrm{Higgs}} \;=\; 2N_fN_c - (N_c^2-1),
\end{equation}
and the gauge-invariant generators (the mesons, with the baryons when $N_f\ge N_c$) coordinatize
exactly this space modulo their classical relations. At $N_f=N_c$, for instance, the two counts agree,
\begin{equation}
\label{eq:V03-higgscount}
 \underbrace{2N_c^2 - (N_c^2-1)}_{\text{moduli-space dim}} = N_c^2+1
 = \underbrace{N_c^2 + 2 - 1}_{\text{$M$ + $B,\widetilde B$ $-$ relation}},
\end{equation}
the $N_f^2 = N_c^2$ meson components plus the two baryons minus the one relation
\eqref{eq:V03-classring}. The chiral ring and the moduli-space dimension agree, a cross-check that
will recur in quiver examples. For the $\mathbb{C}^3/\mathbb{Z}_3$ quiver of \S\ref{sec:V03-amax} the chiral generators are
instead the gauge-invariant closed loops of bifundamentals, $\mathrm{Tr}\,(X_{01}^a X_{12}^b X_{20}^c)$,
the cubic mesonic operators whose ring is the coordinate ring of the singular cone the D-brane probes.

\medskip\noindent\textbf{Non-renormalization.}\enspace
Holomorphy plus the broken flavor symmetries give the non-renormalization theorems. The superpotential
receives no perturbative corrections: treating each coupling as a background chiral spurion and
demanding holomorphy in those spurions, together with the flavor and R-symmetries and the weak-coupling
limit, forbids any perturbative shift of $W$. The holomorphic gauge coupling runs only at one loop in a
holomorphic scheme. The physical, canonically normalized coupling does run to all orders, but its
running is fixed exactly by the NSVZ formula of \S\ref{sec:V03-rsymmetry}, not by an independent
series. These are statements we use throughout; their full proofs are not reproduced here. The lesson
to carry forward is that the holomorphic data, $W$ and $\tau$, are rigid,
and the chiral ring built on them is a reliable, computable invariant of the theory.

\section[Pure super Yang--Mills and the index]{Pure super Yang--Mills: gaugino condensation and the index}
\label{sec:V03-sym}

The simplest $4d\ \mathcal{N}=1$ dynamics has no matter at all: pure $SU(N_c)$ super Yang--Mills, just
the vector multiplet. It is the cleanest place to see $\Lambda$, the R-symmetry, and the index work
together, and it is the worked instance the breaking criterion of \S\ref{sec:V03-susybreaking} rests
on.

\medskip\noindent\textbf{The classical R-symmetry and its anomaly.}\enspace
Classically pure SYM has a $U(1)_R$ acting on the gaugino $\lambda$, which has R-charge one (the
field-strength superfield $W_\alpha$ has R-charge one, the gauge-kinetic chiral superfield
$\mathrm{Tr}\,W_\alpha W^\alpha$ has R-charge two). The Adler--Bell--Jackiw (ABJ) anomaly breaks this
continuous $U(1)_R$ to a discrete subgroup. The anomalous divergence of the R-current is proportional
to $2T(\mathrm{adj}) = 2N_c$ times the instanton density, so a $U(1)_R$ rotation by $\alpha$ shifts the
theta angle by $2N_c\,\alpha$. Rotations that shift $\theta$ by a multiple of $2\pi$ are unbroken,
which leaves
\begin{equation}
\label{eq:V03-Z2Nc}
 U(1)_R \;\longrightarrow\; \mathbb{Z}_{2N_c}.
\end{equation}
This $\mathbb{Z}_{2N_c}$ is an exact symmetry of the quantum theory.

\medskip\noindent\textbf{Gaugino condensation and the $N_c$ vacua.}\enspace
The strong dynamics forms a gaugino condensate, a nonzero vacuum expectation value of the lowest
component of the gauge-kinetic chiral operator,
\begin{equation}
\label{eq:V03-gaugino}
 \langle\, \lambda\lambda \,\rangle_k \;\sim\; \Lambda^3\, e^{\,2\pi i k/N_c},
 \qquad k = 0,1,\dots,N_c-1.
\end{equation}
The operator $\lambda\lambda$ has R-charge two, so a $\mathbb{Z}_{2N_c}$ generator acts on it by
\begin{equation}
\label{eq:V03-Z2Ncphase}
 \lambda\lambda \;\longmapsto\; e^{\,2\pi i\cdot 2/(2N_c)}\,\lambda\lambda \;=\; e^{\,2\pi i/N_c}\,\lambda\lambda,
\end{equation}
which fixes the condensate only for the subgroup $\mathbb{Z}_2\subset\mathbb{Z}_{2N_c}$ (the elements
that fix a charge-two operator), so $\mathbb{Z}_{2N_c}$ is spontaneously broken down to $\mathbb{Z}_2$.
The broken discrete symmetry relates
\begin{equation}
\label{eq:V03-vaccount}
 \frac{|\mathbb{Z}_{2N_c}|}{|\mathbb{Z}_2|} \;=\; \frac{2N_c}{2} \;=\; N_c
\end{equation}
inequivalent vacua, the $N_c$ distinct $N_c$-th roots of $\Lambda^{3N_c}$ displayed in
\eqref{eq:V03-gaugino}.

None of these vacua breaks supersymmetry. The condensate is \emph{not} a supersymmetry-breaking order
parameter: $\lambda\lambda$ is the lowest component of a chiral operator, its vacuum expectation value
is consistent with a zero-energy state, and supersymmetry is unbroken in each of the $N_c$ vacua. The
condensate labels the distinct vacua produced by the discrete R-symmetry breaking; it does not
indicate broken supersymmetry. (Phrasing this as ``an F-flat configuration that breaks an R-symmetry''
is the careful statement; calling the condensate itself a supersymmetry-breaking order parameter is
the trap.)

\medskip\noindent\textbf{The Witten index.}\enspace
The count of supersymmetric vacua is the Witten index,
\begin{equation}
\label{eq:V03-witten}
 \mathrm{Tr}\,(-1)^F \;=\; N_c \;\ne\; 0,
\end{equation}
computed here as $|\mathbb{Z}_{2N_c}|/|\mathbb{Z}_2| = N_c$. The index is a deformation invariant, so
the same number must emerge from a completely different (and weakly coupled) computation. Compactify
the theory on a spatial three-torus of small volume. The gauge field then has flat connections
labelled by commuting holonomies around the three cycles, and for $SU(N_c)$ Witten's count of these
isolated flat-connection vacua, weighted by $(-1)^F$, is exactly $N_c$ (the dual Coxeter number, which
for $SU(N_c)$ equals $N_c$). The weak-coupling count and the strong-coupling gaugino-condensate count
agree, as the deformation invariance of the index demands. A nonzero Witten index forbids dynamical
supersymmetry breaking, so pure super Yang--Mills confines with $N_c$ gaugino-condensate vacua and
unbroken supersymmetry.

\medskip\noindent\textbf{Optional: the glueball superpotential.}\enspace
The same $N_c$ vacua follow from the Veneziano--Yankielowicz effective superpotential for the glueball
chiral superfield $S = -\tfrac{1}{32\pi^2}\,\mathrm{Tr}\,W_\alpha W^\alpha$ (whose lowest component is
$\sim\lambda\lambda$),
\begin{equation}
\label{eq:V03-VY}
 W_{\mathrm{VY}}(S) \;=\; S\left[\,N_c - \ln\!\frac{S^{N_c}}{\Lambda^{3N_c}}\,\right].
\end{equation}
Its F-term equation locates the vacua,
\begin{equation}
\label{eq:V03-VYF}
 \frac{\partial W_{\mathrm{VY}}}{\partial S} = -\ln\!\frac{S^{N_c}}{\Lambda^{3N_c}} = 0
 \;\Longrightarrow\; S^{N_c} = \Lambda^{3N_c}
 \;\Longrightarrow\; \langle S\rangle = \Lambda^3\,e^{2\pi i k/N_c},
\end{equation}
the $N_c$ roots recovering \eqref{eq:V03-gaugino}. Substituting back,
\begin{equation}
\label{eq:V03-VYvalue}
 W_{\mathrm{VY}}(\langle S\rangle) = N_c\langle S\rangle = N_c\,\Lambda^3\,e^{2\pi i k/N_c},
\end{equation}
so the $N_c$ vacua are genuinely distinct (different values of the superpotential), confirming the
discrete-symmetry count. We use it only to locate the vacua; the full
glueball effective action and its derivation are not developed here.

\medskip\noindent\textbf{Tying the SYM scale to SQCD.}\enspace
The pure-SYM scale is not an independent input: it is the $N_f\to 0$ end of the SQCD threshold ladder.
Iterating \eqref{eq:V03-threshold} from $N_f$ flavors down to zero, giving each flavor a mass $m_a$,
\begin{equation}
\label{eq:V03-symscale}
 \Lambda_{\mathrm{SYM}}^{3N_c} \;=\; \Big(\textstyle\prod_{a=1}^{N_f} m_a\Big)\,\Lambda_{N_f}^{3N_c-N_f},
\end{equation}
the exponent on the left being $b_0|_{N_f=0} = 3N_c$. The gaugino condensate of the resulting pure SYM
is then the cube root of \eqref{eq:V03-symscale},
\begin{equation}
\label{eq:V03-symcondensate}
 \langle\lambda\lambda\rangle \;\sim\; \Lambda_{\mathrm{SYM}}^3
 \;=\; \Big(\textstyle\prod_a m_a\Big)^{1/N_c}\,\Lambda_{N_f}^{(3N_c-N_f)/N_c},
\end{equation}
expressed entirely in the SQCD data and the masses. This is the
$\mathcal{N}=1$ holomorphic-scale bookkeeping that the next section's R-symmetry and the duality of
\S\ref{sec:V03-seiberg} both rely on, and it is the construction the scale-matching pointer below
contrasts with the $\mathcal{N}=2$ case.

\medskip\noindent\textbf{A scale-matching pointer, stated carefully.}\enspace
It is tempting to read the $\mathcal{N}=2$ Seiberg--Witten scale of the $4d\ \mathcal{N}=2$ section as this same
$\mathcal{N}=1$ condensate. It is not. The $\mathcal{N}=2$ scale is generated by $\mathcal{N}=2$
dimensional transmutation, the running of the $\mathcal{N}=2$ gauge theory, with its own one-loop
coefficient. The two scales are related only after an $\mathcal{N}=1$ mass deformation: giving the
$\mathcal{N}=2$ adjoint chiral a mass and integrating it out (a threshold matching of the type
\eqref{eq:V03-threshold}, but across the $\mathcal{N}=2\to\mathcal{N}=1$ decoupling) leaves pure
$\mathcal{N}=1$ super Yang--Mills, and the holomorphic matching of scales across that decoupling ties
the two together. The $4d\ \mathcal{N}=2$ section develops the $\mathcal{N}=2$ side. Here we only flag the seam: do
not equate the $\mathcal{N}=2$ transmutation scale with the $\mathcal{N}=1$ condensate before the mass
deformation is performed.

\section{R-symmetry, NSVZ, and the superconformal R}
\label{sec:V03-rsymmetry}

A $U(1)_R$ R-symmetry is part of the superconformal algebra at a fixed point, as Section~1 showed. The
bookkeeping is the fermion R-shift of Section~2. The gauge-kinetic superfield has R-charge two, so the
gaugino fermion has $R=1$. A chiral multiplet whose scalar carries R-charge $R_s$ has its fermion at
\begin{equation}
\label{eq:V03-rshift}
 R_\psi \;=\; R_s - 1.
\end{equation}
Anomaly traces are fermion traces, and forgetting this shift is the most common way to get the wrong
answer. We will use it on every trace in this section.

\medskip\noindent\textbf{The NSVZ beta function.}\enspace
The exact running of the physical (canonically normalized) coupling is the NSVZ beta function,
\begin{equation}
\label{eq:V03-nsvz}
 \beta(g) \;=\; -\frac{g^3}{16\pi^2}\,
 \frac{3\,T(\mathrm{adj}) \;-\; \sum_i T(r_i)\,(1-\gamma_i)}
 {1 \;-\; \dfrac{g^2\,T(\mathrm{adj})}{8\pi^2}},
\end{equation}
with $\gamma_i$ the anomalous dimension of the chiral field in representation $r_i$. The numerator
generalizes $b_0$ of \eqref{eq:V03-b0} by the matter anomalous dimensions; the denominator is the
all-orders correction from the holomorphic-versus-physical rescaling. Its origin is one computation: the
holomorphic coupling runs only at one loop, but passing to the canonically normalized coupling requires
a field rescaling whose Jacobian is anomalous (a Konishi-type anomaly), and that anomaly is exactly the
geometric series $1/(1 - g^2 T(\mathrm{adj})/8\pi^2)$. The one-loop running plus the rescaling anomaly
is the full physical beta function, with no independent higher-loop input. At a fixed point the beta
function vanishes, and since the denominator is finite the numerator vanishes:
\begin{equation}
\label{eq:V03-nsvznum}
 3\,T(\mathrm{adj}) \;=\; \sum_i T(r_i)\,(1-\gamma_i).
\end{equation}
This fixes the $\gamma_i$ at the fixed point, and the fixed-point R-charge of a chiral field follows
from the superconformal relation
\begin{equation}
\label{eq:V03-Rgamma}
 R_i \;=\; \frac{2}{3}\Big(1 + \tfrac12\gamma_i\Big),
\end{equation}
which encodes a chiral primary of dimension
\begin{equation}
\label{eq:V03-Delta}
 \Delta_i \;=\; 1 + \tfrac12\gamma_i \;=\; \tfrac32 R_i \quad\text{at the fixed point.}
\end{equation}
A free field has $\gamma_i = 0$ and $R_i = \tfrac23$; an interacting fixed point shifts $\gamma_i$ away
from zero and $R_i$ away from $\tfrac23$, and \eqref{eq:V03-Rgamma} is the dictionary between the two. Reading \eqref{eq:V03-nsvznum} and
\eqref{eq:V03-Rgamma} together: the NSVZ condition is one equation in the anomalous dimensions, the
$R$-dictionary turns it into one equation in the $R$-charges, and for SQCD it is the same condition as
gauge-R anomaly freedom.

\medskip\noindent\textbf{The SQCD quark R-charge, derived.}\enspace
For $SU(N_c)$ SQCD with $N_f$ flavors the flavor symmetry forces all the quarks to share one R-charge,
$R(Q) = R(\widetilde Q) \equiv R_Q$. The NSVZ condition \eqref{eq:V03-nsvznum} is then equivalent to the
freedom of the mixed gauge-R (ABJ) anomaly, $SU(N_c)^2\,U(1)_R = 0$. That anomaly receives the gaugino
contribution $T(\mathrm{adj})$ at $R_\psi = 1$ and, from each of the $2N_f$ fundamental quark fermions,
$T(\square)$ at $R_\psi = R_Q - 1$:
\begin{equation}
\label{eq:V03-abj}
 T(\mathrm{adj}) \;+\; N_f\,T(\square)\,(R_Q-1) \;+\; N_f\,T(\square)\,(R_{\widetilde Q}-1) \;=\; 0.
\end{equation}
With $T(\mathrm{adj})=N_c$, $T(\square)=\tfrac12$, and $R_Q = R_{\widetilde Q}$, this collapses to
\begin{equation}
\label{eq:V03-abjcollapse}
 N_c \;+\; N_f\,(R_Q-1) \;=\; 0,
\end{equation}
so
\begin{equation}
\label{eq:V03-sqcdR}
 R(Q) \;=\; 1 - \frac{N_c}{N_f},
\end{equation}
the symmetry-determined superconformal R-charge of the SQCD quark, and the quark \emph{fermion} carries
$R_\psi = R(Q)-1 = -N_c/N_f$. This is the data the 't~Hooft anomalies of \S\ref{sec:V03-anomaly} and
the central charges of \S\ref{sec:V03-amax} are built from.

\emph{Sanity check.} The naive guess $R(Q)=\tfrac12$ does \emph{not} solve \eqref{eq:V03-abj} except at
the special point $N_f = 2N_c$,
\begin{equation}
\label{eq:V03-Rqcheck}
 R(Q)\big|_{N_f=2N_c} = 1 - \tfrac{N_c}{2N_c} = \tfrac12, \qquad
 R(Q)\big|_{N_c=3,\,N_f=5} = 1 - \tfrac35 = \tfrac25 \ne \tfrac12.
\end{equation}
And $R(Q)=0$ (a ``free massless quark, no R-charge'') leaves the gaugino's $+N_c$ uncancelled in
\eqref{eq:V03-abjcollapse}. The superconformal R is forced; it is not whatever value looks symmetric.

\medskip\noindent\textbf{What the superconformal R is.}\enspace
The superconformal R is a specific object, not a free choice: it is the one linear combination of the
classical R-symmetry and the abelian flavor symmetries that sits inside the superconformal multiplet.
The naive ultraviolet assignment ($R=\tfrac23$ or equal R-charges) is generally \emph{not} it unless a
symmetry forces it. On a chiral primary the dimension is locked to this R by $\Delta = \tfrac32 R$, the
saturated unitarity bound of Section~1, computed with the superconformal R, never the ultraviolet one.
Symmetry does the work when the candidate R is unique up to constraints, as in SQCD (one flavor
symmetry and one anomaly condition) or a quiver (a discrete symmetry forcing a symmetric assignment).
When the candidate R has a free parameter, as in adjoint SQCD, the selection is the extremization of
\S\ref{sec:V03-amax}, and the answer is generally not a naive guess.

\section{'t~Hooft anomalies and the Konishi anomaly}
\label{sec:V03-anomaly}

The 't~Hooft anomalies of the continuous global symmetries are renormalization-group invariants. They
are computed from the ultraviolet fermion content, they cannot change along the flow, and they must
equal the value computed in any infrared description. Matching them between a proposed ultraviolet and
infrared (or electric and magnetic) frame is therefore a necessary consistency test. It is necessary,
not sufficient: two theories can share every 't~Hooft anomaly and still be different, so anomaly
matching is \emph{evidence} for an identification, never a proof of it. This is the gauge-versus-'t~Hooft
distinction of Section~2 turned into a working tool.

\medskip\noindent\textbf{The SQCD global symmetry and its conventions.}\enspace
For $SU(N_c)$ SQCD with $N_f$ flavors the continuous global symmetry is
\begin{equation}
\label{eq:V03-sqcdsym}
 SU(N_f)_L \;\times\; SU(N_f)_R \;\times\; U(1)_B \;\times\; U(1)_R.
\end{equation}
We declare the representation convention once and keep it for the whole section. The quarks $Q$ are a
fundamental of $SU(N_f)_L$ and a singlet of $SU(N_f)_R$; the antiquarks $\widetilde Q$ are an
antifundamental of $SU(N_f)_R$ and a singlet of $SU(N_f)_L$. Under $U(1)_B$, $Q$ has charge $+1$ and
$\widetilde Q$ has charge $-1$. The R-charges are $R(Q)=R(\widetilde Q)=1-N_c/N_f$, so the quark
fermions carry $R_\psi = -N_c/N_f$ (\eqref{eq:V03-sqcdR}, \eqref{eq:V03-rshift}); the gaugino is at
$R_\psi=1$. The full assignment is collected in Table~\ref{tab:V03-electric}.

\begin{table}[ht]
\centering
\small
\setlength{\tabcolsep}{6pt}
\renewcommand{\arraystretch}{1.3}
\begin{tabular}{@{}lccccc@{}}
\toprule
field & $SU(N_c)$ gauge & $SU(N_f)_L$ & $SU(N_f)_R$ & $U(1)_B$ & $U(1)_R$ \\
\midrule
$Q$ & $\square$ & $\square$ & $\mathbf{1}$ & $+1$ & $1-N_c/N_f$ \\
\midrule
$\widetilde Q$ & $\overline\square$ & $\mathbf{1}$ & $\overline\square$ & $-1$ & $1-N_c/N_f$ \\
\midrule
$\lambda$ (gaugino) & $\mathrm{adj}$ & $\mathbf{1}$ & $\mathbf{1}$ & $0$ & $+1$ \\
\bottomrule
\end{tabular}
\caption{The electric $SU(N_c)$ SQCD matter content and global charges. The quark and antiquark
fermions carry $R_\psi = R(Q)-1 = -N_c/N_f$; the gaugino fermion carries $R_\psi = +1$. The
$SU(N_f)_R$ acts on $\widetilde Q$ as an antifundamental, which fixes the sign of its cubic anomaly.}
\label{tab:V03-electric}
\end{table}

\medskip\noindent\textbf{The cubic flavor anomalies, with the L/R split.}\enspace
Only the $N_c$ colors of a quark are charged under a flavor group, so a cubic flavor anomaly counts
colors. Because $Q$ is a fundamental of $SU(N_f)_L$ and $\widetilde Q$ is an \emph{antifundamental} of
$SU(N_f)_R$, the two cubic anomalies have opposite signs:
\begin{equation}
\label{eq:V03-cubicLR}
 SU(N_f)_L^3 \;=\; +N_c, \qquad SU(N_f)_R^3 \;=\; -N_c,
\end{equation}
each from the $N_c$ colors of the corresponding quark multiplied by the cubic index of a (anti)fundamental
($\pm1$). (Had one conjugated $\widetilde Q$ into a fundamental of $SU(N_f)_R$ instead, both signs would
be $+N_c$; we use the antifundamental convention of Table~\ref{tab:V03-electric} throughout.) Splitting
the two flavor groups is not cosmetic: the sign is exactly what the Seiberg dual must reproduce, and a
single ``$SU(N_f)^3 = N_c$'' obscures which group is which.

\medskip\noindent\textbf{The R-traces, run explicitly.}\enspace
The gravitational and cubic R-anomalies are pure fermion traces over $R_\psi$ and $R_\psi^3$. The
gravitational trace $\mathrm{Tr}\,R$ sums the $N_c^2-1$ gaugino components at $R_\psi=1$ and the $2N_fN_c$
quark fermions at $R_\psi=-N_c/N_f$:
\begin{equation}
\label{eq:V03-TrR}
 \mathrm{Tr}\,R \;=\; (N_c^2-1)\cdot 1 \;+\; 2N_fN_c\left(-\frac{N_c}{N_f}\right)
 \;=\; (N_c^2-1) - 2N_c^2 \;=\; -(N_c^2+1).
\end{equation}
The cubic trace $\mathrm{Tr}\,R^3$ uses the same fermions at the cube of their R-charge:
\begin{equation}
\label{eq:V03-TrR3}
 \mathrm{Tr}\,R^3 \;=\; (N_c^2-1)\cdot 1^3 \;+\; 2N_fN_c\left(-\frac{N_c}{N_f}\right)^3
 \;=\; (N_c^2-1) - \frac{2N_c^4}{N_f^2}.
\end{equation}
The mixed flavor-R trace counts only the quark charged under the flavor group, with the fundamental
index $T(\square)=\tfrac12$:
\begin{equation}
\label{eq:V03-SUNfR}
 SU(N_f)_L^2\,U(1)_R \;=\; N_c\cdot\tfrac12\cdot\left(-\frac{N_c}{N_f}\right)
 \;=\; -\frac{N_c^2}{2N_f},
\end{equation}
and the same value (with the antifundamental of $\widetilde Q$) for $SU(N_f)_R^2\,U(1)_R$. Two more
mixed traces complete the list the duality will test. The flavor-baryon trace counts the colors with
the baryon charge $\pm1$,
\begin{equation}
\label{eq:V03-SUNfB}
 SU(N_f)_L^2\,U(1)_B \;=\; N_c\cdot\tfrac12\cdot(+1) \;=\; \frac{N_c}{2},
\end{equation}
and the cubic-baryon-R trace is
\begin{equation}
\label{eq:V03-B2R}
 U(1)_B^2\,U(1)_R \;=\; 2N_fN_c\cdot(+1)^2\cdot\left(-\frac{N_c}{N_f}\right) \;=\; -2N_c^2,
\end{equation}
summing both quark species (the gaugino is $U(1)_B$-neutral). These are the anomalies the Seiberg dual
of \S\ref{sec:V03-seiberg} must reproduce trace by trace, and the central charges of
\S\ref{sec:V03-amax} are built from \eqref{eq:V03-TrR} and \eqref{eq:V03-TrR3} directly.

\emph{Worked numbers.} At $N_c=3,N_f=6$ the quark fermion R-charge is $-N_c/N_f = -\tfrac12$, so
\begin{align}
\label{eq:V03-TrR-num}
 \mathrm{Tr}\,R\big|_{3,6} &= (9-1) - 2\cdot6\cdot3\cdot\tfrac12 = 8 - 18 = -10 = -(N_c^2+1), \notag\\[2pt]
 \mathrm{Tr}\,R^3\big|_{3,6} &= 8 - 36\cdot\tfrac18 = 8 - \tfrac92 = \tfrac72,
\end{align}
confirming \eqref{eq:V03-TrR} and \eqref{eq:V03-TrR3}. These two numbers feed directly into the central
charges of the next section.

\medskip\noindent\textbf{The Konishi anomaly.}\enspace
The Konishi anomaly is the operator equation that ties the matter sector to the gauge field strength.
Classically the Konishi current $\bar\Phi e^V \Phi$ is conserved up to the superpotential; quantum
mechanically it acquires a gauge-anomaly divergence,
\begin{equation}
\label{eq:V03-konishi}
 \bar D^2\big(\bar\Phi e^V \Phi\big)
 \;=\; \Phi\,\frac{\partial W}{\partial\Phi}
 \;+\; \frac{T(r)}{8\pi^2}\,\mathrm{Tr}\,W_\alpha W^\alpha.
\end{equation}
We normalize $\bar D^2$ so that the classical piece reads $\Phi\,\partial W/\partial\Phi$; its
coefficient is convention-dependent, while the load-bearing \emph{anomaly} coefficient $T(r)/8\pi^2$ is
not. That anomaly coefficient multiplying $\mathrm{Tr}\,W_\alpha W^\alpha$ is the \emph{same} gauge
Dynkin index $T(r)$ that enters the mixed gauge-R (ABJ) anomaly fixing $R(Q)$ above. That shared index is what
makes the Konishi equation consistent with the R-charge bookkeeping: the per-flavor matter contribution
to the gauge-R anomaly is $T(\square)$ times the quark fermion R-charge, with $T(\square)=\tfrac12$, not
$1$. To see the consistency as arithmetic, sum the Konishi equation over the $2N_f$ fundamentals. The
matter contribution to the divergence of the R-current, plus the gaugino's $T(\mathrm{adj})=+N_c$, gives
\begin{equation}
\label{eq:V03-konishisum}
 \underbrace{2N_f\,T(\square)\,R_\psi(Q)}_{\text{matter}} + \underbrace{T(\mathrm{adj})}_{\text{gaugino}}
 \;=\; 2N_f\cdot\tfrac12\cdot\!\left(-\tfrac{N_c}{N_f}\right) + N_c
 \;=\; -N_c + N_c \;=\; 0,
\end{equation}
exactly \eqref{eq:V03-abj}. The Konishi anomaly and the ABJ
anomaly are the same operator statement, written once with a matter superfield and once with the gauge
field strength, and the shared $T(\square)=\tfrac12$ is what makes them agree. The Konishi anomaly feeds
the matter anomalous dimensions and the chiral-ring relations. We state it as the operator equation and
cite its full derivation rather than reproducing it.

\begin{keybox}{Common misconception: anomaly matching is necessary, not a proof}
Equal 't~Hooft anomalies are a \emph{necessary} condition for two theories to flow to the same infrared
physics, never a sufficient one. Two distinct theories can carry identical anomaly polynomials. When a
proposed duality passes the anomaly match, that is real evidence, weighed together with the chiral-ring
and flavor-symmetry matches, but it does not by itself prove the duality. The proof, where one exists,
is a separate argument.
\end{keybox}

\section{a-maximization: the extremization prototype}
\label{sec:V03-amax}

The two conformal central charges of a $4d\ \mathcal{N}=1$ fixed point are built from the same fermion
traces over a trial R-symmetry,
\begin{equation}
\label{eq:V03-ac}
 a(R) = \frac{3}{32}\big(3\,\mathrm{Tr}\,R^3 - \mathrm{Tr}\,R\big), \qquad
 c(R) = \frac{1}{32}\big(9\,\mathrm{Tr}\,R^3 - 5\,\mathrm{Tr}\,R\big),
\end{equation}
with the gaugino at $R=1$ and a chiral fermion at $R_s-1$, the prescription of Section~2. The exact
superconformal R is the trial R that \emph{locally maximizes} $a$. It is $a$ that is extremized, not
$c$, and the extremum is a maximum, not a minimum. This is the prototype of the extremization
through-line of these notes: the $2d$ section does $c$-extremization as the two-dimensional analog of this
section, and the $3d$ section does $F$-maximization as the three-dimensional analog. Both point back
here.

These central charges are not arbitrary combinations: $a$ and $c$ are the coefficients of the Euler and
Weyl-squared terms in the trace anomaly of the stress tensor on a curved background, and supersymmetry
ties them to the 't~Hooft anomalies of the R-current through \eqref{eq:V03-ac}. A single free chiral
multiplet has its fermion at $R = -\tfrac13$, and a free vector multiplet has its gaugino at $R=1$,
giving the standard free-field benchmarks
\begin{align}
\label{eq:V03-freebench}
 \text{chiral:}\quad &\mathrm{Tr}\,R = -\tfrac13,\ \ \mathrm{Tr}\,R^3 = -\tfrac{1}{27}
 \;\Longrightarrow\; a = \tfrac{1}{48},\ c = \tfrac{1}{24}, \notag\\[2pt]
 \text{vector:}\quad &\mathrm{Tr}\,R = 1,\ \ \mathrm{Tr}\,R^3 = 1
 \;\Longrightarrow\; a = \tfrac{3}{16},\ c = \tfrac{1}{8}.
\end{align}
These free benchmarks are how one checks the sign and normalization of any trace. They also exhibit
$a\ne c$ already at free-field level, so the inequality is the rule, not a pathology.

The maximization is constrained, not free. A trial R ranges over the abelian symmetries that respect
gauge-anomaly freedom (the mixed gauge-R, or ABJ, condition $SU(N_c)^2$-$U(1)_R = 0$) and that assign
R-charge two to every superpotential term, together with any flavor symmetry one chooses to impose.
Parametrize the trial R as $R_t = R_0 + \sum_I s_I\,F_I$, with $R_0$ a reference R-charge, $F_I$ the
abelian flavor charges, and $s_I$ the free mixing parameters; the constrained extremization is
\begin{equation}
\label{eq:V03-amaxgen}
 \frac{\partial a}{\partial s_I} \;=\; 0, \qquad
 \frac{\partial^2 a}{\partial s_I\,\partial s_J} \;<\; 0 \ \text{(negative definite)},
\end{equation}
a stationarity equation in the mixing parameters with the Hessian condition selecting the maximum. The
cubic and gravitational 't~Hooft anomalies $\mathrm{Tr}\,R^3$ and $\mathrm{Tr}\,R$ are the \emph{data}
that build $a$ and $c$; they are not constraints that must vanish. One last caveat matters: an
operator whose dimension would fall below the unitarity bound decouples as a free field, and the
maximization must be redone without it. We state that accidental-symmetry correction but do not develop
it in full.

\subsection*{Worked instance: SQCD}

Take $SU(N_c)$ SQCD in the conformal window. Here symmetry already fixes the trial R: one quark
R-charge, one anomaly condition, giving $R(Q) = 1 - N_c/N_f$ as in \eqref{eq:V03-sqcdR}, so the quark
fermion carries $R_\psi = -N_c/N_f$. The two traces are already computed, \eqref{eq:V03-TrR} and
\eqref{eq:V03-TrR3},
\begin{equation}
\label{eq:V03-sqcd-traces}
 \mathrm{Tr}\,R = -(N_c^2+1), \qquad
 \mathrm{Tr}\,R^3 = (N_c^2-1) - \frac{2N_c^4}{N_f^2},
\end{equation}
and \eqref{eq:V03-ac} then gives $a$ and $c$ as explicit rational functions of $(N_c,N_f)$. Substituting
\eqref{eq:V03-sqcd-traces},
\begin{align}
\label{eq:V03-sqcd-ac}
 a(N_c,N_f) &= \frac{3}{32}\!\left[\,3(N_c^2-1) - \frac{6N_c^4}{N_f^2} + (N_c^2+1)\right]
 = \frac{3}{32}\!\left[\,4N_c^2 - 2 - \frac{6N_c^4}{N_f^2}\right]\!, \notag\\[2pt]
 c(N_c,N_f) &= \frac{1}{32}\!\left[\,9(N_c^2-1) - \frac{18N_c^4}{N_f^2} + 5(N_c^2+1)\right]
 = \frac{1}{32}\!\left[\,14N_c^2 - 4 - \frac{18N_c^4}{N_f^2}\right]\!.
\end{align}
These are distinct rational functions: $a\ne c$ in general. Subtracting the two formulas of
\eqref{eq:V03-ac} gives
\begin{equation}
\label{eq:V03-amc}
 a - c \;=\; \tfrac{1}{16}\,\mathrm{Tr}\,R,
\end{equation}
so the central charges coincide precisely when $\mathrm{Tr}\,R=0$ (for example
$\mathcal{N}=4$ super Yang--Mills), a special property, not a generic baseline. For
the explicit case $N_c=3,N_f=6$, with $\mathrm{Tr}\,R = -10$ and $\mathrm{Tr}\,R^3 = \tfrac72$ from
\eqref{eq:V03-TrR-num},
\begin{align}
\label{eq:V03-sqcd-numeric}
 a(3,6) &= \frac{3}{32}\!\left(3\cdot\tfrac72 + 10\right) = \frac{3}{32}\cdot\frac{41}{2} = \frac{123}{64}, \notag\\[2pt]
 c(3,6) &= \frac{1}{32}\!\left(9\cdot\tfrac72 + 50\right) = \frac{1}{32}\cdot\frac{163}{2} = \frac{163}{64}.
\end{align}
Both are finite, positive, and unitary, and $\tfrac{123}{64}\ne\tfrac{163}{64}$ exhibits $a\ne c$ on a
concrete case. Running the same arithmetic across the $N_c=3$ conformal window
($\tfrac92 < N_f < 9$) gives Table~\ref{tab:V03-sqcdac}: $a<c$ in every entry (consistent with
$\tfrac12 \le a/c \le \tfrac32$), and $a/c$ drifts from $0.686$ at the bottom toward $0.798$ at the top,
approaching the free-field ratio as the dynamics weakens. Each number is \eqref{eq:V03-sqcd-ac}
evaluated at the listed $(N_c,N_f)$, computed not quoted.

\begin{table}[ht]
\centering
\small
\setlength{\tabcolsep}{8pt}
\renewcommand{\arraystretch}{1.3}
\begin{tabular}{@{}cccccc@{}}
\toprule
$N_f$ & $\mathrm{Tr}\,R$ & $\mathrm{Tr}\,R^3$ & $a$ & $c$ & $a/c$ \\
\midrule
$5$ & $-10$ & $38/25$ & $273/200 \approx 1.365$ & $199/100 \approx 1.990$ & $0.686$ \\
\midrule
$6$ & $-10$ & $7/2$ & $123/64 \approx 1.922$ & $163/64 \approx 2.547$ & $0.755$ \\
\midrule
$7$ & $-10$ & $230/49$ & $885/392 \approx 2.258$ & $565/196 \approx 2.883$ & $0.783$ \\
\midrule
$8$ & $-10$ & $175/32$ & $2535/1024 \approx 2.476$ & $3175/1024 \approx 3.101$ & $0.798$ \\
\bottomrule
\end{tabular}
\caption{The SQCD central charges $a,c$ across the $N_c=3$ conformal window, computed from
\eqref{eq:V03-sqcd-ac}. The gravitational trace $\mathrm{Tr}\,R = -(N_c^2+1) = -10$ is $N_f$-independent;
$\mathrm{Tr}\,R^3 = 8 - 2\cdot81/N_f^2$ carries the $N_f$ dependence. In every entry $a<c$ and $a/c$
rises toward the free-field value as the window edge is approached.}
\label{tab:V03-sqcdac}
\end{table}

Because the trial R has no free parameter here, there is nothing left to maximize: the
symmetric assignment already is the superconformal R. SQCD shows the machinery on a case where the
answer is rational and symmetry-forced.

\subsection*{Worked instance: adjoint SQCD, the genuine extremization}

Now add one adjoint chiral field $X$ to $SU(N_c)$ with $N_f$ fundamentals (with no superpotential, so
$R_X$ is free; turning on $W=\mathrm{Tr}\,X^{k+1}$ instead fixes $R_X = 2/(k+1)$ by the superpotential
constraint and removes the free parameter, a different problem). The trial R now has a free parameter:
the adjoint R-charge $R_X$ is not fixed by symmetry alone. Only the ABJ condition relates the quark
R-charge to it. The mixed gauge-R anomaly now receives the gaugino, the adjoint chiral $X$ (also index
$T(\mathrm{adj})=N_c$, at $R_\psi = R_X-1$), and the $2N_f$ fundamentals:
\begin{equation}
\label{eq:V03-adjabj}
 N_c \;+\; N_c\,(R_X-1) \;+\; N_f\,(R_Q-1) \;=\; 0
 \quad\Longrightarrow\quad
 R_Q \;=\; 1 - \frac{N_c}{N_f}\,R_X.
\end{equation}
With $R_Q$ slaved to $R_X$, one builds the constrained $a(R_X)$ from the fermion traces. Write
$r_X \equiv R_X - 1$ for the adjoint fermion R-charge and $r_Q \equiv R_Q - 1 = -\tfrac{N_c}{N_f}R_X$
for the quark fermion R-charge, from \eqref{eq:V03-adjabj}. The gravitational trace is
\begin{equation}
\label{eq:V03-adjTrR}
 \mathrm{Tr}\,R \;=\; \underbrace{(N_c^2-1)}_{\text{gaugino}}
 \;+\; \underbrace{(N_c^2-1)\,r_X}_{\text{adjoint }X}
 \;+\; \underbrace{2N_fN_c\,r_Q}_{\text{quarks}},
\end{equation}
and the cubic trace uses the cubes of the same fermion R-charges,
\begin{equation}
\label{eq:V03-adjTrR3}
 \mathrm{Tr}\,R^3 \;=\; (N_c^2-1) \;+\; (N_c^2-1)\,r_X^3 \;+\; 2N_fN_c\,r_Q^3.
\end{equation}
Because $r_Q = -\tfrac{N_c}{N_f}R_X$ is linear in $R_X$ and $r_X = R_X - 1$ is linear in $R_X$, the
trace $\mathrm{Tr}\,R^3$ is a \emph{cubic} polynomial in $R_X$, and so is the trial central charge
\begin{equation}
\label{eq:V03-adja}
 a(R_X) \;=\; \frac{3}{32}\big(3\,\mathrm{Tr}\,R^3 - \mathrm{Tr}\,R\big).
\end{equation}
The stationarity equation
\begin{equation}
\label{eq:V03-adjstationary}
 \frac{da}{dR_X} \;=\; 0
\end{equation}
is therefore a \emph{quadratic} in $R_X$. The bookkeeping is the load-bearing point: the trial $a$ is
cubic, its derivative is quadratic, and the extremum is a root of that quadratic, not of a cubic.
(Calling it ``a root of a cubic'' double-counts a degree; the cubic object is $a$ itself, the equation
to solve is quadratic.) A quadratic with real coefficients has two real roots when its discriminant is
positive; the physical superconformal R is the one with $d^2a/dR_X^2 < 0$ (a local maximum) that
respects the unitarity bounds. The general quadratic root carries a square root of the discriminant,
so $R_X^\star$ is \emph{irrational}, an algebraic number rather than a low-denominator rational. Run
it at $N_c=3,N_f=4$. The gauge-anomaly constraint gives $r_Q = -\tfrac34 R_X$, and assembling
\eqref{eq:V03-adjTrR} and \eqref{eq:V03-adjTrR3} the trial central charge is
\begin{equation}
\label{eq:V03-adjacubic}
 a(R_X) \;=\; -\frac{153}{256}\,R_X^3 \;-\; \frac{27}{4}\,R_X^2 \;+\; \frac{123}{16}\,R_X,
\end{equation}
a genuine cubic. Its derivative is the quadratic
\begin{equation}
\label{eq:V03-adjquad}
 \frac{da}{dR_X} \;=\; -\frac{459}{256}\,R_X^2 \;-\; \frac{27}{2}\,R_X \;+\; \frac{123}{16} \;=\; 0,
\end{equation}
whose two roots are $R_X = -\tfrac{64}{17} \pm \tfrac{4}{51}\sqrt{3001}$. The physical one is
\begin{equation}
\label{eq:V03-adjroot}
 R_X^\star \;=\; -\frac{64}{17} + \frac{4}{51}\sqrt{3001} \;\approx\; 0.5319,
\end{equation}
manifestly irrational (it carries the surd $\sqrt{3001}$), in contrast to the symmetry-forced SQCD
value. The other root, $R_X \approx -8.06$, is unphysical. The second derivative
\begin{equation}
\label{eq:V03-adjhessian}
 a''(R_X^\star) \;=\; -\tfrac{459}{128}\,R_X^\star \;-\; \tfrac{27}{2} \;<\; 0
\end{equation}
confirms it is a maximum, and perturbing $R_X$ either way lowers $a$. The same
$R_X$ does \emph{not} extremize $c(R_X) = \tfrac{1}{32}(9\,\mathrm{Tr}\,R^3 - 5\,\mathrm{Tr}\,R)$ (also
cubic in $R_X$): $dc/dR_X|_{R_X^\star}\ne 0$, the $c$-maximizing root sits at a different value, and
the value of $c$ at the $a$-maximum lies strictly below the $c$-maximum. This is the sharpest
illustration that one maximizes $a$ and not $c$. It is the case where the extremization earns its
name: the answer is selected by the maximum of $a$, not read off a symmetry.

\subsection*{Worked instance: the dP$_0$ quiver}

The $\mathbb{C}^3/\mathbb{Z}_3$ (local $\mathbb{P}^2$, or dP$_0$) orbifold daughter is a standard
three-node $\prod_i U(N_i)$ quiver with nine chiral bifundamentals $X_{i,i+1}^a$ ($a=1,2,3$) and the
cubic superpotential $W = \epsilon_{abc}\,\mathrm{Tr}\,X_{01}^a X_{12}^b X_{20}^c$. Its R-symmetry is
forced two independent ways. The superpotential carries R-charge two, and a cubic word in the
bifundamentals carries R-charge $3R(X)$, so
\begin{equation}
\label{eq:V03-dp0W}
 R(W) = 3\,R(X) = 2 \quad\Longrightarrow\quad R(X) = \tfrac{2}{3}.
\end{equation}
Independently, the per-node mixed gauge-R anomaly cancels at the symmetric assignment. At node $i$ the
fields charged under that gauge factor are its gaugino (index $N$ at $R_\psi=1$) and the six
bifundamentals touching it (three incoming, three outgoing, each contributing $N\cdot\tfrac12$ at
$R_\psi = R(X)-1$):
\begin{equation}
\label{eq:V03-dp0node}
 N \;+\; 6\cdot N\cdot\tfrac12\,(R(X)-1) \;=\; N\big[\,1 + 3(R(X)-1)\,\big] \;=\; 0
 \quad\Longrightarrow\quad R(X) = \tfrac23,
\end{equation}
independent of $N$. Running $a$-maximization confirms that $R(X)=\tfrac23$ is a maximum (the Hessian is
negative), not merely a stationary point. With the R-charge fixed, the central charge is a fermion
trace over the three nodes. Take the gauge group $\prod_{i=0}^{2} SU(N)$ (the interacting factor, the
overall diagonal $U(1)$ decoupled). Each node contributes a gaugino at $R=1$ (dimension $N^2-1$), and
each of the nine bifundamentals contributes a fermion at $R_\psi = R(X)-1 = -\tfrac13$ (dimension
$N^2$):
\begin{equation}
\label{eq:V03-dp0traces}
 \mathrm{Tr}\,R = 3(N^2-1)\cdot 1 + 9N^2\!\left(-\tfrac13\right) = -3, \qquad
 \mathrm{Tr}\,R^3 = 3(N^2-1) + 9N^2\!\left(-\tfrac13\right)^{\!3} = \tfrac83 N^2 - 3,
\end{equation}
and assembling \eqref{eq:V03-ac} the central charges of the dP$_0$ quiver as functions of the rank are
\begin{equation}
\label{eq:V03-dp0ac}
 a \;=\; \tfrac{3}{32}\big(3(\tfrac83 N^2-3) + 3\big) \;=\; \tfrac34 N^2 - \tfrac{9}{16}, \qquad
 c \;=\; \tfrac34 N^2 - \tfrac38,
\end{equation}
with $a\ne c$. (At large $N$ both grow as $\tfrac34 N^2$, the leading planar inheritance from the
$\mathcal{N}=4$ parent; the subleading constants distinguish the daughter.) The numbers displayed here
are the \emph{interacting} central charges with the decoupled diagonal $U(1)$ removed; one must state
which it is, because including the free diagonal $U(1)$ shifts $a$ by the free-vector value
$\tfrac{3}{16}$.
This is the worked link from the general toolkit to a quiver example: we take the standard orbifold
quiver as input and read its superconformal R and central charge off the field theory. We do not
reconstruct the directed McKay quiver or the orbifold projection here; the quiver is only a
field-theory R/anomaly/$a$-maximization fixture.

\subsection*{Worked entry point: the Banks--Zaks fixed point}

The cleanest place to see that the conformal window contains genuine fixed points is its upper edge,
where the dynamics is weakly coupled and the fixed point is perturbatively accessible. Take large
$N_c$ and $N_f$ just below the asymptotic-freedom bound,
\begin{equation}
\label{eq:V03-bzNf}
 N_f \;=\; 3N_c - \epsilon\,N_c, \qquad 0 < \epsilon \ll 1,
\end{equation}
so $b_0 = 3N_c - N_f = \epsilon N_c$ is positive but small. The two-loop beta function
\begin{equation}
\label{eq:V03-bztwoloop}
 \beta \;=\; -\frac{g^3}{16\pi^2}\,b_0 \;-\; \frac{g^5}{(16\pi^2)^2}\,b_1 \;+\; \cdots,
\end{equation}
with the two-loop coefficient $b_1$ negative in this regime, has its two contributions cancel at
\begin{equation}
\label{eq:V03-bzgstar}
 g_*^2 \;=\; -\frac{16\pi^2\,b_0}{b_1} \;\sim\; \frac{\epsilon}{N_c},
\end{equation}
a zero of $\beta$ that is positive and parametrically small. This is the Banks--Zaks fixed point: a
weakly coupled, controllably computable $4d\ \mathcal{N}=1$ superconformal field theory at the top of
the window. The fixed point is found by solving the two-loop beta-function zero, not asserted; at
$N_f$ outside the window the would-be zero \eqref{eq:V03-bzgstar} sits at strong or negative coupling
and is not a controlled fixed point. Because the coupling is small, perturbation theory controls the
anomalous dimensions and correlation functions. The exact infrared R-charge is still the
anomaly/$a$-maximization value
\begin{equation}
\label{eq:V03-bzRexpand}
 R(Q) \;=\; 1-\frac{N_c}{N_f}
 \;=\; \frac23-\frac{\epsilon}{9}+O(\epsilon^2),
\end{equation}
so it approaches the free value $R=\tfrac23$ as $\epsilon\to0$; the weak-coupling expansion is the
controlled realization of that exact R-symmetry, not a separate definition of it. The Banks--Zaks point
is therefore the controlled entry point to the conformal window: it is where $a$-maximization, the NSVZ
condition, and the central charges are all computable in a small parameter, and it anchors the rest of
the window by continuity.

\begin{keybox}{Common misconception: $a$-maximization maximizes $a$, with constraints}
The extremized quantity is $a$, not $c$, and the extremum is a local \emph{maximum}, not a minimum or
an unconstrained extremum. The maximization runs over trial R-symmetries constrained by gauge-anomaly
freedom (the ABJ condition $SU(N_c)^2$-$U(1)_R=0$) and by every superpotential term carrying R-charge
two; the cubic and gravitational 't~Hooft anomalies are the \emph{data} that build the central charges,
not constraints that must vanish. The superconformal R is generally not the naive ultraviolet
assignment: it equals it only when symmetry leaves no free parameter (SQCD, the dP$_0$ quiver), and is
selected by the maximum otherwise (adjoint SQCD, where it is irrational). Operators that hit the
unitarity bound must be decoupled and the maximization redone.
\end{keybox}

\subsection*{The window edges and meson unitarity}

The two edges of the conformal window are not arbitrary. They are fixed by the unitarity bound on
chiral operators that Section~1 established: a gauge-invariant chiral primary obeys
$\Delta \ge 1$, with $\Delta = 1$ exactly for a free field, and a chiral primary saturates
$\Delta = \tfrac32 R$. At the interacting fixed point the quark superfield carries the
anomaly-fixed R-charge
\begin{equation}
\label{eq:V03-RQ}
R(Q)\ =\ 1-\frac{N_c}{N_f},
\end{equation}
so the meson $M = Q\widetilde Q$, a product of two quark superfields, has
\begin{equation}
\label{eq:V03-RM}
R(M)\ =\ 2R(Q)\ =\ 2\Bigl(1-\frac{N_c}{N_f}\Bigr),
\qquad
\Delta(M)\ =\ \frac32\,R(M)\ =\ 3\Bigl(1-\frac{N_c}{N_f}\Bigr).
\end{equation}
Now read off the edges,
\begin{equation}
\label{eq:V03-edges}
 \Delta(M)\big|_{N_f=\frac32 N_c} = 3\big(1-\tfrac23\big) = 1, \qquad
 R(Q)\big|_{N_f=3N_c} = 1 - \tfrac{N_c}{3N_c} = \tfrac23.
\end{equation}
At the lower edge $N_f = \tfrac32 N_c$ the meson has reached the free-field value $\Delta(M)=1$, and
below this edge it would violate the unitarity bound as an interacting operator. Unitarity thus
diagnoses the end of the interacting electric continuation: near and below the lower edge the composite
meson is becoming a free field, and the magnetic dual, where $M$ is an elementary field with its own
kinetic term, is the weakly coupled or infrared-free description that takes over. At the upper edge
$N_f = 3N_c$ the quark reaches its free value $R(Q) = \tfrac23$, matching the weakly coupled ultraviolet
picture where asymptotic freedom is marginal. Both endpoint values \eqref{eq:V03-edges} are computed
symbolically, and they tie the Section~1 unitarity bound directly to the location of the
SQCD window.

\begin{keybox}{Exactly marginal deformations: not every conformal manifold is $\mathcal{N}=2$}
A fixed point can sit on a continuous family of fixed points, a \emph{conformal manifold}, reached by
adding an exactly marginal operator. In $4d\ \mathcal{N}=1$ the classically marginal operators are the
superpotential terms of R-charge two and the gauge-kinetic term, and a combination of their couplings
is exactly marginal when the corresponding beta function vanishes. The Leigh--Strassler analysis shows,
in the standard examples, that this happens along a subspace cut out by the vanishing of a combination
of anomalous dimensions, so the conformal manifold has dimension equal to the number of marginal
couplings minus the number of independent broken global symmetries. This is genuinely $\mathcal{N}=1$ data: it is
not the special geometry of an $\mathcal{N}=2$ Coulomb branch (Section~5), and it has its own marginality
and anomaly constraints. The canonical example is $\mathcal{N}=4$ super Yang--Mills viewed as
$\mathcal{N}=1$, whose three-coupling exactly marginal family contains the $\beta$-deformation. The
proof that the codimension condition is exact, and the structure of these manifolds, are beyond this
crash course; here we only record that $\mathcal{N}=1$ fixed points come in families.
\end{keybox}

\section{Seiberg duality and the conformal window}
\label{sec:V03-seiberg}

The deepest fact about SQCD is that one gauge theory has two ultraviolet descriptions that flow to the
same infrared physics. The electric theory is $SU(N_c)$ with $N_f$ fundamental flavors. The magnetic
dual is $SU(N_f - N_c)$ with dual quarks $q,\widetilde q$, a gauge-singlet meson $M$ (the electric
composite $Q\widetilde Q$, now an elementary field on the magnetic side), and the cubic superpotential
\begin{equation}
\label{eq:V03-seibergW}
 W \;=\; M\, q\, \widetilde q.
\end{equation}
These are different ultraviolet theories (different gauge groups, different matter, an extra gauge
singlet), not two copies of one theory. The claim is an infrared equivalence, supported by four checks.

\begin{figure}[ht]
\centering
\begin{tikzpicture}[scale=1.0]
 % == electric theory ==
 \begin{scope}[shift={(0,0)}]
 \node[gaugenode] (eg) at (0,0) {$N_c$};
 \node[flavnode] (elf) at (-2,0) {$N_f$};
 \node[flavnode] (erf) at (2,0) {$N_f$};
 \draw[bifund] (elf) -- node[above,font=\scriptsize] {$Q$} (eg);
 \draw[bifund] (eg) -- node[above,font=\scriptsize] {$\widetilde Q$} (erf);
 \node[font=\scriptsize] at (-2,-0.75) {$SU(N_f)_L$};
 \node[font=\scriptsize] at (2,-0.75) {$SU(N_f)_R$};
 \node[font=\scriptsize] at (0,-0.75) {$SU(N_c)$};
 \node[font=\footnotesize] at (0,1.4) {electric};
 \end{scope}
 % == duality arrow ==
 \draw[<->,thick,RoyalBlue] (3.1,0) -- node[above,font=\scriptsize,RoyalBlue] {Seiberg} node[below,font=\scriptsize,RoyalBlue] {duality} (4.5,0);
 % == magnetic theory ==
 \begin{scope}[shift={(7.6,0)}]
 \node[gaugenode] (mg) at (0,0) {$\widetilde N_c$};
 \node[flavnode] (mlf) at (-2,0) {$N_f$};
 \node[flavnode] (mrf) at (2,0) {$N_f$};
 % reversed quark arrows
 \draw[bifund] (mg) -- node[above,font=\scriptsize] {$q$} (mlf);
 \draw[bifund] (mrf) -- node[above,font=\scriptsize] {$\widetilde q$} (mg);
 % the gauge-singlet meson line, drawn above
 \draw[bifund] (mlf) to[bend left=32] node[above,font=\scriptsize] {$M$} (mrf);
 \node[font=\scriptsize] at (-2,-0.75) {$SU(N_f)_L$};
 \node[font=\scriptsize] at (2,-0.75) {$SU(N_f)_R$};
 \node[font=\scriptsize] at (0,-0.75) {$SU(N_f{-}N_c)$};
 \node[font=\footnotesize] at (0,1.4) {magnetic};
 \node[font=\scriptsize] at (0,2.05) {$W = M\,q\,\widetilde q$};
 \end{scope}
\end{tikzpicture}
\caption{Seiberg duality as a quiver move. Circles are gauge groups, squares are flavor groups, and
arrows are chiral superfields, with their orientation distinguishing fundamental from
antifundamental representations. The electric $SU(N_c)$ theory has quarks $Q,\widetilde Q$
running through the gauge node. The magnetic dual has gauge group $SU(N_f-N_c)$, the dual quarks
$q,\widetilde q$ with \emph{reversed} arrows, an added gauge-singlet meson $M$ (the elementary image of
$Q\widetilde Q$, drawn as the arc joining the two flavor nodes), and the cubic superpotential
$W=Mq\widetilde q$ \eqref{eq:V03-seibergW}.}
\label{fig:V03-seiberg}
\end{figure}
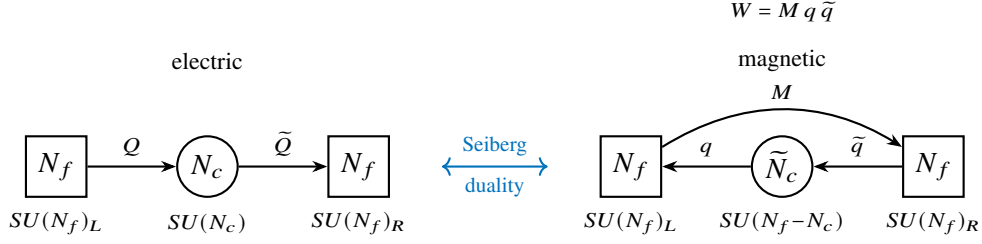

\medskip\noindent\textbf{Check one: the continuous global symmetries match.}\enspace
Both frames carry $SU(N_f)_L \times SU(N_f)_R \times U(1)_B \times U(1)_R$. On the magnetic side the
dual quarks sit at $R(q) = R(\widetilde q) = N_c/N_f$ and the meson at $R_M = 2(1 - N_c/N_f)$, so the
magnetic superpotential $W = Mq\widetilde q$ carries R-charge
\begin{equation}
\label{eq:V03-magWR}
 R_M + R(q) + R(\widetilde q) \;=\; 2\big(1-\tfrac{N_c}{N_f}\big) + 2\tfrac{N_c}{N_f} \;=\; 2,
\end{equation}
as a superpotential must. The dual quark $q$ is an antifundamental of $SU(N_f)_L$ and the meson $M$
carries one $SU(N_f)_L$ fundamental and one $SU(N_f)_R$ antifundamental index.

\medskip\noindent\textbf{Check two: the gauge-invariant chiral operators match.}\enspace
The electric meson $Q\widetilde Q$ maps to the elementary magnetic singlet $M$, and the electric baryons
map to the dual baryons. This is not only a slogan; the conserved charges match by computation. The
electric baryon $B_{\mathrm{el}} \sim Q^{N_c}$ uses $N_c$ electric quarks of baryon number $B(Q)=1$, so
$B(B_{\mathrm{el}}) = N_c$. The dual baryon $b_{\mathrm{mag}} \sim q^{N_f-N_c}$ uses $N_f-N_c$ dual
quarks, whose baryon number is rescaled to $B(q) = N_c/(N_f-N_c)$ precisely so that
\begin{equation}
\label{eq:V03-baryonmap}
B(b_{\mathrm{mag}})\ =\ (N_f-N_c)\,\frac{N_c}{N_f-N_c}\ =\ N_c\ =\ B(B_{\mathrm{el}}).
\end{equation}
The R-charges match by an independent computation. The electric baryon carries
$R(B_{\mathrm{el}}) = N_c\,R(Q) = N_c\bigl(1-N_c/N_f\bigr)$, while the dual quark sits at
$R(q) = 1 - (N_f-N_c)/N_f = N_c/N_f$, so
\begin{equation}
\label{eq:V03-baryonR}
R(b_{\mathrm{mag}})\ =\ (N_f-N_c)\,\frac{N_c}{N_f}\ =\ N_c\Bigl(1-\frac{N_c}{N_f}\Bigr)\ =\ R(B_{\mathrm{el}}).
\end{equation}
Both the baryon number and the R-charge match, each built from independent magnetic data (the rescaled
$B(q)$ and the dual-quark $R(q)$); the equalities are machine-verified. The chiral rings
agree generator by generator.

\medskip\noindent\textbf{Check three: the 't~Hooft anomalies match, trace by trace.}\enspace
The magnetic matter content and global charges are collected in Table~\ref{tab:V03-magnetic}, the
companion to the electric Table~\ref{tab:V03-electric}. The magnetic fermion content is the dual
gaugino (the adjoint of $SU(N_f-N_c)$, dimension $(N_f-N_c)^2-1$, at $R_\psi=1$), the $2N_f$ dual quark
fermions (dimension $N_f-N_c$ each, at $R_\psi = N_c/N_f - 1$), and crucially the $N_f^2$ meson fermions
(at $R_\psi = R_M - 1 = 1 - 2N_c/N_f$). Run the first flavor anomaly.

\begin{table}[ht]
\centering
\small
\setlength{\tabcolsep}{5pt}
\renewcommand{\arraystretch}{1.3}
\begin{tabular}{@{}lccccc@{}}
\toprule
field & $\widetilde N$ gauge & $SU(N_f)_L$ & $SU(N_f)_R$ & $U(1)_B$ & $U(1)_R$ \\
\midrule
$q$ & $\square$ & $\overline\square$ & $\mathbf{1}$ & $+N_c/\widetilde N$ & $N_c/N_f$ \\
\midrule
$\widetilde q$ & $\overline\square$ & $\mathbf{1}$ & $\square$ & $-N_c/\widetilde N$ & $N_c/N_f$ \\
\midrule
$M$ & $\mathbf{1}$ & $\square$ & $\overline\square$ & $0$ & $2(1-N_c/N_f)$ \\
\midrule
$\widetilde\lambda$ (dual gaugino) & $\mathrm{adj}$ & $\mathbf{1}$ & $\mathbf{1}$ & $0$ & $+1$ \\
\bottomrule
\end{tabular}
\caption{The magnetic dual matter content and global charges (writing $\widetilde N \equiv N_f - N_c$
for the dual gauge rank), with superpotential $W = Mq\widetilde q$. The meson $M$ carries one
$SU(N_f)_L$ fundamental and one $SU(N_f)_R$ antifundamental index and is gauge-singlet. The baryon
charge of the dual quark is rescaled so that the dual baryon matches the electric baryon. The meson
fermion ($R_\psi = 1-2N_c/N_f$) is the field that closes every flavor and R anomaly match below.}
\label{tab:V03-magnetic}
\end{table}

Under $SU(N_f)_L$, the dual quark $q$ is an antifundamental (it carries
$-1$ in the cubic index) with $N_f-N_c$ colors, and the meson $M$ is a fundamental with $N_f$ ``colors''
(its $SU(N_f)_R$ index runs over $N_f$ values):
\begin{equation}
\label{eq:V03-magSUNfL3}
 SU(N_f)_L^3\big|_{\mathrm{mag}}
 \;=\; \underbrace{-(N_f-N_c)}_{\text{dual }q}
 \;+\; \underbrace{N_f}_{\text{meson }M}
 \;=\; N_c \;=\; SU(N_f)_L^3\big|_{\mathrm{el}}.
\end{equation}
The meson contribution is what closes the match: drop $M$ and the magnetic side gives $-(N_f-N_c)\ne N_c$.
The meson is essential. The opposite flavor group works the same way with the conjugate
representations,
\begin{equation}
\label{eq:V03-magSUNfR3}
 SU(N_f)_R^3\big|_{\mathrm{mag}}
 \;=\; \underbrace{+(N_f-N_c)}_{\text{dual }\widetilde q}
 \;-\; \underbrace{N_f}_{\text{meson }M}
 \;=\; -N_c \;=\; SU(N_f)_R^3\big|_{\mathrm{el}},
\end{equation}
the sign flipped because $\widetilde q$ is a fundamental and $M$ an antifundamental of $SU(N_f)_R$. The
cubic and gravitational R-traces match as exact identities in $(N_c,N_f)$. Summing the three magnetic
fermion species in Table~\ref{tab:V03-magnetic} (dual gaugino, $2N_f$ dual quarks, $N_f^2$ meson
components) gives
\begin{align}
\label{eq:V03-magTrR}
 \mathrm{Tr}\,R\big|_{\mathrm{mag}}
 &= \big[\widetilde N^2-1\big] + 2N_f\widetilde N\!\left(\tfrac{N_c}{N_f}-1\right)
 + N_f^2\!\left(1-\tfrac{2N_c}{N_f}\right) = -(N_c^2+1), \\[2pt]
\label{eq:V03-magTrR3}
 \mathrm{Tr}\,R^3\big|_{\mathrm{mag}}
 &= \big[\widetilde N^2-1\big] + 2N_f\widetilde N\!\left(\tfrac{N_c}{N_f}-1\right)^{\!3}
 + N_f^2\!\left(1-\tfrac{2N_c}{N_f}\right)^{\!3} = (N_c^2-1) - \frac{2N_c^4}{N_f^2},
\end{align}
with $\widetilde N \equiv N_f - N_c$ the dual rank of Table~\ref{tab:V03-magnetic}. Both sums equal the
electric \eqref{eq:V03-TrR} and \eqref{eq:V03-TrR3} (the algebra is a short exercise in expanding the
cubes, and the meson's $N_f^2$ term is indispensable in \eqref{eq:V03-magTrR3}). The mixed trace matches
as well,
\begin{equation}
\label{eq:V03-magmixed}
 SU(N_f)_L^2\,U(1)_R\big|_{\mathrm{mag}}
 \;=\; (N_f{-}N_c)\cdot\tfrac12\!\left(\tfrac{N_c}{N_f}-1\right)
 + N_f\cdot\tfrac12\!\left(1-\tfrac{2N_c}{N_f}\right)
 \;=\; -\frac{N_c^2}{2N_f},
\end{equation}
the electric value \eqref{eq:V03-SUNfR}, again with the meson's contribution included. We have now run
five explicit traces ($SU(N_f)_L^3$, $SU(N_f)_R^3$, $\mathrm{Tr}\,R$, $\mathrm{Tr}\,R^3$, and
$SU(N_f)_L^2 U(1)_R$), each matching electric to magnetic, and the meson is required in every one that
carries flavor or R charge.

\medskip\noindent\textbf{Check four: decoupling a flavor is an involution.}\enspace
Giving one electric quark a mass, $\Delta W = m\,Q_{N_f}\widetilde Q_{N_f}$, removes a flavor,
$N_f\to N_f-1$, with the holomorphic scales matched by \eqref{eq:V03-threshold}. On the magnetic side
the mass maps (under $M = Q\widetilde Q$) to a linear meson term $\Delta W = m\,M_{N_fN_f}$. Its F-term
$\partial W/\partial M_{N_fN_f} = q_{N_f}\widetilde q_{N_f} + m = 0$ forces a nonzero dual-quark vacuum
expectation value, which Higgses the dual gauge group: $SU(N_f-N_c)\to SU(N_f-1-N_c)$, lowering the
dual rank by one. The magnetic theory after Higgsing is the dual of the $N_f-1$ electric theory, so
dualizing it back returns
\begin{equation}
\label{eq:V03-involution}
 N_f - (N_f - N_c) \;=\; N_c,
\end{equation}
the original electric rank: the duality of the $(N_f-1)$-flavor theory is the original electric theory
with one flavor removed. The duality squares to the identity, a strong internal consistency that no
single anomaly trace could supply. Dualizing twice without removing a flavor likewise returns
$SU(N_f - (N_f-N_c)) = SU(N_c)$, with the meson-of-the-meson identified back as $Q\widetilde Q$.

\medskip\noindent\textbf{The conformal window and the phases.}\enspace
The window of behaviors as $N_f$ varies is the map every later section cites. It is collected in
Table~\ref{tab:V03-window}. The conformal window, where both the electric and magnetic frames flow to
the same interacting fixed point, is
\begin{equation}
\label{eq:V03-window}
 \tfrac{3}{2}\,N_c \;<\; N_f \;<\; 3\,N_c,
\end{equation}
with the upper edge $N_f = 3N_c$ the asymptotic-freedom boundary ($b_0 = 0$) and the lower edge
$N_f = \tfrac32 N_c$ where the magnetic theory loses asymptotic freedom and the dual becomes free
(below it, the magnetic $SU(N_f-N_c)$ is infrared-free with the meson, the ``free magnetic'' phase). The
upper edge is $3N_c$, not $2N_c$: at $N_c=3$ the window is $4.5 < N_f < 9$, so $N_f=5,6,7,8$ are all
conformal. Below the window the behaviors are sharper, and three of them are worth working explicitly:
the runaway, the quantum deformation, and s-confinement.

\subsection*{The Affleck--Dine--Seiberg runaway: $0 < N_f < N_c$}

For $0<N_f<N_c$ there is too little matter to lift the runaway, and a nonperturbative superpotential is
generated on the meson moduli space,
\begin{equation}
\label{eq:V03-ads}
 W_{\mathrm{ADS}} \;=\; (N_c-N_f)\left(\frac{\Lambda^{\,3N_c-N_f}}{\det M}\right)^{1/(N_c-N_f)},
\end{equation}
where $M = Q\widetilde Q$ is the $N_f\times N_f$ meson and the overall coefficient is fixed by the
symmetries up to a convention-dependent constant. The exponent is forced by holomorphy and the
R-symmetry, and that determination is worth doing. The superpotential must carry R-charge two. The
quark fermion R-charge here is $R_\psi(Q) = -N_c/N_f$ (the same formula \eqref{eq:V03-sqcdR}), so the
meson scalar and its determinant carry
\begin{equation}
\label{eq:V03-adsRdetM}
 R(M) = 2\big(1 - \tfrac{N_c}{N_f}\big), \qquad
 R(\det M) = N_f\,R(M) = 2(N_f - N_c),
\end{equation}
$\det M$ being a product of $N_f$ eigenvalues. A power $(\det M)^p$ then has R-charge $2p(N_f-N_c)$,
and combining with $\Lambda$ (R-neutral, mass dimension three per power) the unique holomorphic,
R-charge-two, dimension-three combination is
\begin{equation}
\label{eq:V03-adspower}
 W_{\mathrm{ADS}} \;\propto\; \Lambda^{(3N_c-N_f)/(N_c-N_f)}\,(\det M)^{-1/(N_c-N_f)},
\end{equation}
which is \eqref{eq:V03-ads} up to the constant: the power $1/(N_c-N_f)$ is the unique exponent that
makes $R(W)=2$ and gives $W$ mass dimension three. Take the F-term. On the diagonal slice $M =
v\,\mathds{1}_{N_f}$, $\det M = v^{N_f}$, so
\begin{equation}
\label{eq:V03-adsF}
 \frac{\partial W_{\mathrm{ADS}}}{\partial v}
 \;\propto\; \frac{\partial}{\partial v}\,v^{-N_f/(N_c-N_f)}
 \;\propto\; v^{-N_f/(N_c-N_f)-1}.
\end{equation}
This is a fixed-sign power of $v$: it never vanishes at finite $v$, and it decreases monotonically as
$v\to\infty$. There is no finite stationary point; the F-term is solved only as $v\to\infty$, that is
$\det M\to\infty$. The correct statement is therefore not ``no supersymmetric vacuum survives'' but
\begin{equation}
\label{eq:V03-adsrunaway}
 \text{no \emph{finite} supersymmetric vacuum; the meson runs away to } \det M \to \infty.
\end{equation}

\emph{Sanity check.} Contrast a Wess--Zumino mass term, whose F-term has a genuine finite zero,
\begin{equation}
\label{eq:V03-wzvacuum}
 W = \tfrac12 m v^2 - cv \;\Longrightarrow\; \frac{\partial W}{\partial v} = mv - c = 0
 \;\Longrightarrow\; v = \frac{c}{m}.
\end{equation}
The ADS F-term \eqref{eq:V03-adsF} has no such zero; its monotone fall-off to zero at infinity is the
runaway. A finite zero versus a monotone approach to zero at infinity is exactly the difference between
a vacuum and a runaway.

\emph{Worked case: $SU(3)$ with $N_f = 2$.} Here $3N_c - N_f = 7$ and $N_c - N_f = 1$, so the exponent
$1/(N_c-N_f) = 1$ and the ADS superpotential is a clean simple pole,
\begin{equation}
\label{eq:V03-adsSU3}
 W_{\mathrm{ADS}} \;=\; (N_c-N_f)\,\frac{\Lambda^{7}}{\det M}\ =\ \frac{\Lambda^{7}}{\det M},
\end{equation}
with $M$ the $2\times2$ meson. On the diagonal slice $M = v\,\mathds{1}_2$, $\det M = v^2$, so
\begin{equation}
\label{eq:V03-adsSU3F}
 W_{\mathrm{ADS}} \;\propto\; \Lambda^7 v^{-2}, \qquad
 \frac{\partial W_{\mathrm{ADS}}}{\partial v} \;\propto\; -2\,\Lambda^7 v^{-3}.
\end{equation}
The F-term has no zero at finite $v$ and falls to zero only as $v\to\infty$: the meson runs away,
exactly the general pattern made concrete. (The
runaway and the sign of the exponent are machine-checked across several $(N_c,N_f)$ in the runaway
range.)

\subsection*{The quantum-deformed moduli space: $N_f = N_c$}

At $N_f = N_c$ the matter is just enough to support a moduli space but not a runaway. Classically the
mesons and baryons satisfy
\begin{equation}
\label{eq:V03-classicalconstraint}
 \det M - B\widetilde B \;=\; 0,
\end{equation}
an identity that follows from the epsilon contractions ($M$ is $N_c\times N_c$ here, and $B,\widetilde B$
are the single baryon and antibaryon). Quantum mechanically the constraint is \emph{deformed},
\begin{equation}
\label{eq:V03-quantumconstraint}
 \det M - B\widetilde B \;=\; \Lambda^{2N_c}.
\end{equation}
Read the right-hand side. The exponent is $2N_c = b_0|_{N_f=N_c}$ (with $b_0 = 3N_c - N_f = 2N_c$), the
one-instanton factor: the deformation is generated at one instanton. The symmetry check is immediate.
The anomaly-free R-charge here is $R(Q) = 1 - N_c/N_c = 0$, so $\det M$ and $B\widetilde B$ are both
R-neutral, and $\Lambda^{2N_c}$ (a pure power of the holomorphic scale) is a singlet of
$SU(N_f)_L\times SU(N_f)_R\times U(1)_B\times U(1)_R$: every term in \eqref{eq:V03-quantumconstraint}
carries the same (trivial) global charges, so the deformation is allowed. The physical consequence is
that the \emph{origin is removed}: at $M = B = \widetilde B = 0$ the left side of
\eqref{eq:V03-quantumconstraint} is zero but the right side is $\Lambda^{2N_c}\ne 0$, so the classically
singular origin is not part of the quantum moduli space. Different regions of the quantum moduli
space realize different symmetry-breaking patterns: mesonic regions break the chiral flavor
symmetry, while the baryonic region keeps the nonabelian flavor symmetry but breaks $U(1)_B$. What
the deformation forbids is a single vacuum at $M = B = \widetilde B = 0$ that preserves all the
classical global symmetries at once.

The smallest case is fully explicit. At $N_c=N_f=2$ the meson $M$ is a $2\times2$ matrix and there is
one baryon $B$ and one antibaryon $\widetilde B$, so the constraint reads
\begin{equation}
\label{eq:V03-Nc2constraint}
 M_{11}M_{22} - M_{12}M_{21} - B\widetilde B \;=\; \Lambda^4.
\end{equation}
Two limiting branches illustrate the deformation,
\begin{equation}
\label{eq:V03-Nc2branches}
 \underbrace{\det M = \Lambda^4}_{B=\widetilde B=0,\ \text{chiral symmetry broken}}, \qquad
 \underbrace{B\widetilde B = -\Lambda^4}_{M=0,\ U(1)_B\ \text{broken}},
\end{equation}
the meson branch ($B=\widetilde B=0$) with the quarks condensed and the flavor symmetry broken to a
maximal subgroup, the baryon branch ($M=0$) preserving the nonabelian flavor symmetry. Neither branch
passes through
$M=B=\widetilde B=0$: the quantum constraint physically separates the chiral-symmetry-breaking and
baryon-number-breaking regions and forbids the symmetric origin. This is one of the canonical SQCD
calculations, not just a table row.

\subsection*{S-confinement: $N_f = N_c + 1$}

At $N_f = N_c+1$ the theory confines without breaking chiral symmetry, and the infrared degrees of
freedom are the smooth gauge invariants $M$ (an $(N_c+1)\times(N_c+1)$ meson), $B$, and $\widetilde B$
(the baryon and antibaryon, now $N_c+1$-component flavor vectors). They are governed by a dynamically
generated superpotential
\begin{equation}
\label{eq:V03-sconf}
 W \;=\; \frac{1}{\Lambda^{2N_c-1}}\big(\,B\,M\,\widetilde B - \det M\,\big),
\end{equation}
where the prefactor exponent $2N_c-1 = b_0|_{N_f=N_c+1}$ (with $b_0 = 3N_c - (N_c+1) = 2N_c-1$) again
carries the right instanton scaling. The F-terms of \eqref{eq:V03-sconf} reproduce the classical
constraints. Differentiating with respect to the meson,
\begin{equation}
\label{eq:V03-sconfFM}
 \frac{\partial W}{\partial M_{ij}} \;\propto\; B_i\widetilde B_j - (\mathrm{adj}\,M)_{ji} \;=\; 0,
\end{equation}
where $(\mathrm{adj}\,M)_{ji}$ is the cofactor (the derivative of $\det M$); this is the classical
relation $B_i\widetilde B_j = \mathrm{cofactor}_{ij}(M)$. The baryon F-terms give
\begin{equation}
\label{eq:V03-sconfFB}
 \frac{\partial W}{\partial B_i} \;\propto\; (M\widetilde B)_i = 0, \qquad
 \frac{\partial W}{\partial \widetilde B_j} \;\propto\; (B M)_j = 0.
\end{equation}
The point is not to prove s-confinement in general; it is to show how a single dynamically generated
superpotential encodes the entire set of moduli-space relations. \emph{Sanity check:} dropping the
$\det M$ term from \eqref{eq:V03-sconf} would leave
\begin{equation}
\label{eq:V03-sconfdrop}
 \frac{\partial W}{\partial M_{ij}}\Big|_{\det M\ \mathrm{dropped}} \;\propto\; B_i\widetilde B_j \;\ne\; (\mathrm{adj}\,M)_{ji},
\end{equation}
which is \emph{not} the classical relation; both terms are required for the F-terms to reproduce the
constraints. The two pieces $B M\widetilde B$ and $\det M$ are not independent decorations; together they
are the smooth chiral-ring relation.

\begin{table}[ht]
\centering
\small
\setlength{\tabcolsep}{6pt}
\renewcommand{\arraystretch}{1.25}
\begin{tabular}{@{}>{\raggedright\arraybackslash}p{2.7cm} >{\raggedright\arraybackslash}p{4.5cm} >{\raggedright\arraybackslash}p{4.7cm}@{}}
\toprule
flavor range & infrared behavior & signature \\
\midrule
$0 < N_f < N_c$ & runaway, no finite vacuum & ADS superpotential \eqref{eq:V03-ads}, $\det M\to\infty$\\
\midrule
$N_f = N_c$ & quantum-deformed moduli space & $\det M - B\widetilde B = \Lambda^{2N_c}$, origin removed\\
\midrule
$N_f = N_c + 1$ & s-confinement & smooth origin, dynamical $W$ \eqref{eq:V03-sconf}, F-terms = constraints\\
\midrule
$N_c + 1 < N_f \le \tfrac32 N_c$ & free magnetic & $SU(N_f - N_c)$ infrared-free with meson $M$\\
\midrule
$\tfrac32 N_c < N_f < 3N_c$ & conformal window & both frames flow to one fixed point\\
\midrule
$N_f \ge 3N_c$ & free electric / not asymptotically free & $b_0 \le 0$, no scale generated\\
\bottomrule
\end{tabular}
\caption{The Seiberg-duality map of $SU(N_c)$ SQCD as a function of the flavor number $N_f$. The upper
edge of the conformal window is $3N_c$, the asymptotic-freedom boundary, not $2N_c$. The four checks of
the text are run trace by trace in the magnetic frame for the conformal-window entries, and the three
sub-window phases (runaway, quantum deformation, s-confinement) are worked above.}
\label{tab:V03-window}
\end{table}

\begin{keybox}{Common misconception: Seiberg duality is not an identity of one theory}
The electric $SU(N_c)$ SQCD and the magnetic $SU(N_f-N_c)$ theory with meson $M$ and $W=Mq\widetilde q$
are \emph{different} ultraviolet theories: different gauge groups, different matter, an extra gauge
singlet. They flow to the \emph{same} infrared fixed point in the conformal window. The shared 't~Hooft
anomalies and chiral ring are evidence for that infrared equivalence, not a statement that the two
theories are the same in the ultraviolet. Anomaly matching is necessary, not sufficient.
\end{keybox}

\section{Dynamical supersymmetry breaking and the index}
\label{sec:V03-susybreaking}

Supersymmetry can be broken by the dynamics, but only when the dynamics allows it, and the Witten index
is the obstruction. If $\mathrm{Tr}\,(-1)^F \ne 0$, a nonzero index guarantees a zero-energy state that
supersymmetry pairs nowhere, so it cannot be broken. Pure super Yang--Mills is the worked instance: its
index $\mathrm{Tr}\,(-1)^F = N_c$ of \eqref{eq:V03-witten} is nonzero, so it confines without breaking
supersymmetry. Dynamical breaking therefore requires the index to vanish or to be inapplicable, which
the chiral, runaway, and metastable mechanisms arrange.

\medskip\noindent\textbf{The $3$-$2$ model, at field-content level.}\enspace
The canonical chiral example is the $3$-$2$ model: gauge group $SU(3)\times SU(2)$ with one quark
$Q$ in the $(\mathbf{3},\mathbf{2})$, antiquarks $\bar U,\bar D$ in the $(\overline{\mathbf{3}},
\mathbf{1})$, and a doublet $L$ in the $(\mathbf{1},\mathbf{2})$, with a tree-level superpotential
$W_{\mathrm{tree}} = h\,Q\bar D L$. The matter is chiral (it does not come in vector-like pairs), so
there is no gauge-invariant mass term and no continuous deformation back to a vector-like theory; the
Witten-index argument does not apply, and the door to breaking is open. Without the tree coupling the
model has classical flat directions. The tree superpotential lifts those directions, while the $SU(3)$
dynamics (with $N_f=2 < N_c=3$) generates an ADS-type superpotential of the form \eqref{eq:V03-ads};
the tree and dynamical terms then compete, and the combined F-term equations have no supersymmetric
solution, producing a stable, supersymmetry-breaking vacuum at a calculable scale when $h\ll 1$. The
structural point is that chiral matter evades the index obstruction and the result is genuine (not
metastable) breaking. We treat the detailed dynamics at statement level and cite the original analysis.

\medskip\noindent\textbf{A worked rank-condition fixture: the ISS model.}\enspace
The cleanest worked breaking is metastable, and it is a rank condition. Take $SU(N_c)$ SQCD with massive
flavors in the free magnetic range, $N_c+1 < N_f < \tfrac32 N_c$. Its magnetic dual is an infrared-free
$SU(N_f-N_c)$ theory with dual quarks $q,\widetilde q$, the meson $\Phi$ (an $N_f\times N_f$ matrix,
$\Phi = M/\mu$ canonically normalized), and a superpotential inherited from the electric mass term,
\begin{equation}
\label{eq:V03-iss}
 W \;=\; h\, q\,\Phi\,\widetilde q \;-\; h\,\mu^2\,\mathrm{Tr}\,\Phi.
\end{equation}
The F-term equation for the meson is
\begin{equation}
\label{eq:V03-issF}
 F_\Phi \;=\; h\,\big(q\,\widetilde q - \mu^2\,\mathds{1}_{N_f}\big) \;=\; 0,
\end{equation}
which would require the $N_f\times N_f$ bilinear $q\widetilde q$ to equal $\mu^2\mathds{1}_{N_f}$, a
full-rank matrix. But $q$ and $\widetilde q$ are $(N_f-N_c)\times N_f$ rectangular matrices, so
\begin{equation}
\label{eq:V03-issrank}
 \mathrm{rank}(q\,\widetilde q) \;\le\; N_f-N_c \;<\; N_f.
\end{equation}
The product $q\widetilde q$ cannot reach the rank-$N_f$ matrix $\mu^2\mathds{1}_{N_f}$: at most
$N_f-N_c$ of the $N_f$ diagonal F-term equations can be satisfied, and $N_c$ of them are left
unsatisfied. Diagonalizing, the best one can do is
\begin{equation}
\label{eq:V03-issbest}
 q\widetilde q = \mathrm{diag}(\underbrace{\mu^2,\dots,\mu^2}_{N_f-N_c},\underbrace{0,\dots,0}_{N_c})
 \;\Longrightarrow\; F_\Phi = -h\mu^2 \ \text{on the last } N_c \text{ entries,}
\end{equation}
with $N_f-N_c$ satisfied entries. The scalar potential is the sum of $|F|^2$ over all F-terms, so its
minimum is
\begin{equation}
\label{eq:V03-issV}
 V_{\min} \;=\; \sum_{\text{unsatisfied}} |F_\Phi|^2 \;=\; N_c\,|h\,\mu^2|^2 \;>\; 0,
\end{equation}
strictly positive, the unambiguous signal of broken supersymmetry: the energy of the lowest state is
above zero. This is the rank-condition mechanism: a full-rank F-term equation cannot be solved by a
manifestly lower-rank bilinear, and the rank deficit $N_c$ counts the unsatisfied F-terms, each
contributing $|h\mu^2|^2$ to the vacuum energy. (The would-be Goldstino sits in the
$\mathrm{Tr}\,\Phi$ direction whose F-term \eqref{eq:V03-issF} is the one that cannot be cancelled.)

The breaking here is metastable, not absolute. Supersymmetric vacua do reappear elsewhere in field
space, restored nonperturbatively where the meson runs out to large values, but they are separated from
the rank-condition vacuum by a barrier and it is parametrically long-lived. For a foundations section
this is enough: the rank condition shows \emph{how} the dynamics breaks supersymmetry, and the index
criterion shows \emph{when} it is allowed to.

\emph{Sanity check.} At full rank the obstruction disappears,
\begin{equation}
\label{eq:V03-issfullrank}
 N_c = 0 \;\Longrightarrow\; N_f - N_c = N_f \;\Longrightarrow\; \text{rank deficit } = N_f-(N_f-N_c) = 0,
\end{equation}
so the F-term \eqref{eq:V03-issF} \emph{is} solvable and supersymmetry is unbroken (no gauge group).
Were the bilinear allowed full rank, there would be no breaking; it is the geometry of rectangular
matrices, the rank deficit $N_c>0$, that forces the breaking.

\medskip\noindent\textbf{The supersymmetric index, named.}\enspace
The Witten index has a refined cousin, the supersymmetric (or superconformal) index on
$S^3 \times S^1$. It is a protected counting partition function $\mathrm{Tr}\,(-1)^F x^{\cdots}$ over
states annihilated by a chosen supercharge: it counts short multiplets with signs, is invariant under
continuous deformations, and matches across Seiberg duals, a refinement of the 't~Hooft anomalies. We
name its role and route the computation, the elliptic-gamma-function integral and the proofs of duality
invariance, to the further-reading literature. No index is computed here beyond the Witten count.

\section*{Exit checklist}
\addcontentsline{toc}{subsection}{Exit checklist}
\markboth{Exit checklist}{Exit checklist}

After this section the reader can
\begin{enumerate}
\item identify the multiplets and chiral ring of a $4d\ \mathcal{N}=1$ theory, read the dynamical scale
off $b_0 = 3N_c - \sum_i T(r_i)$, and match scales across a threshold by
$\Lambda_{N_f-1}^{3N_c-(N_f-1)} = m\,\Lambda_{N_f}^{3N_c-N_f}$;
\item compute pure-SYM gaugino condensation, the $\mathbb{Z}_{2N_c}\to\mathbb{Z}_2$ breaking, the $N_c$
vacua, and the Witten index $\mathrm{Tr}\,(-1)^F = N_c$;
\item assign a trial R-symmetry, apply the fermion shift $R_\psi = R_s - 1$, derive $R(Q) = 1-N_c/N_f$
from gauge-R anomaly freedom, and compute the 't~Hooft traces $\mathrm{Tr}\,R = -(N_c^2+1)$,
$\mathrm{Tr}\,R^3 = (N_c^2-1) - 2N_c^4/N_f^2$, $SU(N_f)_L^3 = +N_c$, $SU(N_f)_R^3 = -N_c$;
\item run $a$-maximization to the superconformal R, getting $a(3,6)=123/64$, $c(3,6)=163/64$ for SQCD,
the irrational extremum for adjoint SQCD (a root of the \emph{quadratic} $da/dR_X=0$), and the
symmetry-forced $R(X)=\tfrac23$ for the dP$_0$ quiver, and remember that it is $a$, not $c$, that is
maximized;
\item write the Seiberg dual of SQCD, verify the four checks (including the magnetic $SU(N_f)_L^3 =
-(N_f-N_c)+N_f = N_c$ with the meson essential), work the ADS runaway, the quantum-deformed constraint
$\det M - B\widetilde B = \Lambda^{2N_c}$, and the s-confining superpotential, and place a theory in the
window map of Table~\ref{tab:V03-window};
\item apply the Witten-index criterion for dynamical supersymmetry breaking, work the ISS rank
condition $\mathrm{rank}(q\widetilde q)\le N_f-N_c < N_f$, and state which deep theorems
($a$-maximization, Seiberg duality, the superconformal index) are cited rather than proved.
\end{enumerate}

\bigskip
\section*{Sources and notes}
\addcontentsline{toc}{subsection}{Sources and notes}
\markboth{Sources and notes}{Sources and notes}
{\small

\noindent\textsf{\textcolor{RoyalBlue}{Sources and notes.}}\enspace
This is the anchor dimension section of these notes, the field-theory home of the four-supercharge
holomorphic toolkit.

\medskip\noindent\textsf{\textcolor{RoyalBlue}{\textbf{\S\ref{sec:V03-multiplets}\enspace Multiplets, holomorphy, scale, threshold.}}}\enspace
The chiral/vector multiplets, the holomorphic $\tau$ \eqref{eq:V03-tau}, the one-loop coefficient
$b_0 = 3N_c - \sum_i T(r_i)$ \eqref{eq:V03-b0} ($=3N_c-N_f$ for SQCD, \eqref{eq:V03-b0sqcd}), the
RG-invariant scale $\Lambda^{b_0} = \mu^{b_0}e^{2\pi i\tau}$ \eqref{eq:V03-lambda}, and the threshold
matching $\Lambda_{N_f-1}^{3N_c-(N_f-1)} = m\,\Lambda_{N_f}^{3N_c-N_f}$ \eqref{eq:V03-threshold} (the
exponent shifts by the removed Dynkin index $2T(\square)=1$); the gauge-group-versus-algebra aside.
(\textcite{Tachikawa:2018sae} the holomorphy/$\Lambda$ spine). 

\medskip\noindent\textsf{\textcolor{RoyalBlue}{\textbf{\S\ref{sec:V03-chiralring}\enspace The chiral ring and non-renormalization.}}}\enspace
The chiral ring as gauge-invariant chiral words modulo $\partial W = 0$, the meson \eqref{eq:V03-meson}
and baryon \eqref{eq:V03-baryon} generators and their gradings; the non-renormalization of $W$ and the
one-loop-exact holomorphic gauge coupling. (\textcite{Intriligator:1995au} the SQCD
chiral-ring and non-renormalization review). 

\medskip\noindent\textsf{\textcolor{RoyalBlue}{\textbf{\S\ref{sec:V03-sym}\enspace Pure super Yang--Mills.}}}\enspace
The anomalous $U(1)_R\to\mathbb{Z}_{2N_c}$ \eqref{eq:V03-Z2Nc}, gaugino condensation
$\langle\lambda\lambda\rangle_k \sim \Lambda^3 e^{2\pi ik/N_c}$ \eqref{eq:V03-gaugino} breaking
$\mathbb{Z}_{2N_c}\to\mathbb{Z}_2$, the $N_c$ vacua, the Witten index $\mathrm{Tr}\,(-1)^F = N_c$
\eqref{eq:V03-witten}, and the optional Veneziano--Yankielowicz $W_{\mathrm{VY}}$ \eqref{eq:V03-VY}
recovering $S^{N_c}\sim\Lambda^{3N_c}$; the condensate is unbroken-SUSY zero-energy vacua, not a
SUSY-breaking order parameter; the carefully stated scale-matching pointer to the $4d\ \mathcal{N}=2$ section (the
$\mathcal{N}=2$ Seiberg--Witten scale is $\mathcal{N}=2$ transmutation, related to this condensate only
after an $\mathcal{N}=1$ mass deformation). (\textcite{Witten:1982df} the index;
\textcite{Tachikawa:2018sae} the condensate / vacuum count). 

\medskip\noindent\textsf{\textcolor{RoyalBlue}{\textbf{\S\ref{sec:V03-rsymmetry}\enspace R-symmetry, NSVZ, the superconformal R.}}}\enspace
The fermion R-rule $R_\psi = R_s - 1$ \eqref{eq:V03-rshift} (gaugino $R=1$), the full NSVZ beta function
\eqref{eq:V03-nsvz} and its numerator condition \eqref{eq:V03-nsvznum}, the ABJ
$SU(N_c)^2$-$U(1)_R = 0$ condition \eqref{eq:V03-abj} solving $R(Q) = 1 - N_c/N_f$ \eqref{eq:V03-sqcdR},
and the superconformal R as the constrained abelian combination with $\Delta = \tfrac32 R$ (back to
Section~1). (\textcite{Novikov:1983uc} the NSVZ beta function). 

\medskip\noindent\textsf{\textcolor{RoyalBlue}{\textbf{\S\ref{sec:V03-anomaly}\enspace 't~Hooft matching and the Konishi anomaly.}}}\enspace
't~Hooft anomalies as renormalization-group invariants, the electric SQCD table (Table~\ref{tab:V03-electric}) with the declared $SU(N_f)_R$
antifundamental convention, the L/R-split cubic traces $SU(N_f)_L^3 = +N_c$, $SU(N_f)_R^3 = -N_c$
\eqref{eq:V03-cubicLR}, the explicitly run $\mathrm{Tr}\,R = -(N_c^2+1)$ \eqref{eq:V03-TrR},
$\mathrm{Tr}\,R^3 = (N_c^2-1) - 2N_c^4/N_f^2$ \eqref{eq:V03-TrR3}, $SU(N_f)_L^2 U(1)_R = -N_c^2/(2N_f)$
\eqref{eq:V03-SUNfR}, and the Konishi anomaly \eqref{eq:V03-konishi} with its coefficient equal to the
gauge index feeding the ABJ trace. (\textcite{tHooft:1979rat} anomaly matching;
\textcite{Konishi:1983hf} the Konishi anomaly). 

\medskip\noindent\textsf{\textcolor{RoyalBlue}{\textbf{\S\ref{sec:V03-amax}\enspace a-maximization.}}}\enspace
The central charges $a,c$ \eqref{eq:V03-ac} as fermion traces over a trial R, the local-maximum
selection of the superconformal R (it is $a$, not $c$; the constraints are gauge-anomaly freedom and
$R(W)=2$, the 't~Hooft anomalies being the data, not vanishing constraints); the SQCD rational functions
\eqref{eq:V03-sqcd-ac} and the numeric $a(3,6)=123/64$, $c(3,6)=163/64$ \eqref{eq:V03-sqcd-numeric} with
$a\ne c$; the genuine adjoint-SQCD extremum (the trial $a(R_X)$ cubic so $da/dR_X=0$ is \emph{quadratic};
the maximizing root irrational, Hessian $<0$, $c$ not maximized); and the dP$_0$ symmetric
$R(X)=\tfrac23$ \eqref{eq:V03-dp0W}--\eqref{eq:V03-dp0node}; the Banks--Zaks fixed point (the weakly
coupled entry to the conformal window at $N_f = 3N_c - \epsilon N_c$) is the point where perturbation
theory controls the fixed point and checks the exact anomaly/$a$-maximization R. (\textcite{Intriligator:2003jj} $a$-maximization; \textcite{Banks:1981nn}
the fixed point; \textcite{Lawrence:1998ja} the orbifold-daughter quiver). 

\medskip\noindent\textsf{\textcolor{RoyalBlue}{\textbf{\S\ref{sec:V03-seiberg}\enspace Seiberg duality and the conformal window.}}}\enspace
The magnetic dual $SU(N_f - N_c)$ with meson $M$ and $W = Mq\widetilde q$ \eqref{eq:V03-seibergW}; the
four checks (global symmetries, chiral operators, 't~Hooft anomalies with the explicit magnetic
$SU(N_f)_L^3 = -(N_f-N_c)+N_f = N_c$ \eqref{eq:V03-magSUNfL3} and the matched R-traces
\eqref{eq:V03-magTrR}, the meson essential, flavor-decoupling involution \eqref{eq:V03-involution}); the
conformal window $\tfrac32 N_c < N_f < 3N_c$ \eqref{eq:V03-window}; the worked sub-window phases, the ADS
runaway \eqref{eq:V03-ads}--\eqref{eq:V03-adsrunaway} (no finite vacuum, $\det M\to\infty$, F-term
\eqref{eq:V03-adsF}), the quantum-deformed constraint \eqref{eq:V03-quantumconstraint} (origin removed),
the s-confining superpotential \eqref{eq:V03-sconf} and its F-terms \eqref{eq:V03-sconfFM}--\eqref{eq:V03-sconfFB};
and the full flavor map (Table~\ref{tab:V03-window}). (\textcite{Seiberg:1994pq} the duality +
window; \textcite{Affleck:1983mk} the ADS superpotential). 

\medskip\noindent\textsf{\textcolor{RoyalBlue}{\textbf{\S\ref{sec:V03-susybreaking}\enspace Dynamical SUSY breaking and the index.}}}\enspace
The Witten-index obstruction ($\mathrm{Tr}\,(-1)^F \ne 0 \Rightarrow$ unbroken), worked on pure SYM; the
ISS rank-condition fixture \eqref{eq:V03-iss}--\eqref{eq:V03-issrank} (the F-term
$F_\Phi = h(q\widetilde q - \mu^2\mathds{1}_{N_f})$ unsolvable because
$\mathrm{rank}(q\widetilde q)\le N_f-N_c < N_f$, metastable breaking, SUSY vacua restored
nonperturbatively); the $3$-$2$ model named at statement level; the superconformal index named and
not computed here. (\textcite{Affleck:1984xz} the $3$-$2$ model;
\textcite{Intriligator:2006dd} the metastable vacuum). 

\medskip\noindent\textbf{Stated and cited, not proved here.}\enspace
The $a$-maximization / $a$-theorem / superconformal-R theorem and the accidental-symmetry correction; the Seiberg-duality theorem and the full quantum moduli spaces; the superconformal
index computation, the elliptic-gamma integral, and its duality invariance; and the gauge
global-form / line-operator classification. All are stated and cited here; none is proved from first
principles in this section.
}

\subsection*{Further reading}
\addcontentsline{toc}{subsection}{Further reading}
The exact holomorphic structure of $4d\ \mathcal{N}=1$ gauge theories is developed in the lectures
\textcite{Intriligator:1995au}; exactly marginal deformations and superconformal manifolds in
\textcite{Leigh:1995ep}; the glueball effective superpotential in \textcite{Veneziano:1982ah}. The
central charges $a,c$ and $a$-maximization are treated in \textcite{Anselmi:1997am}. Dynamical
supersymmetry breaking, including the metastable-vacua mechanism, is reviewed in
\textcite{Intriligator:2007cp}. The superconformal index that counts protected operators is introduced
in \textcite{Romelsberger:2005eg,Kinney:2005ej} and developed in \textcite{Dolan:2008qi}.

For modern entry points, \textcite{Gadde:2020yah} gives a graduate-level introduction to the
superconformal index, \textcite{Smilga:2023dlp} reviews Witten-index calculations across
four-dimensional supersymmetric gauge theories, and \textcite{Bajeot:2023gyl} illustrates how
the classical s-confinement and duality toolkit continues to generate new $\mathcal{N}=1$ dynamics.

\section*{References}
\addcontentsline{toc}{subsection}{References}
\markboth{References}{References}
\printbibliography[heading=none]
\end{refsection}
\begin{refsection}\chapter{\texorpdfstring{$2d$}{2d} supersymmetric field theories}
\label{ch:V04}

\noindent\textbf{Guide to this section.}\enspace
Sections~1 and~2 fixed the algebra and the common words, and Section~3 built the
four-supercharge holomorphic world of $4d\ \mathcal{N}=1$ field theory with its extremization
principle. This section builds the two-dimensional world, as a subject to be owned rather than a list
to be surveyed. Two dimensions is where supersymmetry splits by worldsheet chirality, and that split
is the source of everything special here. The section is two complete worlds and a shared interface.
Block A is the balanced four-supercharge $\mathcal{N}=(2,2)$ world: two holomorphic sectors ($W$ and $\widetilde
W$), the gauged linear sigma model as dynamics, and the A/B twists with mirror symmetry. Block B is
the chiral two-supercharge $\mathcal{N}=(0,2)$ world: Fermi multiplets with independent $E/J$ data, holomorphic
bundles, gauge and gravitational anomalies, $c$-extremization, and triality. Block C is the shared
interface: the elliptic genus and a comparison table. Each structure is developed the same way: why it
has to appear, its precise definition, one canonical calculation done in full, what goes wrong if it is
misused. The deep theorems (the phase, triality, $c$-extremization, and elliptic-genus modularity
theorems) are stated and cited; the working calculations are done here. Where the argument needs the chiral ring, the
trial R-symmetry, or the extremization principle, it recalls Section~3 rather than re-deriving it.

\begin{keybox}{What this section delivers}
The $2d$ SUSY kinematics: the $(p,q)$ count, worldsheet chirality, $\mathcal{N}=(2,2)$ versus $\mathcal{N}=(0,2)$, and the
R-symmetries $U(1)_V$, $U(1)_A$ (\S\ref{sec:V04-kinematics}). Block A, the $\mathcal{N}=(2,2)$ world: the two
holomorphic sectors $W$ versus $\widetilde W$ (\S\ref{sec:V04-22data}); the gauged linear sigma model,
the $\mathbb{CP}^{N-1}$ quantum cohomology $\mathbb{C}[\Sigma]/(\Sigma^N - q)$, and the quintic phases
(\S\ref{sec:V04-glsm}); the A/B twists and the Hori--Vafa mirror (\S\ref{sec:V04-twist}). Block B, the
$\mathcal{N}=(0,2)$ world: the $E/J$ scalar potential and closure $\sum_a \mathrm{Tr}(E_a J^a) = 0$
(\S\ref{sec:V04-ej}); bundles and the gauged $\mathcal{N}=(2,2)\to\mathcal{N}=(0,2)$ $E$-term (\S\ref{sec:V04-bundles});
anomalies and $c$-extremization to $c_R = 3k/(k+2)$ (\S\ref{sec:V04-cext}); and triality
(\S\ref{sec:V04-triality}). Block C, the shared interface: the elliptic genus
(\S\ref{sec:V04-ellgenus}) and the $\mathcal{N}=(2,2)$-versus-$\mathcal{N}=(0,2)$ comparison (\S\ref{sec:V04-compare}).
\end{keybox}

\section[Kinematics and R-symmetries]{\texorpdfstring{$2d$}{2d} kinematics and R-symmetries}
\label{sec:V04-kinematics}

Everything special about two dimensions starts from one fact: a spinor in two dimensions can be
Majorana--Weyl. A Majorana--Weyl spinor carries a single real component and a definite worldsheet
chirality, left-moving or right-moving. A supercharge inherits that chirality, and so the amount of
supersymmetry cannot be quoted by a single number as in four dimensions. It is a pair $(p,q)$ with
$p$ left-moving and $q$ right-moving real supercharges, and the total count is the sum
\begin{equation}
\label{eq:V04-pqcount}
 \#\,\text{supercharges} \;=\; p + q,
\end{equation}
the $2d$ Majorana--Weyl reality row of Section~1. This section treats the two ends of the
four-supercharge family. The maximal case is $\mathcal{N}=(2,2)$,
\begin{equation}
\label{eq:V04-22count}
 \mathcal{N}=(2,2): \quad p = 2,\ q = 2, \qquad \#\,\text{supercharges} \;=\; 4,
\end{equation}
the dimensional reduction of $4d\ \mathcal{N}=1$ on a two-torus: the four real supercharges survive,
now sorted into two right-moving and two left-moving,
\begin{equation}
\label{eq:V04-reduction}
 4d\ \mathcal{N}=1 \;\xrightarrow{\ T^2\ }\; 2d\ \mathcal{N}=(2,2), \qquad 4 = 2 + 2,
\end{equation}
the four-supercharge row that reads $4d\ \mathcal{N}=1 \equiv 3d\ \mathcal{N}=2 \equiv 2d\ \mathcal{N}=(2,2)$ on
the Section~1 ladder. The minimal case is $\mathcal{N}=(0,2)$,
\begin{equation}
\label{eq:V04-02count}
 \mathcal{N}=(0,2): \quad p = 0,\ q = 2, \qquad \#\,\text{supercharges} \;=\; 2,
\end{equation}
two real supercharges, all right-moving. This is the chiral, holomorphic half of $\mathcal{N}=(2,2)$: it keeps
the two right-moving supercharges and discards the two left-moving ones. It has no
four-dimensional Lorentz-invariant parent, since there is no way to sort four-dimensional
supercharges into right-movers and left-movers, so its structures are the genuinely two-dimensional
part of the section.

\medskip\noindent\textbf{The section map.}\enspace
The split by chirality organizes the section. Block A is the $\mathcal{N}=(2,2)$ world, four balanced supercharges,
Kähler and complex geometry: the two holomorphic sectors (\S\ref{sec:V04-22data}), the gauged linear
sigma model (\S\ref{sec:V04-glsm}), the A/B twists and mirror symmetry (\S\ref{sec:V04-twist}). Block B
is the $\mathcal{N}=(0,2)$ world, the chiral half with two right-movers and no left-movers, a holomorphic bundle
over a target: $E/J$ data (\S\ref{sec:V04-ej}), bundles and the gauged $E$-term
(\S\ref{sec:V04-bundles}), anomalies and $c$-extremization (\S\ref{sec:V04-cext}), triality
(\S\ref{sec:V04-triality}). Block C is the shared interface: the elliptic genus
(\S\ref{sec:V04-ellgenus}), a $\mathcal{N}=(0,2)$ object that specializes to $\mathcal{N}=(2,2)$, and a comparison table
(\S\ref{sec:V04-compare}). Each structure sits in the world it belongs to: mirror symmetry and the
twists are $\mathcal{N}=(2,2)$-only, triality and $c$-extremization are $\mathcal{N}=(0,2)$-central, the elliptic genus is
genuinely shared.

\medskip\noindent\textbf{The two R-symmetries.}\enspace
Why should two dimensions carry \emph{two} abelian R-symmetries where four dimensions carries one?
Because the two right-moving and two left-moving supercharges can be rotated by two commuting phases,
not one. A $\mathcal{N}=(2,2)$ theory has an R-symmetry group that contains two commuting $U(1)$ factors, the
\emph{vector} R-symmetry $U(1)_V$ and the \emph{axial} R-symmetry $U(1)_A$. The vector R-symmetry
acts with the same sign on the left- and right-moving supercharges; the axial R-symmetry acts with
opposite signs. Writing the two supercharge phases as $Q_\pm \to e^{i\alpha_\pm} Q_\pm$ (the
subscript is the worldsheet chirality), the two combinations are
\begin{equation}
\label{eq:V04-VA}
 U(1)_V: \ \alpha_+ = \alpha_- = \alpha_V, \qquad
 U(1)_A: \ \alpha_+ = -\alpha_- = \alpha_A,
\end{equation}
the vector rotation the diagonal, the axial the anti-diagonal. This is not bookkeeping. The two
R-symmetries distinguish the two topological twists of \S\ref{sec:V04-twist} (A uses $U(1)_V$, B uses
$U(1)_A$), and the axial one is the fragile factor that can be anomalous. Its anomaly is proportional
to the first Chern class of a sigma-model target, so $U(1)_A$ survives precisely on a Calabi--Yau,
which is why the B-twist (and only the B-twist) is gated by the Calabi--Yau condition, and why the
sigma-model coupling can run when only $U(1)_V$ survives.

The two sectors carry opposite axial charge. Under $U(1)_V$ a chiral $\Phi$ and a twisted-chiral
$\Sigma$ both carry their vector R-charge, and $W(\Phi)$, $\widetilde W(\Sigma)$ must each be
$U(1)_V$-neutral as a superspace integral. Under $U(1)_A$ the chiral and twisted-chiral fields carry
opposite sign, because the axial rotation distinguishes $D_+$ from $D_-$. So the two rings of
\S\ref{sec:V04-22data} are graded by different R-symmetries, the algebraic root of mirror symmetry,
which exchanges the two R-symmetries and so the two sectors,
\begin{equation}
\label{eq:V04-mirrorexchange}
 U(1)_V \;\leftrightarrow\; U(1)_A, \qquad
 \Phi \;\leftrightarrow\; \Sigma, \qquad
 W \;\leftrightarrow\; \widetilde W.
\end{equation}
Carry away one fact: two dimensions has two abelian R-symmetries because it has two holomorphic
sectors, and each sector is graded by the R-symmetry its superspace measure respects.

\medskip\noindent\textbf{The multiplet vocabulary.}\enspace
The multiplets are graded by which combinations of the supersymmetry covariant derivatives annihilate
them. Three multiplets carry the $\mathcal{N}=(2,2)$ content.
\begin{itemize}
\item \emph{Chiral} $\Phi$, annihilated by both barred derivatives $\bar D_\pm \Phi = 0$. Content: a
complex scalar $\phi$, a Dirac fermion (one right-mover $\psi_+$ and one left-mover $\psi_-$), and an
auxiliary field. It is the reduction of the $4d\ \mathcal{N}=1$ chiral multiplet of Section~2, and it
carries the ordinary superpotential $W(\Phi)$.
\item \emph{Twisted-chiral} $\Sigma$, annihilated by the \emph{opposite} pair $\bar D_+ \Sigma = D_-
\Sigma = 0$. It has the same field content as a chiral (a complex scalar, a Dirac fermion, an
auxiliary) and the opposite constraint. It has no four-dimensional analog. It is indispensable
because the field strength of a $\mathcal{N}=(2,2)$ vector, $\Sigma = \bar D_+ D_- V$, is twisted-chiral, and its
holomorphic self-coupling is the \emph{twisted} superpotential $\widetilde W(\Sigma)$, developed in
\S\ref{sec:V04-22data}.
\item \emph{Vector} $V$, carrying the gauge field, a Dirac gaugino, and the auxiliary $D$-field. In
two dimensions the gauge field has no propagating polarization, so the physical content of the vector
sits in its twisted-chiral field strength $\Sigma$.
\end{itemize}
The two chiralities are the two constraints
\begin{equation}
\label{eq:V04-22constraints}
 \text{chiral}: \ \bar D_\pm \Phi = 0, \qquad
 \text{twisted-chiral}: \ \bar D_+ \Sigma = D_- \Sigma = 0, \qquad
 \Sigma = \bar D_+ D_- V,
\end{equation}
the opposite pairing $\bar D_+, D_-$ marking the twisted sector. The $\mathcal{N}=(0,2)$ algebra has one barred
right-moving derivative $\bar D_+$, and its three multiplets are graded by worldsheet chirality.
\begin{itemize}
\item \emph{Chiral} $\Phi$, annihilated by $\bar D_+$. Content: a complex scalar $\phi$ and a single
right-moving Weyl fermion $\psi_+$, nothing left-moving.
\item \emph{Fermi} $\Lambda$, the genuinely new object. Content: a single left-moving Weyl fermion
$\lambda_-$ and \emph{no} propagating scalar. It is only constrained chiral, $\bar D_+ \Lambda = E$,
with $E$ a holomorphic function of the chiral fields; $E = 0$ is a free Fermi multiplet, and a
nonzero $E$ is one of the two pieces of $\mathcal{N}=(0,2)$ interaction data of \S\ref{sec:V04-ej}.
\item \emph{Vector} $V$, carrying the gauge field and a left-moving gaugino.
\end{itemize}
The two chiral $\mathcal{N}=(0,2)$ constraints are
\begin{equation}
\label{eq:V04-02constraints}
 \text{chiral}: \ \bar D_+ \Phi = 0, \qquad
 \text{Fermi}: \ \bar D_+ \Lambda = E(\Phi),
\end{equation}
the Fermi only constrained chiral, its right-hand side $E$ the first piece of $\mathcal{N}=(0,2)$ interaction
data. The Fermi multiplet is what makes two dimensions chiral: it supplies a left-moving fermion with
no bosonic partner, so a generic $\mathcal{N}=(0,2)$ spectrum has a right-minus-left imbalance with no analog in
four dimensions.

\medskip\noindent\textbf{The $\mathcal{N}=(2,2)\to\mathcal{N}=(0,2)$ decomposition.}\enspace
Every $\mathcal{N}=(2,2)$ multiplet is a sum of $\mathcal{N}=(0,2)$ multiplets, and the bookkeeping is a clean first check.
Count the content of a $\mathcal{N}=(2,2)$ chiral: one complex scalar (two real bosonic fields), one right-moving
fermion, and one left-moving fermion, written $(\text{boson}, \text{right}, \text{left}) = (2,1,1)$.
A $\mathcal{N}=(0,2)$ chiral has content $(2,1,0)$: the scalar and the right-mover, no left-mover. A $\mathcal{N}=(0,2)$
Fermi has content $(0,0,1)$: the missing left-mover, nothing else. So
\begin{equation}
\label{eq:V04-decompchiral}
\begin{aligned}
 \mathcal{N}=(2,2)\ \text{chiral} &\;=\; \mathcal{N}=(0,2)\ \text{chiral} \;+\; \mathcal{N}=(0,2)\ \text{Fermi}, \\
 (2,1,1) &= (2,1,0) + (0,0,1),
\end{aligned}
\end{equation}
component by component. The Fermi is not optional in the decomposition: dropping it would leave the
$\mathcal{N}=(2,2)$ chiral without its left-moving fermion, breaking the count. Likewise a $\mathcal{N}=(2,2)$ vector splits
as
\begin{equation}
\label{eq:V04-decompvector}
 \mathcal{N}=(2,2)\ \text{vector} \;=\; \mathcal{N}=(0,2)\ \text{vector} \;+\; \mathcal{N}=(0,2)\ \text{adjoint chiral},
\end{equation}
the $\mathcal{N}=(0,2)$ vector carrying the gauge field and the left-moving gaugino, the adjoint $\mathcal{N}=(0,2)$ chiral
carrying the complex scalar and right-moving fermion. This is how any $\mathcal{N}=(2,2)$ theory is read as a
$\mathcal{N}=(0,2)$ theory, and it is exactly why the anomaly cancellations of \S\ref{sec:V04-cext} are automatic
for $\mathcal{N}=(2,2)$ theories: each $\mathcal{N}=(2,2)$ chiral brings a $\mathcal{N}=(0,2)$ chiral and a $\mathcal{N}=(0,2)$ Fermi in matched
pairs, so their chiral contributions cancel term by term. Table~\ref{tab:V04-multiplets} collects the
content.

\begin{table}[ht]
\centering
\small
\setlength{\tabcolsep}{7pt}
\renewcommand{\arraystretch}{1.3}
\begin{tabular}{@{}llccc@{}}
\toprule
supersymmetry & multiplet & complex scalar & right fermion $\psi_+$ & left fermion $\lambda_-$ \\
\midrule
$\mathcal{N}=(2,2)$ & chiral $\Phi$ & $1$ & $1$ & $1$ \\
\midrule
$\mathcal{N}=(2,2)$ & twisted-chiral $\Sigma$ & $1$ & $1$ & $1$ \\
\midrule
$\mathcal{N}=(2,2)$ & vector $V$ ($+\,A_\mu$) & $1$ & $1$ & $1$ \\
\midrule
$\mathcal{N}=(0,2)$ & chiral $\Phi$ & $1$ & $1$ & $0$ \\
\midrule
$\mathcal{N}=(0,2)$ & Fermi $\Lambda$ & $0$ & $0$ & $1$ \\
\midrule
$\mathcal{N}=(0,2)$ & vector $V$ & $0$ & $0$ & $1$ (gaugino) \\
\bottomrule
\end{tabular}
\caption{The $2d$ multiplet content by worldsheet chirality. A $\mathcal{N}=(2,2)$ chiral $= \mathcal{N}=(0,2)$ chiral $+
\mathcal{N}=(0,2)$ Fermi (the right-fermion goes with the scalar, the left-fermion is the Fermi); a $\mathcal{N}=(2,2)$
vector $= \mathcal{N}=(0,2)$ vector $+ \mathcal{N}=(0,2)$ adjoint chiral, and it also carries the gauge field $A_\mu$, which
has no propagating polarization in $2d$. The twisted-chiral $\Sigma$ is the gauge-invariant field
strength of the $\mathcal{N}=(2,2)$ vector: same field content as a chiral, opposite constraint $\bar D_+ \Sigma
= D_- \Sigma = 0$, and it carries the complexified Fayet--Iliopoulos and theta parameter $t = r -
i\theta$ in its twisted superpotential.}
\label{tab:V04-multiplets}
\end{table}

\begin{keybox}{Common misconception: $\mathcal{N}=(0,2)$ is ``the $\mathcal{N}=2$ of two dimensions''}
The pair $\mathcal{N}=(0,2)$ names \emph{two} supercharges, not four. Reading it as ``$\mathcal{N}=2$'' by analogy
with four dimensions, and so expecting four supercharges, confuses it with $\mathcal{N}=(2,2)$, which is the
genuine four-supercharge algebra. The two indices are the left-moving and right-moving counts, so
$\mathcal{N}=(0,2)$ is two right-movers and zero left-movers; the total is $p + q = 2$. Likewise $(4,4)$, with
eight supercharges, is a different and larger algebra, outside this section. The label counts the two
chiralities separately precisely because $2d$ supersymmetry is graded by worldsheet chirality.
\end{keybox}

\bigskip
\begin{center}
\rule{0.4\textwidth}{0.4pt}\\[3pt]
{\large\textsf{\textbf{Block A.\enspace The $\mathcal{N}=(2,2)$ world}}}\\[2pt]
\rule{0.4\textwidth}{0.4pt}
\end{center}
\medskip

\noindent The next three sections build $\mathcal{N}=(2,2)$ dynamics as one coherent world: its two holomorphic
sectors and rings (\S\ref{sec:V04-22data}), the gauged linear sigma model that turns those data into
Calabi--Yau geometry (\S\ref{sec:V04-glsm}), and the two topological twists with mirror symmetry that
compute the two rings and exchange them (\S\ref{sec:V04-twist}). All of it is $\mathcal{N}=(2,2)$-specific; the
$\mathcal{N}=(0,2)$ world of Block B keeps only the pieces that survive discarding two supercharges.

\section{The \texorpdfstring{$\mathcal{N}=(2,2)$}{N=(2,2)} holomorphic sectors and their rings}
\label{sec:V04-22data}

A $4d\ \mathcal{N}=1$ theory has one holomorphic self-coupling, the superpotential. A $2d\ \mathcal{N}=(2,2)$
theory has two, and understanding why is the first genuinely two-dimensional lesson. The reason is
the twisted-chiral multiplet: because the two right-moving and two left-moving supercharges combine in
two inequivalent ways, there are two kinds of holomorphic constraint a multiplet can satisfy, and each
supports its own superspace integral.

\medskip\noindent\textbf{Two holomorphic sectors.}\enspace
A chiral field $\Phi$ obeys $\bar D_\pm \Phi = 0$ and couples through the ordinary superpotential, a
holomorphic function integrated over half of superspace,
\begin{equation}
\label{eq:V04-Wordinary}
 W(\Phi), \qquad \int d^2\theta\; W(\Phi) + \text{c.c.},
\end{equation}
exactly as in four dimensions. A twisted-chiral field $\Sigma$ obeys the opposite constraint $\bar
D_+ \Sigma = D_- \Sigma = 0$ and couples through the \emph{twisted} superpotential, a holomorphic
function integrated over the \emph{other} half of superspace,
\begin{equation}
\label{eq:V04-Wtwisted}
 \widetilde W(\Sigma), \qquad \int d\theta^+ d\bar\theta^-\; \widetilde W(\Sigma) + \text{c.c.}
\end{equation}
The two constraints and the two measures are genuinely separate data. Holding $W$ and $\widetilde W$
apart is the whole reason for the twisted-chiral distinction: a $\mathcal{N}=(2,2)$ gauge theory carries
\emph{both} an ordinary superpotential for its matter and a twisted superpotential for its field
strength, and they never mix.

\medskip\noindent\textbf{FI and theta as twisted-chiral data.}\enspace
Here is the canonical use of the twisted sector. The field strength of a $\mathcal{N}=(2,2)$ vector is the
twisted-chiral $\Sigma = \bar D_+ D_- V$, and its lowest allowed self-coupling is the \emph{linear}
twisted superpotential
\begin{equation}
\label{eq:V04-twistedt}
 \widetilde W(\Sigma) \;=\; t\,\Sigma, \qquad t \;=\; r - i\,\theta.
\end{equation}
The single coefficient $t$ packages two real parameters. Expanding $\int d\theta^+ d\bar\theta^-\,
t\,\Sigma$ into components, the real part gives $r\,D$ (the Fayet--Iliopoulos coupling to the auxiliary
$D$-field), the imaginary part $\theta$ times the field strength,
\begin{equation}
\label{eq:V04-tsplit}
 r \;=\; \mathrm{Re}\,t \quad(\text{FI}), \qquad \theta \;=\; -\mathrm{Im}\,t \quad(\text{theta}).
\end{equation}
So the FI parameter and theta angle are the coefficient of the lowest twisted superpotential, and $t$
is the Kähler modulus labelling the chambers of the gauged linear sigma model in \S\ref{sec:V04-glsm}.
Because the field strength is twisted-chiral, its parameters sit in $\widetilde W$, not $W$.

\medskip\noindent\textbf{Two rings.}\enspace
The two holomorphic sectors have two chiral rings, computed with the two nilpotent supercharges. The
\emph{chiral ring} is the cohomology of the chiral supercharge, the gauge-invariant chiral operators
modulo the F-term relations $\partial W = 0$; the B-twist of \S\ref{sec:V04-twist} computes it, and for
a Landau--Ginzburg model $W(\Phi)$ it is the Jacobi ring
\begin{equation}
\label{eq:V04-jacobiring}
 \mathcal{R}_B \;=\; \mathbb{C}[\Phi_i]\,/\,\big(\partial_i W\big).
\end{equation}
The \emph{twisted-chiral ring} is the cohomology of the other supercharge; the A-twist computes it, and
for a gauge theory it is the quantum-cohomology ring generated by $\Sigma$, worked for $\mathbb{CP}^{N-1}$
in \S\ref{sec:V04-glsm}. Mirror symmetry exchanges the two rings, swapping $W$ with $\widetilde W$ and
complex-structure with Kähler data. That one theory carries both rings, computed by two twists and
swapped by mirror symmetry, is the structural heart of $2d\ \mathcal{N}=(2,2)$ physics.

\medskip\noindent\textbf{A worked chiral ring.}\enspace
The chiral ring is concrete, so it is worth computing once. Take the $A_2$ Landau--Ginzburg model, one
chiral field $\Phi$ with superpotential $W = \Phi^3$. The Jacobi ring \eqref{eq:V04-jacobiring} is
$\mathbb{C}[\Phi]/(\partial_\Phi W)$ with $\partial_\Phi W = 3\Phi^2$, so
\begin{equation}
\label{eq:V04-A2ring}
 \mathcal{R}_B \;=\; \mathbb{C}[\Phi]\,/\,(\Phi^2) \;=\; \{1,\ \Phi\},
\end{equation}
a two-dimensional ring with basis the identity and $\Phi$ (the field $\Phi^2$ is set to zero by the
F-term relation). Its dimension counts the ground states, two, and its grading by the R-charge
$R(\Phi) = 1/3$ reproduces the two chiral primaries of the $A_2$ minimal model. For the general
$A_{k+1}$ model $W = \Phi^{k+2}$ the Jacobi ring is
\begin{equation}
\label{eq:V04-Akring}
 \mathcal{R}_B \;=\; \mathbb{C}[\Phi]\,/\,(\Phi^{k+1}) \;=\; \{1, \Phi, \dots, \Phi^k\},
 \qquad \dim \mathcal{R}_B \;=\; k+1,
\end{equation}
the $k+1$ chiral primaries. This is the ring the B-twist computes, and its central charge
\begin{equation}
\label{eq:V04-Akchat}
 \hat c \;=\; 1 - 2R(\Phi) \;=\; \frac{k}{k+2}
\end{equation}
is exactly the $\hat c$ that reappears in the $c$-extremization of the same model in
\S\ref{sec:V04-cext}, where $c_R = 3\hat c = 3k/(k+2)$. The two computations (the Jacobi ring here,
the anomaly trace there) are two views of one fixed point, and their agreement is a check the reader
can run by hand.

\begin{keybox}{Common misconception: a $\mathcal{N}=(2,2)$ theory has one superpotential}
Carrying the four-dimensional picture wholesale into two dimensions, one expects a single holomorphic
superpotential. A $\mathcal{N}=(2,2)$ theory carries \emph{two} holomorphic data: the ordinary superpotential
$W(\Phi)$ for chiral fields, integrated over $\int d^2\theta$, and the twisted superpotential
$\widetilde W(\Sigma)$ for twisted-chiral fields, integrated over $\int d\theta^+ d\bar\theta^-$. They
are constrained independently and never mix. The Fayet--Iliopoulos and theta parameters are
twisted-chiral data, the coefficient $t = r - i\theta$ of the linear $\widetilde W = t\,\Sigma$; they
do \emph{not} live in the ordinary superpotential. Correspondingly there are two rings (chiral and
twisted-chiral), computed by the B- and A-twists and exchanged by mirror symmetry.
\end{keybox}

\section[The GLSM, from CP\texorpdfstring{$^{N-1}$}{N-1} to the quintic]{The gauged linear sigma model, from CP\texorpdfstring{$^{N-1}$}{N-1} to the quintic}
\label{sec:V04-glsm}

The workhorse construction of $2d\ \mathcal{N}=(2,2)$ theories is the gauged linear sigma model. It is an abelian
(or product-unitary) gauge theory whose low-energy physics computes geometry: in one regime it is a
nonlinear sigma model on a Calabi--Yau, in another a Landau--Ginzburg orbifold, and the quantum
cohomology of a projective target falls out of a single one-loop calculation on its Coulomb branch.
We lead with the simplest such calculation, $\mathbb{CP}^{N-1}$, because it teaches twisted-chiral
fields, the FI/theta parameter, quantum cohomology, the vacuum count, and the A-model at once; then
the quintic as the flagship Calabi--Yau.

\medskip\noindent\textbf{The construction.}\enspace
A $\mathcal{N}=(2,2)$ gauged linear sigma model is a $U(1)^k$ (or $\prod_i U(N_i)$) gauge theory with chiral
matter of integer charges $Q_i^a$, an FI/theta parameter $t_a = r_a - i\theta_a$ for each gauge
factor, and a gauge-invariant superpotential $W$. Its bosonic potential has two pieces. The D-term
sets the moment map of the gauge action equal to the FI parameter,
\begin{equation}
\label{eq:V04-dterm}
 D_a \;=\; \sum_i Q^a_i\,|\phi_i|^2 \;-\; r_a \;=\; 0,
\end{equation}
one equation per gauge factor, and the F-terms impose $\partial W / \partial \phi_i = 0$. The charge
matrix $Q^a_i$ is the same integer data a toric variety is built from on the geometry side, and the
low-energy physics depends on the sign chamber of the FI parameters.

\subsection*{Worked instance: CP\texorpdfstring{$^{N-1}$}{N-1} quantum cohomology from the effective twisted superpotential}

Take the simplest gauged linear sigma model with no superpotential: one $U(1)$ with $N$ chiral fields
$\phi_1, \dots, \phi_N$ of charge $+1$, and FI/theta parameter $t = r - i\theta$. In the $r \gg 0$
chamber the D-term $\sum_i |\phi_i|^2 = r$ modulo the $U(1)$ phase is the projective space
$\mathbb{CP}^{N-1}$, and the theory flows to the sigma model on it. Classically the cohomology ring
of $\mathbb{CP}^{N-1}$ is
\begin{equation}
\label{eq:V04-cpnclassical}
 H^*(\mathbb{CP}^{N-1}) \;=\; \mathbb{C}[x]\,/\,(x^N), \qquad x^N \;=\; 0,
\end{equation}
with $x$ the hyperplane class, $x^N = 0$ because $\mathbb{CP}^{N-1}$ has complex dimension $N-1$. The
gauge theory computes the \emph{quantum} correction, and the computation is one loop on the Coulomb
branch, where the twisted-chiral field strength $\Sigma$ takes a large value.

Give $\Sigma$ a large expectation value. The coupling of a charged chiral to $\Sigma$ is a
\emph{twisted mass},
\begin{equation}
\label{eq:V04-twistedmass}
 m_i \;=\; Q_i\,\Sigma \;=\; \Sigma \quad (\text{charge } +1),
\end{equation}
so each of the $N$ fields gets mass $\Sigma$ and can be integrated out at large $\Sigma$. This
generates a one-loop contribution to the effective twisted superpotential (a massive charged field runs
in a loop and shifts the holomorphic coupling, as in four dimensions); the one-loop integral of a
single field of twisted mass $\Sigma$ is
\begin{equation}
\label{eq:V04-oneloop}
 \delta\widetilde W_{\mathrm{eff}} \;=\; -\,\Sigma\big(\log\Sigma - 1\big),
\end{equation}
the standard $\Sigma\log\Sigma$ running of a $2d$ twisted mass. Summing the $N$ identical
contributions and adding the classical FI term $-t\,\Sigma$ gives
\begin{equation}
\label{eq:V04-cpnWeff}
 \widetilde W_{\mathrm{eff}}(\Sigma) \;=\; -\,t\,\Sigma \;-\; N\,\Sigma\big(\log\Sigma - 1\big),
\end{equation}
the second term the sum of the $N$ one-loop contributions (one per charged chiral). The Coulomb-branch
vacua are the critical points of $\widetilde W_{\mathrm{eff}}$. Differentiating, the $\Sigma\cdot
(1/\Sigma)$ piece of $\partial_\Sigma[\Sigma(\log\Sigma - 1)]$ cancels against the $-1$, leaving
\begin{equation}
\label{eq:V04-cpnvacuum}
 \frac{\partial \widetilde W_{\mathrm{eff}}}{\partial \Sigma} \;=\; -\,t \;-\; N\,\log\Sigma \;=\; 0
 \quad\Longrightarrow\quad \log\Sigma = -\frac{t}{N} \quad\Longrightarrow\quad
 \Sigma^N \;=\; e^{-t} \;=:\; q.
\end{equation}
The vacuum equation is $\Sigma^N = q$ with $q = e^{-t}$ the instanton (Kähler) parameter, and it has
$N$ solutions, the $N$ Nth roots of $q$: the theory has $N$ isolated massive vacua on the Coulomb
branch. Reading \eqref{eq:V04-cpnvacuum} as a ring relation, the twisted-chiral ring generated by
$\Sigma$ is
\begin{equation}
\label{eq:V04-cpnQC}
 \mathcal{R}_A\big(\mathbb{CP}^{N-1}\big) \;=\; \mathbb{C}[\Sigma]\,/\,\big(\Sigma^N - q\big),
\end{equation}
the \emph{quantum cohomology} of $\mathbb{CP}^{N-1}$: the classical ring \eqref{eq:V04-cpnclassical}
with $x^N = 0$ deformed to $x^N = q$, the worldsheet instanton lifting the top power of the hyperplane
class from zero to $q$. As $q \to 0$ ($r \to +\infty$) it degenerates to the classical cohomology, of
dimension $N$ either way, matching $\dim H^*(\mathbb{CP}^{N-1}) = N$. One calculation has delivered the
twisted-chiral $\Sigma$, the parameter $t$ that sets $q$, the quantum cohomology ring, the vacuum count
$N$, and the A-model data, all from one effective twisted superpotential.

The smallest case makes the numbers concrete. For $N = 2$, the sigma model on $\mathbb{P}^1$, the
vacuum equation \eqref{eq:V04-cpnvacuum} is $\Sigma^2 = e^{-t} = q$, so the two vacua sit at
\begin{equation}
\label{eq:V04-p1vacua}
 \Sigma \;=\; \pm\, q^{1/2}, \qquad \mathcal{R}_A(\mathbb{P}^1) \;=\; \mathbb{C}[x]\,/\,(x^2 - q).
\end{equation}
The classical $H^*(\mathbb{P}^1) = \mathbb{C}[x]/(x^2)$ has $x^2 = 0$; the instanton correction lifts it
to $x^2 = q$, the degree-one rational curve contributing $q = e^{-t}$. The two massive vacua are the two
supersymmetric ground states, the Witten index
\begin{equation}
\label{eq:V04-p1index}
 \mathrm{Tr}(-1)^F \;=\; 2 \;=\; \chi(\mathbb{P}^1),
\end{equation}
the Euler characteristic, and the ring has dimension two, matching $\dim H^*(\mathbb{P}^1) = 2$. This is
the entire A-model of $\mathbb{P}^1$ off two lines of a one-loop calculation, reproduced geometrically
by the A-twist of \S\ref{sec:V04-twist}.

\begin{keybox}{Common misconception: the effective twisted superpotential sign is $+N\Sigma(\log\Sigma-1)$}
The one-loop term in \eqref{eq:V04-cpnWeff} carries a \emph{minus} sign, $-N\Sigma(\log\Sigma - 1)$.
This sign is not a convention one is free to flip: it is fixed by the requirement that the vacuum
equation give $\Sigma^N = e^{-t} = q$ with $q$ the instanton factor. The opposite sign
$+N\Sigma(\log\Sigma - 1)$ would give $\Sigma^N = e^{+t}$, inconsistent both with the instanton weight
$q = e^{-t}$ and with the Hori--Vafa mirror of \S\ref{sec:V04-twist}. The calculation that backs this
section carries the sign that reproduces $\Sigma^N = e^{-t}$; a draft that writes the plus sign is
computing the wrong vacuum equation.
\end{keybox}

\medskip\noindent\textbf{The two phases and one RG story.}\enspace
Now add a superpotential and let the target be Calabi--Yau, so the gauged linear sigma model flows to
a conformal field theory. The low-energy physics depends on the FI chamber, and the two chambers
realize two very different-looking descriptions of the \emph{same} infrared theory. For $r \gg 0$ the
D-term forces some positively-charged field nonzero, the origin is excluded, and setting the D-term to
$r$ and dividing by the gauge group is the symplectic reduction
\begin{equation}
\label{eq:V04-gitquotient}
 X_r \;=\; \big\{\,\phi : \textstyle\sum_i Q^a_i |\phi_i|^2 = r_a\,\big\}\big/ \prod_a U(1)_a,
\end{equation}
equivalently the geometric-invariant-theory (GIT) quotient, a toric variety whose fan is read from
the charge matrix. The F-terms of $W$ then cut a hypersurface (or complete intersection) inside
$X_r$: this is the \emph{geometric} phase, a nonlinear sigma model on a Calabi--Yau. For $r \ll 0$
the same D-term forces a negatively-charged field nonzero instead, Higgsing the gauge group to a
finite subgroup and leaving the remaining fields as a Landau--Ginzburg orbifold with superpotential
$W$: the \emph{Landau--Ginzburg} phase.

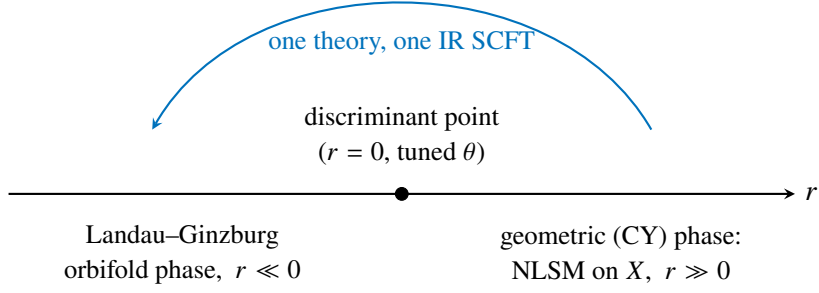
\begin{figure}[ht]
\centering
\begin{tikzpicture}[>=stealth]
 \draw[->,thick] (-5.2,0) -- (5.2,0) node[right] {$r$};
 \filldraw (0,0) circle (2.4pt);
 \node[align=center,above=7pt] at (0,0) {\small discriminant point\\[-1pt] \small ($r=0$, tuned $\theta$)};
 \node[align=center] at (2.9,-0.8) {\small geometric (CY) phase:\\[-1pt] \small NLSM on $X$, \ $r\gg0$};
 \node[align=center] at (-2.9,-0.8) {\small Landau--Ginzburg\\[-1pt] \small orbifold phase, \ $r\ll0$};
 \draw[->,thick,RoyalBlue] (3.3,0.85) to[out=120,in=60] (-3.3,0.85);
 \node[RoyalBlue] at (0,2.0) {\small one theory, one IR SCFT};
\end{tikzpicture}
\caption{The two phases of a Calabi--Yau gauged linear sigma model as the two ends of the quantum
Kähler moduli space, the $t = r - i\theta$ cylinder. For $r \gg 0$ the theory is a nonlinear sigma
model on the Calabi--Yau $X$; for $r \ll 0$ it is a Landau--Ginzburg orbifold. These are two
ultraviolet regimes of one gauged linear sigma model that flows to one infrared superconformal field
theory. The classical $r = 0$ chamber wall sits inside a quantum discriminant locus (here a single
point at $r = 0$ with a tuned theta angle, where a noncompact Coulomb branch opens and a massless
state appears); a generic theta angle steers the path around it, but the smoothness of the
interpolation is a statement about the quantum moduli space, not the classical $r$-axis.}
\label{fig:V04-phases}
\end{figure}

The prose must be careful: the naive picture (``two chambers of the $r$-axis, smoothly connected
except at one point'') is only the classical cartoon. The precise statement has four layers.
Classically the sign of $r$ divides the FI axis into chambers. The FI and theta parameters then
complexify into $t = r - i\theta$, ranging over a cylinder, the \emph{quantum Kähler moduli space}
whose two ends are the classical chambers. That moduli space carries a \emph{discriminant locus}, the
codimension-one set where the theory becomes singular (a noncompact Coulomb branch opens and a new
massless state appears); it is one point for one-parameter models but a genuine subvariety in general.
Finally, the physical statement: the gauged linear sigma model is one theory flowing to one infrared
superconformal field theory in the Calabi--Yau case, and the geometric and Landau--Ginzburg
descriptions are two ultraviolet regimes of that one flow. They are \emph{not} two unrelated theories,
but neither does the classical $r$-axis by itself prove a globally smooth interpolation, which is a
statement about the quantum moduli space and its discriminant. The full analysis (secondary fan,
discriminant, analytic continuation) is the phase theorem, not developed here; the worked computation
here is the quintic (Figure~\ref{fig:V04-phases}).

The discriminant point makes ``two regimes of one flow'' precise. On the Coulomb branch (large
$\Sigma$) the effective twisted superpotential of a Calabi--Yau gauged linear sigma model generically
has no critical point at finite $\Sigma$: the D-term lifts it. At a special $t$ on the discriminant a
would-be massive vacuum becomes massless, a critical point runs off to $\Sigma \to \infty$, and a new
massless degree of freedom appears in the infrared. That is the singular locus: not a thermodynamic
phase boundary but a point where the low-energy theory degenerates because a massless state (a
noncompact Coulomb branch) appears. A generic theta angle steers the straight-line path in $t$ around
it, so the two descriptions are connected by a path along which the theory never degenerates. This is
``no phase transition'': the connection runs through the quantum moduli space, avoiding the
discriminant, a statement about where the theory is singular, not about the sign of classical $r$.

The Calabi--Yau condition on the geometric phase is a condition on the charges. The axial R-symmetry
$U(1)_A$ is non-anomalous, equivalently the target has vanishing first Chern class, precisely when
\begin{equation}
\label{eq:V04-quinticCY}
 \sum_a Q^a \;=\; 0
\end{equation}
for each gauge factor. When \eqref{eq:V04-quinticCY} fails the sigma model still exists, but the FI
parameter runs (the target is Fano or of general type) and it is not the conformal phase this section
exemplifies.

\subsection*{Worked instance: the quintic}

The canonical Calabi--Yau example is the quintic. Take $U(1)$ with five chiral fields $\phi_1, \dots,
\phi_5$ of charge $+1$ and one field $P$ of charge $-5$, with superpotential
\begin{equation}
\label{eq:V04-quinticW}
 W \;=\; P\,G_5(\phi),
\end{equation}
$G_5$ a generic degree-five homogeneous polynomial in the five $\phi_i$. The Calabi--Yau condition
\eqref{eq:V04-quinticCY} holds,
\begin{equation}
\label{eq:V04-quintic}
 \sum_a Q^a \;=\; 5\cdot(+1) + (-5) \;=\; 0,
\end{equation}
so $U(1)_A$ is non-anomalous and the theory flows to a conformal fixed point. Now read the two
chambers. In the $r \gg 0$ chamber the D-term is $\sum_i |\phi_i|^2 - 5|P|^2 = r > 0$, and the F-terms
read
\begin{equation}
\label{eq:V04-quinticFterms}
 \frac{\partial W}{\partial P} = G_5(\phi) = 0, \qquad
 \frac{\partial W}{\partial \phi_i} = P\,\partial_i G_5 = 0 \ \Rightarrow\ P = 0
\end{equation}
($G_5$ generic). With $P = 0$ the D-term reduces to $\sum_i |\phi_i|^2 = r$ modulo the $U(1)$ phase,
the projective space
\begin{equation}
\label{eq:V04-P4}
 \{\,\textstyle\sum_i |\phi_i|^2 = r\,\}\big/ U(1) \;=\; \mathbb{P}^4, \qquad
 \dim_{\mathbb{C}} \mathbb{P}^4 \;=\; 5 - 1 \;=\; 4,
\end{equation}
and the F-term $G_5 = 0$ cuts a degree-five hypersurface inside $\mathbb{P}^4$, of complex dimension
\begin{equation}
\label{eq:V04-quinticdim}
 \dim_{\mathbb{C}}\{G_5 = 0\} \;=\; 4 - 1 \;=\; 3,
\end{equation}
the quintic Calabi--Yau threefold, with central charge $\hat c = \dim_{\mathbb{C}} X = 3$. The
construction is the GIT quotient \eqref{eq:V04-P4} \emph{and then} the F-term cut, not the naive
vanishing locus of $W$ among all six fields: one equation in six variables would give the wrong
dimension. Getting the two steps in the right order is the working discipline of the geometric phase.

In the $r \ll 0$ chamber the D-term $\sum_i |\phi_i|^2 - 5|P|^2 = r < 0$ can only be solved with $P
\neq 0$, at $|P|^2 = -r/5$. A nonzero $P$ Higgses the gauge group down to the stabilizer of a
charge-$-5$ field,
\begin{equation}
\label{eq:V04-Z5}
 U(1) \;\longrightarrow\; \mathbb{Z}_5,
\end{equation}
the fifth roots of unity. The five $\phi_i$ are then a Landau--Ginzburg model with superpotential $W
= G_5(\phi)$ (with $P$ at its VEV) orbifolded by this $\mathbb{Z}_5$. Its central charge is read from
the R-charges, fixed by quasi-homogeneity of $W$: for $W$ to be a good superpotential its superspace
integral must be neutral, so $W$ carries R-charge one; since $W = G_5$ is degree five and homogeneous,
each field carries R-charge $q_i$ with $5\,q_i = 1$, that is $q_i = 1/d = 1/5$ for a degree-$d$
superpotential. A single Landau--Ginzburg field of R-charge $q$ contributes $1 - 2q$ to the central
charge (the standard $\mathcal{N}=2$ minimal-model formula: a free chiral has $q = 0$ and $\hat c = 1$,
and the superpotential lowers it), so the Landau--Ginzburg central charge is the sum
\begin{equation}
\label{eq:V04-lgchat}
 \hat c \;=\; \sum_i (1 - 2 q_i).
\end{equation}
For the quintic, five fields each with $q_i = 1/5$,
\begin{equation}
\label{eq:V04-quinticlgchat}
 \hat c \;=\; 5\left(1 - \tfrac{2}{5}\right) \;=\; 5\cdot\tfrac{3}{5} \;=\; 3,
\end{equation}
exactly the geometric $\hat c = 3$. The two phases share their central charge, as they must if they
are two regimes of one renormalization-group flow. The general pattern is transparent from
\eqref{eq:V04-lgchat}: a degree-$d$ polynomial in $d$ fields gives
\begin{equation}
\label{eq:V04-lgchatgen}
 \hat c \;=\; d\left(1 - \frac{2}{d}\right) \;=\; d - 2,
\end{equation}
which is $\dim_{\mathbb{C}}$ of the degree-$d$ hypersurface in $\mathbb{P}^{d-1}$, a Calabi--Yau
$(d{-}2)$-fold. The same gauged linear sigma model, with $d$ charge-$+1$ fields and one charge-$-d$
field, produces each member,
\begin{equation}
\label{eq:V04-cytable}
 d = 3: \ \hat c = 1\ (\text{elliptic curve}), \qquad
 d = 4: \ \hat c = 2\ (\text{K3}), \qquad
 d = 5: \ \hat c = 3\ (\text{quintic}),
\end{equation}
the charge sum $d\cdot(+1) + (-d) = 0$ ensuring the Calabi--Yau condition in every case, and the
Landau--Ginzburg $\hat c$ matching the geometric dimension throughout.

\begin{keybox}{Common misconception: the two phases are different theories}
The geometric ($r \gg 0$) and Landau--Ginzburg ($r \ll 0$) phases are two ultraviolet regimes of one
gauged linear sigma model, flowing to one infrared superconformal field theory in the Calabi--Yau
case, not distinct theories. The geometric phase is the GIT (symplectic) quotient $\{\sum_i Q^a_i
|\phi_i|^2 = r_a\}/\text{gauge}$ (a toric variety), further cut by the F-terms of $W$, \emph{not} the
naive vanishing locus of $W$ alone (which would give the wrong dimension). The interpolation between
the two ends is smooth in the \emph{quantum Kähler moduli space} (the $t = r - i\theta$ cylinder),
away from its discriminant locus (where a noncompact Coulomb branch opens and a massless state
appears); do not read the classical $r$-axis by itself as proving a globally smooth interpolation. And
the phases exist for any FI chamber: $\sum_a Q^a = 0$ is the Calabi--Yau, conformal, anomaly-free
condition of the intended example, not a prerequisite for a phase to exist.
\end{keybox}

\section{The A and B topological twists and mirror symmetry}
\label{sec:V04-twist}

A $\mathcal{N}=(2,2)$ theory admits two inequivalent topological twists, and they compute the two rings of
\S\ref{sec:V04-22data}. This is $\mathcal{N}=(2,2)$-specific technology, placed here immediately after the gauged
linear sigma model whose rings it computes, not among the shared tools of Block C. The twist mixes the
Lorentz rotation with an R-symmetry, turning one supercharge into a scalar nilpotent operator whose
cohomology is the topological ring. Which R-symmetry is used distinguishes the two models, the $U(1)_V$
versus $U(1)_A$ distinction of \S\ref{sec:V04-kinematics}.

\medskip\noindent\textbf{The A-twist: quantum cohomology.}\enspace
The A-twist uses the \emph{vector} R-symmetry $U(1)_V$. The A-model localizes on holomorphic maps to
the target (the worldsheet instantons) and computes the quantum Kähler data: Gromov--Witten invariants
and the quantum-cohomology ring. For $\mathbb{CP}^{N-1}$ it is exactly the ring of \S\ref{sec:V04-glsm},
\begin{equation}
\label{eq:V04-ARing}
 \mathcal{R}_A\big(\mathbb{CP}^{N-1}\big) \;=\; \mathbb{C}[x]\,/\,\big(x^N - q\big), \qquad q =
 e^{-t},
\end{equation}
the classical $x^N = 0$ deformed to $x^N = q$ by instantons. The A-twist is the twisted-chiral ring at
work, and because $U(1)_V$ is non-anomalous for a broad class of sigma models it exists
\emph{generally}: it does \emph{not} require a Calabi--Yau target. One can A-twist a Fano target and
get sensible quantum cohomology.

\medskip\noindent\textbf{The B-twist: the Jacobi ring.}\enspace
The B-twist uses the \emph{axial} R-symmetry $U(1)_A$. The B-model localizes on constant maps and
computes the classical complex-structure data (periods and variations of Hodge structure), an
instanton-free holomorphic answer, the chiral ring of \S\ref{sec:V04-22data}. For a Landau--Ginzburg
model $W(\Phi)$ it is the Jacobi ring
\begin{equation}
\label{eq:V04-BRing}
 \mathcal{R}_B \;=\; \mathbb{C}[\Phi_i]\,/\,\big(\partial_i W\big),
\end{equation}
the F-term relations of $W$ as a quotient. But $U(1)_A$ is anomalous in general: its anomaly is
controlled by the first Chern class, so the B-twist requires
\begin{equation}
\label{eq:V04-btwistCY}
 c_1(TX) \;=\; 0,
\end{equation}
the Calabi--Yau condition. For a gauged linear sigma model this is the charge condition $\sum_i Q_i =
0$ of \eqref{eq:V04-quinticCY}, because the $U(1)_A$ anomaly is proportional to the sum of the gauge
charges,
\begin{equation}
\label{eq:V04-U1Aanomaly}
 \text{$U(1)_A$ anomaly} \;\propto\; \sum_i Q_i \;\propto\; c_1(TX).
\end{equation}
This gating is asymmetric,
\begin{equation}
\label{eq:V04-twistgate}
 (+1)^5, -5: \ \textstyle\sum_i Q_i = 0 \ (\text{B-twist } \checkmark), \qquad
 (+1)^5, -4: \ \textstyle\sum_i Q_i = 1 \neq 0 \ (\text{no B-twist}),
\end{equation}
the non-Calabi--Yau assignment $(+1)^5, -4$ having anomalous $U(1)_A$, yet it still admits the A-twist
because $U(1)_V$ is non-anomalous regardless. The A-twist is not gated by the Calabi--Yau condition;
the B-twist is.

\subsection*{Mirror symmetry and the Hori--Vafa construction}

Mirror symmetry is a duality between two full $\mathcal{N}=(2,2)$ theories, not a property of their twists. A
$\mathcal{N}=(2,2)$ sigma model on $X$ is physically equivalent, as an untwisted theory, to a $\mathcal{N}=(2,2)$ theory on a
mirror $\hat X$; for a gauged linear sigma model the mirror is the Hori--Vafa dual Landau--Ginzburg /
Toda theory, and needs no twist. The construction is at field level: dualize each charged chiral
$\Phi_i$ to a periodic twisted-chiral $Y_i$, real part $|\phi_i|^2$, shift symmetry the phase of
$\Phi_i$. The dual is a Landau--Ginzburg model of the $Y_i$ with mirror twisted superpotential
\begin{equation}
\label{eq:V04-hvmirror}
 \widetilde W(Y) \;=\; \sum_i e^{-Y_i} \;+\; \sum_a \Sigma_a\Big(\sum_i Q^a_i\, Y_i - t_a\Big),
\end{equation}
the $\Sigma_a$ enforcing the charge constraints $\sum_i Q^a_i Y_i = t_a$. For the $\mathbb{P}^{N-1}$
sigma model (one $U(1)$, $N$ fields of charge $+1$) the mirror is
\begin{equation}
\label{eq:V04-hvcpn}
 \widetilde W \;=\; \sum_{i=1}^{N} e^{-Y_i}, \qquad \sum_{i=1}^{N} Y_i = t,
\end{equation}
whose $N$ critical points $e^{-Y_i} = e^{-t/N}$ reproduce the $N$ ring generators, recovering the
quantum cohomology
\begin{equation}
\label{eq:V04-hvring}
 \mathbb{C}[x]\,/\,(x^N - e^{-t})
\end{equation}
of $\mathbb{P}^{N-1}$ from the \emph{mirror} side, matching \eqref{eq:V04-ARing}. The Kähler parameter
$t$ of $X$ becomes the complex-structure parameter of the mirror: the same $t$ cylinder carries the
mirror map.

The twists are how the duality is most sharply \emph{used}. In the protected topological sector the
A-twist of $X$ and the B-twist of $\hat X$ compute the same correlators, so mirror symmetry exchanges
the two models,
\begin{equation}
\label{eq:V04-mirror}
 A\text{-model}(X) \;\cong\; B\text{-model}(\hat X),
\end{equation}
trading the quantum Kähler data of $X$ for the classical complex-structure data of $\hat X$, and so
$\mathcal{R}_A(X)$ for $\mathcal{R}_B(\hat X)$. The full topological-correlator machinery (Frobenius /
quantum-cohomology structure, variation of Hodge structure) and the mirror theorem are left to
the mirror-theorem literature; this section constructs the untwisted Hori--Vafa mirror, names the two
twists, ties them to the two rings, and gates the B-twist by the Calabi--Yau condition.

\bigskip
\begin{center}
\rule{0.4\textwidth}{0.4pt}\\[3pt]
{\large\textsf{\textbf{Block B.\enspace The $\mathcal{N}=(0,2)$ world}}}\\[2pt]
\rule{0.4\textwidth}{0.4pt}
\end{center}
\medskip

\noindent The next four sections build $\mathcal{N}=(0,2)$ dynamics as the second coherent world. Discarding the
two left-moving supercharges of $\mathcal{N}=(2,2)$ frees the left-moving fermions from their partners, and the
consequences run through the whole part: two independent holomorphic data $(E, J)$ per Fermi and a
scalar potential built from them (\S\ref{sec:V04-ej}); a holomorphic bundle over the target, seen
sharply as a gauged $\mathcal{N}=(2,2)\to\mathcal{N}=(0,2)$ deformation (\S\ref{sec:V04-bundles}); a gauge anomaly that must
cancel, a gravitational anomaly that only matches, and the $c$-extremization principle
(\S\ref{sec:V04-cext}); and an order-three triality (\S\ref{sec:V04-triality}).

\section[\texorpdfstring{$\mathcal{N}=(0,2)$}{N=(0,2)} Fermi multiplets and the E/J data]{\texorpdfstring{$\mathcal{N}=(0,2)$}{N=(0,2)} Fermi multiplets, the E/J data, and the scalar potential}
\label{sec:V04-ej}

Now the genuinely two-dimensional world: two right-moving supercharges, no $\mathcal{N}=(2,2)$ completion. A
$\mathcal{N}=(0,2)$ theory has no single superpotential but two holomorphic data per Fermi multiplet. Why those
data are not free, and what they do to the scalar potential, is the central physics of the $\mathcal{N}=(0,2)$
world. We build it as field theory first, then read off the standard fixture.

\medskip\noindent\textbf{The two data.}\enspace
Each Fermi multiplet $\Lambda_a$ is only constrained chiral,
\begin{equation}
\label{eq:V04-Edatum}
 \bar D_+ \Lambda_a \;=\; E_a(\Phi),
\end{equation}
with $E_a$ a holomorphic function of the $\mathcal{N}=(0,2)$ chiral fields (itself a $\mathcal{N}=(0,2)$ chiral superfield).
The second datum enters the $\mathcal{N}=(0,2)$ superpotential, a Fermi-linear integral over half of the $\mathcal{N}=(2,2)$
measure,
\begin{equation}
\label{eq:V04-Jdatum}
 W_{\mathcal{N}=(0,2)} \;=\; \int d\theta^+\; \sum_a \Lambda_a\, J^a(\Phi) \;+\; \text{c.c.},
\end{equation}
with $J^a$ a second holomorphic function of the chiral fields. The pair $(E_a, J^a)$ is two
independent holomorphic data; neither is more fundamental, and they enter the physics symmetrically, as
the scalar potential and the closure condition below make precise.

\medskip\noindent\textbf{The scalar potential.}\enspace
The physical content of $E$ and $J$ is that they are the two halves of the $\mathcal{N}=(0,2)$ scalar potential.
Component-expanding the Fermi kinetic term $|\bar D_+\Lambda_a|^2 = |E_a|^2 + \cdots$ and the
superpotential \eqref{eq:V04-Jdatum} generates the bosonic potential
\begin{equation}
\label{eq:V04-02potential}
 V \;\supset\; \sum_a \big|E_a(\phi)\big|^2 \;+\; \sum_a \big|J^a(\phi)\big|^2 \;+\; V_D,
\end{equation}
with $V_D$ the gauge D-term potential. This is what makes $E$ and $J$ field theory rather than
bookkeeping. The supersymmetric vacua are the zeros of $V$,
\begin{equation}
\label{eq:V04-02vacua}
 E_a(\phi) \;=\; 0, \qquad J^a(\phi) \;=\; 0, \qquad \text{(plus the D-term equations)},
\end{equation}
for every $a$: the $E_a = 0$ and $J^a = 0$ conditions carve out the $\mathcal{N}=(0,2)$ moduli space, exactly as
the F- and D-term equations do in four dimensions, and both are needed. Without
\eqref{eq:V04-02potential} the pair would be a formal labelling; with it, $E_a$ and $J^a$ are the
objects whose vanishing loci are the vacua.

The two $|{\cdot}|^2$ terms have parallel origins. The $|E_a|^2$ piece is the norm of the Fermi
constraint $\bar D_+ \Lambda_a = E_a$: the bottom component of $E_a$ appears in the Fermi kinetic term
as $|E_a(\phi)|^2$, the $\mathcal{N}=(0,2)$ analog of a $4d$ auxiliary field integrating out. The $|J^a|^2$ piece
comes from the superpotential \eqref{eq:V04-Jdatum}: the $\theta^+$ integral and the auxiliary
component of $\Lambda_a$ leave $|J^a(\phi)|^2$, exactly as a $4d$ superpotential leaves $\sum_i
|\partial_i W|^2$. So a supersymmetric vacuum needs \emph{both} $E_a = 0$ and $J^a = 0$: they are the
two independent contributions to a non-negative potential, and a nonzero value of either breaks
supersymmetry.

\medskip\noindent\textbf{E and J are not exchanged by conjugation.}\enspace
It is tempting to say that $E$ and $J$ are exchanged when the Fermi multiplet is conjugated,
$\Lambda_a \leftrightarrow \bar\Lambda_a$. They are not. The two data enter symmetrically in the
bosonic potential \eqref{eq:V04-02potential} and in the closure condition below, and in some duality
operations one can exchange the E-type and J-type roles. But ordinary Hermitian conjugation
\emph{complex-conjugates} the data,
\begin{equation}
\label{eq:V04-conjnoexchange}
 \Lambda_a \leftrightarrow \bar\Lambda_a: \quad E_a \;\longmapsto\; \bar E_a, \qquad
 J^a \;\longmapsto\; \bar J^a,
\end{equation}
sending $E_a$ to $\bar E_a$ and $J^a$ to $\bar J^a$; it does not literally identify $E$ with $J$. The
symmetry that matters is the symmetric appearance of $|E_a|^2$ and $|J^a|^2$ in the potential and of
$E_a$ and $J^a$ in the trace $\sum_a \mathrm{Tr}(E_a J^a)$, not a claim that conjugation swaps the two
holomorphic functions.

\medskip\noindent\textbf{The closure condition.}\enspace
The two data are not independent as functions. The interaction \eqref{eq:V04-Jdatum} is invariant
under the single right-moving supercharge only if they satisfy the trace condition
\begin{equation}
\label{eq:V04-ejclosure}
 \sum_a \mathrm{Tr}\,\big(E_a\, J^a\big) \;=\; 0,
\end{equation}
the sum over all Fermi multiplets, the trace over gauge indices. This is the $2d$ analog of the trace
condition that makes a $4d$ superpotential gauge invariant, and it is the statement that $\bar D_+$
acting on the interaction closes,
\begin{equation}
\label{eq:V04-barclosure}
 \bar D_+(\Lambda_a J^a) \;=\; E_a J^a + \Lambda_a \bar D_+ J^a,
\end{equation}
so that gauge-invariance plus the chirality of $J^a$ leaves exactly the trace \eqref{eq:V04-ejclosure}
to cancel. It is not automatic: a set of $E$ and $J$ chosen at random will not satisfy it, and a theory
whose data violate it is not a consistent $\mathcal{N}=(0,2)$ theory. In a brick or toric phase each datum splits
into two signed terms, $E_a = E_a^+ - E_a^-$ and $J^a = J^{a,+} - J^{a,-}$, and
\eqref{eq:V04-ejclosure} is the statement that the four products cancel in pairs, the plaquette closure
of the tiling.

\subsection*{Worked instance: the C\texorpdfstring{$^4$}{4} brick and its orbifold}

The cleanest example where the closure \eqref{eq:V04-ejclosure} is a genuine identity, not an
accident, is the $\mathbb{C}^4$ brane-brick model. There the chiral fields are four
matrices $X_1, \dots, X_4$ (the four coordinates of $\mathbb{C}^4$), and the three Fermi data are
commutators
\begin{equation}
\label{eq:V04-ejcommutators}
 E_a \;=\; [\,X_4,\, X_a\,], \qquad J^a \;=\; [\,X_{a+1},\, X_{a+2}\,], \qquad a = 1,2,3,
\end{equation}
indices cyclic. The E- and J-index pairs differ for every $a$, so the two are genuinely distinct
functions. The trace sum is
\begin{equation}
\label{eq:V04-jacobi}
 \sum_{a=1}^{3} \mathrm{Tr}\big([\,X_4, X_a\,]\,[\,X_{a+1}, X_{a+2}\,]\big) \;=\; 0
\end{equation}
identically, for \emph{any} matrices: it is the Jacobi identity, written as a trace. Expanding one
term, $\mathrm{Tr}([X_4, X_1][X_2, X_3])$, and summing the three cyclic images, every monomial
$\mathrm{Tr}(X_i X_j X_k X_l)$ that appears is cancelled by another with the opposite sign. The
matrices genuinely do not commute, so \eqref{eq:V04-jacobi} is a nontrivial cancellation, not a
term-by-term zero. It also singles out the right partner: the wrong pairing, for instance
\begin{equation}
\label{eq:V04-wrongpair}
 \sum_a \mathrm{Tr}\big(E_a\, E_a\big) \;=\; \sum_a \mathrm{Tr}\big([X_4, X_a]^2\big) \;\neq\; 0,
\end{equation}
does not vanish, so the closure is the diagnostic that fixes $J^a$ given $E_a$.

The identity survives orbifolding. The $\mathbb{C}^4/\mathbb{Z}_4$ diagonal orbifold, weights
$(1,1,1,1)$ satisfying the Calabi--Yau condition
\begin{equation}
\label{eq:V04-orbcy}
 1+1+1+1 \;=\; 0 \bmod 4,
\end{equation}
projects the four $X$ matrices to block form and distributes the multiplets over four gauge nodes. The
orbifold count is
\begin{equation}
\label{eq:V04-orbcount}
 \mathbb{C}^4/\mathbb{Z}_4: \quad 4\ \text{nodes}, \quad 16\ \text{chiral}, \quad 12\ \text{Fermi},
\end{equation}
and the projected commutators still obey $\sum_a \mathrm{Tr}(E_a J^a) = 0$, inherited from the parent
Jacobi because the projection commutes with the trace. The Calabi--Yau condition is essential: an
orbifold whose weights do not sum to zero modulo $n$ has inconsistent Fermi charges and does not
descend to a consistent $\mathcal{N}=(0,2)$ theory. This is the field-theory content behind brane-brick
models; the scalar potential \eqref{eq:V04-02potential}, the vacua \eqref{eq:V04-02vacua}, and the
closure \eqref{eq:V04-ejclosure} are its field-theory home.

\begin{keybox}{Common misconception: a $\mathcal{N}=(0,2)$ theory has a single superpotential}
A $\mathcal{N}=(0,2)$ theory carries \emph{two} holomorphic data per Fermi multiplet, the pair $(E_a, J^a)$:
$E_a$ enters $\bar D_+ \Lambda_a = E_a$, and $J^a$ enters $W_{\mathcal{N}=(0,2)} = \int d\theta^+ \sum_a \Lambda_a
J^a$. They are the two halves of the scalar potential $V \supset \sum_a|E_a|^2 + \sum_a|J^a|^2 + V_D$,
so the supersymmetric vacua obey \emph{both} $E_a = 0$ and $J^a = 0$. Supersymmetry of the interaction
is not automatic: it requires $\sum_a \mathrm{Tr}(E_a J^a) = 0$, the $2d$ analog of the $4d$ trace
condition. The two data enter symmetrically, but they are not exchanged by Hermitian conjugation:
conjugation complex-conjugates $E \to \bar E$, $J \to \bar J$, it does not swap the two functions. A
$\mathcal{N}=(2,2)$ theory is the special case in which $J$ and $E$ are fixed by the extended supersymmetry, worked
in \S\ref{sec:V04-bundles}; genuinely $\mathcal{N}=(0,2)$ theories keep $E$ and $J$ independent.
\end{keybox}

\section[Bundles and \texorpdfstring{$\mathcal{N}=(0,2)$}{N=(0,2)} deformations]{\texorpdfstring{$\mathcal{N}=(0,2)$}{N=(0,2)} bundles and deformations of \texorpdfstring{$\mathcal{N}=(2,2)$}{N=(2,2)} GLSMs}
\label{sec:V04-bundles}

The most important reason $\mathcal{N}=(0,2)$ theories matter, beyond quiver technology, is geometric. A $\mathcal{N}=(2,2)$
nonlinear sigma model on a Kähler target $X$ has both its fermions valued in the tangent bundle $TX$,
the left-movers locked to the right-movers by extended supersymmetry. A $\mathcal{N}=(0,2)$ model relaxes that
locking: the right-movers still couple to $TX$, but the left-moving Fermi fermions couple to an
independent holomorphic vector bundle $E \to X$,
\begin{equation}
\label{eq:V04-bundle}
 \mathcal{N}=(2,2)\ \text{NLSM}: \ \psi_\pm \in TX, \qquad\qquad
 \mathcal{N}=(0,2)\ \text{NLSM}: \ \psi_+ \in TX,\ \lambda_- \in E,
\end{equation}
and the $\mathcal{N}=(2,2)$ theory is recovered as the special case $E = TX$. The freedom to choose $E \neq TX$ is
what makes $\mathcal{N}=(0,2)$ models a genuinely larger world (heterotic compactifications, bundle-deformed
Calabi--Yau models). But $E$ is not arbitrary: consistency of the quantum theory demands anomaly
conditions, the geometric avatar of the $2d$ gauge and gravitational anomalies of the next section,
\begin{equation}
\label{eq:V04-bundleanom}
 \mathrm{ch}_2(E) \;=\; \mathrm{ch}_2(TX), \qquad c_1(E) \;\equiv\; c_1(TX) \pmod 2.
\end{equation}
The first is the vanishing of the sigma-model gauge anomaly (the left-moving fermion one-loop trace
matches the right-moving one, degree by degree); the second is the Green--Schwarz condition on the
first Chern class. They are the target image of the chiral fermion traces that \S\ref{sec:V04-cext}
treats as the $2d$ gauge and gravitational anomalies, so the abstract $E/J$ data are the local shadow
of a bundle over a geometry.

\medskip\noindent\textbf{Ungauged versus gauged: where E comes from.}\enspace
The $\mathcal{N}=(2,2)$ locus $E = TX$ locks the two data by extended supersymmetry, but \emph{how} they lock
depends on whether the theory is gauged. For an \emph{ungauged} $\mathcal{N}=(2,2)$ Landau--Ginzburg model with
fields $\Phi_a$ and superpotential $W$, a $\mathcal{N}=(0,2)$ subalgebra splits each chiral into a $\mathcal{N}=(0,2)$ chiral
plus a paired Fermi $\Gamma_a$, with data
\begin{equation}
\label{eq:V04-J22}
 E_a \;=\; 0, \qquad J^a \;=\; \frac{\partial W}{\partial \Phi_a}
 \qquad(\text{ungauged $\mathcal{N}=(2,2)$ LG}),
\end{equation}
so $E$ vanishes and $J$ is the gradient of one superpotential. For a $\mathcal{N}=(2,2)$ Wess--Zumino model with
$\Phi_1, \Phi_2$ and $W = \Phi_1 \Phi_2^2$, read as $\mathcal{N}=(0,2)$,
\begin{equation}
\label{eq:V04-22Jexample}
 J^1 \;=\; \frac{\partial W}{\partial \Phi_1} \;=\; \Phi_2^2, \qquad
 J^2 \;=\; \frac{\partial W}{\partial \Phi_2} \;=\; 2\,\Phi_1 \Phi_2, \qquad E_a = 0,
\end{equation}
and the potential reduces to $|\partial_1 W|^2 + |\partial_2 W|^2$, the ordinary F-term potential. This
is why one superpotential suffices for an ungauged model.

For a \emph{gauged} $\mathcal{N}=(2,2)$ GLSM the story is different, and this is the point of the subsection: $E$
is \emph{not} zero. The extra ingredient is the $\mathcal{N}=(2,2)$ vector, whose $\mathcal{N}=(0,2)$ decomposition
\eqref{eq:V04-decompvector} contains the $\mathcal{N}=(0,2)$ field strength $\Sigma$. A $\mathcal{N}=(2,2)$ chiral $\Phi_i$ of
gauge charge $Q_i$ splits into a $\mathcal{N}=(0,2)$ chiral and a Fermi $\Gamma_i$ whose $E$-term is fixed by the
gauge coupling,
\begin{equation}
\label{eq:V04-gaugedE}
 \bar D_+ \Gamma_i \;=\; E_i \;=\; \sqrt{2}\, Q_i\, \Sigma\, \Phi_i,
\end{equation}
a nonzero $E$ built from the field strength and the charged field. This is the $\mathcal{N}=(0,2)$ statement of the
$2d$ twisted mass $Q_i\Sigma$ of \S\ref{sec:V04-glsm}: the gaugino partner of that mass is exactly the
$E$-term \eqref{eq:V04-gaugedE}. So a gauged $\mathcal{N}=(2,2)$ GLSM, read as $\mathcal{N}=(0,2)$, has $E_i \neq 0$ from the
gauge coupling and $J^{\Gamma_i}$ from the superpotential, and the two data must together satisfy the
closure \eqref{eq:V04-ejclosure}.

\subsection*{Worked instance: the gauged quintic E-term and its closure by Euler's theorem}

The quintic gauged linear sigma model of \S\ref{sec:V04-glsm} makes this concrete, and the closure is
a deep, one-line calculation. The fields are five $\phi_i$ of charge $Q_i = +1$ and one $P$ of charge
$Q_P = -5$, with $W = P\,G_5(\phi)$. Reading it as $\mathcal{N}=(0,2)$, each $\mathcal{N}=(2,2)$ chiral becomes a $\mathcal{N}=(0,2)$
chiral plus a Fermi, and the two data are the gauged $E$-terms \eqref{eq:V04-gaugedE} together with the
superpotential $J$'s,
\begin{align}
\label{eq:V04-quinticEJ}
 E_i &\;=\; \sqrt{2}\,\Sigma\,\phi_i, & E_P &\;=\; -5\sqrt{2}\,\Sigma\,P, \notag\\
 J^i &\;=\; P\,\frac{\partial G_5}{\partial \phi_i}, & J^P &\;=\; G_5(\phi),
\end{align}
the $E$'s carrying the charges $+1$ and $-5$ through \eqref{eq:V04-gaugedE}, the $J$'s the derivatives
of $W = P\,G_5$. This is the correction the earlier ungauged reading misses: for the gauged theory
$E \neq 0$. Now form the closure sum \eqref{eq:V04-ejclosure},
\begin{equation}
\label{eq:V04-quinticclosure}
 \sum_i E_i J^i + E_P J^P
 \;=\; \sqrt{2}\,\Sigma\, P \sum_i \phi_i\,\frac{\partial G_5}{\partial \phi_i}
 \;-\; 5\sqrt{2}\,\Sigma\, P\, G_5
 \;=\; \sqrt{2}\,\Sigma\, P\Big(\sum_i \phi_i\,\partial_i G_5 - 5\,G_5\Big),
\end{equation}
the common factor $\sqrt2\,\Sigma P$ pulled out. The bracket vanishes by Euler's theorem for a
homogeneous function: $G_5$ is homogeneous of degree five, so
\begin{equation}
\label{eq:V04-euler}
 \sum_i \phi_i\,\frac{\partial G_5}{\partial \phi_i} \;=\; 5\, G_5
 \quad\Longrightarrow\quad
 \sum_i E_i J^i + E_P J^P \;=\; 0.
\end{equation}
The closure follows from gauge invariance of the superpotential, equivalently the quasi-homogeneity
identity $\sum_i Q_i \Phi_i \partial_i W = 0$. For the quintic the same integer $5$ appears in three
places: gauge invariance of $W = P G_5$ (the charge $Q_P = -5$ balancing the degree $\deg G_5 = 5$), the
Euler homogeneity $\sum_i \phi_i \partial_i G_5 = 5 G_5$, and the Calabi--Yau charge sum $5(+1) + (-5) = 0$.
A superpotential of the wrong degree is not gauge invariant, and then its $E$/$J$ do not close: with
$Q_P = -5$ fixed but $\deg G_d = d$, the same computation leaves
\begin{equation}
\label{eq:V04-wrongdegree}
 \sum_i E_i J^i + E_P J^P \;=\; \sqrt{2}\,\Sigma\, P\,(d - 5)\,G_d \;\neq\; 0 \qquad (d \neq 5),
\end{equation}
so closure requires $d = 5 = -Q_P$, which is precisely gauge invariance of $W$. In this hypersurface
model that same integer $5$ is also the field-theory shadow of the Calabi--Yau charge assignment, but the
two statements are distinct: $E$/$J$ closure is the supersymmetry / gauge-invariance condition, while the
Calabi--Yau condition is the separate statement that the axial R-symmetry is non-anomalous.

\subsection*{Worked instance: a $\mathcal{N}=(0,2)$ deformation of the quintic GLSM}

The freedom to choose $E \neq TX$ is best seen as a deformation of the $\mathcal{N}=(2,2)$ point just worked. At
the $\mathcal{N}=(2,2)$ (tangent-bundle) locus the Fermi data are the locked pair \eqref{eq:V04-quinticEJ}. A
$\mathcal{N}=(0,2)$ deformation replaces the locked $J^i$ by more general holomorphic functions $J^i = J^i(\phi,
P)$ and deforms the $E_i$ away from \eqref{eq:V04-gaugedE}, subject to two constraints: the closure
$\sum_a \mathrm{Tr}(E_a J^a) = 0$ of \eqref{eq:V04-ejclosure} (supersymmetry) and the bundle-anomaly
conditions \eqref{eq:V04-bundleanom}. The deformed model describes the same Calabi--Yau $X$ but with
the left-moving fermions coupled to a bundle $E \neq TX$: a \emph{bundle-deformed} gauged linear sigma
model. The anomaly conditions bound the family, keeping $\mathrm{ch}_2(E) = \mathrm{ch}_2(TX)$ and
$c_1(E) \equiv c_1(TX) \bmod 2$, so the surviving deformations are those preserving the two
characteristic classes.

The lesson is the shape of the moduli space. The $\mathcal{N}=(2,2)$ locus (complex-structure and Kähler
deformations) sits inside a larger $\mathcal{N}=(0,2)$ moduli space (bundle deformations with $E \neq TX$), the
enlargement the extra freedom in $E$ and $J$ that the anomaly conditions do not kill. So $\mathcal{N}=(0,2)$ models
are the field-theory description of heterotic compactifications on $(X, E)$, their marginal deformations
the bundle moduli. Which deformations survive is controlled here by \eqref{eq:V04-bundleanom} together
with the closure \eqref{eq:V04-ejclosure}.

\section{2d anomalies and c-extremization}
\label{sec:V04-cext}

Chirality makes two dimensions dangerous, and the danger is quantified by chiral anomalies of two
kinds, handled in opposite ways. It is also chirality that lets an extremization principle fix the
exact superconformal R-symmetry and the central charges. We take the anomalies first, then
$c$-extremization, which is built from them.

\medskip\noindent\textbf{The gauge anomaly.}\enspace
A $2d$ chiral fermion in a gauge representation contributes to the gauge two-point anomaly with a sign
set by its worldsheet chirality: right-movers one sign, left-movers the other. Collecting the $\mathcal{N}=(0,2)$
content, the right-moving fermions in chiral multiplets enter with a plus, and the left-moving
fermions in Fermi multiplets and the gaugino (an adjoint left-mover) enter with a minus. The theory is
consistent only if the total vanishes,
\begin{equation}
\label{eq:V04-gaugeanom}
 \mathrm{Tr}_{\text{chiral}} T^A T^B \;-\; \mathrm{Tr}_{\text{Fermi}} T^A T^B \;-\;
 \mathrm{Tr}_{\text{adj}} T^A T^B \;=\; 0.
\end{equation}
On the $\mathbb{C}^4$ brick, where the content per node is four adjoint chirals, three adjoint Fermis,
and one adjoint gaugino, this reads in units of the adjoint index
\begin{equation}
\label{eq:V04-c4anom}
 4\,T(\mathrm{adj}) \;-\; 3\,T(\mathrm{adj}) \;-\; T(\mathrm{adj}) \;=\; 0,
\end{equation}
the clean $4 - 3 - 1 = 0$. The gaugino term is load-bearing: dropping it leaves $4 - 3 = 1 \neq 0$, an
inconsistent theory. For a $\mathcal{N}=(2,2)$ theory the cancellation is automatic, and
\eqref{eq:V04-decompchiral} is why: each $\mathcal{N}=(2,2)$ chiral supplies one $\mathcal{N}=(0,2)$ chiral and one $\mathcal{N}=(0,2)$
Fermi in the same representation, so their traces cancel. For $n$ $\mathcal{N}=(2,2)$ chirals plus one $\mathcal{N}=(2,2)$
vector, the $\mathcal{N}=(0,2)$ counts are $n{+}1$ chirals (the $n$ matter chirals plus the vector's adjoint
chiral), $n$ Fermis, and one vector, so \eqref{eq:V04-gaugeanom} reads
\begin{equation}
\label{eq:V04-22autoanom}
 (n+1) \;-\; n \;-\; 1 \;=\; 0
\end{equation}
identically. The Gadde--Gukov--Putrov node of \S\ref{sec:V04-triality} is the worked nontrivial case: a
$U(N_c)$ gauge group with $N_1$ fundamental chirals, $N_2$ antifundamental chirals, and $N_3$ Fermis is
gauge-anomaly-free precisely when
\begin{equation}
\label{eq:V04-ggpanom}
 N_1 + N_2 - N_3 - 2N_c \;=\; 0,
\end{equation}
fixing the balanced rank $N_c = \tfrac12(N_1 + N_2 - N_3)$, the $2N_c$ the gaugino's adjoint
contribution. At $(N_1, N_2, N_3) = (5,3,4)$ the balanced rank is $N_c = 2$, and $5 + 3 - 4 - 4 = 0$.
Reading the coefficient block by block makes the sign structure explicit: each fundamental chiral
contributes $+T(\square) = +\tfrac12$ (right-moving), each Fermi $-T(\square) = -\tfrac12$
(left-moving), and the gaugino $-T(\mathrm{adj}) = -N_c$ (left-moving adjoint), so the total is
\begin{equation}
\label{eq:V04-ggpcoeff}
 \tfrac12(N_1 + N_2) \;-\; \tfrac12 N_3 \;-\; N_c \;=\; \tfrac12\big(N_1 + N_2 - N_3 - 2N_c\big),
\end{equation}
zero exactly on the balanced node. The rank $N_c$ is not free: it is \emph{determined} by requiring the
gauge anomaly to vanish, which is why $\mathcal{N}=(0,2)$ SQCD has a fixed gauge rank for given flavor content, and
it is this same balancing that the triality of \S\ref{sec:V04-triality} preserves as it permutes the
flavor blocks.

\medskip\noindent\textbf{The gravitational anomaly.}\enspace
The gravitational anomaly is the net worldsheet chirality of the massless fermions,
\begin{equation}
\label{eq:V04-grav}
 c_R - c_L \;=\; \mathrm{Tr}\,\gamma^3,
\end{equation}
where $\gamma^3 = +1$ on a right-mover and $-1$ on a left-mover, so $\mathrm{Tr}\,\gamma^3$ counts
right-movers minus left-movers. Unlike the gauge anomaly, this is generically \emph{nonzero}, and a
nonzero value is allowed and physical. It is a 't~Hooft anomaly of the $2d$ Lorentz (chirality)
symmetry, invariant along the renormalization-group flow, and it must \emph{match} between two
descriptions of the same theory rather than cancel. A content of three right-movers and one left-mover
has $c_R - c_L = 3 - 1 = 2$, a consistent chiral theory. The $\mathbb{C}^4/\mathbb{Z}_4$ brick, with
$16$ right-moving chiral fermions, $12$ left-moving Fermi fermions, and $4$ left-moving gauginos, has
\begin{equation}
\label{eq:V04-gravc4z4}
 c_R - c_L \;=\; 16 - 12 - 4 \;=\; 0,
\end{equation}
a non-chiral theory. The matching statement is the working tool: two descriptions of the same infrared
physics must report the same $c_R - c_L$, one of the invariants the three triality frames share. As a
matching example, two spectra grouped differently but sharing the net chirality,
\begin{align}
\label{eq:V04-gravmatch}
 \text{A:}&\quad 4\ \text{right} + 2\ \text{left}: \quad c_R - c_L = 4 - 2 = 2, \notag\\
 \text{B:}&\quad (3 + 1)\ \text{right} + 2\ \text{left}: \quad c_R - c_L = (3 + 1) - 2 = 2,
\end{align}
report the same $c_R - c_L = 2$: the anomaly sees only the net right-minus-left count, and being
RG-invariant it is the same in the ultraviolet and infrared. This is what makes it a diagnostic: if two
ultraviolet Lagrangians claiming the same infrared theory report different $c_R - c_L$, one is wrong,
and the computation is a one-line count of fermions weighted by $\gamma^3$.

\begin{keybox}{Common misconception: the gravitational anomaly must cancel}
Two dimensions has two distinct chiral-anomaly conditions, handled differently. The \emph{gauge}
anomaly $\mathrm{Tr}_{\text{chiral}} - \mathrm{Tr}_{\text{Fermi}} - \mathrm{Tr}_{\text{adj}} = 0$
must cancel for the theory to exist. The \emph{gravitational} anomaly $c_R - c_L = \mathrm{Tr}\,
\gamma^3$, the net fermion chirality, is generically nonzero; it is a 't~Hooft anomaly of the $2d$
Lorentz symmetry, RG-invariant, and it must \emph{match} between descriptions (for instance the three
triality frames), not cancel. A nonzero $c_R - c_L$ is allowed and physical; an uncancelled
\emph{gauge} anomaly is fatal.
\end{keybox}

\medskip\noindent\textbf{c-extremization.}\enspace
The exact right-moving R-symmetry of a $2d$ $\mathcal{N}=(0,2)$ (or $\mathcal{N}=(2,2)$) fixed point is fixed by an
extremization principle, the two-dimensional member of the extremization through-line whose $4d$
prototype is the $a$-maximization of Section~3. The right-moving central charge of the fixed point is
the 't~Hooft anomaly of its right-moving R-symmetry, so for a trial right-moving R the trial central
charge is
\begin{equation}
\label{eq:V04-cext}
 c_R^{\mathrm{tr}}(R) \;=\; 3\,\mathrm{Tr}\,\big[\gamma^3\, R^2\big],
\end{equation}
a sum over fermions weighted by the chirality $\gamma^3 = \pm 1$ and the square of the R-charge, the
$2d$ analog of the trace that builds $a$ in four dimensions. A trial R ranges over the exact R mixed
with the abelian flavors, $R = R_0 + \sum_F s_F\,F$, and the exact superconformal R is the stationary
point, where the mixed anomaly with each flavor vanishes,
\begin{equation}
\label{eq:V04-krf}
 k_{RF} \;=\; \mathrm{Tr}\,\big[\gamma^3\, R\, F\big] \;=\; 0,
\end{equation}
one equation per flavor. Solving \eqref{eq:V04-krf} gives $R_{\mathrm{exact}}$, evaluating
\eqref{eq:V04-cext} there gives $c_R$, and the left-moving central charge follows from the
gravitational anomaly,
\begin{equation}
\label{eq:V04-cLfromcR}
 c_L \;=\; c_R - \mathrm{Tr}\,\gamma^3.
\end{equation}
The hypotheses match $a$-maximization: the theory must flow to a good candidate infrared fixed point,
the trial R must be anomaly-free (a genuine symmetry), and no accidental symmetries mix into the R at
the fixed point. A formal fermion list violating any of these can be fed through \eqref{eq:V04-cext},
but its output is not the central charge of a unitary theory.

\medskip\noindent\textbf{Stationary, not maximal.}\enspace
Here is the sharp difference from $a$-maximization. In four dimensions the exact R \emph{maximizes} $a$,
and the Hessian is negative definite. In two dimensions the exact R only \emph{extremizes}
$c_R^{\mathrm{tr}}$, and the stationary point is generically a \emph{saddle}. Differentiating
\eqref{eq:V04-cext} twice in the mixing parameters, with $\partial R/\partial s_F = F$, gives the
$\gamma^3$-weighted flavor metric
\begin{equation}
\label{eq:V04-hessian}
 \frac{\partial^2 c_R^{\mathrm{tr}}}{\partial s_F\,\partial s_{F'}}
 \;=\; 6\,\mathrm{Tr}\,\big[\gamma^3\, F\, F'\big],
\end{equation}
and because $\gamma^3$ takes both signs this metric is \emph{indefinite}. To see it on a two-flavor
fixture, take one right-moving chiral ($\gamma^3 = +1$) carrying $F_1 = 1$, $F_2 = 0$ and one
left-moving Fermi ($\gamma^3 = -1$) carrying $F_1 = 0$, $F_2 = 1$. Then $c_R^{\mathrm{tr}}(s_1, s_2) =
3[(+1)(R_0 + s_1)^2 + (-1)(R_0 + s_2)^2]$, and its Hessian is
\begin{equation}
\label{eq:V04-hessworked}
 \frac{\partial^2 c_R^{\mathrm{tr}}}{\partial s_a\,\partial s_b} \;=\; 6\begin{pmatrix} +1 & 0 \\ 0 & -1
 \end{pmatrix}, \qquad \text{eigenvalues } \{+6,\ -6\},
\end{equation}
one positive and one negative, a genuine saddle: the trial central charge increases along the
right-moving flavor and decreases along the left-moving one, so no stationary point is a local maximum.
What $c$-extremization fixes is the stationarity condition and the extremum type, not a maximum.

The saddle-versus-maximum split is structural, not accidental. Four-dimensional $a$-maximization
extremizes a \emph{cubic} anomaly $\mathrm{Tr}\,R^3$ whose Hessian quadratic form is negative definite,
so $a$ is truly maximized. Two-dimensional $c$-extremization extremizes a \emph{quadratic} anomaly
$\mathrm{Tr}[\gamma^3 R^2]$ whose Hessian \eqref{eq:V04-hessian} is weighted by $\gamma^3$: right-moving
flavors give positive curvature, left-moving negative, so the metric is indefinite. The same chirality
that made two dimensions dangerous turns the maximization of four dimensions into an extremization here,
and one must \emph{solve} $k_{RF} = 0$ rather than search for the largest $c_R$.

\subsection*{Worked instance: the $\mathcal{N}=(0,2)$ $A_{k+1}$ minimal model (positive $c_R$)}

The physical worked example is a genuine unitary $\mathcal{N}=(0,2)$ superconformal field theory: the $\mathcal{N}=(0,2)$
description of the $A_{k+1}$ Landau--Ginzburg minimal model. Take one $\mathcal{N}=(0,2)$ chiral $\Phi$ (with
right-moving fermion $\psi_+$, $\gamma^3 = +1$) and one $\mathcal{N}=(0,2)$ Fermi $\Lambda$ (with left-moving
$\lambda_-$, $\gamma^3 = -1$), coupled by
\begin{equation}
\label{eq:V04-mmJ}
 J \;=\; \Phi^{k+1}, \qquad E = 0.
\end{equation}
The superpotential $\int d\theta^+\,\Lambda\, \Phi^{k+1}$ is the $\mathcal{N}=(0,2)$ face of the $\mathcal{N}=(2,2)$
Landau--Ginzburg model with $W = \Phi^{k+2}$. There is no residual abelian flavor to mix here, so the
R-symmetry is fixed by quasi-homogeneity alone and $c$-extremization returns it with no free parameter.
Quasi-homogeneity of $J$ and the superpotential condition $R(\Lambda) + R(J) = 1$ fix the R-charges,
\begin{equation}
\label{eq:V04-mmR}
 R(\phi) \;=\; \frac{1}{k+2}, \qquad
 R(\psi_+) \;=\; R(\phi) - 1 \;=\; -\frac{k+1}{k+2}, \qquad
 R(\lambda_-) \;=\; \frac{1}{k+2},
\end{equation}
the last from $R(J) = (k+1)/(k+2)$. Evaluating the trial central charge \eqref{eq:V04-cext} on these
two fermions,
\begin{equation}
\label{eq:V04-mmcR}
 c_R \;=\; 3\Big[\,(+1)\Big(\tfrac{k+1}{k+2}\Big)^2 \;+\; (-1)\Big(\tfrac{1}{k+2}\Big)^2\,\Big]
 \;=\; 3\cdot\frac{(k+1)^2 - 1}{(k+2)^2} \;=\; \frac{3k}{k+2},
\end{equation}
positive for every $k \geq 1$: a genuine unitary superconformal field theory. It equals $3\hat c$ with
the Landau--Ginzburg central charge $\hat c = k/(k+2) = 1 - 2R(\phi)$, matching \eqref{eq:V04-Akchat}.
The smallest nontrivial case is $k = 2$,
\begin{equation}
\label{eq:V04-mmk2}
 k = 2: \quad R(\phi) = \tfrac14, \quad c_R = \frac{3\cdot 2}{2+2} = \frac{3}{2},
\end{equation}
the $c_R = 3/2$ of the $k = 2$ member of the $\mathcal{N} = 2$ minimal-model series. As $k$ grows,
$c_R = 3k/(k+2) \to 3$ from below. This is the flagship $c$-extremization example: a physical model, a
positive central charge, an R-symmetry fixed by symmetry with no free mixing.

\subsection*{Worked instance: a formal saddle fixture (not a unitary SCFT)}

To exhibit the indefinite Hessian directly one needs a free mixing parameter, and a minimal fixture is
a \emph{formal} fermion list chosen only to expose the saddle, not a unitary theory. Take two
right-moving chiral fermions ($\gamma^3 = +1$) with flavor charges $F = +1$ and $F = +2$, and two
left-moving Fermi fermions ($\gamma^3 = -1$) with $F = -1$ and $F = +1$. The $\gamma^3$-weighted
flavor norm is
\begin{equation}
\label{eq:V04-flavornorm}
 \mathrm{Tr}\,\big[\gamma^3\, F^2\big] \;=\; (1 + 4) - (1 + 1) \;=\; 3 \;\neq\; 0,
\end{equation}
so the mixing genuinely moves the mixed anomaly. Writing $R(s) = R_0 + s\,F$ with a non-extremal seed
$R_0 = 0.2$, the stationarity $k_{RF}(s) = \mathrm{Tr}[\gamma^3 R F] = 0$ is one linear equation,
\begin{equation}
\label{eq:V04-krfsolve}
 k_{RF}(s) \;=\; 0.6 + 3s \;=\; 0 \quad\Longrightarrow\quad s_* \;=\; -0.2,
\end{equation}
with the constant $0.6 = k_{RF}(0)$ the anomaly at the seed and the slope $3$ the flavor norm
\eqref{eq:V04-flavornorm}. At $s_*$ the four R-charges are $R = 0,\, -0.2,\, 0.4,\, 0$, and the trial
central charge there is
\begin{equation}
\label{eq:V04-cRvalue}
 c_R \;=\; 3\big[(0)^2 + (0.2)^2 - (0.4)^2 - (0)^2\big] \;=\; 3(0.04 - 0.16) \;=\; -0.36.
\end{equation}
The negative output is the point: a negative ``central charge'' signals a fermion list that is
\emph{not} a unitary superconformal field theory. The extremization algebra runs on any anomaly data,
but only a genuine candidate fixed point (like the minimal model above) returns a physical $c_R$. This
fixture is here only to display the saddle, never as a physical central charge.

\subsection*{Worked instance: the symmetric brick (R forced by structure)}

The contrast partner is the case where symmetry does the work, the $2d$ analog of the symmetric-R
quiver of Section~3. For the $\mathbb{C}^4/\mathbb{Z}_4$ $\mathcal{N}=(0,2)$ brick the R-charge is forced by the
$E/J$ and superpotential structure with no free mixing. A Fermi obeys $\bar D_+ \Lambda_a = E_a$ with
$E_a$ bilinear in the chirals, so $R(E_a) = 2R_X$ with $R_X$ the common chiral R-charge; the $\bar D_+$
carries R-charge $+1$, so $R(\Lambda_a) = R(E_a) - 1 = 2R_X - 1$. The superpotential $W_{\mathcal{N}=(0,2)} = \int
d\theta^+ \sum_a \Lambda_a J^a$ must carry R-charge one, and $J^a$ is also bilinear, $R(J^a) = 2R_X$,
so
\begin{equation}
\label{eq:V04-RWconstraint}
 R(\Lambda_a) + R(J^a) \;=\; (2R_X - 1) + 2R_X \;=\; 1
 \quad\Longrightarrow\quad R_X \;=\; \tfrac12, \quad R(\Lambda_a) = 0.
\end{equation}
The symmetric chiral R is forced to $R_X = 1/2$ and the Fermi R to zero by the structure alone;
$c$-extremization then confirms this assignment, exactly as $a$-maximization confirmed the symmetric R
of the $4d\ \mathrm{dP}_0$ quiver. When a symmetry pins the R uniquely the extremization is degenerate;
only when a free mixing parameter survives does one actually solve \eqref{eq:V04-krf}.

\section{\texorpdfstring{$\mathcal{N}=(0,2)$}{N=(0,2)} triality}
\label{sec:V04-triality}

Three different $2d$ $\mathcal{N}=(0,2)$ gauge theories can flow to the same infrared $\mathcal{N}=(0,2)$ superconformal field
theory. This is the two-dimensional cousin of Seiberg duality, but with a decisive difference: it is
an \emph{order-three} relation, a triality, not an order-two involution. In the graded-quiver ladder
the $\mathcal{N}=(0,2)$ brick is the $m = 2$ rung, and the duality move is the $n$-ality of order $m + 1 = 3$. We
pin the Gadde--Gukov--Putrov conventions exactly, because the flavor labels are convention-sensitive
and the field-theory content is more than a rank map.

\medskip\noindent\textbf{The Gadde--Gukov--Putrov node in GGP conventions.}\enspace
The family that carries the statement is the $\mathcal{N}=(0,2)$ SQCD node. In the conventions of
Gadde--Gukov--Putrov it is a $U(N_c)$ gauge theory with three flavor blocks,
\begin{equation}
\label{eq:V04-ggpmatter}
\begin{aligned}
 &N_1\ \text{fundamental chirals } P, \qquad
 N_2\ \text{antifundamental chirals } \Phi, \\
 &N_3\ \text{fundamental Fermis } \Gamma,
\end{aligned}
\end{equation}
together with the $E$ and $J$ couplings that consistency requires. Here $N_1$ counts fundamental chiral
multiplets, $N_2$ antifundamental chiral multiplets, and $N_3$ Fermi multiplets, and the $SU(N_c)^2$
gauge anomaly equation is \eqref{eq:V04-ggpanom},
\begin{equation}
\label{eq:V04-ggpanomrestate}
 N_1 + N_2 - N_3 - 2N_c \;=\; 0 \quad\Longrightarrow\quad N_c \;=\; \tfrac12(N_1 + N_2 - N_3),
\end{equation}
each fundamental chiral $+\tfrac12$, each Fermi $-\tfrac12$, the gaugino $-N_c$. The triality move at
the dualized node sends the rank to
\begin{equation}
\label{eq:V04-trialityrank}
 N_c \;\longmapsto\; N_c' \;=\; N_2 - N_c \;=\; \tfrac12(N_2 + N_3 - N_1),
\end{equation}
the value fixed by requiring the dual node anomaly-free in the same convention, and it cyclically
permutes the three flavor roles,
\begin{equation}
\label{eq:V04-trialityperm}
 (N_1, N_2, N_3) \;\longmapsto\; (N_2, N_3, N_1)
\end{equation}
(fundamental chirals $\to$ antifundamental chirals $\to$ Fermis $\to$ fundamental chirals). The rank
map alone, $N_c \mapsto N_2 - N_c$, is an involution ($N_2 - (N_2 - N_c) = N_c$), an order-two 2-cycle
just like Seiberg duality; it is the flavor permutation that lifts the combined move to order three.

\medskip\noindent\textbf{The dual matter content.}\enspace
Triality is not just rank arithmetic. After dualizing the $U(N_c)$ node to $U(N_c')$ with $N_c' = N_2 -
N_c$, the dual matter is
\begin{itemize}
\item the permuted flavor blocks of \eqref{eq:V04-trialityperm}: $N_2$ fundamental chirals, $N_3$
antifundamental chirals, and $N_1$ fundamental Fermis of the dual gauge group $U(N_c')$;
\item the \emph{meson} chiral multiplets $M = \Phi P$ (gauge singlets built from the original quarks),
which appear as elementary fields in the dual, exactly as in $4d$ Seiberg duality;
\item the additional gauge-singlet \emph{Fermi} multiplets required to write the dual $E$ and $J$
couplings, which pair the mesons to the dual quarks so that $\sum \mathrm{Tr}(E J) = 0$ closes in the
dual frame;
\item the flip fields (singlet chirals) whose $J$-terms remove the operators that became redundant, the
$\mathcal{N}=(0,2)$ counterpart of the $4d$ Seiberg meson superpotential $W = M q \tilde q$.
\end{itemize}
The mesons and singlet Fermis are essential: without them the dual reproduces neither the operator
spectrum nor the elliptic genus and central charges. A statement of triality giving only the rank map
is incomplete; the relation is the full matter content plus the shared anomalies, central charges, and
elliptic genus. Concretely, dualizing the first frame of the orbit below, $(N_1, N_2, N_3) = (5,3,4)$ at
$U(2)$, gives the dual frame
\begin{equation}
\label{eq:V04-dualframe}
 U(N_c') = U(N_2 - N_c) = U(1), \qquad
 (N_1', N_2', N_3') = (3,4,5), \qquad
 N_M = N_1 N_2 = 15,
\end{equation}
a $U(1)$ gauge theory with the permuted quark blocks $(3,4,5)$, the $N_M = 15$ mesons $M = \Phi P$ as
singlet chiral fields, and the singlet Fermi multiplets carrying the dual $J$-couplings $\Gamma\,(M -
\Phi P)$ that enforce the meson definition. The rank dropped from $2$ to $1$, but the operator content did
not shrink: the mesons that were composite in the first frame are elementary in the second, exactly as
in $4d$ Seiberg duality. This is why triality is a relation between full theories, not a rank
recursion.

\medskip\noindent\textbf{The order-three return, checked in all three frames.}\enspace
Apply the combined move to an explicit asymmetric balanced triple, $(N_1, N_2, N_3) = (5,3,4)$, with
$N_c = \tfrac12(5 + 3 - 4) = 2$. Recomputing the rank from \eqref{eq:V04-ggpanomrestate} after each
flavor permutation,
\begin{align}
\label{eq:V04-trialityranks}
 (5,3,4): \ &N_c = \tfrac12(5+3-4) = 2, \quad 5+3-4-2\cdot 2 = 0, \notag\\
 (3,4,5): \ &N_c = \tfrac12(3+4-5) = 1, \quad 3+4-5-2\cdot 1 = 0, \notag\\
 (4,5,3): \ &N_c = \tfrac12(4+5-3) = 3, \quad 4+5-3-2\cdot 3 = 0,
\end{align}
each frame separately gauge-anomaly-free by \eqref{eq:V04-ggpanomrestate}. Tracking $(N_1, N_2, N_3;
N_c)$ through the three steps,
\begin{equation}
\label{eq:V04-trialityorbit}
 (5,3,4;\,2) \;\to\; (3,4,5;\,1) \;\to\; (4,5,3;\,3) \;\to\; (5,3,4;\,2),
\end{equation}
the intermediate ranks $2, 1, 3$ genuinely different and the third application returning the original
data, $T^3 = \mathds{1}$. The orbit has length three, not two. Without the flavor permutation the bare
rank recursion would close at two; the permutation is what makes it a triality. When the flavor numbers
coincide the node is self-dual,
\begin{equation}
\label{eq:V04-selfdual}
 N_1 = N_2 = N_3 \;\Longrightarrow\; N_2 = 2N_c, \quad N_c' = N_2 - N_c = N_c,
\end{equation}
the permutation the identity and the rank fixed: a triality fixed point of order one, a single
self-dual frame.

\medskip\noindent\textbf{The evidence.}\enspace
At toolkit depth the load-bearing evidence is anomaly and index matching. Each frame is separately
gauge-anomaly-free \eqref{eq:V04-ggpanomrestate}, and the three share the global symmetry, the
gravitational anomaly $c_R - c_L$, the central charges from $c$-extremization, and the elliptic genus of
\S\ref{sec:V04-ellgenus}. The elliptic genus is the sharpest: a whole function of the fugacities, its
equality across the three frames pins the full protected spectrum. As a cheap necessary check one can
also form a permutation-symmetric combination of the flavor numbers, the toy quantity
\begin{equation}
\label{eq:V04-flavorinv}
 I \;=\; N_1 N_2 + N_2 N_3 + N_3 N_1 - \tfrac12\big(N_1^2 + N_2^2 + N_3^2\big),
\end{equation}
symmetric in $(N_1, N_2, N_3)$ and so unchanged by the permutation \eqref{eq:V04-trialityperm}. This is
a toy symmetric check, not a named central-charge or anomaly formula: it is invariant precisely because
it is a symmetric function of the flavor multiset, so it is only bookkeeping. On the orbit the three
frames each give
\begin{equation}
\label{eq:V04-flavorinvworked}
 I \;=\; (15 + 12 + 20) - \tfrac12(25 + 9 + 16) \;=\; 47 - 25 \;=\; 22,
\end{equation}
the same value (the same multiset reordered), while a mismatched $(6,3,4)$ gives $I = 54 - 30.5 = 23.5
\neq 22$ and is not a frame. The check does real work only in the negative direction (rejecting a
wrong-multiset candidate); the genuine relation is the anomaly, central-charge, and elliptic-genus
matches, not a symmetric polynomial in the ranks. The full spectrum-reconstructed match of $c_R - c_L$
(needing the complete singlet and meson content) and the proof that the three theories share an
infrared fixed point are cited rather than proved here; the exact GGP conventions, the dual matter
content, the order-three return, and the anomaly matches are what is established.

\begin{table}[ht]
\centering
\small
\setlength{\tabcolsep}{9pt}
\renewcommand{\arraystretch}{1.3}
\begin{tabular}{@{}ccccc@{}}
\toprule
frame & $N_1$ & $N_2$ & $N_3$ & $N_c$ \\
\midrule
$1$ & $5$ & $3$ & $4$ & $2$ \\
\midrule
$2$ (dualize) & $3$ & $4$ & $5$ & $1$ \\
\midrule
$3$ (dualize) & $4$ & $5$ & $3$ & $3$ \\
\midrule
$1$ (return) & $5$ & $3$ & $4$ & $2$ \\
\bottomrule
\end{tabular}
\caption{One $\mathcal{N}=(0,2)$ triality orbit in Gadde--Gukov--Putrov conventions. Each step applies the rank
map $N_c \mapsto N_2 - N_c$ together with the flavor cyclic permutation $(N_1, N_2, N_3) \mapsto (N_2,
N_3, N_1)$, and generates the dual matter (permuted quark blocks plus mesons and singlet Fermis). The
intermediate ranks $2, 1, 3$ differ and the third step returns the original data ($T^3 =
\mathds{1}$): the orbit has length three, not two.}
\label{tab:V04-triality}
\end{table}

\begin{keybox}{Common misconception: $\mathcal{N}=(0,2)$ triality is a $2d$ Seiberg duality}
Seiberg duality in four dimensions is an order-two relation: dualizing twice returns the original
theory. $\mathcal{N}=(0,2)$ triality is order \emph{three}. The rank map $N_c \mapsto N_2 - N_c$ alone is an
involution, but the triality is the rank map \emph{together with} the flavor cyclic permutation
$(N_1, N_2, N_3) \mapsto (N_2, N_3, N_1)$, and the combined move closes only after three applications,
$T^3 = \mathds{1}$. The three frames are three \emph{different} ultraviolet gauge theories, with dual
matter (permuted quarks plus mesons and singlet Fermis), that flow to the same infrared physics, and
the relation is checked by the anomaly, central-charge, and elliptic-genus matches across all three.
\end{keybox}

\bigskip
\begin{center}
\rule{0.4\textwidth}{0.4pt}\\[3pt]
{\large\textsf{\textbf{Block C.\enspace The shared protected interface}}}\\[2pt]
\rule{0.4\textwidth}{0.4pt}
\end{center}
\medskip

\noindent The last two sections are the genuinely shared technology. The elliptic genus
(\S\ref{sec:V04-ellgenus}) is a $\mathcal{N}=(0,2)$ object that specializes to $\mathcal{N}=(2,2)$ and is the sharpest
protected check for both worlds, including the triality of Block B and the GLSM phases of Block A. A
final comparison table (\S\ref{sec:V04-compare}) reads the two worlds side by side.

\section{The elliptic genus}
\label{sec:V04-ellgenus}

The sharpest protected observable of a $2d$ $\mathcal{N}=(0,2)$ theory is its elliptic genus, the supersymmetric
partition function on a two-torus. It is the two-dimensional analog of the $4d$ superconformal index of
Section~3: a flavored, signed counting function, invariant under continuous deformations, that matches
across dualities. It is a $\mathcal{N}=(0,2)$ object that specializes to $\mathcal{N}=(2,2)$, which is why it is the shared
tool that closes the section rather than belonging only to Block A or Block B.

\medskip\noindent\textbf{A protected index.}\enspace
The elliptic genus is the torus trace, in the Ramond--Ramond sector, of $(-1)^F$ weighted by the left-
and right-moving Hamiltonians and the flavor fugacities,
\begin{equation}
\label{eq:V04-ellgenus}
 Z_{T^2}(\tau, z, \dots) \;=\; \mathrm{Tr}_{RR}\,\big[(-1)^F\, q^{H_L}\, \bar q^{H_R}\, y^{J}\,
 \cdots\big], \qquad q = e^{2\pi i\tau},\ y = e^{2\pi i z}.
\end{equation}
Why is it protected? It counts only states annihilated by one right-moving supercharge, with signs.
The right-moving oscillators come in boson-fermion pairs that cancel in the graded trace, so every
state with $H_R > 0$ drops out and the result is holomorphic in $\tau$ (meromorphic in $z$), depending
on $\tau$ and $z$ but not $\bar\tau$. That cancellation is why the genus is an index, not a full
partition function. For a $\mathcal{N}=(0,2)$ SCFT it is a weak Jacobi form in $(\tau, z)$ whose modular weight and
fugacity index are fixed by the anomaly data. It is RG-invariant, and one of the quantities that must
agree across the three triality frames.

Place it among the indices already known. The Witten index $\mathrm{Tr}(-1)^F$ counts ground states
with signs, one number. The elliptic genus refines it: it keeps the dependence on $\tau$ (through
$q^{H_L}$) and on the fugacities $z$, so it is a whole function, and it counts the entire tower of
right-moving-supersymmetric states graded by left-moving energy and flavor charge. Its two limits
recover the coarser indices,
\begin{equation}
\label{eq:V04-genuslimits}
 \begin{aligned}
 Z_{T^2}\big|_{q \to 0}
   &= \chi_y\text{-genus of the target} &&(\text{compact sigma model}),\\
 Z_{T^2}\big|_{q \to 0,\,y\to 1}
   &= \mathrm{Tr}(-1)^F &&(\text{Witten index}),
 \end{aligned}
\end{equation}
the first keeping the right-moving ground states while retaining the flavor grading, the second
forgetting that grading and reducing to one signed count. At generic $q$ the elliptic genus is richer
than either limit.
Because it is a weak Jacobi form in $\tau$ and $z$, matching it across a duality is far stronger than
matching one number: two theories can share a Witten index yet differ in genus, and equality of the
genus pins the full protected spectrum. This is why it is the sharpest triality test in
\S\ref{sec:V04-triality}.

\medskip\noindent\textbf{The free-field building blocks.}\enspace
The genus of an interacting theory is assembled from free-field factors, one per multiplet, and the
two $\mathcal{N}=(0,2)$ multiplets contribute reciprocal factors. A free $\mathcal{N}=(0,2)$ chiral of fugacity $z$ contributes
\begin{equation}
\label{eq:V04-chiralfactor}
 Z_{\Phi} \;=\; \frac{i\,\eta(\tau)}{\theta_1(\tau, z)},
\end{equation}
$\theta_1$ in the \emph{denominator}, a pole reflecting the boson zero modes. A free $\mathcal{N}=(0,2)$ Fermi of
fugacity $z$ contributes
\begin{equation}
\label{eq:V04-fermifactor}
 Z_{\Lambda} \;=\; \frac{i\,\theta_1(\tau, z)}{\eta(\tau)},
\end{equation}
$\theta_1$ in the \emph{numerator}, a zero reflecting the left-moving fermion zero mode. With
\emph{matched} fugacity and R-shift assignments the two are reciprocal up to the phase,
\begin{equation}
\label{eq:V04-reciprocal}
 Z_{\Phi}\, Z_{\Lambda} \;=\; \frac{i\,\eta}{\theta_1}\cdot\frac{i\,\theta_1}{\eta} \;=\; i^2 \;=\;
 -1,
\end{equation}
$\theta_1$ and $\eta$ cancelling. This is a useful matched-argument cancellation, but it is only a
\emph{mnemonic}, not the general genus of a $\mathcal{N}=(2,2)$ chiral. In the
$\mathcal{N}=(2,2)\to\mathcal{N}=(0,2)$ decomposition the chiral and Fermi factors usually carry
R/flavor-shifted arguments, so their product is the standard nontrivial theta-function ratio; only when
the arguments are deliberately matched do the pole and zero cancel as in \eqref{eq:V04-reciprocal}. The
universal statement is the \emph{placement}: chiral is $\theta_1$-in-the-denominator, Fermi is
$\theta_1$-in-the-numerator, never swapped; for noncompact free chirals the boson zero modes add
another subtlety.

The $q$-expansions make the matched-argument reciprocity concrete. In a schematic oscillator limit with
the zero-mode prefactors suppressed, the chiral factor is the inverse product (the right-moving boson
tower) and the Fermi factor is the product itself (the alternating single left-moving fermion sum),
\begin{align}
\label{eq:V04-qexpansions}
 Z_{\Phi} &\;\sim\; \frac{1}{\prod_{n\ge 1}(1 - q^n)} \;=\; 1 + q + 2q^2 + 3q^3 + 5q^4 + \cdots,
 \notag\\[2pt]
 Z_{\Lambda} &\;\sim\; \prod_{n\ge 1}(1 - q^n) \;=\; 1 - q - q^2 + q^5 + \cdots,
\end{align}
the chiral coefficients the partition numbers $p(n) = 1,1,2,3,5,\dots$, the Fermi coefficients the
pentagonal-number signs of Euler's identity. Their product is $1$ order by order,
\begin{equation}
\label{eq:V04-qproduct}
 Z_{\Phi}\, Z_{\Lambda} \;\sim\; \big(1 + q + 2q^2 + \cdots\big)\big(1 - q - q^2 + \cdots\big)
 \;=\; 1 + 0\cdot q + 0\cdot q^2 + \cdots \;=\; 1,
\end{equation}
the reciprocity read off the series. This computation is deliberately limited: it says that a chiral
pole and a Fermi zero cancel when their arguments match. It does \emph{not} say that the matter
contribution of every $\mathcal{N}=(2,2)$ theory is trivial. A $\mathcal{N}=(2,2)$ Calabi--Yau sigma model has a full
$q$-dependent elliptic genus, a weak Jacobi form; its $q\to0$ limit is the geometric $\chi_y$-genus, and
its specialization to $y=1$ gives the coarser signed index.

\medskip\noindent\textbf{The gauge-theory genus and its localization.}\enspace
For a gauge theory the elliptic genus is a contour integral over the gauge-holonomy fugacities $u$, the
one-loop determinants of the matter and gauge multiplets assembled into a meromorphic form and
integrated by the Jeffrey--Kirwan prescription,
\begin{equation}
\label{eq:V04-JK}
 Z_{T^2} \;=\; \frac{1}{|W|}\oint_{\mathrm{JK}} \prod_{a} \frac{du_a}{2\pi i\, u_a}\;
 Z_{\text{vec}}(u,\tau)\;\prod_{\text{matter}} Z_{\Phi/\Lambda}(u, z, \tau),
\end{equation}
the chiral \eqref{eq:V04-chiralfactor} and Fermi \eqref{eq:V04-fermifactor} factors of the charged
matter dressed with the gauge fugacities, $Z_{\text{vec}}$ the Cartan factor, $|W|$ the Weyl-group
order. The Jeffrey--Kirwan residue prescription selects which poles contribute (a chamber-dependent
rule resolving the flat-direction ambiguity), and its output is a weak Jacobi form. We name the
skeleton and cite the localization proof, the JK contour rule, and the modularity theorem.

\medskip\noindent\textbf{Modularity is gated by the gauge anomaly.}\enspace
The genus is modular only on a consistent theory. Under $\tau \to -1/\tau$ the $\theta_1$ and $\eta$
pick up known factors that combine into a weak-Jacobi-form transformation only if the anomalous phases
cancel, precisely when the gauge anomaly \eqref{eq:V04-gaugeanom} vanishes,
\begin{align}
\label{eq:V04-modulargate}
 \text{gauge anomaly} = 0 \ &\Longrightarrow\ Z_{T^2} \ \text{a weak Jacobi form}, \notag\\
 \text{gauge anomaly} \neq 0 \ &\Longrightarrow\ \text{no modular } Z_{T^2}.
\end{align}
On the $\mathbb{C}^4$ brick ($4 - 3 - 1 = 0$) the genus is a weak Jacobi form of fixed weight and index;
on a gauge-anomalous combination ($5 - 3 - 1 = 1 \neq 0$) the $\theta_1$ factors do not close under
modular transformations and there is no modular genus. So an anomaly-free spectrum is a prerequisite for
the genus to be a triality invariant.

\section[Comparison]{\texorpdfstring{$\mathcal{N}=(2,2)$}{N=(2,2)} versus \texorpdfstring{$\mathcal{N}=(0,2)$}{N=(0,2)}: comparison}
\label{sec:V04-compare}

With both worlds built, a side-by-side comparison makes the architecture explicit.
Table~\ref{tab:V04-compare} reads the two worlds together:
the balanced four-supercharge $\mathcal{N}=(2,2)$ theory with its two holomorphic sectors and Kähler geometry,
against the chiral two-supercharge $\mathcal{N}=(0,2)$ theory with its independent $E/J$ data and holomorphic
bundle.

\begin{table}[ht]
\centering
\small
\setlength{\tabcolsep}{5pt}
\renewcommand{\arraystretch}{1.35}
\begin{tabular}{@{}p{0.22\textwidth}p{0.35\textwidth}p{0.35\textwidth}@{}}
\toprule
feature & $\mathcal{N}=(2,2)$ & $\mathcal{N}=(0,2)$ \\
\midrule
supercharges & $4$ (two right, two left) & $2$ (two right) \\
\midrule
holomorphic data & $W(\Phi)$ and $\widetilde W(\Sigma)$, two sectors & $E_a$ and $J^a$, two independent
data per Fermi \\
\midrule
geometric meaning & Kähler / complex geometry, GLSM phases & holomorphic bundle $E \to X$, heterotic
data \\
\midrule
anomaly role & axial anomaly gates the B-twist / CY condition & gauge anomaly must cancel, gravitational
anomaly only matches \\
\midrule
extremization & inherited as the special case $E = TX$ & $c$-extremization central, $c_R = 3\,\mathrm{Tr}
[\gamma^3 R^2]$ \\
\midrule
protected index & elliptic genus specializes ($\chi_y$-genus) & elliptic genus natural, a weak Jacobi
form \\
\midrule
duality & mirror symmetry ($A(X) \cong B(\hat X)$) & triality (order three) \\
\midrule
construction use & GLSM / mirror / Calabi--Yau phases & brane-brick models / $\mathcal{N}=(0,2)$ quivers \\
\bottomrule
\end{tabular}
\caption{The two $2d$ four-supercharge and two-supercharge worlds side by side. The $\mathcal{N}=(2,2)$ theory is
recovered from $\mathcal{N}=(0,2)$ by locking $E = TX$; the genuinely $\mathcal{N}=(0,2)$ physics is the freedom $E \neq TX$,
the independent $E/J$ data, and the order-three triality. The elliptic genus is the shared protected
object, a $\mathcal{N}=(0,2)$ weak Jacobi form that specializes to the $\mathcal{N}=(2,2)$ geometric index.}
\label{tab:V04-compare}
\end{table}

\section*{Exit checklist}
\addcontentsline{toc}{subsection}{Exit checklist}
\markboth{Exit checklist}{Exit checklist}

After this section the reader can
\begin{enumerate}
\item place a $2d$ theory on the supercharge ladder ($\mathcal{N}=(2,2) = 4$, $\mathcal{N}=(0,2) = 2$ real supercharges by $p
+ q$), identify the $\mathcal{N}=(2,2)$ chiral, twisted-chiral, and vector multiplets and the $\mathcal{N}=(0,2)$ chiral,
Fermi, and vector multiplets, decompose $\mathcal{N}=(2,2)\ \text{chiral} = \mathcal{N}=(0,2)\ \text{chiral} + \mathcal{N}=(0,2)\
\text{Fermi}$ component by component, and name the vector and axial R-symmetries $U(1)_V$, $U(1)_A$;
\item (Block A) hold the two $\mathcal{N}=(2,2)$ holomorphic sectors apart (the ordinary $W(\Phi)$ over $\int
d^2\theta$ and the twisted $\widetilde W(\Sigma)$ over $\int d\theta^+ d\bar\theta^-$), read the
FI/theta $t = r - i\theta$ as the coefficient of the linear $\widetilde W = t\Sigma$, and name the
chiral (Jacobi) and twisted-chiral (quantum-cohomology) rings;
\item (Block A) compute the $\mathbb{CP}^{N-1}$ quantum cohomology from the effective twisted
superpotential $\widetilde W_{\mathrm{eff}} = -t\Sigma - N\Sigma(\log\Sigma - 1)$, get $\Sigma^N =
e^{-t} = q$ and the ring $\mathbb{C}[\Sigma]/(\Sigma^N - q)$ with $N$ vacua, and work the quintic in
both chambers (the $\mathbb{P}^4$ quotient of dimension $4$, the $G_5 = 0$ cut to $3$, the
$U(1)\to\mathbb{Z}_5$ Higgsing at $|P|^2 = -r/5$, the matched $\hat c = 3$) without confusing $W = 0$
with the quotient-then-cut construction;
\item (Block A) name the A-twist ($U(1)_V$, quantum cohomology $\mathbb{C}[x]/(x^N - q)$, not gated by
the CY condition) and the B-twist ($U(1)_A$, the Jacobi ring $\mathbb{C}[\Phi_i]/(\partial_i W)$, gated
by $c_1(TX) = 0$), and read the Hori--Vafa mirror exchanging them at the level of the untwisted
theories;
\item (Block B) write the $\mathcal{N}=(0,2)$ scalar potential $V \supset \sum_a|E_a|^2 + \sum_a|J^a|^2 + V_D$,
read its supersymmetric vacua $E_a = 0$, $J^a = 0$, verify the closure $\sum_a \mathrm{Tr}(E_a J^a) =
0$ (the Jacobi identity on the $\mathbb{C}^4$ brick), and state correctly that $E$ and $J$ enter
symmetrically but are \emph{not} exchanged by Hermitian conjugation (which complex-conjugates the data);
\item (Block B) distinguish an ungauged $\mathcal{N}=(2,2)$ LG model ($E = 0$, $J = \partial W$) from a gauged
$\mathcal{N}=(2,2)$ GLSM ($E_i = \sqrt2\, Q_i\, \Sigma\, \Phi_i \neq 0$), and verify the gauged quintic closure
$\sum_i E_i J^i + E_P J^P = \sqrt2\,\Sigma P(\sum_i \phi_i\partial_i G_5 - 5 G_5) = 0$ by Euler homogeneity and
gauge invariance (in this hypersurface the same degree-five charge assignment is also the Calabi--Yau
condition), plus the bundle interpretation ($E = TX$ for $\mathcal{N}=(2,2)$, $E \neq
TX$ for $\mathcal{N}=(0,2)$, with $\mathrm{ch}_2(E) = \mathrm{ch}_2(TX)$, $c_1(E) \equiv c_1(TX) \bmod 2$);
\item (Block B) check the $2d$ gauge anomaly ($4 - 3 - 1 = 0$) and the gravitational anomaly $c_R - c_L
= \mathrm{Tr}\,\gamma^3$ (matched, not cancelled), and run $c$-extremization: assemble $c_R^{\mathrm{tr}}
= 3\mathrm{Tr}[\gamma^3 R^2]$, solve $k_{RF} = 0$, get the positive $c_R = 3k/(k+2)$ of the $\mathcal{N}=(0,2)$
$A_{k+1}$ minimal model, read $c_L = c_R - \mathrm{Tr}\,\gamma^3$, and recognize the saddle (indefinite
Hessian) and the negative-output formal fixture for what they are;
\item (Block B) follow $\mathcal{N}=(0,2)$ triality in exact GGP conventions (the $U(N_c)$ node with $(N_1, N_2,
N_3)$, the rank map $N_c \mapsto N_2 - N_c$, the flavor permutation, the dual matter with mesons and
singlet Fermis), verify the order-three return $T^3 = \mathds{1}$, and match anomalies across the three
frames, with the elliptic genus the sharpest check and the symmetric flavor quantity a toy consistency
test;
\item (Block C) define the elliptic genus as a protected index (right-movers cancel), assign the
chiral ($i\eta/\theta_1$) and Fermi ($i\theta_1/\eta$) factors (never swapped, reciprocal only for
matched fugacities), write the Jeffrey--Kirwan localization skeleton, gate its modularity on the gauge
anomaly, and read the $\mathcal{N}=(2,2)$-versus-$\mathcal{N}=(0,2)$ comparison table.
\end{enumerate}

\bigskip
\section*{Sources and notes}
\addcontentsline{toc}{subsection}{Sources and notes}
\markboth{Sources and notes}{Sources and notes}
{\small

\noindent\textsf{\textcolor{RoyalBlue}{Sources and notes.}}\enspace
This is the two-dimensional dimension section of these notes, the low-supersymmetry holomorphic and
extremization partner of the $4d\ \mathcal{N}=1$ anchor.

\medskip\noindent\textsf{\textcolor{RoyalBlue}{\textbf{\S\ref{sec:V04-kinematics}\enspace Kinematics and R-symmetries.}}}\enspace
The $(p,q)$ supercharge count \eqref{eq:V04-pqcount} with $\mathcal{N}=(2,2) = 4$ \eqref{eq:V04-22count} and $\mathcal{N}=(0,2)
= 2$ \eqref{eq:V04-02count}; the vector and axial
R-symmetries $U(1)_V$, $U(1)_A$ \eqref{eq:V04-VA} (the axial anomaly $\propto c_1$); the $\mathcal{N}=(2,2)$
chiral, twisted-chiral, and vector multiplets, the $\mathcal{N}=(0,2)$ chiral, Fermi, and vector, and the
$\mathcal{N}=(2,2)\to\mathcal{N}=(0,2)$ decompositions \eqref{eq:V04-decompchiral}, \eqref{eq:V04-decompvector}
(Table~\ref{tab:V04-multiplets}); the section map into the three blocks. (\textcite{Witten:1993yc} the $\mathcal{N}=(2,2)$ multiplet grammar; \textcite{Hori:2003ic} the $2d$ multiplet
spine). 

\medskip\noindent\textsf{\textcolor{RoyalBlue}{\textbf{\S\ref{sec:V04-22data}\enspace Block A: the $\mathcal{N}=(2,2)$ holomorphic sectors and rings.}}}\enspace
The two holomorphic data of a $\mathcal{N}=(2,2)$ theory: the ordinary $W(\Phi)$ over $\int d^2\theta$
\eqref{eq:V04-Wordinary} and the twisted $\widetilde W(\Sigma)$ over $\int d\theta^+ d\bar\theta^-$
\eqref{eq:V04-Wtwisted}, with the linear $\widetilde W = t\Sigma$, $t = r - i\theta$
\eqref{eq:V04-twistedt} carrying the FI/theta parameter as twisted-chiral data
\eqref{eq:V04-tsplit}; the chiral (Jacobi) ring $\mathcal{R}_B = \mathbb{C}[\Phi_i]/(\partial_i W)$
\eqref{eq:V04-jacobiring} and the twisted-chiral (quantum-cohomology) ring, exchanged by mirror
symmetry, with the worked $A_{k+1}$ Jacobi ring \eqref{eq:V04-A2ring}--\eqref{eq:V04-Akchat}. This is (the superpotential) with (the FI/theta $t$). (\textcite{Witten:1993yc} the $\mathcal{N}=(2,2)$ grammar and FI/theta; \textcite{Hori:2003ic} the two rings).

\medskip\noindent\textsf{\textcolor{RoyalBlue}{\textbf{\S\ref{sec:V04-glsm}\enspace Block A: the GLSM, $\mathbb{CP}^{N-1}$, the quintic.}}}\enspace
The gauged linear sigma model, the D-term $\sum_i Q^a_i |\phi_i|^2 = r_a$ \eqref{eq:V04-dterm}, the
$\mathbb{CP}^{N-1}$ effective twisted superpotential $\widetilde W_{\mathrm{eff}} = -t\Sigma -
N\Sigma(\log\Sigma - 1)$ \eqref{eq:V04-cpnWeff}, the vacuum equation $\partial_\Sigma \widetilde
W_{\mathrm{eff}} = -t - N\log\Sigma = 0 \Rightarrow \Sigma^N = e^{-t} = q$ \eqref{eq:V04-cpnvacuum}, and
the quantum cohomology $\mathbb{C}[\Sigma]/(\Sigma^N - q)$ with $N$ vacua \eqref{eq:V04-cpnQC}; the FI
chambers, the quantum Kähler moduli space with its discriminant locus (a massless state / noncompact
Coulomb branch, not a shrinking curve), the geometric (GIT quotient / Calabi--Yau) and Landau--Ginzburg
phases as two regimes of one RG flow, and the CY condition $\sum_a Q^a = 0$ \eqref{eq:V04-quinticCY};
the quintic walked ($U(1)$ with $(+1)^5, -5$, $W = P G_5$ \eqref{eq:V04-quinticW}, the $\mathbb{P}^4$
quotient of dimension $4$ \eqref{eq:V04-P4}, the $G_5 = 0$ cut to $3$ \eqref{eq:V04-quinticdim}, the
$U(1) \to \mathbb{Z}_5$ Higgsing at $|P|^2 = -r/5$ \eqref{eq:V04-Z5}, and $\hat c = 3$ geometric and LG
\eqref{eq:V04-lgchat}--\eqref{eq:V04-quinticlgchat}). (\textcite{Witten:1993yc} the phases,
the FI chambers, the quintic; \textcite{Hori:2000kt} the effective twisted superpotential and mirror;
\textcite{Witten:1993jg} the LG central charge). 

\medskip\noindent\textsf{\textcolor{RoyalBlue}{\textbf{\S\ref{sec:V04-twist}\enspace Block A: the A and B twists and mirror symmetry.}}}\enspace
The two $\mathcal{N}=(2,2)$ topological twists: the A-twist ($U(1)_V$, holomorphic maps, quantum cohomology
$\mathcal{R}_A = \mathbb{C}[x]/(x^N - q)$ \eqref{eq:V04-ARing} tied to \S\ref{sec:V04-glsm}, NOT gated
by the CY condition) and the B-twist ($U(1)_A$, constant maps, the Jacobi ring $\mathcal{R}_B =
\mathbb{C}[\Phi_i]/(\partial_i W)$ \eqref{eq:V04-BRing}, requires $c_1(TX) = 0$ \eqref{eq:V04-btwistCY},
i.e. $\sum_i Q_i = 0$ for a GLSM \eqref{eq:V04-U1Aanomaly}); mirror symmetry as a duality of the full
untwisted $\mathcal{N}=(2,2)$ theories, realized by the Hori--Vafa GLSM mirror
\eqref{eq:V04-hvmirror}--\eqref{eq:V04-hvring}, exchanging $A(X) \cong B(\hat X)$ \eqref{eq:V04-mirror}
and so $\mathcal{R}_A(X) \cong \mathcal{R}_B(\hat X)$, the twists how the duality is used, not a
prerequisite. (\textcite{Witten:1993yc} the twists and the CY condition; \textcite{Hori:2000kt}
the mirror realization). 

\medskip\noindent\textsf{\textcolor{RoyalBlue}{\textbf{\S\ref{sec:V04-ej}\enspace Block B: the $\mathcal{N}=(0,2)$ E/J data and scalar potential.}}}\enspace
The two holomorphic data $\bar D_+ \Lambda_a = E_a$ \eqref{eq:V04-Edatum} and $W_{\mathcal{N}=(0,2)} = \int
d\theta^+ \sum_a \Lambda_a J^a$ \eqref{eq:V04-Jdatum}, the scalar potential $V \supset \sum_a|E_a|^2 +
\sum_a|J^a|^2 + V_D$ \eqref{eq:V04-02potential} with the SUSY vacua $E_a = 0$, $J^a = 0$
\eqref{eq:V04-02vacua}; the precise statement that $E$ and $J$ enter symmetrically but conjugation
complex-conjugates the data rather than exchanging them \eqref{eq:V04-conjnoexchange} (must-fix); the
closure $\sum_a \mathrm{Tr}(E_a J^a) = 0$ \eqref{eq:V04-ejclosure} as the Jacobi identity on the
$\mathbb{C}^4$ brick commutators \eqref{eq:V04-ejcommutators}--\eqref{eq:V04-jacobi} (with the wrong
pairing nonzero \eqref{eq:V04-wrongpair}), inherited by the $\mathbb{C}^4/\mathbb{Z}_4$ orbifold
(4/16/12 \eqref{eq:V04-orbcount}). This is in its $\mathcal{N}=(0,2)$ form (REUSED, no new
convention). (\textcite{Distler:1993mk} the $\mathcal{N}=(0,2)$ $E/J$ / Landau--Ginzburg structure;
\textcite{Franco:2015tya} the brick $E/J$ and $\sum \mathrm{Tr}(EJ) = 0$). 

\medskip\noindent\textsf{\textcolor{RoyalBlue}{\textbf{\S\ref{sec:V04-bundles}\enspace Block B: $\mathcal{N}=(0,2)$ bundles and the gauged E-term.}}}\enspace
The bundle interpretation ($E = TX$ for $\mathcal{N}=(2,2)$, an independent bundle for $\mathcal{N}=(0,2)$, with
$\mathrm{ch}_2(E) = \mathrm{ch}_2(TX)$, $c_1(E) \equiv c_1(TX) \bmod 2$
\eqref{eq:V04-bundle}--\eqref{eq:V04-bundleanom}); the ungauged $\mathcal{N}=(2,2)$ LG specialization $E = 0$, $J^a
= \partial W/\partial\Phi_a$ \eqref{eq:V04-J22}--\eqref{eq:V04-22Jexample}; the correction that a GAUGED
$\mathcal{N}=(2,2)$ GLSM has $E_i = \sqrt2\, Q_i\, \Sigma\, \Phi_i \neq 0$ \eqref{eq:V04-gaugedE} (must-fix), with
the worked gauged quintic $E_i = \sqrt2\,\Sigma\phi_i$, $E_P = -5\sqrt2\,\Sigma P$, $J^i = P\partial_i
G_5$, $J^P = G_5$ \eqref{eq:V04-quinticEJ} and its closure $\sum_i E_i J^i + E_P J^P = \sqrt2\,\Sigma
P(\sum_i \phi_i\partial_i G_5 - 5 G_5) = 0$ by gauge invariance / quasi-homogeneity \eqref{eq:V04-quinticclosure}--\eqref{eq:V04-euler}
(in this hypersurface the same degree-five charge assignment is also the Calabi--Yau condition), and the $\mathcal{N}=(0,2)$ deformation with $E \neq TX$. This is
in its $\mathcal{N}=(0,2)$ form (REUSED). (\textcite{Distler:1993mk} the $\mathcal{N}=(0,2)$ LG / GLSM $E$-term
and bundle anomalies; \textcite{Witten:1993yc} the gauged linear sigma model). 

\medskip\noindent\textsf{\textcolor{RoyalBlue}{\textbf{\S\ref{sec:V04-cext}\enspace Block B: anomalies and c-extremization.}}}\enspace
The gauge anomaly $\mathrm{Tr}_{\text{chiral}} - \mathrm{Tr}_{\text{Fermi}} - \mathrm{Tr}_{\text{adj}}
= 0$ \eqref{eq:V04-gaugeanom} (the brick $4-3-1 = 0$ \eqref{eq:V04-c4anom}, automatic for $\mathcal{N}=(2,2)$
\eqref{eq:V04-22autoanom}, the GGP node $N_1+N_2-N_3-2N_c = 0$ \eqref{eq:V04-ggpanom}) and the
gravitational anomaly $c_R - c_L = \mathrm{Tr}\,\gamma^3$ \eqref{eq:V04-grav} (generically nonzero, the
$\mathbb{C}^4/\mathbb{Z}_4$ $16 - 12 - 4 = 0$ \eqref{eq:V04-gravc4z4}, matched, not cancelled);
$c$-extremization $c_R^{\mathrm{tr}} = 3\mathrm{Tr}[\gamma^3 R^2]$ \eqref{eq:V04-cext}, the stationarity
$k_{RF} = 0$ \eqref{eq:V04-krf}, $c_L = c_R - \mathrm{Tr}\,\gamma^3$ \eqref{eq:V04-cLfromcR}, a
STATIONARY point (a saddle, the indefinite Hessian \eqref{eq:V04-hessian}--\eqref{eq:V04-hessworked}),
NOT a maximum, with the hypotheses stated (candidate IR SCFT, anomaly-free trial R, no accidental
symmetries); the POSITIVE flagship $\mathcal{N}=(0,2)$ $A_{k+1}$ minimal model $c_R = 3k/(k+2)$ ($k=2 \Rightarrow
c_R = 3/2$) \eqref{eq:V04-mmJ}--\eqref{eq:V04-mmk2}, the deliberately-formal saddle fixture $c_R =
-0.36$ \eqref{eq:V04-flavornorm}--\eqref{eq:V04-cRvalue} (NOT a unitary SCFT, demoted), and the
symmetric brick $R_X = 1/2$ \eqref{eq:V04-RWconstraint}. (\textcite{Benini:2012cz}
$c$-extremization; \textcite{Benini:2013cda} the companion; \textcite{Distler:1993mk} the $\mathcal{N}=(0,2)$ gauge
anomaly; \textcite{Gadde:2013lxa} the GGP node anomaly). 

\medskip\noindent\textsf{\textcolor{RoyalBlue}{\textbf{\S\ref{sec:V04-triality}\enspace Block B: $\mathcal{N}=(0,2)$ triality.}}}\enspace
The order-three duality (the $m=2$ rung of the graded-quiver ladder signposted) in exact GGP
conventions; the $U(N_c)$ node with $N_1$ fundamental chirals, $N_2$ antifundamental chirals, $N_3$
fundamental Fermis \eqref{eq:V04-ggpmatter}, balanced rank $N_c = \tfrac12(N_1 + N_2 - N_3)$
\eqref{eq:V04-ggpanomrestate}, the rank map $N_c \mapsto N_2 - N_c$ \eqref{eq:V04-trialityrank} with the
flavor permutation \eqref{eq:V04-trialityperm}, the dual matter (permuted quarks, mesons $M = \Phi P$,
singlet Fermis), the order-three return $(5,3,4) \to (3,4,5) \to (4,5,3) \to (5,3,4)$
\eqref{eq:V04-trialityorbit}, $T^3 = \mathds{1}$ (Table~\ref{tab:V04-triality}). The load-bearing
evidence is the anomaly and elliptic-genus matches across the three frames; the symmetric flavor
quantity $I$ \eqref{eq:V04-flavorinv} is a toy consistency check (a permutation-symmetric polynomial in
the ranks, useful only to reject a wrong-multiset candidate), NOT a named central-charge or anomaly
formula. (\textcite{Gadde:2013lxa} the $2d\ \mathcal{N}=(0,2)$ triality, the rank map, the dual matter).

\medskip\noindent\textsf{\textcolor{RoyalBlue}{\textbf{\S\ref{sec:V04-ellgenus}\enspace Block C: the elliptic genus.}}}\enspace
The torus trace $Z_{T^2} = \mathrm{Tr}_{RR}[(-1)^F q^{H_L} \bar q^{H_R} y^J \cdots]$
\eqref{eq:V04-ellgenus}, a protected index (right-movers cancel) and a weak Jacobi form for a $\mathcal{N}=(0,2)$
SCFT; the $q\to0$ limit to the $\chi_y$-genus and the further $y\to1$ specialization to the Witten
index \eqref{eq:V04-genuslimits}; the free chiral factor $i\eta/\theta_1$ (DENOMINATOR)
\eqref{eq:V04-chiralfactor} and the free Fermi factor $i\theta_1/\eta$ (NUMERATOR)
\eqref{eq:V04-fermifactor}, reciprocal only for matched arguments \eqref{eq:V04-reciprocal} (a
mnemonic, not the general $\mathcal{N}=(2,2)$ chiral genus: R/flavor shifts usually make a nontrivial
theta-function ratio, and noncompact free chirals have subtle zero modes), with $q$-expansions
$1/\prod(1-q^n)$ and $\prod(1-q^n)$ \eqref{eq:V04-qexpansions}; the Jeffrey--Kirwan localization
skeleton \eqref{eq:V04-JK} (named, not derived); modularity gated by the gauge anomaly
\eqref{eq:V04-modulargate}. (\textcite{Benini:2013nda} and \textcite{Benini:2013xpa} the
elliptic genus by localization, the JK residue, the Jacobi-form structure). 

\medskip\noindent\textsf{\textcolor{RoyalBlue}{\textbf{\S\ref{sec:V04-compare}\enspace Block C: the $\mathcal{N}=(2,2)$-versus-$\mathcal{N}=(0,2)$ comparison.}}}\enspace
The side-by-side comparison table (Table~\ref{tab:V04-compare}) reading the two worlds together
(supercharges, holomorphic data, geometric meaning, anomaly role, extremization, protected index,
duality, construction use). A summary section, no new claim; it recapitulates the structures of
Blocks A and B.

\medskip\noindent\textbf{Stated and cited, not proved here.}\enspace
The $c$-extremization theorem, the $2d$ c-theorem, the accidental-symmetry / unitarity-bound
corrections, and the proof of $\mathcal{N}=(0,2)$ triality beyond the necessary anomaly / index matches;
the gauged-linear-sigma-model phase theorem (the full quantum-Kähler-moduli / secondary-fan /
discriminant analysis), the elliptic-genus localization / Jeffrey--Kirwan / modularity / duality
machinery, and the A/B-twist topological-correlator / mirror-theorem machinery. All
are stated and, where a computation is possible at toolkit depth, run on fixtures; their proofs are
left to the cited literature.
}

\subsection*{Further reading}
\addcontentsline{toc}{subsection}{Further reading}
The gauged linear sigma model and its phases are developed in \textcite{Witten:1993yc}; the original
geometric mirror statement is \textcite{Greene:1990ud}, and quantum cohomology of toric targets is in
\textcite{Morrison:1994fr}. The $tt^*$ geometry and the classification of two-dimensional
$\mathcal{N}=2$ theories appear in \textcite{Cecotti:1991me,Vafa:1990mu}, and the Bethe/gauge
correspondence behind the vacuum equations in \textcite{Nekrasov:2009uh,Nekrasov:2009ui}. For $(0,2)$
models see \textcite{Silverstein:1994ih}; the two-dimensional conformal minimal models are in
\textcite{Friedan:1983xq}, with the standard CFT reference \textcite{DiFrancesco:1997nk}.

For modern extensions, global forms, discrete theta angles, and nonabelian mirrors of pure
$\mathcal{N}=(2,2)$ gauge theories are analyzed in \textcite{Gu:2020ivl}, new
$\mathcal{N}=(0,2)$ dualities and their elliptic-genus tests in \textcite{Sacchi:2020pet},
and the related decomposition phenomenon of two-dimensional gauge theories is reviewed in
\textcite{Sharpe:2022ene}.

\section*{References}
\addcontentsline{toc}{subsection}{References}
\markboth{References}{References}
\printbibliography[heading=none]
\end{refsection}
\begin{refsection}\chapter{\texorpdfstring{$4d$ $\mathcal{N}=2$}{4d N=2} supersymmetric field theories}
\label{ch:V05}

\noindent\textbf{Guide to this section.}\enspace
Sections~1 and~2 fixed the algebra and the common words, and Section~3 built the
four-dimensional $\mathcal{N}=1$ world. This section builds the four-dimensional $\mathcal{N}=2$
world: eight real supercharges, the Coulomb-branch special geometry, and Seiberg--Witten theory.
It is a foundations section. It states the special structures of $4d\ \mathcal{N}=2$ field theory
and cites the deep proofs without reproducing them. It writes everything in the $\mathcal{N}=1$
language of Section~3, so the multiplet, chiral-ring, R-symmetry, and anomaly machinery is recalled,
not re-derived. By the end you can read an $\mathcal{N}=2$ theory in $\mathcal{N}=1$ language,
separate its Coulomb (special-K\"ahler) branch from its Higgs (hyperk\"ahler) branch, write the
Seiberg--Witten periods and the central charge $Z=a\,n_e+a_D\,n_m$ with $M=\sqrt{2}\,|Z|$, and
locate a theory's massless BPS points from where its curve degenerates.

\begin{keybox}{What this section delivers}
The $\mathcal{N}=2$ multiplets in $\mathcal{N}=1$ language, the canonical coupling
$W_{\mathcal{N}=2}=\sqrt{2}\,\widetilde{Q}\,\Phi\,Q+m\,\widetilde{Q}Q$, and the
$SU(2)_R\times U(1)_R$ that rotates the eight supercharges (\S\ref{sec:V05-multiplets}); the
low-energy electromagnetic duality, the rank-$r$ charge lattice with its Dirac pairing, and the
duality group $Sp(2r,\mathbb{Z})$ with $SL(2,\mathbb{Z})=Sp(2,\mathbb{Z})$ at rank one
(\S\ref{sec:V05-emduality}); the Coulomb branch and its rigid special geometry, fixed by one
holomorphic prepotential, with the one-loop derivatives worked out (\S\ref{sec:V05-coulomb}); the
Seiberg--Witten curve, differential, periods, and the BPS central charge, walked end to end on pure
$SU(2)$ with the discriminant, the singularities, the charges, and the ordered monodromy product,
followed by the local monopole/dyon frames (\S\ref{sec:V05-sw}); matter deformations, the flavor
decoupling relation, and BPS wall-crossing discipline (\S\ref{sec:V05-matter}); the hyperk\"ahler
Higgs branch with a worked $U(1)$ quotient and mixed branches (\S\ref{sec:V05-higgs});
$4d\ \mathcal{N}=2$ S-duality, the $N_f=4$ marginal coupling, and $\mathrm{Spin}(8)$ triality, stated
as four-dimensional facts (\S\ref{sec:V05-classS}); the Lagrangian $\mathcal{N}=2$ conformal central
charges $a=(5n_v+n_h)/24$, $c=(2n_v+n_h)/12$, with the $SU(2)$, $N_f=4$, the $\mathcal{N}=4$, and the
$SU(N)$, $N_f=2N$ fixtures and the $2a-c$ Coulomb-dimension relation (\S\ref{sec:V05-centralcharges});
and the Argyres--Douglas non-Lagrangian fixed
point with its fractional Coulomb-branch dimension, closed by an $\mathcal{N}=2$ SCFT data card
(\S\ref{sec:V05-ad}).
\end{keybox}

\section{\texorpdfstring{$\mathcal{N}=2$}{N=2} multiplets in \texorpdfstring{$\mathcal{N}=1$}{N=1} language}
\label{sec:V05-multiplets}

A four-dimensional $\mathcal{N}=2$ theory carries eight real supercharges. Section~1 records this
count as the $8Q$ row of the dimensional ladder, the same amount as $5d\ \mathcal{N}=1$ and
$3d\ \mathcal{N}=4$. Eight supercharges are twice the four of the $4d\ \mathcal{N}=1$ theories of
Section~3, so every $\mathcal{N}=2$ multiplet splits into a pair of $\mathcal{N}=1$ multiplets
under the $\mathcal{N}=1$ subalgebra. This is the working description used throughout the section.
It is a chart on the local grammar, not a claim that the second supersymmetry has gone away.

There are two basic multiplets. The $\mathcal{N}=2$ \emph{vector multiplet} decomposes, under
the $\mathcal{N}=1$ subalgebra, into an $\mathcal{N}=1$ vector multiplet $V$ and an $\mathcal{N}=1$
adjoint chiral multiplet $\Phi$,
\begin{equation}
\label{eq:V05-vector}
\text{$\mathcal{N}=2$ vector}\ =\ \underbrace{V}_{\text{$\mathcal{N}=1$ vector}}
\ \oplus\ \underbrace{\Phi}_{\text{$\mathcal{N}=1$ adjoint chiral}}.
\end{equation}
It contains a gauge field $A_\mu$, a complex adjoint scalar $\phi$ (the scalar of $\Phi$), and two
Weyl gauginos, the gaugino of $V$ together with the fermion of $\Phi$. The two gauginos are not
independent under the second supersymmetry. They form a doublet of the $SU(2)_R$ that rotates them
into each other. The $\mathcal{N}=2$ \emph{hypermultiplet} decomposes into two $\mathcal{N}=1$
chiral multiplets in conjugate representations,
\begin{equation}
\label{eq:V05-hyper}
\text{$\mathcal{N}=2$ hyper}\ =\ Q\ \oplus\ \widetilde{Q},
\qquad Q\in \mathbf{R},\quad \widetilde{Q}\in\overline{\mathbf{R}}.
\end{equation}
It carries four real scalars, the components of $Q$ and $\widetilde{Q}$, which assemble into a
quaternion, and two Weyl fermions. The four scalars are again an $SU(2)_R$ doublet of complex pairs.
The names $V$, $\Phi$, $Q$, $\widetilde{Q}$ are exactly the $\mathcal{N}=1$ multiplets of Section~3.
What is new is the pairing rule, and the R-symmetry that enforces it.

The R-symmetry of $4d\ \mathcal{N}=2$ is the master-table row of Section~1,
\begin{equation}
\label{eq:V05-Rgroup}
R\ =\ SU(2)_R\times U(1)_R,
\qquad
\dim R\ =\ 3+1\ =\ 4,
\qquad
\operatorname{rank} R\ =\ 1+1\ =\ 2.
\end{equation}
The $SU(2)_R$ rotates the two supercharges of each chirality into one another, and the $U(1)_R$
acts on the adjoint scalar $\phi$ and grades the chiral ring. This R-group is small. It is not
$SU(3)$, whose $\dim SU(3)=8$ is too large to act on a doublet. That confusion, $SU(2)_R\times
U(1)_R$ misread as $SU(3)$, is the first thing to avoid. The eight-supercharge count, the
decompositions, and the rank and dimension of the R-group are all read off the Section~1 ladder.

An $\mathcal{N}=2$ gauge theory is then built from these pieces: the vector multiplets gauge a group
$G$, the hypermultiplets sit in a representation of $G$, and the second supersymmetry fixes the
superpotential and gauge kinetic term in terms of $\mathcal{N}=1$ data. The new physics is not the
field content but the eight-supercharge structure the $\mathcal{N}=1$ language misses: the rigid
special geometry of the Coulomb branch, the curve that solves it, and the BPS central charge.

\medskip\noindent\textbf{The canonical $\mathcal{N}=2$ coupling, in $\mathcal{N}=1$ words.}\enspace
The pairing rule also pins down the interactions. In $\mathcal{N}=1$ superspace, the adjoint chiral
$\Phi$ couples to a hypermultiplet $(Q,\widetilde{Q})$ through a superpotential the second
supersymmetry fixes completely,
\begin{equation}
\label{eq:V05-n2coupling}
W_{\mathcal{N}=2}\ =\ \sqrt{2}\,\widetilde{Q}\,\Phi\,Q\ +\ m\,\widetilde{Q}\,Q.
\end{equation}
Three facts make this $\mathcal{N}=2$ rather than a generic $\mathcal{N}=1$ superpotential. First,
$\Phi$ sits in the adjoint of $G$ and contracts the gauge indices of $Q\in\mathbf{R}$ and
$\widetilde{Q}\in\overline{\mathbf{R}}$ into a singlet, the only cubic coupling of these three
multiplets,
\begin{equation}
\label{eq:V05-singlet}
\mathbf{R}\otimes\mathrm{adj}\otimes\overline{\mathbf{R}}\ \supset\ \mathbf{1}.
\end{equation}
Second, the cubic coefficient is not free: the second supersymmetry rotates the gaugino of $V$ into
the fermion of $\Phi$, tying the gauge coupling $g$ to the Yukawa coupling in
\eqref{eq:V05-n2coupling}. The factor $\sqrt{2}$ is the normalization that makes the central charge
come out as the fixed $Z=a\,n_e+a_D\,n_m$ of \S\ref{sec:V05-sw}, not an adjustable parameter.
Third, the mass term is the unique $\mathcal{N}=2$-preserving mass, giving equal masses to the two
$\mathcal{N}=1$ chirals as the $SU(2)_R$ demands,
\begin{equation}
\label{eq:V05-equalmass}
m_Q\ =\ m_{\widetilde{Q}}\ =\ m.
\end{equation}
A generic $\mathcal{N}=1$ superpotential spoils all three and breaks the second supersymmetry.
Reading \eqref{eq:V05-n2coupling} as a free $\mathcal{N}=1$ choice is the trap: the
eight-supercharge structure fixes it, and that rigidity is what makes $\mathcal{N}=2$ theories
solvable.

\section{Electromagnetic duality and the charge lattice}
\label{sec:V05-emduality}

On a generic vacuum the adjoint scalar $\phi$ acquires a VEV, the gauge group breaks to its Cartan
$U(1)^r$, and the low-energy theory is abelian. An abelian gauge theory in four dimensions carries
both electric and magnetic charges and has an electromagnetic-duality symmetry that exchanges them.
This is the structure the rest of the section solves.

The low-energy coupling is packaged into one complex number, the complexified coupling
\begin{equation}
\label{eq:V05-tau}
\tau\ =\ \frac{\theta}{2\pi}\ +\ \frac{4\pi i}{g^2},
\qquad \operatorname{Im}\tau\ =\ \frac{4\pi}{g^2}\ >\ 0,
\end{equation}
whose real part is the theta-angle and whose imaginary part is the inverse gauge coupling squared.
The positivity of $\operatorname{Im}\tau$ is the positivity of the gauge kinetic term. The
electromagnetic-duality group is $SL(2,\mathbb{Z})$, generated by two transformations,
\begin{equation}
\label{eq:V05-ST}
S=\begin{pmatrix} 0 & -1\\ 1 & 0\end{pmatrix}:\ \tau\mapsto -\frac{1}{\tau},
\qquad
T=\begin{pmatrix} 1 & 1\\ 0 & 1\end{pmatrix}:\ \tau\mapsto \tau+1.
\end{equation}
The generator $S$ exchanges electric and magnetic charges and inverts the coupling, $T$ is the
periodicity of the theta-angle $\theta\mapsto\theta+2\pi$. Each acts on $\tau$ by the M\"obius
transformation of its matrix and on the charge vector by the matrix itself,
\begin{equation}
\label{eq:V05-mobius}
\tau\ \mapsto\ \frac{a\tau+b}{c\tau+d},
\qquad
\begin{pmatrix} n_m\\ n_e\end{pmatrix}\ \mapsto\
\begin{pmatrix} a & b\\ c & d\end{pmatrix}\begin{pmatrix} n_m\\ n_e\end{pmatrix}.
\end{equation}
Both generators have determinant one, and they obey the defining relations of $SL(2,\mathbb{Z})$,
\begin{equation}
\label{eq:V05-STrelations}
\det S\ =\ \det T\ =\ 1,
\qquad
S^2\ =\ -\mathds{1},
\qquad
(ST)^3\ =\ -\mathds{1},
\end{equation}
the small algebraic check that the duality is well defined.

The magnetically charged states are solitons: on the Coulomb branch the broken gauge symmetry
supports a 't Hooft--Polyakov monopole, a finite-energy solution of one unit of magnetic charge,
while a purely electric state is a quantum of the unbroken $U(1)$. The duality $S$ exchanges these
two kinds of object. That a soliton and an elementary excitation can be exchanged by a symmetry is
the deep content of electromagnetic duality, and the reason a strong-coupling $\mathcal{N}=2$
description can look unlike its weak-coupling Lagrangian.

\medskip\noindent\textbf{The rank-$r$ charge lattice and $Sp(2r,\mathbb{Z})$.}\enspace
Rank one is the simplest case, not the general one. A theory of rank $r$ breaks to $U(1)^r$, and a
BPS state carries $r$ electric and $r$ magnetic charges. The charge vector lives in the
electromagnetic lattice $\Gamma_{\mathrm{em}}\cong\mathbb{Z}^{2r}$,
\begin{equation}
\label{eq:V05-chargevec}
\gamma\ =\ (n_{e,I},\,n_m^I)\ \in\ \Gamma_{\mathrm{em}},
\qquad I=1,\dots,r.
\end{equation}
The lattice carries a canonical antisymmetric integer pairing, the \emph{Dirac pairing},
\begin{equation}
\label{eq:V05-dirac2}
\langle\gamma,\gamma'\rangle\ =\ n_{e,I}\,n_m^{\prime I}\ -\ n_m^I\,n'_{e,I}\ =\ \gamma^{T} J\,\gamma',
\qquad J=\begin{pmatrix} 0 & \mathds{1}_r\\ -\mathds{1}_r & 0\end{pmatrix}.
\end{equation}
The pairing is the integer counting the relative electric-magnetic charge of two states,
\begin{equation}
\label{eq:V05-diracprops}
\langle\gamma,\gamma'\rangle\ =\ -\langle\gamma',\gamma\rangle,
\qquad
\langle\gamma,\gamma\rangle\ =\ 0,
\qquad
\langle e_I,m^J\rangle\ =\ \delta_I^{\,J},
\end{equation}
antisymmetric, self-pairing zero, basic electric against basic magnetic pairing to $\pm1$. (The
check code verifies these from $J$ directly.) We fix the ordering once: the prose labels every
charge electric-first, $\gamma=(n_{e,I},n_m^I)$, following \eqref{eq:V05-Zgamma}. A duality matrix
acting on the period vector transforms a fixed charge by the dual (contragredient) action, so the
central charge $Z_\gamma$ stays invariant; \S\ref{sec:V05-sw} acts the matrices on the period
column $(a_D,a)^T$ and states its paired charge ordering $(n_m,n_e)$ at the point of use. The
low-energy data are packaged in the \emph{period vector} $\Pi=(a_D^I,a_I)$, built from the special
coordinates and their duals, and the central charge of a state of charge $\gamma$ is the symplectic
contraction
\begin{equation}
\label{eq:V05-Zgamma}
Z_\gamma\ =\ n_{e,I}\,a^I\ +\ n_m^I\,a_{D,I}\ +\ \text{(flavor-mass terms)},
\end{equation}
the rank-$r$ generalization of the rank-one $Z=a\,n_e+a_D\,n_m$ of \S\ref{sec:V05-sw} (the flavor
masses are added in \S\ref{sec:V05-matter}). The duality group is the group of lattice
automorphisms that preserve the Dirac pairing, the integer symplectic group
\begin{equation}
\label{eq:V05-sp2r}
M\in Sp(2r,\mathbb{Z}):\qquad M^{T} J\, M\ =\ J,\qquad M\in\mathrm{GL}(2r,\mathbb{Z}).
\end{equation}
At rank one a $2\times2$ integer matrix preserves $J$ exactly when its determinant is one, so the
symplectic group collapses to the modular group,
\begin{equation}
\label{eq:V05-sp2sl2}
Sp(2,\mathbb{Z})\ =\ SL(2,\mathbb{Z}),
\end{equation}
recovering the $S,T$ generators above. At higher rank a determinant-one integer matrix need
\emph{not} be symplectic, so $\det M=1$ alone is not the rank-$r$ duality condition; the check code
confirms both facts. The trap is to let the rank-one $SL(2,\mathbb{Z})$ language look universal:
the low-energy duality group of a rank-$r$ theory is $Sp(2r,\mathbb{Z})$, and $SL(2,\mathbb{Z})$ is
its rank-one face.

\begin{keybox}{Three dualities that wear the same symbol}
The letter ``S-duality'' and the group $SL(2,\mathbb{Z})$ appear in three distinct places in this
section, and conflating them is the standing error of the section.
\begin{enumerate}
\item \emph{Low-energy electromagnetic duality} on the abelian Coulomb branch
(\S\ref{sec:V05-emduality}): the group $Sp(2r,\mathbb{Z})$ acting on the lattice
$\Gamma_{\mathrm{em}}$ and the period vector $\Pi$, $SL(2,\mathbb{Z})$ at rank one.
\item \emph{Monodromy} of the period vector around singular loci of the Coulomb branch
(\S\ref{sec:V05-sw}): a specific \emph{subgroup} of the low-energy duality group, generated by the
degenerations of the Seiberg--Witten geometry (for pure $SU(2)$, the $M_m,M_d,M_\infty$ below).
\item \emph{Exact S-duality} of a full superconformal theory (\S\ref{sec:V05-classS}): a duality of
the \emph{interacting} theory at its marginal coupling, e.g.\ $SU(2)$ with $N_f=4$, carrying extra
flavor data ($\mathrm{Spin}(8)$ triality) and global-form caveats.
\end{enumerate}
The first is a property of the Cartan theory, the second is a property of the curve, and the third
is a property of the SCFT. They are related but not the same.
\end{keybox}

This layer comes first because it gives the curve's monodromies a home: the Seiberg--Witten solution
attaches to each vacuum a coupling $\tau$ and a charge vector, and carrying the vacuum around a
singular point acts on the pair by an $Sp(2r,\mathbb{Z})$ matrix (at rank one, $SL(2,\mathbb{Z})$).
Those monodromy matrices are exactly the electromagnetic-duality transformations of this section.

\section{The Coulomb branch and rigid special geometry}
\label{sec:V05-coulomb}

The Coulomb branch is the family of vacua on which the adjoint scalars are turned on. Their VEVs
break the gauge group to its Cartan $U(1)^r$, so the branch has complex dimension the rank,
\begin{equation}
\label{eq:V05-coulombdim}
\dim_{\mathbb{C}}\mathcal{M}_C\ =\ r.
\end{equation}
It is parametrized not by the gauge-dependent bare VEVs but by the gauge-invariant Casimirs built
from $\phi$, the special coordinates of the abelian theory. For $SU(2)$ there is one such invariant,
\begin{equation}
\label{eq:V05-ucoord}
u\ =\ \langle \operatorname{Tr}\phi^2\rangle,
\end{equation}
and the Coulomb branch is the complex $u$-plane. Calling it ``the space of scalar VEVs'' is
imprecise: the bare VEV $\langle\phi\rangle$ is gauge-dependent, while the branch is coordinatized
by the gauge-invariant Casimirs and, locally, by the special coordinates $a,a_D$ the prepotential
below builds. The Coulomb branch is not a generic space of scalar expectation values; it carries a
rigid local special geometry, the data Seiberg--Witten theory solves.

The low-energy effective Lagrangian on the Coulomb branch is fixed by a single holomorphic function,
the \emph{prepotential} $\mathcal{F}(a)$, where $a^I$ are the special coordinates of the $r$ abelian
vector multiplets. From $\mathcal{F}$ one builds the dual special coordinates, the low-energy
coupling matrix, and the metric,
\begin{equation}
\label{eq:V05-special}
a_{D,I}\ =\ \frac{\partial\mathcal{F}}{\partial a^I},
\qquad
\tau_{IJ}\ =\ \frac{\partial^2\mathcal{F}}{\partial a^I\,\partial a^J}\ =\ \frac{\partial a_{D,I}}{\partial a^J},
\qquad
ds^2\ =\ \operatorname{Im}\tau_{IJ}\,da^I\,d\bar a^J.
\end{equation}
The second derivative is the period matrix $\tau_{IJ}$, the rank-$r$ coupling whose rank-one entry
is the $\tau$ of \eqref{eq:V05-tau}, now holomorphic in the position on the branch. The two ways of
writing it agree by construction. The metric is $\operatorname{Im}\tau_{IJ}$, its positivity the
positivity of the gauge kinetic term. This is what \emph{rigid special geometry} means: one
holomorphic prepotential controls the special coordinates, their duals, the coupling matrix, and the
metric at once. The prepotential is only \emph{locally} defined, since a duality transformation
changes $(a,a_D)$ and with it $\mathcal{F}$; the globally defined object is the period vector
$\Pi=(a_{D,I},a^I)$, transforming under $Sp(2r,\mathbb{Z})$, of which $\mathcal{F}$ is a
frame-dependent potential. From here on, rank one.

A model makes the structure concrete and checkable. At weak coupling the one-loop running of the
$SU(2)$ theory gives the prepotential
\begin{equation}
\label{eq:V05-Fweak}
\mathcal{F}(a)\ =\ \frac{i}{2\pi}\,a^2\,\ln\frac{a^2}{\Lambda^2},
\end{equation}
with $\Lambda$ the dynamical scale. Now differentiate it, not just name the rule. The first
derivative is the dual special coordinate,
\begin{equation}
\label{eq:V05-aDoneloop}
a_D\ =\ \frac{d\mathcal{F}}{da}\ =\ \frac{i}{\pi}\,a\!\left(\ln\frac{a^2}{\Lambda^2}+1\right),
\end{equation}
where the $+1$ comes from differentiating the prefactor $a^2$ and is load-bearing: it is the
constant the period integral reproduces. The second derivative is the running coupling,
\begin{equation}
\label{eq:V05-tauoneloop}
\tau\ =\ \frac{d^2\mathcal{F}}{da^2}\ =\ \frac{da_D}{da}\ =\ \frac{i}{\pi}\!\left(\ln\frac{a^2}{\Lambda^2}+3\right),
\end{equation}
and the two routes to $\tau$ agree as they must. The check code carries out this differentiation
symbolically. The physically meaningful content is the logarithm: its coefficient $i/\pi$ is the
asymptotic-freedom slope fixed by the one-loop beta function, and the metric is positive in the
weak-coupling region,
\begin{equation}
\label{eq:V05-imtaupos}
\operatorname{Im}\tau\ =\ \frac1\pi\,\ln\frac{a^2}{\Lambda^2}\ >\ 0
\qquad\text{at large }a.
\end{equation}
The additive constants ($+1$ in $a_D$, $+3$ in $\tau$) depend on the definition of $\Lambda$ and
the scheme; the discriminating data are the logarithm and its slope, not the shift. A prepotential
giving $\operatorname{Im}\tau<0$ would signal a wrong-sign kinetic term, and the off-by-one-derivative
misidentification $\tau=d\mathcal{F}/da$ must be rejected; the check code rejects both. The Coulomb
branch is special-K\"ahler, with a running coupling and monodromies; a Higgs branch is
hyperk\"ahler, with neither.

\medskip\noindent\textbf{The higher-rank atlas: $SU(N)$ Coulomb branches.}\enspace
The worked theory below is $SU(2)$, but the structure is general. An $SU(N)$ gauge theory breaks on
its Coulomb branch to $U(1)^{N-1}$, so its rank is $r=N-1$. The gauge-invariant coordinates are the
Casimirs of the adjoint scalar,
\begin{equation}
\label{eq:V05-casimirs}
u_k\ =\ \langle\operatorname{Tr}\phi^k\rangle,\qquad k=2,3,\dots,N,
\qquad
\#\{u_k\}\ =\ N-1\ =\ r,
\end{equation}
their number matching the rank. (These are the $\mathcal{N}=2$ avatars of the $\operatorname{Tr}\phi^k$
chiral-ring generators of the $\mathcal{N}=1$ section.) The low-energy theory
is an abelian gauge theory of rank $r$, with a period vector of $2r$ components $(a_{D,I},a^I)$, a
charge lattice $\Gamma_{\mathrm{em}}\cong\mathbb{Z}^{2r}$ carrying the Dirac pairing
\eqref{eq:V05-dirac2}, and a duality group $Sp(2r,\mathbb{Z})$. For $N=2$ this is $r=1$, one coordinate
$u=u_2$, and $Sp(2,\mathbb{Z})=SL(2,\mathbb{Z})$, the case the rest of the section solves in full. The
higher-rank Seiberg--Witten curves (hyperelliptic of genus $r$, with their period matrices) belong to
the class S and geometric-engineering literature; this section solves only the rank-one curve, but the
rank-$r$ scaffolding above is what every higher example specializes.
Table~\ref{tab:V05-rankatlas} is the bookkeeping.

\begin{table}[ht]
\centering
\small
\setlength{\tabcolsep}{8pt}
\renewcommand{\arraystretch}{1.3}
\begin{tabular}{@{}lccc@{}}
\toprule
$G$ & rank $r$ & Coulomb coordinates & period-vector length $2r$ \\
\midrule
$SU(2)$ & $1$ & $u_2$ & $2$ \\
\midrule
$SU(3)$ & $2$ & $u_2,u_3$ & $4$ \\
\midrule
$SU(N)$ & $N-1$ & $u_2,\dots,u_N$ & $2(N-1)$ \\
\bottomrule
\end{tabular}
\caption{The $SU(N)$ Coulomb-branch atlas. The rank is $r=N-1$, the Casimir coordinates are
$u_k=\langle\operatorname{Tr}\phi^k\rangle$ for $k=2,\dots,N$, and the duality group is
$Sp(2r,\mathbb{Z})$. The section solves the $SU(2)$ ($r=1$) row in full; the rest fix the notation for
higher-rank Seiberg--Witten systems.}
\label{tab:V05-rankatlas}
\end{table}

\section{Seiberg--Witten theory and the central charge}
\label{sec:V05-sw}

The prepotential is holomorphic but not single-valued: around the singular points of the Coulomb
branch it has the $SL(2,\mathbb{Z})$ monodromies of \S\ref{sec:V05-emduality}. The Seiberg--Witten
solution encodes the exact, non-perturbative $\mathcal{F}$ in geometry. Over each point $u$ it
places an auxiliary Riemann surface $\Sigma_u$, the \emph{Seiberg--Witten curve}, carrying a
meromorphic one-form $\lambda_{\mathrm{SW}}$, the \emph{Seiberg--Witten differential}. The special
coordinate and its dual are the periods of $\lambda_{\mathrm{SW}}$ over a symplectic basis $(A,B)$
of one-cycles,
\begin{equation}
\label{eq:V05-period}
a(u)\ =\ \oint_{A}\lambda_{\mathrm{SW}},
\qquad
a_D(u)\ =\ \oint_{B}\lambda_{\mathrm{SW}}.
\end{equation}
The curve is auxiliary: not the spacetime, not the moduli space, but a bookkeeping geometry whose
periods compute $(a,a_D,\tau)$ and whose degenerations locate the massless states.

The BPS central charge follows from the periods. A state of electric and magnetic charges
$(n_e,n_m)$ has central charge and BPS mass
\begin{equation}
\label{eq:V05-centralcharge}
Z\ =\ a\,n_e\ +\ a_D\,n_m,
\end{equation}
\begin{equation}
\label{eq:V05-bpsmass}
M\ =\ \sqrt{2}\,|Z|\ =\ \sqrt{2}\,\big|\,a\,n_e+a_D\,n_m\,\big|.
\end{equation}
This is the normalization convention adopted from the Seiberg--Witten solution and already used in
Section~1. The factor $\sqrt{2}$ is part of the convention; dropping it to write
$M=|Z|$ is wrong in this normalization. The three species of state read off the charges,
\begin{equation}
\label{eq:V05-states}
\text{W-boson }(1,0):\ M=\sqrt2\,|a|;
\qquad
\text{monopole }n_m\neq0;
\qquad
\text{dyon }n_e n_m\neq0.
\end{equation}
Where a one-cycle of $\Sigma_u$ pinches, that is, where two branch points collide and the
discriminant vanishes, the corresponding period vanishes and the BPS state of that charge becomes
massless. The $SL(2,\mathbb{Z})$ monodromy around that point is the duality transformation the
massless state generates.

\medskip\noindent\textbf{Pure $SU(2)$, worked end to end.}\enspace
The cleanest instance is $SU(2)$ with one vector multiplet and no hypermultiplets. The Coulomb
branch is the $u$-plane of \eqref{eq:V05-ucoord}. The microscopic $U(1)_R$ is broken by the
anomaly to $\mathbb{Z}_8$, and since $u$ has R-charge four, a residual $\mathbb{Z}_2$ acts on the
branch as $u\mapsto -u$. The Seiberg--Witten curve, in the standard form, is
\begin{equation}
\label{eq:V05-su2curve}
y^2\ =\ (x^2-\Lambda^4)\,(x-u),
\end{equation}
a family of elliptic curves over the $u$-plane. (The strong-coupling-symmetric form
$y^2=(x^2-u)^2-\Lambda^4$ is an alternative, birationally equivalent normalization of the same
Seiberg--Witten family; the worked discriminant check below uses the cubic form \eqref{eq:V05-su2curve}.) An elliptic curve $y^2=p(x)$ with $p$ a cubic is a double cover
of the $x$-sphere, branched over the three roots of $p$ together with the point at infinity. Reading
off the cubic on the right of \eqref{eq:V05-su2curve}, the three finite branch points are
\begin{equation}
\label{eq:V05-branchpts}
x\ =\ +\Lambda^2,\qquad x\ =\ -\Lambda^2,\qquad x\ =\ u,
\end{equation}
the first two fixed by the scale and the third moving with the position $u$ on the Coulomb branch.
The curve degenerates, and a one-cycle pinches, exactly when the moving root meets a fixed root,
\begin{equation}
\label{eq:V05-collide}
u\ =\ \pm\Lambda^2.
\end{equation}

\begin{figure}[ht]
\centering
\begin{tikzpicture}[scale=1.0]
 % == (a) smooth torus ==
 \begin{scope}[shift={(0,0)}]
 \draw[thick] (0,0) ellipse (1.5 and 0.95);
 % the hole (two arcs, front rim lower, back rim upper)
 \draw[thick] (-0.62,0.08).. controls (-0.32,-0.22) and (0.32,-0.22).. (0.62,0.08);
 \draw[thick] (-0.48,-0.04).. controls (-0.28,0.14) and (0.28,0.14).. (0.48,-0.04);
 % alpha cycle: around the tube (right side)
 \draw[RoyalBlue,thick] (1.18,0) ellipse (0.2 and 0.42);
 \node[RoyalBlue,font=\scriptsize] at (1.75,0.5) {$\alpha$};
 % beta cycle: around the hole
 \draw[purple,thick,dashed] (0,0) ellipse (1.02 and 0.6);
 \node[purple,font=\scriptsize] at (0,-0.78) {$\beta$};
 \node[font=\footnotesize] at (0,-1.35) {(a) generic $u$: smooth torus};
 \end{scope}
 % == arrow ==
 \draw[->,thick,RoyalBlue] (2.6,0) -- node[above,font=\scriptsize,RoyalBlue] {$u\to\pm\Lambda^2$} (4.0,0);
 % == (b) pinched (nodal) torus: the tube collapses to a node on the right ==
 \begin{scope}[shift={(6.7,0)}]
 % outer profile: from the node on the right, around the left lobe, back to the node
 \draw[thick] (1.3,0).. controls (1.1,0.92) and (-1.3,0.95).. (-1.5,0)
.. controls (-1.3,-0.95) and (1.1,-0.92).. (1.3,0);
 % inner (hole) profile: ENDS at the SAME node, so the tube thickness -> 0 there (pinch)
 \draw[thick] (1.3,0).. controls (1.0,0.26) and (-0.32,0.42).. (-0.6,0)
.. controls (-0.32,-0.42) and (1.0,-0.26).. (1.3,0);
 % the node where the vanishing one-cycle has shrunk to a point
 \fill[red] (1.3,0) circle (2pt);
 \draw[red,thick,->] (2.45,0.85).. controls (1.9,0.55).. (1.42,0.06);
 \node[red,font=\scriptsize,align=center] at (2.55,1.12) {node:\\[-1pt] cycle $\to 0$};
 \node[font=\footnotesize] at (0,-1.35) {(b) $u=\pm\Lambda^2$: the one-cycle pinches};
 \end{scope}
\end{tikzpicture}
\caption{The Seiberg--Witten curve \eqref{eq:V05-su2curve} is a torus fibered over the $u$-plane.
At a generic $u$ (a) it is smooth, with a basis of one-cycles $\alpha,\beta$ whose periods are
$a_D=\oint_\alpha\lambda_{\mathrm{SW}}$ and $a=\oint_\beta\lambda_{\mathrm{SW}}$. At $u=\pm\Lambda^2$ (b)
two branch points collide and a one-cycle pinches: the period of the vanishing cycle goes to zero, so
the BPS state with that charge becomes massless. The vanishing cycle is a different one-cycle at each
point, the monopole cycle (period $a_D$) at $+\Lambda^2$ and the dyon cycle (period $-a+a_D$) at
$-\Lambda^2$. The curve is an auxiliary Riemann surface, not the spacetime geometry.}
\label{fig:V05-torus}
\end{figure}
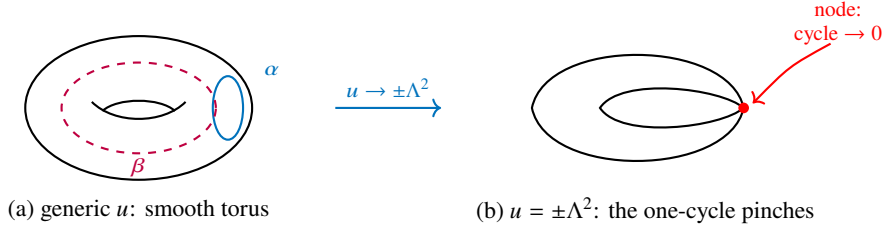 The invariant
statement is that the discriminant of the cubic in $x$ vanishes. Computing it (the check code does
this with \texttt{sympy.discriminant}) gives, up to a $u$-independent constant,
\begin{equation}
\label{eq:V05-disc}
\Delta(u)\ \propto\ (u^2-\Lambda^4)^2\ =\ (u-\Lambda^2)^2\,(u+\Lambda^2)^2,
\end{equation}
a product of \emph{double} zeros at the two finite points
\begin{equation}
\label{eq:V05-singpts}
u\ =\ +\Lambda^2 \qquad\text{and}\qquad u\ =\ -\Lambda^2,
\end{equation}
and at no other finite value, since $\Delta(0)\neq0$: the origin is not singular. There is also the
semiclassical singularity at $u=\infty$, where the theory is weakly coupled. So there are exactly
two strong-coupling singularities, not one. The $\mathbb{Z}_2$ symmetry $u\mapsto -u$ relates the
two finite points but does not collapse them: the discriminant \eqref{eq:V05-disc} is even in $u$
with two roots, not one. The check code reproduces \eqref{eq:V05-disc} symbolically, confirms the
double roots at $\pm\Lambda^2$, and rejects the scale-free curve $y^2=(x^2-u)^2$, whose
$x$-discriminant is identically zero and isolates nothing: the scale $\Lambda$ splits the branch
points and creates the monopole and dyon vacua. The multiplicity two is a discriminant-convention
artifact; the reduced singular locus $u^2-\Lambda^4=0$ is two points, each carrying one light BPS
hypermultiplet (monopole at $+\Lambda^2$, dyon at $-\Lambda^2$).

At each singular point a BPS state becomes massless, and its central charge follows from
\eqref{eq:V05-centralcharge},
\begin{equation}
\label{eq:V05-masslessZ}
\begin{aligned}
u=+\Lambda^2:&\ \ (n_e,n_m)=(0,1),\ \ Z=a_D; \\
u=-\Lambda^2:&\ \ (n_e,n_m)=(-1,1),\ \ Z=-a+a_D.
\end{aligned}
\end{equation}
The monopole period $a_D\to0$ at $+\Lambda^2$ and the dyon combination $-a+a_D\to0$ at $-\Lambda^2$.
The masslessness is the period degeneration, read through the central charge.

Carrying the special coordinates around each singular point produces an $SL(2,\mathbb{Z})$
monodromy. With the monodromies acting on the column vector $(a_D,a)^T$, the three are
\begin{equation}
\label{eq:V05-monodromy}
M_m=\begin{pmatrix} 1 & 0\\ -2 & 1\end{pmatrix},
\qquad
M_d=\begin{pmatrix} -1 & 2\\ -2 & 3\end{pmatrix},
\qquad
M_\infty=\begin{pmatrix} -1 & 2\\ 0 & -1\end{pmatrix},
\end{equation}
the monopole monodromy $M_m$ at $u=+\Lambda^2$, the dyon monodromy $M_d$ at $u=-\Lambda^2$, and the
semiclassical monodromy $M_\infty$ at $u=\infty$. Each lies in $SL(2,\mathbb{Z})$,
\begin{equation}
\label{eq:V05-monodets}
\det M_m\ =\ \det M_d\ =\ \det M_\infty\ =\ 1,
\end{equation}
as \S\ref{sec:V05-emduality} requires. The consistency condition is that a large loop enclosing
both finite singularities is homotopic to the loop at infinity, giving the ordered product
$M_\infty=M_m M_d$. Multiplying with $M_m$ acting first,
\begin{equation}
\label{eq:V05-product}
M_m\,M_d\ =\ \begin{pmatrix} 1 & 0\\ -2 & 1\end{pmatrix}\begin{pmatrix} -1 & 2\\ -2 & 3\end{pmatrix}
\ =\ \begin{pmatrix} -1 & 2\\ 0 & -1\end{pmatrix}\ =\ M_\infty,
\end{equation}
where the lower-left entry is $(-2)(-1)+(1)(-2)=0$, the relation closes exactly. The order matters:
the reversed product is
\begin{equation}
\label{eq:V05-wrongorder}
M_d\,M_m\ =\ \begin{pmatrix} -1 & 2\\ -2 & 3\end{pmatrix}\begin{pmatrix} 1 & 0\\ -2 & 1\end{pmatrix}
\ =\ \begin{pmatrix} -5 & 2\\ -8 & 3\end{pmatrix}\ \neq\ M_\infty.
\end{equation}
This product is the backbone of the solution: the strong-coupling physics, two massless states at
finite $u$, is consistent with the weak-coupling one-loop monodromy at infinity.

\begin{figure}[ht]
\centering
\begin{tikzpicture}[scale=1.0]
 % big loop at infinity (drawn first, behind everything)
 \draw[purple,thick,dashed,-{Stealth[length=2mm]}] (0,0)++(14:3.3 and 1.8) arc (14:374:3.3 and 1.8);
 \node[purple,font=\scriptsize] at (2.6,1.42) {$M_\infty$};
 % the u-plane axes
 \draw[->] (-3.8,0) -- (4.0,0) node[right,font=\scriptsize] {$\operatorname{Re}u$};
 \draw[->] (0,-2.05) -- (0,2.35) node[above,font=\scriptsize] {$\operatorname{Im}u$};
 % singular points
 \fill (1.7,0) circle (1.8pt);
 \fill (-1.7,0) circle (1.8pt);
 % monodromy loop around +Lambda^2 (monopole)
 \draw[RoyalBlue,thick,-{Stealth[length=2mm]}] (1.7,0)++(-30:0.5 and 0.42) arc (-30:305:0.5 and 0.42);
 \node[RoyalBlue,font=\scriptsize] at (1.7,0.82) {$M_m$};
 % monodromy loop around -Lambda^2 (dyon)
 \draw[Green!55!black,thick,-{Stealth[length=2mm]}] (-1.7,0)++(-30:0.5 and 0.42) arc (-30:305:0.5 and 0.42);
 \node[Green!55!black,font=\scriptsize] at (-1.7,0.82) {$M_d$};
 % point labels, two lines, clearly below the loops
 \node[font=\scriptsize] at (1.7,-0.76) {$+\Lambda^2$};
 \node[font=\scriptsize] at (1.7,-1.12) {(monopole)};
 \node[font=\scriptsize] at (-1.7,-0.76) {$-\Lambda^2$};
 \node[font=\scriptsize] at (-1.7,-1.12) {(dyon)};
\end{tikzpicture}
\caption{The Coulomb-branch $u$-plane with its three singularities: the monopole point $u=+\Lambda^2$
(monodromy $M_m$), the dyon point $u=-\Lambda^2$ (monodromy $M_d$), and the semiclassical point
$u=\infty$ (monodromy $M_\infty$). A large loop enclosing both finite singularities is homotopic to the
loop at infinity, so the monodromies obey the ordered product $M_\infty=M_m M_d$
\eqref{eq:V05-product}, the consistency condition that ties the strong-coupling physics to the
one-loop running at weak coupling.}
\label{fig:V05-uplane}
\end{figure}
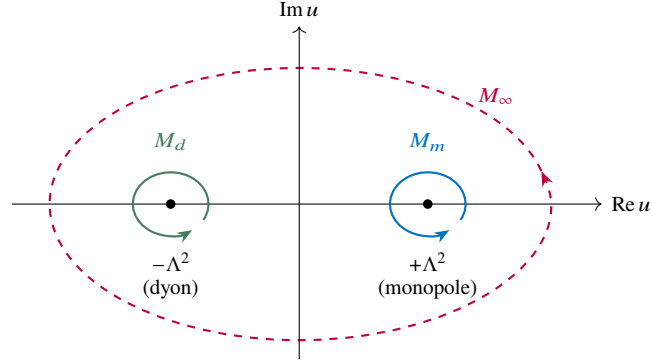 The matrices act on the period column $(a_D,a)^T$, whose symplectically paired charge
column is ordered $(n_m,n_e)$, while the prose labels charges $(n_e,n_m)$ following
\eqref{eq:V05-centralcharge}. The discriminant, the two singular points, the determinants, and the
ordered product \eqref{eq:V05-product} are all checked symbolically against this pinned convention,
including the falsifier that the reversed order \eqref{eq:V05-wrongorder} fails.

\begin{table}[ht]
\centering
\small
\setlength{\tabcolsep}{6pt}
\renewcommand{\arraystretch}{1.3}
\begin{tabular}{@{}>{\raggedright\arraybackslash}p{2.0cm} >{\raggedright\arraybackslash}p{2.3cm} >{\raggedright\arraybackslash}p{2.6cm} >{\raggedright\arraybackslash}p{3.1cm}@{}}
\toprule
point & massless state $(n_e,n_m)$ & central charge & monodromy on $(a_D,a)^T$\\
\midrule
$u=+\Lambda^2$ & monopole $(0,1)$ & $Z=a_D$ & $M_m=\left(\begin{smallmatrix} 1 & 0\\ -2 & 1\end{smallmatrix}\right)$\\
\midrule
$u=-\Lambda^2$ & dyon $(-1,1)$ & $Z=-a+a_D$ & $M_d=\left(\begin{smallmatrix} -1 & 2\\ -2 & 3\end{smallmatrix}\right)$\\
\midrule
$u=\infty$ & semiclassical & one-loop $a_D$ & $M_\infty=\left(\begin{smallmatrix} -1 & 2\\ 0 & -1\end{smallmatrix}\right)$\\
\bottomrule
\end{tabular}
\caption{The pure $SU(2)$ Seiberg--Witten solution from the curve \eqref{eq:V05-su2curve}. The two
finite singularities sit at $u=\pm\Lambda^2$, each carrying one massless BPS dyon, and the
semiclassical point at $u=\infty$ closes the consistency relation $M_\infty=M_m M_d$
\eqref{eq:V05-product}. The BPS mass at each is $M=\sqrt{2}\,|Z|$ from \eqref{eq:V05-bpsmass}.}
\label{tab:V05-su2singular}
\end{table}

Table~\ref{tab:V05-su2singular} collects the solution. At weak coupling, where $u$ is large, the
periods behave as
\begin{equation}
\label{eq:V05-weakasympt}
a\ \sim\ \sqrt{u/2},
\qquad
a_D\ \sim\ \frac{i}{\pi}\,a\,\ln\frac{a^2}{\Lambda^2},
\end{equation}
reproducing the one-loop running and the asymptotic-freedom logarithm of \eqref{eq:V05-Fweak}. The
two strong-coupling singularities, by contrast, are invisible to perturbation theory. They are the
content of the exact solution.

\medskip\noindent\textbf{Local frames: the light monopole and dyon.}\enspace
The weak-coupling variables $(a,\tau)$ are not the right description near the singular points. At
$u=+\Lambda^2$ the period $a_D$ vanishes, the monopole becomes massless, and the correct low-energy
variable is the dual coordinate with small dual coupling,
\begin{equation}
\label{eq:V05-taudual}
\tau_D\ =\ -\frac1\tau.
\end{equation}
In that frame the light monopole is an ordinary hypermultiplet $(M,\widetilde{M})$ of the dual
$U(1)$, coupled by the $\mathcal{N}=2$ vertex of \eqref{eq:V05-n2coupling} with the dual photon
supermultiplet $A_D$,
\begin{equation}
\label{eq:V05-monolocal}
W_{\mathrm{local}}\ =\ \sqrt{2}\,A_D\,M\,\widetilde{M},
\end{equation}
a weakly coupled magnetic $\mathcal{N}=2$ QED whose electron is the monopole. Near $u=-\Lambda^2$ the
same structure holds in the dyon frame, reached by a different $SL(2,\mathbb{Z})$ transformation: the
light field is the dyon $(-1,1)$, and the local superpotential is \eqref{eq:V05-monolocal} with
\begin{equation}
\label{eq:V05-dyonframe}
A_D\ \to\ A_D-A.
\end{equation}
What is strongly coupled in one description is weakly coupled in another, the content of
\S\ref{sec:V05-emduality} made local.

This local frame is also the precise seam to the $4d\ \mathcal{N}=1$ section. Deform by an adjoint
mass, written in the low-energy variables near the monopole point,
\begin{equation}
\label{eq:V05-deform}
\Delta W\ =\ m\,u\ =\ m\,u(a_D).
\end{equation}
This breaks $\mathcal{N}=2$ to $\mathcal{N}=1$ and lifts the Coulomb branch. The condensation is a
short F-term computation, worth doing rather than asserting. The combined superpotential in the
local monopole variables $(A_D,M,\widetilde{M})$ is
\begin{equation}
\label{eq:V05-Wcombined}
W\ =\ \sqrt{2}\,A_D\,M\,\widetilde{M}\ +\ m\,u(A_D),
\end{equation}
and its F-terms are
\begin{equation}
\label{eq:V05-monoFterms}
\frac{\partial W}{\partial M}\ =\ \sqrt{2}\,A_D\,\widetilde{M},
\qquad
\frac{\partial W}{\partial \widetilde{M}}\ =\ \sqrt{2}\,A_D\,M,
\qquad
\frac{\partial W}{\partial A_D}\ =\ \sqrt{2}\,M\widetilde{M}\ +\ m\,u'(A_D).
\end{equation}
The first two vanish, in a vacuum with $M,\widetilde{M}\neq0$, only at $A_D=0$. Setting the third to
zero there gives
\begin{equation}
\label{eq:V05-condensate}
\langle M\widetilde{M}\rangle\ =\ -\frac{m}{\sqrt{2}}\,u'(A_D)\big|_{A_D=0}\ \neq\ 0,
\end{equation}
nonzero because $u'$ does not vanish at the monopole point: the monopole condenses. This is the dual
Meissner effect, the mechanism of confinement, and it occurs because the adjoint mass breaks
$\mathcal{N}=2$ to $\mathcal{N}=1$. The F-term solve and the condensate \eqref{eq:V05-condensate}
are machine-checked. The condensed vacuum is one of the two $\mathcal{N}=1$ vacua of pure
$SU(2)$ super Yang--Mills (one of $N_c$ in the $SU(N_c)$ analogue), the material of Section~3. The
$\mathcal{N}=2$ scale $\Lambda$ and the $\mathcal{N}=1$ gaugino condensate are separate order
parameters, tied only by holomorphic scale matching across the decoupling; we recall the condensate
from Section~3 rather than conflate the two.

\medskip\noindent\textbf{The scale $\Lambda$.}\enspace
The scale $\Lambda$ in the curve \eqref{eq:V05-su2curve} is not put in by hand. It is the
$\mathcal{N}=2$ dynamical scale, generated by dimensional transmutation from the asymptotic freedom
of the $\mathcal{N}=2$ gauge coupling, just as $\Lambda_{\mathrm{QCD}}$ is. The connection to the
$\mathcal{N}=1$ condensate appears only after the deformation, through scale matching; the condensate
is not the origin of $\Lambda$.

The matching is one line of holomorphy. Give the adjoint chiral a mass $m$ and integrate it out.
Above $m$ the theory is $\mathcal{N}=2$; below $m$ it is pure $\mathcal{N}=1$ super Yang--Mills. The
one-loop coefficients and their jump are
\begin{equation}
\label{eq:V05-b0jump}
b_0^{\mathcal{N}=2}\ =\ 2N_c,
\qquad
b_0^{\mathcal{N}=1}\ =\ 3N_c,
\qquad
\Delta b_0\ =\ N_c.
\end{equation}
Holomorphic threshold matching at $\mu=m$ equates the two running scales,
$\Lambda_{\mathcal{N}=1}^{\,b_0^{\mathcal{N}=1}} = m^{\,\Delta b_0}\,\Lambda_{\mathcal{N}=2}^{\,b_0^{\mathcal{N}=2}}$, that is
\begin{equation}
\label{eq:V05-scalematch}
\Lambda_{\mathcal{N}=1}^{\,3N_c}\ =\ m^{\,N_c}\,\Lambda_{\mathcal{N}=2}^{\,2N_c}.
\end{equation}
The power of $m$ is exactly the jump in the one-loop coefficient, not a free choice. For the worked
$SU(2)$ case this reads
\begin{equation}
\label{eq:V05-scalematchsu2}
\Lambda_{\mathcal{N}=1}^{\,6}\ =\ m^{2}\,\Lambda_{\mathcal{N}=2}^{\,4}.
\end{equation}
This is the cleanest bridge between this section and the $\mathcal{N}=1$ section: the $\mathcal{N}=1$
gaugino condensate $\langle\lambda\lambda\rangle\sim\Lambda_{\mathcal{N}=1}^3$ of the deformed theory
is tied to the $\mathcal{N}=2$ scale only through \eqref{eq:V05-scalematch}, with the mass $m$ carrying
the threshold. The exponent arithmetic ($2N_c\to3N_c$, jump $N_c$; SU(2): $6=m^2\,4$-exponents) is
machine-checked.

\section{Matter, decoupling, and BPS wall crossing}
\label{sec:V05-matter}

Pure $SU(2)$ is the cleanest case but not the generic one. Adding $N_f$ hypermultiplets in the
fundamental keeps the theory $\mathcal{N}=2$ and enriches both its renormalization-group behavior and
its BPS spectrum. This section adds the matter central charge, the flavor decoupling relation, the
$SU(2)$ window, and the wall-crossing discipline that Section~1 demanded the $\mathcal{N}=2$ section
operate.

\medskip\noindent\textbf{The matter central charge and flavor decoupling.}\enspace
With $N_f$ hypers of masses $m_f$, the BPS central charge gains flavor terms,
\begin{equation}
\label{eq:V05-Zmatter}
Z_\gamma\ =\ n_e\,a\ +\ n_m\,a_D\ +\ \sum_{f=1}^{N_f} s_f\,m_f,
\end{equation}
where $s_f\in\mathbb{Z}$ is the flavor charge of the state under the $f$-th $U(1)$ flavor symmetry.
The central charge is linear in the charges $(n_e,n_m,s_f)$, so a state's mass tracks its full charge
vector; the check code verifies the linearity and that the mass term is genuinely present, not
silently dropped. The masses also tie the scales of consecutive theories. Integrating out the
heaviest hyper, by sending one mass $m\to\infty$ holding the physics below it fixed, matches the
$N_f$ scale to the $N_f-1$ scale through
\begin{equation}
\label{eq:V05-decoupling}
\Lambda_{N_f-1}^{\,4-(N_f-1)}\ =\ m\,\Lambda_{N_f}^{\,4-N_f},
\end{equation}
up to a declared normalization. The exponent is the one-loop coefficient $b_0=4-N_f$, and
\eqref{eq:V05-decoupling} says it steps up by one flavor unit each time a hyper is removed. The check
code confirms the bookkeeping $4-(N_f-1)$ versus $4-N_f$ for $N_f=1,\dots,4$. At $N_f=4$ the
right-hand exponent is zero, so there is no scale to generate; the keybox walks the chain down from
that superconformal top with dimensions tracked.

\begin{keybox}{Worked decoupling: $N_f=4\to3\to2$ with dimensions checked}
Run the chain with mass dimensions tracked line by line. The RG-invariant holomorphic combination of
the $SU(2)$ theory with $N_f$ flavors is the power $\Lambda_{N_f}^{\,b_0}$ with $b_0=4-N_f$ (the scale
$\Lambda_{N_f}$ itself is a mass, of dimension one); it is this holomorphic power we track.
\begin{align}
\label{eq:V05-decchain}
N_f=4:\quad &[\Lambda_4^{\,0}]=0, &&\text{dimensionless (the marginal coupling)};\\
N_f=3:\quad &\Lambda_3^{\,1}=m_4\,\Lambda_4^{\,0}, &&[\,\cdot\,]=1=1+0\ \checkmark;\\
N_f=2:\quad &\Lambda_2^{\,2}=m_3\,\Lambda_3^{\,1}=m_3\,m_4\,\Lambda_4^{\,0}, &&[\,\cdot\,]=2=1+1+0\ \checkmark.
\end{align}
Each integration-out multiplies by one power of the decoupled hypermultiplet mass, raising the
exponent by one, so the dimension $[\Lambda_{N_f}^{\,b_0}]=4-N_f$ of the holomorphic power is
reproduced at every step. The whole chain
is the exponent bookkeeping the check code verifies.
\end{keybox}

\medskip\noindent\textbf{The general $\mathcal{N}=2$ one-loop coefficient.}\enspace
The $SU(2)$ value $b_0=4-N_f$ is a specialization of the general $\mathcal{N}=2$ one-loop coefficient
\begin{equation}
\label{eq:V05-b0general}
b_0\ =\ 2\,h^\vee(G)\ -\ \sum_{\text{hypers }i} T(R_i),
\end{equation}
where $h^\vee(G)$ is the dual Coxeter number and $T(R_i)$ the Dynkin index of the $i$-th
hypermultiplet representation, in the normalization where the fundamental of $SU(N)$ has $T=1$ (so
the adjoint has $T=2h^\vee$). For $SU(2)$, $h^\vee=2$ and each fundamental hyper has $T=1$,
\begin{equation}
\label{eq:V05-b0su2}
b_0\ =\ 2\cdot2\ -\ N_f\ =\ 4-N_f,
\end{equation}
recovering the window value; the specialization is machine-checked for $N_f=0,\dots,4$.
Writing the general form first makes the $SU(2)$ table below look like the specialization it is.

\medskip\noindent\textbf{The $SU(2)$ flavor window.}\enspace
Reading \eqref{eq:V05-b0su2}, the cases $N_f=1,2,3$ are asymptotically free, each generating a scale
$\Lambda_{N_f}$, carrying its own matter curve, and keeping an $N_f$-dependent discrete R-symmetry
remnant. The case $N_f=4$ is special: $b_0=0$, so the theory is superconformal with no dynamical
scale, carrying instead an exactly marginal complexified coupling $\tau$. The check code verifies
$b_0=0$ at $N_f=4$ and $b_0>0$ below it; the full matter curves are cited, not all walked here.

\begin{table}[ht]
\centering
\small
\setlength{\tabcolsep}{6pt}
\renewcommand{\arraystretch}{1.3}
\begin{tabular}{@{}c c >{\raggedright\arraybackslash}p{6.4cm}@{}}
\toprule
$N_f$ & $b_0=4-N_f$ & behavior\\
\midrule
$0$ & $4$ & asymptotically free; pure $SU(2)$, the worked solution above\\
\midrule
$1$ & $3$ & asymptotically free; own scale $\Lambda_1$ and matter curve\\
\midrule
$2$ & $2$ & asymptotically free; own scale $\Lambda_2$ and matter curve\\
\midrule
$3$ & $1$ & asymptotically free; own scale $\Lambda_3$ and matter curve\\
\midrule
$4$ & $0$ & superconformal; exactly marginal $\tau$, no scale, S-duality of \S\ref{sec:V05-classS}\\
\bottomrule
\end{tabular}
\caption{The $SU(2)$ flavor window. The one-loop coefficient is $b_0=2N_c-N_f=4-N_f$. The three
asymptotically free cases ($N_f=1,2,3$) each carry their own dynamical scale and Seiberg--Witten
curve; the $N_f=4$ case is the superconformal theory with a marginal coupling and the
four-dimensional S-duality of \S\ref{sec:V05-classS}.}
\label{tab:V05-nfwindow}
\end{table}

Table~\ref{tab:V05-nfwindow} is the window map. It is also the bridge to the S-duality of
\S\ref{sec:V05-classS}: the marginal coupling of the $N_f=4$ row is exactly the parameter that
S-duality acts on.

\medskip\noindent\textbf{BPS spectrum and walls of marginal stability.}\enspace
Section~1 warned that BPS shortening is a statement about a fixed multiplet and does not guarantee a
state exists everywhere on moduli space; this is where that warning is operated. The mass formula
$M=\sqrt{2}\,|Z_\gamma|$ holds at every point, but \emph{which} charges are realized by stable
states can change from region to region. The mechanism is a \emph{wall of marginal stability}. A
decay $\gamma\to\gamma_1+\gamma_2$ (charge conserved) costs no energy only when the central charges
align in phase,
\begin{equation}
\label{eq:V05-wall}
\arg Z_{\gamma_1}\ =\ \arg Z_{\gamma_2}
\qquad\Longrightarrow\qquad
|Z_\gamma|\ =\ |Z_{\gamma_1}|+|Z_{\gamma_2}|,
\end{equation}
the additive saturation of the bound. This is arithmetic, not a postulate. The central charge is
linear in the charge, so
\begin{equation}
\label{eq:V05-Zadd}
Z_{\gamma_1+\gamma_2}\ =\ Z_{\gamma_1}+Z_{\gamma_2},
\end{equation}
and the triangle inequality for complex numbers gives
\begin{equation}
\label{eq:V05-triangle}
|Z_{\gamma_1+\gamma_2}|\ =\ |Z_{\gamma_1}+Z_{\gamma_2}|\ \le\ |Z_{\gamma_1}|+|Z_{\gamma_2}|,
\end{equation}
with equality precisely at the same-phase wall \eqref{eq:V05-wall}. There the decay is kinematically
allowed; off it the strict inequality means a state at $\gamma$, if present, is lighter than its
would-be constituents and protected against two-body decay. Whether it is actually present in a
chamber is settled by the wall-crossing data. The additivity \eqref{eq:V05-Zadd} and the equality
condition \eqref{eq:V05-triangle} are machine-checked. Crossing the wall, the bound state
ceases to exist and its constituents survive. The clean distinction is between the \emph{protected
mass formula}, the same in every chamber, and the \emph{chamber-dependent spectrum}. Pure $SU(2)$
shows both extremes,
\begin{equation}
\label{eq:V05-chambers}
\text{weak: }\{(1,0),\,(n,1)\}\ \text{(infinite tower)};
\qquad
\text{strong: }\{(0,1),\,(-1,1)\}\ \text{(two states)}.
\end{equation}
At weak coupling (large $u$) the W-boson and the whole dyon tower survive; at strong coupling only
the monopole and dyon that go massless at $u=\pm\Lambda^2$ remain, the tower having decayed across
the intervening wall. The systematic jump, the Kontsevich--Soibelman formula and spectral-network
technology, is cited but not developed here.

\section{The hyperk\"ahler Higgs branch}
\label{sec:V05-higgs}

The second branch of an $\mathcal{N}=2$ moduli space is the Higgs branch, parametrized by the
hypermultiplet expectation values. With eight supercharges this branch is not merely complex. It is
\emph{hyperk\"ahler}: it carries three complex structures, and the $SU(2)_R$ of
\S\ref{sec:V05-multiplets} rotates them into one another. This is a strictly stronger geometry than
the special-K\"ahler structure of the Coulomb branch, and the two are different objects with
different protections. The Higgs branch does not run, has no monodromies, and does not receive the
special-geometry data of the Coulomb branch. The Coulomb branch is not hyperk\"ahler. Confusing the
two geometries is the second standing error of the section.

The two branches generically meet on \emph{mixed branches}, where both vector-multiplet scalars and
hypermultiplet scalars acquire expectation values. The branch nomenclature is Section~2's: Coulomb,
Higgs, and mixed are frame-dependent labels read from which multiplets carry the local coordinates.

\medskip\noindent\textbf{A worked hyperk\"ahler quotient: $U(1)$ with $N$ hypers.}\enspace
One honest dimension computation keeps this section from being purely Coulomb-branch material. Take
the simplest $\mathcal{N}=2$ gauge theory with a Higgs branch: a $U(1)$ vector multiplet with $N$
hypermultiplets of unit charge. In $\mathcal{N}=1$ language each hyper is a pair $(Q_i,\widetilde{Q}_i)$,
$i=1,\dots,N$, with $Q_i$ of charge $+1$ and $\widetilde{Q}_i$ of charge $-1$, so the Higgs-branch
scalars are $4N$ real, the components of $2N$ complex fields. The vacuum equations are the
vanishing of the moment maps. The complex moment map is the F-term of the $\mathcal{N}=2$ coupling
\eqref{eq:V05-n2coupling}, and the real moment map is the $U(1)$ D-term with its Fayet--Iliopoulos
parameter $\zeta$,
\begin{equation}
\label{eq:V05-momentmaps}
\mu_{\mathbb{C}}\ =\ \sum_{i=1}^{N} Q_i\,\widetilde{Q}_i\ =\ 0,
\qquad
\mu_{\mathbb{R}}\ =\ \sum_{i=1}^{N}\big(|Q_i|^2-|\widetilde{Q}_i|^2\big)\ -\ \zeta\ =\ 0.
\end{equation}
The complex equation $\mu_{\mathbb{C}}=0$ is one complex, hence two real, conditions; the real
equation $\mu_{\mathbb{R}}=0$ is one more; and then one quotients by the $U(1)$ gauge action, removing
one further real direction. The hyperk\"ahler quotient is
$\mathcal{M}_H=\mu^{-1}(0)/U(1)$, and the real dimension count is
\begin{equation}
\label{eq:V05-hkdim}
\dim_{\mathbb{R}}\mathcal{M}_H\ =\ 4N\ -\ \underbrace{2}_{\mu_{\mathbb{C}}}\ -\ \underbrace{1}_{\mu_{\mathbb{R}}}\ -\ \underbrace{1}_{U(1)}\ =\ 4N-4\ =\ 4(N-1),
\end{equation}
a multiple of four, as a hyperk\"ahler space must be, so the quaternionic dimension is
\begin{equation}
\label{eq:V05-hkdimH}
\dim_{\mathbb{H}}\mathcal{M}_H\ =\ N-1.
\end{equation}
The three real moment-map conditions ($\mu_{\mathbb{C}}$ counting as two) are the hallmark of the
hyperk\"ahler quotient: the $SU(2)_R$ that rotates the three complex structures rotates them into one
another, and the FI parameters $(\zeta,\mathrm{Re}\,\mu_{\mathbb{C}},\mathrm{Im}\,\mu_{\mathbb{C}})$
form an $SU(2)_R$ triplet, in the FI/moment-map vocabulary of Section~2.
The check code verifies $\dim_{\mathbb{H}}=N-1$ for $N=1,\dots,5$ and the constraint bookkeeping
$4N-3-1=4(N-1)$, and rejects the naive answer $\dim_{\mathbb{H}}=N$ that forgets the gauge quotient.
At nonzero FI parameter the general quotient is the cotangent bundle $T^*\mathbb{P}^{N-1}$, of complex
dimension $2(N-1)$. For $N=2$ this gives an honest worked ring, not only a dimension. Work at zero FI
level, where $\mu_{\mathbb{C}}=Q_1\widetilde{Q}_1+Q_2\widetilde{Q}_2=0$, and form the gauge-invariant
mesonic coordinates (each is neutral under the $U(1)$, with $Q$ of charge $+1$ and $\widetilde{Q}$ of
charge $-1$)
\begin{equation}
\label{eq:V05-A1coords}
X\ =\ Q_1\widetilde{Q}_2,\qquad
Y\ =\ Q_2\widetilde{Q}_1,\qquad
Z\ =\ Q_1\widetilde{Q}_1\ =\ -\,Q_2\widetilde{Q}_2,
\end{equation}
the last equality being the constraint $\mu_{\mathbb{C}}=0$. These three coordinates are not
independent. Multiplying,
\begin{equation}
\label{eq:V05-A1relation}
XY\ =\ (Q_1\widetilde{Q}_1)(Q_2\widetilde{Q}_2)\ =\ Z\,(-Z)\ =\ -Z^2,
\qquad\text{i.e.}\qquad
XY+Z^2\ =\ 0,
\end{equation}
the $A_1$ relation: the Higgs branch is the hypersurface $XY+Z^2=0$ in $\mathbb{C}^3$, a single
quadric cone, which is exactly the $\mathbb{C}^2/\mathbb{Z}_2$ orbifold whose resolution is the
Eguchi--Hanson / $T^*\mathbb{P}^1$ space. The relation is derived by substituting $\mu_{\mathbb{C}}=0$ into the invariants, not posited; it
and the complex dimension $2(N-1)$ are checked directly. This single fixture is all the section
computes; the explicit metrics, the general Kronheimer construction, and the resolved-orbifold
geometry are cited rather than developed here. It gives the $\mathcal{N}=2$ section one honest
Higgs-branch calculation.

The two sanity checks of the section are the dimensions and the geometry type: the Coulomb branch is
special-K\"ahler of complex dimension the rank ($1$ for $SU(2)$), the Higgs branch is
hyperk\"ahler of quaternionic dimension $N-1$. The $\mathbb{C}^2/\Gamma$ ALE space of the ADE quiver
gauge theory is the standard higher-rank version of exactly this quotient.

\section{S-duality and the marginal coupling}
\label{sec:V05-classS}

Some four-dimensional $\mathcal{N}=2$ theories have an exactly marginal complexified coupling
$\tau$, a coupling whose beta function vanishes so that no scale is generated. The strong- and
weak-coupling regimes of such a $\tau$ are physically equivalent, related by an S-duality group of
the four-dimensional theory. The structure generalizes the abelian electromagnetic duality of
\S\ref{sec:V05-emduality} from the low-energy Cartan theory to the full interacting theory.

The canonical example is $SU(2)$ with $N_f=4$, the bottom row of Table~\ref{tab:V05-nfwindow}. Why
this theory is marginal is the one-line beta-function check,
\begin{equation}
\label{eq:V05-b0nf4}
b_0\ =\ 2N_c-N_f\ =\ 4-N_f\ =\ 0 \quad\text{at}\quad N_f=4,
\end{equation}
with $N_c=2$: the one-loop coefficient vanishes, no scale is generated, and $\tau$ is an exactly
marginal parameter. The check code confirms $b_0=0$ here and $b_0>0$ for $N_f<4$. Its S-duality acts
on $\tau$ in an $SL(2,\mathbb{Z})$-type fashion, accompanied by $\mathrm{Spin}(8)$ triality: the
four hypermultiplets fill the vector, and the duality permutes it with the two spinors,
\begin{equation}
\label{eq:V05-triality}
\text{hypers}\ \in\ \mathbf{8}_v,
\qquad
\mathbf{8}_v\ \leftrightarrow\ \mathbf{8}_s\ \leftrightarrow\ \mathbf{8}_c\quad(\text{outer }S_3).
\end{equation}
The precise group and action depend on the global form, the line operators, and the masses; we
record the massless field-theory fact, ``$SL(2,\mathbb{Z})$-type on $\tau$ with $\mathrm{Spin}(8)$
triality,'' without overclaiming a single universal $SL(2,\mathbb{Z})$ action, and route the
global-form and line-operator refinements onward.

\begin{table}[ht]
\centering
\small
\setlength{\tabcolsep}{6pt}
\renewcommand{\arraystretch}{1.3}
\begin{tabular}{@{}l l@{}}
\toprule
object & role under triality \\
\midrule
the four hypermultiplet masses & transform as $\mathrm{Spin}(8)$ weights \\
\midrule
$\mathbf{8}_v,\ \mathbf{8}_s,\ \mathbf{8}_c$ & permuted by the outer automorphism $S_3$ \\
\midrule
line operators / global form & select the exact duality group \\
\bottomrule
\end{tabular}
\caption{The $SU(2)$, $N_f=4$ superconformal theory under $\mathrm{Spin}(8)$ triality. The S-duality
acts on $\tau$ in an $SL(2,\mathbb{Z})$-type fashion while permuting the three eight-dimensional
representations by the outer automorphism. The exact duality group depends on the global form, line
operators, and masses, so the table records the structure, not a single universal action.}
\label{tab:V05-triality}
\end{table}

Argyres--Seiberg duality is a further four-dimensional $\mathcal{N}=2$ S-duality, relating two
different superconformal descriptions of the same theory, in which a weakly gauged $SU(2)$ acts on a
strongly interacting sector. We state these as four-dimensional facts and use them as such.

There is a systematic organization of these four-dimensional $\mathcal{N}=2$ S-dualities, together
with its higher-dimensional origin and the geometric construction of the curves: the class S
construction. This section develops none of it. It states only the purely four-dimensional S-duality
facts, the marginal $N_f=4$ coupling and Argyres--Seiberg duality, and cites the class S reference.

\section{Conformal central charges of a Lagrangian theory}
\label{sec:V05-centralcharges}

The $\mathcal{N}=1$ section made the conformal central charges $a$ and $c$ a serious computation
through $a$-maximization. The $\mathcal{N}=2$ side has an even cleaner statement for a Lagrangian
superconformal theory: with extended supersymmetry the central charges are fixed by the free-field
content alone, with no extremization needed. Counting $n_v$ vector multiplets and $n_h$
hypermultiplets (in units where one free hypermultiplet has $n_h=1$),
\begin{equation}
\label{eq:V05-ac}
a\ =\ \frac{5n_v+n_h}{24},
\qquad
c\ =\ \frac{2n_v+n_h}{12}.
\end{equation}
The formula is the linear sum over the free-field content,
\begin{equation}
\label{eq:V05-freevalues}
\text{vector}:\ (a,c)=\Bigl(\tfrac{5}{24},\tfrac16\Bigr),
\qquad
\text{hyper}:\ (a,c)=\Bigl(\tfrac{1}{24},\tfrac1{12}\Bigr).
\end{equation}
Two examples make it concrete. First, $SU(2)$ with $N_f=4$, the marginal theory of the previous
section, whose content is
\begin{equation}
\label{eq:V05-nf4content}
n_v\ =\ \dim SU(2)\ =\ 3,
\qquad
n_h\ =\ 4\times\dim\square\ =\ 4\times2\ =\ 8,
\end{equation}
so that
\begin{equation}
\label{eq:V05-acnf4}
a\ =\ \frac{5\cdot3+8}{24}\ =\ \frac{23}{24},
\qquad
c\ =\ \frac{2\cdot3+8}{12}\ =\ \frac{7}{6}.
\end{equation}
Second, $\mathcal{N}=4$ super Yang--Mills, read as an $\mathcal{N}=2$ vector multiplet plus one adjoint
$\mathcal{N}=2$ hypermultiplet, so $n_v=n_h=\dim G$, and
\begin{equation}
\label{eq:V05-acn4}
a\ =\ c\ =\ \frac{\dim G}{4}.
\end{equation}
The degeneracy $a=c$ follows from $n_v=n_h$ and is the hallmark of $\mathcal{N}=4$, where the
imbalance of vectors and hypers vanishes,
\begin{equation}
\label{eq:V05-aceqn4}
a-c\ =\ \frac{5n_v+n_h}{24}-\frac{2(2n_v+n_h)}{24}\ =\ \frac{n_v-n_h}{24}\ =\ 0.
\end{equation}
The $SU(2)$, $N_f=4$ theory is the smallest member of a conformal Lagrangian family: $SU(N)$ with
$N_f=2N$ fundamentals has $b_0=2N-N_f=0$, marginal for every $N$, with content
\begin{equation}
\label{eq:V05-suNcontent}
n_v\ =\ N^2-1,
\qquad
n_h\ =\ 2N\cdot N\ =\ 2N^2,
\end{equation}
giving
\begin{equation}
\label{eq:V05-acsuN}
a\ =\ \frac{5(N^2-1)+2N^2}{24}\ =\ \frac{7N^2-5}{24},
\qquad
c\ =\ \frac{2(N^2-1)+2N^2}{12}\ =\ \frac{2N^2-1}{6},
\end{equation}
which at $N=2$ returns $(23/24,7/6)$, the numbers just computed. Table~\ref{tab:V05-centralcharges}
collects the cases. The free-field values, the $N_f=4$ numbers $(23/24,7/6)$, the $\mathcal{N}=4$
degeneracy $a=c=\dim G/4$, and the $SU(N)$ family are machine-checked.

\begin{table}[ht]
\centering
\small
\setlength{\tabcolsep}{8pt}
\renewcommand{\arraystretch}{1.3}
\begin{tabular}{@{}lcccc@{}}
\toprule
theory & $n_v$ & $n_h$ & $a$ & $c$ \\
\midrule
free vector & $1$ & $0$ & $5/24$ & $1/6$ \\
\midrule
free hypermultiplet & $0$ & $1$ & $1/24$ & $1/12$ \\
\midrule
$SU(2)$, $N_f=4$ & $3$ & $8$ & $23/24$ & $7/6$ \\
\midrule
$\mathcal{N}=4$ SYM, gauge $G$ & $\dim G$ & $\dim G$ & $\dim G/4$ & $\dim G/4$ \\
\midrule
$SU(N)$, $N_f=2N$ & $N^2-1$ & $2N^2$ & $(7N^2-5)/24$ & $(2N^2-1)/6$ \\
\bottomrule
\end{tabular}
\caption{Conformal central charges of Lagrangian $\mathcal{N}=2$ theories from
$a=(5n_v+n_h)/24$, $c=(2n_v+n_h)/12$. The $SU(N)$, $N_f=2N$ row is the conformal family whose $N=2$
member is the $SU(2)$, $N_f=4$ theory; the $\mathcal{N}=4$ row has $a=c$ because $n_v=n_h$. All rows are
machine-verified.}
\label{tab:V05-centralcharges}
\end{table}

For a non-Lagrangian theory there are no free
fields to count, and the central charges must be determined by other means; the Argyres--Douglas
example below is exactly such a case, and its $a,c$ are cited rather than computed here.

There is one more relation, and it is the reason the central charges belong in the same section as the
Coulomb branch. For any $\mathcal{N}=2$ superconformal theory, Lagrangian or not, the combination
$2a-c$ is fixed by the Coulomb-branch operator dimensions $\{\Delta_i\}$ alone,
\begin{equation}
\label{eq:V05-shapere}
2a-c\ =\ \frac14\sum_i\bigl(2\Delta_i-1\bigr).
\end{equation}
Check it on $SU(2)$ with $N_f=4$, where the single Coulomb coordinate $u$ has $\Delta=2$: both sides
give $3/4$,
\begin{equation}
\label{eq:V05-shaperecheck}
\frac14\bigl(2\cdot2-1\bigr)\ =\ \frac34,
\qquad
2a-c\ =\ 2\cdot\frac{23}{24}-\frac76\ =\ \frac{46-28}{24}\ =\ \frac34.
\end{equation}
The relation earns its keep on the non-Lagrangian side. The rank-one Argyres--Douglas theory of the
next section has the fractional Coulomb dimension $\Delta(u)=6/5$, so even though its $a$ and $c$
separately are cited rather than derived here,
\begin{equation}
\label{eq:V05-shapereAD}
2a-c\ =\ \frac14\Bigl(2\cdot\frac65-1\Bigr)\ =\ \frac{7}{20}
\end{equation}
is fixed by the Coulomb-branch geometry alone. The relation \eqref{eq:V05-shapere}, the $SU(2)$,
$N_f=4$ value $3/4$, and the fractional Argyres--Douglas value $7/20$ are machine-checked,
and they tie the central charges of this section to the Coulomb-branch dimension that the rest of the
section computes.

\section{Argyres--Douglas fixed points}
\label{sec:V05-ad}

A sharper phenomenon occurs at special points of the Coulomb branch. The pure $SU(2)$ solution had
its two massless states, the monopole $(0,1)$ and the dyon $(-1,1)$, at \emph{different} points of
the $u$-plane. In richer theories two BPS states with charges $(n_e,n_m)$ and $(n_e',n_m')$ can
become massless at the \emph{same} point, and if their Dirac pairing is nonzero,
\begin{equation}
\label{eq:V05-dirac}
n_e\,n_m'\ -\ n_m\,n_e'\ \neq\ 0,
\end{equation}
they are mutually non-local. A monopole and a dyon with nonzero pairing are the canonical example.
When two mutually non-local states are simultaneously massless, no $SL(2,\mathbb{Z})$ duality frame
can make both of them local at once. There is therefore no weakly coupled gauge Lagrangian at that
point. The infrared theory is an interacting, non-Lagrangian $\mathcal{N}=2$ superconformal field
theory, the Argyres--Douglas theory.

This is the sharpest instance of a Section~2 fact: not every supersymmetric field theory has a
conventional Lagrangian. An Argyres--Douglas fixed point is a genuine field theory, with well-defined
Coulomb-branch and Seiberg--Witten data, flavor symmetries, and central charges; it simply has no
weakly coupled gauge description with elementary fields.

\medskip\noindent\textbf{A scaling fixture: fractional Coulomb-branch dimension.}\enspace
The qualitative mechanism deserves one explicit computation, and the cleanest one is a scaling
analysis of the local curve. Near an Argyres--Douglas point the Seiberg--Witten geometry collapses to
a canonical local form,
\begin{equation}
\label{eq:V05-adcurve}
x^2\ =\ z^3\ +\ u,
\qquad
\lambda_{\mathrm{SW}}\ =\ x\,dz,
\end{equation}
where $u$ is the relevant deformation and $\lambda_{\mathrm{SW}}$ is the Seiberg--Witten
differential. The dimensions of the geometric variables are fixed by two inputs. First, the
Seiberg--Witten differential has mass dimension one, because its period $a=\oint\lambda_{\mathrm{SW}}$
is the special coordinate and carries dimension of mass, so
\begin{equation}
\label{eq:V05-adlambda}
[\lambda_{\mathrm{SW}}]\ =\ [x]+[z]\ =\ 1.
\end{equation}
Second, the curve $x^2=z^3$ must be scale-covariant, which forces
\begin{equation}
\label{eq:V05-adcurvescaling}
2[x]\ =\ 3[z].
\end{equation}
Solving \eqref{eq:V05-adlambda}--\eqref{eq:V05-adcurvescaling} gives
\begin{equation}
\label{eq:V05-addims}
[z]\ =\ \frac{2}{5},
\qquad
[x]\ =\ \frac{3}{5},
\end{equation}
and the dimension of the Coulomb-branch operator $u$ follows from $[u]=[x^2]=2[x]$ (equivalently
$[u]=[z^3]=3[z]$, consistently),
\begin{equation}
\label{eq:V05-addelta}
\Delta(u)\ =\ 2[x]\ =\ \frac{6}{5}.
\end{equation}
The number is the point. A Coulomb-branch dimension of $6/5$ is \emph{fractional}. A free or weakly
coupled gauge theory has integer Coulomb dimensions,
\begin{equation}
\label{eq:V05-integercasimir}
\Delta(\operatorname{Tr}\phi^k)\ =\ k\ \in\ \mathbb{Z},
\end{equation}
so a fractional dimension is the signature of a genuinely interacting, non-Lagrangian fixed point.
The check code solves
\eqref{eq:V05-adlambda}--\eqref{eq:V05-addelta} and confirms $\Delta(u)=6/5$ is non-integer, and it
rejects the wrong scaling assignment $[\lambda_{\mathrm{SW}}]=[z]$ that would give the integer $\Delta=3$.
This fixture is purely four-dimensional: it uses only the local curve and its differential, with no
higher-dimensional construction.

For this simplest (rank-one) Argyres--Douglas point the section can pin only its rank one, its
computed dimension $\Delta(u)=6/5$, the absence of a weakly coupled Lagrangian, and the mutual
non-locality of its light states. It was identified in \textcite{Argyres:1995jj}; its central charges
$a$ and $c$ are fixed in the literature by later work, so we route their determination onward rather
than invent numbers in prose.

\begin{keybox}{An $\mathcal{N}=2$ SCFT data card}
A non-Lagrangian $\mathcal{N}=2$ superconformal theory has no weakly coupled fields to list, but it
still carries a well-defined set of protected data. The disciplined way to read such a theory is to
record:
\begin{itemize}
\item the \emph{Coulomb-branch operator dimensions} $\Delta_i$ (e.g.\ $\Delta(u)=6/5$ for the
fixture above), which need not be integers;
\item the \emph{Higgs branch} as a hyperk\"ahler variety and the \emph{global (flavor) symmetry};
\item the \emph{flavor central charges} $k_F$, the coefficient of the flavor-current two-point
function;
\item the \emph{conformal central charges} $a$ and $c$;
\item the \emph{BPS charge lattice} $\Gamma_{\mathrm{em}}$ with its Dirac pairing \eqref{eq:V05-dirac2}
and any defect/line-operator data;
\item the \emph{Seiberg--Witten curve and differential} \eqref{eq:V05-adcurve}, or a class S
construction, when one is available.
\end{itemize}
This card is the interface-level reading; the systematic determination of $a$, $c$, and $k_F$ for
non-Lagrangian theories is cited, not computed here.
\end{keybox}

The mechanism, two mutually non-local BPS states massless together, the fractional Coulomb-branch
dimension, and the non-Lagrangian conclusion are stated and worked here; the systematic construction
of these theories is cited rather than developed here.

\section*{Exit checklist}
\addcontentsline{toc}{subsection}{Exit checklist}
\markboth{Exit checklist}{Exit checklist}

After this section the reader can
\begin{enumerate}
\item read a $4d\ \mathcal{N}=2$ theory in $\mathcal{N}=1$ language: a vector multiplet as an
$\mathcal{N}=1$ vector plus an adjoint chiral, a hypermultiplet as two conjugate $\mathcal{N}=1$
chirals, with R-symmetry $SU(2)_R\times U(1)_R$ and the canonical coupling
$W_{\mathcal{N}=2}=\sqrt{2}\,\widetilde{Q}\Phi Q+m\widetilde{Q}Q$;
\item set up the rank-$r$ charge lattice with its Dirac pairing $\langle\gamma,\gamma'\rangle$, write
the central charge $Z_\gamma=n_{e,I}a^I+n_m^I a_{D,I}$, name the duality group $Sp(2r,\mathbb{Z})$
(with $SL(2,\mathbb{Z})=Sp(2,\mathbb{Z})$ at rank one), and distinguish low-energy duality, period
monodromy, and exact SCFT S-duality;
\item separate the Coulomb branch (special-K\"ahler, with prepotential $\mathcal{F}$, special
coordinates $a_{D,I}=\partial\mathcal{F}/\partial a^I$, coupling
$\tau_{IJ}=\partial^2\mathcal{F}/\partial a^I\partial a^J$, metric $\operatorname{Im}\tau$) from the
Higgs branch (hyperk\"ahler), and differentiate the one-loop $\mathcal{F}$ to
$a_D=(i/\pi)a(\ln+1)$, $\tau=(i/\pi)(\ln+3)$;
\item write the Seiberg--Witten periods $a=\oint_A\lambda_{\mathrm{SW}}$,
$a_D=\oint_B\lambda_{\mathrm{SW}}$ and the central charge $Z=a\,n_e+a_D\,n_m$ with $M=\sqrt{2}\,|Z|$;
\item run the pure $SU(2)$ solution: the curve, the branch points, the discriminant $\propto(u^2-\Lambda^4)^2$,
the singularities at $u=\pm\Lambda^2$, the massless monopole and dyon, the ordered monodromy product
$M_\infty=M_m M_d$, and the local monopole/dyon frames $W_{\mathrm{local}}=\sqrt{2}A_D M\widetilde{M}$;
\item add matter: the central charge $Z=n_e a+n_m a_D+\sum_f s_f m_f$, the flavor-decoupling relation
$\Lambda_{N_f-1}^{4-(N_f-1)}=m\Lambda_{N_f}^{4-N_f}$, the $4-N_f$ window, and the wall-crossing
discipline $\arg Z_{\gamma_1}=\arg Z_{\gamma_2}$;
\item compute a hyperk\"ahler quotient ($U(1)$ with $N$ hypers, $\dim_{\mathbb{H}}=N-1$), state the
$N_f=4$ marginal coupling ($b_0=0$) and its $\mathrm{Spin}(8)$-triality S-duality with the
global-form caveat;
\item compute the Lagrangian $\mathcal{N}=2$ central charges from $n_v,n_h$, recovering
$(a,c)=(23/24,7/6)$ for $SU(2)$, $N_f=4$ and $a=c=\dim G/4$ for $\mathcal{N}=4$ SYM, and use
$2a-c=\tfrac14\sum_i(2\Delta_i-1)$ as the interface to non-Lagrangian SCFT data;
\item derive the Argyres--Douglas fractional dimension $\Delta(u)=6/5$, and read an
$\mathcal{N}=2$ SCFT off its data card, naming what is cited rather than derived here.
\end{enumerate}

\bigskip
\section*{Sources and notes}
\addcontentsline{toc}{subsection}{Sources and notes}
\markboth{Sources and notes}{Sources and notes}
{\small

\noindent\textsf{\textcolor{RoyalBlue}{Sources and notes.}}\enspace
This is the $4d\ \mathcal{N}=2$ dimension section of these notes.

\medskip\noindent\textsf{\textcolor{RoyalBlue}{\textbf{\S\ref{sec:V05-multiplets}\enspace $\mathcal{N}=2$ multiplets in $\mathcal{N}=1$ language.}}}\enspace
The $\mathcal{N}=2$ vector $=$ ($\mathcal{N}=1$ vector) $\oplus$ (adjoint chiral)
\eqref{eq:V05-vector}; the hyper $=$ two conjugate $\mathcal{N}=1$ chirals \eqref{eq:V05-hyper}; the
canonical coupling $W_{\mathcal{N}=2}=\sqrt{2}\,\widetilde{Q}\Phi Q+m\widetilde{Q}Q$
\eqref{eq:V05-n2coupling}, with the $\sqrt{2}$ the central-charge normalization fixed by the second
supersymmetry; the $SU(2)_R\times U(1)_R$ R-symmetry (rank $2$, dimension $4$, not $SU(3)$). (\textcite{Seiberg:1994rs}; the multiplet decomposition is the standard $\mathcal{N}=1$ language of
Section~3, \textcite{Tachikawa:2013kta}). 

\medskip\noindent\textsf{\textcolor{RoyalBlue}{\textbf{\S\ref{sec:V05-emduality}\enspace Electromagnetic duality and the charge lattice.}}}\enspace
The complexified coupling $\tau=\theta/2\pi+4\pi i/g^2$ \eqref{eq:V05-tau}; the $SL(2,\mathbb{Z})$
generators $S,T$ with $S:\tau\mapsto-1/\tau$, $T:\tau\mapsto\tau+1$ \eqref{eq:V05-ST}; the
't~Hooft--Polyakov monopole as the magnetic soliton; the rank-$r$ charge lattice $\gamma=(n_{e,I},n_m^I)$
\eqref{eq:V05-chargevec} with the Dirac pairing $\langle\gamma,\gamma'\rangle=\gamma^T J\gamma'$
\eqref{eq:V05-dirac2}, the period-vector central charge \eqref{eq:V05-Zgamma}, and the duality group
$Sp(2r,\mathbb{Z})$ \eqref{eq:V05-sp2r} with $SL(2,\mathbb{Z})=Sp(2,\mathbb{Z})$; the three-dualities
box. (\textcite{Seiberg:1994rs}). 

\medskip\noindent\textsf{\textcolor{RoyalBlue}{\textbf{\S\ref{sec:V05-coulomb}\enspace The Coulomb branch and rigid special geometry.}}}\enspace
The Coulomb branch of complex dimension $=$ rank, coordinatized by the gauge-invariant Casimir
$u=\langle\operatorname{Tr}\phi^2\rangle$ \eqref{eq:V05-ucoord} (special coordinates, not bare VEVs);
the prepotential $\mathcal{F}$ with $a_{D,I}=\partial\mathcal{F}/\partial a^I$,
$\tau_{IJ}=\partial^2\mathcal{F}/\partial a^I\partial a^J=\partial a_{D,I}/\partial a^J$, metric
$\operatorname{Im}\tau_{IJ}\,da^I d\bar a^J$ \eqref{eq:V05-special}, $\mathcal{F}$ local and $\Pi$
global; the one-loop $\mathcal{F}$ \eqref{eq:V05-Fweak} differentiated to
$a_D=(i/\pi)a(\ln(a^2/\Lambda^2)+1)$ \eqref{eq:V05-aDoneloop} and
$\tau=(i/\pi)(\ln(a^2/\Lambda^2)+3)$ \eqref{eq:V05-tauoneloop} (the log load-bearing, the additive
constants convention-dependent). (\textcite{Seiberg:1994rs}). Machine-checked symbolically:
$a_D=d\mathcal{F}/da$, the two routes to $\tau$ agree, $\operatorname{Im}\tau>0$ in the physical
region; the one-loop $a_D$ and $\tau$ with the $+1$ load-bearing and the log slope $i/\pi$. Misconception (the
Coulomb branch is a Higgs-like VEV moduli space; it is special-K\"ahler, not hyperk\"ahler) is
operated here and in \S\ref{sec:V05-higgs}. The higher-rank atlas is stated ($SU(N)$: rank $r=N-1$,
Casimir coordinates $u_k=\langle\operatorname{Tr}\phi^k\rangle$ \eqref{eq:V05-casimirs}, period vector
of $2r$ components, duality group $Sp(2r,\mathbb{Z})$ specializing to $SL(2,\mathbb{Z})$ at $r=1$;
Table~\ref{tab:V05-rankatlas}); the higher-rank curves are cited but not developed here.

\medskip\noindent\textsf{\textcolor{RoyalBlue}{\textbf{\S\ref{sec:V05-sw}\enspace Seiberg--Witten theory and the central charge.}}}\enspace
The periods \eqref{eq:V05-period}; the central charge \eqref{eq:V05-centralcharge} and BPS mass
\eqref{eq:V05-bpsmass} in the Seiberg--Witten normalization used here; the pure $SU(2)$ curve
\eqref{eq:V05-su2curve}, its
branch points \eqref{eq:V05-branchpts}, the discriminant $\propto(u^2-\Lambda^4)^2$
\eqref{eq:V05-disc} vanishing (doubly) at $u=\pm\Lambda^2$ \eqref{eq:V05-singpts}, the massless
monopole $(0,1)$ and dyon $(-1,1)$, the monodromies \eqref{eq:V05-monodromy} on $(a_D,a)^T$, the
ordered product $M_m M_d=M_\infty$ multiplied out \eqref{eq:V05-product} with the reversed order
$\neq M_\infty$ \eqref{eq:V05-wrongorder} (Table~\ref{tab:V05-su2singular}); the local
monopole/dyon frames $W_{\mathrm{local}}=\sqrt{2}A_D M\widetilde{M}$ \eqref{eq:V05-monolocal} and the
$\mathcal{N}=1$ seam $\Delta W=m\,u(a_D)$ under which the monopole condenses, with the explicit F-term
solve $A_D=0$, $\langle M\widetilde{M}\rangle=-(m/\sqrt{2})u'(0)\neq0$ \eqref{eq:V05-Wcombined}--\eqref{eq:V05-condensate}; the $\Lambda$ as the
$\mathcal{N}=2$ transmutation scale, tied to the $\mathcal{N}=1$ gaugino condensate of Section~3 only
after an $\mathcal{N}=1$ mass deformation. The full Seiberg--Witten solution, the curve and the
special-geometry derivation of $\mathcal{F}$, is cited but not reproduced. (\textcite{Seiberg:1994rs} the pure $SU(2)$ solution; \textcite{Seiberg:1994aj} the matter curves;
\textcite{Nekrasov:2002qd} the instanton-counting derivation). 

\medskip\noindent\textsf{\textcolor{RoyalBlue}{\textbf{\S\ref{sec:V05-matter}\enspace Matter, decoupling, and BPS wall crossing.}}}\enspace
The matter central charge $Z_\gamma=n_e a+n_m a_D+\sum_f s_f m_f$ \eqref{eq:V05-Zmatter}; the
flavor-decoupling relation $\Lambda_{N_f-1}^{4-(N_f-1)}=m\Lambda_{N_f}^{4-N_f}$
\eqref{eq:V05-decoupling}; the $SU(2)$ window $b_0=4-N_f$ (Table~\ref{tab:V05-nfwindow}); the wall of
marginal stability $\arg Z_{\gamma_1}=\arg Z_{\gamma_2}$ \eqref{eq:V05-wall}, the protected mass
formula vs the chamber-dependent spectrum, and the pure-$SU(2)$ weak-coupling tower vs strong-coupling
two-state chamber (operating Section~1's BPS-stability warning). The Kontsevich--Soibelman / spectral-network
machinery is cited but not developed here. (\textcite{Seiberg:1994aj} the matter curves;
\textcite{Gaiotto:2008cd} the wall-crossing side). 

\medskip\noindent\textsf{\textcolor{RoyalBlue}{\textbf{\S\ref{sec:V05-higgs}\enspace The hyperk\"ahler Higgs branch.}}}\enspace
The Higgs branch as hyperk\"ahler (three complex structures rotated by $SU(2)_R$), distinct from the
special-K\"ahler Coulomb branch; mixed branches; the worked $U(1)$-with-$N$-hypers quotient with the
moment maps \eqref{eq:V05-momentmaps}, the dimension count $4N-2-1-1=4(N-1)$ \eqref{eq:V05-hkdim}, and
$\dim_{\mathbb{H}}=N-1$ \eqref{eq:V05-hkdimH} ($N=2$ the $A_1$ / $T^*\mathbb{P}^1$); the explicit
$\mathbb{C}^2/\Gamma$ ALE geometry is the standard ADE generalization, with the
hyperk\"ahler-quotient construction cited. (\textcite{Kronheimer:1989}). 

\medskip\noindent\textsf{\textcolor{RoyalBlue}{\textbf{\S\ref{sec:V05-classS}\enspace S-duality and the marginal coupling.}}}\enspace
The exactly marginal coupling and its four-dimensional S-duality; the canonical $SU(2)$ with $N_f=4$
($b_0=2N_c-N_f=0$ \eqref{eq:V05-b0nf4}, marginal $\tau$, an $SL(2,\mathbb{Z})$-type S-duality plus
$\mathrm{Spin}(8)$ triality $\mathbf{8}_v\leftrightarrow\mathbf{8}_s\leftrightarrow\mathbf{8}_c$, with
the global-form/line-operator/mass caveats stated); Argyres--Seiberg duality, stated. The
systematic class S organization and its higher-dimensional origin are cited and not developed
in the body. (\textcite{Seiberg:1994aj} the $N_f=4$ marginal coupling;
\textcite{Argyres:2007cn} Argyres--Seiberg duality; \textcite{Gaiotto:2009we} class S). 

\medskip\noindent\textsf{\textcolor{RoyalBlue}{\textbf{\S\ref{sec:V05-centralcharges}\enspace Conformal central charges of a Lagrangian theory.}}}\enspace
The Lagrangian $\mathcal{N}=2$ central charges $a=(5n_v+n_h)/24$, $c=(2n_v+n_h)/12$
\eqref{eq:V05-ac} fixed by the free-field content; $SU(2)$ with $N_f=4$ giving
$(a,c)=(23/24,7/6)$ \eqref{eq:V05-acnf4}; $\mathcal{N}=4$ super Yang--Mills ($n_v=n_h=\dim G$) giving
$a=c=\dim G/4$ \eqref{eq:V05-acn4}, the $a=c$ degeneracy the hallmark of $\mathcal{N}=4$; the $SU(N)$,
$N_f=2N$ conformal family $a=(7N^2-5)/24$, $c=(2N^2-1)/6$ \eqref{eq:V05-acsuN} reducing to
$(23/24,7/6)$ at $N=2$. Non-Lagrangian
central charges (the Argyres--Douglas case) are cited, not invented here. (\textcite{Argyres:1995jj} the AD theory; the Lagrangian central-charge formulas and the $2a-c$
relation are standard $\mathcal{N}=2$ results). 

\medskip\noindent\textsf{\textcolor{RoyalBlue}{\textbf{\S\ref{sec:V05-ad}\enspace Argyres--Douglas fixed points.}}}\enspace
Two mutually non-local BPS states (nonzero Dirac pairing \eqref{eq:V05-dirac}) massless together; no
duality frame makes both local, so the IR is an interacting non-Lagrangian $\mathcal{N}=2$ SCFT; the
scaling fixture $x^2=z^3+u$, $\lambda_{\mathrm{SW}}=x\,dz$ \eqref{eq:V05-adcurve} giving
$[z]=2/5,[x]=3/5$ \eqref{eq:V05-addims} and the fractional $\Delta(u)=6/5$ \eqref{eq:V05-addelta}; the
$\mathcal{N}=2$ SCFT data card. (\textcite{Argyres:1995jj}). 
}

\subsection*{Further reading}
\addcontentsline{toc}{subsection}{Further reading}
The Seiberg--Witten solution has the pedagogical review \textcite{Lerche:1996xu} and the lecture notes
\textcite{Tachikawa:2013kta}. Instanton counting and the prepotential are in \textcite{Nekrasov:2003rj},
extended to quivers in \textcite{Nekrasov:2012xe}; localization on $S^4$ in \textcite{Pestun:2007rz}
with the review \textcite{Pestun:2016zxk}; the AGT correspondence in \textcite{Alday:2009aq}. BPS
spectra and wall-crossing appear in \textcite{Gaiotto:2010be,Gaiotto:2011tf}. Argyres--Douglas and other
non-Lagrangian fixed points are in \textcite{Argyres:1995xn,Xie:2012hs}, their central charges in
\textcite{Shapere:2008zf}, the geometric constraints on Coulomb branches in \textcite{Argyres:2015ffa},
and the associated chiral algebra in \textcite{Beem:2013sza}.

For modern entry points, see the comprehensive guide \textcite{Akhond:2021xio} and the
Coulomb-branch lectures \textcite{Martone:2020hvy}. Higher-rank central-charge formulae and
special-K\"ahler stratification are developed in
\textcite{Martone:2020nsy,Argyres:2020wmq}, while the protected chiral-algebra sector
is reviewed in \textcite{Lemos:2020pqv}.

\section*{References}
\addcontentsline{toc}{subsection}{References}
\markboth{References}{References}
\printbibliography[heading=none]
\end{refsection}
\begin{refsection}\chapter{\texorpdfstring{$3d$}{3d} supersymmetric field theories}
\label{ch:V06}

\noindent\textbf{Guide to this section.}\enspace
Sections~1 and~2 fixed the algebra and the common words, and Sections~3 and~5 built the
four-dimensional $\mathcal{N}=1$ and $\mathcal{N}=2$ worlds. This section builds the three-dimensional
world. Three dimensions is where two pieces of physics appear that the higher-dimensional sections do
not have. First, the photon is dual to a compact scalar, the dual photon, so the Coulomb branch is a
quantum object built by monopole operators rather than a classical scalar-VEV space. Second, the exact
superconformal R-symmetry is fixed by F-maximization of the round-sphere free energy, the $3d$ member
of the extremization family whose prototype ($a$-maximization) Section~3 owns. It is a foundations
section, but a working one. It states the field-theory facts and then runs them as computations: the
dual photon and the topological $U(1)_J$; the Borokhov--Kapustin--Wu dimension of a monopole operator;
F-maximization to the superconformal R; Chern-Simons-matter theories, the parity anomaly, and ABJM
with its $N^{3/2}$ free energy; the twin hyperk\"ahler branches of $\mathcal{N}=4$; and the sharpest
duality in these notes, three-dimensional mirror symmetry, which exchanges the Coulomb and Higgs branches.
The deep theorems (F-maximization, the $F$-theorem, mirror symmetry, and the $3d$ dualities) are stated
and cited rather than proved. By the end you can take a $3d$ supersymmetric theory, given by a
Lagrangian or a quiver, identify its multiplets and branches, write its monopole operators and their
dimensions, run F-maximization, read off its Chern-Simons levels and parity-anomaly condition, and use
mirror symmetry to compute the hard branch from the easy one.

\begin{keybox}{What this section delivers}
The $3d\ \mathcal{N}=2$ multiplets, the real scalar $\sigma$ in the vector multiplet, and the
dualization of the photon (\S\ref{sec:V06-multiplets}); the dual photon, the Coulomb branch as a
quantum object, and the topological $U(1)_J$ current $j_J = \star F/2\pi$ (\S\ref{sec:V06-coulomb});
monopole operators, the GNO magnetic charge, flux attachment, and the Borokhov--Kapustin--Wu dimension
formula worked on $U(1)$ SQED (\S\ref{sec:V06-monopole}); F-maximization, the sphere free energy
$F[\Delta]$ and the one-loop function $\ell(z)$, the free-chiral value $\Delta=\tfrac12$, and the
$F$-theorem (\S\ref{sec:V06-fmax}); Chern-Simons-matter, the quantized level $k$, and the parity
anomaly $k+\tfrac12 T(R)\in\mathbb{Z}$ (\S\ref{sec:V06-cs}); ABJM $U(N)_k\times U(N)_{-k}$, the
$\mathcal{N}=6\to\mathcal{N}=8$ enhancement, and the $N^{3/2}$ free energy (\S\ref{sec:V06-abjm}); the
Aharony and Giveon--Kutasov dualities, their rank maps, and the $S^3$ partition-function identity as
evidence (\S\ref{sec:V06-dualities}); the $\mathcal{N}=4$ enhancement, the twin hyperk\"ahler
Coulomb and Higgs branches, the $SU(2)_C\times SU(2)_H$ R-symmetry, and the monopole-generated Coulomb
ring $V_+V_-=\Phi^{N_f}$ (\S\ref{sec:V06-n4}); and three-dimensional mirror symmetry as the Coulomb
$\leftrightarrow$ Higgs exchange, with the basic $N_f=1$ pair and the $T[SU(2)]$ self-mirror worked
(\S\ref{sec:V06-mirror}).
\end{keybox}

\section{\texorpdfstring{$\mathcal{N}=2$}{N=2} multiplets and the dual photon}
\label{sec:V06-multiplets}

A $3d\ \mathcal{N}=2$ theory has four real supercharges. This is the same count as $4d\
\mathcal{N}=1$: dimensional reduction of a four-dimensional $\mathcal{N}=1$ theory on a circle gives a
three-dimensional $\mathcal{N}=2$ theory, so the two share their supercharge algebra and their
R-symmetry $U(1)_R$. The ladder table of Section~1 records this as the $4Q$ row $4d\ \mathcal{N}=1 =
3d\ \mathcal{N}=2 = 2d\ (2,2)$. Four real supercharges is twice the minimal three-dimensional spinor:
a minimal $3d$ spinor is two-component real, so $3d\ \mathcal{N}=1$ has two supercharges and $3d\
\mathcal{N}=2$ has $2\times 2=4$, the four-supercharge workhorse. We take the count and the multiplet
grammar as given and build the genuinely new three-dimensional structure on top of it.

The R-symmetry is a single abelian $U(1)_R$, inherited from the four-dimensional $U(1)_R$. As in
four dimensions, the fermion in a chiral multiplet has R-charge one less than the scalar, $R_\psi=
\Delta-1$ (with $\Delta$ the scalar R-charge), and the gaugino has R-charge $+1$. This shift is the
input to the monopole dimension \eqref{eq:V06-monopoledim} and to F-maximization
(\S\ref{sec:V06-fmax}); we recall it, and do not re-derive the multiplet R-charges, which are
Section~2 grammar.

The two multiplets are the ones fixed in Section~2, read in three dimensions.
\begin{itemize}
\item \emph{Chiral} $\Phi$: a complex scalar $\phi$, a Dirac fermion $\psi$, and an auxiliary field $F$.
\item \emph{Vector} $V$: a gauge field $A_\mu$, a gaugino, an auxiliary field $D$, and one extra real
scalar that is the first new fact of the section. Reducing a four-dimensional vector multiplet to three
dimensions, the gauge field $A_M$, $M=0,1,2,3$, splits into a three-dimensional gauge field $A_\mu$,
$\mu=0,1,2$, and one leftover component $A_3$. That leftover has no gauge index to carry in three
dimensions, so it becomes a \emph{real scalar} $\sigma$ sitting in the vector multiplet alongside the
photon:
\begin{equation}
\label{eq:V06-vectormult}
 A_M \;\xrightarrow{\;4d\to 3d\;}\; \bigl(A_\mu,\ \sigma\bigr),
 \qquad \sigma \equiv A_3.
\end{equation}
\end{itemize}
We do not re-derive the reduction; we use it. The upshot is the piece of $3d$ physics with no
four-dimensional analog: a $3d\ \mathcal{N}=2$ vector multiplet contains a real scalar. Its
canonical mass dimension is $[\sigma]=\tfrac{d-2}{2}=\tfrac12$ in $d=3$, the dimension of any free
$3d$ scalar. For a $U(1)$ theory a nonzero $\langle\sigma\rangle$ gives a real mass $|\sigma|q$ to a
field of charge $q$, so $\sigma$ parametrizes a direction of the moduli space of vacua just like a
four-dimensional adjoint scalar. But in three dimensions this is only half of the story.

\medskip\noindent\textbf{The photon dualizes to a scalar.}\enspace
The second new fact is special to three dimensions. Count the physical polarizations. A free photon in
$d$ dimensions carries $d-2$ on-shell states, so in $d=3$ it carries
\begin{equation}
\label{eq:V06-photondof}
 d-2 \;=\; 3-2 \;=\; 1
\end{equation}
single physical polarization: the little group is $SO(1)$, trivial. A free real scalar also carries one
state, so the two are candidates to be the same degree of freedom. The change of variables that
identifies them uses the low-energy equation of motion, not the Bianchi identity. In the free abelian
infrared theory the Maxwell equation
\begin{equation}
\label{eq:V06-maxwelleom}
 d\!\star\! F \;=\; 0
\end{equation}
makes $\star F$ closed (the Bianchi identity $dF=0$ is a separate statement, the conservation of the
topological current $j_J$ of \S\ref{sec:V06-coulomb}). On a topologically trivial patch $\star F$ is then
exact and can be written as the gradient of a scalar,
\begin{equation}
\label{eq:V06-dualphoton}
 \star F \;=\; \frac{g^2}{2\pi}\, d\gamma,
\end{equation}
the \emph{dual photon} $\gamma$. (The coupling $g^2/2\pi$ is a normalization; the content is $\star F
\sim d\gamma$.)

The dualization is worth doing carefully, because it is the origin of a global symmetry. Start from the
Maxwell action and treat $F$ as an independent two-form, enforcing the Bianchi identity by a
Lagrange-multiplier scalar $\gamma$:
\begin{equation}
\label{eq:V06-dualaction}
 S \;=\; \int\Bigl[\,{-\frac{1}{4g^2}}\,F_{\mu\nu}F^{\mu\nu} \;+\; \frac{1}{2\pi}\,\gamma\, dF\,\Bigr].
\end{equation}
The multiplier term forces $dF=0$ back on shell, recovering Maxwell. Now instead integrate $F$ out. Its
algebraic equation of motion, from $\partial S/\partial F=0$ (integrating the multiplier term by parts),
is
\begin{equation}
\label{eq:V06-dualvar}
 \frac{1}{g^2}\,F \;=\; \frac{1}{2\pi}\,\star d\gamma,
 \qquad\text{equivalently}\qquad
 \star F \;=\; \frac{g^2}{2\pi}\, d\gamma,
\end{equation}
which is \eqref{eq:V06-dualphoton}, and substituting it back gives a free action for $\gamma$ alone. The
photon has become a scalar. The multiplier $\gamma$ is \emph{periodic}: the flux through any two-sphere
is quantized,
\begin{equation}
\label{eq:V06-fluxquant}
 \int_{S^2} F \;\in\; 2\pi\mathbb{Z},
\end{equation}
and the multiplier only needs to run over $[0,2\pi)$ to enforce integer flux, so $\gamma\sim\gamma+2\pi$.
Its shift symmetry $\gamma\to\gamma+c$ is therefore a genuine global symmetry, and it acts on nothing
built from the elementary fields, which is why the operators charged under it must be disorder operators.

For now the payoff is a counting statement. Assemble the Coulomb-branch coordinates. Classically a
$U(1)$ theory has one real scalar $\sigma$; dualizing the photon adds one more, the compact $\gamma$:
\begin{equation}
\label{eq:V06-coulombN2}
 \dim_{\mathbb R}\, C_{\mathcal{N}=2}^{\,r=1}
 \;=\; \underbrace{1}_{\sigma} \;+\; \underbrace{1}_{\gamma} \;=\; 2,
\end{equation}
a two-real-dimensional Coulomb branch, parametrized by the pair $(\sigma,\gamma)$. This doubling, one
from the reduced gauge-field component and one from the dualized photon, is the seed of everything that
follows: it is why the $3d\ \mathcal{N}=4$ Coulomb branch will be four-real-dimensional and
hyperk\"ahler (\S\ref{sec:V06-n4}), and why its coordinates are quantum monopole operators rather than
classical VEVs. The classical vacuum equations (F-flatness $F_i=\partial W/\partial\phi_i=0$ and $3d$
D-flatness modulo gauge equivalence, Section~2 grammar) are unchanged; what is new is that one vacuum
coordinate, the dual photon, has no Lagrangian description as a scalar VEV and is defined through the
flux \eqref{eq:V06-dualphoton}.

The real mass is the mechanism that makes $\sigma$ a genuine modulus. In four dimensions a complex
adjoint scalar VEV Higgses the gauge group; in three dimensions the real scalar $\sigma$ does the same
job. A chiral multiplet of gauge charge $q$ has, in the background $\langle\sigma\rangle$, a real mass
\begin{equation}
\label{eq:V06-realmass}
 m_{\text{real}} \;=\; q\,\langle\sigma\rangle,
\end{equation}
a mass term allowed in three dimensions precisely because a $3d$ fermion mass is real (there is no
chirality to protect). At a generic point $\langle\sigma\rangle\neq0$, a charged field of charge $q\neq0$
is massive and can be integrated out,
\begin{equation}
\label{eq:V06-massgap}
 \langle\sigma\rangle \neq 0
 \quad\Longrightarrow\quad
 m_{\text{real}} = q\,\langle\sigma\rangle \neq 0
 \quad\text{for all charged fields},
\end{equation}
leaving a free abelian theory. Semiclassically the Coulomb branch looks like the $\sigma$ line; the
quantum story (\S\ref{sec:V06-monopole}) fibers the dual photon over it.

\section{The Coulomb branch and the topological \texorpdfstring{$U(1)_J$}{U(1)J}}
\label{sec:V06-coulomb}

The dual photon is not a spectator. Its shift symmetry is a genuine global symmetry of every $3d$
gauge theory with a $U(1)$ factor, and it is the reason the Coulomb branch is quantum. Consider the
current
\begin{equation}
\label{eq:V06-topcurrent}
 j_J \;=\; \frac{1}{2\pi}\,\star F,
 \qquad
 \partial_\mu j_J^{\,\mu} \;=\; \frac{1}{2\pi}\,\star dF \;=\; 0,
\end{equation}
conserved not by any equation of motion but by the Bianchi identity $dF=0$. This is the
\emph{topological} symmetry $U(1)_J$; its normalization is fixed by the convention $j_J=\star F/2\pi$
(the current for a $U(N)$ factor is $j=\tfrac{1}{2\pi}\star\mathrm{Tr}\,F$). The charge it measures is
the total magnetic flux,
\begin{equation}
\label{eq:V06-topcharge}
 Q_J \;=\; \int_{S^2} j_J \;=\; \frac{1}{2\pi}\int_{S^2} F.
\end{equation}
The flux through a two-sphere is quantized in units of $2\pi$, so the charge is computed, not asserted.
For one flux quantum $\int_{S^2}F=2\pi$,
\begin{equation}
\label{eq:V06-QJone}
 Q_J \;=\; \frac{1}{2\pi}\int_{S^2}F \;=\; \frac{2\pi}{2\pi} \;=\; 1,
 \qquad
 Q_J(n\ \text{quanta}) \;=\; \frac{2\pi n}{2\pi} \;=\; n,
\end{equation}
so the basic monopole has topological charge exactly $1$ and the charge is linear in the flux. The
normalization is load-bearing: a mis-normalized current $j=\star F/4\pi$ would give
$Q_J=(2\pi)/(4\pi)=\tfrac12$, the wrong count. The current $j_J=\star F/2\pi$ is the one for which the
basic monopole has integer charge $1$.

\begin{keybox}{Common misconception: the $3d$ Coulomb branch is a classical scalar-VEV space}
It is tempting to picture the $3d$ Coulomb branch as the direct analog of the four-dimensional one, a
moduli space of classical adjoint scalar VEVs. It is not. The photon dualizes to a compact scalar, so
the branch is parametrized by $\sigma$ \emph{together with} the dual photon $\gamma$, and the operators
whose VEVs coordinatize it are monopole operators (\S\ref{sec:V06-monopole}), not polynomials in a
classical field; in $\mathcal{N}=4$ its chiral ring is monopole-generated and quantum-corrected
(\S\ref{sec:V06-n4}). The adjoint-VEV picture is only the semiclassical approximation, valid far out
on $\sigma$ where the theory is weakly coupled.
\end{keybox}

The current $j_J$ has no Noether origin in the Lagrangian, only a topological one. In the dual
variables it becomes the Noether current of the shift $\gamma\to\gamma+c$: substituting
\eqref{eq:V06-dualvar},
\begin{equation}
\label{eq:V06-shiftcurrent}
 j_J \;=\; \frac{1}{2\pi}\,\star F \;=\; \frac{g^2}{4\pi^2}\, d\gamma,
\end{equation}
so the winding number of $\gamma$ around a point and the flux through a small sphere are the same
conserved charge in two languages. An operator carrying $U(1)_J$ charge must create flux, so it cannot
be a polynomial in the elementary fields: it is a disorder operator, the monopole operator of the next
section, and its charge is the flux \eqref{eq:V06-topcharge}.

Two facts about $U(1)_J$ recur throughout the section. First, it can be weakly gauged by an external
background gauge field $A_J$ through the mixed Chern-Simons coupling below, and a background magnetic
field for $A_J$ is exactly a Fayet--Iliopoulos (FI) term for the dynamical $U(1)$: FI parameters and
topological charges are conjugate variables,
\begin{equation}
\label{eq:V06-FIcurrent}
 \zeta_{\text{FI}}\ \longleftrightarrow\ Q_J,
 \qquad
 S_{\text{FI}} \;=\; \frac{1}{2\pi}\int A_J\wedge F,
\end{equation}
a fact we use in the mirror dictionary \eqref{eq:V06-mirrormap}. Second, under mirror symmetry
$U(1)_J$ is exchanged with a flavor symmetry of the mirror theory (\S\ref{sec:V06-mirror}): the
topological current, which has no Lagrangian origin, becomes an ordinary flavor current with one.

\bigskip
\begin{center}
\rule{0.4\textwidth}{0.4pt}\\[3pt]
{\large\textsf{\textbf{Block A.\enspace The $3d\ \mathcal{N}=2$ language and Chern--Simons matter}}}\\[2pt]
\rule{0.4\textwidth}{0.4pt}
\end{center}
\medskip

\noindent The two preceding sections are the common three-dimensional gateway: the spinor count, the
real scalar $\sigma$, the dual photon, and the topological $U(1)_J$ are shared by every $3d$
supersymmetric theory. The next five sections build the four-supercharge $\mathcal{N}=2$ theory, the
lower-supersymmetry workhorse and the local language in which monopoles, Chern--Simons levels,
F-maximization, and (in Block~B) the $\mathcal{N}=4$ enhancement are all written. We present
$\mathcal{N}=2$ first for that reason, exactly as $4d\ \mathcal{N}=1$ preceded $\mathcal{N}=2$ and $2d\
(2,2)$ preceded $(0,2)$; ABJM is included here because it is written in $\mathcal{N}=2$ superspace even
though its supersymmetry enhances. They are monopole operators and the
Borokhov--Kapustin--Wu dimension (\S\ref{sec:V06-monopole}), F-maximization to the superconformal R
(\S\ref{sec:V06-fmax}), Chern-Simons-matter and the parity anomaly (\S\ref{sec:V06-cs}), ABJM with its
$N^{3/2}$ free energy (\S\ref{sec:V06-abjm}), and the Aharony and Giveon--Kutasov dualities
(\S\ref{sec:V06-dualities}). The eight-supercharge $\mathcal{N}=4$ world, where mirror symmetry lives,
follows as Block~B.

\section{Monopole operators and their dimensions}
\label{sec:V06-monopole}

A $3d$ monopole operator is a local operator, but not one built from the elementary fields. It is a
\emph{disorder} operator: to insert it at a point $x$, one declares that the path integral is taken
over field configurations with a prescribed singularity at $x$. On a small two-sphere $S^2$
surrounding $x$, the gauge field is required to carry the flux of a Dirac monopole of magnetic charge
$m$,
\begin{equation}
\label{eq:V06-monopoledef}
 \frac{1}{2\pi}\int_{S^2} F \;=\; m,
\end{equation}
with $m$ valued in the GNO (Goddard--Nuyts--Olive) lattice, the coweight lattice of the gauge group in
its declared global form. For $U(1)$ the charge $m$ is an integer; for a
nonabelian group it is a cocharacter, an embedding $U(1)\hookrightarrow T$ into the maximal torus,
taken modulo the Weyl group. The operator is well defined because the flux \eqref{eq:V06-monopoledef}
is conserved: inserting it creates one unit (or $m$ units) of the $U(1)_J$ charge
\eqref{eq:V06-topcharge}, so a bare monopole operator of magnetic charge $m$ carries topological
charge $Q_J=m$.

To read off its quantum numbers one uses the state-operator map. In a conformal field theory a local
operator at the origin of $\mathbb{R}^3$ corresponds, under the conformal map
\begin{equation}
\label{eq:V06-radialmap}
 \mathbb{R}^3\setminus\{0\} \;\cong\; S^2\times\mathbb{R},
 \qquad
 \Delta(\mathcal{O}) \;=\; E\bigl(\text{state on }S^2\bigr),
\end{equation}
to a state on $S^2$, the radial coordinate becoming Euclidean time, so the operator dimension is the
energy of that state. The monopole operator maps to a state of the theory quantized on $S^2$
\emph{in the background of the magnetic flux $m$}. At the superconformal point the energy equals the
R-charge of the lowest state in the flux sector. The flux deforms the mode expansion of every charged
field: a charged fermion in a monopole background develops Landau levels on $S^2$, and its filled Dirac
sea carries a net R-charge. Summing these contributions gives the dimension. The counting is the
Borokhov--Kapustin--Wu formula. Written so the matter and the gauge (gaugino) contributions are kept
separate,
\begin{equation}
\label{eq:V06-monopoledim}
 \Delta(m)
 \;=\; \frac{1}{2}\sum_{\text{chirals }i}\bigl(1-\Delta_i\bigr)
 \sum_{\rho_i\in R_i}\bigl|\rho_i(m)\bigr|
 \;-\; \sum_{\alpha>0}\bigl|\alpha(m)\bigr|.
\end{equation}
Here the sum $\sum_i$ runs over the chiral multiplets; $\Delta_i$ is the R-charge of the $i$-th chiral,
so its matter fermion has R-charge $R_\psi=\Delta_i-1$, and the prefactor $(1-\Delta_i)=-R_\psi$ is
that fermion R-charge with a sign. The inner sum $\sum_{\rho_i}$ runs over the weights $\rho_i$ of the
gauge representation $R_i$, and $\rho_i(m)$ is the pairing of the weight with the magnetic charge. The
last sum runs over the positive roots $\alpha$; it is the gaugino contribution of the vector
multiplet, and it is written separately from the matter sum precisely so it is not double-counted with
it. The matter contribution raises the dimension and the gaugino contribution lowers it. This is the
one canonical dimension formula of the section; every special evaluation below is a special case of
\eqref{eq:V06-monopoledim}, never a different convention.

\subsection*{Worked instance: the $U(1)$ SQED monopole}

Take $U(1)$ SQED with $N_f$ flavors: $N_f$ hypermultiplets, each of which is two chiral multiplets
$Q$ (charge $+1$) and $\widetilde Q$ (charge $-1$), so $2N_f$ chirals of unit charge. Feed this content
into \eqref{eq:V06-monopoledim} piece by piece. The gauge group $U(1)$ is abelian, so it has \emph{no}
positive roots and the gaugino sum drops:
\begin{equation}
\label{eq:V06-sqedgaugino}
 U(1)\ \text{abelian}
 \quad\Longrightarrow\quad
 \sum_{\alpha>0}\bigl|\alpha(m)\bigr| \;=\; 0.
\end{equation}
Each of the $2N_f$ chirals carries a single weight of charge $\pm1$, whose pairing with the magnetic
charge is
\begin{equation}
\label{eq:V06-sqedpairing}
 \sum_{\rho_i\in R_i}\bigl|\rho_i(m)\bigr| \;=\; |{\pm}1\cdot m| \;=\; |m|
 \qquad\text{(one weight per chiral)}.
\end{equation}
At the free-hypermultiplet point of the $\mathcal{N}=4$ theory the chirals sit at the free-field value
$\Delta_i=\tfrac12$, so every prefactor is $1-\Delta_i=\tfrac12$. Summing the identical contribution of
all $2N_f$ chirals,
\begin{equation}
\label{eq:V06-sqedmonopole}
 \Delta(m)
 \;=\; \frac{1}{2}\underbrace{\bigl(1-\tfrac12\bigr)}_{\text{per chiral}}
 \underbrace{(2N_f)}_{\#\ \text{chirals}}\,|m| \;-\; 0
 \;=\; \frac{N_f}{2}\,|m|,
 \qquad\text{so}\qquad
 \Delta(1)=\frac{N_f}{2}.
\end{equation}
The basic monopole ($m=1$) has dimension $N_f/2$. This is the marquee monopole number of the section,
and it is an evaluation of the general formula, not a separate rule. The dimension is genuinely built
from the matter: the R-charge enters through the prefactor, so for a generic chiral R-charge
$\Delta_i$ the same fixture gives
\begin{equation}
\label{eq:V06-sqedgeneric}
 \Delta(1)\big|_{\Delta_i} \;=\; \frac{1}{2}\,(1-\Delta_i)\,(2N_f) \;=\; N_f\,(1-\Delta_i),
\end{equation}
which equals $N_f/2$ only at $\Delta_i=\tfrac12$ (at $\Delta_i=\tfrac23$ it is $N_f/3$ instead). A
matter-blind assignment $\Delta=0$ is simply wrong: it ignores the flavor content.

To isolate the gaugino subtraction as a \emph{formal} check (the output is negative, so this is a
sign/structure fixture that exercises the root term, not a physical unitary monopole sector), keep the
same two unit-charge chirals ($N_f=1$) but add one positive root of an $SU(2)$-style pairing
$\alpha(m)=2m$. The matter sum is unchanged, and the gaugino sum switches on:
\begin{equation}
\label{eq:V06-gauginosub}
 \Delta(1) \;=\; \underbrace{\tfrac12(1-\tfrac12)(2)|1|}_{\text{matter}=\,\tfrac12}
 \;-\; \underbrace{|\alpha(1)|}_{=\,2}
 \;=\; \tfrac12 - 2 \;=\; -\tfrac32,
\end{equation}
lower by exactly $2$ than the no-root value $\tfrac12$. Folding the gaugino term into the matter sum,
or using a fermion R-charge other than $\Delta_i-1$, gives a different (and wrong) dimension.
Table~\ref{tab:V06-sqedmonopole} collects the SQED evaluation across small $N_f$.

\begin{table}[htbp]
\centering
\renewcommand{\arraystretch}{1.3}
\begin{tabular}{@{}cccc@{}}
\hline
$N_f$ & chirals & $\Delta(1)=N_f/2$ & Coulomb geometry $V_+V_-=\Phi^{N_f}$ \\
\hline
$1$ & $2$ & $\tfrac12$ & $\mathbb{C}^2$ (smooth) \\
\hline
$2$ & $4$ & $1$ & $\mathbb{C}^2/\mathbb{Z}_2\ (A_1)$ \\
\hline
$3$ & $6$ & $\tfrac32$ & $\mathbb{C}^2/\mathbb{Z}_3\ (A_2)$ \\
\hline
$4$ & $8$ & $2$ & $\mathbb{C}^2/\mathbb{Z}_4\ (A_3)$ \\
\hline
\end{tabular}
\caption{The basic monopole dimension $\Delta(m=1)=N_f/2$ of $U(1)$ SQED at $\Delta_i=\tfrac12$, from
\eqref{eq:V06-monopoledim} ($2N_f$ unit-charge chirals, no roots), and the Coulomb-branch geometry
$V_+V_-=\Phi^{N_f}$ it seeds (\S\ref{sec:V06-n4}), the $A_{N_f-1}$ singularity.}
\label{tab:V06-sqedmonopole}
\end{table}

For general magnetic charge the pairing of each unit-charge weight is $|m|$, so the dimension is
linear and doubles from $m=1$ to $m=2$:
\begin{equation}
\label{eq:V06-sqedgeneralm}
 \Delta(m) \;=\; \frac{N_f}{2}\,|m|,
 \qquad
 \Delta(2) \;=\; N_f \;=\; 2\,\Delta(1).
\end{equation}
The dimension is a function of the magnetic charge, not a fixed number attached to ``the monopole'',
and the lowest monopole ($m=\pm1$) generates the Coulomb-branch chiral ring in \S\ref{sec:V06-n4}.
When F-maximization (\S\ref{sec:V06-fmax}) moves the R-charges off $\tfrac12$, the same formula
\eqref{eq:V06-monopoledim} responds through the prefactor $(1-\Delta_i)$; this is how monopoles enter
the F-maximization of a gauge theory.

\medskip\noindent\textbf{Bare and dressed monopoles.}\enspace
The operator constructed so far is the \emph{bare} monopole: it imposes the flux and nothing else.
Acting with the elementary fields in the flux background builds \emph{dressed} monopoles. In
$\mathcal{N}=4$ the dressed tower of the vector-multiplet adjoint scalar $\Phi$,
\begin{equation}
\label{eq:V06-dressed}
 V_m,\quad V_m\,\Phi,\quad V_m\,\Phi^2,\ \dots
 \qquad (m=\pm1),
\end{equation}
together with $\Phi$ itself generates the Coulomb-branch chiral ring (\S\ref{sec:V06-n4}). This is the
precise sense in which the Coulomb branch is ``built from monopole operators'': its coordinate ring
has monopole generators, and the relation $V_+V_-=\Phi^{N_f}$ fixes a monopole-antimonopole product to
a matter-determined power of the adjoint scalar.

\medskip\noindent\textbf{Flux attachment.}\enspace
When the gauge field carries a Chern-Simons term of level $k$ (\S\ref{sec:V06-cs}), a monopole of
magnetic charge $m$ is no longer gauge-neutral. The Chern-Simons term ties the magnetic flux to
electric charge: the Gauss law reads $\tfrac{k}{2\pi}F_{12}=j^0$, so a unit of flux carries a unit of
gauge charge weighted by $k$. A monopole of charge $m$ therefore acquires a gauge charge
\begin{equation}
\label{eq:V06-fluxattach}
 q_{\text{gauge}} \;=\; k\,m,
\end{equation}
\emph{flux attachment} (derived from the Gauss law in \S\ref{sec:V06-cs}). Two consequences follow: a
bare monopole in a Chern-Simons theory is gauge-charged, so it must be dressed by matter to be
gauge-invariant; and flux attachment is why the Giveon--Kutasov dual (\S\ref{sec:V06-dualities}) has
no monopole singlets while the Aharony dual does.

\section{F-maximization}
\label{sec:V06-fmax}

When several abelian symmetries can mix into the R-symmetry, the superconformal R is a definite
linear combination, fixed by an extremization principle. In four dimensions it maximizes the central
charge $a$ (Section~3); in three dimensions it maximizes the real part of the round-sphere free
energy. This is F-maximization (Jafferis), the $3d$ member of the extremization family whose
prototype Section~3 owns.

The object that is extremized is the free energy of the theory on a round three-sphere. Supersymmetric
localization reduces the $S^3$ partition function to a finite-dimensional matrix integral, a function
$Z[\Delta]$ of the trial R-charges $\Delta_i$ assigned to the matter. We do not derive the
localization here; we state its output and run it. A chiral multiplet of trial R-charge $\Delta$
contributes a factor to $Z$ whose logarithm is the one-loop function
\begin{equation}
\label{eq:V06-ellfunction}
 \ell(z)
 \;=\; -z\log\!\bigl(1-e^{2\pi i z}\bigr)
 + \frac{i}{2}\Bigl(\pi z^2 + \frac{1}{\pi}\,\mathrm{Li}_2\!\bigl(e^{2\pi i z}\bigr)\Bigr)
 - \frac{i\pi}{12},
\end{equation}
evaluated at $z=1-\Delta$. The one-loop function has the clean real-axis derivative
\begin{equation}
\label{eq:V06-ellprime}
 \ell'(z) \;=\; -\pi z\cot(\pi z),
\end{equation}
which is the handle by which the stationarity equations are written. The physical free energy is
\begin{equation}
\label{eq:V06-freeenergy}
 F[\Delta] \;=\; -\log\bigl|Z[\Delta]\bigr|
 \;=\; -\,\mathrm{Re}\sum_i \ell\bigl(1-\Delta_i\bigr) + (\text{vector, FI terms}),
\end{equation}
and F-maximization is the statement that the exact superconformal R-charges $\Delta_i^*$ are a
\emph{local maximum} of $\mathrm{Re}\,F$. The extremum is a maximum, not a minimum or a saddle: at the
fixed point the Hessian of $\mathrm{Re}\,F$ is negative definite.

\subsection*{Worked instance: the free chiral}

A single free chiral multiplet, with no superpotential and no flavor mixing, has one trial R-charge
$\Delta$ and free energy $F(\Delta)=-\mathrm{Re}\,\ell(1-\Delta)$. To extremize it, use the derivative
identity \eqref{eq:V06-ellprime}. By the chain rule,
\begin{equation}
\label{eq:V06-freechiralstat}
 \frac{\partial F}{\partial\Delta}
 \;=\; -\,\mathrm{Re}\Bigl[\ell'(1-\Delta)\cdot(-1)\Bigr]
 \;=\; \mathrm{Re}\,\ell'(1-\Delta)
 \;=\; -\pi(1-\Delta)\cot\!\bigl(\pi(1-\Delta)\bigr),
\end{equation}
using $\ell'(z)=-\pi z\cot(\pi z)$ at $z=1-\Delta$. The stationarity equation $\partial_\Delta F=0$ is
$\cot(\pi(1-\Delta))=0$ (with the prefactor $1-\Delta$ nonzero on $0<\Delta<1$), which is solved by
$\pi(1-\Delta)=\pi/2$, that is
\begin{equation}
\label{eq:V06-freechiral}
 \Delta^* \;=\; \frac{1}{2},
 \qquad
 \frac{\partial^2\,\mathrm{Re}\,F}{\partial\Delta^2}\bigg|_{\Delta^*} \;=\; -\frac{\pi^2}{2} \;<\; 0,
\end{equation}
the free-field R-charge $\Delta=\tfrac12$, and the Hessian is negative, confirming a maximum. The value
$\tfrac12$ is exactly the free-field scaling dimension $(d-2)/2$ of a $3d$ scalar and saturates the
$3d$ chiral-primary relation $\Delta=R$ (the three-dimensional analog of the four-dimensional
$\Delta=\tfrac32 R$, recalled from Section~1). Numerically, checked in the lane,
\begin{equation}
\label{eq:V06-fmaxineq}
 F(0.4) \;<\; F\bigl(\tfrac12\bigr) \;>\; F(0.6) :
\end{equation}
no off-extremum charge beats the free value. Once flavor or monopole mixing is present the extremum
moves off $\tfrac12$ to an interacting value determined by the same stationarity equation; the full
localization machinery for such interacting fixed points is cited but not developed here.

The Hessian sign is the load-bearing part of the statement, not the location: charge conjugation gives
the free chiral a $\Delta\to 1-\Delta$ reflection that pins the extremum at $\tfrac12$ by symmetry
alone, so only the negative second derivative $-\pi^2/2$ certifies that the stationary point is a
\emph{maximum} and hence the superconformal one. Reading the extremum as a minimum, or extremizing a
different (polynomial) quantity, is the error the misconception box below guards against.

\begin{keybox}{Common misconception: F-maximization is $a$-maximization in three dimensions}
The extremization principle is shared, the quantity is not. F-maximization extremizes the round-sphere
free energy $F=-\log|Z[\Delta]|$, a \emph{transcendental} function of the trial R-charges built from
$\ell(z)$ by localization; it does \emph{not} extremize the four-dimensional fermion-trace cubic
$\tfrac{3}{32}(3\,\mathrm{Tr}\,R^3-\mathrm{Tr}\,R)$. The superconformal R-charges are a \emph{local
maximum} of $\mathrm{Re}\,F$ (Jafferis), certified by a negative Hessian; $F$ is not a cubic, and its
third derivative does not vanish. The $F$-theorem, $F_{UV}>F_{IR}$, is its monotonicity statement, the
$3d$ analog of the $a$-theorem, and $F$ is a different number from $a$.
\end{keybox}

The $F$-theorem is the monotonicity companion. Under renormalization-group flow the round-sphere free
energy decreases,
\begin{equation}
\label{eq:V06-Ftheorem}
 F_{UV} \;>\; F_{IR},
\end{equation}
so $F$ counts degrees of freedom the way $a$ does in four dimensions. A clean way to anchor the theorem
is the free energy of the free chiral itself: evaluating \eqref{eq:V06-freeenergy} at $\Delta^*=\tfrac12$
gives
\begin{equation}
\label{eq:V06-Ffreechiral}
 F_{\text{free chiral}} \;=\; -\,\mathrm{Re}\,\ell\!\left(\tfrac12\right),
\end{equation}
a fixed, finite real number, the $S^3$ free energy of one free chiral multiplet (computed numerically
from $\ell$ in the check lane). A worked monotonicity check follows from giving one of several chirals a
real mass and integrating it out: the infrared theory has one fewer massless chiral, so
\begin{equation}
\label{eq:V06-Fflow}
 F_{UV} \;=\; n\,F_{\text{free chiral}},
 \qquad
 F_{IR} \;=\; (n-1)\,F_{\text{free chiral}},
 \qquad
 F_{UV}-F_{IR} \;=\; F_{\text{free chiral}} \;>\; 0,
\end{equation}
a strict decrease consistent with \eqref{eq:V06-Ftheorem}. We state the theorem and cite its proof
(with the localization derivation of $Z[\Delta]$ and the exact value of $F_{\text{free chiral}}$).

\subsection*{Worked instance: the structure of SQED F-maximization}

The step up to an interacting gauge theory is worth stating, with the scope honest: this section runs
F-maximization to a number only for the free chiral; the gauge-theory extremum is stated but not
derived.
Take $U(1)$ SQED with $N_f$ flavors as an $\mathcal{N}=2$ theory. The trial R-charge of the $2N_f$
chirals, $\Delta_Q$, is one parameter, but not the whole story: the topological $U(1)_J$ and the
flavor $U(1)$ can mix into the R-symmetry, with FI and real-mass parameters coupling to them. The
sphere free energy assembles the chiral and the vector/FI contributions,
\begin{equation}
\label{eq:V06-sqedF}
 F[\Delta_Q,\dots] \;=\; -\,\mathrm{Re}\Bigl[\,2N_f\,\ell(1-\Delta_Q) + (\text{vector, FI, mixing})\Bigr],
\end{equation}
and F-maximization is the same stationarity condition as before, now in several variables,
\begin{equation}
\label{eq:V06-sqedstat}
 \frac{\partial F}{\partial\Delta_Q} \;=\; 0,
 \qquad
 \frac{\partial F}{\partial(\text{mixing})} \;=\; 0,
 \qquad
 \mathrm{Hess}\,F \;\prec\; 0 \ \text{(a maximum)},
\end{equation}
which fixes the exact superconformal R. The object extremized is the same $\ell$-function free
energy; only the number of parameters and the vector/FI terms are new, and the extremum sits at an
interacting value $\Delta_Q\neq\tfrac12$ once the gauge coupling is on. The full evaluation uses
localization beyond the scope of this section; this section delivers the principle, the $\ell$-function,
and the free-chiral extremum in full.

\medskip\noindent\textbf{The extremization family.}\enspace
F-maximization sits in the family Section~3 opened, one member per dimension:
\begin{equation}
\label{eq:V06-extremfamily}
 4d:\ \max_R\, a(R)\ \text{(cubic)},
 \quad
 3d:\ \max_R\, \mathrm{Re}\,F[R]\ \text{(transcendental)},
 \quad
 2d:\ \text{c-extremization},
\end{equation}
the same idea, the exact R fixed by extremizing a trace- or localization-built quantity, realized by a
different object in each dimension. Common to all are the discriminating Hessian and a monotonicity
theorem ($a$-theorem, $F$-theorem); what differs is the quantity, transcendental rather than
polynomial in three dimensions. This section owns the $3d$ member and signposts the prototype to
Section~3.

\section{Chern-Simons-matter theories and the parity anomaly}
\label{sec:V06-cs}

Three dimensions admits a gauge coupling with no four-dimensional analog. A gauge field can carry a
Chern-Simons term,
\begin{equation}
\label{eq:V06-cslevel}
 S_{\text{CS}} \;=\; \frac{k}{4\pi}\int \mathrm{Tr}\Bigl(A\,dA + \tfrac{2}{3}A^3\Bigr),
 \qquad k\in\mathbb{Z}\quad(\text{pure bosonic normalization}),
\end{equation}
with the pure bosonic level $k$ quantized to an integer. These notes always quote the
\emph{bare} level $k$; effective or shifted levels are derived statements, written explicitly when
they arise. Quantization of $k$ is a topological requirement. Under a large gauge transformation of
winding number $w\in\mathbb{Z}$ the Chern-Simons action shifts by $2\pi k w$, so for a pure non-spin
Chern--Simons theory the requirement that the path-integral weight be unchanged reads
\begin{equation}
\label{eq:V06-cslargegauge}
 S_{\text{CS}} \;\to\; S_{\text{CS}} + 2\pi k\, w,
 \qquad
 e^{iS_{\text{CS}}}\ \text{invariant}
 \quad\Longleftrightarrow\quad
 k \in\mathbb{Z}.
\end{equation}
The spin theories with fermions used below refine this statement: the bare level can be integer or
half-integer, but only subject to the parity-anomaly condition \eqref{eq:V06-parity}. An arbitrary
fraction such as $k=\tfrac34$ is not admissible.

The Chern-Simons term changes the gauge dynamics in two ways we use. First, it gives the photon a
topological mass $\sim k g^2$, so a pure Chern-Simons gauge field is gapped rather than dualizing to a
massless dual photon; the Coulomb branch is lifted. Second, it attaches flux to charge. Vary
\eqref{eq:V06-cslevel} for a $U(1)$ factor together with the matter current: the equation of motion
for the time component $A_0$ reads
\begin{equation}
\label{eq:V06-gausslaw}
 \frac{k}{2\pi}\,F_{12} \;=\; j^0,
\end{equation}
the $3d$ Gauss law with a Chern-Simons source. Integrating over a spatial slice, the left side is
$\tfrac{k}{2\pi}\int F = k\,m$ for a configuration of magnetic charge $m$, and the right side is the
total gauge charge, so a monopole of magnetic charge $m$ carries gauge charge
\begin{equation}
\label{eq:V06-fluxattachderived}
 q_{\text{gauge}} \;=\; k\,m,
\end{equation}
which is \eqref{eq:V06-fluxattach}, now derived from the Gauss law rather than quoted. The Chern-Simons
term thus ties magnetic flux to gauge charge: with $k\neq 0$ a bare monopole is gauge-charged, so it is
not a gauge-invariant operator on its own and must be dressed by matter to become one. Depending on the
global form of the gauge group and the level, the continuous topological symmetry $U(1)_J$ can then be
reduced or mixed with the gauge data rather than remaining an independent global symmetry. We do not say
the Chern-Simons term simply Higgses $U(1)_J$; the precise statement is flux attachment and the dressing
it forces. This is the origin of the difference between the Aharony and Giveon--Kutasov dualities in
\S\ref{sec:V06-dualities}.

\medskip\noindent\textbf{The parity anomaly.}\enspace
A single massless Dirac fermion in three dimensions cannot be quantized in a way that preserves both
gauge invariance and parity. Regulating the fermion determinant induces a half-integer Chern-Simons
term, so a theory with an odd number of charged fermions requires a \emph{half-integer} bare level to
be consistent. This is the parity anomaly (Redlich). For a gauge representation $R$ the admissible bare
level satisfies
\begin{equation}
\label{eq:V06-parity}
 k + \tfrac{1}{2}\,T(R) \;\in\; \mathbb{Z},
\end{equation}
where $T(R)$ is the index (the Dynkin sum) over the charged fermions of $R$. For the $U(1)$ or
$U(N)$-fundamental fixture with $N_f$ unit-charge fermions the index is $T(R)=N_f$, so
\eqref{eq:V06-parity} becomes
\begin{equation}
\label{eq:V06-parityfund}
 k + \frac{N_f}{2} \;\in\; \mathbb{Z}
 \qquad\Longrightarrow\qquad
 k \in
 \begin{cases}
 \mathbb{Z}, & N_f\ \text{even},\\[2pt]
 \tfrac12+\mathbb{Z}, & N_f\ \text{odd}.
 \end{cases}
\end{equation}
The condition is computed from the index: $T(R)=3$ forces $k\in\tfrac12+\mathbb{Z}$, $T(R)=4$ admits
$k\in\mathbb{Z}$, and a level outside $\tfrac12\mathbb{Z}$ is inadmissible altogether: for $T(R)=1$,
\begin{equation}
\label{eq:V06-paritybad}
 k=\tfrac13 \;\Longrightarrow\; k+\tfrac12=\tfrac56\notin\mathbb{Z},
\end{equation}
so $k=\tfrac13$ is forbidden. The $N_f/2$ form \eqref{eq:V06-parityfund} is the fundamental special
case of the general representation-dependent statement \eqref{eq:V06-parity};
Table~\ref{tab:V06-parity} tabulates small $N_f$.

\begin{table}[htbp]
\centering
\renewcommand{\arraystretch}{1.3}
\begin{tabular}{@{}cccc@{}}
\hline
$N_f$ & $T(R)=N_f$ & $k+N_f/2\in\mathbb{Z}$ requires & admissible bare $k$ \\
\hline
$1$ & $1$ & $k\in\tfrac12+\mathbb{Z}$ & half-integer \\
\hline
$2$ & $2$ & $k\in\mathbb{Z}$ & integer \\
\hline
$3$ & $3$ & $k\in\tfrac12+\mathbb{Z}$ & half-integer \\
\hline
$4$ & $4$ & $k\in\mathbb{Z}$ & integer \\
\hline
\end{tabular}
\caption{The parity anomaly on the $U(1)$/$U(N)$-fundamental fixture. The index is $T(R)=N_f$, and the
admissible bare Chern-Simons level satisfies $k+N_f/2\in\mathbb{Z}$: half-integer for odd $N_f$,
integer for even $N_f$.}
\label{tab:V06-parity}
\end{table}

\section{ABJM and the \texorpdfstring{$N^{3/2}$}{N-3/2} free energy}
\label{sec:V06-abjm}

The canonical Chern-Simons-matter theory is ABJM (Aharony--Bergman--Jafferis--Maldacena). It is a
\begin{equation}
\label{eq:V06-abjmgroup}
 U(N)_k \times U(N)_{-k}
\end{equation}
gauge theory (the first factor at level $+k$, the second at $-k$) with two bifundamental
hypermultiplets in the conjugate representations
\begin{equation}
\label{eq:V06-abjmmatter}
 A_i \in (\mathbf{N},\overline{\mathbf{N}}),
 \qquad
 B_j \in (\overline{\mathbf{N}},\mathbf{N}),
 \qquad i,j=1,2,
\end{equation}
coupled through a quartic superpotential. It is $\mathcal{N}=6$ superconformal for generic $k$, and its
supersymmetry enhances to $\mathcal{N}=8$ at the special levels $k=1,2$. The two Chern-Simons levels are
opposite in sign, and each must independently satisfy the parity anomaly \eqref{eq:V06-parity} for the
bifundamental matter,
\begin{equation}
\label{eq:V06-abjmlevels}
 \bigl(k,\,-k\bigr),
 \qquad
 k + \tfrac12\,T(R) \in\mathbb{Z},
\end{equation}
which is automatic for the equal-and-opposite bifundamental content. ABJM is the worldvolume theory of
$N$ M2-branes probing the orbifold $\mathbb{C}^4/\mathbb{Z}_k$; here it is the marquee
Chern--Simons-matter fixture.

The reason it is the marquee example is a single number, the growth of its free energy with the rank
$N$. The round-sphere free energy of ABJM, computed by localization and its M-theory limit
(Drukker--Marino--Putrov), scales as
\begin{equation}
\label{eq:V06-abjmF}
 F_{S^3} \;=\; \frac{\pi\sqrt{2}}{3}\, k^{1/2}\, N^{3/2}.
\end{equation}
The exponent is $3/2$, not the naive $2$ of a gauge theory with $O(N^2)$ matrix degrees of freedom;
the $N^{3/2}$ growth is the field-theory signature of $N$ M2-branes, the microscopic count of degrees
of freedom on the M2-brane worldvolume. Logarithmic differentiation reads the exponent off directly,
\begin{equation}
\label{eq:V06-abjmexp}
 \frac{\mathrm{d}\log F}{\mathrm{d}\log N} \;=\; \frac{3}{2},
\end{equation}
the power of $N$. As a concrete evaluation, at $k=2$, $N=4$,
\begin{equation}
\label{eq:V06-abjmval}
 F(2,4) \;=\; \frac{\pi\sqrt2}{3}\,\sqrt{2}\cdot 4^{3/2} \;=\; \frac{\pi\sqrt2}{3}\,\sqrt2\cdot 8
 \;=\; \frac{16\pi}{3}.
\end{equation}
The sharpest way to see the exponent is the ratio of free energies at two ranks, which cancels the
overall constant:
\begin{equation}
\label{eq:V06-abjmratio}
 \frac{F(k,4)}{F(k,1)} \;=\; \frac{4^{3/2}}{1^{3/2}} \;=\; 8
 \quad\text{($N^{3/2}$ reading)},
 \qquad\text{versus}\qquad
 4^2 \;=\; 16 \quad\text{(wrong $N^2$)}.
\end{equation}
Table~\ref{tab:V06-abjm} shows the ratio growing as $N^{3/2}$, not $N^2$, the M2-brane distinction. The
localization derivation of the ABJM matrix model and its M-theory limit (the constant
$\tfrac{\pi\sqrt2}{3}$ and the $N^{3/2}$ law) is cited but not derived here.

\begin{table}[htbp]
\centering
\renewcommand{\arraystretch}{1.3}
\begin{tabular}{@{}cccc@{}}
\hline
$N$ & $F(k,N)/F(k,1)=N^{3/2}$ & the $N^{3/2}$ reading & the $N^2$ reading (wrong) \\
\hline
$1$ & $1$ & $1$ & $1$ \\
\hline
$2$ & $2^{3/2}=2\sqrt2$ & $\approx 2.83$ & $4$ \\
\hline
$4$ & $4^{3/2}=8$ & $8$ & $16$ \\
\hline
$9$ & $9^{3/2}=27$ & $27$ & $81$ \\
\hline
\end{tabular}
\caption{The ABJM free-energy ratio $F(k,N)/F(k,1)$, which cancels the overall constant and isolates
the $N$-scaling. The M2-brane $N^{3/2}$ growth (third column) is well below the naive gauge-theory
$N^2$ (fourth column); the two diverge sharply with $N$, so the exponent is a decisive check.}
\label{tab:V06-abjm}
\end{table}

The $N^{3/2}$ scaling is the field-theory shadow of the M-theory origin. The moduli space of vacua
and the supersymmetry both track the orbifold:
\begin{equation}
\label{eq:V06-abjmmoduli}
 \mathcal{M} \;=\; \mathrm{Sym}^N\bigl(\mathbb{C}^4/\mathbb{Z}_k\bigr),
 \qquad
 \mathcal{N} \;=\;
 \begin{cases}
 8, & k=1,2,\\[2pt]
 6, & k\geq 3,
 \end{cases}
\end{equation}
the configuration space of $N$ indistinguishable M2-branes on the orbifold, with the level $k$ the
order of the orbifold action. At $k=1$ the orbifold is trivial ($\mathbb{C}^4$) and the theory is the
maximally supersymmetric worldvolume theory of M2-branes in flat space; at $k=2$ the $\mathbb{Z}_2$
still preserves $\mathcal{N}=8$. The $N^{3/2}$ number is the one to carry away.

\begin{keybox}{Common misconception: the ABJM gauge group is $SU(4)$ with an effective level}
Two errors travel together here. First, the $\mathcal{N}=6$ R-symmetry is $SU(4)$, and it acts on the
matter, but it is \emph{not} the gauge group: the ABJM gauge group is $U(N)_k\times U(N)_{-k}$, with
the declared global quotient as global-form data, and the ABJM matter weights $(1,1,1,1)/k$ lie in
$U(4)$, not $SU(4)$. Second, these notes quote the \emph{bare} level $k$. The parity anomaly
\eqref{eq:V06-parity} is a quantization condition on the bare level (the $k+N_f/2\in\mathbb{Z}$ form
is the fundamental special case, not the general law), a constraint, not a redefinition. The global
form (the diagonal $U(1)$, the $\mathbb{Z}_k$ quotient) affects the line and monopole spectrum and
the $\mathbb{C}^4/\mathbb{Z}_k$ orbifold statement; the full line-operator classification is not
developed here.
\end{keybox}

\section{The Aharony and Giveon--Kutasov dualities}
\label{sec:V06-dualities}

Three-dimensional $\mathcal{N}=2$ gauge theories have their own Seiberg-type infrared dualities, and
the monopole operators of \S\ref{sec:V06-monopole} make them differ from the four-dimensional Seiberg
duality of Section~3 in a way worth stating precisely. There are two, distinguished by whether a
Chern-Simons term is present.

\medskip\noindent\textbf{Aharony duality (no Chern-Simons term).}\enspace
A $U(N_c)$ theory with $N_f$ flavors (and no Chern-Simons term) is infrared-dual to a $U(N_f-N_c)$
theory with $N_f$ dual flavors, a gauge-singlet meson $M$, and, crucially, two extra gauge-singlet
\emph{monopole operators} $V_\pm$, all coupled through a superpotential. The rank map is
\begin{equation}
\label{eq:V06-aharony}
 U(N_c) \;\longleftrightarrow\; U(N_f-N_c),
\end{equation}
and the monopole singlets $V_\pm$ are the three-dimensional feature with no four-dimensional analog:
the electric theory's monopole operators are dynamical, so the magnetic dual carries them as explicit
singlet fields. The map is an involution, as any duality must be:
\begin{equation}
\label{eq:V06-aharonyinv}
 U(N_c)\ \longrightarrow\ U(N_f-N_c)\ \longrightarrow\ U\bigl(N_f-(N_f-N_c)\bigr) \;=\; U(N_c).
\end{equation}

\medskip\noindent\textbf{Giveon--Kutasov duality (Chern-Simons level $k$).}\enspace
Turn on a Chern-Simons level $k$. A $U(N_c)_k$ theory with $N_f$ flavors is dual to a
\begin{equation}
\label{eq:V06-gk}
 U(N_c)_k \;\longleftrightarrow\; U\bigl(N_f+|k|-N_c\bigr)_{-k}
\end{equation}
theory with $N_f$ flavors and a meson, but \emph{no} monopole singlets. The Chern-Simons term lifts the
monopoles by flux attachment \eqref{eq:V06-fluxattachderived}: a bare monopole is no longer
gauge-neutral, so it does not survive as a light singlet, and the dual has none. The dual level flips
sign to $-k$, and the rank picks up the $|k|$ shift. Re-dualizing returns $U(N_c)_k$, an involution,
and the $k\to 0$ limit is the Aharony map,
\begin{equation}
\label{eq:V06-gkreduce}
 U\bigl(N_f+|k|-N_c\bigr)_{-k}\ \xrightarrow{\ k\to 0\ }\ U(N_f-N_c),
\end{equation}
with the monopole singlets reappearing, so the two dualities are consistent limits of one another.

The rank arithmetic on examples: for Aharony $U(2)$ with $N_f=5$, the dual and its re-dual are
\begin{equation}
\label{eq:V06-aharonyex}
 U(2) \;\xrightarrow{\ N_f=5\ }\; U(N_f-N_c)=U(3)
 \;\xrightarrow{\ N_f=5\ }\; U(5-3)=U(2),
\end{equation}
back to the start. For Giveon--Kutasov take $U(2)_1$ with $N_f=3$: with the $|k|$ shift and the
level flip,
\begin{equation}
\label{eq:V06-gkex}
 U(2)_1 \;\xrightarrow{\ N_f=3\ }\; U(3+1-2)_{-1}=U(2)_{-1}
 \;\xrightarrow{\ N_f=3\ }\; U(3+1-2)_{1}=U(2)_{1},
\end{equation}
again an involution (the second step uses level $|{-1}|=1$). Dropping the $|k|$ shift (writing
$U(N_f-N_c)_{-k}$) breaks the involution; Table~\ref{tab:V06-rankmaps} collects the worked ranks.

\begin{table}[htbp]
\centering
\renewcommand{\arraystretch}{1.3}
\begin{tabular}{@{}ccccc@{}}
\hline
Duality & $(N_c,N_f,k)$ & dual rank & re-dual & involutive? \\
\hline
Aharony & $(2,5,0)$ & $U(3)$ & $U(2)$ & yes \\
\hline
Aharony & $(3,7,0)$ & $U(4)$ & $U(3)$ & yes \\
\hline
Giveon--Kutasov & $(2,3,1)$ & $U(2)_{-1}$ & $U(2)_{1}$ & yes \\
\hline
Giveon--Kutasov & $(1,4,2)$ & $U(5)_{-2}$ & $U(1)_{2}$ & yes \\
\hline
\end{tabular}
\caption{Worked rank maps. Aharony ($k=0$): $U(N_c)\to U(N_f-N_c)$. Giveon--Kutasov ($k\neq0$):
$U(N_c)_k\to U(N_f+|k|-N_c)_{-k}$. Every map is involutive: re-dualizing returns $U(N_c)$ (at the
original level $k$ for Giveon--Kutasov).}
\label{tab:V06-rankmaps}
\end{table}

\begin{table}[htbp]
\centering
\renewcommand{\arraystretch}{1.35}
\begin{tabular}{@{}p{0.20\linewidth}p{0.30\linewidth}p{0.42\linewidth}@{}}
\hline
Duality & Rank map & Extra content \\
\hline
Aharony ($k=0$) & $U(N_c)\to U(N_f-N_c)$ & meson $M$ and monopole singlets $V_\pm$ \\
\hline
Giveon--Kutasov ($k\neq 0$) & $U(N_c)_k\to U(N_f+|k|-N_c)_{-k}$ & meson $M$; no monopole singlets (lifted by the CS term) \\
\hline
\end{tabular}
\caption{The two $3d\ \mathcal{N}=2$ Seiberg-type dualities. Both rank maps are involutive
(re-dualizing returns $N_c$), and the Giveon--Kutasov map reduces to the Aharony map at $k=0$. The
monopole singlets are the intrinsically three-dimensional content: dynamical in Aharony, lifted by the
Chern-Simons term in Giveon--Kutasov.}
\label{tab:V06-dualities}
\end{table}

\begin{keybox}{Common misconception: the $3d$ dualities are the reduction of $4d$ Seiberg duality}
Reducing four-dimensional Seiberg duality on a circle does not give the three-dimensional dualities on
the nose. The reduction-and-flow generates extra monopole superpotential terms, so the Aharony dual
carries explicit monopole singlets $V_\pm$ that the naive four-dimensional map has no room for; and a
Chern-Simons level changes the rank map entirely, $U(N_c)_k\to U(N_f+|k|-N_c)_{-k}$, which has no
four-dimensional counterpart. The same warning applies, even more sharply, to mirror symmetry
(\S\ref{sec:V06-mirror}), which has no four-dimensional analog at all.
\end{keybox}

The Aharony dual superpotential is where the monopole singlets earn their place. The magnetic theory
is $U(N_f-N_c)$ with dual quarks $q,\widetilde q$, the meson $M$ (a gauge singlet with the flavor
quantum numbers of the electric meson $Q\widetilde Q$), and the two monopole singlets $V_\pm$, coupled
by
\begin{equation}
\label{eq:V06-aharonyW}
 W \;=\; M\,q\,\widetilde q \;+\; V_+\,\widetilde V_- \;+\; V_-\,\widetilde V_+,
\end{equation}
where $\widetilde V_\pm$ are the monopole operators of the \emph{magnetic} gauge group. The first term
is the meson coupling of four-dimensional Seiberg duality (Section~3); the last two tie the electric
monopole singlets to the magnetic monopoles and are the intrinsically three-dimensional part: dropping
them would leave extra massless singlets and spoil the match. The field-theory content of the duality
is the operator map, each gauge-invariant chiral operator of the electric theory matched to one of the
magnetic theory, collected in Table~\ref{tab:V06-opmap}.

\begin{table}[htbp]
\centering
\renewcommand{\arraystretch}{1.35}
\begin{tabular}{@{}p{0.42\linewidth}p{0.50\linewidth}@{}}
\hline
Electric object ($U(N_c)$, $N_f$ flavors) & Aharony magnetic object ($U(N_f-N_c)$) \\
\hline
meson $Q\widetilde Q$ & gauge singlet $M$ \\
\hline
monopoles $V_\pm$ & gauge singlets $V_\pm$ (dynamical, coupled in $W$) \\
\hline
magnetic monopoles $\widetilde V_\pm$ & lifted by $W\supset V_+\widetilde V_- + V_-\widetilde V_+$ \\
\hline
FI parameter / $U(1)_J$ topological & a flavor symmetry acting on $M$, $V_\pm$ \\
\hline
\end{tabular}
\caption{The Aharony operator map. The electric mesons $Q\widetilde Q$ become the singlet $M$; the
electric monopole operators $V_\pm$ appear as explicit singlets in the dual (the intrinsically $3d$
content), coupled through the dual superpotential \eqref{eq:V06-aharonyW}; and the topological $U(1)_J$
maps to a flavor symmetry of the dual. Reducing $4d$ Seiberg duality has no room for the monopole
singlets, which is why the $3d$ duality is not that reduction.}
\label{tab:V06-opmap}
\end{table}

\subsection*{Worked instance: the evidence on a small rank}

The evidence for these dualities is the three-dimensional analog of Section~3's four checks, named on
a concrete pair: $U(1)$ with $N_f$ flavors (Aharony), dual to $U(N_f-1)$ with the meson and monopole
singlets. \emph{Global symmetries:} both sides carry
\begin{equation}
\label{eq:V06-dualglobal}
 SU(N_f)\times SU(N_f)\times U(1)\times U(1)_J,
\end{equation}
with the electric $U(1)_J$ acting as a flavor symmetry on the dual mesons and monopoles.
\emph{Chiral operators:} $Q\widetilde Q\mapsto M$ and $V_\pm\mapsto V_\pm$ of
\eqref{eq:V06-aharonyW}, a map with no four-dimensional analog. \emph{Parity anomaly:} the bare-level
condition \eqref{eq:V06-parity} matches, since the fermion index is duality-invariant. \emph{The $S^3$
partition function:} the two round-sphere partition functions are equal \emph{as functions} of the
real masses $m_a$ and the FI parameter $\zeta$,
\begin{equation}
\label{eq:V06-S3identity}
 Z^{S^3}_{\text{electric}}(m_a;\zeta) \;=\; Z^{S^3}_{\text{magnetic}}(m_a;\zeta)
 \qquad\text{(under the duality parameter map)},
\end{equation}
an identity of matrix integrals, the strongest $3d$ check because it is an exact equality of
computable functions rather than a match of discrete data. The identity and the proofs of both
dualities are cited but not proved here.

\bigskip
\begin{center}
\rule{0.4\textwidth}{0.4pt}\\[3pt]
{\large\textsf{\textbf{Block B.\enspace The $3d\ \mathcal{N}=4$ world}}}\\[2pt]
\rule{0.4\textwidth}{0.4pt}
\end{center}
\medskip

\noindent Doubling to eight supercharges opens the geometric, mirror world of three dimensions.
Three-dimensional $\mathcal{N}=4$ mirror symmetry was one of the earliest sharp understandings of $3d$
supersymmetric dynamics: it exchanges the Higgs and Coulomb branches of a pair of theories, and thereby
makes a quantum, monopole-generated Coulomb branch computable from the classical Higgs branch of the
mirror. Where Block~A treats $\mathcal{N}=2$ as the lower-supersymmetry workhorse (monopoles,
F-maximization, Chern-Simons-matter and ABJM, the Aharony and Giveon--Kutasov dualities), Block~B is
the eight-supercharge world: two hyperk\"ahler branches, the $SU(2)_C\times SU(2)_H$ R-symmetry that
acts on them separately, and mirror symmetry as a central part of the subject rather than an appendix.
The two sections build the twin branches (\S\ref{sec:V06-n4}) and then the mirror exchange that swaps
them (\S\ref{sec:V06-mirror}).

\section[The \texorpdfstring{$\mathcal{N}=4$}{N=4} enhancement]{The \texorpdfstring{$\mathcal{N}=4$}{N=4} enhancement: two branches}
\label{sec:V06-n4}

Doubling the supersymmetry to $3d\ \mathcal{N}=4$ gives eight real supercharges, the same count as
$4d\ \mathcal{N}=2$: dimensional reduction of a four-dimensional $\mathcal{N}=2$ theory on a circle
gives a three-dimensional $\mathcal{N}=4$ theory, the $8Q$ row of the Section~1 ladder table. The
enhancement changes the R-symmetry qualitatively. Where $\mathcal{N}=2$ has a single $U(1)_R$, the
$\mathcal{N}=4$ theory has
\begin{equation}
\label{eq:V06-rsymmetry}
 SU(2)_C \times SU(2)_H \;=\; \mathrm{Spin}(4),
\end{equation}
two commuting $SU(2)$ factors (Section~1 master table): $\mathrm{so}(4)=\mathrm{su}(2)\oplus
\mathrm{su}(2)$, rank $1+1=2$, dimension $3+3=6$. The two factors act separately on the two kinds of
branch. This is not a single $SU(2)_R$; the two-factor structure is the whole point, and it is what
mirror symmetry will exchange.

\medskip\noindent\textbf{The multiplet content, read in $\mathcal{N}=2$ language.}\enspace
The cleanest way to see the multiplet content is to decompose each $\mathcal{N}=4$ multiplet into
$\mathcal{N}=2$ pieces, exactly parallel to Section~5's decomposition of $4d\ \mathcal{N}=2$ into
$\mathcal{N}=1$. There are two decompositions:
\begin{equation}
\label{eq:V06-n4decompvec}
 \text{$\mathcal{N}=4$ vector}
 \;=\; \underbrace{\text{$\mathcal{N}=2$ vector}}_{A_\mu,\ \sigma,\ \text{gaugino}}
 \;+\; \underbrace{\text{$\mathcal{N}=2$ adjoint chiral }\Phi}_{\text{2 real scalars}},
\end{equation}
\begin{equation}
\label{eq:V06-n4decomphyp}
 \text{$\mathcal{N}=4$ hyper}
 \;=\; \underbrace{\text{$\mathcal{N}=2$ chiral }Q}_{(R)}
 \;+\; \underbrace{\text{$\mathcal{N}=2$ chiral }\widetilde Q}_{(\overline{R})},
\end{equation}
a pair of $\mathcal{N}=2$ chirals in conjugate representations. Read off the vector-multiplet scalars
from \eqref{eq:V06-n4decompvec}: the $\mathcal{N}=2$ vector supplies the one real scalar $\sigma$ of
\eqref{eq:V06-vectormult}, and the adjoint chiral $\Phi$ supplies two more real scalars, so the
$\mathcal{N}=4$ vector multiplet carries
\begin{equation}
\label{eq:V06-tripletcount}
 \underbrace{1}_{\sigma\ (\mathcal{N}=2\ \text{vector})}
 \;+\; \underbrace{2}_{\Phi\ (\mathcal{N}=2\ \text{adjoint chiral})}
 \;=\; 3\ \text{real scalars},
\end{equation}
a \emph{triplet} of $SU(2)_C$. The hypermultiplet's four real scalars (two complex, one each in $Q$ and
$\widetilde Q$) form a doublet of doublets under $SU(2)_H$. On the abelian Coulomb branch the photon
again dualizes to the compact $\gamma$, which is \emph{not} one of the three triplet scalars: it is
the separate fourth real coordinate of the local Coulomb geometry. For a rank-$1$ ($U(1)$) theory the
count is
\begin{equation}
\label{eq:V06-coulombdim}
 \underbrace{3}_{SU(2)_C\text{ triplet }\sigma_a} \;+\; \underbrace{1}_{\text{dual photon }\gamma}
 \;=\; 4 \ \text{real} \;=\; 1 \ \text{quaternionic},
\end{equation}
four real coordinates assembling one quaternionic (hyperk\"ahler) dimension: locally a hyperk\"ahler
$\mathbb{C}^2$. For gauge rank $r$ the generic abelian Coulomb branch has real dimension
\begin{equation}
\label{eq:V06-coulombrankr}
 \dim_{\mathbb R} C \;=\; \underbrace{3r}_{\text{triplet scalars}} + \underbrace{r}_{\text{dual photons}}
 \;=\; 4r \;=\; 4\dim_{\mathbb H}C,
 \qquad \dim_{\mathbb H} C = r,
\end{equation}
the $3r$ triplet scalars plus the $r$ dual photons, assembling $r$ quaternionic coordinates. The
factor of $4$ is the hyperk\"ahler signature: a hyperk\"ahler manifold has real dimension a multiple
of $4$. Compactifying the $4d\ \mathcal{N}=2$ Coulomb branch (complex dimension $r$, real $2r$) on a
circle adds the $r$ Wilson lines and the $r$ dual photons, promoting the $2r$-real special-K\"ahler
branch to this $4r$-real hyperk\"ahler one; the dual photon is the piece with no four-dimensional
counterpart.

The two R-symmetry factors act on the two branches separately, the structural fact mirror symmetry
exploits:
\begin{equation}
\label{eq:V06-raction}
 SU(2)_C:\ C\ \text{rotated},\ H\ \text{fixed};
 \qquad
 SU(2)_H:\ H\ \text{rotated},\ C\ \text{fixed}.
\end{equation}
The $SU(2)_C$ factor rotates the three triplet scalars of the vector multiplet, the $SU(2)_H$ factor
the hypermultiplet scalars, and the branch each acts on is identified by which factor moves it.
Because the two factors are on an equal footing in the algebra, exchanging them is a symmetry of the
abstract structure, and mirror symmetry (\S\ref{sec:V06-mirror}) realizes that exchange by a pair of
physical theories.

Both branches are hyperk\"ahler. The Higgs branch is the hyperk\"ahler quotient of the hypermultiplet
scalars by the gauge group, and it is classically exact, precisely the object Section~5 built for the
four-dimensional $\mathcal{N}=2$ theory; for $U(1)$ with $N_f$ hypers its quaternionic dimension is
\begin{equation}
\label{eq:V06-higgsdim}
 \dim_{\mathbb H} H \;=\; \frac{4N_f - 3 - 1}{4} \;=\; N_f - 1,
\end{equation}
the $4N_f$ real hypermultiplet scalars, minus the $3$ real moment-map (triplet D-term) constraints,
minus the $1$ gauge direction, all over $4$. The Coulomb branch is also hyperk\"ahler, but
\emph{quantum-corrected} and monopole-built: the dual photon supplies the extra real direction
\eqref{eq:V06-coulombdim} that completes the local geometry to a quaternionic space, promoting the
special-K\"ahler Coulomb branch of Section~5's $4d\ \mathcal{N}=2$ world to a hyperk\"ahler one while
the Higgs branch is unchanged. For $U(1)$ with $N_f$ hypers both branches turn out to be $A$-type
singularities, which is exactly why $T[SU(2)]$ (the $N_f=2$ case) can be self-mirror: its Coulomb
$A_1$ and its Higgs $A_1$ coincide.

\begin{keybox}{Common misconception: the $\mathcal{N}=4$ Coulomb branch is special-K\"ahler}
The four-dimensional $\mathcal{N}=2$ Coulomb branch is special-K\"ahler; the three-dimensional
$\mathcal{N}=4$ Coulomb branch is \emph{hyperk\"ahler}. It is not the four-dimensional special-K\"ahler
object reduced naively. The dual photon supplies the fourth real coordinate
\eqref{eq:V06-coulombdim}, promoting the local geometry to quaternionic, and the branch is
quantum-corrected and monopole-built rather than governed by a classical prepotential. Nor is the
R-symmetry a single $SU(2)_R$: it is $SU(2)_C\times SU(2)_H$, two factors acting on the two branches
separately, and this two-factor structure is what mirror symmetry exchanges.
\end{keybox}

\subsection*{Worked instance: the $U(1)$ Coulomb-branch chiral ring}

The monopole operators do more than carry topological charge; in $\mathcal{N}=4$ they generate the
Coulomb-branch chiral ring. For $U(1)$ with $N_f$ flavors the ring is generated by the two bare
monopoles $V_+$ (charge $+1$) and $V_-$ (charge $-1$) together with the vector-multiplet complex scalar
$\Phi$ (the complexified $\sigma$). The product $V_+V_-$ has vanishing topological charge, so it is
a genuine function of $\Phi$; its R-charge is fixed by the monopole dimensions. Using the SQED value
$\Delta(\pm1)=N_f/2$ from \eqref{eq:V06-sqedmonopole},
\begin{equation}
\label{eq:V06-ringRcharge}
 R(V_+V_-) \;=\; R(V_+)+R(V_-) \;=\; \frac{N_f}{2}+\frac{N_f}{2} \;=\; N_f,
 \qquad
 R(\Phi) \;=\; 1,
\end{equation}
so $V_+V_-$ has the R-charge of $\Phi^{N_f}$. The counting is the matter zero-mode counting: each of the
$N_f$ charged hypermultiplets contributes a fermion zero mode in the monopole background, and the
$N_f$ zero modes are what promote a single power of $\Phi$ to the $N_f$-th power. The ring relation
is therefore
\begin{equation}
\label{eq:V06-coulombring}
 V_+ V_- \;=\; \Phi^{N_f}.
\end{equation}
In coordinates $(x,y,z)=(V_+,V_-,\Phi)$ this is $xy=z^{N_f}$, the equation of the $A_{N_f-1}$
singularity, the orbifold $\mathbb{C}^2/\mathbb{Z}_{N_f}$. The orbifold order is $N_f$, read off the
exponent of $z$ in the ring relation, not asserted: the $\mathbb{Z}_{N_f}$ acts on $\mathbb{C}^2=
\{(u,v)\}$ and its invariants satisfy the relation identically,
\begin{equation}
\label{eq:V06-orbaction}
\begin{aligned}
 &(u,v)\to(\omega u,\omega^{-1}v),\quad \omega=e^{2\pi i/N_f},
 \qquad x=u^{N_f},\ y=v^{N_f},\ z=uv \\[2pt]
 &\Longrightarrow\quad xy = u^{N_f}v^{N_f} = (uv)^{N_f} = z^{N_f}.
\end{aligned}
\end{equation}
Two special cases anchor the mirror computation of the next section. For $N_f=1$ and $N_f=2$,
\begin{equation}
\label{eq:V06-ringcases}
\begin{aligned}
 N_f=1:&\quad V_+V_-=\Phi \;\Rightarrow\; xy=z,\quad \text{smooth }\mathbb{C}^2, \\[2pt]
 N_f=2:&\quad V_+V_-=\Phi^2 \;\Rightarrow\; xy=z^2,\quad \mathbb{C}^2/\mathbb{Z}_2=A_1.
\end{aligned}
\end{equation}
The $N_f=1$ ring solves for $z=xy$, a smooth $\mathbb{C}^2$ in coordinates $(x,y)$; the $N_f=2$ ring
$xy=z^2$ is the $A_1$ singular point, \emph{not} smooth, and getting the order wrong by one
misidentifies the geometry. This ring is the quantum geometry that mirror symmetry will map to a
classical Higgs geometry; the general machinery (abelianization, Hilbert series) is cited but not
developed here.

\section{Three-dimensional mirror symmetry}
\label{sec:V06-mirror}

Three-dimensional mirror symmetry is the sharpest duality in these notes. It is a duality of
$\mathcal{N}=4$ theories that \emph{exchanges the two branches}: for a theory $T$ and its mirror
$T^\vee$,
\begin{equation}
\label{eq:V06-mirror}
 C(T)\ \cong\ H(T^\vee),
 \qquad
 H(T)\ \cong\ C(T^\vee),
\end{equation}
the Coulomb branch of one theory equal to the Higgs branch of the other and vice versa (the branch
frame is always carried: $C(T^\vee)\cong H(T)$ is a claim, not a renaming). Under
the exchange the two R-symmetry factors swap, $SU(2)_C\leftrightarrow SU(2)_H$, and the exchange
dictionary swaps the data that source the two branches:
\begin{equation}
\label{eq:V06-mirrormap}
 \text{FI parameters}\ \longleftrightarrow\ \text{hypermultiplet masses},
 \qquad
 U(1)_J\ \longleftrightarrow\ \text{a flavor symmetry}.
\end{equation}
Written as a single line, with the branch frame carried and the FI parameter $\zeta_T$ and mass
$m_{T^\vee}$ named on the two theories, the mirror dictionary is
\begin{equation}
\label{eq:V06-mirrordict}
\begin{aligned}
 &\text{Coulomb branch of }T \ \leftrightarrow\ \text{Higgs branch of }T^\vee, \\
 &\zeta_T \leftrightarrow m_{T^\vee},\qquad U(1)_{J,T}\leftrightarrow U(1)_{F,T^\vee}.
\end{aligned}
\end{equation}
The dictionary is an involution: applying it twice returns each item to itself. Its computational
power is the asymmetry of difficulty between the branches: the Coulomb branch is quantum and
monopole-generated, the Higgs branch a classical hyperk\"ahler quotient, so mirror symmetry turns a
hard monopole computation on one side into an easy polynomial computation on the other.

\subsection*{Worked instance: the basic pair and the self-mirror}

Two pairs make the exchange concrete, and each is worked by computing the two branches independently and
only then matching them.

\medskip\noindent\emph{The basic pair: $U(1)$ with $N_f=1$ against a free hyper.}\enspace
Compute the four branch dimensions. The gauge theory has a one-dimensional (quaternionic) Coulomb branch
and a point Higgs branch,
\begin{equation}
\label{eq:V06-basicdims}
 \dim_{\mathbb H} C\bigl(U(1),N_f{=}1\bigr) = 1,
 \qquad
 \dim_{\mathbb H} H\bigl(U(1),N_f{=}1\bigr) = N_f-1 = 0,
\end{equation}
while the free hyper has a one-dimensional Higgs branch (its four real scalars) and no Coulomb branch.
The geometries match crosswise. From the ring \eqref{eq:V06-coulombring} with $N_f=1$ the Coulomb branch
is the smooth $\mathbb{C}^2$, which equals the free hyper's Higgs branch:
\begin{equation}
\label{eq:V06-basicpair}
\begin{aligned}
 C\bigl(U(1),\,N_f=1\bigr) &\;=\; \mathbb{C}^2 \;=\; H(\text{free hyper}), \\[2pt]
 H\bigl(U(1),\,N_f=1\bigr) &\;=\; \{\cdot\} \;=\; C(\text{free hyper}),
\end{aligned}
\end{equation}
the quantum Coulomb branch of the gauge theory equal to the classical Higgs branch of the free hyper,
and the trivial branches likewise exchanged.

\medskip\noindent\emph{The self-mirror: $U(1)$ with $N_f=2$, the theory $T[SU(2)]$.}\enspace
Now both branches are one-dimensional. The Coulomb branch is read from the ring, the Higgs branch from
the hyperk\"ahler quotient (Section~5):
\begin{equation}
\label{eq:V06-tsu2dims}
 C\bigl(U(1),N_f{=}2\bigr) \;\overset{\eqref{eq:V06-coulombring}}{=}\; \mathbb{C}^2/\mathbb{Z}_2 = A_1,
 \qquad
 H\bigl(U(1),N_f{=}2\bigr) \;\overset{\dim_{\mathbb H}=N_f-1=1}{=}\; \mathbb{C}^2/\mathbb{Z}_2 = A_1.
\end{equation}
Both branches are the same $A_1$ singularity $\mathbb{C}^2/\mathbb{Z}_2$, so mirror symmetry maps the
theory to itself:
\begin{equation}
\label{eq:V06-selfmirror}
 C\bigl(T[SU(2)]\bigr) \;=\; H\bigl(T[SU(2)]\bigr) \;=\; \mathbb{C}^2/\mathbb{Z}_2,
 \qquad T[SU(2)]^\vee = T[SU(2)].
\end{equation}
The branch geometries here are computed, the Coulomb branch from the monopole ring and the Higgs branch
from the hyperk\"ahler quotient, and only then matched; they are not asserted equal.

\begin{table}[htbp]
\centering
\renewcommand{\arraystretch}{1.35}
\begin{tabular}{@{}p{0.24\linewidth}p{0.24\linewidth}p{0.24\linewidth}p{0.18\linewidth}@{}}
\hline
Theory $T$ & Coulomb $C(T)$ & Higgs $H(T)$ & Mirror $T^\vee$ \\
\hline
$U(1)$, $N_f=1$ & $\mathbb{C}^2$ & $\{\cdot\}$ & free hyper \\
\hline
$U(1)$, $N_f=2$ ($T[SU(2)]$) & $\mathbb{C}^2/\mathbb{Z}_2$ & $\mathbb{C}^2/\mathbb{Z}_2$ & itself \\
\hline
\end{tabular}
\caption{The basic mirror pairs. For $N_f=1$ the gauge theory's quantum Coulomb branch $\mathbb{C}^2$
equals the free hypermultiplet's classical Higgs branch $\mathbb{C}^2$. For $N_f=2$, $T[SU(2)]$ is
self-mirror, both branches the $A_1$ singularity $\mathbb{C}^2/\mathbb{Z}_2$. The Coulomb branches are
read off the ring $V_+V_-=\Phi^{N_f}$; the Higgs branches from the hyperk\"ahler quotient.}
\label{tab:V06-mirror}
\end{table}

\begin{keybox}{The $3d\ \mathcal{N}=4$ deformation dictionary}
The two deformation parameters resolve the two branches, and the assignment is the opposite of the naive
guess. In a $3d\ \mathcal{N}=4$ gauge theory,
\begin{equation*}
 \text{FI } \zeta \ \text{resolves } \mathbf{Higgs},
 \qquad
 \text{real mass } m \ \text{resolves } \mathbf{Coulomb}.
\end{equation*}
Mirror symmetry exchanges the two deformations along with the two branches:
\begin{equation*}
 \zeta(T)\ \longleftrightarrow\ m(T^\vee),
 \qquad
 H(T)\ \longleftrightarrow\ C(T^\vee).
\end{equation*}
So the FI parameter of $T$ resolves the Higgs branch of $T$, which is the Coulomb branch of $T^\vee$,
resolved there by the mirror \emph{mass}. Do \emph{not} say the FI parameter resolves the Coulomb branch
of the same theory: it resolves that theory's Higgs branch (equivalently, the mirror's Coulomb branch).
\end{keybox}

The exchange dictionary makes the match more than a coincidence of geometries, though the basic pair
realizes it degenerately. Under \eqref{eq:V06-mirrormap} the FI parameter $\zeta$ of the gauge theory
(the background conjugate to $U(1)_J$, by \eqref{eq:V06-FIcurrent}) maps to a real mass of the free
hyper; that mass deforms the free hyper's \emph{Higgs} branch, the branch mirror to the gauge theory's
Coulomb branch,
\begin{equation}
\label{eq:V06-basicdeform}
 \zeta_{U(1),N_f=1}
 \;\xrightarrow{\ \text{mirror}\ }\;
 m_{\text{free hyper}}:\ \ \text{deforms}\ H(\text{free hyper}) = C\bigl(U(1),N_f{=}1\bigr).
\end{equation}
Because the free hyper has no nontrivial Coulomb branch, this pair is not a useful example of a
nontrivial Coulomb-branch \emph{resolution}. The clean nontrivial case is the self-mirror $T[SU(2)]$
below, where an FI parameter resolves one $A_1$ branch, a real mass resolves the mirror $A_1$ branch,
and the self-mirror symmetry exchanges the two.
The symmetries map the same way,
\begin{equation}
\label{eq:V06-basicsym}
 U(1)_J(T)\ \longleftrightarrow\ U(1)_{\text{flavor}}(T^\vee),
 \qquad
 SU(2)_C(T)\ \longleftrightarrow\ SU(2)_H(T^\vee),
\end{equation}
the topological current with no Lagrangian origin becoming an ordinary flavor current with a manifest
Noether one. Every piece of data that sources one branch on one side sources the exchanged branch on
the other, which is what promotes the geometric match to a duality.

For the self-mirror $T[SU(2)]$ the dictionary closes on itself: at $N_f=2$ the topological symmetry
enhances, and the exchange of the flavor $SU(2)$ (Higgs side, rotating the two hypers) with it,
\begin{equation}
\label{eq:V06-tsu2exchange}
 SU(2)_{\text{flavor}}
 \ \longleftrightarrow\
 SU(2)_J\ (\text{topological, enhanced at }N_f{=}2),
\end{equation}
is a symmetry of the theory itself, the two factors on an equal footing.

Beyond these worked pairs, the systematic construction of mirror pairs, by brane realizations (the
Hanany--Witten setup) and magnetic quivers, is cited but not developed here; the proof of mirror
symmetry is also cited rather than reproduced. This section delivers the statement, the exchange dictionary, and the two
worked pairs.

\section*{Exit checklist}
\addcontentsline{toc}{subsection}{Exit checklist}
\markboth{Exit checklist}{Exit checklist}

After this section the reader can
\begin{enumerate}
\item identify the $3d\ \mathcal{N}=2$ multiplets, name the real scalar $\sigma$ in the vector
multiplet, dualize the photon to the compact dual photon $\gamma$ ($\star F\sim d\gamma$), and state
that a rank-$1$ Coulomb branch is two-real-dimensional;
\item write the topological current $j_J=\star F/2\pi$, compute the basic-monopole $U(1)_J$ charge
$Q_J=1$ from the flux quantum $\int_{S^2}F=2\pi$, and explain why the Coulomb branch is a quantum
object built by monopole operators, not a classical scalar-VEV space;
\item define a monopole operator by its GNO magnetic charge $m$, apply the Borokhov--Kapustin--Wu
formula $\Delta(m)=\tfrac12\sum_i(1-\Delta_i)\sum_{\rho_i}|\rho_i(m)|-\sum_{\alpha>0}|\alpha(m)|$,
recover the SQED value $\Delta(1)=N_f/2$ as its special case at $\Delta_i=\tfrac12$, and attach flux
by $q_{\text{gauge}}=km$;
\item run F-maximization, extremize $\mathrm{Re}\,F=-\log|Z[\Delta]|$ built from the one-loop function
$\ell(z)$ ($\ell'(z)=-\pi z\cot\pi z$), get the free-chiral value $\Delta^*=\tfrac12$ with a negative
Hessian (a maximum), and state the $F$-theorem $F_{UV}>F_{IR}$;
\item write a Chern-Simons term with pure bosonic integer level, then impose the spin-theory parity
anomaly $k+\tfrac12 T(R)\in\mathbb{Z}$ (allowing integer or half-integer bare $k$, and reducing to
$k+N_f/2\in\mathbb{Z}$ on the fundamental fixture), and
recall that ABJM is $U(N)_k\times U(N)_{-k}$ with the $\mathcal{N}=6\to\mathcal{N}=8$ enhancement at
$k=1,2$ and free energy $F=\tfrac{\pi\sqrt2}{3}k^{1/2}N^{3/2}$, the $N^{3/2}$ M2-brane scaling;
\item write the Aharony rank map $U(N_c)\to U(N_f-N_c)$ (with monopole singlets) and the
Giveon--Kutasov map $U(N_c)_k\to U(N_f+|k|-N_c)_{-k}$ (no monopole singlets), check both are
involutive, and name the $S^3$ partition-function identity as the strongest evidence;
\item state the $3d\ \mathcal{N}=4$ R-symmetry $SU(2)_C\times SU(2)_H$, count the four real Coulomb
coordinates (three triplet scalars plus the dual photon) into one quaternionic dimension, explain why
both branches are hyperk\"ahler while the $4d\ \mathcal{N}=2$ Coulomb branch is only special-K\"ahler,
and read off the Coulomb ring $V_+V_-=\Phi^{N_f}=\mathbb{C}^2/\mathbb{Z}_{N_f}$;
\item use three-dimensional mirror symmetry as the Coulomb $\leftrightarrow$ Higgs exchange, apply the
dictionary (FI $\leftrightarrow$ mass, $U(1)_J\leftrightarrow$ flavor), work the basic $N_f=1$ pair
($C=\mathbb{C}^2=H$ of the free hyper) and the $T[SU(2)]$ self-mirror (both branches
$\mathbb{C}^2/\mathbb{Z}_2$), and state which deep theorems (F-maximization, the $F$-theorem, mirror
symmetry, the $3d$ dualities) are cited rather than proved.
\end{enumerate}

\bigskip
\section*{Sources and notes}
\addcontentsline{toc}{subsection}{Sources and notes}
\markboth{Sources and notes}{Sources and notes}
{\small

\noindent\textsf{\textcolor{RoyalBlue}{Sources and notes.}}\enspace
This is the three-dimensional dimension section of these notes, the Coulomb-branch, more-rigid partner of
the $4d\ \mathcal{N}=2$ section.

\medskip\noindent\textsf{\textcolor{RoyalBlue}{\textbf{\S\ref{sec:V06-multiplets}\enspace Multiplets and the dual photon.}}}\enspace
The $3d\ \mathcal{N}=2$ four-supercharge count as the reduction of $4d\ \mathcal{N}=1$, the vector
multiplet's real scalar $\sigma=A_3$ \eqref{eq:V06-vectormult}, the dualization of the photon
$\star F\sim d\gamma$ \eqref{eq:V06-dualphoton}, and the two-real-dimensional rank-$1$ Coulomb branch.
(\textcite{Aharony:1997bx} the $3d$ Coulomb-branch spine). 

\medskip\noindent\textsf{\textcolor{RoyalBlue}{\textbf{\S\ref{sec:V06-coulomb}\enspace The Coulomb branch and $U(1)_J$.}}}\enspace
The topological current $j_J=\star F/2\pi$ \eqref{eq:V06-topcurrent}, conserved by the Bianchi
identity, its charge the flux \eqref{eq:V06-topcharge}, and the basic-monopole charge $Q_J=1$; the FI
parameter as the $U(1)_J$ background and the $U(1)_J\leftrightarrow$ flavor mirror exchange previewed.
(\textcite{Aharony:1997bx} the topological symmetry). 

\medskip\noindent\textsf{\textcolor{RoyalBlue}{\textbf{\S\ref{sec:V06-monopole}\enspace Monopole operators and their dimensions.}}}\enspace
The monopole as a GNO-flux disorder operator \eqref{eq:V06-monopoledef}, the topological charge $Q_J=m$,
the Borokhov--Kapustin--Wu dimension \eqref{eq:V06-monopoledim} with the gaugino term kept separate,
the worked SQED $\Delta(1)=N_f/2$ \eqref{eq:V06-sqedmonopole} as a special case, and flux attachment
$q_{\text{gauge}}=km$ \eqref{eq:V06-fluxattach}. (\textcite{Borokhov:2002ib} the monopole
dimension by radial quantization; \textcite{Borokhov:2003yu} the follow-up). 

\medskip\noindent\textsf{\textcolor{RoyalBlue}{\textbf{\S\ref{sec:V06-fmax}\enspace F-maximization.}}}\enspace
The sphere free energy $F=-\log|Z[\Delta]|$ built from the one-loop function $\ell(z)$
\eqref{eq:V06-ellfunction} with $\ell'(z)=-\pi z\cot\pi z$ \eqref{eq:V06-ellprime}, the free energy
\eqref{eq:V06-freeenergy}, the free-chiral extremum $\Delta^*=\tfrac12$ with Hessian $-\pi^2/2<0$
\eqref{eq:V06-freechiral} (a maximum, the $3d$ $\Delta=R$ chiral-primary relation), the signpost to
$a$-maximization (Section~3), and the $F$-theorem $F_{UV}>F_{IR}$. (\textcite{Jafferis:2010un} F-maximization; \textcite{Kapustin:2009kz} the $S^3$ localization;
\textcite{Pufu:2016zxm} the $F$-theorem review). 

\medskip\noindent\textsf{\textcolor{RoyalBlue}{\textbf{\S\ref{sec:V06-cs}\enspace Chern-Simons-matter and the parity anomaly.}}}\enspace
The Chern-Simons action \eqref{eq:V06-cslevel} with integer pure bosonic level, the spin-theory
refinement by the parity anomaly $k+\tfrac12 T(R)\in\mathbb{Z}$ \eqref{eq:V06-parity} (so the bare
level may be integer or half-integer, not arbitrary fractional), the topological photon mass, and flux
attachment, reducing to $k+N_f/2\in\mathbb{Z}$ on the fundamental fixture. (\textcite{Redlich:1983dv} the parity anomaly). 

\medskip\noindent\textsf{\textcolor{RoyalBlue}{\textbf{\S\ref{sec:V06-abjm}\enspace ABJM and the $N^{3/2}$ free energy.}}}\enspace
ABJM $U(N)_k\times U(N)_{-k}$ \eqref{eq:V06-abjmgroup} with bifundamental hypers, $\mathcal{N}=6$
superconformal, the $\mathcal{N}=6\to\mathcal{N}=8$ enhancement at $k=1,2$, the M2-brane at
$\mathbb{C}^4/\mathbb{Z}_k$, and the free energy $F=\tfrac{\pi\sqrt2}{3}k^{1/2}N^{3/2}$
\eqref{eq:V06-abjmF} with the $N^{3/2}$ M2-brane scaling. (\textcite{Aharony:2008ug} ABJM;
\textcite{Drukker:2010nc} the $N^{3/2}$ free energy). 

\medskip\noindent\textsf{\textcolor{RoyalBlue}{\textbf{\S\ref{sec:V06-dualities}\enspace Aharony and Giveon--Kutasov dualities.}}}\enspace
The Aharony rank map $U(N_c)\to U(N_f-N_c)$ with monopole singlets \eqref{eq:V06-aharony}, the
Giveon--Kutasov map $U(N_c)_k\to U(N_f+|k|-N_c)_{-k}$ without them \eqref{eq:V06-gk}
(Table~\ref{tab:V06-dualities}), the involutions, the $k\to0$ reduction, and the evidence (global
symmetries, the chiral-operator/monopole map, parity matching, and the $S^3$ partition-function
identity). (\textcite{Aharony:1997gp} Aharony duality; \textcite{Giveon:2008zn}
Giveon--Kutasov duality). 

\medskip\noindent\textsf{\textcolor{RoyalBlue}{\textbf{\S\ref{sec:V06-n4}\enspace The $\mathcal{N}=4$ twin hyperk\"ahler branches.}}}\enspace
The $8Q$ count as the reduction of $4d\ \mathcal{N}=2$, the R-symmetry $SU(2)_C\times SU(2)_H=
\mathrm{Spin}(4)$ \eqref{eq:V06-rsymmetry}, the vector multiplet's three triplet scalars with the dual
photon the separate fourth Coulomb coordinate \eqref{eq:V06-coulombdim}, both branches hyperk\"ahler
(Coulomb quantum/monopole-built, not special-K\"ahler), and the Coulomb ring $V_+V_-=\Phi^{N_f}=
\mathbb{C}^2/\mathbb{Z}_{N_f}$ \eqref{eq:V06-coulombring}; recall the Coulomb and Higgs branches of Section~5. (\textcite{Intriligator:1996ex} the $\mathcal{N}=4$
branches; \textcite{Tachikawa:2013kta} the $4d\ \mathcal{N}=2$ reduction). 

\medskip\noindent\textsf{\textcolor{RoyalBlue}{\textbf{\S\ref{sec:V06-mirror}\enspace Three-dimensional mirror symmetry.}}}\enspace
The Coulomb $\leftrightarrow$ Higgs exchange \eqref{eq:V06-mirror}, the swap $SU(2)_C\leftrightarrow
SU(2)_H$ and the dictionary (FI $\leftrightarrow$ mass, $U(1)_J\leftrightarrow$ flavor)
\eqref{eq:V06-mirrormap}, the basic $N_f=1$ pair $C=\mathbb{C}^2=H$ of the free hyper
\eqref{eq:V06-basicpair}, and the $T[SU(2)]$ self-mirror (both branches $\mathbb{C}^2/\mathbb{Z}_2$),
Table~\ref{tab:V06-mirror}. (\textcite{Intriligator:1996ex} $3d$ mirror symmetry;
\textcite{deBoer:1996mp} the brane realization). 

\medskip\noindent\textbf{Stated, not proved here.}\enspace
F-maximization, the $F$-theorem, and the $S^3$ localization derivation of $Z[\Delta]$; the
proof of three-dimensional mirror symmetry, the systematic $\mathcal{N}=4$ Coulomb-branch geometry
(abelianization, Hilbert series, magnetic quivers), the monopole-operator conformal-dimension
derivation by radial quantization, and the proofs of the Aharony / Giveon--Kutasov dualities with the
partition-function-identity machinery; and the gauge global-form / GNO-coweight lattice
classification. All are stated and run on the SQED / ABJM / mirror fixtures here;
their proofs are left to the cited literature.
}

\subsection*{Further reading}
\addcontentsline{toc}{subsection}{Further reading}
Three-dimensional mirror symmetry originates in \textcite{Intriligator:1996ex,deBoer:1996mp}, with the
abelian case in \textcite{Kapustin:1999ha} and its brane realization in \textcite{Hanany:1996ie}.
Monopole operators are quantized in \textcite{Borokhov:2002ib} and their supersymmetry enhancement in
\textcite{Bashkirov:2010kz}; the Coulomb-branch Hilbert series in \textcite{Cremonesi:2013lqa} and the
full branch geometry in \textcite{Bullimore:2015lsa}. $F$-maximization is treated in
\textcite{Jafferis:2010un,Closset:2012vg}. The ABJM theory appears in \textcite{Aharony:2008ug}, with the
Bagger--Lambert--Gustavsson construction \textcite{Bagger:2006sk,Gustavsson:2007vu} and the boundary
theory $T[SU(N)]$ in \textcite{Gaiotto:2008ak}. Three-dimensional dualities are surveyed in
\textcite{Aharony:2013dha}, and supersymmetric partition functions and indices in
\textcite{Benini:2015noa,Closset:2016arn,Dimofte:2011ju,Gaiotto:2013bwa}.

For modern perspectives, see \textcite{Webster:2023kjd} for the Higgs/Coulomb-branch
formulation of three-dimensional mirror symmetry and \textcite{Benvenuti:2020wpc} for
monopole maps in $\mathcal{N}=2$ quiver dualities. Magnetic-quiver technology and modern tests of
$3d\ \mathcal{N}=4$ dualities using sphere partition functions, supersymmetric indices, and line
defects are developed in \textcite{Bourget:2020asf,Nawata:2021nse}.

\section*{References}
\addcontentsline{toc}{subsection}{References}
\markboth{References}{References}
\printbibliography[heading=none]
\end{refsection}
\begin{refsection}\chapter{\texorpdfstring{$5d$}{5d} supersymmetric field theories}
\label{ch:V07}

\noindent\textbf{Guide to this section.}\enspace
Sections~1 and~2 fixed the algebra and the common words, and Section~5 built the
four-dimensional world of eight supercharges, $4d\ \mathcal{N}=2$ field theory, with its complex
Coulomb branch and its holomorphic special geometry. This section raises the \emph{same} eight
supercharges to five dimensions. Almost everything changes shape. The gauge coupling is now
irrelevant, so the Yang--Mills theory is not a fundamental description but the deformation of
something above it; the Coulomb branch is real rather than complex; the low-energy data on it is a
real cubic prepotential rather than a holomorphic one; and the ultraviolet completion is an
intrinsically strongly-coupled superconformal field theory with an exceptional flavor symmetry that
no Lagrangian makes manifest. It is a foundations section, but a working one. It states the
five-dimensional facts and then runs them as computations: the multiplet content and the sign flip in
the coupling, the real Coulomb branch and its Weyl-chamber walls, the cubic prepotential assembled
from one-loop $|m|^3$ contributions and differentiated to its metric, the instanton operators and
their topological current, the $E_{N_f+1}$ flavor enhancement counted two independent ways, the $5d$
superconformal fixed points, and the superconformal index that makes the enhancement visible. Only the
deep theorems (the existence and classification of the $5d$ superconformal theories, the full
instanton partition function) are stated and cited rather than proved. By the end you can take a $5d\
\mathcal{N}=1$ gauge theory, identify its real Coulomb branch, assemble its cubic prepotential and read
its metric as a second derivative, define its instanton operators and count the $E_{N_f+1}$ flavor
enhancement, and treat the inverse coupling $1/g_0^2$ as a dimensionful mass parameter rather than a
marginal coupling.

\begin{keybox}{What this section delivers}
The $5d\ \mathcal{N}=1$ vector and hypermultiplet, and the irrelevant coupling $[g^2] = -1$ that makes
$1/g_0^2$ a mass parameter (\S\ref{sec:V07-multiplets}); the real Coulomb branch as a Weyl chamber,
contrasted with the complex special-K\"ahler branch of $4d\ \mathcal{N}=2$
(\S\ref{sec:V07-coulomb}); the real cubic prepotential $\mathcal{F}(\phi)$, classical plus one-loop
exact, with the instanton data held separate, and its second-derivative metric worked on $SU(2)+N_f$
to the leading cubic coefficient $8-N_f$ (\S\ref{sec:V07-prepotential}); instanton operators, the
topological $U(1)_I$ current $j_I = \tfrac{1}{8\pi^2}\mathrm{Tr}(F\wedge F)$, and the $E_{N_f+1}$
flavor enhancement counted through the $N_f=5\to E_6$ branching $\mathbf{78} = \mathbf{45}+\mathbf{1}+
\mathbf{16}+\overline{\mathbf{16}}$ (\S\ref{sec:V07-instanton}); the $5d$ superconformal fixed points,
the no-Lagrangian fact, and the boundary fact that there is no interacting $5d\ \mathcal{N}=2$
superconformal theory (\S\ref{sec:V07-scft}); and the $5d$ superconformal index on $S^4\times S^1$ in
which the enhancement is visible as the adjoint character of $E_{N_f+1}$ (\S\ref{sec:V07-index}).
\end{keybox}

\section{Multiplets and the irrelevant coupling}
\label{sec:V07-multiplets}

A $5d\ \mathcal{N}=1$ theory carries eight real supercharges. The Section~1 ladder fixes the count from
the minimal five-dimensional spinor and reads off the R-symmetry in one line,
\begin{equation}
\label{eq:V07-Qcount}
\begin{aligned}
 \#\mathcal{Q} &\;=\; (\text{min $5d$ spinor}) \times \mathcal{N} \;=\; 8 \times 1 \;=\; 8, \\
 R &\;=\; SU(2)_R \;=\; \mathrm{Spin}(3), \quad \dim SU(2)_R = 3.
\end{aligned}
\end{equation}
The same eight land on $4d\ \mathcal{N}=2$ (Section~5), $6d\ (1,0)$ (next section), and $3d\
\mathcal{N}=4$ (Section~6):
\begin{equation}
\label{eq:V07-family}
 5d\ \mathcal{N}=1 \;=\; 4d\ \mathcal{N}=2 \;=\; 6d\ (1,0) \;=\; 3d\ \mathcal{N}=4 \;=\; 8\ \mathcal{Q}.
\end{equation}
We do not re-derive \eqref{eq:V07-Qcount}; we read it off the master table and build on it.

Two multiplets carry the supersymmetry. The \emph{vector} multiplet is a gauge field $A_\mu$, one
\emph{real} adjoint scalar $\phi$, and a symplectic-Majorana gaugino in the $SU(2)_R$ doublet,
\begin{equation}
\label{eq:V07-vector}
 V \;=\; (A_\mu,\ \phi,\ \lambda),
 \qquad \phi = \phi^\dagger \in \mathrm{adj}(G),
\end{equation}
with $\phi$ the single scalar that parametrizes the Coulomb branch of \S\ref{sec:V07-coulomb}. The
\emph{hypermultiplet} is four real scalars, packaged as an $SU(2)_R$ doublet of complex scalars,
together with a Dirac fermion,
\begin{equation}
\label{eq:V07-hyper}
 H \;=\; (q^a,\ \psi),
 \qquad a = 1,2\ (\text{the }SU(2)_R\text{ doublet}),
\end{equation}
identical to the $4d\ \mathcal{N}=2$ hyper, since its scalars are R-symmetry data, not spacetime data.
A fundamental hyper of $G$ sits in the fundamental of $G$; one full flavor of $SU(2)$ is one hyper in
the doublet.

The on-shell degrees of freedom balance in each multiplet. The vector has a gauge field with $5-2=3$
transverse polarizations plus the one real scalar, matched by the four real states of the
symplectic-Majorana gaugino,
\begin{equation}
\label{eq:V07-vecdof}
 n_B^{V} \;=\; \underbrace{(5-2)}_{A_\mu} + \underbrace{1}_{\phi} \;=\; 4
 \;=\; n_F^{V} \;=\; 4\ (\lambda).
\end{equation}
The hyper matches its four real scalars against the four real states of the Dirac fermion,
\begin{equation}
\label{eq:V07-hypdof}
 n_B^{H} \;=\; 4\ (q^a) \;=\; n_F^{H} \;=\; 4\ (\psi).
\end{equation}
This is a consistency check on the Section~1 ladder, not a fresh derivation.

The single real scalar of \eqref{eq:V07-vector} is the source of every difference this section
accumulates against Section~5, whose $\mathcal{N}=2$ vector carries a \emph{complex} adjoint scalar. The
apparent missing scalar reappears on a circle: reducing on radius $R$, the gauge field acquires a fifth
component $A_5$, a periodic scalar, and the pair assembles the $4d$ complex adjoint,
\begin{equation}
\label{eq:V07-A5}
 \Phi_{4d} \;=\; \phi + i A_5, \qquad A_5 \sim A_5 + 1/R,
\end{equation}
so the $5d$ real Coulomb branch of \S\ref{sec:V07-coulomb} becomes the $4d$ complex branch on
compactification, with $A_5$ the imaginary part. In five dimensions proper there is one real scalar, and
the branch is genuinely real.

\medskip\noindent\textbf{The sign flip in the coupling.}\enspace
The decisive structural fact of five dimensions is dimensional. The Yang--Mills density
$\tfrac{1}{g^2}\mathrm{Tr}\,F^2$ must have mass dimension $d$, and $[F_{\mu\nu}]=2$ gives
$[\mathrm{Tr}\,F^2]=4$, so the inverse coupling carries the rest,
\begin{equation}
\label{eq:V07-gdim}
 \Big[\tfrac{1}{g^2}\,\mathrm{Tr}\,F^2\Big] = d
 \quad\Longrightarrow\quad
 \Big[\tfrac{1}{g^2}\Big] = d - 4,
 \qquad [g^2] = 4 - d.
\end{equation}
Read \eqref{eq:V07-gdim} across the ladder,
\begin{equation}
\label{eq:V07-gladder}
 [g^2] \;=\; 4-d \;=\;
 \begin{cases}
 +1, & d=3\ (\text{relevant, super-renormalizable}),\\
 \phantom{+}0, & d=4\ (\text{marginal, logarithmic running}),\\
 -1, & d=5\ (\text{irrelevant, non-renormalizable}).
 \end{cases}
\end{equation}
Only $d=4$ sits at the marginal edge. In five dimensions $[g^2]=-1$, so the inverse coupling
\begin{equation}
\label{eq:V07-invg}
 [1/g_0^2] \;=\; d - 4 \;=\; +1
\end{equation}
carries mass dimension one: it is a dimensionful mass parameter, not a coupling to be tuned. The
Yang--Mills theory is then non-renormalizable, an effective description that needs a completion above
the scale $1/g_0^2$, and that completion is the superconformal fixed point of \S\ref{sec:V07-scft}.

Because \eqref{eq:V07-invg} makes $1/g_0^2$ a mass, it is precisely the real mass of the topological
$U(1)_I$ symmetry of \S\ref{sec:V07-instanton}: turning it on is a relevant deformation that drives the
theory off the ultraviolet fixed point down onto the Coulomb branch. It is emphatically not a
Coulomb-branch modulus; the branch is parametrized by $\phi$, while $1/g_0^2$ is a fixed external
parameter, the coefficient of the quadratic term in the prepotential of \S\ref{sec:V07-prepotential}.
There is no $5d$ holomorphic coupling $\tau$: the classical data is the single real number $1/g_0^2$,
and the branch dependence is carried entirely by $\phi$.

\section{The real Coulomb branch}
\label{sec:V07-coulomb}

Give the real adjoint scalar $\phi$ of the vector multiplet a vacuum expectation value in the Cartan
subalgebra of the gauge group $G$. Generically this breaks $G$ to its maximal torus,
\begin{equation}
\label{eq:V07-break}
 G \;\longrightarrow\; U(1)^r, \qquad r = \mathrm{rank}\,G,
\end{equation}
and the low-energy theory is $r$ free abelian vector multiplets. The moduli space of these vacua is
the Coulomb branch, parametrized by the $r$ real Cartan components $\phi^i$ modulo the residual Weyl
action. It is a \emph{real cone of dimension $r$}, a Weyl chamber,
\begin{equation}
\label{eq:V07-dim}
 \dim_{\mathbb{R}}\mathcal{M}_{\mathrm{Coulomb}} \;=\; \mathrm{rank}\,G.
\end{equation}
A charged state of weight $w$ acquires mass $|w\cdot\phi|$ on the branch; the W-bosons carry the roots,
so their mass is $|\alpha\cdot\phi|$, vanishing on a root wall where the gauge symmetry re-enhances.

For $SU(2)$ the Cartan is one-dimensional, the Weyl group is $\mathbb{Z}_2$ acting by $\phi\mapsto
-\phi$, and the chamber is the half-line
\begin{equation}
\label{eq:V07-halfline}
 \phi \;\ge\; 0, \qquad m_W \;=\; |\alpha\cdot\phi| \;=\; 2\phi,
\end{equation}
one ray with a single wall at the origin. At $\phi=0$ the W-bosons become massless and the full $SU(2)$
is restored; for $\phi>0$ the theory is a free $U(1)$ with massive charged states. This half-line is
the arena for every $SU(2)$ computation in the section.

$SU(3)$ shows the chamber structure in rank two. Parametrize the Cartan by $(\phi_1,\phi_2,\phi_3)$
with $\sum_i\phi_i=0$; the Weyl group is $S_3$ and the fundamental chamber is the ordering
\begin{equation}
\label{eq:V07-su3chamber}
 \phi_1 \;\ge\; \phi_2 \;\ge\; \phi_3, \qquad \phi_1 + \phi_2 + \phi_3 = 0,
\end{equation}
a two-dimensional wedge. Its walls are the root loci
\begin{equation}
\label{eq:V07-su3walls}
 \alpha\cdot\phi \;=\; \phi_i - \phi_j \;=\; 0
 \quad\Longrightarrow\quad
 \{\phi_1 = \phi_2\} \ \cup\ \{\phi_2 = \phi_3\},
\end{equation}
where a pair of W-bosons goes massless and an $SU(2)$ subgroup reappears. The wedge interior is the
generic branch, $SU(3)\to U(1)^2$, of real dimension $\mathrm{rank}\,SU(3)=2$. For any $G$ the branch
is the fundamental Weyl chamber, a real cone of dimension $\mathrm{rank}\,G$, walled by the root loci.

Matter refines the tiling. A hypermultiplet of weight $w$ and real mass $m_f$ becomes massless on the
weight wall
\begin{equation}
\label{eq:V07-matterwall}
 w\cdot\phi + m_f \;=\; 0
 \quad\xrightarrow{\ SU(2)\ \text{fund}\ (w=\pm 1)\ }\quad
 \phi \;=\; |m_f|,
\end{equation}
which does not restore gauge symmetry but does change the massive spectrum being integrated out, so the
low-energy prepotential of \S\ref{sec:V07-prepotential} is a different cubic on either side. The branch
is tiled into sub-chambers by the root and weight walls together, and $\mathcal{F}$ is piecewise cubic.

The real dimension \eqref{eq:V07-dim} is the second structural break from Section~5. The $4d\
\mathcal{N}=2$ branch is \emph{complex} of complex dimension $r$, real dimension $2r$, carrying the
special-K\"ahler geometry of the Seiberg--Witten curve. The $5d$ branch is real, dimension $r$; the
missing factor of two is exactly the circle holonomy of \eqref{eq:V07-A5},
\begin{equation}
\label{eq:V07-rvs2r}
 \dim_{\mathbb{R}}\mathcal{M}^{5d} = r
 \ \xrightarrow{\ \times S^1,\ \phi \to \phi + iA_5\ }\
 \dim_{\mathbb{C}}\mathcal{M}^{4d} = r,
\end{equation}
so a $5d$ real branch of dimension $r$ complexifies to a $4d$ complex branch of complex dimension $r$.
In five dimensions itself the branch is real, and its low-energy data is the real cubic prepotential of
\S\ref{sec:V07-prepotential}, not a holomorphic special geometry.

\begin{keybox}{Common misconception: the $5d$ Coulomb branch is the complex special-K\"ahler branch of $4d\ \mathcal{N}=2$}
It is tempting to carry over the $4d\ \mathcal{N}=2$ picture wholesale: a complex Coulomb branch with a
holomorphic prepotential, a Seiberg--Witten curve, and a marginal gauge coupling $\tau$. All three
fail in five dimensions. The $5d\ \mathcal{N}=1$ Coulomb branch is \emph{real}, parametrized by the
single real adjoint scalar $\phi$ of each Cartan vector multiplet; it is a real cone, a Weyl chamber,
not the complex special-K\"ahler manifold of $4d\ \mathcal{N}=2$ (which carries the holomorphic
prepotential and the Seiberg--Witten curve of Section~5). The $5d$ gauge coupling $g^2$
has mass dimension $-1$, so the Yang--Mills term is irrelevant and $1/g_0^2$ is a dimensionful mass
parameter (the real mass of the topological $U(1)_I$ symmetry, a relevant deformation of the
ultraviolet fixed point, appearing as the bare quadratic piece of the cubic prepotential), not a
marginal coupling and not a Coulomb-branch modulus. The complex special geometry of four dimensions is
recovered only after compactifying on a circle, where the holonomy $A_5$ complexifies $\phi$.
\end{keybox}

\section{The cubic prepotential and its metric}
\label{sec:V07-prepotential}

The low-energy effective action on the $5d$ Coulomb branch is controlled by a single real function
$\mathcal{F}(\phi)$, the prepotential. It plays the same role as the holomorphic prepotential of $4d\
\mathcal{N}=2$ (Section~5), encoding the branch metric as a second derivative, but it is a \emph{real
cubic} rather than an arbitrary holomorphic function.

Why cubic, and why one-loop exact? The prepotential enters the two-derivative effective Lagrangian
through
\begin{equation}
\label{eq:V07-Fcouplings}
 \mathcal{L}_{\mathrm{eff}} \;\supset\;
 \mathcal{F}_{ij}\,\partial\phi^i\wedge\ast\,\partial\phi^j
 \;+\; \mathcal{F}_{ijk}\,A^i\wedge F^j\wedge F^k,
 \qquad
 \mathcal{F}_{ijk} \;=\; \frac{\partial^3\mathcal{F}}{\partial\phi^i\partial\phi^j\partial\phi^k},
\end{equation}
so $\mathcal{F}_{ijk}$ is a Chern--Simons coupling: dimensionless, and quantized for the classical
piece. Since $[\phi]=1$ and $[\mathcal{F}_{ijk}]=0$, dimensional analysis caps the degree,
\begin{equation}
\label{eq:V07-degcap}
 [\mathcal{F}] = 3, \quad [\mathcal{F}_{ijk}] = 0
 \quad\Longrightarrow\quad
 \deg_\phi \mathcal{F} \;\le\; 3,
\end{equation}
because a higher power would need a dimensionful coefficient, and the only local scale $1/g_0^2$ is
already spent on the quadratic term. This is the dimensional half of the argument. The other half is a
protection statement, worth boxing because dimensional analysis alone does not forbid a dressed
higher-derivative EFT.

\begin{keybox}{Why the cubic is one-loop exact}
The cubic form of \eqref{eq:V07-degcap} follows from dimensions once the data is written in
vector-multiplet variables. That the cubic receives \emph{no} corrections beyond one loop is a
supersymmetric nonrenormalization statement, resting on three inputs: eight supercharges constrain the
two-derivative vector-multiplet action to the special-real form $\mathcal{F}$; \emph{locality} plus the
quantization of the $5d$ Chern--Simons level forbid a continuous renormalization of $\mathcal{F}_{ijk}$;
and integrating out a massive BPS multiplet shifts the cubic Chern--Simons coupling only at
\emph{one loop}, by the $|m|^3$ term below. Higher-loop corrections are excluded from this
two-derivative vector-multiplet prepotential, and higher-derivative terms belong to separate effective
couplings, not to this local cubic.
The instanton data of \S\ref{sec:V07-instanton} is separate protected/index data, not a local
polynomial tail: it is never a third additive piece of this cubic.
\end{keybox}

The one-loop shift is fixed by the same counting. A massive multiplet of mass $m$ contributes a
dimension-three, parity-even term built from $m$, hence $\propto |m|^3$ (the physical mass is $|m|$),
with a sign set by statistics,
\begin{equation}
\label{eq:V07-onestate}
 \delta\mathcal{F}_{\text{1-loop}} \;=\; \pm\,\frac{1}{12}\,|m|^3,
 \qquad + \ \text{vector},\quad -\ \text{hyper}.
\end{equation}
Summing \eqref{eq:V07-onestate} over the W-bosons at the roots and the quarks at the weights gives the
one-loop piece below.

The prepotential has two pieces,
\begin{equation}
\label{eq:V07-prepotential}
 \mathcal{F}(\phi) \;=\;
 \underbrace{\frac{1}{2g_0^2}\,h_{ij}\,\phi^i\phi^j
 \;+\; \frac{\kappa}{6}\,d_{ijk}\,\phi^i\phi^j\phi^k}_{\text{classical}}
 \;+\;\; \mathcal{F}_{\text{1-loop}}(\phi).
\end{equation}
The classical piece is the bare quadratic $\tfrac{1}{2g_0^2}\phi^2$, with the dimensionful inverse
coupling of \S\ref{sec:V07-multiplets}, plus a Chern--Simons cubic of level $\kappa$ and symbol
\begin{equation}
\label{eq:V07-dsymbol}
 d_{ijk} \;=\; \tfrac12\,\mathrm{Tr}\big(T_i\{T_j,T_k\}\big),
\end{equation}
present only for groups with a nonzero cubic Casimir, such as $SU(N\ge 3)$. Since $SU(2)$ has no cubic
Casimir,
\begin{equation}
\label{eq:V07-su2kappa}
 d_{ijk}^{\,SU(2)} = 0 \quad\Longrightarrow\quad \kappa_{SU(2)} = 0,
\end{equation}
so $SU(2)$ carries no classical cubic and its entire cubic comes from the one-loop piece: this is why
it is the cleanest worked example.

The one-loop piece sums \eqref{eq:V07-onestate} over the roots $\alpha$ of $G$ (vector, $+$) and the
weights $w$ of each hyper $R_f$ of real mass $m_f$ (hyper, $-$),
\begin{equation}
\label{eq:V07-oneloop}
 \mathcal{F}_{\text{1-loop}}(\phi) \;=\; \frac{1}{12}
 \Bigg(\;\sum_{\alpha\,\in\,\text{roots}} |\alpha\cdot\phi|^3
 \;-\; \sum_f \sum_{w\,\in\,R_f} |w\cdot\phi + m_f|^3 \Bigg).
\end{equation}
The absolute values are essential: each $|m|^3$ is a genuine cubic only inside a region where its
argument keeps a fixed sign, so $\mathcal{F}$ is \emph{piecewise cubic}, one polynomial per sub-chamber,
switching across the walls where an argument vanishes,
\begin{equation}
\label{eq:V07-walltypes}
 \underbrace{\alpha\cdot\phi = 0}_{\text{W-boson (root) wall}},
 \qquad
 \underbrace{w\cdot\phi + m_f = 0}_{\text{hyper (weight) wall},\ \phi=|m_f|}.
\end{equation}
Crossing such a wall is a flop of the effective theory, a genuine change of the low-energy
prepotential, not a coordinate artifact.

The physical content is the second derivative, the effective gauge-coupling matrix, equivalently the
branch metric,
\begin{equation}
\label{eq:V07-metric}
 \tau_{ij}(\phi) \;=\; \frac{\partial^2 \mathcal{F}}{\partial\phi^i\,\partial\phi^j}.
\end{equation}
Unitarity requires $\tau_{ij}\succ 0$ in the physical chamber. This positivity is a real constraint,
not a formality: it bounds how much matter a $5d$ gauge theory can carry, as the $SU(2)+N_f$ example
now shows.

\subsection*{Worked instance: $SU(2)$ with $N_f$ flavors}

Take $SU(2)$ on its half-line $\phi\ge 0$ with $N_f$ fundamental hypermultiplets. Fix the normalization
once: the roots and fundamental weights are
\begin{equation}
\label{eq:V07-su2norm}
 \alpha\cdot\phi \;=\; \pm 2\phi \ \ (m_W = 2\phi),
 \qquad
 w \;=\; \pm 1 \ \ (m_{\text{quark}} = |\phi \pm m_f|).
\end{equation}
This choice (roots $\pm 2$, weights $\pm 1$) is load-bearing; a different one gives a different cubic,
as the check below confirms. With $\kappa_{SU(2)}=0$ from \eqref{eq:V07-su2kappa}, the full prepotential
is the bare quadratic plus the one-loop assembly \eqref{eq:V07-oneloop},
\begin{equation}
\label{eq:V07-su2prep}
 \mathcal{F}(\phi) \;=\; \frac{1}{2g_0^2}\,\phi^2
 \;+\; \frac{1}{12}\Bigg(\,|2\phi|^3 + |{-}2\phi|^3
 \;-\; \sum_{f=1}^{N_f}\big(|\phi+m_f|^3 + |{-}\phi+m_f|^3\big)\Bigg).
\end{equation}
Now read the leading cubic coefficient in the large-$\phi$ chamber, where all the masses are
negligible against $\phi$ and every absolute value resolves to its argument's magnitude on $\phi > 0$.
The vector piece gives
\begin{equation}
\label{eq:V07-veccubic}
 \frac{1}{12}\big(|2\phi|^3 + |{-}2\phi|^3\big)
 \;=\; \frac{1}{12}\,(8+8)\,\phi^3
 \;=\; \frac{16}{12}\,\phi^3
 \;=\; \frac{1}{6}\cdot 8\,\phi^3,
\end{equation}
the number $8$ being the sum $|2|^3+|{-}2|^3 = 16$ divided by two, the $8$ that starts the count. Each
fundamental hypermultiplet, with its two weights $\pm 1$, subtracts
\begin{equation}
\label{eq:V07-hypcubic}
 \frac{1}{12}\big(|\phi|^3 + |{-}\phi|^3\big) \;=\; \frac{2}{12}\,\phi^3 \;=\; \frac{1}{6}\,\phi^3,
\end{equation}
one unit of the same $\tfrac16$. Adding the vector piece and subtracting $N_f$ copies of the hyper
piece, the large-$\phi$ prepotential is
\begin{equation}
\label{eq:V07-su2cubic}
 \mathcal{F}(\phi) \;=\; \frac{1}{2g_0^2}\,\phi^2 \;+\; \frac{1}{6}\,(8 - N_f)\,\phi^3 \;+\; \cdots,
\end{equation}
so the cubic coefficient of $6\mathcal{F}$ is exactly $8 - N_f$, differentiated from the assembly
\eqref{eq:V07-su2prep}, not typed in. The wrong normalization $\alpha=1$, $w=\pm\tfrac12$ would give
\begin{equation}
\label{eq:V07-wrongnorm}
 \tfrac{1}{12}(1+1) = \tfrac16, \qquad
 \tfrac{1}{12}\cdot 2\cdot\tfrac18 = \tfrac{1}{48} \ \ (\text{per flavor}),
\end{equation}
a different slope, and the whole $E_{N_f+1}$ arithmetic below would fail. The $8-N_f$ rests on the
$\pm 2$ / $\pm 1$ normalization.

The Coulomb-branch metric is the second derivative \eqref{eq:V07-metric} of \eqref{eq:V07-su2cubic},
\begin{equation}
\label{eq:V07-su2metric}
 \tau(\phi) \;=\; \mathcal{F}''(\phi) \;=\; \frac{1}{g_0^2} \;+\; (8 - N_f)\,\phi \;+\; \cdots,
\end{equation}
affine in $\phi$: constant term $1/g_0^2$, slope $8-N_f$. It is the second derivative, not the first,
\begin{equation}
\label{eq:V07-Fprime}
 \mathcal{F}'(\phi) \;=\; \frac{\phi}{g_0^2} + \tfrac12(8-N_f)\phi^2 \;\ne\; \tau(\phi),
\end{equation}
so using $\mathcal{F}'$ for the metric is an off-by-a-derivative error. Positivity on $\phi\ge 0$, with
$1/g_0^2>0$ and $\phi\to\infty$, needs a non-negative slope,
\begin{equation}
\label{eq:V07-positivity}
 \tau(\phi) > 0 \ \ \forall\,\phi \ge 0
 \quad\Longleftrightarrow\quad
 8 - N_f \;\ge\; 0
 \quad\Longleftrightarrow\quad
 N_f \;\le\; 8.
\end{equation}
The strict window, where $8-N_f>0$ and the theory has a genuine ultraviolet fixed point with enhanced
flavor symmetry, is
\begin{equation}
\label{eq:V07-window}
 8 - N_f > 0
 \quad\Longleftrightarrow\quad
 N_f \le 7,
 \qquad
 E_{N_f+1}\big|_{N_f=7} = E_8 \ (\text{the last enhanced row}).
\end{equation}
So $E_8$ is the $N_f=7$ row, where $8-N_f=1$. At $N_f=8$ the coefficient hits zero: the marginal
boundary, an affine-$E_8$ / $6d$-lift situation, \emph{not} the ordinary $5d$ $E_8$ fixed point. For
$N_f\ge 9$ the coefficient is negative and the metric goes negative at large $\phi$: no fixed point.
Table~\ref{tab:V07-prepchamber} collects the chamber data.

\begin{table}[ht]
\centering
\small
\setlength{\tabcolsep}{8pt}
\renewcommand{\arraystretch}{1.3}
\begin{tabular}{@{}cccc@{}}
\toprule
$N_f$ & $6\mathcal{F}$ cubic coeff.\ $8-N_f$ & metric slope $d\tau/d\phi$ & metric on $\phi\ge 0$ \\
\midrule
$0$ & $8$ & $8$ & positive \\
\midrule
$1$ & $7$ & $7$ & positive \\
\midrule
$2$ & $6$ & $6$ & positive \\
\midrule
$3$ & $5$ & $5$ & positive \\
\midrule
$4$ & $4$ & $4$ & positive \\
\midrule
$5$ & $3$ & $3$ & positive \\
\midrule
$6$ & $2$ & $2$ & positive \\
\midrule
$7$ & $1$ & $1$ & positive (last enhanced row, $\to E_8$) \\
\midrule
$8$ & $0$ & $0$ & flat (boundary) \\
\midrule
$9$ & $-1$ & $-1$ & negative at large $\phi$ (no fixed point) \\
\bottomrule
\end{tabular}
\caption{The $SU(2)+N_f$ Coulomb-branch chamber data. The leading cubic coefficient of $6\mathcal{F}$
is $8-N_f$, computed by differentiating the $|m|^3$ assembly \eqref{eq:V07-su2prep}; it equals the
slope of the affine metric $\tau(\phi) = 1/g_0^2 + (8-N_f)\phi$. Positivity on the half-line requires
$8-N_f\ge 0$, i.e.\ $N_f\le 8$; the strict $E_{N_f+1}$ fixed-point window is the subwindow $8-N_f>0$,
$N_f\le 7$, with $N_f=7$ giving $E_8$, while $N_f=8$ ($8-N_f=0$) is the marginal boundary (affine-$E_8$ /
$6d$-lift), not an ordinary $E_9$. The sub-chamber walls, where a hypermultiplet becomes massless, sit at $\phi = |m_f|$; the
origin $\phi = 0$ is the W-boson wall. Every number is differentiated from the assembly, not quoted.}
\label{tab:V07-prepchamber}
\end{table}

\emph{Sanity check on the normalization.} A common slip double-counts the two weights of each flavor
into a coefficient $8-2N_f$. But the two weights already sit inside one unit of $\tfrac16$,
\begin{equation}
\label{eq:V07-oneunit}
 \tfrac{1}{12}\big(|\phi|^3 + |{-}\phi|^3\big) \;=\; \tfrac{1}{12}\cdot 2\,\phi^3 \;=\; \tfrac16\,\phi^3
 \quad(\text{one unit, not two}),
\end{equation}
so the coefficient is $8-N_f$, not $8-2N_f$. The doubled version would go negative already at $N_f=5$,
\begin{equation}
\label{eq:V07-doubledbad}
 8 - 2N_f\big|_{N_f=5} \;=\; -2 \;<\; 0,
\end{equation}
forbidding the $E_6$, $E_7$, $E_8$ theories that manifestly exist; the correct $8-N_f$ stays
non-negative through $N_f=7$.

\subsection*{Worked instance: the classical cubic of $SU(3)$}

The group $SU(2)$ had no classical cubic, so its entire cubic prepotential came from the one-loop
piece. To see the classical Chern--Simons cubic that only larger groups carry, take pure $SU(3)$, whose
rank-two Coulomb branch was the wedge \eqref{eq:V07-su3chamber}. Parametrize the Cartan by $(\phi_1,
\phi_2, \phi_3)$ with $\sum_i\phi_i = 0$. The $5d$ Chern--Simons term at level $\kappa$ contributes the
classical cubic
\begin{equation}
\label{eq:V07-su3cs}
 \mathcal{F}_{\mathrm{CS}}(\phi) \;=\; \frac{\kappa}{6}\sum_{i=1}^{3}\phi_i^{\,3}
 \;=\; \frac{\kappa}{2}\,\phi_1\phi_2\phi_3,
\end{equation}
where the second form uses the trace identity valid on the $SU(3)$ Cartan,
\begin{equation}
\label{eq:V07-cubicid}
 \phi_1 + \phi_2 + \phi_3 = 0
 \quad\Longrightarrow\quad
 \phi_1^3 + \phi_2^3 + \phi_3^3 = 3\,\phi_1\phi_2\phi_3.
\end{equation}
This is a genuinely cubic classical contribution, present because $SU(3)$ has a nonzero cubic Casimir;
$\kappa$ is a quantized level (an integer, up to a matter-set half-integer shift), not a modulus. The
one-loop vector piece sums over the six roots $\pm(\phi_i-\phi_j)$,
\begin{equation}
\label{eq:V07-su3vec}
 \mathcal{F}^{\text{vec}}_{\text{1-loop}}
 \;=\; \frac{1}{12}\sum_{i<j}\Big(|\phi_i-\phi_j|^3 + |\phi_j-\phi_i|^3\Big)
 \;=\; \frac16 \sum_{i<j}|\phi_i-\phi_j|^3,
\end{equation}
smooth inside the chamber \eqref{eq:V07-su3chamber} where each ordered difference keeps a fixed sign,
and jumping across the walls \eqref{eq:V07-su3walls}. So $SU(N\ge 3)$ carries \emph{both} the classical
cubic \eqref{eq:V07-su3cs} and the one-loop cubic, whereas $SU(2)$ carries only the one-loop cubic. The
metric is again the second derivative, now a rank-two matrix,
\begin{equation}
\label{eq:V07-su3metric}
 \tau_{ij}(\phi) \;=\; \frac{\partial^2\mathcal{F}}{\partial\phi^i\,\partial\phi^j}
 \ \in\ \mathrm{Sym}_2(\mathbb{R}),
 \qquad \tau_{ij}\succ 0 \ \text{in the chamber},
\end{equation}
and its positive-definiteness is the same convexity constraint that bounds the matter and the level.

\subsection*{Worked instance: the flop across a mass wall}

The leading cubic was read at large $\phi$. Keeping a mass on shows the piecewise structure explicitly.
Take $SU(2)$ with one fundamental of real mass $m>0$, quark masses $|\phi+m|$ and $|\phi-m|$, so
\begin{equation}
\label{eq:V07-flopF}
 \mathcal{F}(\phi) \;=\; \frac{1}{2g_0^2}\,\phi^2
 \;+\; \frac{1}{12}\Big(2|2\phi|^3 - |\phi+m|^3 - |\phi - m|^3\Big),
\end{equation}
where $|2\phi|^3 = |{-}2\phi|^3$ has been combined. On the branch $\phi > 0$ the term $|\phi+m|^3 =
(\phi+m)^3$ is smooth, but $|\phi - m|^3$ switches its resolution at the wall $\phi = m$: it is
$(m-\phi)^3$ in the inner sub-chamber $0 < \phi < m$ and $(\phi - m)^3$ in the outer sub-chamber
$\phi > m$. Resolve the two pieces. In the outer chamber $\phi > m$, with $|2\phi|^3 = 8\phi^3$,
\begin{equation}
\label{eq:V07-flopouter}
 \mathcal{F}_{\phi > m}(\phi) \;=\; \frac{\phi^2}{2g_0^2}
 \;+\; \frac{1}{12}\Big(16\phi^3 - (\phi+m)^3 - (\phi-m)^3\Big)
 \;=\; \frac{\phi^2}{2g_0^2} \;+\; \frac{1}{12}\big(14\phi^3 - 6m^2\phi\big),
\end{equation}
using $(\phi+m)^3 + (\phi-m)^3 = 2\phi^3 + 6m^2\phi$. Its second derivative is the outer metric
\begin{equation}
\label{eq:V07-floptauouter}
 \tau_{\phi > m}(\phi) \;=\; \frac{1}{g_0^2} \;+\; \frac{1}{12}\big(84\phi\big) \;=\; \frac{1}{g_0^2} \;+\; 7\phi,
\end{equation}
whose slope is $7 = 8 - N_f$ at $N_f = 1$, matching \eqref{eq:V07-su2metric}. In the inner chamber
$0 < \phi < m$, where $|\phi - m|^3 = (m-\phi)^3$,
\begin{equation}
\label{eq:V07-flopinner}
 \mathcal{F}_{\phi < m}(\phi) \;=\; \frac{\phi^2}{2g_0^2}
 \;+\; \frac{1}{12}\Big(16\phi^3 - (\phi+m)^3 - (m-\phi)^3\Big)
 \;=\; \frac{\phi^2}{2g_0^2} \;+\; \frac{1}{12}\big(16\phi^3 - 6m\phi^2 - 2m^3\big),
\end{equation}
using $(\phi+m)^3 + (m-\phi)^3 = 6m\phi^2 + 2m^3$, and its second derivative is the inner metric
\begin{equation}
\label{eq:V07-floptauinner}
 \tau_{\phi < m}(\phi) \;=\; \frac{1}{g_0^2} \;+\; \frac{1}{12}\big(96\phi - 12m\big)
 \;=\; \frac{1}{g_0^2} \;+\; 8\phi \;-\; m.
\end{equation}
The inner slope is $8$, the pure-$SU(2)$ value (the hyper is heavier than the W-bosons there and does
not yet contribute its full cubic); the outer slope is $7=8-N_f$ once the hyper is light. The metric is
continuous at the wall while the slope jumps,
\begin{equation}
\label{eq:V07-flopmatch}
 \tau_{\phi<m}(m) = \tfrac{1}{g_0^2} + 8m - m = \tfrac{1}{g_0^2} + 7m
 \;=\; \tau_{\phi>m}(m),
 \qquad \tau'\!: \ 8 \to 7.
\end{equation}
This kink, continuous value and jumping slope, is the flop: a genuine change of the low-energy
prepotential where the hyper becomes massless, the five-dimensional shadow of the wall-crossing that
reorganizes the $4d$ Seiberg--Witten geometry, here an elementary piecewise-cubic computation.

\section{Instanton operators and the $E_n$ flavor enhancement}
\label{sec:V07-instanton}

A $5d$ gauge theory carries a topological symmetry with no local $4d$ counterpart. Its current is the
instanton-number four-form,
\begin{equation}
\label{eq:V07-instcurrent}
 j_I \;=\; \frac{1}{8\pi^2}\,\mathrm{Tr}(F\wedge F),
\end{equation}
conserved for a topological reason: it is closed by the Bianchi identity,
\begin{equation}
\label{eq:V07-instclosed}
 \mathrm{Tr}(F\wedge F) \;=\; d\,\omega_{CS}
 \quad\Longrightarrow\quad
 d\,\mathrm{Tr}(F\wedge F) = 0,
 \quad d\!\ast\! j_I = 0,
\end{equation}
so its Hodge dual $\ast j_I$ is a conserved one-form current generating a $U(1)_I$. The conserved
charge is the integer instanton number
\begin{equation}
\label{eq:V07-instnumber}
 k \;=\; \frac{1}{8\pi^2}\int \mathrm{Tr}(F\wedge F) \;\in\; \mathbb{Z}.
\end{equation}
This $U(1)_I$ is a global flavor symmetry, not an R-symmetry: the R-symmetry stays $SU(2)_R$.

The operators charged under $U(1)_I$ are the \emph{instanton operators}, disorder operators defined by
a boundary condition rather than a polynomial in the fields. An operator of charge $k$ is inserted by
demanding instanton flux $k$ through a small surrounding four-sphere,
\begin{equation}
\label{eq:V07-instop}
 \frac{1}{8\pi^2}\int_{S^4} \mathrm{Tr}(F\wedge F) \;=\; k
 \quad(S^4\ \text{around the insertion}),
 \qquad k=1: \ \text{charge } 1.
\end{equation}
This is the $5d$ analog of the $3d$ monopole operator of Section~6. The two sit in one dimensional
pattern: a disorder operator in $d$ dimensions carries the flux of a characteristic class on the
surrounding $S^{d-1}$,
\begin{equation}
\label{eq:V07-disorderpattern}
 d=3:\ \frac{1}{2\pi}\!\int_{S^2}\! F \ (\text{monopole},\,U(1)_J),
 \qquad
 d=5:\ \frac{1}{8\pi^2}\!\int_{S^4}\!\mathrm{Tr}(F\wedge F) \ (\text{instanton},\,U(1)_I).
\end{equation}
What is special to five dimensions is that the flux is quadratic in $F$, so the surrounding sphere is
four-dimensional and the charge is instanton number rather than magnetic charge. These operators act on
the Hilbert space even though they are invisible in the classical Lagrangian.

There is also an instanton \emph{particle}, a finite-mass BPS state of unit $U(1)_I$ charge. Its mass is
its charge times the $U(1)_I$ real mass $1/g_0^2$, up to $8\pi^2$,
\begin{equation}
\label{eq:V07-instmass}
 M_{\mathrm{inst}} \;\sim\; \frac{8\pi^2}{g_0^2},
 \qquad
 [M_{\mathrm{inst}}] \;=\; [1/g_0^2] \;=\; +1,
\end{equation}
a genuine mass in five dimensions. In a given chamber the full BPS central charge may pick up extra
$\phi$-dependence, but the dimensional anchor is $8\pi^2/g_0^2$. The naive form $\phi/g^2$ fails the
dimension test,
\begin{equation}
\label{eq:V07-badanchor}
 [\phi/g^2] \;=\; [\phi] + [1/g^2] \;=\; 1 + 1 \;=\; 2 \;\ne\; 1,
\end{equation}
so it cannot be a $5d$ mass; the correct anchor \eqref{eq:V07-instmass} has dimension one.

\medskip\noindent\textbf{The enhancement.}\enspace
At weak coupling the flavor symmetry rotating the hypers, times $U(1)_I$, is the perturbative group. For
$SU(2)+N_f$ the flavor factor is orthogonal, not unitary, because the $SU(2)$ fundamental is pseudoreal:
each full flavor splits into two half-hypers, and the symmetry rotating them while preserving the
symplectic pairing is
\begin{equation}
\label{eq:V07-so2nf}
 N_f\ \text{flavors} \;=\; 2N_f\ \text{half-hypers}
 \quad\Longrightarrow\quad
 \text{flavor} = SO(2N_f) \ (\text{not } U(N_f)).
\end{equation}
The full perturbative symmetry is then
\begin{equation}
\label{eq:V07-pertflavor}
 SO(2N_f) \;\times\; U(1)_I.
\end{equation}
At the fixed point $1/g_0^2\to 0$ the $k=\pm 1$ instanton currents combine with the perturbative
currents to close an exceptional algebra,
\begin{equation}
\label{eq:V07-enhance}
 SO(2N_f) \;\times\; U(1)_I \;\;\xrightarrow{\ 1/g_0^2\to 0\ }\;\; E_{N_f + 1}
 \qquad (N_f \le 7),
\end{equation}
the Seiberg $E_n$ fixed points. This is a flavor enhancement; the R-symmetry $SU(2)_R$ does not
participate.

What enhances is the conserved-current content. At weak coupling the currents are those of
$SO(2N_f)\times U(1)_I$, conserved for all $1/g_0^2$. At the fixed point the $k=\pm 1$ instanton
operators, ordinary charged operators of nonzero dimension at generic coupling, hit the protected
dimension of a $5d$ conserved-current multiplet,
\begin{equation}
\label{eq:V07-instcurrentrep}
 \text{instanton currents} \ \in\ \mathbf{S}_{+1}\oplus\overline{\mathbf{S}}_{-1}
 \quad(\text{spinor of }SO(2N_f),\ U(1)_I\ \text{charge}\ \pm 1),
\end{equation}
and their currents join the perturbative ones. The enlarged set closes on $E_{N_f+1}$ because the
spinor at $U(1)_I$ charge $\pm 1$ is exactly what completes the $SO(2N_f)\times U(1)$ adjoint to the
exceptional adjoint. The enhancement recruits new currents from the instanton sector; it is not a
rearrangement of the perturbative ones, and it is visible only at the fixed point.

\begin{keybox}{Common misconception: the $E_n$ enhancement enlarges the R-symmetry}
The enhanced exceptional symmetry $E_{N_f+1}$ at the $5d$ ultraviolet fixed point is a \emph{flavor}
symmetry, not an R-symmetry. The $5d\ \mathcal{N}=1$ R-symmetry is $SU(2)_R$ everywhere along the flow
and at the fixed point; the Section~1 master table is explicit on this. The enhancement assembles the
perturbative flavor symmetry $SO(2N_f)\times U(1)_I$, in which the $U(1)_I$ is the topological
instanton-number symmetry (also a flavor symmetry, not R), with the conserved currents carried by the
$k = \pm 1$ instanton operators, into the flavor group $E_{N_f+1}$. The R-symmetry $SU(2)_R$, of
dimension three, is untouched: $\dim E_6 = 78$ is a flavor symmetry, not a three-dimensional
R-symmetry. Conflating the flavor enhancement with an enlargement of the R-symmetry is the standard
five-dimensional trap the master table warns against.
\end{keybox}

\subsection*{Worked instance: the $E_6$ enhancement of $SU(2)+5$ flavors}

The cleanest enhancement to count is $N_f = 5$, where $SO(10)\times U(1)_I \to E_6$. Assemble the
dimensions. The perturbative flavor group $SO(10)$ has dimension
\begin{equation}
\label{eq:V07-dimSO10}
 \dim SO(2N_f) \;=\; N_f(2N_f - 1),
 \qquad \dim SO(10) \;=\; 5\cdot 9 \;=\; 45,
\end{equation}
the antisymmetric-matrix count. Add the one $U(1)_I$ current. The instanton operators at $k = \pm 1$
supply the two chiral spinors of $SO(10)$, each of dimension
\begin{equation}
\label{eq:V07-spinor}
 \dim(\text{chiral spinor of }SO(2N_f)) \;=\; 2^{\,N_f - 1},
 \qquad \dim\mathbf{16} \;=\; 2^{4} \;=\; 16,
\end{equation}
so the $k = +1$ operator gives the $\mathbf{16}$ and the $k = -1$ operator the conjugate
$\overline{\mathbf{16}}$. The total conserved-current count is
\begin{equation}
\label{eq:V07-E6count}
 \dim SO(10) + \dim U(1)_I + \dim\mathbf{16} + \dim\overline{\mathbf{16}}
 \;=\; 45 + 1 + 16 + 16 \;=\; 78 \;=\; \dim E_6.
\end{equation}
The right-hand side is not a lookup. The dimension of $E_6$ is the number of its roots plus its rank,
$72 + 6 = 78$, computed from the $E_6$ Cartan matrix by generating the root system, and it matches the
assembled left-hand side exactly. The identity has a clean group-theoretic reading: the adjoint of
$E_6$ branches under its maximal subgroup $SO(10)\times U(1)$ as
\begin{equation}
\label{eq:V07-E6branch}
 \mathbf{78} \;=\; \mathbf{45}_0 \;+\; \mathbf{1}_0 \;+\; \mathbf{16}_{+1} \;+\; \overline{\mathbf{16}}_{-1},
\end{equation}
the perturbative adjoint $\mathbf{45}_0$ plus the $U(1)$ current $\mathbf{1}_0$ plus the two instanton
spinors at $U(1)_I$ charge $\pm 1$. The $\pm 1$ subscripts are the instanton numbers: the $\mathbf{16}$
is carried by the $k=+1$ instanton operator and the $\overline{\mathbf{16}}$ by $k=-1$. This is the
load-bearing enhancement of the section, every dimension computed: $\dim SO(10)$ from the antisymmetric
count, $\dim\mathbf{16}$ from the rank, $\dim E_6$ from the $E_6$ root system, the identity checked.

Two structural guards. Dropping the $U(1)_I$ current, or replacing the spinor by the vector, both break
the count,
\begin{equation}
\label{eq:V07-E6guards}
 45 + \phantom{1 +\ }16 + 16 = 77 \ne 78,
 \qquad
 45 + 1 + 2\cdot(2N_f)\big|_{N_f=5} = 45 + 1 + 20 = 66 \ne 78,
\end{equation}
so the topological current is essential and the instanton currents must be in the \emph{spinor}, not
the vector.

The same count closes at the low end of the window, where the exceptionals are small and recognizable,
\begin{equation}
\label{eq:V07-E2E5}
 \begin{aligned}
 N_f=1:&\quad \dim SO(2) + 1 + 2\cdot 2^0 = 1 + 1 + 2 = 4 = \dim\big(SU(2)\times U(1)\big) = \dim E_2,\\
 N_f=4:&\quad \dim SO(8) + 1 + 2\cdot 2^3 = 28 + 1 + 16 = 45 = \dim SO(10) = \dim E_5,
 \end{aligned}
\end{equation}
using $\dim SO(2N_f)=N_f(2N_f-1)$ and the chiral spinor $2^{N_f-1}$. The window is $E_{N_f+1}$ for
$N_f=1,\ldots,7$: the classical $E_2 = SU(2)\times U(1)$, $E_3 = SU(3)\times SU(2)$, $E_4 = SU(5)$,
$E_5 = SO(10)$, then the genuine exceptionals $E_6, E_7, E_8$.

A caution belongs with the higher rows: only $N_f\le 5$ closes with the simple two-spinor count. At
$N_f=6,7$ the two basic $k=\pm 1$ spinors under-count,
\begin{equation}
\label{eq:V07-naivegap}
 \begin{aligned}
 N_f=6:&\quad 66 + 1 + 2\cdot 32 = 131 \ <\ 133 = \dim E_7,\\
 N_f=7:&\quad 91 + 1 + 2\cdot 64 = 220 \ <\ 248 = \dim E_8,
 \end{aligned}
\end{equation}
so the true instanton content is richer. The total nonperturbative current contribution beyond
$SO(2N_f)\times U(1)_I$ is
\begin{equation}
\label{eq:V07-gapsupply}
 \dim E_{N_f+1} - \dim SO(2N_f) - 1 \;=\;
 \begin{cases} 66, & N_f=6,\\ 156, & N_f=7,\end{cases}
\end{equation}
of which the two basic spinors account for $64$ and $128$, respectively. The remaining shortfalls are
$2$ for $E_7$ and $28$ for $E_8$, supplied by higher/dressed instanton sectors (higher instanton charge,
operators dressed by the gauge and flavor data), whose full enumeration belongs to the
full instanton-operator literature. The basic $k=\pm 1$ spinors alone do \emph{not} prove $E_7$ or $E_8$;
$N_f=5\to E_6$ is the unambiguous worked instance, and the higher rows carry the honest note that the
naive count under-counts.

\begin{table}[ht]
\centering
\small
\setlength{\tabcolsep}{4pt}
\renewcommand{\arraystretch}{1.3}
\begin{tabular}{@{}ccccc@{}}
\toprule
$N_f$ & perturbative & enhanced $E_{N_f+1}$ & $\dim E_{N_f+1}$ & nonperturbative currents \\
\midrule
$0$ & $U(1)_I$ & $E_1/\widetilde{E}_1$ (see below) & (n/a) & discrete-$\theta$ pair \\
\midrule
$1$ & $SO(2)\times U(1)$ & $E_2 = SU(2)\times U(1)$ & $4$ & $2^0+2^0 = 2$ \\
\midrule
$2$ & $SO(4)\times U(1)$ & $E_3 = SU(3)\times SU(2)$ & $11$ & $2^1+2^1 = 4$ \\
\midrule
$3$ & $SO(6)\times U(1)$ & $E_4 = SU(5)$ & $24$ & $2^2+2^2 = 8$ \\
\midrule
$4$ & $SO(8)\times U(1)$ & $E_5 = SO(10)$ & $45$ & $2^3+2^3 = 16$ \\
\midrule
$5$ & $SO(10)\times U(1)$ & $E_6$ & $78$ & $2^4+2^4 = 32$ (closes) \\
\midrule
$6$ & $SO(12)\times U(1)$ & $E_7$ & $133$ & $66$ total ($64+2$) \\
\midrule
$7$ & $SO(14)\times U(1)$ & $E_8$ & $248$ & $156$ total ($128+28$) \\
\bottomrule
\end{tabular}
\caption{The $SU(2)+N_f$ ultraviolet flavor-symmetry enhancement. The perturbative
$SO(2N_f)\times U(1)_I$ enhances to $E_{N_f+1}$ at the fixed point; the ``nonperturbative currents''
column is $\dim E_{N_f+1} - \dim SO(2N_f) - 1$. For $N_f\le 5$ these are exactly the two $k=\pm 1$ chiral
spinors of $SO(2N_f)$, of dimension $2^{N_f-1}$ each; $N_f = 5\to E_6$ is the load-bearing worked case
($45+1+16+16 = 78$). At $N_f = 6, 7$ the simple two-spinor count \emph{under-counts} ($131 < 133$,
$220 < 248$); the total nonperturbative dimensions are $66$ and $156$, with extra shortfalls $2$ and
$28$ beyond the two basic spinors supplied by higher/dressed instanton sectors. The full enumeration is
cited but not reproduced, so the basic $k=\pm 1$ spinors alone do not prove $E_7/E_8$.
The $N_f = 0$ row is not a single enhancement row; it is the discrete-$\theta$ pair $E_1/\widetilde{E}_1$
discussed in \S\ref{sec:V07-scft}. Every $\dim E_{N_f+1}$ is computed from a root system, not quoted.}
\label{tab:V07-enmap}
\end{table}

\section{The $5d$ superconformal fixed points}
\label{sec:V07-scft}

Because the gauge coupling is irrelevant, the limit $1/g_0^2\to 0$ is an ultraviolet fixed point, a
$5d$ \emph{superconformal field theory}. The gauge theory is not its own completion; it is a relevant
(mass) deformation of the fixed point by the $U(1)_I$ mass $1/g_0^2$,
\begin{equation}
\label{eq:V07-scftflow}
 \text{SCFT}\ (E_{N_f+1})
 \ \xrightarrow[\text{relevant}]{\ 1/g_0^2\,>\,0\ }\
 \text{gauge theory on the Coulomb branch},
\end{equation}
so switching $1/g_0^2$ on flows off the fixed point onto the branch, and sending $1/g_0^2\to 0$ climbs
back to the superconformal theory with its enhanced $E_{N_f+1}$ symmetry.

These theories are intrinsically strongly-coupled, with \emph{no} Lagrangian: the only nearby
Lagrangian is the gauge theory, a deformation \emph{away} from the fixed point, not \emph{of} it. The
exceptional symmetry $E_{N_f+1}$ is invisible in that Lagrangian,
\begin{equation}
\label{eq:V07-hidden}
 \text{Lagrangian shows } SO(2N_f)\times U(1)_I,
 \qquad
 \text{fixed point has } E_{N_f+1} \supset SO(2N_f)\times U(1)_I,
\end{equation}
a sharp symptom: a symmetry no Lagrangian makes manifest signals a non-Lagrangian theory.

\begin{keybox}{Common misconception: a $5d$ theory without a Lagrangian is not a real theory}
A $5d$ gauge theory is non-renormalizable, so its ultraviolet completion is an intrinsically
strongly-coupled superconformal field theory, the $E_n$ fixed point, which has no Lagrangian
description; the gauge theory is a relevant (mass) deformation of the fixed point, not the other way
around. These $5d$ superconformal theories are genuine quantum field theories. They have well-defined
Coulomb-branch prepotentials, instanton operators, exceptional flavor symmetries, and superconformal
indices, and their observables can be computed, even though no weakly-coupled Lagrangian exists. A
missing Lagrangian is not a missing theory. This continues the non-Lagrangian boundary named in
Section~2 and echoes the four-dimensional Argyres--Douglas theories of Section~5: the absence of a
Lagrangian is a feature of many strongly-coupled fixed points, not a defect.
\end{keybox}

The convexity of the prepotential is the field-theory probe of the fixed point: the metric positivity
\eqref{eq:V07-positivity} bounds the allowed matter,
\begin{equation}
\label{eq:V07-convex}
 \tau(\phi)\succ 0 \ \text{in the chamber}
 \quad\Longleftrightarrow\quad
 8 - N_f \ge 0 \quad (N_f\le 8).
\end{equation}
Metric nonnegativity gives only $N_f\le 8$; the ordinary $E_{N_f+1}$ SCFT window is the strict subwindow
$8-N_f>0$, $N_f\le 7$ (with $N_f=7$ giving $E_8$), while $N_f=8$ ($8-N_f=0$) is the affine-$E_8$ /
$6d$-lift boundary, not an ordinary $5d$ $E_9$ fixed point.
The prepotential does not prove the fixed point exists, but it detects the obstruction when the theory
is too heavily matter-loaded to have one. The sequence runs $E_{N_f+1}$ for $N_f=1,\ldots,7$, up to
$E_8$ at $N_f=7$; the existence and classification are cited but not proved here.

\medskip\noindent\textbf{The $N_f = 0$ caveat.}\enspace
Pure $SU(2)$ is not a single row. Its instanton sectors are classified by a $\mathbb{Z}_2$,
\begin{equation}
\label{eq:V07-pi4}
 \pi_4(SU(2)) \;=\; \mathbb{Z}_2
 \quad\Longrightarrow\quad
 \theta \in \{0,\pi\},
\end{equation}
a discrete theta angle. The two choices give different theories,
\begin{equation}
\label{eq:V07-E1pair}
 \theta = 0: \ U(1)_I \to SU(2)_I \ (E_1),
 \qquad
 \theta = \pi: \ \text{only } U(1)_I \ (\widetilde{E}_1),
\end{equation}
with a further mass deformation reaching the rank-zero $E_0$. Conventions for the lower $E_0/E_1/E_2$
rows vary, so we keep $N_f=0$ compact as this discrete-theta pair and lean on $N_f=5,6,7\to E_6,E_7,E_8$
as the unambiguous anchors.

\medskip\noindent\textbf{No $5d\ \mathcal{N}=2$ superconformal algebra.}\enspace
The Nahm classification of Section~1 permits a $5d$ superconformal algebra for exactly one amount of
supersymmetry,
\begin{equation}
\label{eq:V07-F4}
 \text{$5d$ SCA}: \ F(4), \quad \mathcal{N}=1 \ (8\,\mathcal{Q}),
\end{equation}
the exceptional superalgebra $F(4)$. The $5d\ \mathcal{N}=2$ theory has twice as many supercharges,
above the ceiling,
\begin{equation}
\label{eq:V07-noN2}
 5d\ \mathcal{N}=2 = 16\,\mathcal{Q} \;>\; 8\,\mathcal{Q}
 \quad\Longrightarrow\quad
 \text{no interacting $5d\ \mathcal{N}=2$ SCFT},
\end{equation}
so maximal $5d$ super Yang--Mills has no fixed point of its own; its completion is the $6d\ (2,0)$
theory on a circle, developed in the next section. We state this boundary fact and recall the
classification from Section~1, folding it into the same consistency check as the supercharge count.

\subsection*{Geometric pictures}

Everything here is field theory, but two geometric pictures help organize the same data (we prove
neither construction here). Two constructions realize the $5d$
superconformal theories: the $(p,q)$ five-brane web in type IIB string theory, and geometric
engineering by M-theory on a local Calabi--Yau threefold. In both, the three field-theory objects of
this section become geometry,
\begin{equation}
\label{eq:V07-webdict}
 \begin{aligned}
 \text{Coulomb metric / prepotential } \mathcal{F}
 &\ \longleftrightarrow\ \text{web face areas / CY triple intersection},\\
 \text{instanton particle}
 &\ \longleftrightarrow\ \text{wrapped brane (M2 on a curve)},\\
 \text{enhanced } E_{N_f+1}
 &\ \longleftrightarrow\ \text{collapsed curves / $7$-branes}.
 \end{aligned}
\end{equation}
So the cubic prepotential, the instanton particles, and the $E_n$ enhancement are the same three
objects defined field-theoretically above, now with a geometric picture. This is intuition, not a
derivation: the section uses only the field-theory toolkit.

\section{The $5d$ superconformal index}
\label{sec:V07-index}

The enhanced $E_{N_f+1}$ flavor symmetry of a $5d$ superconformal theory is not manifest in any
Lagrangian, so one needs a protected observable to see it. That observable is the $5d$ superconformal
index. It is a partition function on $S^4 \times S^1$, a weighted trace over the states of the theory
on $S^4$,
\begin{equation}
\label{eq:V07-index}
 \mathcal{I} \;=\; \mathrm{Tr}\,(-1)^F\, x^{\,2(E + j)}\,\prod_a u_a^{\,F_a},
\end{equation}
where the trace runs over states annihilated by a chosen supercharge, $x$ grades by the energy $E$ and
an $SU(2)$ angular momentum $j$ compatible with $SU(2)_R$, and $u_a$ are fugacities for the flavor and
instanton charges $F_a$. By radial quantization, states on $S^4$ correspond to local operators and the
$S^1$ is the trace, so the index counts operators weighted by charge. Only short (BPS) multiplets
contribute, and pairs that could lift cancel in the $(-1)^F$ sum, so the index is deformation-invariant.
In particular it is blind to $1/g_0^2$,
\begin{equation}
\label{eq:V07-indexinvariant}
 \frac{\partial\mathcal{I}}{\partial(1/g_0^2)} \;=\; 0
 \quad\Longrightarrow\quad
 \mathcal{I}_{\text{gauge theory}} \;=\; \mathcal{I}_{\text{SCFT}},
\end{equation}
so the weakly-coupled gauge theory and its strongly-coupled fixed point share one index. A quantity
computed in the gauge theory by summing over instanton sectors thus equals a quantity of the fixed
point, which is how a gauge-theory computation sees the enhanced symmetry of a theory with no
Lagrangian.

The enhancement is visible at the first nontrivial order. The coefficient of $x^2$ in the index, the
conserved-current contribution, counts the conserved currents of the theory, refined by their flavor
and instanton charges. Computed in the gauge theory it has two parts: the perturbative currents, which
assemble the adjoint character of $SO(2N_f)\times U(1)_I$, and the one-instanton contribution, the
$k = \pm 1$ sector, which supplies the spinor characters. Adding them,
\begin{equation}
\label{eq:V07-indexchar}
 \chi_{\mathrm{adj}}(E_{N_f+1})
 \;=\; \chi_{\mathrm{adj}}(SO(2N_f)) \;+\; \chi_{U(1)_I}
 \;+\; \chi^{(k=+1)}_{\mathrm{spinor}} \;+\; \chi^{(k=-1)}_{\mathrm{spinor}},
\end{equation}
the conserved-current order reorganizes into the adjoint character of $E_{N_f+1}$. This simple form
(the perturbative adjoint, the $U(1)_I$ current, and the $k=\pm 1$ spinor characters) closes the
$E_{N_f+1}$ adjoint through $N_f\le 5$; for $N_f=6,7$ it under-counts, and higher/dressed instanton
sectors complete the $E_7$ and $E_8$ adjoints (\S\ref{sec:V07-instanton}), an enumeration cited but
not reproduced here. Setting all
fugacities to one evaluates each character at the identity, giving the dimension identity, exactly the
count \eqref{eq:V07-E6count} of \S\ref{sec:V07-instanton} now from the index side,
\begin{equation}
\label{eq:V07-indexdim}
 \dim E_6 \;=\; \dim SO(10) + 1 + \dim\mathbf{16} + \dim\overline{\mathbf{16}}
 \;=\; 45 + 1 + 16 + 16 \;=\; 78.
\end{equation}
The two routes agree. Dropping the instanton characters breaks it,
\begin{equation}
\label{eq:V07-indexnoinst}
 \dim SO(10) + 1 \;=\; 45 + 1 \;=\; 46 \;\ne\; 78,
\end{equation}
so the instanton operators are as essential to the index as to the symmetry count.

There is no $5d\ \mathcal{N}=2$ superconformal algebra, so this index story is an $\mathcal{N}=1$-only
object: the superconformal index counts states annihilated by a supercharge of the $5d$ superconformal
algebra $F(4)$, which exists only for eight supercharges. The full index machinery, the
instanton-partition-function integrand that produces \eqref{eq:V07-index} order by order and the
higher-order coefficients beyond the conserved-current order, is a substantial computation cited but
not developed here; the load-bearing content is the conserved-current order and its $E_{N_f+1}$
reorganization, worked at $N_f = 5$.

\section*{Exit checklist}
\addcontentsline{toc}{subsection}{Exit checklist}
\markboth{Exit checklist}{Exit checklist}

After this section the reader can
\begin{enumerate}
\item identify the $5d\ \mathcal{N}=1$ vector multiplet (gauge field, one real adjoint scalar $\phi$,
symplectic-Majorana gaugino, $SU(2)_R$) and hypermultiplet (four real scalars, Dirac fermion), read
off the eight-supercharge count, and derive $[g^2] = 4-d = -1$ from
$[\tfrac{1}{g^2}\mathrm{Tr}\,F^2] = d$, so that $1/g_0^2$ is a dimensionful mass parameter, not a
marginal coupling;
\item recognize the real Coulomb branch as a Weyl chamber of real dimension $\mathrm{rank}\,G$ (the
half-line $\phi\ge 0$ for $SU(2)$), and distinguish it from the complex special-K\"ahler branch of
$4d\ \mathcal{N}=2$;
\item assemble the cubic prepotential $\mathcal{F}(\phi) = \tfrac{1}{2g_0^2}\phi^2 +
\tfrac{1}{12}(\sum_{\mathrm{roots}}|\alpha\cdot\phi|^3 - \sum_{\mathrm{hypers}}|w\cdot\phi+m|^3)$ from
the root/weight $|m|^3$ contributions, hold the instanton data separate from this classical plus
one-loop cubic, differentiate the $SU(2)+N_f$ assembly to the leading cubic coefficient $8-N_f$ of
$6\mathcal{F}$, read the metric $\tau(\phi) = \mathcal{F}''(\phi) = 1/g_0^2 + (8-N_f)\phi$ as a second
derivative, and locate the sub-chamber walls at $\phi = |m_f|$;
\item verify metric positivity on the branch for $N_f\le 8$ and the strict $E_{N_f+1}$ window
$N_f\le 7$, and reject the doubled coefficient $8-2N_f$ that would forbid $E_6, E_7, E_8$;
\item define the topological $U(1)_I$ current $j_I = \tfrac{1}{8\pi^2}\mathrm{Tr}(F\wedge F)$ and the
instanton operators as disorder operators of charge $k$ on a surrounding $S^4$, anchor the instanton
mass at $M_{\mathrm{inst}}\sim 8\pi^2/g_0^2$ (dimension one, rejecting the dimension-two $\phi/g^2$),
and count the $E_{N_f+1}$ flavor enhancement through $\dim E_6 = \dim SO(10) + 1 + 16 + 16 = 78$ with
the branching $\mathbf{78} = \mathbf{45}_0 + \mathbf{1}_0 + \mathbf{16}_{+1} + \overline{\mathbf{16}}_{-1}$,
remembering that $E_{N_f+1}$ is a flavor symmetry, not the R-symmetry $SU(2)_R$;
\item explain why $1/g_0^2\to 0$ is a $5d$ superconformal fixed point with no Lagrangian description
(the $E_n$ theories), read the same enhancement off the conserved-current order of the $5d$
superconformal index on $S^4\times S^1$, and state the boundary fact that there is no interacting
$5d\ \mathcal{N}=2$ superconformal theory (sixteen supercharges sits above the eight-supercharge
ceiling of the unique $5d$ superconformal algebra $F(4)$), with the fixed-point existence, the full
instanton partition function, and the $6d$ origin cited rather than proved here.
\end{enumerate}

\bigskip
\section*{Sources and notes}
\addcontentsline{toc}{subsection}{Sources and notes}
\markboth{Sources and notes}{Sources and notes}
{\small

\noindent\textsf{\textcolor{RoyalBlue}{Sources and notes.}}\enspace
This is the first of the two higher-dimension sections in these notes, the field-theory home of the
five-dimensional eight-supercharge toolkit.

\medskip\noindent\textsf{\textcolor{RoyalBlue}{\textbf{\S\ref{sec:V07-multiplets}\enspace Multiplets and the irrelevant coupling.}}}\enspace
The $5d\ \mathcal{N}=1$ vector (one real adjoint scalar, $SU(2)_R$) and hypermultiplet (four real
scalars, Dirac fermion), the eight-supercharge count, and the sign flip $[g^2] = 4-d = -1$
\eqref{eq:V07-gdim} making $1/g_0^2$ a dimensionful mass parameter (the $U(1)_I$ real mass), contrasted
with the marginal $4d$ value and with the $4d\ \mathcal{N}=2$ complex adjoint of
Section~5. (\textcite{Seiberg:1996bd} the $5d$ fixed points and multiplet
structure). 

\medskip\noindent\textsf{\textcolor{RoyalBlue}{\textbf{\S\ref{sec:V07-coulomb}\enspace The real Coulomb branch.}}}\enspace
The real adjoint vev breaking $G\to U(1)^r$ \eqref{eq:V07-break}, the real Coulomb branch of dimension
$\mathrm{rank}\,G$ \eqref{eq:V07-dim} (the $SU(2)$ half-line $\phi\ge 0$ \eqref{eq:V07-halfline}), and
the contrast with the complex special-K\"ahler branch of $4d\ \mathcal{N}=2$ (Section~5,
recovered on a circle by $A_5$ complexifying $\phi$). (\textcite{Seiberg:1996bd};
\textcite{Intriligator:1997pq} the Coulomb-branch structure). 

\medskip\noindent\textsf{\textcolor{RoyalBlue}{\textbf{\S\ref{sec:V07-prepotential}\enspace The cubic prepotential and its metric.}}}\enspace
The two-piece prepotential \eqref{eq:V07-prepotential} (classical $\tfrac{1}{2g_0^2}\phi^2$ plus the
$SU(N\ge 3)$ Chern--Simons cubic, $SU(2)$ having none; the one-loop $|m|^3$ assembly
\eqref{eq:V07-oneloop}, piecewise cubic), classical plus one-loop exact with the instanton data held
separate; the metric as the second derivative $\tau_{ij} = \partial^2\mathcal{F}$
\eqref{eq:V07-metric}; the $SU(2)+N_f$ assembly \eqref{eq:V07-su2prep} differentiated to the leading
cubic coefficient $8-N_f$ of $6\mathcal{F}$ \eqref{eq:V07-su2cubic}, the affine metric $\tau(\phi) =
1/g_0^2 + (8-N_f)\phi$ \eqref{eq:V07-su2metric}, positivity for $N_f\le 8$ and the strict window
$N_f\le 7$ \eqref{eq:V07-positivity}, and the chamber walls at $\phi=|m_f|$
(Table~\ref{tab:V07-prepchamber}). (\textcite{Seiberg:1996bd}; \textcite{Intriligator:1997pq} the full cubic prepotential and
Coulomb-branch phases). 

\medskip\noindent\textsf{\textcolor{RoyalBlue}{\textbf{\S\ref{sec:V07-instanton}\enspace Instanton operators and the $E_n$ enhancement.}}}\enspace
The topological current $j_I = \tfrac{1}{8\pi^2}\mathrm{Tr}(F\wedge F)$ \eqref{eq:V07-instcurrent} and
the integer instanton number \eqref{eq:V07-instnumber}, the instanton operators as disorder operators
of charge $k$ on a surrounding $S^4$ \eqref{eq:V07-instop} (the $5d$ analog of the $3d$ monopole
operator), the instanton-particle mass anchor $M_{\mathrm{inst}}\sim 8\pi^2/g_0^2$
\eqref{eq:V07-instmass} (dimension one), the perturbative $SO(2N_f)\times U(1)_I$ enhancing to
$E_{N_f+1}$ \eqref{eq:V07-enhance}, and the worked $N_f=5\to E_6$ count $45+1+16+16 = 78$
\eqref{eq:V07-E6count} with the adjoint branching \eqref{eq:V07-E6branch} (Table~\ref{tab:V07-enmap};
the naive two-spinor count honestly under-counting at $N_f = 6, 7$). (\textcite{Lambert:2014jna} the $5d$ instanton operators; \textcite{Tachikawa:2015mha} the
instanton-current symmetry enhancement; \textcite{Seiberg:1996bd} the $E_n$ fixed points).

\medskip\noindent\textsf{\textcolor{RoyalBlue}{\textbf{\S\ref{sec:V07-scft}\enspace The $5d$ superconformal fixed points.}}}\enspace
The $1/g_0^2\to 0$ ultraviolet limit as a $5d$ superconformal theory, the gauge theory as a relevant
(mass) deformation, the no-Lagrangian fact, the prepotential convexity as the field-theory probe (the
$8-N_f\ge 0$ bound $=$ the $E_{N_f+1}$ window), the $N_f = 0$ discrete-$\theta$ pair
$E_1/\widetilde{E}_1$ ($\pi_4(SU(2)) = \mathbb{Z}_2$), and the boundary fact that there is no
interacting $5d\ \mathcal{N}=2$ superconformal theory (the unique $5d$ superconformal algebra is
$F(4)$ at eight supercharges; maximal $5d$ SYM UV-completes to $6d\ (2,0)$ on a circle, forward-pointed
to Section~8).
(\textcite{Seiberg:1996bd} the fixed-point existence; \textcite{Cordova:2016emh} the
superconformal-algebra catalog / $F(4)$). 

\medskip\noindent\textsf{\textcolor{RoyalBlue}{\textbf{\S\ref{sec:V07-index}\enspace The $5d$ superconformal index.}}}\enspace
The index on $S^4\times S^1$ \eqref{eq:V07-index} as a protected, deformation-invariant counting
partition function, the enhancement visible at the conserved-current ($x^2$) order as the adjoint
character of $E_{N_f+1}$ \eqref{eq:V07-indexchar}, the index-side dimension identity $\dim E_6 =
45+1+16+16 = 78$ \eqref{eq:V07-indexdim} agreeing with the symmetry-bookkeeping route, and the
$\mathcal{N}=1$-only nature of the index (no $5d\ \mathcal{N}=2$ superconformal algebra). (\textcite{Kim:2012gu} the $5d$ superconformal index with enhanced
$E_n$ symmetry). 

\medskip\noindent\textsf{\textcolor{RoyalBlue}{\textbf{Stated and cited, not proved here.}}}\enspace
The existence and classification of the $5d$ superconformal fixed points (the $E_n$ sequence, the
higher-rank $5d$ theories, the convexity / gauge-theory-phase classification, the string / M-theory /
brane-web construction that establishes existence) are cited but not proved. The
full $5d$ instanton-operator technology (the Nekrasov instanton partition function, the higher-order
superconformal-index evaluation, the proof of the symmetry enhancement beyond the conserved-current
count, and the brane-web realization of instanton particles) is cited but not developed here.
No $5d$ holographic dual, $S^5$ free energy, or $5d$ central-charge technology is developed
beyond naming it. The $6d$ origin of the $5d$ $E_n$ theories on a circle, and the completion of maximal
$5d$ SYM to $6d\ (2,0)$, are forward-pointed to Section~8. The $(p,q)$
five-brane webs and local-Calabi--Yau geometries that engineer $5d$ superconformal theories and realize
the cubic prepotential, the instanton particles, and the $E_n$ enhancement, are cited as geometric
pictures rather than proved here.
}

\subsection*{Further reading}
\addcontentsline{toc}{subsection}{Further reading}
Five-dimensional superconformal fixed points and their exceptional flavor symmetry are in
\textcite{Seiberg:1996bd}; the geometric and extremal-transition picture in
\textcite{Morrison:1996xf,Intriligator:1997pq}; the $(p,q)$ five-brane webs in
\textcite{Aharony:1997ju,Aharony:1997bh}. The systematic geometric classification of $5d$ superconformal
theories is developed in \textcite{Jefferson:2018irk,Closset:2020scj}.

Recent classification and compactification perspectives on $5d$ SCFTs are developed in
\textcite{Bhardwaj:2019jtr,Bhardwaj:2019fzv}; the global form of flavor symmetry, instantonic
symmetry, higher-form symmetry, and their anomalies are analyzed in
\textcite{BenettiGenolini:2020doj,Apruzzi:2021vcu}.

\section*{References}
\printbibliography[heading=none]
\end{refsection}
\begin{refsection}\chapter{\texorpdfstring{$6d$}{6d} supersymmetric field theories}
\label{ch:V08}

\noindent\textbf{Guide to this section.}\enspace
Sections~1, 2, 3, 5, and~7 fixed the algebra, the words, and the four- and five-dimensional worlds. This
section builds the six-dimensional world, the top of the dimension ladder before the pure
super-Yang--Mills interface of Section~9. Six dimensions carries three structures no lower-dimensional
section has met: a chiral two-form and the tensor branch it lives on, an anomaly in an eight-form that
must Green--Schwarz factorize, and the sharpest non-Lagrangian theory in these notes, the $6d\ \mathcal{N}=(2,0)$ theory
of coincident M5-branes. It is a working foundations section: it states the $6d$ facts and runs them as
computations, the self-dual-tensor degree count, a worked anomaly eight-form and its factorization, and
the $\mathcal{N}=(2,0)$ anomaly $a\sim N^3$. The deep results, the anomaly-polynomial derivation, the $6d$ SCFT
classification, and the construction of the $\mathcal{N}=(2,0)$ theory, are stated and cited rather
than proved here.

\begin{keybox}{What this section delivers}
The $6d\ \mathcal{N}=(1,0)$ multiplets (vector, hyper, tensor with its anti-self-dual two-form) and the on-shell
count that closes the tensor multiplet (\S\ref{sec:V08-multiplets}); the tensor branch, the tensor scalar
as the inverse gauge coupling, and the tensionless string at the origin (\S\ref{sec:V08-tensorbranch});
the anomaly eight-form $I_8$, its $\hat A(R)\,\mathrm{ch}(F)$ blocks, and the irreducible
$\mathrm{tr}\,F^4$ obstruction, worked on the rank-one $E$-string (\S\ref{sec:V08-anomaly}); the
Green--Schwarz--Sagnotti--West factorization $I_8=\tfrac12\Omega_{ij}X_4^iX_4^j$ and why the tensor
branch is required (\S\ref{sec:V08-gs}); the four Weyl coefficients $(a,c_1,c_2,c_3)$, matched on the
branch, with $6d$ beside the extremization family (\S\ref{sec:V08-acanomaly}); the $6d\ \mathcal{N}=(2,0)$ theory,
its absence of a Lagrangian, the $(\mathbb{R}^5)^{N-1}/S_N$ branch, and the marquee $a\sim N^3$
(\S\ref{sec:V08-twozero}); and $6d\ \mathcal{N}=(1,1)$ maximal super-Yang--Mills as the Lagrangian contrast
(\S\ref{sec:V08-oneone}).
\end{keybox}

\medskip\noindent\textbf{The $6d$ supersymmetry data, recalled once.}\enspace
Six-dimensional supersymmetry comes in three amounts, and Section~1's master table fixed each. A minimal
$6d$ chiral (Weyl) spinor supercharge carries eight real components, so the three cases are
\begin{equation}
\label{eq:V08-Qcount}
 \mathcal{N}=(1,0):\ 8, \qquad \mathcal{N}=(2,0):\ 16, \qquad \mathcal{N}=(1,1):\ 16
\end{equation}
real supercharges. The minimal case $6d\ \mathcal{N}=(1,0)$ has $R$-symmetry $SU(2)_R$ and sits at the top of the
eight-supercharge ladder,
\begin{equation}
\label{eq:V08-ladder}
 6d\ \mathcal{N}=(1,0) \;\equiv\; 5d\ \mathcal{N}=1 \;\equiv\; 4d\ \mathcal{N}=2 \;\equiv\; 3d\ \mathcal{N}=4
 \qquad(8\ \text{supercharges}),
\end{equation}
the sibling family of Sections~5 and~7. The two maximal cases each add a second supercharge; their
chirality choice fixes the $R$-symmetry,
\begin{equation}
\label{eq:V08-Rsym}
\begin{aligned}
 \mathcal{N}=(1,0):&\ SU(2)_R, \\
 \mathcal{N}=(2,0):&\ USp(4)_R=Spin(5)_R, \\
 \mathcal{N}=(1,1):&\ Spin(4)_R=SU(2)^2.
\end{aligned}
\end{equation}
The $\mathcal{N}=(2,0)$ pair has the same chirality (larger $R$-symmetry, $SO(5)$); the $\mathcal{N}=(1,1)$ pair has opposite
chirality ($SO(4)$). We do not re-derive \eqref{eq:V08-Qcount}--\eqref{eq:V08-Rsym}; they are the
Section~1 oracle, and \eqref{eq:V08-Qcount} is machine checked against the ladder. We also use
Section~2's grammar once and do not re-teach it: multiplets are supersymmetry representations, moduli
spaces split into branches named for the multiplet whose scalars parametrize them, gauge anomalies
cancel while 't~Hooft anomalies are matched, and a theory can be a field theory without a Lagrangian.
Every one of those three returns in a sharper $6d$ form below.

\bigskip
\begin{center}
\rule{0.4\textwidth}{0.4pt}\\[3pt]
{\large\textsf{\textbf{Block A.\enspace The $6d\ \mathcal{N}=(1,0)$ world}}}\\[2pt]
\rule{0.4\textwidth}{0.4pt}
\end{center}
\medskip

\noindent The first five sections build $6d\ \mathcal{N}=(1,0)$ dynamics: the multiplets and the chiral tensor, the tensor branch, the anomaly polynomial, its Green--Schwarz factorization, and the conformal anomalies.

\section{Multiplets and the self-dual tensor}
\label{sec:V08-multiplets}

A $6d\ \mathcal{N}=(1,0)$ theory is assembled from three multiplets, the third with no lower-dimensional analog.
\begin{itemize}
\item \emph{Vector}: a gauge field $A_\mu$ and a gaugino, with \emph{no scalar} (unlike $4d$/$5d$, no
adjoint Coulomb-branch scalar in $6d$).
\item \emph{Hyper}: four real scalars (an $SU(2)_R$ doublet, as in four and five dimensions) and a
fermion, the $6d$ matter multiplet.
\item \emph{Tensor}: one real scalar $\phi$, an antisymmetric two-form $B^-_{\mu\nu}$ with anti-self-dual
$H^-=dB^-$, and a fermion. Its chiral two-form is what makes six dimensions special.
\end{itemize}
The vector and hyper are familiar; the tensor is the new object. Table~\ref{tab:V08-multiplets} collects
all three alongside the maximal cases.

Six is the lowest dimension hosting a chiral two-form. The field strength $H=dB$ is a three-form, and in
$d=6$ a three-form is middle-dimensional, $\star:\Omega^3\to\Omega^{6-3}=\Omega^3$, so $H=\pm\star H$
makes sense. We take the $\mathcal{N}=(1,0)$ two-form anti-self-dual and the $\mathcal{N}=(2,0)$ (and supergravity) tensor
self-dual,
\begin{equation}
\label{eq:V08-selfdual}
 \mathcal{N}=(1,0):\ H^- = -\star H^-, \qquad \mathcal{N}=(2,0):\ H^+ = +\star H^+.
\end{equation}
Only the sign is a convention; the physical content is that the two-form is chiral.
We fix the $\mathcal{N}=(1,0)$ sign as in \eqref{eq:V08-selfdual} so that the tensor-branch gauge kinetic term
$\phi\,\mathrm{tr}\,F^2$ of \S\ref{sec:V08-tensorbranch} comes out with the right sign.

The R-symmetry $SU(2)_R$ organizes the fermions. The gaugino and the tensor fermion are $SU(2)_R$
doublets (each a symplectic-Majorana--Weyl spinor), the bosons $A_\mu$, $B_{\mu\nu}$, $\phi$ are
singlets, and only the hypermultiplet has $SU(2)_R$-charged bosons,
\begin{equation}
\label{eq:V08-Rcharges}
 \text{doublets: }\lambda,\ \psi_T;\quad
 \text{singlets: }A_\mu,\ B_{\mu\nu},\ \phi;\quad
 \text{hyper: }(\mathbf 2,\mathbf 2)\text{ of }SU(2)_R\times SU(2)_F.
\end{equation}
The four hyper scalars form two $SU(2)_R$ doublets (a half-hypermultiplet is one doublet, for a
pseudoreal representation), the same $8$-supercharge bookkeeping as the $4d\ \mathcal{N}=2$ and
$5d\ \mathcal{N}=1$ hypermultiplets of Sections~5 and~7. The new element is the tensor multiplet.

That $H=\pm\star H$ is a real condition, not a vacuous one, rests on a signature check. The square of the
Hodge star on a $k$-form in $d$ Lorentzian dimensions is $(-1)^{k(d-k)}s$ with $s=-1$; on a $6d$
three-form,
\begin{equation}
\label{eq:V08-starsq}
 \star^2\big|_{k=3,\,d=6} \;=\; (-1)^{k(d-k)}\,s \;=\; (-1)^{3\cdot 3}\,(-1) \;=\; (-1)^{9}\,(-1) \;=\; +1.
\end{equation}
Because $\star^2=+1$ the projectors $\tfrac12(1\pm\star)$ are genuine and $H=\pm\star H$ has nontrivial
real solutions. In Euclidean signature $s=+1$ gives $\star^2=-1$, the real condition collapses, and the
chiral tensor is a Lorentzian object. Equation \eqref{eq:V08-starsq} is machine checked.

\subsection*{Worked instance: the tensor-multiplet degree-of-freedom count}

The tensor multiplet must close the supersymmetry algebra, so its on-shell bosonic and fermionic
degrees of freedom must match. Counting them is the smallest piece of tensor-multiplet arithmetic, and
it feeds directly into the anomaly polynomial of \S\ref{sec:V08-anomaly}. Start with the two-form off
shell: an antisymmetric $B_{\mu\nu}$ in six dimensions has
\begin{equation}
\label{eq:V08-offshell}
 \binom{6}{2} \;=\; 15
\end{equation}
independent components. Gauge redundancy and the equations of motion cut this to physical states,
counted cleanest in the massless little group $SO(4)=SU(2)\times SU(2)$. The three-form field strength
splits self-dual and anti-self-dual, sitting in the $(\mathbf 3,\mathbf 1)$ and $(\mathbf 1,\mathbf 3)$,
so a generic two-form carries both while self-duality keeps one half,
\begin{equation}
\label{eq:V08-fulldof}
 (\mathbf{3},\mathbf{1})\oplus(\mathbf{1},\mathbf{3}) = 3+3 = 6
 \ \xrightarrow{\ H^-=-\star H^-\ }\
 (\mathbf{1},\mathbf{3}) = 3.
\end{equation}
The fermion is a symplectic-Majorana--Weyl spinor; a $6d$ Weyl spinor has $8$ real off-shell components
and the reality condition keeps $8/2=4$ on-shell,
\begin{equation}
\label{eq:V08-selfdof}
 n_{\mathrm{fermi}} \;=\; \tfrac12\cdot 8 \;=\; 4.
\end{equation}
Adding the real scalar $\phi$ to the tensor's $3$ states, the counts balance,
\begin{equation}
\label{eq:V08-susybalance}
 n_{\mathrm{bose}} \;=\; \underbrace{3}_{B^-}+\underbrace{1}_{\phi} \;=\; 4 \;=\;
 \underbrace{4}_{\psi_T} \;=\; n_{\mathrm{fermi}},
\end{equation}
so the tensor multiplet closes the eight-supercharge algebra. The little-group dimensions in
\eqref{eq:V08-fulldof}--\eqref{eq:V08-selfdof} are computed, not asserted, and \eqref{eq:V08-susybalance}
is machine checked; the falsifier that a self-dual two-form has six on-shell states is rejected.

The same counting fixes the other two multiplets, so the anomaly assembly of \S\ref{sec:V08-anomaly} has
all three. The vector has a gauge field ($6-2=4$ states, an $SO(4)$ vector) plus a gaugino; the hyper
has four real scalars plus a Weyl fermion:
\begin{equation}
\label{eq:V08-all3dof}
 \text{vector}:\ 4_B=4_F,\qquad \text{hyper}:\ 4_B=4_F,\qquad \text{tensor}:\ (3+1)_B=4_F.
\end{equation}
Each $\mathcal{N}=(1,0)$ multiplet carries $(4,4)$ on-shell states, the smallest $8$-supercharge representations;
the tensor is the one whose bosonic $4$ splits $3+1$ rather than $4+0$, and that split feeds the
self-dual-tensor term into the anomaly polynomial. The three states are a chiral two-form, the origin of
every $6d$-specific structure below: no covariant Lagrangian (\S\ref{sec:V08-twozero}), a self-dual
string (\S\ref{sec:V08-tensorbranch}), a Hirzebruch term in $I_8$ (\S\ref{sec:V08-anomaly}).

\begin{table}[ht]
\centering
\small
\setlength{\tabcolsep}{6pt}
\renewcommand{\arraystretch}{1.3}
\begin{tabular}{@{}lll>{\raggedright\arraybackslash}p{42mm}@{}}
\toprule
supersymmetry & supercharges & R-symmetry & multiplets \\
\midrule
$6d\ \mathcal{N}=(1,0)$ & $8$ & $SU(2)_R$ & vector; hyper; tensor (anti-self-dual $B^-$) \\
\midrule
$6d\ \mathcal{N}=(2,0)$ & $16$ & $USp(4)_R$ & tensor (self-dual $B^+$, five scalars) \\
\midrule
$6d\ \mathcal{N}=(1,1)$ & $16$ & $Spin(4)_R$ & vector, four scalars: maximal SYM \\
\bottomrule
\end{tabular}
\caption{The $6d$ supersymmetries, their real supercharge counts, R-symmetries (the Section~1
master-table $6d$ row), and multiplet content. The R-symmetries are $SU(2)_R$ for $\mathcal{N}=(1,0)$,
$USp(4)_R=Spin(5)_R$ for $\mathcal{N}=(2,0)$, and $Spin(4)_R=SU(2)\times SU(2)$ for $\mathcal{N}=(1,1)$. The $\mathcal{N}=(1,0)$ vector has
no scalar; the tensor multiplet's two-form is chiral. The $\mathcal{N}=(2,0)$ tensor is self-dual with five
$SO(5)_R$ scalars; $\mathcal{N}=(1,1)$ maximal SYM is a Lagrangian gauge theory with four scalars. The supercharge
counts are machine checked against the Section~1 ladder.}
\label{tab:V08-multiplets}
\end{table}

\section{The tensor branch}
\label{sec:V08-tensorbranch}

A $6d\ \mathcal{N}=(1,0)$ theory with $n_T$ tensor multiplets has $n_T$ real tensor scalars $\phi^i$, whose vacuum
expectation values parametrize the \emph{tensor branch}, the $6d$ member of Section~2's branch
vocabulary. Its real dimension is exactly the number of tensor multiplets,
\begin{equation}
\label{eq:V08-branchdim}
 \dim_{\mathbb{R}}\mathcal{M}_{\mathrm{tensor}} \;=\; n_T,
\end{equation}
machine checked for $n_T=1,2,3,5$. The interacting SCFT sits at the origin $\langle\phi^i\rangle=0$;
generic $\phi^i$ move onto the branch, where the theory is weakly coupled or free.

The key mechanism is that the tensor scalar \emph{is} the gauge coupling. On the branch the gauge kinetic
term is controlled by $\phi^i$, so the inverse coupling is a modulus,
\begin{equation}
\label{eq:V08-tensorbranch}
 \mathcal{L}_{\mathrm{gauge}} \;\supset\; \phi^i\,\mathrm{tr}\,F_i^2,\qquad
 \frac{1}{g_i^2} \;\sim\; \langle\phi^i\rangle,\qquad
 \langle\phi^i\rangle\to\infty:\ g_i\to0,\quad \langle\phi^i\rangle\to0:\ g_i\to\infty.
\end{equation}
This is unlike lower dimensions, where the coupling is a fixed input; here it runs from free
($\langle\phi\rangle\to\infty$) to the strongly-coupled origin. The inverse-coupling identity is machine
checked, and it is why the tensor branch, not a Coulomb branch, carries the coupling data.

Charged under $B^i$ is a \emph{self-dual string}, a one-dimensional soliton whose tension is the tensor
scalar, tensionless at the origin,
\begin{equation}
\label{eq:V08-stringtension}
 T_i \;\sim\; \langle\phi^i\rangle \;\xrightarrow{\ \langle\phi^i\rangle\to0\ }\; 0.
\end{equation}
A tensionless string is a new kind of light degree of freedom, not a particle, and its appearance is the
hallmark of an interacting $6d$ fixed point: the origin contains light string excitations no Lagrangian
in the tensor-branch fields can capture. It is the $6d$ counterpart of a massless BPS particle on a
lower-dimensional Coulomb branch, one dimension bigger. Equation \eqref{eq:V08-stringtension} is machine
checked; the string dynamics is cited but not developed here.

\begin{figure}[ht]
\centering
\begin{tikzpicture}[scale=1.0]
 % the tensor branch as a ray, phi from 0 (origin/SCFT) outward
 \draw[thick,-{Stealth[length=2.4mm]}] (0,0) -- (7.2,0) node[right] {$\langle\phi\rangle$};
 \filldraw[RoyalBlue] (0,0) circle (2.6pt);
 \node[below=5pt,RoyalBlue] at (0,0) {\small origin: SCFT};
 \node[above=8pt] at (0,0.05) {\small $T\to0$};
 % a generic tensor-branch point
 \filldraw (4.4,0) circle (1.8pt);
 \node[below=5pt] at (4.4,0) {\small tensor branch};
 \node[above=8pt] at (4.4,0) {\small $T\sim\langle\phi\rangle$};
 % the self-dual string tension growing: schematic segments of increasing length
 \draw[very thick,gray] (0.0,0.9) -- (0.45,0.9);
 \node[gray,right] at (0.55,0.9) {\tiny (short string, light)};
 \draw[very thick,gray] (4.1,0.9) -- (5.4,0.9);
 \node[gray,right] at (5.5,0.9) {\tiny (long string, heavy)};
\end{tikzpicture}
\caption{The $6d\ \mathcal{N}=(1,0)$ tensor branch, schematic. The real tensor scalar $\langle\phi\rangle$ runs
outward from the origin. On the branch the gauge coupling is set by $1/g^2\sim\langle\phi\rangle$ and a
self-dual string (charged under $B$) has tension $T\sim\langle\phi\rangle$. At the origin (blue) the
string becomes tensionless and the interacting superconformal field theory appears. The figure is a
one-dimensional cartoon of an $n_T$-dimensional branch; the string, not a particle, is the light object
at the origin.}
\label{fig:V08-tensorbranch}
\end{figure}
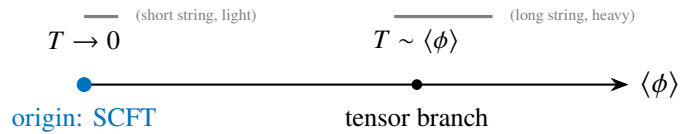

\begin{keybox}{Common misconception: the tensor branch is the Coulomb branch under another name}
It is tempting to call the $6d$ tensor branch a Coulomb branch, since both are moduli spaces carrying
the low-energy coupling data, but the difference is structural. The $4d/5d$ Coulomb branch is
parametrized by \emph{vector-multiplet} scalars, and the $6d\ \mathcal{N}=(1,0)$ vector has \emph{no scalar at all},
so there is no $6d$ Coulomb branch of that type. The moduli space carrying the special $6d$ data is
parametrized by \emph{tensor-multiplet} scalars, and it is the tensor branch. Branch names attach to the
theory's own multiplets (Section~2); on the tensor branch the scalar plays the inverse-coupling role
\eqref{eq:V08-tensorbranch} and a self-dual string, not a particle, goes light at the origin.
\end{keybox}

A purely gravitational constraint ties $n_T$ to the spectrum. The pure-gravitational anomaly must itself
be Green--Schwarz-removable, which coupled to gravity becomes the $6d$ gravitational-anomaly condition
\begin{equation}
\label{eq:V08-gravcond}
 n_H \;-\; n_V \;+\; 29\,n_T \;=\; 273,
\end{equation}
one of the sharpest constraints on $6d$ spectra and the reason $n_T$ is not free; the coupling to gravity
is not developed here.

The tensor branch is where the anomaly technology lives. The origin SCFT has an $I_8$ hard to compute
directly; on the branch the theory is free, so its $I_8$ is a sum of free-field blocks. Since $I_8$ is an
RG invariant, the matching
\begin{equation}
\label{eq:V08-branchmatch}
 I_8^{\mathrm{SCFT}}\big|_{\text{origin}} \;=\; I_8\big|_{\text{tensor branch}}
\end{equation}
makes it computable and reads off both the gauge-anomaly cancellation of \S\ref{sec:V08-anomaly} and the
conformal anomalies of \S\ref{sec:V08-acanomaly}. F-theory constructions build such tensor branches
geometrically; the classification is cited but not reproduced here.

\section{The anomaly polynomial}
\label{sec:V08-anomaly}

A chiral theory in six dimensions has gauge, gravitational, mixed, and R-symmetry anomalies, and the
right object to organize them is a polynomial. The anomaly-inflow formalism packages the anomaly of a
$d$-dimensional theory into a characteristic class of degree $d+2$; a one-loop anomaly in $2n$ dimensions
comes from an $(n+1)$-gon diagram, so
\begin{equation}
\label{eq:V08-Idplus2}
 d=2:\ I_4\ (\text{$2$-gon}),\qquad
 d=4:\ I_6\ (\text{triangle}),\qquad
 d=6:\ I_8\ (\text{box}),
\end{equation}
and in six dimensions the whole analysis is arithmetic on the eight-form $I_8$, generated by a box
(four-point) diagram. The anomaly is recovered from $I_8$ by the Stora--Zumino descent chain. Because
$I_8$ is a closed, gauge-invariant characteristic class it is locally exact, and one descends by two
steps,
\begin{equation}
\label{eq:V08-descent}
 I_8 \;=\; dI_7^{(0)}, \qquad
 \delta_\lambda I_7^{(0)} \;=\; dI_6^{(1)}, \qquad
 \delta_\lambda \log Z \;=\; 2\pi i\int_{M_6} I_6^{(1)},
\end{equation}
the Chern--Simons seven-form $I_7^{(0)}$, then its gauge variation
($\lambda$ the gauge/Lorentz parameter) descending to the six-form $I_6^{(1)}$, and finally
$\int_{M_6}I_6^{(1)}$ as the anomalous variation of the effective action. Each step lowers the degree by
one, $8\to7\to6$, landing on a six-form over spacetime; the bookkeeping is machine checked.
The entire anomaly is packaged in the single polynomial $I_8$, and the physics questions become algebra
on its coefficients.

\begin{keybox}{Common misconception: the $6d$ anomaly is a triangle diagram}
In four dimensions the anomaly comes from a triangle (three-point) diagram and the polynomial is the
six-form $I_6$. Six dimensions is not the same: a $d$-dimensional anomaly is a degree-$(d+2)$ class, so
the $6d$ anomaly is the \emph{eight-form} $I_8$, generated by a \emph{box} (four-point) diagram, per the
$2n$-dimensional $(n+1)$-gon pattern of \eqref{eq:V08-Idplus2}. The consequence is that $6d$
cancellation is richer than the $4d$ cubic vanishing: an irreducible $\mathrm{tr}\,F^4$ piece must cancel
group-theoretically \emph{and} the residual must factorize for Green--Schwarz inflow to absorb it. The
$6d$ tool is the eight-form polynomial and its factorization, not a single triangle.
\end{keybox}

\medskip\noindent\textbf{The building blocks.}\enspace
The anomaly polynomial is assembled additively from the chiral field content, each field contributing
with its chirality sign. A chiral Weyl fermion in a representation $R$ of the gauge or flavor group
contributes the degree-eight part of the index-density form
\begin{equation}
\label{eq:V08-inflow}
 \big[\hat A(R)\,\mathrm{ch}(F)\big]_{8},
\end{equation}
the entry-9 inflow building block, $\hat A(R)$ the Dirac genus and $\mathrm{ch}(F)$ the Chern character,
the subscript $8$ selecting the eight-form; a chiral tensor contributes a signature (Hirzebruch) term.
Expanding the two factors,
\begin{equation}
\label{eq:V08-Ahatch}
 \hat A \;=\; 1 \;-\; \frac{p_1}{24} \;+\; \frac{7p_1^2-4p_2}{5760} \;+\;\dots, \qquad
 \mathrm{ch}(F) \;=\; \mathrm{rk} \;+\; \mathrm{tr}\,e^{iF/2\pi}\big|_{\ge2},
\end{equation}
the degree-eight product collects gravitational, mixed, and gauge pieces,
\begin{equation}
\label{eq:V08-A8sources}
 [\hat A\,\mathrm{ch}]_8 \;\supset\;
 \underbrace{p_1^2,\,p_2}_{\hat A|_8},\quad
 \underbrace{(\mathrm{tr}\,F^2)\,p_1}_{\text{deg-4}\,\times\,\text{deg-4}},\quad
 \underbrace{\mathrm{tr}\,F^4,\,(\mathrm{tr}\,F^2)^2}_{\mathrm{ch}|_8},
\end{equation}
plus the $c_2(R)$ monomials from the same expansion on the $SU(2)_R$ background. The index-theorem
derivation is cited but not reproduced here. The full degree-eight basis, built from $F$, $p_1,p_2$, and $c_2(R)$, is
\begin{equation}
\label{eq:V08-monobasis}
 \mathrm{tr}\,F^4,\quad (\mathrm{tr}\,F^2)^2,\quad (\mathrm{tr}\,F^2)\,p_1,\quad p_1^2,\quad p_2,\quad
 c_2(R)^2,\quad c_2(R)\,p_1,\quad c_2(R)\,(\mathrm{tr}\,F^2),
\end{equation}
and assembling $I_8$ means computing each rational coefficient from the field content.

The gauge dependence enters through group-theory index data. For a chiral fermion in a representation
$R$, the quartic invariants reduce to a reference (the fundamental, or the smallest exceptional
representation) by
\begin{equation}
\label{eq:V08-indexdata}
 \mathrm{tr}_R F^2 \;=\; A_R\,\mathrm{tr}\,F^2, \qquad
 \mathrm{tr}_R F^4 \;=\; B_R\,\mathrm{tr}\,F^4 \;+\; C_R\,(\mathrm{tr}\,F^2)^2,
\end{equation}
with $A_R$ the Dynkin index and $B_R,C_R$ the quartic index data of $R$. The coefficient of the
irreducible $\mathrm{tr}\,F^4$ in $I_8$ is then a signed sum over the chiral matter, the gaugino minus
the hypermultiplets,
\begin{equation}
\label{eq:V08-Bsum}
 \big[I_8\big]_{\mathrm{tr}\,F^4} \;\propto\; B_{\mathrm{adj}} \;-\; \sum_f n_f\,B_{R_f},
\end{equation}
while the reducible $(\mathrm{tr}\,F^2)^2$ collects the $C_R$ and $A_R$ pieces. When the group has no
independent quartic Casimir, $\mathrm{tr}\,F^4$ itself reduces to $(\mathrm{tr}\,F^2)^2$ and $B_R\to0$;
the analysis splits into a $B_R$ (irreducible) part that must vanish and a $C_R/A_R$ (reducible) part
that must factorize.

\medskip\noindent\textbf{The irreducible $\mathrm{tr}\,F^4$ obstruction.}\enspace
The single-trace quartic $\mathrm{tr}\,F^4$ is \emph{irreducible}: it is not a product of two
degree-four forms, so it has no home in the factorized $\tfrac12\Omega_{ij}X_4^iX_4^j$ (a sum of
squares) and Green--Schwarz inflow cannot absorb it. Consistency therefore forces its coefficient to
\emph{vanish on its own}, from the group theory of the matter, the six-dimensional analog of the $4d$
cubic gauge anomaly. Whether a group carries an independent quartic Casimir decides the story,
\begin{equation}
\label{eq:V08-quarticlist}
 \text{has one: }SU(N\ge4),\ SO(N\ge8),\ Sp(N);\quad
 \text{none: }SU(2),SU(3),G_2,F_4,E_6,E_7,E_8.
\end{equation}
For the second list $\mathrm{tr}\,F^4$ reduces to $(\mathrm{tr}\,F^2)^2$. The sharpest case is $E_8$,
where
\begin{equation}
\label{eq:V08-e8identity}
 \mathrm{tr}_{\mathbf{248}}F^4 \;=\; \frac{1}{100}\,\big(\mathrm{tr}_{\mathbf{248}}F^2\big)^2,
\end{equation}
so an $E_8$ theory has no irreducible quartic at all: any would-be $\mathrm{tr}\,F^4$ folds, with the
constant $\tfrac{1}{100}$, into the reducible $(\mathrm{tr}\,F^2)^2$ slot. The identity
\eqref{eq:V08-e8identity} and the reduction it drives are machine checked, along with the falsifier that
a group with an independent quartic Casimir, $SU(5)$, cannot fold its $\mathrm{tr}\,F^4$ and must cancel
it on its own.

\subsection*{Worked instance: the anomaly eight-form of the rank-one $E$-string}

The pinned fixture carried through this section and the next is the \emph{rank-one $E$-string} theory:
one tensor multiplet ($n_T=1$), no gauge algebra on the tensor branch, and an $E_8$ flavor symmetry.
Its anomaly-polynomial coefficients are the modern values of Ohmori, Shimizu, Tachikawa, and Yonekura
(OSTY), the standard $6d$ Green--Schwarz source. The $E$-string is the cleanest genuine
example: a single tensor, a single exceptional flavor group, and a residual anomaly that factorizes with
one degree-four form.

First the group theory. Because the flavor group is $E_8$, the $E_8$ identity
\eqref{eq:V08-e8identity} applies, and the irreducible $\mathrm{tr}\,F_{E_8}^4$ coefficient in the
assembled $I_8$ is
\begin{equation}
\label{eq:V08-trF4vanishes}
 \big[I_8\big]_{\mathrm{tr}\,F^4} \;=\; 0.
\end{equation}
There is no independent quartic Casimir for $E_8$ to obstruct the factorization; the whole $E_8$
dependence collapses into $\mathrm{tr}\,F_{E_8}^2$ and its square. This vanishing is machine checked,
and it is the reason the $E$-string is Green--Schwarz consistent. One caution: here $E_8$ is a
\emph{background flavor} symmetry, so its terms in $I_8$ are 't~Hooft anomalies, i.e.\ invariant data,
not gauge-consistency obstructions. Green--Schwarz cancellation is a requirement for \emph{dynamical}
gauge fields; for the $E_8$ flavor field the factorization is just the way inflow packages the anomaly
and the tensor-branch matching data.

With the quartic gone, the reducible part of $I_8$ is built from the six product-of-degree-four
monomials of \eqref{eq:V08-monobasis}. What factorizes into a single perfect square is the
\emph{residual} obtained after subtracting the free tensor/hyper contributions and the pure
gravitational $p_2$ piece from the full OSTY polynomial; we carry only this residual,
\begin{equation}
\label{eq:V08-estringX4}
 I_8^{\mathrm{res}} \;=\; \tfrac12\,X_4\,X_4, \qquad
 X_4 \;=\; -\tfrac14\,p_1 \;-\; \tfrac14\,\mathrm{tr}\,F_{E_8}^2 \;+\; 2\,c_2(R),
\end{equation}
one tensor ($\Omega=1$) with a single degree-four $X_4$, in the pinned OSTY normalization. This is not
the \emph{full} $E$-string $I_8$: the complete OSTY expression carries a gravitational $p_2$ term and
free-field pieces outside any $X_4X_4$ square, which we do not expand here. We defer the solution for $X_4$ to
\S\ref{sec:V08-gs}; here we run the assembly \emph{forward}, squaring the pinned $X_4$ and comparing term
by term against OSTY. Expanding $\tfrac12 X_4^2$,
\begin{equation}
\label{eq:V08-I8coeffs}
 I_8^{\mathrm{res}} \;=\; \tfrac{1}{32}\,p_1^2 \;+\; \tfrac{1}{32}\,(\mathrm{tr}\,F_{E_8}^2)^2 \;+\; 2\,c_2(R)^2
 \;+\; \tfrac{1}{16}\,(\mathrm{tr}\,F_{E_8}^2)\,p_1 \;-\; \tfrac12\,c_2(R)\,p_1
 \;-\; \tfrac12\,c_2(R)\,(\mathrm{tr}\,F_{E_8}^2),
\end{equation}
where each square coefficient is $\tfrac12$ times the square of an $X_4$ coefficient and each cross
coefficient is a product,
\begin{equation}
\label{eq:V08-sqcross}
 \tfrac12(-\tfrac14)^2=\tfrac{1}{32},\quad \tfrac12(2)^2=2;\qquad
 (-\tfrac14)(-\tfrac14)=\tfrac{1}{16},\quad (2)(-\tfrac14)=-\tfrac12.
\end{equation}
Every coefficient in \eqref{eq:V08-I8coeffs} matches the pinned OSTY residual, machine checked term by
term (compared against the OSTY table, not self-declared; a wrong $X_4$ dropping the $E_8$ term is
rejected); Table~\ref{tab:V08-anomaly} records their origin as squares and products of the $X_4$ data.
The cross coefficients are the load-bearing anti-vacuous check: they are not free but the \emph{products}
of the $X_4$ coefficients, so an $I_8$ whose cross terms miss those products is not a perfect square.
That conspiracy is Green--Schwarz factorization, the next section.

\subsection*{The gauge-anomaly cancellation condition}

In a $6d\ \mathcal{N}=(1,0)$ theory \emph{with} a gauge group $\mathfrak{g}$ carrying an independent quartic Casimir,
the tr\,$F^4$ vanishing \eqref{eq:V08-Bsum} becomes a constraint on the charged matter, the working tool
the reader most often needs. With hypermultiplets in representations $R_f$ of multiplicity $n_f$,
\begin{equation}
\label{eq:V08-gaugecancel}
 B_{\mathrm{adj}} \;-\; \sum_f n_f\,B_{R_f} \;=\; 0
\end{equation}
must hold identically ($B_R$ the quartic index of \eqref{eq:V08-indexdata}, hypers entering opposite to
the gaugino), the direct $6d$ analog of the $4d$ cubic-anomaly cancellation. When $\mathfrak{g}$ has no
independent quartic Casimir, \eqref{eq:V08-gaugecancel} is vacuous and the whole gauge anomaly is left to
Green--Schwarz. When it does hold, the residual reducible anomaly ($A_R$, $C_R$, gravitational,
R-symmetry) must still factorize, fixing $n_T$ and the lattice $\Omega_{ij}$. Both steps are necessary;
the machine-checked $SU(5)$ falsifier (an independent quartic that cannot fold) is the statement that
\eqref{eq:V08-gaugecancel} is not automatic.

A concrete count makes this tangible. For $\mathfrak{su}(N)$ ($N\ge4$), with the fundamental normalized
to $B_{\square}=1$, the quartic indices are
\begin{equation}
\label{eq:V08-suNindices}
 B_{\mathrm{adj}} \;=\; 2N, \qquad
 B_{\square} \;=\; 1, \qquad
 B_{\mathrm{antisym}} \;=\; N-8,
\end{equation}
so with $n_{\square}$ fundamental and $n_{\mathrm{a}}$ antisymmetric hypermultiplets
\eqref{eq:V08-gaugecancel} reads
\begin{equation}
\label{eq:V08-suNcancel}
 2N \;-\; n_{\square} \;-\; n_{\mathrm{a}}\,(N-8) \;=\; 0,
\end{equation}
a Diophantine constraint on the matter with two standard solutions,
\begin{equation}
\label{eq:V08-suNsolutions}
 n_{\mathrm{a}}=0:\ n_{\square}=2N;\qquad
 n_{\mathrm{a}}=1:\ n_{\square}=2N-(N-8)=N+8.
\end{equation}
The first, $\mathfrak{su}(N)$ with $2N$ fundamentals, is the $6d$ cousin of ``vector-like matter is
anomaly-free''; the second, one antisymmetric plus $N+8$ fundamentals, is the content on the tensor
branch of a well-known $6d$ SCFT family. Either way the residual reducible anomaly must then factorize
with a suitable $n_T$, the step the F-theory engines solve geometrically.

\begin{table}[ht]
\centering
\small
\setlength{\tabcolsep}{8pt}
\renewcommand{\arraystretch}{1.4}
\begin{tabular}{@{}lcl@{}}
\toprule
monomial & coefficient in $I_8$ & origin in $\tfrac12 X_4^2$ \\
\midrule
$p_1^2$ & $\tfrac{1}{32}$ & $\tfrac12(-\tfrac14)^2$ (square) \\
\midrule
$(\mathrm{tr}\,F_{E_8}^2)^2$ & $\tfrac{1}{32}$ & $\tfrac12(-\tfrac14)^2$ (square) \\
\midrule
$c_2(R)^2$ & $2$ & $\tfrac12(2)^2$ (square) \\
\midrule
$(\mathrm{tr}\,F_{E_8}^2)\,p_1$ & $\tfrac{1}{16}$ & $(-\tfrac14)(-\tfrac14)$ (cross) \\
\midrule
$c_2(R)\,p_1$ & $-\tfrac12$ & $(2)(-\tfrac14)$ (cross) \\
\midrule
$c_2(R)\,(\mathrm{tr}\,F_{E_8}^2)$ & $-\tfrac12$ & $(2)(-\tfrac14)$ (cross) \\
\midrule
$\mathrm{tr}\,F_{E_8}^4$ & $0$ & irreducible: cancels by $E_8$ group theory \\
\bottomrule
\end{tabular}
\caption{The residual anomaly eight-form $I_8$ of the rank-one $E$-string, coefficient by coefficient,
with each entry's origin as a square or a cross product of the $X_4=-\tfrac14 p_1-\tfrac14\mathrm{tr}\,
F_{E_8}^2+2c_2(R)$ coefficients. The three cross products are the nontrivial factorization checks: they
are fixed once the squares fix $X_4$, so they are not free. The irreducible $\mathrm{tr}\,F^4$ vanishes
because $E_8$ has no independent quartic Casimir. Every entry is machine checked against the pinned OSTY
values.}
\label{tab:V08-anomaly}
\end{table}

\section{Green--Schwarz--Sagnotti--West factorization}
\label{sec:V08-gs}

Once the irreducible $\mathrm{tr}\,F^4$ piece cancels, the residual anomaly is a reducible eight-form.
For the theory to be consistent this residual must \emph{factorize} into a sum of squares of
degree-four forms, one per tensor multiplet,
\begin{equation}
\label{eq:V08-gsfactor}
 I_8 \;=\; \tfrac12\,\Omega_{ij}\,X_4^i\,X_4^j,
\end{equation}
where $i,j=1,\dots,n_T$ run over the tensor multiplets, $\Omega_{ij}$ is a symmetric integer matrix
(the Dirac-pairing lattice metric of the self-dual string charges), and each
\begin{equation}
\label{eq:V08-X4general}
 X_4^i \;=\; \tfrac14\,a^i\,p_1 \;+\; \sum_a b^i_a\,\tfrac14\,\mathrm{tr}_a F^2 \;+\; \dots
\end{equation}
is a degree-four form carrying the Green--Schwarz coefficients $a^i$ (the gravitational coefficient)
and $b^i_a$ (the gauge/flavor coefficients). Factorizability is the condition that makes Green--Schwarz
inflow possible. Under a gauge or Lorentz transformation the two-form $B^i$ is postulated to shift,
$\delta B^i\sim X_2^{i,(1)}$ (the descent partner of $X_4^i$), so that the \emph{classical} coupling
\begin{equation}
\label{eq:V08-gscoupling}
 S \;\supset\; \sum_{i,j}\Omega_{ij}\int B^i\wedge X_4^j
\end{equation}
has a gauge variation that is exactly the one-loop anomaly with the opposite sign. To see the
cancellation as bookkeeping, apply the descent \eqref{eq:V08-descent} to the factorized polynomial. The
one-loop anomalous variation of the effective action is $\delta_\lambda\log Z=2\pi i\int_{M_6}I_6^{(1)}$
with $I_6^{(1)}$ descended from $I_8=\tfrac12\Omega_{ij}X_4^iX_4^j$. The classical coupling
\eqref{eq:V08-gscoupling}, with the postulated shift $\delta_\lambda B^i=\Omega^{ij}X_{2,j}^{(1)}$ (the
descent partner $X_4^j=dX_3^{(0)j}$, $\delta_\lambda X_3^{(0)j}=dX_2^{(1)j}$), varies into
\begin{equation}
\label{eq:V08-inflowcancel}
 \delta_\lambda\Big(\sum_{i,j}\Omega_{ij}\int_{M_6} B^i\wedge X_4^j\Big)
 \;=\; \sum_{i,j}\Omega_{ij}\int_{M_6}\big(\delta_\lambda B^i\big)\wedge X_4^j
 \;=\; -\,2\pi i\int_{M_6} I_6^{(1)},
\end{equation}
equal and opposite to the one-loop anomaly, so the total is anomaly-free. The mechanism is the
six-dimensional descendant of the $10d$ Green--Schwarz cancellation, the multi-tensor version due to
Sagnotti and to Schwarz and West; the counterterm is classical, the coupling of the dynamical tensor
$B^i$ to the anomaly factor $X_4^j$.

This is the deepest reason the tensor branch is not optional. A theory whose residual does \emph{not}
factorize as \eqref{eq:V08-gsfactor} admits no local counterterm and is inconsistent: the tensor
multiplets supply the $B^i$ and the coupling \eqref{eq:V08-gscoupling}, so $n_T$ and $\Omega_{ij}$ are
fixed by anomaly consistency. Every interacting $6d\ \mathcal{N}=(1,0)$ SCFT therefore has a tensor branch, demanded
by Green--Schwarz.

\subsection*{Worked instance: factorizing the $E$-string residual}

Return to the rank-one $E$-string, now solving \eqref{eq:V08-gsfactor} from the residual $I_8^{\mathrm{res}}$
of \eqref{eq:V08-I8coeffs} rather than assuming the answer. With one tensor multiplet we take
$\Omega=1$ (a single self-dual string of unit charge, the unimodular $1\times1$ lattice) and seek a
single $X_4=\alpha\,p_1+\beta\,\mathrm{tr}\,F_{E_8}^2+\gamma\,c_2(R)$. The three square coefficients fix
the magnitudes,
\begin{equation}
\label{eq:V08-squares}
 \tfrac12\alpha^2 = \tfrac{1}{32},\qquad
 \tfrac12\beta^2 = \tfrac{1}{32},\qquad
 \tfrac12\gamma^2 = 2,
\end{equation}
so $\alpha=\pm\tfrac14$, $\beta=\pm\tfrac14$, $\gamma=\pm2$. Fixing the signs by the OSTY convention
(the two Pontryagin/gauge coefficients negative, the R-symmetry coefficient positive) gives
\begin{equation}
\label{eq:V08-X4solved}
 \Omega \;=\; 1,\qquad
 X_4 \;=\; -\tfrac14\,p_1 \;-\; \tfrac14\,\mathrm{tr}\,F_{E_8}^2 \;+\; 2\,c_2(R),
\end{equation}
which are the Green--Schwarz coefficients $a=-1$ (from $X_4\supset\tfrac14 a\,p_1$), $b_{E_8}=1$, and
$c_2(R)$-coefficient $2$. So far this only used the three square coefficients; the factorization is
\emph{nontrivial} because the same $\alpha,\beta,\gamma$ must reproduce the three \emph{cross}
coefficients of $I_8$:
\begin{equation}
\label{eq:V08-crosschecks}
 \alpha\beta \;=\; \tfrac{1}{16},\qquad
 \gamma\alpha \;=\; -\tfrac12,\qquad
 \gamma\beta \;=\; -\tfrac12,
\end{equation}
matching the $(\mathrm{tr}\,F^2)p_1$, $c_2(R)p_1$, and $c_2(R)(\mathrm{tr}\,F^2)$ coefficients of
\eqref{eq:V08-I8coeffs} exactly. All three hold with the solved values, so the residual genuinely
factorizes, and reassembling $\tfrac12\Omega X_4^2$ reproduces the pinned OSTY residual on every
monomial. The full solve-and-reassemble is machine verified: $X_4$ is \emph{solved} from the residual,
the three cross relations \eqref{eq:V08-crosschecks} are an independent test, and the reassembly is
compared against the pinned OSTY data. The falsifiers bite, a stray $p_2$ or a nonzero irreducible
$\mathrm{tr}\,F^4$ has no $X_4X_4$ home and does not factorize, and dropping the overall $\tfrac12$ or
the $E_8$ term fails the compare. The $E$-string passes: its consistency is the tensor multiplet doing
its Green--Schwarz job.

\section{Conformal anomalies on the tensor branch}
\label{sec:V08-acanomaly}

A $6d$ superconformal field theory has a Weyl anomaly, and unlike four dimensions it carries \emph{four}
independent coefficients. On a curved background the trace of the stress tensor is the six-dimensional
Euler density $E_6$ plus three Weyl-tensor invariants $I_1,I_2,I_3$,
\begin{equation}
\label{eq:V08-weyl}
 \langle T^\mu_{\ \mu}\rangle \;=\; a\,E_6 \;+\; c_1\,I_1 \;+\; c_2\,I_2 \;+\; c_3\,I_3.
\end{equation}
There is no single $4d$-style $c$: six dimensions has three degree-six Weyl invariants where four
dimensions had one Weyl-squared term, so $(a,c_1,c_2,c_3)$ is the central-charge data. The free
multiplets fix the normalization; their $(a,c_i)$ contributions are the benchmarks, exactly as in
Section~3, with two structural facts,
\begin{equation}
\label{eq:V08-cfacts}
 \frac{c_1}{c_2}\bigg|_{\text{hyper}} \neq \frac{c_1}{c_2}\bigg|_{\text{tensor}}
 \ \Rightarrow\ \text{no universal }c;\qquad
 a\ \text{monotone (the $a$-theorem)},\quad c_i\ \text{not}.
\end{equation}
The three $c_i$ are genuinely independent (hyper and tensor have different ratios), while only $a$ is
universal and monotone. The $\mathcal{N}=(2,0)$ per-rank benchmark used below is the decomposition
\begin{equation}
\label{eq:V08-2020decomp}
 \mathcal{N}=(2,0)\ \text{tensor} \;=\; \mathcal{N}=(1,0)\ \text{tensor} \;\oplus\; \mathcal{N}=(1,0)\ \text{hyper}.
\end{equation}

These coefficients are read off the anomaly polynomial by matching on the tensor branch. The polynomial
$I_8$ is an RG invariant, the same at the origin and on the branch (a free collection of tensor, vector,
and hypermultiplets), so
\begin{equation}
\label{eq:V08-matching}
 I_8^{\mathrm{SCFT}} \;=\; \sum_{\text{free fields}}I_8^{\mathrm{free}}
 \;+\; I_8^{\text{origin}},
\end{equation}
with $I_8^{\text{origin}}$ the interacting piece localized at the origin. This plays the organizing role
that a-maximization plays in $4d\ \mathcal{N}=1$, but it is an equation, not an extremization.

The extraction works because supersymmetry ties the anomalies to $I_8$: the R-symmetry sits in the
stress-tensor multiplet, so the $c_2(R),p_1,p_2$ coefficients of the gravitational/R part
\begin{equation}
\label{eq:V08-I8grav}
 I_8 \;\supset\; \alpha\,c_2(R)^2 \;+\; \beta\,c_2(R)\,p_1 \;+\; \gamma\,p_1^2 \;+\; \delta\,p_2
\end{equation}
map linearly to the conformal anomalies. The Córdova--Dumitrescu--Intriligator (CDI) relation for the
Euler anomaly is
\begin{equation}
\label{eq:V08-cdimap}
 a \;=\; \tfrac{16}{7}\big(\alpha-\beta+\gamma\big) \;+\; \tfrac{6}{7}\,\delta,\qquad
 c_i \;=\; \lambda_{c_i}\big(\alpha,\beta,\gamma,\delta\big),
\end{equation}
the Euler row a fixed linear functional of $(\alpha,\beta,\gamma,\delta)$, normalized so a free
$\mathcal{N}=(2,0)$ tensor has $a=1$; the $c_i$ rows and the full $\mathcal{N}=(2,0)$
$(\alpha,\beta,\gamma,\delta)$ evaluation are cited but not derived. We run the Euler row on the
$\mathcal{N}=(2,0)$ case in the next section. The point is that one computes $I_8$ once, and the same
data that ran the Green--Schwarz factorization delivers the conformal anomalies.

\begin{keybox}{Common misconception: $6d$ has a trace-extremization principle like a-maximization}
The extremization through-line in these notes runs a-maximization ($4d\ \mathcal{N}=1$), $c$-extremization
($2d$), and $F$-maximization ($3d$), each selecting the superconformal R by extremizing an anomaly-trace
function over trial symmetries. Six dimensions hosts no such member: the $(a,c_i)$ are fixed directly by
anomaly-polynomial \emph{factorization} and \emph{matching} on the tensor branch, not by a variational
principle, and there is no $6d$ trace whose extremum is the superconformal R. Six dimensions sits
\emph{beside} the a-max/$c$-ext/$F$-max family; the closest analog is the matching
\eqref{eq:V08-matching}, an equation, not an extremum.
\end{keybox}

There is a $6d$ $a$-theorem: along a flow between $6d$ SCFTs the Euler coefficient decreases,
\begin{equation}
\label{eq:V08-atheorem}
 a_{\mathrm{UV}} \;>\; a_{\mathrm{IR}},
\end{equation}
the cousin of the four-dimensional $a$-theorem. The flow to the tensor branch is itself a $6d$ flow with
$a_{\mathrm{SCFT}}>a_{\text{free branch}}$; we state \eqref{eq:V08-atheorem} and route its proof to
the cited literature. The worked number is the $\mathcal{N}=(2,0)$ Euler
coefficient $a\sim N^3$, computed next.

\bigskip
\begin{center}
\rule{0.4\textwidth}{0.4pt}\\[3pt]
{\large\textsf{\textbf{Block B.\enspace The $6d\ \mathcal{N}=(2,0)$ world}}}\\[2pt]
\rule{0.4\textwidth}{0.4pt}
\end{center}
\medskip

\noindent The maximal $\mathcal{N}=(2,0)$ theory is the non-Lagrangian contrast: an ADE-classified interacting fixed point with no Lagrangian, its M5-brane origin, and the marquee conformal anomaly $a\sim N^3$.

\section{The maximal \texorpdfstring{$\mathcal{N}=(2,0)$}{N=(2,0)} theory}
\label{sec:V08-twozero}

The maximal supersymmetric fixed point of six dimensions is the $6d\ \mathcal{N}=(2,0)$ theory: sixteen real
supercharges (two same-chirality $6d$ Weyl supercharges), R-symmetry $USp(4)_R=Spin(5)_R$, and a single
$\mathcal{N}=(2,0)$ \emph{tensor} multiplet,
\begin{equation}
\label{eq:V08-twozerocontent}
 B^+\ (H^+=+\star H^+),\qquad 5\ \text{scalars in the }\mathbf 5\text{ of }SO(5)_R,\qquad \text{fermions},
\end{equation}
the five scalars being transverse position moduli and $B^+$ the maximal-symmetry version of the $\mathcal{N}=(1,0)$
chiral tensor. The interacting theories are classified by a simply-laced (ADE) algebra
$\mathfrak{g}\in\{A_n,D_n,E_6,E_7,E_8\}$, one per algebra, with tensor branch and strings
\begin{equation}
\label{eq:V08-adebranch}
 \mathcal{M} \;=\; (\mathbb{R}^5\otimes\mathfrak{h})/W,\qquad
 \mathrm{rank}=\mathrm{rank}\,\mathfrak{g},\qquad
 \text{tensionless strings }\leftrightarrow\ \text{roots}.
\end{equation}
The $A_{N-1}$ series is $N$ coincident M5-branes in flat space; the $D,E$ series need an orbifold or
M-theory orientifold, which we cite rather than construct. The ADE pattern is the same simply-laced structure
that classified the $C^2/\Gamma$ singularities and the $4d\ \mathcal{N}=2$ gauge symmetries. We work the
$A_{N-1}$ marquee; the classification proof is cited.

\medskip\noindent\textbf{M5-branes and the rank.}\enspace
A single M5-brane carries one \emph{free} abelian $\mathcal{N}=(2,0)$ tensor, its center-of-mass sector. Stacking
$N$, the interacting theory is the \emph{relative} theory after decoupling that free tensor,
\begin{equation}
\label{eq:V08-rank}
 \mathrm{rank}\big(A_{N-1}\big) \;=\; N-1,\qquad
 N=1:\ \text{rank }0\ (\text{free tensor}),\qquad
 A_1:\ N=2.
\end{equation}
This is a common place to slip: a single M5 is the free tensor, not an interacting theory, and the
rank-one \emph{interacting} theory $A_1$ is \emph{two} coincident M5s. Equation \eqref{eq:V08-rank} is
machine checked, with the falsifier that a single M5 is $A_1$ rejected.

\medskip\noindent\textbf{The absence of a Lagrangian.}\enspace
The obstruction to a $\mathcal{N}=(2,0)$ Lagrangian is the self-dual tensor itself. The naive kinetic term
\emph{vanishes} on a self-dual $H$,
\begin{equation}
\label{eq:V08-nolag}
 \int H\wedge\star H \;=\; \int H\wedge H \;=\; 0 \qquad (H=\star H\ \text{up to signs}),
\end{equation}
and imposing self-duality as an equation of motion from a Lorentz-covariant action is obstructed. Even
the free $\mathcal{N}=(2,0)$ tensor is non-Lagrangian in the strict covariant sense; the interacting $A_{N-1}$ theory
has no Lagrangian at all without auxiliary or non-covariant machinery. This is the sharpest instance
here of a field theory with no Lagrangian.

\begin{keybox}{Common misconception: no Lagrangian means it is not a field theory}
The $\mathcal{N}=(2,0)$ theory has no covariant Lagrangian, and it is tempting to conclude it is not a field theory.
That is wrong. It is a bona fide local quantum field theory: a stress tensor, R-symmetry $USp(4)_R$, a
moduli space, BPS self-dual strings, a well-defined $I_8$, and conformal anomalies $(a,c_i)$ with the
sharp $a\sim N^3$ below. The absence of a Lagrangian is a statement about \emph{presentation}, not
existence. This is the $6d$ sharpening of Section~2's non-Lagrangian grammar, and the partner of the
$4d\ \mathcal{N}=2$ Argyres--Douglas fact of Section~5: there the obstruction was mutually non-local BPS
states, here the self-dual tensor.
\end{keybox}

\medskip\noindent\textbf{The tensor branch.}\enspace
The five scalars of each tensor multiplet are the transverse M5-brane positions in $\mathbb{R}^5$.
Separating $N$ coincident branes gives
\begin{equation}
\label{eq:V08-twozerobranch}
 \mathcal{M}_{A_{N-1}} \;=\; \big(\mathbb{R}^5\big)^{N-1}\big/S_N,
\end{equation}
the $N$ positions modulo the overall center of mass (leaving $N-1$ factors) and modulo the permutation
$S_N$ exchanging identical branes; its rank $N-1$ matches \eqref{eq:V08-rank}. At a generic point the
theory is $N-1$ free $\mathcal{N}=(2,0)$ tensors; at the origin the interacting SCFT appears. The self-dual strings
are M2-branes stretched between the M5s, with tension proportional to the separation, so they become
tensionless as the branes come together, exactly the mechanism of \S\ref{sec:V08-tensorbranch}.

\begin{figure}[ht]
\centering
\begin{tikzpicture}[scale=1.0]
 % separated M5-branes (vertical bars) with an M2 stretched between two of them
 \foreach \x/\lab in {0/1,1.1/2,2.6/3} {
 \draw[very thick,RoyalBlue] (\x,-0.9) -- (\x,0.9);
 }
 \node[below=2pt] at (0,-0.9) {\small M5};
 \node[below=2pt] at (1.1,-0.9) {\small M5};
 \node[below=2pt] at (2.6,-0.9) {\small M5};
 % M2 stretched between brane 2 and 3 = a self-dual string of tension ~ separation
 \fill[gray!30] (1.1,-0.25) rectangle (2.6,0.25);
 \node at (1.85,0) {\small M2};
 \node at (1.85,1.2) {\tiny $T\sim$ separation};
 \draw[|-|] (1.1,-0.55) -- (2.6,-0.55);
 % arrow to the coincident (origin) configuration
 \draw[-{Stealth[length=2.4mm]}] (3.4,0) -- (4.6,0);
 \node[above] at (4.0,0.05) {\tiny origin};
 % coincident stack
 \draw[very thick,RoyalBlue] (5.4,-0.9) -- (5.4,0.9);
 \draw[very thick,RoyalBlue] (5.5,-0.9) -- (5.5,0.9);
 \draw[very thick,RoyalBlue] (5.6,-0.9) -- (5.6,0.9);
 \node[below=2pt] at (5.5,-0.9) {\small $N$ M5};
 \node at (5.5,1.2) {\tiny $T\to0$: SCFT};
\end{tikzpicture}
\caption{The $\mathcal{N}=(2,0)$ $A_{N-1}$ tensor branch as M5-brane transverse positions (here $N=3$), schematic.
Separated M5-branes give $N-1$ free tensor multiplets; an M2-brane stretched between two M5s is a
self-dual string with tension proportional to the separation. As the branes coincide (right) the
stretched M2 becomes a tensionless string and the interacting superconformal field theory of rank $N-1$
appears at the origin. The five transverse directions are drawn as one.}
\label{fig:V08-m5}
\end{figure}
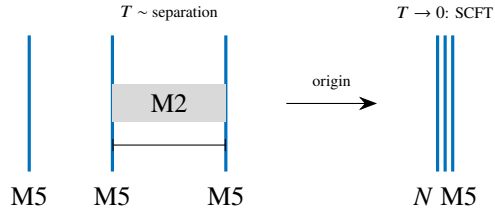

\medskip\noindent\textbf{The $\mathcal{N}=(2,0)$ anomaly polynomial.}\enspace
Although the $\mathcal{N}=(2,0)$ theory has no Lagrangian, its anomaly polynomial is completely known, the clearest
sign that it is a genuine field theory. For the $A_{N-1}$ theory the eight-form is
(Harvey--Minasian--Moore, Intriligator)
\begin{equation}
\label{eq:V08-I820}
 I_8^{(2,0)}[A_{N-1}] \;=\; (N-1)\,I_8^{\mathrm{free}} \;+\; \frac{N(N^2-1)}{24}\,p_2(N_R),
\end{equation}
with $I_8^{\mathrm{free}}$ the free-tensor eight-form (the $N-1$ Cartan multiplets) and $p_2(N_R)$ the
Pontryagin class of the $SO(5)_R$ normal bundle. Two features are load-bearing,
\begin{equation}
\label{eq:V08-cubicvanish}
 N(N^2-1) \;=\; N^3-N \quad(\text{cubic in }N),\qquad
 N(N^2-1)\big|_{N=1} \;=\; 0 \quad(\text{rank }0),
\end{equation}
the cubic coefficient being the anomaly-polynomial origin of the $N^3$ and its vanishing at $N=1$ the
free single-M5 limit \eqref{eq:V08-rank}. The $a$ and $c_i$ are read off \eqref{eq:V08-I820} by the
matching of \S\ref{sec:V08-acanomaly}, and both inherit the cubic growth.

\subsection*{Worked instance: the \texorpdfstring{$\mathcal{N}=(2,0)$}{N=(2,0)} conformal anomaly $a\sim N^3$}

The marquee number of the $\mathcal{N}=(2,0)$ theory is its Euler conformal anomaly, growing as the \emph{cube} of
$N$, the sharpest ``non-Lagrangian but field-theoretic'' number in these notes: a theory with no Lagrangian
has an exact central charge. The CDI relation \eqref{eq:V08-cdimap}, evaluated on the
$\mathcal{N}=(2,0)$ anomaly data, gives their closed form for the Euler anomaly of the type-$g$ theory,
$a=\tfrac{16}{7}\,h^\vee_g d_g + r_g$ (CDI, Table~1). For $A_{N-1}=\mathfrak{su}(N)$ the dual Coxeter
number, dimension, and rank are $(h^\vee,d,r)=(N,\,N^2-1,\,N-1)$, so this splits into $(N-1)$ free
tensors plus an interacting piece with the Harvey--Minasian--Moore cubic structure,
\begin{equation}
\label{eq:V08-aN}
 a(N) \;=\; \tfrac{16}{7}\,h^\vee d + r \;=\; (N-1)\,a_{\mathrm{free}} \;+\; a_{\mathrm{int}}(N), \qquad
 a_{\mathrm{int}}(N) \;=\; \tfrac{16}{7}\,\big(N^3-N\big),
\end{equation}
in the normalization $a_{\mathrm{free}}=1$ per free $\mathcal{N}=(2,0)$ tensor (the CDI convention); with
that normalization fixed, the interacting coefficient $\tfrac{16}{7}$ is the CDI value, not a free
choice. The convention-independent content is the cubic growth $a\sim N^3$ and the tensor-branch
decomposition into $(N-1)$ free tensors plus the interacting $\tfrac{16}{7}(N^3-N)$; the displayed $a(N)$ is
thus the \emph{full} $A_{N-1}$ anomaly, the tensor-branch free tensors together with the interacting
piece, not the interacting-relative theory with the center-of-mass tensor removed. Expanding,
\begin{equation}
\label{eq:V08-aNexpanded}
 a(N) \;=\; \tfrac{16}{7}\,N^3 \;+\; N \;-\; \tfrac{16}{7}\,N \;-\; 1
 \;=\; \tfrac{16}{7}\,N^3 \;-\; \tfrac97\,N \;-\; 1,
\end{equation}
a genuine cubic whose degree is read off the polynomial, not hard-coded. The limits and the leading
behaviour pin the physics,
\begin{equation}
\label{eq:V08-aNvals}
 a(1)=0,\qquad a(2)=\tfrac{128}{7}-\tfrac{18}{7}-1=\tfrac{103}{7},\qquad a(N)\sim\tfrac{16}{7}\,N^3,
\end{equation}
$a(1)=0$ the free rank-zero limit, $a(2)=\tfrac{103}{7}$ the $A_1$ (two-M5) value, and $\tfrac{16}{7}N^3$
the large-$N$ leading term. This beats the naive $N^2$ of a gauge theory,
\begin{equation}
\label{eq:V08-N3overN2}
 \lim_{N\to\infty}\frac{a(N)}{N^3}=\tfrac{16}{7}\ (\text{finite}),\qquad
 \lim_{N\to\infty}\frac{a(N)}{N^2}=\infty,
\end{equation}
so the $\mathcal{N}=(2,0)$ theory has parametrically more degrees of freedom than any Lagrangian gauge theory of
comparable size. The exact coefficient is the CDI Table-1 value $a=\tfrac{16}{7}h^\vee d+r$; the engine
reconstructs this polynomial and machine-checks its degree ($3$), its leading coefficient
($\tfrac{16}{7}$, discriminated against the wrong $\tfrac43$), the values $a(1)=0$ and
$a(2)=\tfrac{103}{7}$, and the $N^3$-over-$N^2$ growth by symbolic limits. The falsifiers reject an
$N^2$ scaling, a vanishing leading coefficient, and a single M5 mis-identified as $A_1$.

The $N^3$ is what makes the $\mathcal{N}=(2,0)$ theory strange. An ordinary $SU(N)$ gauge theory counts
$\dim\mathfrak{su}(N)\sim N^2$ gluons, so every large-$N$ central charge grows as $N^2$; the $\mathcal{N}=(2,0)$
theory has no gluons, no coupling, no Lagrangian, yet grows one power faster, both $a$ and the $c_i$
sharing the cubic behaviour of \eqref{eq:V08-I820}. The same $N^3$ appears on the gravity side as the
free energy of $N$ M5-branes in eleven-dimensional supergravity (Henningson--Skenderis), the holographic
$N^3$ that first signalled M5-branes have no weakly-coupled field-theory description. The field-theory
computation reproduces it from the anomaly polynomial alone. The construction, tensionless strings, and
M-theory origin are cited rather than developed here.

\medskip\noindent\textbf{Compactification and the class S pointer.}\enspace
The $\mathcal{N}=(2,0)$ theory is the geometric origin of the $4d\ \mathcal{N}=2$ S-dualities of Section~5. Because
it is conformal it has no intrinsic scale, so reducing $A_{N-1}$ on a two-torus $T^2$ of complex
structure $\tau$ gives a $4d$ theory depending on $T^2$ only through $\tau$,
\begin{equation}
\label{eq:V08-T2reduction}
 A_{N-1}\ \text{on}\ T^2_\tau \;\longrightarrow\; 4d\ \mathcal{N}=4\ \mathfrak{su}(N),\qquad
 \tau=\text{coupling},\quad SL(2,\mathbb{Z})_{T^2}=\text{S-duality}.
\end{equation}
Montonen--Olive duality is then the statement that two $SL(2,\mathbb{Z})$-related tori are the same
torus, invisible from six dimensions, and it exists only because the parent has no coupling of its own.
Compactifying instead on a punctured Riemann surface $\mathcal{C}_{g,n}$ gives the class S theories,
\begin{equation}
\label{eq:V08-classS}
 A_{N-1}\ \text{on}\ \mathcal{C}_{g,n} \;\longrightarrow\; 4d\ \mathcal{N}=2,\qquad
 \text{couplings}=\text{moduli},\quad
 \text{S-duality}=\text{MCG},
\end{equation}
with the Seiberg--Witten curve of Section~5 a branched $N$-fold cover of $\mathcal{C}_{g,n}$. We develop
none of the construction; we have stated the $\mathcal{N}=(2,0)$ field-theory facts
the class S program reads from.

\bigskip
\begin{center}
\rule{0.4\textwidth}{0.4pt}\\[3pt]
{\large\textsf{\textbf{Block C.\enspace Maximal super-Yang--Mills}}}\\[2pt]
\rule{0.4\textwidth}{0.4pt}
\end{center}
\medskip

\noindent The $\mathcal{N}=(1,1)$ maximal super-Yang--Mills is the Lagrangian sixteen-supercharge contrast, the non-renormalizable uplift of $4d\ \mathcal{N}=4$ that links to the Section~9 super-Yang--Mills ladder.

\section{The \texorpdfstring{$\mathcal{N}=(1,1)$}{N=(1,1)} maximal super-Yang--Mills}
\label{sec:V08-oneone}

The other sixteen-supercharge theory is $6d\ \mathcal{N}=(1,1)$, and it is the opposite of $\mathcal{N}=(2,0)$ in the one way
that matters here: it \emph{is} a Lagrangian gauge theory. It has R-symmetry
$Spin(4)_R=SU(2)\times SU(2)$ and is maximal $6d$ super-Yang--Mills, the uplift of $4d\ \mathcal{N}=4$,
\begin{equation}
\label{eq:V08-oneonecontent}
 \text{adjoint: }A_\mu,\ 4\ \text{scalars},\ \text{fermions};\qquad
 \text{manifest }\tfrac{1}{g^2}\,\mathrm{tr}\,F^2\ \text{action}.
\end{equation}
But six is above the critical dimension for Yang--Mills. Since $\mathrm{tr}\,F^2$ has mass dimension four
and a $6d$ Lagrangian density must have dimension six,
\begin{equation}
\label{eq:V08-g2dim}
 \Big[\tfrac{1}{g^2}\Big] \;=\; 6-4 \;=\; 2, \qquad\text{hence}\qquad [g^2] \;=\; -2,
\end{equation}
the theory is non-renormalizable, an effective description below a cutoff, with \emph{no} interacting
$6d$ fixed point of its own; its UV completion is stringy (a little-string theory). So $\mathcal{N}=(1,1)$ is the
Lagrangian sixteen-supercharge row and $\mathcal{N}=(2,0)$ the non-Lagrangian one.

One identification must be resisted. The $6d\ \mathcal{N}=(1,1)$ SYM is \emph{not} the $\mathcal{N}=(2,0)$ theory on a circle,
\begin{equation}
\label{eq:V08-circlereduce}
  \begin{aligned}
  \mathcal{N}=(2,0)\ \text{on}\ S^1_R
    &\;\longrightarrow\; 5d\ \mathcal{N}=2\ \text{maximal SYM},\\
  g_5^2 &\sim R,\qquad 1/g_5^2\sim 1/R,\\
  B^+ &\to A_\mu^{(5)},
  \end{aligned}
\end{equation}
one dimension \emph{down}, with the Kaluza--Klein and wrapped-string modes the $5d$ instanton tower. That
reduction gives $5d$, not $6d\ \mathcal{N}=(1,1)$; the two are distinct theories.

The $\mathcal{N}=(1,1)$ SYM is the $6d$ entry of the sixteen-supercharge SYM ladder that Section~9 continues,
$6d\ \mathcal{N}=(1,1)\to 7d\to\dots\to 10d$. In $d$ dimensions the coupling dimension is
\begin{equation}
\label{eq:V08-ladderdim}
 \Big[\tfrac{1}{g^2}\Big]=d-4,\qquad [g^2]=4-d,
\end{equation}
marginal at $d=4$ (the conformal $4d\ \mathcal{N}=4$) and increasingly irrelevant above, as
Table~\ref{tab:V08-symladder} records. We state $\mathcal{N}=(1,1)$ as that contrast; the ladder is developed in
Section~9.

\begin{table}[ht]
\centering
\small
\setlength{\tabcolsep}{8pt}
\renewcommand{\arraystretch}{1.3}
\begin{tabular}{@{}lcc>{\raggedright\arraybackslash}p{52mm}@{}}
\toprule
dimension & supersymmetry & $[g^2]=4-d$ & status \\
\midrule
$4d$ & $\mathcal{N}=4$ & $0$ & marginal: superconformal \\
\midrule
$5d$ & $\mathcal{N}=2$ & $-1$ & irrelevant: UV completion is $6d\ \mathcal{N}=(2,0)$ on $S^1$ \\
\midrule
$6d$ & $\mathcal{N}=(1,1)$ & $-2$ & irrelevant: UV completion stringy (little string) \\
\midrule
$7d$--$9d$ & maximal & $-3$ to $-5$ & irrelevant: no interacting fixed point \\
\midrule
$10d$ & $\mathcal{N}=1$ & $-6$ & low-energy limit of the open superstring \\
\bottomrule
\end{tabular}
\caption{The sixteen-supercharge maximal super-Yang--Mills ladder. The gauge coupling has dimension
$[g^2]=4-d$, marginal only in four dimensions ($4d\ \mathcal{N}=4$). Above four dimensions maximal SYM
is non-renormalizable with no interacting fixed point of its own; $6d\ \mathcal{N}=(1,1)$ is one rung, distinct from
the $\mathcal{N}=(2,0)$ theory (which on a circle gives the $5d$ row, not the $6d$ one). Section~9 develops the
ladder; the coupling dimensions here are the elementary count $[1/g^2]=d-4$.}
\label{tab:V08-symladder}
\end{table}

\medskip\noindent\textbf{The $6d\to5d$ ladder bridge.}\enspace
The $\mathcal{N}=(1,0)$ theories close the dimension ladder downward too. Reducing $6d\ \mathcal{N}=(1,0)$ on a circle gives
$5d\ \mathcal{N}=1$ (Section~7),
\begin{equation}
\label{eq:V08-10to5}
 (\phi,B^-)\ \to\ (\phi_5,A_\mu^{(5)}),\qquad
 \text{self-dual string on }S^1\ \to\ 5d\ \text{instanton}\ (U(1)_I),
\end{equation}
the tensor scalar and two-form descending to the $5d$ vector multiplet and the wrapped string to the
$5d$ instanton tower whose charge is the Section~7 topological $U(1)_I$. The $6d\ \mathcal{N}=(1,0)$ SCFTs are thus
the parents of the $5d\ \mathcal{N}=1$ theories. F-theory classifications of $6d$ SCFTs use the same
tensor-branch, anomaly-polynomial, and Green--Schwarz data emphasized here. We state the bridge as a
pointer, not a worked reduction.

\section*{Exit checklist}
\addcontentsline{toc}{subsection}{Exit checklist}
\markboth{Exit checklist}{Exit checklist}

After this section the reader can
\begin{enumerate}
\item identify the $6d\ \mathcal{N}=(1,0)$ multiplets (vector with no scalar, hyper, tensor with an anti-self-dual
two-form), count the on-shell degrees of freedom of a self-dual two-form as the
$\dim(\mathbf{1},\mathbf{3})=3$ of the little group $SO(4)$, and check that the tensor multiplet closes
supersymmetry as $3+1=4$ bosonic $=4$ fermionic;
\item describe the tensor branch, its dimension $n_T$, the tensor scalar as the inverse gauge coupling
$1/g^2\sim\langle\phi\rangle$, and the self-dual string that becomes tensionless at the origin, and
distinguish it from the (absent) $6d$ Coulomb branch;
\item write the $6d$ anomaly as the eight-form $I_8$, assemble it from the $\hat A(R)\,\mathrm{ch}(F)$
building blocks and the self-dual-tensor term, and impose the irreducible $\mathrm{tr}\,F^4$
cancellation (vanishing for $E_8$ by $\mathrm{tr}\,F^4=\tfrac{1}{100}(\mathrm{tr}\,F^2)^2$);
\item run the Green--Schwarz--Sagnotti--West factorization $I_8=\tfrac12\Omega_{ij}X_4^iX_4^j$ on the
rank-one $E$-string, extract $\Omega=1$ and $X_4=-\tfrac14 p_1-\tfrac14\mathrm{tr}\,F_{E_8}^2+2c_2(R)$,
verify the three cross relations, and explain why the tensor branch is required for consistency;
\item state the four $6d$ Weyl-anomaly coefficients $(a,c_1,c_2,c_3)$, read them off the anomaly
polynomial by matching on the tensor branch, and place $6d$ beside the a-max / $c$-ext / $F$-max
extremization family rather than inside it;
\item describe the $6d\ \mathcal{N}=(2,0)$ theory (self-dual tensor multiplet, $USp(4)_R$, the ADE list), explain
why it has no Lagrangian yet is a field theory, count its rank as $N-1$ (a single M5 is the free tensor,
$A_1$ is two M5s), and compute its conformal anomaly $a(N)=\tfrac{16}{7} N^3-\tfrac97 N-1\sim N^3$;
\item contrast $6d\ \mathcal{N}=(1,1)$ maximal super-Yang--Mills (Lagrangian, $[g^2]=-2$, no interacting fixed
point) with $\mathcal{N}=(2,0)$, distinguish it from the $\mathcal{N}=(2,0)$ theory on a circle (which is $5d$ maximal SYM), and
state which deep results (the anomaly-polynomial derivation, the $6d$ SCFT classification, the $\mathcal{N}=(2,0)$
construction) are cited rather than proved.
\end{enumerate}

\bigskip
\section*{Sources and notes}
\addcontentsline{toc}{subsection}{Sources and notes}
\markboth{Sources and notes}{Sources and notes}
{\small

\noindent\textsf{\textcolor{RoyalBlue}{Sources and notes.}}\enspace
This is the six-dimensional dimension section of these notes, the top of the dimension ladder before the
pure super-Yang--Mills interface of Section~9.

\medskip\noindent\textsf{\textcolor{RoyalBlue}{\textbf{\S\ref{sec:V08-multiplets}\enspace Multiplets and the self-dual tensor.}}}\enspace
The $6d\ \mathcal{N}=(1,0)$ vector (no scalar), hyper, and tensor multiplets (Table~\ref{tab:V08-multiplets}); the
anti-self-dual two-form $H^-=-\star H^-$ \eqref{eq:V08-selfdual}, the Lorentzian signature check
$\star^2=+1$ on a $6d$ three-form \eqref{eq:V08-starsq}, the off-shell count $\binom{6}{2}=15$
\eqref{eq:V08-offshell}, the on-shell self-dual count $\dim(\mathbf{1},\mathbf{3})=3$ (half the generic
$6$) and the fermion $\tfrac12\cdot8=4$ \eqref{eq:V08-fulldof}--\eqref{eq:V08-selfdof}, and the
tensor-multiplet SUSY balance $3+1=4=4$ \eqref{eq:V08-susybalance}. (\textcite{Cordova:2016emh} the $6d$ multiplet
structure). 

\medskip\noindent\textsf{\textcolor{RoyalBlue}{\textbf{\S\ref{sec:V08-tensorbranch}\enspace The tensor branch.}}}\enspace
The tensor branch of real dimension $n_T$ \eqref{eq:V08-branchdim}, the gauge kinetic term
$\phi^i\,\mathrm{tr}\,F_i^2$ and the tensor scalar as the inverse gauge coupling
$1/g^2\sim\langle\phi\rangle$ \eqref{eq:V08-tensorbranch}, and the self-dual string with tension
$T\sim\langle\phi\rangle$ becoming tensionless at the origin
\eqref{eq:V08-stringtension}. (\textcite{Seiberg:1996qx} the interacting $6d$ fixed points and
the tensionless string; \textcite{Heckman:2013pva} the F-theory tensor-branch classification, the engine
home). 

\medskip\noindent\textsf{\textcolor{RoyalBlue}{\textbf{\S\ref{sec:V08-anomaly}\enspace The anomaly polynomial.}}}\enspace
The $6d$ anomaly as the eight-form $I_8$, generated by a box diagram, the $(n+1)$-gon pattern
\eqref{eq:V08-Idplus2} with the descent chain $8\to7\to6$; the
$\hat A(R)\,\mathrm{ch}(F)$ building blocks \eqref{eq:V08-inflow} and the degree-eight monomial basis
\eqref{eq:V08-monobasis}; the irreducible $\mathrm{tr}\,F^4$ obstruction and the $E_8$ identity
$\mathrm{tr}\,F^4=\tfrac{1}{100}(\mathrm{tr}\,F^2)^2$ \eqref{eq:V08-e8identity}; and the worked assembly
of the rank-one $E$-string $I_8$ \eqref{eq:V08-I8coeffs} with $[I_8]_{\mathrm{tr}\,F^4}=0$
\eqref{eq:V08-trF4vanishes} compared to the pinned OSTY values. (\textcite{Schwarz:1995zw} the
$6d$ $\mathrm{tr}\,F^4$ / factorization conditions; \textcite{Ohmori:2014kda} the OSTY anomaly-polynomial
table, the entry-9 named $6d$ source). 

\medskip\noindent\textsf{\textcolor{RoyalBlue}{\textbf{\S\ref{sec:V08-gs}\enspace Green--Schwarz--Sagnotti--West factorization.}}}\enspace
The factorization $I_8=\tfrac12\Omega_{ij}X_4^iX_4^j$ \eqref{eq:V08-gsfactor} with the degree-four form
$X_4^i$ \eqref{eq:V08-X4general}, the $B\wedge X_4$ inflow coupling $\sum\Omega_{ij}\int B^i\wedge X_4^j$
\eqref{eq:V08-gscoupling}, and the tensor branch as required for consistency; the worked $E$-string
factorization solving $\Omega=1$ and $X_4$ \eqref{eq:V08-X4solved} from the square coefficients
\eqref{eq:V08-squares}, with the three cross relations \eqref{eq:V08-crosschecks} and the reassembly
against pinned OSTY data. (\textcite{Green:1984sg} the $10d$ Green--Schwarz mechanism;
\textcite{Sagnotti:1992qw} the multi-tensor $6d$ factorization; \textcite{Ohmori:2014kda} the pinned
$E$-string coefficients). 

\medskip\noindent\textsf{\textcolor{RoyalBlue}{\textbf{\S\ref{sec:V08-acanomaly}\enspace Conformal anomalies on the tensor branch.}}}\enspace
The four $6d$ Weyl-anomaly coefficients $(a,c_1,c_2,c_3)$ \eqref{eq:V08-weyl} (one Euler $a$, three Weyl
$c_i$, no single $4d$-style $c$), read off the anomaly polynomial by matching on the tensor branch
\eqref{eq:V08-matching}; the $6d$ $a$-theorem $a_{\mathrm{UV}}>a_{\mathrm{IR}}$ stated and cited; and $6d$ placed beside the extremization family. (\textcite{Cordova:2016xhm} the
$6d$ Weyl-anomaly $(a,c_i)$ structure; \textcite{Cordova:2015fha} the $6d$ $a$-theorem on the tensor
branch). 

\medskip\noindent\textsf{\textcolor{RoyalBlue}{\textbf{\S\ref{sec:V08-twozero}\enspace The $\mathcal{N}=(2,0)$ theory.}}}\enspace
The $6d\ \mathcal{N}=(2,0)$ tensor multiplet ($USp(4)_R$, self-dual $B^+$, five $SO(5)_R$ scalars) and the ADE
list; the M5 rank bookkeeping $\mathrm{rank}(A_{N-1})=N-1$ \eqref{eq:V08-rank} (single M5 free, $A_1$ two
M5s); the absence of a Lagrangian (the self-dual tensor obstruction); the tensor branch
$(\mathbb{R}^5)^{N-1}/S_N$ \eqref{eq:V08-twozerobranch}; the marquee conformal anomaly
$a(N)=\tfrac{16}{7} N^3-\tfrac97 N-1\sim N^3$ \eqref{eq:V08-aN}--\eqref{eq:V08-aNvals}; and the class S /
pointer. (\textcite{Harvey:1998bx} the $\mathcal{N}=(2,0)$ anomaly polynomial and the $N^3$ scaling;
\textcite{Cordova:2015fha} the $a=\tfrac{16}{7}h^\vee d+r$ Euler-anomaly relation and Table-1 value;
\textcite{Intriligator:2000eq} the anomaly matching and rank corrections; \textcite{Henningson:1998gx}
the gravity-dual $N^3$ cross-check). 

\medskip\noindent\textsf{\textcolor{RoyalBlue}{\textbf{\S\ref{sec:V08-oneone}\enspace The $\mathcal{N}=(1,1)$ maximal super-Yang--Mills.}}}\enspace
$6d\ \mathcal{N}=(1,1)$ as Lagrangian maximal SYM ($Spin(4)_R$, the uplift of $4d\ \mathcal{N}=4$), non-renormalizable
with $[g^2]=-2$ \eqref{eq:V08-g2dim} and no interacting $6d$ fixed point; the warning that it is not the
$\mathcal{N}=(2,0)$ theory on a circle (that reduction is $5d$ maximal SYM); the $6d\to5d$ ladder bridge (tensor
$\to$ $5d$ vector, wrapped string $\to$ instanton tower); and the forward pointer to the Section~9
sixteen-supercharge SYM ladder. (\textcite{Witten:1995zh} the $6d$ string dynamics and the
little-string / $\mathcal{N}=(2,0)$ context). Statement-level aside; the $16$Q count is the Section~1 oracle and the
SYM-ladder bookkeeping is Section~9's; this is a stated contrast and bridge.

\medskip\noindent\textbf{Stated, not proved here.}\enspace
The anomaly-polynomial building-block derivation, the general Green--Schwarz-factorization theorem, the
full $6d$ anomaly-inflow argument, and the $6d$ $a$-theorem; the construction of the interacting
$\mathcal{N}=(2,0)$ theory, the tensionless-string dynamics, the M5-brane / M-theory origin, and the class S
compactification of $\mathcal{N}=(2,0)$ on a punctured surface; and
the F-theory tensor-branch classification of $6d\ \mathcal{N}=(1,0)$ SCFTs. All are stated and cited
here; their proofs and constructions are outside the scope of these notes.
}

\subsection*{Further reading}
\addcontentsline{toc}{subsection}{Further reading}
The $\mathcal{N}=(2,0)$ theory was identified in \textcite{Witten:1995zh}, with tensionless strings from
\textcite{Strominger:1995ac}. Its anomaly polynomial and $N^3$ scaling are in
\textcite{Harvey:1998bx,Yi:2001bz}, the general six-dimensional case in \textcite{Ohmori:2014kda}, and
the $E$-string in \textcite{Ohmori:2014pca}. The Euler anomaly and the six-dimensional $a$-theorem are
in \textcite{Cordova:2015fha}. The classification of $6d\ \mathcal{N}=(1,0)$ theories and conformal
matter is in \textcite{Heckman:2013pva,DelZotto:2014hpa,Bhardwaj:2015xxa}, reviewed in
\textcite{Heckman:2018jxk}.

More recent refinements of the $6d$ classification appear in \textcite{Bhardwaj:2019hhd}; modern
treatments of higher-group symmetry and anomaly matching in six dimensions are
\textcite{Cordova:2020tij,Apruzzi:2021mlh}, while the bordism-based status of global gauge anomalies
is clarified in \textcite{Davighi:2020kok}.

\section*{References}
\addcontentsline{toc}{subsection}{References}
\markboth{References}{References}
\printbibliography[heading=none]
\end{refsection}
\begin{refsection}\chapter{\texorpdfstring{$7d$ to $10d$}{7d to 10d} supersymmetric field theories}
\label{ch:V09}

\noindent\textbf{Guide to this section.}\enspace
Sections~3 through~8 taught the rich low dimensions, each with its own dimension-specific machinery:
the holomorphic workhorse, the gauged linear sigma model, special geometry, the monopole operator, the
cubic prepotential, the tensor branch. This final dimension section caps the ladder. Above six
dimensions, at sixteen supercharges, there is exactly one multiplet and one parent theory, and the
whole story is that one object seen at four spacetime dimensions plus the bookkeeping that ties them
together. The parent is $10d\ \mathcal{N}=1$ super Yang--Mills, the theory in the maximal dimension in
which a super Yang--Mills theory exists at all. The section is still the lightest of the dimension
sections, but its one calculation is worked in full: we list the parent multiplet, reduce it on a
torus one circle at a time, and display every step of the $10d \to 9d \to 8d \to 7d$ descent, the
index splits, the fermion decompositions, the action-term bookkeeping, and the coupling dimensions,
at a level a strong undergraduate can follow line by line. Along the way the R-symmetry appears as
the rotation group of the reduced directions and the Coulomb branch opens up with the reduced
scalars. We close with the one structural fact that removes this whole range from the extremization
through-line: there is no interacting superconformal field theory above six dimensions. By the end
you can take a $7d$-$10d$ sixteen-supercharge SYM theory, identify its multiplet as the reduced
$10d$ vector, run the reduction yourself, read off its R-symmetry as $SO(10-d)$ with the $7d$ cover
to $SU(2)$, and count its Coulomb-branch real dimension $(10-d)\,r$, while knowing that $10d$ is the
non-renormalizable parent whose ultraviolet home is string or M-theory.

\begin{keybox}{What this section delivers}
The $10d\ \mathcal{N}=1$ super Yang--Mills parent: the vector multiplet listed field by field, the
on-shell counting chain $64 \to 32 \to 16 \to 8$ and the degree-of-freedom match $8 = 8$, the
Lagrangian, the coupling dimension $[g^2]=-6$, and the corrected reason there is no $11d$ SYM (a
Majorana $16$ against a gauge-field $9$) (\S\ref{sec:V09-tenD}); the reduction ladder $10d \to 7d$
with each step displayed in full, the index splits, the $45$-component field-strength bookkeeping,
the fermion decompositions $16 = 16 \times 1 = 8 + 8 = 8 \times 2$, the action splits, and the
itemized $9d$, $8d$, $7d$ vector multiplets (\S\ref{sec:V09-ladder}); the R-symmetry ladder
$SO(10-d)$ with the uniform spin-cover rule and the $7d$ $SO(3)\to SU(2)_R$ cover
(\S\ref{sec:V09-rsymmetry}); the Coulomb branch that opens only on reduction, its commuting-vev flat
directions, and its real dimension $(10-d)\,r$ (\S\ref{sec:V09-coulomb}); the per-dimension catalog
with the $8d$ F-theory and $7d$ M-theory engine intuitions (\S\ref{sec:V09-rows}); and the statement
that no $\ge 7d$ superconformal field theory exists at sixteen supercharges
(\S\ref{sec:V09-noscft}).
\end{keybox}

\medskip\noindent
One paragraph of recall fixes the setting; we do not re-derive it. Section~1's master table assigns
the sixteen-supercharge column its four high-dimensional rows: $10d\ \mathcal{N}=1$ SYM at the top,
then $9d$, $8d$, $7d\ \mathcal{N}=1$, all carrying $16$ real supercharges, with R-symmetries none,
none, $U(1)_R$, and $SU(2)_R$ reading down. The count $16$ is preserved at every step of the
reduction (the Section~1 ladder map), and the $32Q$ ceiling above $10d$ forces a graviton, which is
why $10d$ is the maximal SYM dimension and the top of this section. We take the supercharge
counting, the reduction map, and the R-symmetry rows from Section~1 and spend this section deriving,
at working depth, exactly the numbers Section~1 quotes.

\section{\texorpdfstring{$10d$ $\mathcal{N}=1$}{10d N=1} super Yang--Mills: the parent}
\label{sec:V09-tenD}

The maximal-dimension pure super Yang--Mills theory is $10d\ \mathcal{N}=1$ SYM. Its entire field
content is one multiplet, which we list explicitly rather than describe in passing:

\begin{itemize}
\item \emph{The $10d\ \mathcal{N}=1$ vector multiplet} (sixteen supercharges; the unique multiplet
without gravity in $10d$). Constraint: the gaugino is Majorana--Weyl with the fixed chirality
$\Gamma_{11}\lambda = +\lambda$ (the heterotic and type~I gaugino chirality of Section~1). Field
content: a gauge field $A_M$, $M = 0, 1, \dots, 9$, in the adjoint of the gauge group $G$, and the
adjoint gaugino $\lambda$; no scalars, no other fermions. Role: this one object, in $10d$ or
reduced, is every theory in this section.
\end{itemize}

\noindent
The Lagrangian is the minimal gauge-covariant action for these two fields,
\begin{equation}
\label{eq:V09-lagrangian}
 \mathcal{L}
 \;=\; -\frac{1}{4 g^2}\,\mathrm{Tr}\,F_{MN} F^{MN}
 \;+\; \frac{i}{2 g^2}\,\mathrm{Tr}\,\bar\lambda\,\Gamma^M D_M \lambda,
\end{equation}
with $F_{MN} = \partial_M A_N - \partial_N A_M - i[A_M, A_N]$ the field strength and $D_M$ the
adjoint gauge-covariant derivative, in the mostly-plus signature of Section~1. The two terms are the
unique gauge-invariant kinetic terms, and supersymmetry fixes their relative coefficient. We do not
vary the action here; Section~2 fixed the multiplet grammar. What we do check is the one algebraic
consistency condition any supersymmetric theory must satisfy: on-shell bosonic and fermionic degrees
of freedom must match.

\medskip\noindent\textbf{The on-shell degree-of-freedom match.}\enspace
A massless gauge field in $D$ spacetime dimensions has $D - 2$ physical transverse polarizations,
after gauge-fixing and the equations of motion remove the timelike and longitudinal modes. At
$D = 10$,
\begin{equation}
\label{eq:V09-gaugedof}
 \#\,A_M \big|_{\mathrm{on\text{-}shell}} \;=\; D - 2 \;=\; 8.
\end{equation}
The gaugino count is a four-step chain, and it is worth displaying every step once. A Dirac spinor
of $\mathrm{Spin}(1,9)$ has $2^{\lfloor 10/2 \rfloor} = 32$ complex components, that is $64$ real
ones. The Majorana condition is a reality condition and halves the real count. The Weyl condition
$\Gamma_{11}\lambda = +\lambda$ halves it again (the two conditions are compatible precisely in
$D = 2 \bmod 8$, so at $D = 10$ in this section's range):
\begin{equation}
\label{eq:V09-spinorchain}
 64 \ \xrightarrow{\ \text{Majorana}\ }\ 32
 \ \xrightarrow{\ \text{Weyl}\ }\ 16
 \qquad \text{real off-shell components.}
\end{equation}
On-shell, the Dirac equation halves the count once more:
\begin{equation}
\label{eq:V09-mwdof}
 \#\,\lambda \big|_{\mathrm{on\text{-}shell}} \;=\; \frac{16}{2} \;=\; 8.
\end{equation}
The bosonic and fermionic counts agree,
\begin{equation}
\label{eq:V09-dofmatch}
 8 \;=\; 8,
\end{equation}
the smallest sixteen-supercharge consistency check and the anchor for everything downstream. Neither
$8$ is put in by hand: the polarization count is $D - 2$ and the spinor count is the chain
\eqref{eq:V09-spinorchain} halved, both evaluated at $D = 10$. The same two counting functions carry
the match down the reduction ladder, where the gauge field loses polarizations and scalars appear to
replace them.

\medskip\noindent\textbf{Why there is no $11d$ super Yang--Mills.}\enspace
The match is genuinely a statement about $D$, and the neighboring dimension shows it. One must be
careful with the formulas here: $11$ is odd, so there is no Weyl condition and the Majorana--Weyl
chain \eqref{eq:V09-spinorchain} does not apply. The correct $11d$ counting starts from the same
$2^{\lfloor 11/2 \rfloor} = 32$ complex Dirac components; the minimal $11d$ spinor is Majorana, with
$32$ real components, and the Dirac equation halves it on-shell:
\begin{equation}
\label{eq:V09-11dspinor}
 64 \ \xrightarrow{\ \text{Majorana}\ }\ 32
 \ \xrightarrow{\ \text{on-shell}\ }\ 16.
\end{equation}
A putative $11d$ SYM multiplet would have to match these $16$ fermionic states against a gauge
field alone, which supplies only
\begin{equation}
\label{eq:V09-11dmismatch}
 D - 2 \;=\; 9 \;\ne\; 16,
\end{equation}
and a pure gauge multiplet has no scalars available to make up the difference. So the minimal $11d$
multiplet is not a gauge multiplet at all: it is the \emph{supergravity} multiplet, whose counting
does close,
\begin{equation}
\label{eq:V09-11dsugra}
 \underbrace{44}_{g_{MN}} \;+\; \underbrace{84}_{C_{MNP}} \;=\; 128
 \;=\; \underbrace{128}_{\psi_M},
\end{equation}
the graviton ($\tfrac{1}{2}(D-2)(D-1) - 1 = 44$), the three-form ($\binom{D-2}{3} = 84$), and the
Majorana gravitino ($(D-3) \times 16 = 128$) at $D = 11$. Eleven dimensions is the home of
supergravity, forced by the $32Q$ ceiling of Section~1, and $10d$ is the ceiling for a gauge
theory. The degree-of-freedom arithmetic and the supercharge ceiling agree: $10d\ \mathcal{N}=1$
SYM is the top.

\medskip\noindent\textbf{The coupling is irrelevant.}\enspace
The Yang--Mills coupling $g^2$ carries a mass dimension. With the gauge field at its geometric
dimension $[A] = 1$, so $[F] = 2$ and $[\mathrm{Tr}\,F^2] = 4$, the requirement that
$\int d^d x\,\mathcal{L}$ be dimensionless fixes the prefactor:
\begin{equation}
\label{eq:V09-actiondim}
 \Big[\frac{1}{g^2}\Big] + 4 \;=\; d
 \qquad\Longrightarrow\qquad
 \Big[\frac{1}{g^2}\Big] \;=\; d - 4,
\end{equation}
that is,
\begin{equation}
\label{eq:V09-couplingdim}
 [g^2] \;=\; 4 - d.
\end{equation}
At $d = 10$ this is
\begin{equation}
\label{eq:V09-tenDcoupling}
 [g^2] \;=\; 4 - 10 \;=\; -6 \;<\; 0.
\end{equation}
A negative coupling dimension means the gauge coupling is irrelevant: the theory is
non-renormalizable, a low-energy effective theory rather than an ultraviolet-complete field theory.
The coupling itself sets the cutoff scale,
\begin{equation}
\label{eq:V09-cutoff}
 \Lambda \;\sim\; (g^2)^{1/(4-d)} \;=\; (g_{10}^2)^{-1/6} \quad \text{at } d = 10,
\end{equation}
above which new degrees of freedom must enter,
and in $10d$ those degrees of freedom are string oscillator modes: $10d\ \mathcal{N}=1$ SYM is the
field-theory content of the heterotic and type~I gauge sector, and its ultraviolet completion is
string theory, not a field-theory fixed point. We state the non-renormalizability, name the
completion, and route the completion itself; the section is precisely the field-theory
interface that these string constructions read.

\medskip\noindent\textbf{No Coulomb branch, no continuous R-symmetry.}\enspace
Two absences at the parent level shape the rest of the section. First, there is no Coulomb branch. A
Coulomb branch is parametrized by the vacuum expectation values of adjoint scalars in the vector
multiplet, but the $10d$ vector multiplet has no scalar: there is nothing to give a vacuum
expectation value, so the moduli space is a point. Second, there is no continuous R-symmetry. The
R-symmetry of a reduced theory will turn out to be the rotation group of the reduced directions, but
at $d = 10$ there are no reduced directions: the transverse rotation group is $SO(10 - 10) = SO(0)$,
the trivial group. Both absences are undone by reduction, and understanding why is the content of
the next three sections. In $10d$ the theory is rigid: one multiplet, one coupling, no moduli, no
continuous R-symmetry.

\begin{keybox}{Worked instance: the $10d$ degree-of-freedom match}
The $10d\ \mathcal{N}=1$ vector multiplet carries, on-shell, a gauge field with $D - 2 = 8$
transverse polarizations and a Majorana--Weyl gaugino with $64 \to 32 \to 16 \to 8$ real components
after the Majorana, Weyl, and on-shell halvings, so bosons match fermions, $8 = 8$. The match fails
at $D = 11$: the minimal spinor there is Majorana with $32$ real components, giving $16$ on-shell
fermionic states against only $9$ gauge polarizations and no scalars, so the minimal $11d$ multiplet
is the supergravity multiplet \eqref{eq:V09-11dsugra}, not a SYM multiplet. The identity
$(d-2) + (10-d) = 8$ carries the bosonic total unchanged down the reduction ladder.
\end{keybox}

\section{The reduction ladder \texorpdfstring{$10d \to 7d$}{10d to 7d}}
\label{sec:V09-ladder}

Every sixteen-supercharge super Yang--Mills theory below $10d$ is a dimensional reduction of the
$10d$ parent. This is the central calculation of the section, and we run it slowly: first the general
mechanism, then the $10d \to 9d$ step displayed in full as the template, then the $9d \to 8d$ and
$8d \to 7d$ steps at the same level of explicitness but more compactly, each step ending with the
resulting vector multiplet listed item by item. Nothing beyond elementary index bookkeeping is
needed to follow any of it.

\medskip\noindent\textbf{The mechanism: zero modes on a torus.}\enspace
Reduction means the simplest possible operation. Place the $10d$ theory on the product spacetime
\begin{equation}
\label{eq:V09-torus}
 \mathbb{R}^{1, d-1} \times T^{10-d},
 \qquad
 x^m \;\simeq\; x^m + 2\pi R_m, \quad m = d, \dots, 9,
\end{equation}
a $d$-dimensional Minkowski factor times a torus of the remaining $10 - d$ directions. Every field
can be Fourier-expanded along each circle,
\begin{equation}
\label{eq:V09-fourier}
 \Phi(x^\mu, x^m)
 \;=\; \sum_{n \in \mathbb{Z}} \Phi^{(n)}(x^\mu)\, e^{i n x^m / R_m},
\end{equation}
and a mode with Fourier number $n$ carries a Kaluza--Klein mass set by the circle size,
\begin{equation}
\label{eq:V09-kkmass}
 m_{\mathrm{KK}} \;=\; \frac{|n|}{R_m}.
\end{equation}
Reduction keeps only the constant modes $n = 0$, the fields that do not vary along the torus, which
is the same as imposing $\partial_m = 0$ on everything. The supersymmetry parameter $\epsilon$ is
also a constant spinor, untouched by the truncation, so all $16$ real supercharges survive: the
reduced theory carries the same $16$ supercharges as the parent at every step (the Section~1
reduction map, recalled, not re-derived).

\medskip\noindent\textbf{The gauge-field split.}\enspace
The $10d$ gauge field $A_M$ has its vector index split by the reduction into a $d$-dimensional part
and a transverse part,
\begin{equation}
\label{eq:V09-split}
 A_M \;\longrightarrow\; \big(\,A_\mu,\ \phi^m\,\big),
 \qquad \mu = 0, \dots, d-1, \quad m = d, \dots, 9.
\end{equation}
The $d$ components $A_\mu$ remain a $d$-dimensional gauge field. The $10 - d$ components
$\phi^m \equiv A_m$ no longer carry a $d$-dimensional vector index; from the $d$-dimensional point
of view each is a scalar. The gauge transformations that survive the truncation are the
$x^m$-independent ones, and under them the two pieces transform differently,
\begin{equation}
\label{eq:V09-adjoint}
 A_\mu \;\to\; U A_\mu U^{-1} + i\, U \partial_\mu U^{-1},
 \qquad
 \phi^m \;\to\; U \phi^m U^{-1} :
\end{equation}
$A_\mu$ as a gauge field, each $\phi^m$ as an adjoint scalar. So reduction produces
\begin{equation}
\label{eq:V09-scalarcount}
 \#\,\text{adjoint scalars} \;=\; 10 - d.
\end{equation}
These $\phi^m$ are the seed of everything the parent lacked: the R-symmetry will rotate them, and
their vacuum expectation values will parametrize the Coulomb branch.

\medskip\noindent\textbf{The field strength and the action, in general.}\enspace
With $\partial_m = 0$, the mixed and transverse components of $F_{MN}$ collapse to covariant
derivatives and commutators of the scalars:
\begin{equation}
\label{eq:V09-Fcomponents}
 F_{\mu m} \;=\; \partial_\mu \phi^m - i[A_\mu, \phi^m] \;=\; D_\mu \phi^m,
 \qquad
 F_{mn} \;=\; -\,i\,[\phi^m, \phi^n].
\end{equation}
Substituting into the gauge kinetic term and separating the index sums
($F_{MN}F^{MN} = F_{\mu\nu}F^{\mu\nu} + 2 F_{\mu m}F^{\mu m} + F_{mn}F^{mn}$) gives the master
split, with repeated $m, n$ summed:
\begin{equation}
\label{eq:V09-mastersplit}
 -\frac{1}{4 g^2}\mathrm{Tr}\,F_{MN}F^{MN}
 =
 -\frac{1}{4 g^2}\mathrm{Tr}\,F_{\mu\nu}F^{\mu\nu}
 - \frac{1}{2 g^2}\mathrm{Tr}\,D_\mu \phi^m D^\mu \phi^m
 + \frac{1}{4 g^2}\mathrm{Tr}\,[\phi^m, \phi^n][\phi^m, \phi^n].
\end{equation}
Each term has a clear origin: the first is the $d$-dimensional gauge kinetic term; the second is the
scalar kinetic term, born from the mixed components $F_{\mu m} = D_\mu\phi^m$ (the factor $2$ from
the two index orders becomes the standard $\tfrac12$); the third is born from the purely transverse
components $F_{mn} = -i[\phi^m,\phi^n]$ and is the scalar potential. For Hermitian scalars the
commutator is anti-Hermitian, so $\mathrm{Tr}\,[\phi^m,\phi^n]^2 \le 0$ and the last term is minus a
non-negative potential; \S\ref{sec:V09-coulomb} takes it up. The reduction also preserves the
on-shell degree-of-freedom count at every step: the gauge field contributes $d - 2$ transverse
polarizations and the scalars contribute $10 - d$, so the bosonic total is
\begin{equation}
\label{eq:V09-dofpreserved}
 (d - 2) \;+\; (10 - d) \;=\; 8,
\end{equation}
independent of $d$, matched at every row by the $8$ on-shell components of the reduced gaugino. This
is the sense in which the whole ladder is one multiplet: the same $8 + 8$ on-shell degrees of
freedom, packaged differently at each dimension. We now watch the packaging change, one circle at a
time.

\subsection*{Worked instance: the \texorpdfstring{$10d \to 9d$}{10d to 9d} step, in full}

Compactify the single direction $x^9$ on a circle of radius $R_9$ and keep the zero modes,
$\partial_9 = 0$.

\medskip\noindent\emph{(a) The index split.}\enspace The ten components of $A_M$ split as
\begin{equation}
\label{eq:V09-9dsplit}
 A_M \;\longrightarrow\; \big(A_\mu,\ \phi^9\big),
 \qquad \mu = 0, \dots, 8, \quad \phi^9 \equiv A_9,
\end{equation}
a $9d$ gauge field plus one adjoint scalar: the scalar count is $10 - 9 = 1$. It is instructive to
watch all $\binom{10}{2} = 45$ independent components of the antisymmetric $F_{MN}$ land somewhere:
\begin{equation}
\label{eq:V09-9dcount}
 45 \;=\; \underbrace{36}_{F_{\mu\nu}} \;+\; \underbrace{9}_{F_{\mu 9}}
 \;+\; \underbrace{0}_{F_{mn}},
\end{equation}
$\binom{9}{2} = 36$ gauge components, $9 \times 1 = 9$ mixed components, and no purely transverse
pair, because a single reduced direction admits no antisymmetric pair. Nothing is lost and nothing
is added; the components are relabeled.

\medskip\noindent\emph{(b) The mixed field strength is the scalar's covariant derivative.}\enspace
With $\partial_9 = 0$,
\begin{equation}
\label{eq:V09-9dF}
 F_{\mu 9} \;=\; \partial_\mu A_9 - \partial_9 A_\mu - i[A_\mu, A_9]
 \;=\; \partial_\mu \phi^9 - i[A_\mu, \phi^9] \;=\; D_\mu \phi^9,
\end{equation}
the middle term dying by the zero-mode condition. This is the general statement
\eqref{eq:V09-Fcomponents} caught in the act.

\medskip\noindent\emph{(c) The action splits.}\enspace Substituting into the kinetic term,
\begin{equation}
\label{eq:V09-9dFsq}
 \mathrm{Tr}\,F_{MN}F^{MN}
 \;=\; \mathrm{Tr}\,F_{\mu\nu}F^{\mu\nu} \;+\; 2\,\mathrm{Tr}\,D_\mu\phi^9 D^\mu\phi^9,
\end{equation}
with no potential term: one scalar has no partner to fail to commute with ($\binom{1}{2} = 0$
commutator pairs, the $0$ of \eqref{eq:V09-9dcount}). The $x^9$ integral in the action contributes
the circle volume, $\int d^{10}x = 2\pi R_9 \int d^9 x$, so
\begin{equation}
\label{eq:V09-9daction}
 S \;=\; -\frac{2\pi R_9}{4 g_{10}^2} \int d^9 x\;
 \mathrm{Tr}\Big( F_{\mu\nu}F^{\mu\nu} + 2\, D_\mu\phi^9 D^\mu\phi^9 \Big) \;+\; \dots
\end{equation}

\medskip\noindent\emph{(d) The coupling and its dimension.}\enspace Reading off the prefactor of the
$9d$ gauge kinetic term defines the $9d$ coupling, and the circle length shifts the dimension by
exactly one:
\begin{equation}
\label{eq:V09-9dcoupling}
 \frac{1}{g_9^2} \;=\; \frac{2\pi R_9}{g_{10}^2}
 \qquad\Longrightarrow\qquad
 [g_9^2] \;=\; [g_{10}^2] + 1 \;=\; -5 \;=\; 4 - 9,
\end{equation}
consistent with the general dimension count \eqref{eq:V09-couplingdim} run directly at $d = 9$. The
coupling is still irrelevant; descending has made it less negative by one unit, not positive.

\medskip\noindent\emph{(e) The gaugino reduces and a Yukawa coupling appears.}\enspace The same
zero-mode condition collapses the gaugino's kinetic term. Since
$D_9 \lambda = -i[\phi^9, \lambda]$,
\begin{equation}
\label{eq:V09-9dyukawa}
 \frac{i}{2 g_{10}^2}\,\mathrm{Tr}\,\bar\lambda\,\Gamma^M D_M \lambda
 \;\longrightarrow\;
 \frac{i}{2 g_9^2}\,\mathrm{Tr}\,\bar\lambda\,\Gamma^\mu D_\mu \lambda
 \;+\; \frac{1}{2 g_9^2}\,\mathrm{Tr}\,\bar\lambda\,\Gamma^9 [\phi^9, \lambda],
\end{equation}
a $9d$ kinetic term plus the scalar--gaugino Yukawa coupling that supersymmetry demands. As a
representation of the $9d$ Lorentz group, the $16$ real components of $\lambda$ simply restrict:
the transverse group $\mathrm{Spin}(1)$ is trivial, so
\begin{equation}
\label{eq:V09-9dspinor}
 \mathbf{16} \;\longrightarrow\; \mathbf{16} \otimes \mathbf{1},
 \qquad 16 \;=\; 16 \times 1,
\end{equation}
one Majorana spinor of $\mathrm{Spin}(1,8)$, which is exactly the minimal $9d$ spinor: $16$ real
components off-shell, $8$ on-shell.

\medskip\noindent\emph{(f) The tally.}\enspace The on-shell bookkeeping closes:
\begin{equation}
\label{eq:V09-9dtally}
 \underbrace{(9 - 2)}_{A_\mu} + \underbrace{1}_{\phi^9} \;=\; 7 + 1 \;=\; 8
 \;=\; \underbrace{16/2}_{\lambda},
\end{equation}
and the Coulomb branch, once we open it in \S\ref{sec:V09-coulomb}, will have real dimension
$(10-9)\,r = r$. The result of the step, listed:

\begin{itemize}
\item \emph{The $9d\ \mathcal{N}=1$ vector multiplet} (sixteen supercharges; the $10d$ vector on one
circle). Field content: gauge field $A_\mu$, $\mu = 0, \dots, 8$ ($7$ on-shell polarizations); $1$
real adjoint scalar $\phi^9$; one Majorana gaugino of $\mathrm{Spin}(1,8)$ with $16$ real components
off-shell, $8$ on-shell. R-symmetry: none (the transverse $SO(1)$ is trivial; the scalar is an
$SO(1)$ singlet). Role: the one-scalar bookkeeping row between the parent and the first R-symmetric
rows.
\end{itemize}

\subsection*{Worked instance: the \texorpdfstring{$9d \to 8d$}{9d to 8d} step}

Compactify $x^8$ on a circle of radius $R_8$; equivalently, run the parent on $T^2$. The same six
displays repeat with one new phenomenon: the first commutator.

\medskip\noindent\emph{(a, b) Split and components.}\enspace The $9d$ gauge field splits again,
\begin{equation}
\label{eq:V09-8dsplit}
 A_{\hat\mu} \;\longrightarrow\; \big(A_\mu,\ \phi^8\big),
 \qquad \hat\mu = 0, \dots, 8, \quad \mu = 0, \dots, 7,
\end{equation}
so the accumulated scalars are $(\phi^8, \phi^9)$, count $10 - 8 = 2$, an $SO(2)$ vector.
Relative to the $10d$ parent the $45$ components of $F_{MN}$ now sit as
\begin{equation}
\label{eq:V09-8dcount}
 45 \;=\; \underbrace{28}_{F_{\mu\nu}} \;+\; \underbrace{16}_{F_{\mu m}}
 \;+\; \underbrace{1}_{F_{89}},
\end{equation}
$\binom{8}{2} = 28$ gauge components, $8 \times 2 = 16$ covariant derivatives, and, for the first
time, one purely transverse component,
\begin{equation}
\label{eq:V09-8dcomm}
 F_{89} \;=\; -\,i\,[\phi^8, \phi^9].
\end{equation}

\medskip\noindent\emph{(c) The first potential.}\enspace That single component feeds the master
split \eqref{eq:V09-mastersplit} its first commutator-squared term, minus the potential
\begin{equation}
\label{eq:V09-8dpot}
 V \;=\; -\frac{1}{2 g_8^2}\,\mathrm{Tr}\,[\phi^8, \phi^9]^2
 \;=\; \frac{1}{2 g_8^2}\,\mathrm{Tr}\Big( [\phi^8, \phi^9]^\dagger [\phi^8, \phi^9] \Big)
 \;\ge\; 0,
\end{equation}
non-negative because the commutator of Hermitian matrices is anti-Hermitian. At $9d$ there was no
potential at all; from $8d$ down, supersymmetric vacua must make the commutators vanish, the seed of
the Coulomb-branch analysis of \S\ref{sec:V09-coulomb}.

\medskip\noindent\emph{(d) The complex scalar.}\enspace The transverse rotation $SO(2)$ rotates
$(\phi^8, \phi^9)$ as a doublet, which packages naturally as one complex adjoint scalar of definite
charge,
\begin{equation}
\label{eq:V09-8dcomplex}
 \phi \;\equiv\; \phi^8 + i\,\phi^9,
 \qquad \phi \;\to\; e^{i\alpha}\,\phi \quad \text{under } SO(2) = U(1)_R,
\end{equation}
the charge-$1$ complex scalar of the $8d$ row (its conjugate carries charge $-1$).

\medskip\noindent\emph{(e) The fermion split.}\enspace Under
$\mathrm{Spin}(1,7) \times \mathrm{Spin}(2)$ the $\mathbf{16}$ decomposes into two Weyl blocks,
\begin{equation}
\label{eq:V09-8dspinor}
 \mathbf{16} \;\longrightarrow\; \mathbf{8}_s^{+1/2} \,\oplus\, \mathbf{8}_c^{-1/2},
 \qquad 16 \;=\; 8 + 8,
\end{equation}
the $10d$ Weyl condition locking the $8d$ chirality to the $\mathrm{Spin}(2) = U(1)_R$ charge
$\pm\tfrac12$. The two blocks are exchanged by complex conjugation, so together they form one
Majorana gaugino of $\mathrm{Spin}(1,7)$, the minimal $8d$ spinor: $16$ real components off-shell,
$8$ on-shell.

\medskip\noindent\emph{(f) Coupling and tally.}\enspace The second circle shifts the coupling
dimension by one more unit,
\begin{equation}
\label{eq:V09-8dcoupling}
 \frac{1}{g_8^2} \;=\; \frac{2\pi R_8}{g_9^2},
 \qquad
 [g_8^2] \;=\; -5 + 1 \cdot 1 \;=\; -4 \;=\; 4 - 8,
\end{equation}
and the on-shell tally closes again,
\begin{equation}
\label{eq:V09-8dtally}
 \underbrace{(8-2)}_{A_\mu} + \underbrace{2}_{\phi^8,\,\phi^9} \;=\; 6 + 2 \;=\; 8.
\end{equation}

\begin{itemize}
\item \emph{The $8d\ \mathcal{N}=1$ vector multiplet} (sixteen supercharges; the $10d$ vector on
$T^2$). Field content: gauge field $A_\mu$, $\mu = 0, \dots, 7$ ($6$ on-shell polarizations); $2$
real adjoint scalars $(\phi^8, \phi^9)$, the vector of $SO(2)$, equivalently one complex scalar
$\phi = \phi^8 + i\phi^9$ of $U(1)_R$ charge $1$; one Majorana gaugino of $\mathrm{Spin}(1,7)$,
decomposing as the Weyl pair $\mathbf{8}_s^{+1/2} \oplus \mathbf{8}_c^{-1/2}$ under $U(1)_R$, $16$
real off-shell, $8$ on-shell. R-symmetry: $U(1)_R = SO(2)$. Role: the first R-symmetric row; the
field-theory home of the F-theory 7-brane gauge theory (\S\ref{sec:V09-rows}).
\end{itemize}

\subsection*{Worked instance: the \texorpdfstring{$8d \to 7d$}{8d to 7d} step}

Compactify $x^7$ on a circle of radius $R_7$; cumulatively, the parent on $T^3$.

\medskip\noindent\emph{(a, b) Split and components.}\enspace
\begin{equation}
\label{eq:V09-7dsplit}
 A_{\hat\mu} \;\longrightarrow\; \big(A_\mu,\ \phi^7\big),
 \qquad \hat\mu = 0, \dots, 7, \quad \mu = 0, \dots, 6,
\end{equation}
with accumulated scalars $(\phi^7, \phi^8, \phi^9)$, count $10 - 7 = 3$, a vector of $SO(3)$. The
$45$ components of the parent field strength now split as
\begin{equation}
\label{eq:V09-7dcount}
 45 \;=\; \underbrace{21}_{F_{\mu\nu}} \;+\; \underbrace{21}_{F_{\mu m}}
 \;+\; \underbrace{3}_{F_{mn}},
\end{equation}
$\binom{7}{2} = 21$, $7 \times 3 = 21$, and $\binom{3}{2} = 3$ transverse pairs, which feed the
potential all three of its commutators,
\begin{equation}
\label{eq:V09-7dpairs}
 V \;=\; \frac{1}{2 g_7^2} \sum_{m < n}
 \mathrm{Tr}\Big( [\phi^m, \phi^n]^\dagger [\phi^m, \phi^n] \Big),
 \qquad (m,n) \in \{ (7,8),\ (8,9),\ (7,9) \}.
\end{equation}

\medskip\noindent\emph{(c) The fermion split.}\enspace Under
$\mathrm{Spin}(1,6) \times \mathrm{Spin}(3)$ the $\mathbf{16}$ factorizes,
\begin{equation}
\label{eq:V09-7dspinor}
 \mathbf{16} \;\longrightarrow\; (\mathbf{8},\, \mathbf{2}),
 \qquad 16 \;=\; 8 \times 2,
\end{equation}
the Dirac spinor $\mathbf{8}$ of $\mathrm{Spin}(1,6)$ times the doublet $\mathbf{2}$ of
$\mathrm{Spin}(3) = SU(2)$. Counted naively this is $8 \times 2 = 16$ \emph{complex} components; a
symplectic-Majorana condition, a reality condition that uses the doublet index, halves it back to
$16$ real components. The $7d$ gaugino is therefore an $SU(2)_R$ doublet $\lambda^i$, $i = 1, 2$,
with $16$ real components off-shell and $8$ on-shell, and the doublet index is the first sighting of
the $7d$ R-symmetry acting on fermions through the \emph{cover} of $SO(3)$
(\S\ref{sec:V09-rsymmetry}).

\medskip\noindent\emph{(d) Coupling and tally.}\enspace
\begin{equation}
\label{eq:V09-7dcoupling}
 \frac{1}{g_7^2} \;=\; \frac{2\pi R_7}{g_8^2},
 \qquad
 [g_7^2] \;=\; -4 + 1 \;=\; -3 \;=\; 4 - 7,
\end{equation}
\begin{equation}
\label{eq:V09-7dtally}
 \underbrace{(7-2)}_{A_\mu} + \underbrace{3}_{\phi^m} \;=\; 5 + 3 \;=\; 8.
\end{equation}

\medskip\noindent\emph{(e) The terminal Lagrangian, assembled.}\enspace Collecting every piece the
three steps produced gives the full $7d\ \mathcal{N}=1$ SYM Lagrangian, the worldvolume theory the
$7d$ engine row reads:
\begin{equation}
\label{eq:V09-7dlagrangian}
\begin{aligned}
 \mathcal{L}_7 \;=\;
 &-\frac{1}{4 g_7^2}\mathrm{Tr}\,F_{\mu\nu}F^{\mu\nu}
 - \frac{1}{2 g_7^2}\mathrm{Tr}\,D_\mu\phi^m D^\mu\phi^m
 + \frac{1}{4 g_7^2}\mathrm{Tr}\,[\phi^m, \phi^n][\phi^m, \phi^n] \\
 &+ \frac{i}{2 g_7^2}\mathrm{Tr}\,\bar\lambda\,\Gamma^\mu D_\mu \lambda
 + \frac{1}{2 g_7^2}\mathrm{Tr}\,\bar\lambda\,\Gamma^m [\phi^m, \lambda],
\end{aligned}
\end{equation}
written in $10d$ spinor notation ($\lambda$ the reduced gaugino components, $\Gamma^m$ the
surviving internal gamma matrices); regrouping the fermions into the $SU(2)_R$ doublet is the split
\eqref{eq:V09-7dspinor}. Every coefficient is the master split \eqref{eq:V09-mastersplit} plus the
Yukawa pattern of \eqref{eq:V09-9dyukawa}, now with all three scalars.

\begin{itemize}
\item \emph{The $7d\ \mathcal{N}=1$ vector multiplet} (sixteen supercharges; the $10d$ vector on
$T^3$). Field content: gauge field $A_\mu$, $\mu = 0, \dots, 6$ ($5$ on-shell polarizations); $3$
real adjoint scalars $(\phi^7, \phi^8, \phi^9)$, the vector of $SO(3)$, equivalently a triplet of
$SU(2)_R$; one symplectic-Majorana gaugino doublet $\lambda^i$ of $\mathrm{Spin}(1,6)$, the
$(\mathbf{8}, \mathbf{2})$ of $\mathrm{Spin}(1,6) \times SU(2)_R$, $16$ real off-shell, $8$
on-shell. R-symmetry: $SU(2)_R$, the spinor cover of the transverse $SO(3)$. Role: the terminal row
of this section; the field-theory home of M-theory on ADE singularities (\S\ref{sec:V09-rows}).
\end{itemize}

\medskip\noindent\textbf{Composing the steps.}\enspace
The three circles compose into one statement: reducing directly on the torus $T^{10-d}$ divides
the parent coupling by the torus volume,
\begin{equation}
\label{eq:V09-composedcoupling}
 \frac{1}{g_d^2} \;=\; \frac{\mathrm{Vol}\,(T^{10-d})}{g_{10}^2},
 \qquad
 [g_d^2] \;=\; -6 + (10 - d) \;=\; 4 - d,
\end{equation}
with $\mathrm{Vol}\,(T^{10-d}) = \prod_m 2\pi R_m$ carrying mass dimension $-(10-d)$. The one-step
relations \eqref{eq:V09-9dcoupling}, \eqref{eq:V09-8dcoupling}, \eqref{eq:V09-7dcoupling} are this
formula peeled one circle at a time.

\begin{keybox}{The reduction dictionary at a glance}
Per step $d+1 \to d$: split the gauge field \eqref{eq:V09-split}; the new scalar's kinetic term is
the mixed field strength \eqref{eq:V09-Fcomponents}; commutators of accumulated scalars feed the
potential \eqref{eq:V09-mastersplit}; the circle volume shifts $[g^2]$ up by one
\eqref{eq:V09-composedcoupling}; the $16$ gaugino components repackage under
$\mathrm{Spin}(1,d-1) \times \mathrm{Spin}(10-d)$; and the bosonic tally $(d-2) + (10-d) = 8$
closes. Everything else in the section reads off this dictionary.
\end{keybox}

\medskip\noindent\textbf{The four rows, collected.}\enspace
Reading the three steps top-down gives the four rows this section owns, collected in
Table~\ref{tab:V09-content}: $10d\ \mathcal{N}=1 \to 9d\ \mathcal{N}=1 \to 8d\ \mathcal{N}=1 \to
7d\ \mathcal{N}=1$, with $10 - d = 0, 1, 2, 3$ adjoint scalars. The descent continues below $7d$ to
$6d\ (1,1)$, $5d\ \mathcal{N}=2$, and $4d\ \mathcal{N}=4$, all still sixteen-supercharge reductions
of the same parent, but those rows belong to their own flavor-family sections (the $6d$ neighbor of
Section~8, the $5d$ neighbor of Section~7, the $4d$ neighbor of Section~5, where $4d\ \mathcal{N}=4$
is the maximal $4d$ superconformal theory, an interacting fixed point that does exist below seven
dimensions). This section stops the descent at $7d$.

\begin{table}[ht]
\centering
\small
\setlength{\tabcolsep}{5pt}
\renewcommand{\arraystretch}{1.3}
\begin{tabular}{@{}cccp{44mm}c@{}}
\toprule
$d$ & gauge field & scalars $10{-}d$ & gaugino ($16$ real, decomposed) & bosonic dof \\
\midrule
$10$ & $A_\mu$ ($8$ polns) & $0$ & Majorana--Weyl, $\Gamma_{11}\lambda = +\lambda$ & $8 + 0 = 8$ \\
\midrule
$9$ & $A_\mu$ ($7$ polns) & $1$ & Majorana of $\mathrm{Spin}(1,8)$:\newline $16 = 16 \times 1$ & $7 + 1 = 8$ \\
\midrule
$8$ & $A_\mu$ ($6$ polns) & $2$ & Weyl pair $\mathbf{8}_s^{+1/2} \oplus \mathbf{8}_c^{-1/2}$:\newline $16 = 8 + 8$ & $6 + 2 = 8$ \\
\midrule
$7$ & $A_\mu$ ($5$ polns) & $3$ & sympl.\ Majorana $(\mathbf{8}, \mathbf{2})$:\newline $16 = 8 \times 2$ & $5 + 3 = 8$ \\
\bottomrule
\end{tabular}
\caption{The $10d \to 7d$ reduction ladder. Each row is the same sixteen-supercharge vector
multiplet: the gauge field keeps $d-2$ on-shell polarizations, the $10-d$ lost components become
adjoint scalars, the $16$ gaugino components repackage under
$\mathrm{Spin}(1,d-1) \times \mathrm{Spin}(10-d)$, and the bosonic total $(d-2)+(10-d)=8$ is fixed,
matched by $8$ on-shell gaugino components at every row. Every entry is a function of $d$, not a
per-row input.}
\label{tab:V09-content}
\end{table}

\medskip\noindent\textbf{The coupling-dimension summary.}\enspace
One further table (Table~\ref{tab:V09-coupling}) collects the numbers that distinguish the rows. Its
central point is that the non-renormalizability is uniform down the whole ladder, not a $10d$-only
affliction: each circle raises $[g^2]$ by exactly one unit (the circle volume in
\eqref{eq:V09-9dcoupling}, \eqref{eq:V09-8dcoupling}, \eqref{eq:V09-7dcoupling}), and starting from
$-6$ three steps cannot reach zero,
\begin{equation}
\label{eq:V09-couplingladder}
 [g^2] \;=\; 4 - d \;=\; -6,\ -5,\ -4,\ -3
 \qquad \text{for } d = 10,\ 9,\ 8,\ 7,
\end{equation}
strictly negative for all four rows. Every $\ge 7d$ sixteen-supercharge SYM theory is
non-renormalizable, an effective theory whose ultraviolet completion is a string or M-theory
construction, not a field-theory fixed point. What grows as one descends is not the coupling
behavior but the matter and moduli data: the scalar count $10 - d$, the R-symmetry $SO(10-d)$, and
the Coulomb-branch real dimension $(10-d)\,r$ all increase as the number of reduced directions
grows. The next two sections derive the R-symmetry column and the Coulomb-branch column.

\begin{table}[ht]
\centering
\small
\setlength{\tabcolsep}{8pt}
\renewcommand{\arraystretch}{1.3}
\begin{tabular}{@{}ccccc@{}}
\toprule
$d$ & scalars $10-d$ & $[g^2]=4-d$ & R-symmetry $SO(10-d)$ & Coulomb dim $(10-d)\,r$ \\
\midrule
$10$ & $0$ & $-6$ & $SO(0)$ (none) & $0$ \\
\midrule
$9$ & $1$ & $-5$ & $SO(1)$ (none) & $r$ \\
\midrule
$8$ & $2$ & $-4$ & $SO(2)=U(1)_R$ & $2r$ \\
\midrule
$7$ & $3$ & $-3$ & $SO(3)$, cover $SU(2)_R$ & $3r$ \\
\bottomrule
\end{tabular}
\caption{The per-row summary of the sixteen-supercharge SYM ladder, $r = \mathrm{rank}\,G$. The
coupling dimension $[g^2]=4-d$ is strictly negative for every row, so the non-renormalizability is
uniform down the whole $\ge 7d$ ladder, not $10d$-only. The scalar count, R-symmetry, and
Coulomb-branch dimension all grow as $10-d$ increases. Each entry is computed from $d$ (and, for the
last column, the rank).}
\label{tab:V09-coupling}
\end{table}

\medskip\noindent\textbf{The reduction is what the ladder reads from.}\enspace
The value of this bookkeeping is that it \emph{is} the sixteen-supercharge ladder that Section~1's
master table quotes. Section~1 tabulates the R-symmetries and
the supercharge counts; this section shows they are consequences of one operation applied to one
parent. A reader handed a $9d$, $8d$, or $7d\ \mathcal{N}=1$ SYM theory can now reconstruct its
multiplet content from the row alone, and, more to the point, can re-derive it: every display in the
three worked steps used nothing but the zero-mode rule $\partial_m = 0$ and index bookkeeping.

\section{The R-symmetry ladder \texorpdfstring{$SO(10-d)$}{SO(10-d)}}
\label{sec:V09-rsymmetry}

The R-symmetry of the $d$-dimensional theory is not an extra input. It is the rotation group of the
$10 - d$ reduced directions. The reduction singled out a torus $T^{10-d}$; rotating those transverse
directions into each other is a symmetry of the $d$-dimensional theory (nothing in
$\mathbb{R}^{1,d-1}$ sees the rotation), and it acts on the $10 - d$ adjoint scalars $\phi^m$
exactly as a vector of that rotation group, because the $\phi^m$ inherited their index $m$ from the
reduced directions. The supercharges, which are spinors of the $10d$ Lorentz group, transform under
the reduced rotations as spinors, so this rotation group is the R-symmetry: it acts on the
supercharges. The R-symmetry is therefore
\begin{equation}
\label{eq:V09-rsymmetry}
 \mathcal{R}(d) \;=\; SO(10 - d)
\end{equation}
at the algebra level. The representation assignments are uniform down the whole ladder, and it is
worth displaying the rule once, in one line:
\begin{equation}
\label{eq:V09-covernote}
 \phi^m \in \text{vector of } SO(10-d),
 \qquad
 \lambda,\ Q \in \text{spinor of } \mathrm{Spin}(10-d),
\end{equation}
scalars see the rotation group, fermions see its spin cover, with the low-rank covers
$\mathrm{Spin}(2) = U(1)$ and $\mathrm{Spin}(3) = SU(2)$. Reading the algebra data down the ladder
makes the pattern concrete. The dimension and rank of $so(n)$ are
\begin{equation}
\label{eq:V09-sorankdim}
 \dim\,so(n) \;=\; \frac{n(n-1)}{2},
 \qquad
 \mathrm{rank}\,so(n) \;=\; \Big\lfloor \frac{n}{2} \Big\rfloor,
\end{equation}
so with $n = 10 - d$ we get, for $d = 10, 9, 8, 7$,
\begin{equation}
\label{eq:V09-rladder}
 so(0),\quad so(1),\quad so(2),\quad so(3),
 \qquad
 \dim = 0,\ 0,\ 1,\ 3.
\end{equation}
The two top rows are trivial. At $d = 10$ the transverse rotation group is $SO(0)$, the trivial
group, which is why the parent has no continuous R-symmetry. At $d = 9$ the transverse group is
$SO(1)$, again trivial (a single direction has no rotations), so the $9d$ row also has no continuous
R-symmetry; the one scalar of the $9d$ row is just bookkeeping, carried but not rotated. The
R-symmetry turns on at $d = 8$.

\medskip\noindent\textbf{The $8d$ row: $U(1)_R$.}\enspace
At $d = 8$ there are two reduced directions, so the transverse rotation group is $SO(2)$, the
rotations of a plane. This is an abelian group, isomorphic to $U(1)$, with rank $1$ and dimension
$1$:
\begin{equation}
\label{eq:V09-8dR}
 \mathcal{R}(8) \;=\; SO(2) \;\cong\; U(1)_R,
 \qquad (\mathrm{rank}, \dim) = (1, 1).
\end{equation}
The single generator rotates the two scalars $\phi^8, \phi^9$ into each other, equivalently rotates
the complex combination $\phi = \phi^8 + i\phi^9$ of \eqref{eq:V09-8dcomplex} by a phase, and acts
on the gaugino's two Weyl blocks with the charges $\pm\tfrac12$ of \eqref{eq:V09-8dspinor}: the
half-integer fermion charges against the integer scalar charge are precisely the spin-cover rule
\eqref{eq:V09-covernote} at work. The $8d\ \mathcal{N}=1$ SYM theory carries $U(1)_R$, and, as the
catalog of \S\ref{sec:V09-rows} records, this is the field-theory home of the F-theory 7-brane gauge
theory (whose worldvolume is $7 + 1 = 8$-dimensional).

\medskip\noindent\textbf{The $7d$ row: $SU(2)_R$, the marquee cover.}\enspace
At $d = 7$ there are three reduced directions, so the transverse rotation group is $SO(3)$, with
rank $1$ and dimension $3$. But the R-symmetry group is not literally $SO(3)$. The R-symmetry is the
automorphism group acting on the supercharges, and the supercharges are spinors; spinors transform
under the double cover of the rotation group, which for $SO(3)$ is
\begin{equation}
\label{eq:V09-7dcover}
 SU(2) \;=\; \mathrm{Spin}(3) \;\longrightarrow\; SO(3),
 \qquad
 so(3) \;\cong\; su(2) \ \text{as Lie algebras},
\end{equation}
the isomorphism holding at the level of the Lie algebra (both have rank $1$ and dimension $3$) while
the groups differ by the $\mathbb{Z}_2$ center. The genuine R-symmetry of $7d\ \mathcal{N}=1$ SYM is
the spinor cover,
\begin{equation}
\label{eq:V09-7dR}
 \mathcal{R}(7) \;=\; SU(2)_R,
\end{equation}
the master-table row ``$SO(3)$ transverse, double-covered by $SU(2)$.'' This is the section's
marquee structural statement: the transverse rotation group of the reduced dimensions becomes the
R-symmetry, and the cover to $SU(2)$ passes exactly at $7d$, where $SO(3)$ first appears. The three
adjoint scalars $\phi^7, \phi^8, \phi^9$ form a vector of $SO(3)$, equivalently a triplet of
$SU(2)_R$; the gaugino forms the doublet $\lambda^i$ that the $8d \to 7d$ step
\eqref{eq:V09-7dspinor} already displayed. The same rule keeps working below this section's range,
which is a useful cross-check against a row the reader already owns: continuing the reduction to
$d = 4$ gives six reduced directions and
\begin{equation}
\label{eq:V09-spin6}
 \mathrm{Spin}(6) \;=\; SU(4),
 \qquad
 so(6) \;\cong\; su(4), \quad (\mathrm{rank}, \dim) = (3, 15),
\end{equation}
the $SU(4)_R$ of $4d\ \mathcal{N}=4$ SYM quoted in Sections~1 and~5, produced by the identical
transverse-rotation logic.

\begin{keybox}{Common misconception: the $7d\ \mathcal{N}=1$ R-symmetry is $SO(3)$}
The transverse rotation group of the three reduced directions is $SO(3)$, and it rotates the three
adjoint scalars $\phi^m$ as a vector. But the R-symmetry is the automorphism of the
\emph{supercharges}, and the supercharges are spinors: they transform under the double cover
$SU(2)=\mathrm{Spin}(3)$, not $SO(3)$. So the $7d$ R-symmetry is $SU(2)_R$. The cover is physical,
not a labeling choice, because the gaugino sits in the spinor (doublet) representation, so the
genuine R-symmetry group is $SU(2)$. The same pattern is why $4d\ \mathcal{N}=4$ has
$SU(4)_R=\mathrm{Spin}(6)$ and not $SO(6)$: whenever the reduced rotation group acts on spinors, the
R-symmetry is its spinor cover.
\end{keybox}

\section{The Coulomb branch on reduction}
\label{sec:V09-coulomb}

The parent has no Coulomb branch because it has no adjoint scalar. Reduction produces $10 - d$
adjoint scalars $\phi^m$, and their vacuum expectation values open a Coulomb branch. We work out
which configurations are supersymmetric vacua and count the resulting moduli.

\medskip\noindent\textbf{The flat-direction condition.}\enspace
The scalars $\phi^m$ inherit their interactions from the $10d$ gauge kinetic term: the last term of
the master split \eqref{eq:V09-mastersplit} is minus the scalar potential, which the $9d \to 8d$
step derived with its coefficient \eqref{eq:V09-8dpot}. In general,
\begin{equation}
\label{eq:V09-potential}
 V(\phi) \;=\; -\frac{1}{2 g_d^2} \sum_{m < n} \mathrm{Tr}\,[\phi^m, \phi^n]^2
 \;=\; \frac{1}{2 g_d^2} \sum_{m < n}
 \mathrm{Tr}\Big( [\phi^m, \phi^n]^\dagger [\phi^m, \phi^n] \Big)
 \;\ge\; 0,
\end{equation}
a sum of squared commutator norms, non-negative because each commutator of Hermitian matrices is
anti-Hermitian. A supersymmetric vacuum is a zero-energy configuration, $V = 0$, which because each
term is a norm forces every commutator to vanish:
\begin{equation}
\label{eq:V09-flatdir}
 [\phi^m, \phi^n] \;=\; 0 \qquad \text{for all } m, n.
\end{equation}
The flat directions are the \emph{commuting} configurations of the $10 - d$ adjoint scalars: sets of
matrices that can be simultaneously diagonalized. Simultaneously diagonalizable adjoint matrices
lie, up to a gauge rotation, in a common Cartan subalgebra. So the vacuum moduli are the $10 - d$
scalars each taking a value in the Cartan of $G$, modulo the residual gauge identifications, which
for the Cartan is the Weyl group $W_G$. The Coulomb branch is the real space
\begin{equation}
\label{eq:V09-coulombspace}
 \mathcal{M}_{\mathrm{Coulomb}} \;=\; \mathbb{R}^{(10-d)\,r} \big/ W_G,
 \qquad r = \mathrm{rank}\,G,
\end{equation}
each of the $10 - d$ scalars contributing its $r$ Cartan directions.

\medskip\noindent\textbf{The dimension, and the generic breaking.}\enspace
The real dimension of the Coulomb branch is therefore
\begin{equation}
\label{eq:V09-coulombdim}
 \dim_{\mathbb{R}}\,\mathcal{M}_{\mathrm{Coulomb}} \;=\; (10 - d)\,r,
\end{equation}
the number of reduced directions times the rank. At a generic point the adjoint scalars have
distinct Cartan eigenvalues and Higgs the gauge group to its maximal torus,
\begin{equation}
\label{eq:V09-higgsing}
 G \;\longrightarrow\; U(1)^r,
 \qquad
 \#\,\text{massive gauge bosons} \;=\; \dim G - r
 \;\;\big(= N^2 - N \ \text{for } SU(N)\big).
\end{equation}
The mechanism is visible in the commutators. A Cartan-valued vev acts diagonally on the root
generators $E_\alpha$,
\begin{equation}
\label{eq:V09-rootaction}
 [\phi^m, E_\alpha] \;=\; \alpha(a_m)\, E_\alpha
 \qquad \text{for } \phi^m \text{ in the Cartan with eigenvalue data } a_m,
\end{equation}
so the off-Cartan modes are charged under the surviving $U(1)^r$ and pick up masses proportional to
the vevs through the commutator terms of \eqref{eq:V09-mastersplit}, while the $r$ Cartan modes
commute with everything and stay massless: exactly the $\dim G - r$ count of
\eqref{eq:V09-higgsing}. At the parent $d = 10$ the
formula \eqref{eq:V09-coulombdim} gives $(10 - 10)\,r = 0$: no Coulomb branch, consistent with the
absence of any adjoint scalar. The dimension grows as one descends, since $10 - d$ grows: $r$ at
$9d$, $2r$ at $8d$, $3r$ at $7d$. This is the last column of Table~\ref{tab:V09-coupling}.

\medskip\noindent\textbf{The branch is real.}\enspace
The Coulomb branch here is a \emph{real} moduli space, a space of commuting Hermitian matrices
modulo Weyl. This is the high-dimensional analog of the $5d\ \mathcal{N}=1$ real Coulomb branch of
Section~7, where the single real vector-multiplet scalar likewise takes real Cartan values. It is
\emph{not} the complex special-K\"ahler Coulomb branch of $4d\ \mathcal{N}=2$ (Section~5), whose
scalar is a complex combination protected into a rigid special geometry, and it is \emph{not} a
hyperk\"ahler Higgs branch: a Higgs branch needs hypermultiplets, and pure super Yang--Mills has
none. The reduced scalars are real adjoint fields, and their moduli space is flat and real, the
simplest kind of Coulomb branch.

\subsection*{Worked instance: the $7d\ \mathcal{N}=1$ $SU(N)$ Coulomb branch}

Take $7d\ \mathcal{N}=1$ SYM with gauge group $SU(N)$. The reduced-direction count is $10 - 7 = 3$,
so there are three adjoint scalars, the $SO(3)$ vector, equivalently the $SU(2)_R$ triplet. The rank
of $SU(N)$ is $r = N - 1$. The Coulomb-branch real dimension is
\begin{equation}
\label{eq:V09-7dSUN}
 \dim_{\mathbb{R}}\,\mathcal{M}_{\mathrm{Coulomb}}
 \;=\; (10 - 7)(N - 1)
 \;=\; 3(N - 1),
\end{equation}
three real Cartan directions per unbroken $U(1)$. For $SU(2)$, $r = 1$ and the branch is
$3(2-1) = 3$-dimensional: three real scalars, the $SU(2)_R$ triplet of the single Cartan direction.
For $SU(3)$, $r = 2$ and the branch is $3(3-1) = 6$-dimensional. A generic point Higgses $SU(N)$ to
its maximal torus $U(1)^{N-1}$, giving mass to $N^2 - N$ gauge bosons ($2$ for $SU(2)$, $6$ for
$SU(3)$, per \eqref{eq:V09-higgsing}).

The flat-direction condition is concrete at $SU(2)$. Write the three adjoint scalars as $2 \times 2$
traceless Hermitian matrices, and align them all with the Cartan generator $\sigma_3$:
\begin{equation}
\label{eq:V09-su2config}
 \phi^m \;=\; a_m\,\sigma_3, \qquad m = 7, 8, 9, \quad a_m \in \mathbb{R}.
\end{equation}
Every commutator $[\phi^m, \phi^n] = a_m a_n [\sigma_3, \sigma_3] = 0$ vanishes, the potential
$V = 0$, and the configuration is a supersymmetric vacuum: three real moduli $a_7, a_8, a_9$,
matching $3(N-1) = 3$. The root action \eqref{eq:V09-rootaction} is explicit here: with
$\sigma_\pm = (\sigma_1 \pm i\sigma_2)/2$,
\begin{equation}
\label{eq:V09-su2root}
 [\phi^m, \sigma_\pm] \;=\; \pm\, 2 a_m\, \sigma_\pm,
\end{equation}
the two off-Cartan generators are charged and become the two massive gauge bosons of
\eqref{eq:V09-higgsing}. If instead two of the scalars point along non-commuting generators, say
$\phi^7 = \sigma_1$ and $\phi^8 = \sigma_2$, then
\begin{equation}
\label{eq:V09-su2comm}
 [\sigma_1, \sigma_2] \;=\; 2i\,\sigma_3 \;\ne\; 0,
\end{equation}
the potential is strictly positive, and the configuration is \emph{not} a vacuum. Only the
commuting, Cartan-aligned configurations survive, exactly the $\mathbb{R}^{3(N-1)}/W$ count.

\begin{keybox}{Common misconception: high-dimensional SYM has a rank-dimensional Coulomb branch}
$10d\ \mathcal{N}=1$ SYM has \emph{no} Coulomb branch at all: the gauge field has no adjoint-scalar
superpartner. The Coulomb branch appears only on dimensional reduction, where the $10-d$ gauge-field
components along the reduced directions become $10-d$ adjoint scalars. Its real dimension is
therefore $(10-d)\,r$, the number of reduced directions times the rank, which \emph{grows} as you
descend, not a fixed $r$. And it is a \emph{real} moduli space of commuting Hermitian matrices, the
high-dimensional analog of the $5d$ real Coulomb branch, not the complex special-K\"ahler branch of
$4d\ \mathcal{N}=2$ and not a hyperk\"ahler Higgs branch (pure SYM has no hypermultiplets). The flat
directions are the simultaneously diagonalizable configurations $[\phi^m,\phi^n]=0$, Higgsing
$G \to U(1)^r$ at a generic point.
\end{keybox}

\section{The per-dimension catalog and geometric origins}
\label{sec:V09-rows}

The four rows can now be read off with their R-symmetries and their string or M-theory homes.
Table~\ref{tab:V09-rows} collects them. The R-symmetry entry in each row is the grounding
computation of \S\ref{sec:V09-rsymmetry}, $SO(10-d)$ with the $7d$ cover; the final column records a
standard geometric origin or use, not a construction developed here.

\begin{table}[ht]
\centering
\small
\setlength{\tabcolsep}{5pt}
\renewcommand{\arraystretch}{1.3}
\begin{tabular}{@{}ccp{22mm}p{56mm}@{}}
\toprule
$d$ & R-symmetry & content & geometric origin / use \\
\midrule
$10$ & none ($SO(0)$) & $A_\mu$, $0$ scalars & heterotic / type~I gauge sector \\
\midrule
$9$ & none ($SO(1)$) & $A_\mu$, $1$ scalar & the one-scalar bookkeeping row \\
\midrule
$8$ & $U(1)_R$ ($SO(2)$) & $A_\mu$, $2$ scalars & F-theory 7-brane gauge theory (worldvolume $8d$) \\
\midrule
$7$ & $SU(2)_R$ ($SO(3)$ cover) & $A_\mu$, $3$ scalars & M-theory on ADE / $G_2$-holonomy singularities \\
\bottomrule
\end{tabular}
\caption{The per-dimension catalog of the $7d$-$10d$ sixteen-supercharge SYM rows. The R-symmetry is
$SO(10-d)$ (trivial at $9d$/$10d$, $U(1)_R$ at $8d$, the $SU(2)_R$ cover at $7d$); the content is
the $10d$ vector reduced, $A_\mu$ plus $10-d$ adjoint scalars; the final column names standard
string- or M-theory contexts in which the row appears. These are pointers, not constructions here.}
\label{tab:V09-rows}
\end{table}

\noindent
The $10d$ row is the gauge sector of the heterotic string and of type~I: the $10d\ \mathcal{N}=1$
vector multiplet, with $\Gamma_{11}\lambda = +\lambda$, is precisely the open-string or heterotic
gaugino and gauge field, and its non-renormalizability is completed by the string oscillator tower.
The two lower rows each deserve a short paragraph of intuition, because they are the reason a
field-theory reader will meet these rows at all.

\medskip\noindent\textbf{Why the $8d$ row matters: F-theory 7-branes.}\enspace
A 7-brane has a $7 + 1 = 8$-dimensional worldvolume, so its gauge theory lives in the $8d$ row (not
the $7d$ row: the ``7'' counts spatial worldvolume dimensions, an off-by-one this row-labeling
prevents). The brane sits at a point in the two transverse directions, and the row's field content
is the geometry of that statement:
\begin{equation}
\label{eq:V09-8dengine}
 \phi \;=\; \phi^8 + i\,\phi^9 \;\;\longleftrightarrow\;\; z_\perp,
 \qquad
 U(1)_R \;=\; SO(2)_\perp,
\end{equation}
the complex adjoint scalar is the brane's complex transverse position, and the $U(1)_R$ we derived
is the rotation of the transverse plane. In F-theory the 7-brane's gauge group is read off from the
singularity type of an elliptic fibration; what the construction consumes from this section is
exactly the $8d$ row of Table~\ref{tab:V09-rows}: multiplet, $U(1)_R$, and the $2r$-dimensional
Coulomb branch of transverse positions.

\medskip\noindent\textbf{Why the $7d$ row matters: M-theory on ADE singularities.}\enspace
A $7d\ \mathcal{N}=1$ gauge theory arises on the singular locus of an ADE-fibered M-theory
background: eleven dimensions minus the four of the ADE space leave a $7d$ worldvolume. The three
adjoint scalars are the three normal deformations of the singular locus,
\begin{equation}
\label{eq:V09-7dengine}
 (\phi^7,\ \phi^8,\ \phi^9) \;\;\longleftrightarrow\;\; \vec{x}_\perp \in \mathbb{R}^3,
 \qquad
 SU(2)_R \;=\; \mathrm{Spin}(3)_\perp,
\end{equation}
and the $SU(2)_R$ triplet structure of the Coulomb branch is the rotation of that normal
$\mathbb{R}^3$. The $G_2$-holonomy compactifications that engineer $4d$ physics pass through this
$7d$ SYM. In each case the construction reads its multiplet, its R-symmetry, and its Coulomb branch
off this section; this section does not build the construction.

\section{No superconformal field theory above six dimensions}
\label{sec:V09-noscft}

The final structural fact removes this whole dimension range from the extremization through-line
that organized Sections~3, 4, and 6. There is no interacting superconformal field theory in
$d \ge 7$ at sixteen supercharges. Consequently there is nothing here to extremize: no
$a$-maximization, no $c$-extremization, no $F$-maximization, and no conformal central charges $a$,
$c$ to compute, because there is no conformal fixed point to compute them at.

Two independent recalled facts force the absence. First, the Nahm classification of Section~1 (its
superconformal-algebra section) permits interacting superconformal algebras only up to $d = 6$:
above six dimensions there is no superconformal algebra with a finite-dimensional R-symmetry to host
an interacting fixed point. This is a statement about the allowed algebras, blind to any particular
Lagrangian, and it closes off $d \ge 7$ before dynamics is even considered. Second, the dimensional
analysis of this section, run row by row in \S\ref{sec:V09-ladder}, gives
\begin{equation}
\label{eq:V09-noscftcoupling}
 [g^2] \;=\; 4 - d \;\le\; -3 \;<\; 0
 \qquad \text{for every } d \ge 7 :
\end{equation}
the gauge coupling is irrelevant, so the theory is non-renormalizable, and its only would-be
ultraviolet completion is gravitational (a string or M-theory embedding), not a field-theory fixed
point. The two arguments agree. The Nahm result says no superconformal algebra exists to sit at; the
coupling analysis says the theory itself is not ultraviolet-complete as a field theory. Either way,
the $\ge 7d$ sixteen-supercharge SYM theories are effective field theories, not conformal field
theories, and they sit outside the extremization through-line.

This is why the section has no extremization member and closes the dimensional ladder. The rich
extremization structure of the lower dimensions, the $a$-maximization anchor of Section~3 and its
$2d$ and $3d$ analogs, has no counterpart here because it presupposes a fixed point that does not
exist at these dimensions. Below seven dimensions the maximal-supercharge theories do reach
interacting fixed points ($4d\ \mathcal{N}=4$ is the standard example, the maximal $4d$
superconformal theory); at and above seven dimensions they do not. Section~1's $32Q$ ceiling caps
the ladder above $10d$ with supergravity, and this section's $11d$ counting
\eqref{eq:V09-11dmismatch} is the same ceiling seen in degrees of freedom; Nahm's $d = 6$ ceiling
caps the superconformal theories below; and this section fills the $7d$-$10d$ gap between them with
the non-renormalizable pure-SYM interface, which is exactly the field theory the string and M-theory
constructions start from.

\section*{Exit checklist}
\addcontentsline{toc}{subsection}{Exit checklist}
\markboth{Exit checklist}{Exit checklist}

After this section the reader can
\begin{enumerate}
\item identify $10d\ \mathcal{N}=1$ super Yang--Mills as the maximal-dimension parent, list its
vector multiplet ($A_M$, Majorana--Weyl gaugino, no scalars), run the counting chain
$64 \to 32 \to 16 \to 8$ against the $D - 2 = 8$ polarizations to verify $8 = 8$, and compute the
coupling dimension $[g^2] = 4 - 10 = -6 < 0$ (non-renormalizable);
\item explain why there is no $11d$ SYM \emph{without} misusing the even-dimensional Majorana--Weyl
formula: the $11d$ minimal spinor is Majorana with $32$ real components, giving $16$ on-shell
fermionic states against $9$ gauge polarizations and no available scalars, so the minimal $11d$
multiplet is the supergravity multiplet, $44 + 84 = 128 = 128$;
\item run each reduction step $d+1 \to d$ explicitly: split the gauge field with its index ranges,
track the $45$ field-strength components ($36+9+0$, $28+16+1$, $21+21+3$), derive
$F_{\mu m} = D_\mu\phi^m$ and $F_{mn} = -i[\phi^m,\phi^n]$ from the zero-mode rule, split the action
into gauge, scalar-kinetic, and commutator-potential terms, and shift the coupling dimension by one
unit per circle ($-6, -5, -4, -3$);
\item decompose the $16$ gaugino components at each row, $16 = 16 \times 1$ ($9d$ Majorana),
$16 = 8 + 8$ (the $8d$ Weyl pair with $U(1)_R$ charges $\pm\tfrac12$), $16 = 8 \times 2$ (the $7d$
symplectic-Majorana $SU(2)_R$ doublet), and verify the preserved totals $(d-2) + (10-d) = 8 = 16/2$;
\item state each reduced vector multiplet as a list (gauge field, $10-d$ real scalars, fermion
content, R-symmetry representations: scalars in the vector of $SO(10-d)$, fermions under
$\mathrm{Spin}(10-d)$, with $\mathrm{Spin}(2) = U(1)$ and $\mathrm{Spin}(3) = SU(2)$);
\item count the Coulomb branch that opens on reduction from the commuting-vev flat directions
$[\phi^m, \phi^n] = 0$, giving the real moduli space $\mathbb{R}^{(10-d)\,r}/W_G$ of dimension
$(10-d)\,r$ ($3(N-1)$ for $7d\ SU(N)$, $0$ at $d = 10$), with $\dim G - r$ gauge bosons massive at a
generic point;
\item place each row in its catalog with its geometric origin or use (heterotic / type~I at $10d$, the F-theory
7-brane theory at $8d$ with the complex transverse position $\phi = \phi^8 + i\phi^9$, M-theory on
ADE / $G_2$ singularities at $7d$ with the scalars as normal deformations), knowing these are
pointers rather than constructions developed here;
\item state that there is no interacting superconformal field theory at $d \ge 7$ with sixteen
supercharges, from the Nahm classification and the $[g^2] \le -3 < 0$ dimensional analysis, so no
extremization runs here and this section closes the dimensional ladder.
\end{enumerate}

\bigskip
\section*{Sources and notes}
\addcontentsline{toc}{subsection}{Sources and notes}
\markboth{Sources and notes}{Sources and notes}
{\small

\noindent\textsf{\textcolor{RoyalBlue}{Sources and notes.}}\enspace
This is the terminal, lightest dimension section of these notes: the high-dimensional
sixteen-supercharge pure-SYM interface, one parent ($10d\ \mathcal{N}=1$ SYM) and three reduction
rows, with the $10d \to 9d \to 8d \to 7d$ reduction bookkeeping worked step by step at
undergraduate-followable depth.

\medskip\noindent\textsf{\textcolor{RoyalBlue}{\textbf{\S\ref{sec:V09-tenD}\enspace The $10d$ parent.}}}\enspace
The itemized $10d\ \mathcal{N}=1$ vector multiplet ($A_M$, Majorana--Weyl gaugino), the Lagrangian
\eqref{eq:V09-lagrangian}, the counting chain $64 \to 32 \to 16$ \eqref{eq:V09-spinorchain} and the
on-shell dof match $D-2 = 8$ \eqref{eq:V09-gaugedof} against $16/2 = 8$ \eqref{eq:V09-mwdof} giving
$8 = 8$ \eqref{eq:V09-dofmatch}, the coupling dimension $[g^2] = 4 - 10 = -6 < 0$
\eqref{eq:V09-actiondim}--\eqref{eq:V09-tenDcoupling} with the cutoff scale
$(g_{10}^2)^{-1/6}$ \eqref{eq:V09-cutoff} (non-renormalizable), the corrected $11d$
argument (the $11d$ minimal spinor is Majorana, $32$ real $\to$ $16$ on-shell
\eqref{eq:V09-11dspinor}, against $9$ polarizations \eqref{eq:V09-11dmismatch}; the minimal $11d$
multiplet is the supergravity multiplet $44 + 84 = 128 = 128$ \eqref{eq:V09-11dsugra}), and the
absence of a Coulomb branch or continuous R-symmetry at $d = 10$. (\textcite{Brink:1976bc} the $10d$ SYM and its reduction; \textcite{Gliozzi:1976qd} the dof match;
\textcite{Cremmer:1978km} the $11d$ supergravity multiplet). 

\medskip\noindent\textsf{\textcolor{RoyalBlue}{\textbf{\S\ref{sec:V09-ladder}\enspace The reduction ladder.}}}\enspace
The zero-mode reduction on $T^{10-d}$ \eqref{eq:V09-torus}--\eqref{eq:V09-kkmass}, the gauge-field
split $A_M \to (A_\mu, \phi^m)$ \eqref{eq:V09-split} with the surviving gauge transformations
\eqref{eq:V09-adjoint} and the scalar count $10 - d$
\eqref{eq:V09-scalarcount}, the general component formulas $F_{\mu m} = D_\mu\phi^m$,
$F_{mn} = -i[\phi^m,\phi^n]$ \eqref{eq:V09-Fcomponents}, the master action split
\eqref{eq:V09-mastersplit}, the preserved bosonic dof $(d-2)+(10-d) = 8$ \eqref{eq:V09-dofpreserved};
the three worked steps: $10d \to 9d$ in full
(\eqref{eq:V09-9dsplit}--\eqref{eq:V09-9dtally}: the $45 = 36 + 9 + 0$ component split, $F_{\mu 9} =
D_\mu\phi^9$, the action split, the coupling relation $1/g_9^2 = 2\pi R_9/g_{10}^2$ with $[g_9^2] =
-5$, the Yukawa coupling, the fermion restriction $16 = 16 \times 1$), $9d \to 8d$
(\eqref{eq:V09-8dsplit}--\eqref{eq:V09-8dtally}: $45 = 28 + 16 + 1$, the first commutator $F_{89}$
and potential, the complex scalar, the Weyl-pair split $16 = 8 + 8$, $[g_8^2] = -4$), and $8d \to
7d$ (\eqref{eq:V09-7dsplit}--\eqref{eq:V09-7dtally}: $45 = 21 + 21 + 3$, the three commutator pairs,
the symplectic-Majorana split $16 = 8 \times 2$, $[g_7^2] = -3$, the assembled terminal Lagrangian
\eqref{eq:V09-7dlagrangian}); the composed torus-volume coupling relation
\eqref{eq:V09-composedcoupling}; the itemized $9d$/$8d$/$7d$ vector
multiplets; the field-content table (Table~\ref{tab:V09-content}) and the coupling-dimension summary
$[g^2] = 4 - d = -6, -5, -4, -3$ \eqref{eq:V09-couplingladder} (Table~\ref{tab:V09-coupling}, the
non-renormalizability uniform down the ladder). (\textcite{Brink:1976bc} the dimensional
reduction; \textcite{Seiberg:1997ax} the sixteen-supercharge rows). 

\medskip\noindent\textsf{\textcolor{RoyalBlue}{\textbf{\S\ref{sec:V09-rsymmetry}\enspace The R-symmetry ladder.}}}\enspace
The R-symmetry as the transverse rotation group $SO(10-d)$ \eqref{eq:V09-rsymmetry}, the uniform
representation rule (scalars in the vector of $SO(10-d)$, fermions under $\mathrm{Spin}(10-d)$, with
$\mathrm{Spin}(2) = U(1)$ and $\mathrm{Spin}(3) = SU(2)$) \eqref{eq:V09-covernote}, the $so(n)$
rank/dimension formulas \eqref{eq:V09-sorankdim} giving the ladder $so(0), so(1), so(2), so(3)$ of
dimension $0, 0, 1, 3$ \eqref{eq:V09-rladder}, the $8d$ $SO(2) = U(1)_R$ \eqref{eq:V09-8dR} with the
$\pm\tfrac12$ fermion charges, the $7d$ $SO(3) \to SU(2)_R$ spinor cover
\eqref{eq:V09-7dcover}--\eqref{eq:V09-7dR}, and the below-the-ladder cross-check
$\mathrm{Spin}(6) = SU(4)$ at $(\mathrm{rank}, \dim) = (3, 15)$ \eqref{eq:V09-spin6}. (\textcite{Seiberg:1997ax} theories with
sixteen supercharges in various dimensions). 

\medskip\noindent\textsf{\textcolor{RoyalBlue}{\textbf{\S\ref{sec:V09-coulomb}\enspace The Coulomb branch on reduction.}}}\enspace
The scalar potential with its derived coefficient
$V = -\tfrac{1}{2g^2}\sum_{m<n}\mathrm{Tr}[\phi^m,\phi^n]^2 \ge 0$ \eqref{eq:V09-potential}
(inherited from the master split \eqref{eq:V09-mastersplit}; the sign from the anti-Hermiticity of
the commutator), the flat-direction / commuting-vev condition $[\phi^m,\phi^n]=0$
\eqref{eq:V09-flatdir}, the real moduli space $\mathbb{R}^{(10-d)r}/W_G$ \eqref{eq:V09-coulombspace}
of dimension $(10-d)\,r$ \eqref{eq:V09-coulombdim}, the generic Higgsing $G \to U(1)^r$ with $\dim G
- r$ massive gauge bosons \eqref{eq:V09-higgsing} through the root action
\eqref{eq:V09-rootaction}, the real-branch type (vs the complex /
hyperk\"ahler branches), and the worked $7d\ SU(N)$ count $3(N-1)$ \eqref{eq:V09-7dSUN} ($SU(2)$:
$3$; $SU(3)$: $6$) with the explicit Cartan configuration \eqref{eq:V09-su2config}, the root
eigenvalues $[\phi^m, \sigma_\pm] = \pm 2a_m \sigma_\pm$ \eqref{eq:V09-su2root}, and the
non-commuting rejection $[\sigma_1,\sigma_2] = 2i\sigma_3$ \eqref{eq:V09-su2comm}. (\textcite{Seiberg:1997ax} the sixteen-supercharge Coulomb branch). 

\medskip\noindent\textsf{\textcolor{RoyalBlue}{\textbf{\S\ref{sec:V09-rows}\enspace The per-dimension catalog.}}}\enspace
The per-row catalog (Table~\ref{tab:V09-rows}) of R-symmetry, content, and geometric origin/use: $10d$
heterotic / type~I gauge sector; $9d$ the one-scalar bookkeeping row; $8d\ U(1)_R$ the F-theory
7-brane gauge theory (worldvolume $8d$; the complex scalar as the transverse position
\eqref{eq:V09-8dengine}); $7d\ SU(2)_R$ M-theory on ADE / $G_2$-holonomy singularities (the three
scalars as normal deformations \eqref{eq:V09-7dengine}). (\textcite{Vafa:1996xn} F-theory
/ the 7-brane; \textcite{Acharya:2001gy} the $G_2$ / $7d$ M-theory row). 

\medskip\noindent\textsf{\textcolor{RoyalBlue}{\textbf{\S\ref{sec:V09-noscft}\enspace No $\ge 7d$ superconformal field theory.}}}\enspace
The statement that no interacting superconformal field theory exists in $d \ge 7$ at sixteen
supercharges, hence no $a$-maximization, no $c$-extremization, no $F$-maximization, and no central
charges to extremize, from the two recalled reasons: the Nahm classification (Section~1's
superconformal-algebra section, interacting superconformal algebras only up to $d=6$) and the
$[g^2] = 4 - d \le -3 < 0$ dimensional analysis \eqref{eq:V09-noscftcoupling} (non-renormalizable, a
gravitational not a field-theory completion). (\textcite{Nahm:1977tg} the
superconformal-algebra classification, recalled from Section~1). This is a statement-level recall:
the absence follows from the Section~1 Nahm result and the $[g^2] < 0$ analysis above, so it carries no separate check.

\medskip\noindent\textbf{Stated, not proved here.}\enspace
The ultraviolet completion of $\ge 7d$ pure super Yang--Mills (non-renormalizable, $[g^2] < 0$, no
field-theory fixed point): the heterotic / type~I gauge sector at $10d$, F-theory 7-branes at $8d$
(worldvolume $7 + 1 = 8d$), and M-theory on ADE / $G_2$-holonomy singularities at $7d$. The
lower sixteen-supercharge rows ($6d\ (1,1)$, $5d\ \mathcal{N}=2$, $4d\ \mathcal{N}=4$) are pointed
at their flavor-family sections (Sections~8, 7, 5) and not developed here. All are stated and named
here; their constructions are outside the scope of these notes.
}

\subsection*{Further reading}
\addcontentsline{toc}{subsection}{Further reading}
Ten-dimensional super Yang--Mills and its dimensional reduction are in
\textcite{Brink:1976bc,Scherk:1979zr}; eleven-dimensional supergravity in
\textcite{Cremmer:1978km,Cremmer:1979up}, with the membrane perspective of \textcite{Townsend:1995kk}
and the Kaluza--Klein review \textcite{Duff:1986hr}. The role of these maximal theories as limits of
string and M-theory dynamics is discussed in \textcite{Witten:1995ex}, and the M-theory realization of
gauge dynamics on $G_2$ and ADE singularities in \textcite{Acharya:1998pm,Atiyah:2001qf}.

For modern viewpoints beyond flat-space component reduction, the twists of SYM in dimensions
$2\leq d\leq10$ are classified in \textcite{Elliott:2020ecf}, curved-space $7d$ localization is
developed in \textcite{Iakovidis:2020znp}, and the $7d$ gauge-theory description of local $G_2$
backgrounds is studied in \textcite{Barbosa:2019bgh}. The pure-spinor organization of
ten-dimensional on-shell data and its field-theory limit are reviewed in \textcite{Mafra:2022wml}.

\section*{References}
\addcontentsline{toc}{subsection}{References}
\markboth{References}{References}
\printbibliography[heading=none]
\end{refsection}

\end{document}